# The Hierarchic Theory of Liquids and Solids.

## Computerized Applications for Ice, Water, and Biosystems

Alex Kaivarainen

# CONTENTS

# PREFACE

For many years our Laboratory of molecular biophysics has been involved in the study of solvent dependent large-scale dynamics of proteins determined by relative thermal mobility of their domains and subunits. The role of large-scale dynamics in the mechanism of protein function, the signal transmission, allosteric effects and other water dependent effects in protein solutions have been investigated.

The modified for this goal physical methods, like NMR, EPR (spin-label), microcalorimetry, spectroscopy, light scattering, refractometry and others were used. A number of new phenomena in physics of biopolymers have been discovered (Kaivarainen, 1985; 1989; 1995, 2001, 2003).

*The most important of them are following:*

1. The ability of proteins to change the bulk water dynamics and thermodynamic activity, as a result of large-scale pulsations of their big inter-domain and inter-subunit cavities, accompanied by assembly $\rightleftharpoons$ disassembly (flickering) of water clusters in these cavities and exchange of this water molecules with bulk water;

2. Solvent - mediated remote interaction between different kinds of proteins in the process of their large - scale dynamics (flexibility) change, induced, respectively, by ligand binding to the active sites, by temperature or by variation of solvent composition;

3. Solvent-mediated distant interaction between protein and cells, accompanied by cells swelling or shrinking, correlated with change of protein flexibility and water activity, enhancing or triggering the passive osmosis via membranes of cells.

A new kind of interaction of water clusters, containing 30 - 70 molecules, with the open inter-domain and inter-subunit cavities of macromolecules/proteins, named *clusterphilic interaction,* was introduced (Kaivarainen, 1985, 1995, 2001). Such interaction can be considered, as the intermediate one between the hydrophobic and hydrophilic ones. It follows from our dynamic model of protein behavior in water, that intramolecular *clusterphilic interaction* stands for remote signal transmission, allosteric properties in multi-domain and oligomeric proteins. It is a consequence of high sensitivity of clusters stability to perturbation of such protein cavities geometry, induced by the ligand binding. Stabilization or destabilization of water clusters in cavities shifts the dynamic equilibrium $\mathbf{B} \rightleftharpoons \mathbf{A}$ between the open (B) and closed (A) states of protein cavities to the left or right, correspondingly. As far the *assembly* $\rightleftharpoons$ *disassembly* of water clusters in cavities represent mesoscopic 1st order phase transitions, the functionally important changes of proteins configuration, accompanied the shift of $\mathbf{B} \rightleftharpoons \mathbf{A}$ equilibrium need very small change of free energy: $\Delta G = \Delta H - T\Delta S \simeq 0$. This change easily can be provided by binding of ligands to the active sites of proteins (Kaivarainen, 1985, 2001).

The intermolecular *clusterphilic interactions* are crucial in the interfacial effects and thixotropic structure formation in colloid systems (Kaivarainen, 1995; 2003).

The part of this research activity was summarized in book of this author: "Solvent - dependent flexibility of proteins and principles of their function", D Reidel Pub Co., 1985, ISBN: 9027715343.

The development of new Hierarchic theory of condensed matter was started by this author in 1986. This work was stimulated by the understanding that the progress in biophysics is not possible without the detailed and quantitative description of water physical properties on mesoscopic and macroscopic level. The existing theories of liquid state was not enough deep and general for this end.

One of the results of application of created computer program (pCAMP, copyrighted in USA in 1997), based on our Hierarchic theory of matter (Kaivarainen, 1995, 2001, 2003), was the discovering of molecular *mesoscopic Bose condensation (mBC)* in the ice and in liquid water at the ambient temperatures (even around $36^0$C) in form of coherent molecular clusters, named the *primary librational effectons.*

The evidences where obtained, using computer simulations, that just the dimensions and dynamics of these water clusters represent the crucial factors in evolution of biopolymer's spatial and dynamic structure.

Our Hierarchic theory of condensed matter got a lot of convincing computerized verifications on examples of water and ice from comparison of calculated and experimental physical parameters, like heat capacity, thermal conductivity, surface tension, vapor pressure, viscosity and self-diffusion. The new quantitative theories of refraction index, Brillouin light scattering, Mössbauer effect and others, based on the same hierarchical model, are also in good correspondence with experiment.

Because of numerous anomalies, water is a good system for testing of new theories of condensed matter. One may anticipate, that if the theory works well quantitatively for such complicated systems, as water and ice, it must be valid for the other liquids, glasses or crystals also.

# Acknowledgments

A significant part of Hierarchic theory of condensed matter was worked out at Universities of Turku and Joensuu (Finland) during 1991-2003. I got a good friendly support from Professors Rauno Hamalainen, Erkki Soini and Jyrki Kauppinen.

I am grateful for invitation to the Institute of Atomic and Molecular Physics (C.N.R., Pisa, Italy), to Drs. Vaselly Moreno and Giuseppe Salvetti for their kindness, making it possible the development of my ideas during spring and summer 1992 in Pisa.

The progress in Hierarchic concept of matter was achieved in time of my visit to Research Institute for Electronic Science (Hokkaido University, Sapporo, Japan), granted by Scandinavia-Japan Sasakawa Foundation in 1993.

The first version of Hierarchic model of consciousness, including quantum and classical stages, was written during my few month long stay (October-December, 1994) at the Advanced Biotechnology Laboratory (University of Arizona, Tucson, USA). I am grateful to Prof. Stuart Hameroff from this University for his hospitality.

Special thanks and regards to Gary and Mary Goodman of Avondale, Arizona, USA, for my English language edition, making this material more perceptible.

I would like to express my deep gratitude also to my friends and colleagues Drs. Andrei Goryunov, Galina Sukhanova and Ludmila Fradkova from my former Laboratory of Molecular Biophysics in Petrozavodsk at Scientific Center of Russian Academy of Science for different kind of support in the course of this long-term and multistage work.

# SUMMARY

Basically new quantitative Hierarchic theory of matter general for liquids and solids is presented in this book. This theory is proved to be more advanced than well known basic models of Einstein and Debye of condensed matter and can be reduced to them only after number of simplifications.

In our approach the condensed matter is considered as a system of three-dimensional (3D) superposition of standing waves of following types:

a) the most probable de Broglie waves, related to molecules translation and librations in composition of condensed matter;
b) the acoustic waves (thermal phonons);
c) the electromagnetic waves (IR photons).

The existence of ambient (high-T) *mesoscopic* Bose condensation (mBC) in composition of liquids and solids in form of coherent molecular clusters with volume of 3D de Broglie waves of molecules, has been discovered as a result of computer simulations, using software, based on Hierarchic theory (copyright, 1997, USA, Kaivarainen). The *macroscopic* Bose condensation, accompanied the origination of superfluidity and superconductivity, can be achieved only at very low temperatures.

Strongly interrelated collective excitations (quasiparticles), named the *effectons, convertons, transitons and deformons and their different combinations* are introduced in hierarchic model of condensed matter. They represent a mesoscopic scale of matter, intermediate between microscopic and macroscopic ones.

The *effectons* (tr and lb) are formed by three-dimensional (3D) superposition of standing de Broglie waves of molecules or atoms, related to their translations (tr) and librations (lb). Two possible thermal in-phase (acoustic, a) or counterphase (optic, b) oscillations in the volume of effectons are coherent. The primary effectons represent a high-temperature mesoscopic Bose-condensation. The primary effectons, with external resulting momentum equal to zero, are coherent clusters, approximated by parallelepiped with edges, determined by three most probable de Broglie standing wave length of molecules, normal to each other. The properties of *secondary effectons* are the result of averaging of all effectons with nonzero external momentum, using Bose-Einstein statistics.

The *convertons* (ac and bc) represent the interconversions between similar states (a or b) of translational and librational effectons $[tr \rightleftharpoons lb]_{ac,bc}$. As far the former effectons are much smaller, than the latter, the convertons can be considered as a flickering clusters, exciting the acoustic waves (thermal phonons) of corresponding frequency and phase velocity in condensed matter.

The *transitons* represent $(a \rightleftharpoons b)_{tr,lb}$ quantum transitions of the translational and librational effectons between acoustic (a) and optic (b) states, which differs only with their potential energies. The kinetic energy and momentum of these two states are equal. Such transitions of primary and secondary effectons (primary and secondary transitons) are accompanied by emission $\rightleftharpoons$ absorption of IR photons and thermal phonons, correspondingly.

The *deformons* of two types: primary (electromagnetic) and secondary (acoustic), are formed as a result of 3D superposition of three standing IR photons and phonons, radiated by primary and secondary *transitons*, correspondingly.

All possible combinations of the above 4 basic excitations lead to hierarchical system of 4! = 24 quasiparticles, describing virtually all physical properties of any condensed matter.

The formulas obtained in hierarchic theory allow to calculate the internal energy, heat capacity, vapor pressure, surface tension, thermal conductivity, viscosity, self-diffusion, solvent activity and a lot of other important parameters for liquids and solids on the same theoretical background. Hierarchic theories of refraction index, Brillouin light scattering and Mössbauer effect also has been developed. The new state equation, mechanisms of osmosis and self-organization in colloid and biological systems are proposed. A new scenarios of turbulence, superfluidity and superconductivity are worked out.

The following four input experimental parameters of matter at the same temperature and pressure are necessary for calculation of the above listed parameters, using created computer program: "Comprehensive analyzer of Matter Properties (pCAMP):
   - Sound velocity;
   - Density;
   - Refraction index;
   - Positions of translational and librational bands in oscillatory spectra.

The coincidence between our theory and available from literature experimental data for ice and water in temperature interval $(0-373)^0 K$ is very good even without adjustable parameters. It is a first quantitative theory enable to explain all known temperature anomalies in ice and water. *It reveals a true mechanism of first and second order phase transitions and describes them quantitatively.*

Solvent-dependent principles of proteins small-scale and large-scale dynamics and their function on examples of antibodies, hemoglobin, albumin, enzymes, myosin are discussed in this book in the framework of hierarchic theory. A new water-dependent mechanism of muscle contraction and cancer emergency is described also.

*A new model of elementary act of perception and memory (Cycle of Mind) is proposed*. It interrelates the dynamics of coherent water clusters (mesoscopic Bose condensate) in the interior of microtubules with their reversible disassembly, gel-sol transition in the nerve cells, accompanied by pulsation of cells volume and synaptic reorganization on the surface of dendrites. This process involve the oscillation between mesoscopic and macroscopic entanglement between microtubules, stimulated by IR photons exchange between water clusters.

*Audience:* The described comprehensive Hierarchic theory of condensed matter and theory based computer program provide a possibility for quantitative analysis of wide range of phenomena, related to condensed matter physics and biophysics, thermodynamics, nanotechnology, nanobiology, self-organization in colloids and biosystems, interaction of light with matter, etc. Our book can be useful for professionals and students of corresponding profiles, looking for novel and general solutions of the old problems.

A special computer program, named Comprehensive Analyzer of Matter Properties (pCAMP), based on Hierarchic theory of condensed matter, allows very detailed quantitative analysis (more than 300 physical parameters) of any types of solids and liquids: crystals, glasses, solutions, colloid & biological systems. The pCAMP also can be used effectively for monitoring of materials and crystals technology, for preparation of ultra-dilute solutions (liquid and solid), for investigation of superconductivity, superfluidity and turbulence. The demonstrational version of pCAMP on examples of water and ice can be downloaded from the front page of author's site: **web.petrsu.ru/~alexk**

# Introduction

A quantitative theory of the liquid state, as well as a general theory of condensed matter, has not been available until now, thus presenting a fundamental problem. The solution of this fundamental problem is crucial for different branches of science and technology. The existing solid-state theories do not allow easy extrapolation to liquids. Widely used molecular dynamics methods for computer simulations of condensed matter properties are based on the classical approach and harmonic approximation.

Understanding of the hierarchic organization of matter is essential. It involves a mesoscopic bridge between microscopic and macroscopic physics, and between liquids and solids. There is strong evidence that the dynamics of many classes of molecules in solids and liquids are anharmonic and do not follow the classical Maxwell-Boltzmann distribution. This implies, then, that only a quantum approach, used in presented in this book model, is suitable for the elaboration of a general theory of condensed matter.

The Langevin equation is not able to describe the non-Markovian, non-Gaussian behavior of liquids (Ferrario et al., 1985). The latter reflects a nonlinear anharmonic interaction between molecules (Bertolini et al., 1992). These nonlinear phenomena are interrelated to the *self-organization* of matter in space and time and represent a much debated problem in modern science. The pioneering contributions to the problem's solution was made by I. Prigogine, P. Glansdorf, J. Nicolis and A. Babloyantz, and are based on *non-equilibrium thermodynamics*, as well as the concept of *dissipative structures*.

Classical examples of self-organization and dissipative structures are hydrodynamic convectional instabilities, brought to light by Benar and by Tailor. The Belousov-Zhabotinsky oscillatory reaction reflects the possibility of macroscopic self-organization in chemical processes. These phenomena are also clearly useful for a proper understanding of evolution in biological systems. The books by M. Eigen, P. Schuster and P. Vinkler are devoted to the emergence of the genetic code. In 1960, H. Haken introduced the term "*synergism*" to describe cooperative interaction in different types of open systems: physical, chemical, biological and even social systems, and the spontaneous formation of new macroscopic structures.

The mathematical description of self-organization processes is formally rather well-developed in terms both of differential equations and of a theory of bifurcation. However, the physical underpinnings of these self-organizing processes remains unclear. The appearance of oscillations and "memory" – slow relaxation of non-equilibrium states – in liquids and solids is an additional class of poorly understood phenomena.

The link between macro- and micro-physics is actual and important. Such an attempt has been made by Fröhlich, who introduced the notion of *single coherently excited polar mode*. Corresponding collective excitations have the properties of a Bose condensate and may be characterized by three features (Fröhlich, 1988):

1) they are relatively stable but far from thermal equilibrium
2) they exhibit non-trivial order
3) they have extraordinary dielectric properties

The highly polar mode can be described in terms of "optical" counterphase thermal vibrations of molecules, with frequencies: $\varpi_j$ ($j = 1, 2, 3..$). As a result of nonlinear effects, the lowest mode frequency $\varpi_1$ differs from zero. In this case, the *long range Coulomb* interaction establishes that the polar modes extend over relatively large (mesoscopic) volumes of matter. However, *the corresponding coherent structures are not true Bose condensates*. The particles are not unified by sufficiently long de Broglie wave with length, exceeding average distance between vibrating particles, as it takes a place in superfluidity and superconductivity (see eq.1).

The Fröhlich' approach to long-range interactions is valid only for *quasi-one-dimensional* chains of identical polarizable molecules interacting via dipole-dipole forces (Tuszynski, 1985). This approach is not valid for description of 3D coherent clusters of molecules in state of mesoscopic Bose condensation, introduced in our Hierarchic theory.

A trend in bioscience, called "Nanobiology", is related to understanding the principles of Life on mesoscopic level - intermediate between microscopic and macroscopic ones. The crucial role of water clusters on this level in state of mesoscopic Bose Condensation (mBC) will be demonstrated in the current work.

Stimulating this interest in the biophysics of water was the comprehensive collection of reviews edited by Felix Franks (1975, 1982). Interesting ideas, concerning the role of water clusters in biopolymers and cell architecture have been developed by John Watterson (1991, 1993).

It becomes evident that it is not possible to understand the emergence and existence of LIFE without a deep understanding of the physical properties of WATER.

The Hierarchic Theory of matter, general for solids and liquids, has been developed by this author during a couple of the last decades. The first related paper was published in J. Mol.Liquids, 1989, v.41, p. 53-84 under the title: "Theory of condensed state as a hierarchical system of quasi-particles formed by phonons and three-dimensional de Broglie waves of molecules. Application the theory for description of thermodynamic properties of water and ice."

This new theory proceeds from the fact that two main types of molecular heat dynamics, translational (tr) and librational (lb) oscillations, can not be described by Maxwell-Boltzmann distributions because of the anharmonicity of these oscillations.

The correct starting point for theory is that the most probable momentums ($p$) determine how the most probable de Broglie wave length ($\lambda_B = h/p = \mathbf{v}_{ph}/\nu_B$) of vibrating molecules is related to its frequency ($\nu_B$) and phase velocity ($\mathbf{v}_{ph}$).

Solids and liquids are considered in this model as a hierarchical system of collective excitations - metastable quasi-particles of four types: *effectons*, *transitons*, *convertons* and *deformons*. These types are all interrelated with one-another.

The *effectons* (**tr and lb**) are formed by the superposition of three-dimensional (3D) standing de Broglie waves of molecules, related to their translations and librations, while the *deformons* are formed by the 3D superposition of photons (primary deformons) and phonons (secondary deformons). The emission / absorption of photons is a result of quantum beats of *primary* effectons between the *acoustic (a) and optic (b)* states. Likewise, the emission-absorption of *phonons* is a result of quantum beats of *secondary* effectons between the *acoustic ($\bar{a}$) and optic ($\bar{b}$)* states.

The oscillations of molecules in the a-state of primary effectons have an analogy with in-phase "acoustic" modes, and, in the b-phase, with counterphase "optic" modes. The kinetic energies of these two modes (a and b) are equal – in contrast to their potential energies. The total energies of both states are far from thermal equilibrium energy, i.e. both states exist in conditions suitable for the formation of dissipative structures.

The *transitons* represent the intermediate - transition states of the effectons between the *acoustic (a)* and *optic (b)* states.

The *convertons* are responsible for inter-conversions between primary translational and librational effectons $[tr \Leftrightarrow lb]_{a,b}$, being in the same states ($a$ or $b$).

The *effectons are quantum collective excitations*. They have the properties of a partly degenerate Bose-gas *(mesoscopic Bose condensate)* and obey the Bose-Einstein distribution.

Primary and secondary deformons originate and annihilate as a result of ($b \to a$) and ($a \to b$) quantum transitions of primary and secondary effectons, respectively. Such an energy exchange between different quasi-particles, even under constant external conditions, reflects a dynamic equilibrium between the subsystems of the *effectons, convertons, transitons and deformons*.

When the effecton's shape, approximated by parallelepiped with edges (1,2,3), determined by the length of most probable de Broglie waves of molecules, exceeds the distances between molecules:

$$\left[\lambda_B = \frac{h}{m\mathbf{v}}\right]_{1,2,3} > (V_0/N_0)^{1/3} \qquad 1$$

then the *coherent* molecular clusters (effectons) originate as a result of the mesoscopic Bose-condensation (mBC). The effectons represent the self-organization of matter on the mesoscopic - nanoscale level.

The coherence of thermal oscillations in the volume of effectons, together with the stability of these clusters as the ambient temperature molecular mesoscopic Bose-condensate (mBC), are the result of local and distant Van der Waals interactions between molecules and unification of all molecules in the volume of mBC by the same wave function.

The assembly of the effectons side-by-side and formation of polyeffectons is also possible due to tunneling processes.

In liquids such as water, all hydrogen bonds between molecules which compose the primary effectons are saturated – similar to an ideal ice crystalline structure. This corresponds to a minimum of molecular thermal mobility and the maximum de Broglie wave length. These quantum effects, related to the mesoscopic Bose condensation of molecules which form the primary effectons, are responsible for the stabilization of clusters in both (a) and (b) states of the effectons. The Frölich's polar mode, if existing, may reflect only one special case of thermal coherent dynamics, corresponding to the optical (b) -state of primary effectons.

The interaction between atoms and molecules in condensed matter is much stronger, and the thermal mobility / momentum much less, than those in the gas phase. This implies that the temperature of the Bose condensation, when condition (1) starts to operate, should be much higher in solids and liquids, than in the gas phase. The weaker is the interaction between molecules, the lower is the temperature of their Bose condensation.

In 1997, Steven Chu, Claude Cohen-Taaoudji and William D. Philips shared the Nobel price in Physics for their work in the 1980s, involving laser- beam cooling atoms down to temperature close to absolute zero. The next stage of such efforts was a creation of a Bose-condensate in a gas of rubidium atoms by Anderson et al. (Science 269:198, 1995). Their work was confirmed in 1995 by Ketterle's group in MIT, and later by a few other groups, showing the Bose-Einstein condensate in gas of neutral atoms, such as sodium (MIT), and lithium (Rice University) at very low temperatures, less than $1^0K$. For a review of this problem see the article by E. Cornel in Scientific American, March, 1998.

Under these particular conditions, the number of atoms in the state of Bose condensation was about 20,000, and the dimension of the corresponding cluster was almost macroscopic – about 15 micrometers. As a comparison, the number of water molecules in a primary librational effecton (*mesoscopic Bose condensate, mBC*) at the freezing point ($273^0K$) – calculated using our computer program (see chapter 6) is only 280. Furthermore, the linear dimension is just about 20 Å (see Figure 7.).

The primary *transitons* and *convertons,* introduced in our approach, have common features with the coherent dissipative structures – a concept introduced by Chatzidimitrov-Dreisman and Brändas in 1989. Such structures were predicted on the basis of a complex scaling method and Prigogine's theory of star-unitary transformations.

Estimated from the Uncertainty Principle, the minimum boson's "degrees of freedom" ($n_{\min}$) in these spontaneous coherent structures is equal to:

$$n_{\min} \geq \tau(2\pi k_B T/h) \qquad 2$$

$$or: \quad \hbar \frac{1}{\tau} \geq \frac{k_B T}{n_{\min}} \qquad 2a$$

where $\tau$ is the relaxation time of the given coherent-dissipative structure.

For example, if $\tau \simeq 10^{-12} sec$, corresponds to the life-time of excitation of a molecular system by an infrared photon, then at $T = 300^0 K$ one obtains $n_{\min} \simeq 250$. This implies that at least 250 degrees of freedom associated with the number of molecules act coherently and produce the IR photon absorption /emission phenomenon. The traditional concept of an oscillating individual molecular dipole, as a source of photons, is replaced by the notion of a coherent molecular group quantum transitions.

A quantum field approach to the description of biosystems, using the ideas of spontaneous symmetry breaking and massless Goldstone boson origination, has been developed by H. Umesavawa's group (1965, 1974, 1982) and an Italian group, consisting of Del Guidice, Doglia, Milani, Vitello and others (1982, 1984, 1985, 1990, 1991). This approach has some common features with the description of the emission ⇌ absorption of photons and phonons in the process of ($a \rightleftharpoons b$) transitions of primary and secondary effectons.

The presented here Hierarchic Theory of matter unifies and extends two basic, well-known and familiar models of the condensed state:

a) the Einstein model of atoms of matter, as independent quantum oscillators (discrete model)

b) the Debye model of matter, which takes into account only the collective phenomena - phonons (continuous model)

Among earlier models of the liquid state, the model of "flickering clusters" by Frank and Wen (1957) is the closest of all to the Hierarchic model. This model, general for liquids and solids, is described in detail below. The model relates the microscopic molecular dynamics of condensed matter to its macroscopic properties via *mesoscopic matter* organization – corresponding to *clusters* and *domains*.

These new physical ideas and theories require a new terminology. This fact is a reason why the reader can feel a certain discomfort when reading the first chapters of this book. To alleviate this discomfort, a special glossary with explanations of notions and terms, is introduced, and is presented below.

**The New Notions and Definitions Introduced in the Hierarchic Model of Condensed Matter**

*1. The most probable de Broglie wave.*

In condensed matter, the dynamics of particles can be characterized by thermal anharmonic oscillations of two types: *translational* (**tr**) and *librational* (**lb**).

The length of the most probable de Broglie wave of the corresponding vibrating molecule, atom or group of atoms in condensed matter may be estimated in two ways:

$$(\lambda_{1,2,3} = h/m\mathbf{v}_{gr}^{1,2,3} = \mathbf{v}_{ph}^{1,2,3}/\nu_B^{1,2,3})_{tr,lb} \qquad 3$$

where the most probable momentum $p^{1,2,3} = m\mathbf{v}_{gr}^{1,2,3}$ is equal to the product of the particle's mass ($m$) and its most probable group velocity ($\mathbf{v}_{gr}^{1,2,3}$). The de Broglie wave length may also be evaluated as the ratio of its most probable phase velocity ($\mathbf{v}_{ph}^{1,2,3}$) to the most probable frequency ($\nu_B^{1,2,3}$).

The indices (1, 2, 3) correspond to selected directions of motion in 3D space, relating to the main axes of the molecule's symmetry and their tensor of polarizability. In the case where the molecular motion is anisotropic, we have the inequalities:

$$\lambda_B^{(1)} \neq \lambda_B^{(2)} \neq \lambda_B^{(3)} \qquad 4$$

Due to the anharmonicity. of oscillations and non-classical, non-Maxwellian behavior of molecules in condensed matter, we have the critical condition for obtaining the most probable kinetic energy of these molecules:

$$T_{\text{kin}} = m\mathbf{v}^2/2 < kT/2 \qquad 5$$

$$\left(T_{\text{kin}} < V\right)_{tr,lb}$$

These equations imply that the kinetic energy of molecules $(T_{\text{kin}})_{tr,lb}$ in condensed matter is less than their potential energy $(V)_{tr,lb}$. Consequently, the most probable de Broglie wave length of the corresponding fraction of molecules of condensed matter is larger than the classical length:

$$\lambda_B^{1,2,3} > h/(mk_BT)^{1/2} = \lambda_T \qquad 6$$

where $\lambda_T$ is the de Broglie thermal wave, following directly from the classical Maxwell-Boltzmann distribution.

*2. The most probable (primary) effectons (tr and lb).*

These new types of quasi-particles (excitations) are represented by 3D superposition of three most probable standing de Broglie waves of the component molecules. The shape of these primary effectons can be approximated by a parallelepiped, with the length of each of its three edges determined by the length of the most probable de Broglie wave.

The volume of primary effectons is equal to:

$$V_{ef} = (9/4\pi)\lambda_1\lambda_2\lambda_3. \qquad 7$$

The number of molecules or atoms forming the effectons is:

$$n_m = (V_{ef})/(V_0/N_0), \qquad 8$$

where $V_0$ and $N_0$ are the molar volume and Avogadro's number, respectively. The number $n_m$ increases with decreasing temperature and may reach as many as hundreds or even thousands of molecules.

In liquids, the primary effectons represent flickering clusters; in solids they are presented by domains or microcrystalites.

The thermal oscillations of molecules in the volume of the corresponding effectons are indeed synchronized. This implies the coherence of the most probable de Broglie wave of molecules, and the unification of their wave functions. We consider the *primary effectons, as a mesoscopic Bose condensate* (mBC) of molecules in condensed matter. The primary effectons correspond to the ground state of the Bose-condensate with the packing number $n_p = 0$, where the resulting external momentum of mBC is equal to zero.

Primary effectons, as clusters of coherent molecules, represent a high temperature mesoscopic Bose condensate (mBC) with quantum properties. On the other hand, if the volume of primary effectons is less than the volume occupied by a single molecule $V_{ef} \leq (V_0/N_0)$, this means that the classical description of the behavior of the system applies in this case.

*3. "Acoustic" (a) and "optical" (b) states of primary effectons.*

The "acoustic" a-state of the effectons is in a dynamic state when molecules, or other particles making up the effectons, oscillate in the same phase ( i.e. without changing the distance between them).

The "optic" b-state of the effectons is in a dynamic state when the particles oscillate in a counterphase manner with periodic change of the distance between particles. This state of primary effectons has common features with Frölich's mode.

The kinetic energies of "acoustic" (a) and "optical" (b) states are equal $[T^a_{kin} = T^b_{kin}]$ – in contrast to their potential energies, which are not equal: $[V^a < V^b]$. This implies that the most probable momentum values in (a) and (b) states and, consequently, the de Broglie wave length and spatial dimensions of the effectons in both of the states are the same. The energy of inter-molecular interaction (Van der Waals, Coulomb, hydrogen bonds etc.) in the a-state is larger than that in b-state. The molecular polarizability in a-state is also bigger than in b-state. This means that the dielectric properties of matter may change as a result of a shift of $(a \Leftrightarrow b)_{tr,lb}$ equilibrium of the effectons leftward or rightward.

*4. Primary transitons (tr and lb).*

Primary transitons represent intermediate transition states between (a) and (b) modes of primary effectons (translational and librational), i.e. the process of quantum beats between these two states. The primary transitons (tr and lb) radiate $(b \rightarrow a)_{tr,lb}$ or absorb $(a \rightarrow b)_{tr,lb}$ the IR photons corresponding to translational and librational bands in oscillatory spectra. The volumes of primary transitons and primary effectons coincide. (See Table 1.)

*5. Primary electromagnetic and acoustic deformons (tr, lb).*

Electromagnetic primary deformons are introduced as quantum collective excitations, representing a 3D superposition of three standing electromagnetic waves, each normal to the other. The IR photons both originate and annihilate (radiated and absorbed) as a result of $(a \Leftrightarrow b)_{tr,lb}$ transitions of primary effectons, i.e. following the corresponding primary transitons. Such quantum transitions are not accompanied by a density fluctuation, but only by the change of polarizability and dipole momenta of the given molecules.

Electromagnetic deformons appear as a result of the superposition of 3 pairs of photons with different selected directions $(1,2,3)$ in the same space volume. It is assumed that each of these three pairs of photons forms a standing wave in condensed matter.

The linear dimension of each of the three edges of any given primary deformon is determined by the wave length of three super-imposed standing EM waves - photons:

$$\lambda^{1,2,3} = \left[\frac{1}{n\tilde{\nu}}\right]^{1,2,3}_{tr,lb} \qquad 9$$

where: $n$ is the refraction index and $(\tilde{\nu})_{tr,lb}$ is the wave number of the translational or librational band. These quasi-particles / quantum excitations are the largest ones and are responsible for the long-range space-time correlations in liquids and solids.

In the case where $(b \to a)_{tr,lb}$ transitions of primary effectons are accompanied by density fluctuations, these fluctuations are followed by emission of phonons instead of photons. This may happen when the primary effectons are the part of the volume of macro- and super-effectons (see below). Primary acoustic deformons of hypersonic frequency may originate or annihilate in such a way. But for independent primary effectons the probability of emission of photons during $(b \to a)_{tr,lb}$ transition without density fluctuation is much higher than that of phonons.

There exists a coherent electromagnetic radiation (termed super-radiation by Dicke in 1954) which is a result of synhronized self-correlated change of many molecular dipole moments and polarization in the volume of coherent molecular clusters. These molecular clusters represent primary effectons - 3D standing molecular de Broglie waves (mBC), each containing $N \gg 1$ molecules.

In the case of this super-radiance, the collective transition time is less than the transition time of an isolated molecule. However, the intensity of the super-radiance

$$I \sim N \times (h\nu/\tau) \sim N^2, where \quad \tau \sim 1/N \qquad 10$$

is much larger than the intensity derived from the very same number of independent molecules: $I \sim N \times (h\nu/T_1) \sim N$.

The $(b \to a)$ transition time of coherent molecular clusters (the primary effectons) has a reverse dependence on this number (the number being $N \sim 1/\tau$) The relaxation transitional time for independent atoms or molecules ($T_1$) is independent of the number $N$.

The greater proportion of this energy is radiated in the directions of the "most elongated volume" of asymmetric coherent cluster.

The phenomenon of super-radiance has been utilized in the design of high-powered masers, as well as in lasers with ultra-short pulses. The super-radiance is thus seen as a natural consequence of the Hierarchic theory of condensed matter. It is hard to reveal such laser properties of condensed matter experimentally due to fast dissipation of EM radiation because of absorption and fluctuations.

*6. Secondary effectons (tr and lb).*

In contrast to primary effectons, secondary effectons are conventional. They are the result of averaging of the frequencies, length and energies of the "acoustic" (a) and "optical" (b) states of a large number of effectons with packing numbers $n_P > 0$, and with nonzero resulting-external momentum, using quantum statistical methods.

In order to obtain the average energies and frequencies of such states, the Bose-Einstein distribution was used, with the condition $T < T_0$ ($T_0$ is a temperature of degeneration, which coincides with a temperature of first-order phase transition, like boiling and melting. It follows from the results of our computer simulations that under these conditions, the chemical potential $\mu = 0$, and Bose-Einstein distribution turns to form of the Planck equation/distribution.

*7. Secondary transitons (tr and lb).*

Secondary transitons, like primary ones, represent an intermediate transition state between the acoustic ($\bar{a}$) and optic ($\bar{b}$) states of secondary effectons - both the translational and the librational types. Secondary transitons are responsible for radiation and absorption of phonons. As well as secondary effectons, secondary transitons are the conventional collective excitations, as far they are the result of the energies and frequencies averaging, using Bose-Einstein statistics. The volumes of secondary transitons and secondary effectons *are equal.*

*8. Secondary "acoustic" deformons (tr and lb).*

This type of quasi-particles is also conventional – as a result of 3D superposition of averaged thermal phonons. These conventional phonons originate and annihilate in a process of $(\bar{a} \rightleftharpoons \bar{b})_{1,2,3}$ thermally activated transitions of secondary conventional effectons (translational and librational). The secondary $(\bar{a} \Leftrightarrow \bar{b})_{tr,lb}$ transitions are accompanied by the fluctuation of density and acoustic waves excitation (thermal phonons).

*9. Convertons* (**tr** $\Leftrightarrow$ **lb**)

The above excitations are introduced in our model as the *interconversions* between translational and librational primary effectons. The $(a_{tr} \rightleftharpoons a_{lb})$ *convertons (acon)* correspond to transitions between the *(a)* states of these two kind of the effectons and $(b_{tr} \rightleftharpoons b_{lb})$ *convertons (bcon)* - to transitions between their *(b)* -states. The dimensions of translational primary effectons are much less than the librational ones, and so the *convertons* could be considered as the dissociation and association of the primary librational effectons, representing in this consideration the *flickering clusters*. Both the convertons, $(a_{tr} \rightleftharpoons a_{lb})$ types as well as the $(b_{tr} \rightleftharpoons b_{lb})$ types, are accompanied by density fluctuation; this excites the phonons with corresponding frequencies in the surrounding medium.

*10. The types of [lb/tr] con-deformons, induced by corresponding $(a_{tr} \rightleftharpoons a_{lb})_{def}$ and $(b_{tr} \rightleftharpoons b_{lb})_{def}$ convertons.*

Three-dimensional (3D) superposition of acoustic waves (phonons), emitted $\rightleftharpoons$ absorbed by two types of convertons: *acon* and *bcon*, represent in our model the acoustic $(a_{tr}/a_{lb})_{def}$ and $(b_{tr}/b_{lb})_{def}$ *types of [lb/tr] con-deformons.* Unlike the *effective secondary deformons, resulting from averaging, using Bose-Einstein statistics,* discussed above, the *con-deformons of both kinds are real collective excitations, i.e. not the result of averaging, using quantum statistics*.

*11. The Macro-convertons and [lb/tr] macrocon-deformons.*

The simultaneous excitation of the $(a_{tr} \rightleftharpoons a_{lb})$ and $(b_{tr} \rightleftharpoons b_{lb})$ types of [lb/tr] convertons in the volume of primary librational effectons leads to the origination of a large fluctuation, similar to a cavitational fluctuation. These types of flickering quasi-particles are termed *Macro-convertons*.
In turn, such fluctuations induce high frequency thermal phonons in the surrounding medium. The 3D-superposition of these phonons forms what are called *macrocon-deformons*.

*12. Macro-effectons (tr and lb).*

*The macro-effectons* (in A and B states)$_{tr,lb}$ are the result of collective *simultaneous* excitations of the primary and secondary effectons in the $[A \sim (a,\bar{a})]_{tr,lb}$ and $[B \sim (b,\bar{b})]_{tr,lb}$ states in the volume of primary electromagnetic translational and librational deformons, respectively. This correlation of similar primary and secondary states results in significant deviations from thermal equilibrium. The both: A and B states of macro-effectons (tr and lb) may be considered as a large correlated thermal fluctuations of condensed matter.

*13. Macro-deformons or macro-transitons (tr and lb).*

This type of conventional quasi-particle is considered to be the transitional state – between A and B states of previously described *macro-effectons* The $(A \to B)_{tr,lb}$ and $(B \to A)_{tr,lb}$ transitons are represented by the coherent transitions of primary and secondary effectons *in the volume of primary electromagnetic deformons* - of translational or librational types. The $(A \to B)_{tr,lb}$ transition of macro-effectons is accompanied by the simultaneous absorption of IR photons, and thermal phonons. The rest of the energy of the $(B \to A)_{tr,lb}$ transition transforms to the energy of density fluctuation, as well as entropy, within the volume of the macro-effectons. This is a dissipative process. The large fluctuations of density during $(A \Leftrightarrow B)_{tr,lb}$ transitions of macro-effectons, named *macro-deformons* are responsible for the Raleigh central component in the Brillouin spectra of light scattering (See Chapter 9). Translational and librational macro-deformons also contribute to the viscosity (Chapter 11).

The volumes of macro-transitons (macro-deformons) and macro-effectons coincide and are equal to that of translational or librational primary electromagnetic deformons.

*14. Super-effectons.*

This mixed type of conventional quasi-particles is composed of translational and librational macro-effectons correlated in space and time in the volume of superimposed electromagnetic primary deformons (translational and librational – simultaneously). Like macro-effectons, the super-effectons may exist in the ground $(A_S^*)$ and excited $(B_S^*)$ states representing strong deviations from thermal equilibrium.

*15. Super-deformons or super-transitons.*

These collective excitations have the lowest probability as compared with other quasi-particles of our model because of high activation threshold. Similar by nature to macro-deformons, super-deformons represent an intermediate $(A_S^* \Leftrightarrow B_S^*)$ transitional state of super-effectons. In the course of these transitions, the translational and librational macro-effectons undergo simultaneous transitions of the following sort:

$$[(A \Leftrightarrow B)_{tr} \text{ and } (A \Leftrightarrow B)_{lb}]$$

The $(A_S^* \to B_S^*)$ transition of super-effectons may be accompanied by the absorption of translational and librational photons simultaneously. The reverse $(B_S^* \to A_S^*)$ transition may accompanied by the same photonic super-radiation as $b \to a$ transition of the primary effectons. If this process occurs without photonic radiation, it represents a large *cavitational fluctuation.* Such a process plays an important role in sublimation, evaporation and boiling as confirmed by our theory based computer program.

The dynamic equilibrium of water molecule dissociation, possible in the process of cavitational fluctuation can be presented as follows:

$$H_2O \rightleftharpoons H^+ + HO^-$$

The corresponding constant of equilibrium should be dependent on the equilibrium constant for super-transitons: $K_{B_S^* \rightleftharpoons A_S^*}$. It is a consequence of our model that $A_S^* \rightleftharpoons B_S^*$ cavitational fluctuations of super-effectons can be accompanied by the activation of a partial dissociation of water molecules.

In contrast to primary and secondary *transitons* and *deformons*, the volumes of [*macro-transitons* and *macro-deformons*]$_{tr,lb}$ as well as [*super-transitons* and *super-deformons*] coincide. Super-transitons and macro-transitons have the properties of dissipative systems. Such types of large collective excitations represent dynamic processes in the volumes of corresponding, primary, electromagnetic deformons - pulsing 3D infrared photons (superposition of translational and librational deformons).

Considering the *translational* deformons (primary, secondary and macro-deformons), one must keep in mind that the *librational* type of modes may remain unperturbed. As well, (vice versa) in the case of librational deformons, translational modes may remain unchanged. Only the excitations of the convertons and super-effectons are accompanied by the inter-conversions between the translational and librational modes, that is, between translational and librational effectons.

*Interrelation between quasi-particles forming solids and liquids.*

**The Hierarchic Model includes 24 types of quasi-particles** (**Table 1**):

$$\begin{bmatrix} 4 - \textit{Effectons} \\ 4 - \textit{Transitons} \\ 4 - \textit{Deformons} \end{bmatrix} \text{ translational and librational, including primary and secondary} \qquad \text{I}$$

$$\begin{bmatrix} 2 - [lb/tr] \textit{ convertons } (a_{tr} \rightleftharpoons a_{lb}) \textit{ and } (b_{tr} \rightleftharpoons b_{lb}) \\ 2 - [lb/tr] \textit{ con-deformons } (a_{tr} \rightleftharpoons a_{lb}) \textit{ and } (b_{tr} \rightleftharpoons b_{lb}) \\ 1 - [lb/tr]\text{-macroconverton } [(a_{tr} \rightleftharpoons a_{lb}) + (b_{tr} \rightleftharpoons b_{lb})] \\ 1 - [lb/tr] \textit{ macrocon-deformon } [(a_{tr} \rightleftharpoons a_{lb}) + (b_{tr} \rightleftharpoons b_{lb})] \end{bmatrix} \text{ the set of interconvertions between translational and librational primary effectons} \qquad \text{II}$$

$$\begin{bmatrix} 2 - \textit{macro-effectons} \\ 2 - \text{Macrocon-deformon} \end{bmatrix} \text{ translational and librational (spatially separated)} \qquad \text{III}$$

$$\begin{bmatrix} 1 - \text{Super-effecton} \\ 1 - \text{Super-deformon} \end{bmatrix} \text{ translational + librational (superposition of } \textit{tr} \text{ and } \textit{lb} \text{ effectons and deformons in the same volume)} \qquad \text{IV}$$

Each level in the hierarchy of quasi-particles (**I** - **IV**) introduced in our model is based on the principle of correlation in space and time. All of these quasi-particles are constructed on the same physical principles as 3D -superposition of different types of standing waves in the same volume.

These considerations imply that condensed matter can be handled as a system of quasi-particles of 24 types. The developed computer program evaluate the properties of each of these excitation.

Since each of the effecton's types, *tr* and *lb*, macro-effectons and super-effectons [*tr/lb*] has two states (acoustic and optic) the total number of excitations increases to

$$N_{ex} = 31$$

Primary and secondary deformons have the properties of electromagnetic and acoustic Goldstone bosons or Goldstone modes with zero mass and zero resulting spin. Zero resulting spin is a consequence of the opposite orientation of spins ($S = +1$ and $S = -1$) in each pair of the standing photons or phonons, forming primary and secondary deformons.

Three types of standing waves are included in our model:

- de Broglie waves of particles

- acoustic waves (thermal phonons)
- electromagnetic waves (IR photons, corresponding to translations and librations of molecules).

The Hierarchical Theory also describes the transition from the ORDER (primary effectons, transitons and deformons) to the CHAOS (macro- and super-deformons). It is important, however, that, in accordance with the terminology of this model, thermal CHAOS is "organized" when considered in terms of hierarchical superposition of definite types of *ordered* quantum excitations. *This means that the essential dynamics of condensed matter only "appear" to be chaotic in nature, being really complex.* The Hierarchic Model makes it possible to analyze the so-called "hidden order" of condensed matter.

The correlation between remote quasi-particles is provided mainly by the *electromagnetic primary deformons* - the largest ones being those that involve a large number of primary and secondary effectons. The volume of primary deformons [tr and lb] can be conventionally subdivided into two equal parts, within the nodes of 3D standing IR electromagnetic waves. The half-length, characterizing the linear dimension of primary effectons is related to the photon frequency ($v_p = c\tilde{v}_p$) and its wave number ($\tilde{v}_p$) and refraction index of the medium ($n$) in the following formulae:

$$\lambda_p/2 = \frac{c}{2nv_p} = \frac{1}{2n\tilde{v}_p} \qquad 11$$

For librational primary deformons in water ($\tilde{v}_p \simeq 700\ cm^{-1}$) this dimension is equal to $(\lambda_p/2)_{lb} = 5\mu = 5 \times 10^4 \text{Å}$. For translational primary deformons ($\tilde{v}_p \simeq 200\ cm^{-1}$) it is $(\lambda_p/2)_{tr} = 17\mu = 1.7 \times 10^5 \text{Å}$.

Consequently, the number of the effectons (primary and secondary) and molecules, in each of two parts of primary deformons is equal. However, their dynamics are orchestrated in such a way, that when one half of the effectons in the volume of a large primary deformon undergoes an $(A \to B)_{tr,lb}$ transition, the other half of the effectons undergoes the opposite $(B \to A)_{tr,lb}$ transition. These processes compensate each other, due to energy exchanges between two parts of primary deformons by means of the IR photons and phonons. This internal dynamic equilibrium makes it possible to consider macro-effectons and macro-deformons as isolated mesoscopic systems. A similarly-orchestrated dynamic equilibrium also obtains for super-effectons and super-deformons.

The increase or decrease in the concentration of primary deformons is directly related to the shift of $(a \Leftrightarrow b)_{tr,lb}$ equilibrium of the primary effectons leftward ($\Leftarrow$) or rightward ($\Rightarrow$) respectively. This shift, in turn, leads to corresponding changes in the energies and concentrations of secondary effectons, deformons and, consequently, to that of super- and macro-deformons. This mechanism provides the feedback reaction between subsystems of effectons and deformons, necessary for long-range self-organization in macroscopic volumes of condensed matter.

The correlated $(a \Leftrightarrow b)_{tr,lb}$ equilibrium shifts of all types of effectons in the volume of primary deformons (tr or lb) will have an influence on the deviations of dielectric properties and the refraction index of matter, because the polarizability of molecules in the ground state of effectons (a) is higher than in the excited state (b). According to our theories of thermal conductivity, the viscosity and self-diffusion parameters, following from our approach, will also change correspondingly.

This scenario gives a mechanism of correlation between microscopic (molecular), mesoscopic (cluster) and macroscopic levels of matter.

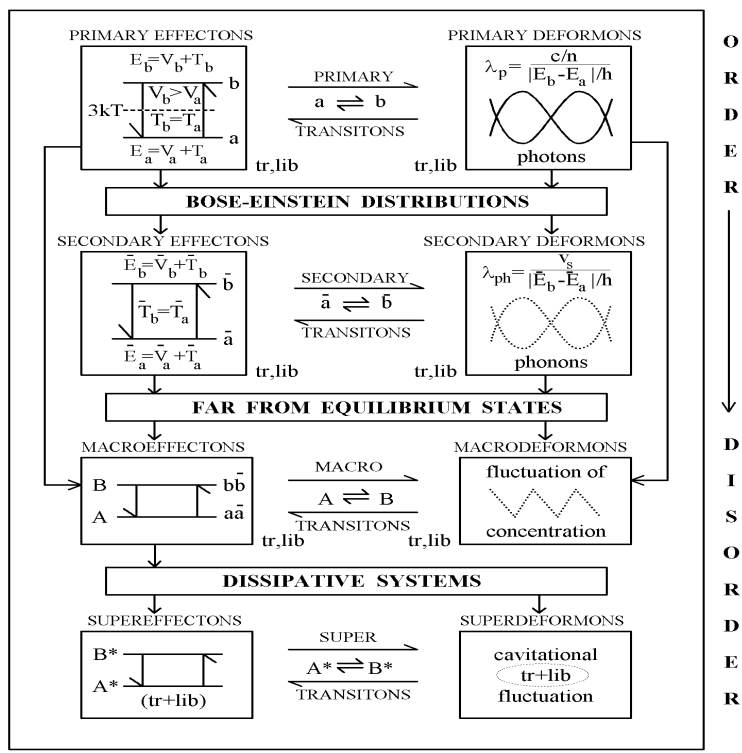

**Table 1**. Schematic representation of the 18 types of quasi-particles of condensed matter as a hierarchical dynamic system, based on the effectons, transitons and deformons. The total number of quasi-particles, introduced in the Hierarchic Theory is 24. See description of hierarchical system/model above, represented by diagram levels I - IV).
The additional six collective excitations, related to *convertons* (interconversions between primary librational and translational effectons and their derivatives) are not represented in this table.

The instability of macro-system, like water, may arise from competition between the discrete - quantum and continuous - thermal energy distributions on mesoscopic scale of coherent molecular clusters. This instability, leading to macroscopic anharmonic oscillations could be stimulated by -'coherent external factors, like geophysical factors, cosmic radiation, gravitational waves, etc.

In certain conditions the primary poly-effectons and coherent superclusters can originate. The filaments, named *poly-effectons* or chain-effectons, are the result of spontaneous polymerization of primary effectons - "side by side" due to distant Van-der-Waals interaction and the presence of Josephson junctions. Such a "polymers" may form 3D net. In liquids it has the properties of an unstable macroscopic Bose condensate, associating and dissociating with certain period. In the closed form the primary poly-effectons can be relatively stable. The "strings" of poly-effectons with macroscopic length can present a superfluid component in $He^2$ (See chapter 12.).

The total internal energy of condensed matter as a system of different 3D standing waves is determined by the sum of contributions of all types of quasi-particles with due regard for their own energy, concentration, and probability of excitation (see section 6.3.2).

The calculated contributions of all members of the following list to the internal energy of matter are very small. This list is as follows: *super-* and *macro-*effectons and corresponding *super-* and *macro-*deformons, as well as poly-effectons and coherent superclusters. These contributions are small (as they are) due to the fact that the concentration of the afore-mentioned particles is much lower than other collective excitations. However, these collective excitations play important roles in *viscosity*, *self-diffusion* and *vapor pressure*, as will be shown in this book (chapter 11).

# Chapter 1
# Theoretical background of the Hierarchic Model
## 1.1. General notions

X-ray data reveal that not only solids, but also a liquids (if the time-frame as short enough) can be modeled as a system of elementary cells having the form of a parallelepiped with edge lengths $(\vec{a}_1, \vec{a}_2, \vec{a}_3)$, and volume:

$$\mathbf{v}_0 = a_1 \times [a_2 a_3] \qquad 1.1$$

The number of either atoms or molecules in this volume is termed the basis (Kittel, 1978; Ashkroft and Mermin, 1976; Blakemore, 1988). The indices (1, 2, 3) denote the main symmetry axes. The basis can be represented by an atom, a molecule or even a group of atoms, arranged in a certain manner relative to one-another.

The *primitive basis* is the minimal number of atoms or molecules corresponding to a primitive (minimum) elementary cell.

The basis vectors of the first Brillouin zone (inverse elementary cell) are introduced from (1.1) as:

$$\begin{aligned} \vec{c}_1 &= [\vec{a}_2 \vec{a}_3]/\mathbf{v}_0 \\ \vec{c}_2 &= [\vec{a}_3 \vec{a}_1]/\mathbf{v}_0 \\ \vec{c}_3 &= [\vec{a}_1 \vec{a}_2]/\mathbf{v}_0 \end{aligned} \qquad 1.2$$

These basis vectors have the dimension of wave numbers: $k = (2\pi/\lambda) \equiv 1/L (cm^{-1})$. They form the first Brillouin zone of the volume:

$$c_1 c_2 c_3 = \frac{1}{\mathbf{v}_0} \qquad 1.3$$

In the limiting case, when the primitive basis contains only one molecule, its volume is:

$$\mathbf{v}_0^{\min} = V_0/N_0 \qquad 1.4$$

where $V_0$ and $N_0$ are the molar volume Avogadro's number, respectively.

The so called *acoustic (in-phase)* and *optical (counter-phase)* oscillations appear if the primitive basis contains two or more particles (atoms or molecules of different or equal masses), which are capable of relative thermal displacements.

For each kind of displacements (longitudinal or transversal with respect to direction of wave propagation), the dependence of the angular frequency of the particle's oscillation ($\omega$) on the wave vector ($k$), has both acoustic and optical branches. The analysis of the simple case of normal oscillations for a linear two-atom chain, yields two corresponding solutions (Blakemore, 1985):

$$\omega^2 = \mu(\tfrac{1}{m} + \tfrac{1}{M}) \pm \mu\left[(\tfrac{1}{m} + \tfrac{1}{M})^2 - \tfrac{4\sin^2(ka)}{mM}\right]^{1/2} \qquad 1.5$$

where $\mu \approx m(\mathbf{v}_s/l)^2$ is the inter-atom rigidity coefficient; $\mathbf{v}_s$ is the velocity of sound; $l$ is the distance between two types of atoms: with masses $m$ and $M$, where $(m < M)$.

The negative sign in (1.5) and the smaller values of the frequencies of oscillation ($\omega_a$) correspond to the acoustic aspect, while the positive sign correspond to frequencies $\omega_b$ – associated with the *optical branch* ($\omega_b \gg \omega_a$).

The analysis of (1.5) shows that the maximal frequency of acoustic oscillations occurs when $\sin(ka) = 1$:

$$\omega_a^{\max} = (2\mu/M)^{1/2} \qquad 1.6$$

and the maximal frequency of the optical aspect, occurs when $\sin(ka) = 0$, is

$$\omega_b^{max} = \left[2\mu(\frac{1}{m} + \frac{1}{M})\right]^{1/2} \qquad 1.7$$

The minimal frequency of optical oscillations, for the case, when $M \to \infty$, (See Figure1b):

$$\omega_b^{min} = (2\mu/m)^{1/2} \qquad 1.8$$

It is important to bear in mind that the waves corresponding to the frequencies $\omega_a$ and $\omega$, are *standing waves* (Kittel, 1975; Ashkroft and Mermin, 1976).

The notions of acoustic and optical modes also hold for 3D crystals with a multiatom basis. The number of transversal modes in this case is just twice the number of longitudinal modes.

If the volume of the medium contains $pN$ atoms, and $N$ is the number of basis in this volume, then the following set of modes appears:

$$\left.\begin{array}{l} N \text{ longitudinal acoustic modes;} \\ 2N \text{ transversal acoustic modes;} \\ (p-1)N \text{ longitudinal optical modes;} \\ 2(p-1)N \text{ transversal optical modes.} \end{array}\right\} \qquad 1.9$$

If a basis cell or a primitive cell consists of two particles (atoms or molecules), then $p = 2$. Therefore, it follows from (1.9) that the number of acoustic and optical modes is equal. *The same situation occurs, when each thermal modes of p-atoms in a basis (longitudinal and transversal) are synchronized or coherent in both the optical and acoustic states.*

The *acoustic* oscillations *(a)* represent the oscillations of particles in a primitive cell in the same phase, when the centers of their masses move together and the distance between them is constant.

The *optical* oscillations *(b)* of particles, on the contrary, are *counter-phasic* – that is, the center of mass within the cell unit is kept fixed.

It is assumed in our model, that the beats between the coherent (a) and (b) interacting anharmonic modes $[a \Leftrightarrow b]$ are accompanied by emission or absorption of IR photons or phonons with frequency:

$$\omega_p = \omega_b - \omega_a \qquad 1.10$$

Figure 1a shows the oscillations of particles in the acoustic *(a)* and optical *(b)* branches (Ashkroft and Mermin, 1976).

In the case of in-phase acoustic oscillations, the displacements of various particles are equal:

$$u = v \qquad 1.11$$

For optical oscillations, the following relation between mass and displacements is valid:

$$uM = -\mathbf{v}m$$

$$\text{or} \qquad 1.12$$

$$u/\mathbf{v} = -m/M$$

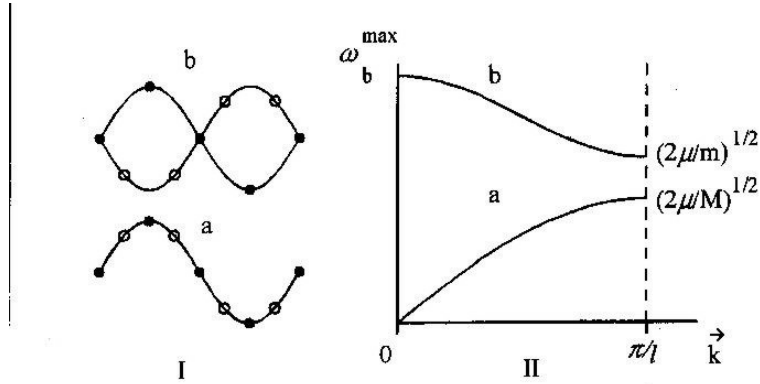

**Figure 1**. (I) The transversal optical *(b)* and transversal acoustic *(a)* waves of *equal length* for the case of a linear two-atom chain. (white dots and black dots) denote two kinds of atoms with masses (m) and (M) and the displacement values: (**v**) and (**u**) respectively; (II) Optical and acoustic phonon branches of the dispersion dependence for two atoms chain; [$l$ - is the lattice constant].

It follows from eqs. (1.6) - (1.8) that the intermediate frequency range:

$$\omega_b > \omega > \omega_a \qquad 1.13$$

is prohibited. If the acoustic and optical modes are purely harmonic, they do not interact, and, as a result, the beats between them are not possible. However, in real condensed matter these oscillations of atoms and molecules are always anharmonic, and so the quantum beats between the acoustic (a) and optical (b) modes should, in fact, exist.

### 1.2. **The main statements of the hierarchic model**

**1**. Because of the anharmonicity of the acoustic (a) and optical (b) modes of the primary effectons and corresponding interaction between them, the quantum $a \Leftrightarrow b$ transitions (quantum beats) with absorption and radiation of photons is existing.

These quantum beats may occur between one longitudinal acoustic $(a_1)$ and one longitudinal optical $(b_1)$ mode, in this manner:

$$a_1 \Leftrightarrow b_1 \qquad 1.14$$

As well, quantum beats may occur between *two* transversal acoustic $(a_2$ and $a_3)$ and two transversal optical $(b_2$ and $b_3)$ modes:

$$a_2 \Leftrightarrow b_2, \qquad 1.15$$
$$a_3 \Leftrightarrow b_3 \qquad 1.15a$$

It is important to note that these three type of transitions are simultaneous. In conditions of equilibrium, when the external source of optical or acoustic signals is absent, the classification of thermal modes – as longitudinal and transversal – is conventional.

The number of acoustic and optical modes is the same (3), if the oscillation of all p-atoms in the basis are coherent in both optical and acoustic dynamic states. The remnant modes are in state of degeneration.

As described in the Introduction, the coherent clusters of molecules, representing primary effectons, are the result of 3D superposition of standing de Broglie waves in any type of condensed matter in liquid or solid phase. It is known that the short-range order, peculiar to crystals, is also retained after *solid* → *liquid* phase transition.

**2**. The acoustic *(a)* and optical *(b)* states of the effectons differ only in their potential energies,

while the most probable values of momentums and kinetic energies of the particles in these states are, in fact, equal. It means that the de Broglie wave length of (a) and (b) states are also equal:

$$\lambda^a_{1,2,3} = \lambda^b_{1,2,3} \qquad 1.16$$

$$\lambda^a_{1,2,3} = h/P^a_{1,2,3} = h/m(\mathbf{v}^{1,2,3}_{gr})^a \qquad 1.16a$$

$$\lambda^b_{1,2,3} = (\mathbf{v}^a_{ph}/\nu_a)_{1,2,3} = (\mathbf{v}^b_{ph}/\nu_b)_{1,2,3} \qquad 1.16b$$

Here $m$ is the mass of the particle, $(P_{1,2,3} = m\mathbf{v}^{1,2,3}_{gr})$ are the projections of the most probable particle momentum onto the axes (1, 2, 3); $(\mathbf{v}^{1,2,3}_{gr})$ are the projections of the most probable group velocities of the particle de Broglie wave onto the axes (1, 2, 3); $(\mathbf{v}^{a,b}_{ph}$ and $\nu^{a,b})_{1,2,3}$ are the most probable phase velocity and frequency of de Broglie waves in the same selected directions in (a) and (b) states. Computer calculations show that the value of the velocity of sound ($\mathbf{v}_s$ of phonons) is between phase velocities $\mathbf{v}^a_{ph}$ and $\mathbf{v}^b_{ph}$ of the effectons states (See Chapter 6. Figure 2.). The length of the de Broglie waves also differs from that of phonons with the same frequency. This means that we can not use the notion of phonons for a description of the effectons. The physical "contexts" of phonons and de Broglie waves are different.

The de Broglie wave length in selected directions [1, 2, 3] of the most probable (primary) effectons and of mean (secondary) effectons are not equal in general case because of possible spatial anisotropy of translations and librations:

$$[\lambda^{a,b}_1 \neq \lambda^{a,b}_2 \neq \lambda^{a,b}_3]_{tr,lb} \qquad 1.17$$

$$[\bar{\lambda}^{a,b}_1 \neq \bar{\lambda}^{a,b}_2 \neq \bar{\lambda}^{a,b}_3]_{tr,lb} \qquad 1.18$$

The momentums and the corresponding de Broglie wave lengths $(\lambda = h/p)_{tr,lb}$ are a consequence of two basic thermal oscillations: *translations* (tr) and *librations* (lb).

It is implied, therefore, that *translations* are related to pure translational displacements of particles of condensed matter, while *librations* are linked with limited rotational-translational displacements. The *small* translational momentum of the libration determines a larger librational de Broglie wave length than that of pure translations: $(\lambda_{lb} = h/p_{lb} > \lambda_{tr} = h/p_{tr})$. Hence, the size of librational effectons normally exceeds that of translational effectons.

It is known (Prochorov, 1990), that the states of any system, minimizing the uncertainty relation, when:

$$\left[\Delta p \, \Delta x \sim \hbar \quad \text{and} \quad \Delta x = L \sim \hbar/\Delta p\right]_{1,2,3} \qquad 1.18a$$

are considered as a quantum *coherent states*.

**3**. The 3D superposition of three most probable standing de Broglie waves $(\lambda_1, \lambda_2, \lambda_3)$ leads to origination of metastable coherent excitations (quasi-particles), defined as the most probable - primary effectons. These effectons can be subdivided to translational or librational types.

Depending on the type of coherent oscillations of all molecules in the volume of primary effectons: acoustic or optical types, the effectons may be in either the (a) or (b) states, respectively. The primary effectons can be approximated by parallelepiped with the length of its edges determined by the set of superimposed de Broglie wave lengths (eq. 1.17).

When the de Broglie waves lengths of different molecules/atoms forming the same effecton are equal, the differences in their masses are compensated for by the differences in their most probable group velocities. The momentums and de Broglie waves in such a case are all equal. See equation. (1.12).

The resulting momentum of primary effectons of a *3D* superposition of standing de Broglie waves in optic-like (b) state is zero, and that in the acoustic-like (a) state is oscillating around zero. The subsystem of primary effectons can be considered as a weakly interacting Bose particles, tending without external mechanical perturbations, like shaking to reversible assembly.

The degree of the Bose-degeneration is proportional to the fraction of molecules in the volume of primary effectons of matter. Degeneration in liquids grows up at lowering temperature and make a jump up as a result of (liquid → solid) phase transition.

It is known from the Bose-Einstein theory of condensation, developed by London in 1938 that **if** the *degeneration factor:*

$$\lambda = \exp(\mu/kT) \qquad 1.19$$

is close to unit at a low chemical potential value:

$$\mu \ll kT \qquad 1.20$$

then the contribution of bosons with the resulting external momentum $P_{ef} \simeq 0$, (similar to the case of primary effectons) cannot be neglected, when calculating internal energy of condensed matter. We assume in our Hierarchic theory, that the condition (1.20) is valid.

Partial (mesoscopic) Bose-Einstein condensation leads to the coherence of vibrations of molecules and atoms forming primary effectons - coherent molecular clusters in the acoustic (a) and optic (b) states. The de Broglie wave length and dimensions of the effectons are the same in both (a) and (b) states. Primary effectons are thus described by the wave functions which are coherent in the volume of the effecton.

Despite the partial Bose-condensation, the effectons with nonzero momentum do exist. They are termed *secondary* effectons. These effectons obey the Bose-Einstein statistics and can be analyzed using it.

The sizes of primary effectons determine the mesoscopic scale of condensed matter organization. According to the model, the domains, nodes, crystallites and clusters observed in solid bodies and in liquid crystals, polymers and biopolymers - can represent the primary translational and librational effectons. Stability of molecular/atomic/ionic clusters (effectons) is provided by distant Van der Waals interaction in conditions of 3D standing de Broglie waves of the coherent particles of clusters.

**4**. The gas → liquid phase transition is accompanied by degeneration of translational degrees of freedom, when the chemical potential $\mu \to 0$ (eq.1.19). Under these conditions the de Broglie wave length, provided by librations, begins to exceed the mean distances between molecules in the liquid phase.

This means that the temperature of the phase transition [*gas → liquid*] coincides with the degeneration temperature ($T_0$). The changes of the effectons volume and shape in three dimensional (3D) space are related to corresponding changes in the momentum space of molecules.

The macroscopic Bose-condensation, in accordance to our theory of superconductivity and superfluidity (Chapter 12), occurs under conditions, where the $[a \Leftrightarrow b]$ equilibrium of primary and secondary effectons strongly shifts to (a) - state and (b) - state becomes thermally inaccessible. Consequently, the probability of photons and phonons excitation and energy dissipation tends to zero. At the same time the de Broglie wave length of molecules tends to its macroscopic value.

For quantum systems at temperatures higher than the degeneration temperature $T_0$ ($T > T_0$), when the chemical potential has a negative value, (as when $(\mu_i = \partial G_i/\partial n_i) < 0$ ) the mean number of Bose-particles ($\bar{n}_i$) in state ($i$) is determined by Bose- Einstein distribution:

$$\bar{n}_i = \frac{1}{\exp[(\epsilon_i - \mu_i)/kT] - 1} \qquad 1.21$$

where $\epsilon_i$ is the energy of the particle in state ($i$).

For real condensed matter, $\epsilon_i \gg \mu \ll kT$ (See eq. 1.20).

The Bose-Einstein statistics, in contrast to the Maxwell-Boltzmann statistics, are applied to the indistinguishable Bose - particles with zero or integer spin values. The Fermi - Dirac distribution is valid for systems of indistinguishable particles with a semi-integer spin obeying the Pauli principle.

In the case of condensed matter at the temperature:
$$0 < T < [T_0 \cong T_c]$$

$N^*$ particles of Bose condensate have a zero momentum (Landau, Lifshitz, 1976):
$$N^* \simeq N\left[1 - (T/T_0)^{3/2}\right] \qquad 1.22$$
where $N$ is the total number of particles in a system.

This equation is in accordance with our Hierarchic Model. For the most probable kinetic energy of a quasi ideal Bose-gas of particles with mass ($m$) and velocity ($\mathbf{v}$) we can write, using Maxwell distribution:
$$\frac{m\mathbf{v}^2}{2} = \frac{\hbar^2}{2mL_0^2} = kT_0 \qquad 1.23$$
where $L_0 = \hbar/m\mathbf{v} = \lambda_B/2\pi = 1/k_B$; $\lambda_0$ and $k_0$ are the most probable de Broglie wave length and wave number of the particles forming primary effectons with degeneration temperature ($T_0$).

At a temperature lower than $T_0$ ($T < T_0$), the following consequence of (eq. 1.23) holds:
$$L_0^3/L_T^3 = (T/T_0)^{3/2} \qquad 1.24$$
On the other hand, the left part of this equation reflects the ratio of the density of primary effectons ($n_{ef}^T$ and $n_{ef}^0$) at two temperatures:
$$L_0^3/L_T^3 = n_{ef}^T/n_{ef}^0 \qquad 1.25$$
Combining eqs. (1.25), (1.24) and (1.22), we obtain the number of particles forming primary effectons ($N_{ef} = N^*$) in condensed matter:
$$N_{ef} = N \cong \left[1 - \frac{n_{ef}^T}{n_{ef}^0}\right] = \left[1 - (T/T_0)^{3/2}\right] \qquad 1.26$$

Clearly, when $T \to 0$ and $L_T \to \infty$, the corresponding concentration of primary effectons: ($n_{ef}^T \simeq 1/L_T^3) \to 0$, because their volume ($V_T \sim L_T$) tends to infinity. Consequently, $N_{ef} \to N$, i.e. all the particles of matter tend to make up one macroscopic primary effecton in the basic *(a)*-state containing all of these $N$ particles. This condition corresponds to formation of macroscopic Bose condensate, like superconductivity or superfluidity.

At the given temperature $[0 < T < T_0]$, the number and energy of the remaining particles $[N - N^*]$ forming effectons with nonzero quantum momentums, obey the Bose-Einstein distribution (eq. 1.21). Under these conditions the chemical potential in (1.21) is close to zero ($\mu_i \simeq 0$) (London, 1938).

Under these conditions, the Bose-Einstein distribution (1.21) for the mean number of particles in state (i) coincides with the Planck formula for the mean quantum number ($\bar{n}$) of harmonic quantum oscillators with quantized energy
$$E_i = h\nu_i = h\nu_0(n_i + 1/2) \qquad 1.27$$
$$n_i = \bar{n} = \frac{1}{\exp(E_i/kT) - 1} \qquad 1.27a$$
Using this approximation, the corresponding mean energy ($\bar{E}_{a,b}$) and the mean frequency ($\bar{\nu}_{a,b}$) of the $\bar{a}$ and $\bar{b}$ states of secondary effectons, can be calculated as:

$$\bar{E}_{a,b} = \bar{n}_{a,b} \, h\nu_{a,b} = \qquad (1.28)$$

$$h\bar{\nu}_{a,b} = \frac{h\nu_{a,b}}{\exp(h\nu_{a,b}/kT) - 1} = h\frac{\bar{\mathbf{v}}_{ph}^{a,b}}{\bar{\lambda}_{a,b}} \qquad (1.28a)$$

where: $\nu_{a,b} = \mathbf{v}_{ph}^{a,b}/\lambda_{a,b}$ are the frequencies of de Broglie waves of molecules forming primary effectons in a and b states; $(\mathbf{v}_{ph}^{a,b}, \bar{\mathbf{v}}_{ph}^{a,b})$ and $(\lambda_{a,b}, \bar{\lambda}_{a,b})$ are the phase velocities and length of de Broglie waves forming primary and secondary effectons, respectively.

The secondary (or mean) effectons, formed by 3D superposition of mean de Broglie waves in the directions $(1, 2, 3)$: $\bar{\lambda}_1^{a,b}, \bar{\lambda}_2^{a,b}, \bar{\lambda}_3^{a,b}$ are conventional in contrast to real – primary or most probable primary – effectons. Secondary effectons are the result of averaging of distribution of quasi-particles by energies and momentums. Like primary effectons, they can exist in two $\bar{a}$ and $\bar{b}$ states and are approximated by a parallelepiped with dimensions determined from (1.18).

Similar to primary effectons (see eq.1.16), the dimensions of secondary effectons, determined by averaged momentums (kinetic energies) are the same in $(\bar{a})$ and $(\bar{b})$ states:

$$\bar{\lambda}_{1,2,3}^{a} = \bar{\lambda}_{1,2,3}^{b} \qquad (1.29)$$

$$\bar{\lambda}_{1,2,3}^{a} = h/\bar{P}_{1,2,3}^{a} = \frac{\bar{\mathbf{v}}_{ph}^{a}}{\bar{\lambda}_a} \qquad (1.29a)$$

$$\bar{\lambda}_{1,2,3}^{b} = h/\bar{P}_{1,2,3}^{b} = \frac{\bar{\mathbf{v}}_{ph}^{b}}{\bar{\lambda}_b} \qquad (1.29b)$$

Primary and secondary effectons are subdivided in terms of their dynamic nature into two classes: *translational* (tr) and *librational* (lb).

**5**. The stability of coherent effectons are determined mainly by the *additive dispersion interactions, provided by unification and synhronization of molecules in state of mesoscopic Bose condensation*. However, other types of distant Van der Waals interactions and hydrogen bonds can contribute to stability of these cluster either. The transitions between the acoustic (a) and optical (b) states of primary effectons are accompanied by changing the molecular polarizability and – in some cases – by changing of the dipole moments. However, the fluctuation of density of molecules does not accompany the $a \Leftrightarrow b$ transitions of primary effectons, i.e. these beats can not activate the phonons in the medium. Only the IR photons are emitted or absorbed as a result of $(b \to a)$ or $(a \to b)$ transitions of primary effectons, respectively.

On the other hand, similar $(\bar{a} \Leftrightarrow \bar{b})_{tr,lb}$ transitions of conventional secondary effectons are accompanied by fluctuation of density in their volume, origination and annihilation of thermal phonons and their propagation (with the velocity of sound) in the medium.

The energy and frequency of the acoustic and optic states of primary and secondary effectons, as well as the frequency of transitions between these states are determined by the energy difference of these states. In the general case thee energies differ with respect to each of three directions $(1, 2, 3)$ along the main axes of the effectons.

For the energy of IR photons, radiated and absorbed by primary effectons in 3 selected directions, we have:

$$\left[\Delta E_{1,2,3}^{a \Leftrightarrow b} = (E_b - E_a)_{1,2,3} = h(\nu_b - \nu_a)_{1,2,3} = \right.$$
$$\left. = (V_b - V_a)_{1,2,3} = h\nu_p^{1,2,3} \right]_{tr,lb} \qquad (1.30)$$

and for the averaged energy of thermal phonons radiated and absorbed by secondary effectons. we have:

$$\left[ \Delta \bar{E}_{1,2,3}^{a \Leftrightarrow b} = |\bar{E}_b - \bar{E}_a|_{1,2,3} = |\bar{V}_b - \bar{V}_a|_{1,2,3} = h\nu_{ph}^{1,2,3} \right]_{tr,lb} \quad 1.31$$

Consequently, the energies of photons $E_p = h\nu_p$ and phonons $E_{ph} = h\nu_{ph}$ are determined by the difference in potential energy ($V_{a,b}$) of the optic and acoustic states only, because, according to the model, the kinetic energies ($T^{a,b}$) of both states are equal:

$$\left( T_k^a = T_k^b \right)_{tr,lb}^{1,2,3} \quad \text{and} \quad \left( \bar{T}_k^a = \bar{T}_k^b \right)_{tr,lb}^{1,2,3} \quad 1.32$$

According to the theory of *dispersion* and *induction* interactions, the change in the potential energy of intermolecular interactions ($V_{dis}$ and $V_i$) must be accompanied by change of polarizability of molecules ($\alpha$), dipole moment ($\mu_c$) at permanent distances (r) between them:

$$\left[ V_{dis} = -\frac{3}{4}\frac{\alpha^2}{r^6}h\nu_0 \simeq -\frac{2}{3}\frac{\alpha^2}{r^6}I_0 \right]^{1,2,3} \quad 1.33$$

$$\left[ V_i = -\frac{2\alpha\mu_c^2}{r^6} \right]^{1,2,3} \quad 1.34$$

where $I_0$ is the ionization potential of a molecule:

$$I_0 \simeq h\nu_0/2 \quad 1.35$$

and

$$\mu_c = el \quad 1.36$$

is the constant dipole moment; $(e)$ is the electron charge; $l$ is a distance between weight centers of the charges.

As was mentioned above, the distant Van der Waals interaction and quantum phenomena, generated by corpuscle-wave pulsation of elementary particles which make up atoms and molecules (Kaivarainen, 2006), may be responsible for stabilization of the effectons as a coherent clusters.

If the transitions between two states of the effectons (*a* and *b*) are followed not only by a change in the *tensor* of polarizability $(\alpha)^{1,2,3}$ but also by change of the *tensor* of dipole moment of these molecules $(\mu^a \neq \mu^b)^{1,2,3}$ at a constant mean distance (r), then the given transitions should be accompanied by the radiation or absorption of photons in 3 selected directions with frequencies: $\nu_p^{(1)}, \nu_p^{(2)}, \nu_p^{(3)}$ – as determined by eq.(1.30).

According to the model, the translational and librational bands in the IR-spectrum of matter ( i.e. water and ice) with wave numbers: $(\tilde{\nu}_p)_{tr}$ and $(\tilde{\nu}_p)_{lb}$ correspond to such $(a \Leftrightarrow b)^{1,2,3}$ transitions of primary translational and librational effectons.

On the other hand, the $(\bar{a} \Leftrightarrow \bar{b})_{tr,lb}^{1,2,3}$ transitions of the *secondary* effectons are accompanied by density fluctuations with change of the mean distance (r) between molecules. The energy of such transitions (excitations, termed *secondary transitons* - see the Introduction) is transmitted to the surrounding medium in the form of *thermal phonons* (eq. 1.31).

The influence of molecules on the properties of the surrounding medium, when they oscillate in the "acoustic" or "optical" states of the effectons, is itself much less than the effects of $(\bar{a} \Leftrightarrow \bar{b})$ transitions. The frequency and amount of *super-radiated* IR photons, which originate and annihilate in the process of coherent $(a \Leftrightarrow b)_{tr,lb}$ transitions of *primary* effectons, should be much bigger than that, excited by optical modes of molecules in *(b)* state of primary effectons (tr or lb).

**6**. The 3D superposition of standing photons emitted due to the $(b \to a)_{tr,lb}$ transitions of primary effectons forms quasi-particles, named the *"most probable"* (primary) electromagnetic deformons (tr and lib).

The dynamics of molecules in the volume of primary (EM) deformons are correlated. Emission of photons as a result of collective spontaneous transitions clusters of coherent molecules or atoms is known as *super-radiance* (Dicke, 1954). In contrast to radiation of isolated atoms or molecules in the absence of coherency, the time of this phenomenon of super-radiance [$\tau \sim 1/N$] of the primary effecton, containing $N$ molecules has an inverse dependence on the $N$.

The intensity of super-radiation is proportional to $N^2$ and can exceed regular radiation of noncoherent molecules by a few orders in magnitude. If the primary effectons are approximated by parallelepipeds, then the greater part of this energy (the number of IR photons) will be radiated in two directions, corresponding to the most elongated edges (Andreev *et al.*, 1988.)

The sizes of primary deformons are determined by the wavelengths of IR photons in selected directions (1,2,3), related to the frequency of primary effectons transitions (eq.1.31):

$$\left[ \lambda_p^{1,2,3} = \frac{c}{n\nu_p} = \frac{c}{nc\tilde{\nu}_p} = \frac{c}{n\tilde{\nu}_p} \right]_{tr,lb}^{1,2,3} \qquad 1.37$$

where: $c$ is the speed of light; $\tilde{\nu}_p$ is the wave number of the oscillatory spectra, and $n$ is the refraction index of the given sample.

The *phonons* appear as a direct consequence of transitions between conventional $\bar{a}$ and $\bar{b}$ states of conventional *secondary* effectons. Their 3D superposition forms secondary acoustic deformons (tr and lb).

*All secondary quasi-particles are conventional*. They are products of averaging, using quantum statistics, in contrast to real primary quasi-particles. Because photons and phonons are bosons, the subsystem of deformons have the properties of Bose gas, like a subsystem of effectons.

**7**. The dynamic equilibrium between the four basic types of quasi-particles (effectons, convertons, transitons and deformons) is very sensitive to various external fields: the acoustic, electromagnetic, etc.

The *competition* between two types of energy distribution: *the continuous* heat distribution and *the discrete*, quantum-mechanical distribution of quasi-particles can induce instability and the long period macroscopic oscillations in condensed matter, accompanied by counterphase changes of the effectons and deformons subsystems energies. It happens in such a way that the total internal energy of matter remains permanent. Such oscillations can be accompanied by slow periodic changes of equilibrium between the "acoustic" and "optic" states of primary and secondary effectons, macro-effectons, and super-effectons (see Table 1 of Introduction).

*Another important optic collective phenomena except super-radiation is "bistability"*. The bistability is possible in materials capable to saturation of the absorption or to nonlinear dispersion. Bistability occur at certain *critical* intensity, when one input beam induces two output stable beams. The analogy between bistability and phase transitions is very deep. In both cases the change of parameter of order, soft mode shift and spontaneous symmetry breach occur.

The resulting dipole moments of *(a)* and *(b)* states of the effectons are not equal $(\mu^a \neq \mu^b)^{1,2,3}$. Furthermore, light can shift the equilibrium between populations of two corresponding energetic levels; the light itself may induce the change of macroscopic static polarization. This effect depends on the intensity of light (density of IR photons) for shifting ($a \Leftrightarrow b$) equilibrium to the right.

The equilibrium could be *bistable*, depending in turn, on saturation effect and cooperative interaction between molecules in the volume of the effectons (*i.e.* by means of acting Lorenz field and Bose condensation).

The light-stimulated phase transitions can be induced not only by an external source of electromagnetic energy, but also self-induced by an internal field in the form of primary electromagnetic deformons (tr and lb). Self-induced bistability in laser mediums, like that seen in the ruby crystal, is related to the "pike regime" of generation. The "*pike regime*" of light radiation is represented by the raw of short momentums (peaks). The main conditions for the emergence of such a regime is a high cooperativeness of the medium, nonequilibrium properties and strong distant interactions between molecules.

The *self-induced bistability and the pike regime* are interrelated with the oscillation of $[a \rightleftharpoons b]$

equilibrium of primary effectons; they should be accompanied by oscillations of the dielectric permeability and matter transparency with the same frequency.

Usually, calculation of the interaction between molecules requires special potentials: the Lennard-Jones, Backingham one, etc. are used. In our hierarchic approach, the intermolecular interactions are taken into account by sizes of the each type of quasi-particles (collective excitations) and their concentrations. The stronger are interactions between molecules of matter and the smaller is their kinetic energy and momentum, the higher is degree of Bose-condensation, reflected by increasing of dimensions of primary effectons.

It will be shown that the Hierarchic Theory is universal, interrelating the micro- and macro-levels in liquids and solids, both the quantum and classical parameters. Theory provides a new explanation of, as well as a quantitative analysis of, a wide range of physical phenomena, making them unified.

# Chapter 2

# Properties of de Broglie waves

### 2.1. Parameters of individual de Broglie waves

The detailed dynamic mechanism of *Corpuscle ⇌ Wave* duality and the nature of de Broglie wave are described in the paper of this author: " Unified Theory of Bivacuum, Particles Duality, Fields & Time. New Bivacuum Mediated Interaction, Overunity Devices, Cold Fusion & Nucleosynthesis", http://arxiv.org/pdf/physics/0207027 (Kaivarainen, 2006).

The well-known de Broglie relation, interrelating wave and corpuscular properties of particle has the simple form:

$$\vec{p} = \hbar\vec{k} = h/\vec{\lambda}_B \quad \text{or} \quad 2.1$$

$$\vec{p} = \hbar/\vec{L}_B = m_V^+ \vec{v}_{gr} \quad 2.1a$$

where $\vec{k} = 2\pi/\vec{\lambda} = 1/\vec{L}_B$ is the wave number, with wavelength $\vec{\lambda}_B = 2\pi\vec{L}_B$, $\vec{p}$ is the momentum of particle with actual mass $(m_V^+)$ and group velocity $(\mathbf{v}_{gr})$, and $\hbar = h/2\pi$ is the Planck constant.

Each particle can be represented as wave packet with group velocity:

$$\mathbf{v}_{gr} = \left(\frac{d\omega_b}{dk}\right)_0 \quad 2.2$$

and phase velocity:

$$\mathbf{v}_{ph} = \frac{\omega_b}{k} \quad 2.3$$

where: $\omega_B$ is the angle frequency of de Broglie wave determining the total energy of the de Broglie wave: $(E_B = \hbar\omega_B)$.

The total energy is equal to the sum of kinetic $(T_k)$ and potential $(V_B)$ energies and is related to relativistic particle's mass and the product of it's phase and group velocities $(\mathbf{v}_{gr}\mathbf{v}_{ph} = \mathbf{c}^2)$ as follows (Grawford, 1973):

$$E_B^{ext} = \hbar\omega_B^{ext} = (T_k + V_B)^{ext} \quad 2.4$$

*or* :

$$E_B^{ext} = \frac{(\hbar k)^2}{2m} + V_B = m_V^+(\mathbf{v}_{gr}\mathbf{v}_{ph})^{ext} = m_V^+ \mathbf{c}^2 \quad 2.4a$$

where (m) is the particle mass; (c) is the velocity of light.

The dispersion relation of the de Broglie waves leading from (eq.2.4) is:

$$\omega_B^{ext} = \frac{\hbar k^2}{2m_V^+} + \frac{V_B}{\hbar} \quad 2.5$$

$$= \frac{\hbar}{2m_V^+ L_B^2} + \frac{V_B}{\hbar} \quad 2.5a$$

Substituting (eq. 2.5) into (2.2), we derive the expression for the group velocity of de Broglie wave, which is equal to the velocity of the particle *per se* (v):

$$\mathbf{v}_{gr}^{ext} = \frac{\hbar k}{m_V^+} \quad or\ : \quad 2.6$$

$$\mathbf{v}_{gr}^{ext} = \frac{p}{m_V^+} = \mathbf{v} \quad 2.6a$$

The external phase velocity of the de Broglie wave taking into account (2.3) and (2.5):

$$\mathbf{v}_{ph}^{ext} = \frac{\omega_B^{ext}}{k} = \frac{\hbar k}{2m_V^+} + \frac{V_B}{\hbar k} = \frac{\mathbf{v}_{gr}}{2} + \frac{V_B}{m_V^+ \mathbf{v}_{gr}} \qquad 2.7$$

Because in the general case, the phase and group velocities of de Broglie waves are not equal to one-another, the corresponding wave packets "diffuse" with time, unless some kind of reverse process is present.

The constant value of de Broglie wave external energy ($E_B$) in (eq. 2.4) implies the counterphase oscillations of instantaneous parameters $m_V^{+t}$ and $\mathbf{v}_{gr}^t$ in the course of time, so that their product, phase velocity $\mathbf{v}_{ph}^t$ and de Broglie wave length $\lambda_B = h/(m_V^{+t} \mathbf{v}_{gr}^t)$ all remain constant:

$$E_B^{ext} = m_V^{+t}\left(\mathbf{v}_{gr}^t \mathbf{v}_{ph}^t\right)^{ext} = h(\mathbf{v}_{ph}/\lambda_B) = \text{const} \qquad 2.8$$

In the particular case, when the external potential and kinetic energies of the de Broglie waves are equal:

$$V_B = T_k = \frac{(\hbar k)^2}{2m_V^+} = \frac{m_V^+ \mathbf{v}_{gr}^2}{2} \qquad 2.9$$

we derive from (2.7) the condition of de Broglie wave harmonization:

$$\mathbf{v}_{ph}^{ext} = \mathbf{v}_{gr}^{ext} \qquad 2.10$$

The formula for calculating phase velocity (see eq. 2.7) with constant potential $V = const$ can be generalized, as well, in the case of a spatially inhomogeneous potential $V = V(\vec{r})$, where $\vec{r}$ is the spatial coordinate:

$$2\left(\frac{\mathbf{v}_{ph}}{\mathbf{v}_{gr}}\right) - 1 = \frac{V(\vec{r})}{T_k(\vec{r})} \qquad 2.11$$

We can write down this equation as:

$$\left(\frac{\mathbf{v}_{ph}}{\mathbf{v}_{gr}}\right)^{ext} = \frac{T_k + V(\vec{r})}{2T_k(\vec{r})} = \frac{E_B}{2T_k(\vec{r})} \qquad 2.12$$

For a free particle or an ideal gas molecule, when the potential energy is equal to zero:

$$V = 0 \quad \text{and} \quad E_B = T_k \qquad 2.14$$

we derive from (2.11):

$$\mathbf{v}_{ph} = \frac{1}{2}\mathbf{v}_{gr} \qquad 2.15$$

The de Broglie formula (2.1) contains information only about the kinetic energy of an individual particle:

$$T_k = h^2/2m\lambda_B^2 = \hbar^2/2mL_B^2 \qquad 2.16$$

where the parameter $L_B = \lambda_B/2\pi$ for a particle in the field of central forces, like electrostatic and gravitational ones has a meaning of a circular orbit radius equal to radius of standing de Broglie wave. Therefore, we can term $L_B$ as the radius of de Broglie wave.

As an approximation to the harmonic oscillator the average kinetic energy of the particle is:

$$T_{\text{kin}} = V = \frac{1}{2}E_B = \frac{1}{2}kT = \frac{1}{2}m\mathbf{v}_{gr}^2 \qquad 2.17$$

For this case we can simply calculate the de Broglie wave's length as:

$$\lambda_B = h/m\mathbf{v}_{gr} = h/(kTm)^{1/2} \qquad 2.18$$

However, for real condensed matter, the harmonic approximation is not valid – formula (2.18) does not work. In the gas phase, $\lambda_B$ is usually much less than the average distance between molecules, hence the molecules can be considered as free, rather than bound.

That said, it was shown that at temperatures close to absolute zero, in samples of sodium gas,

rubidium, etc. atoms, Bose condensates also occur (Ketterle, 1995).

In liquids and solids the following condition should exist:

$$\lambda_B^{(1)} \lambda_B^{(2)} \lambda_B^{(3)} \geq V_m = V_0/N_0 \qquad 2.19$$

where $V_0$ and $N_0$ are the molar volume and Avogadro's number.

Eq.(2.19) implies that the volume of primary effectons exceeds that occupied by one molecule **of** condensed matter.

Under this condition $\lambda_B^{1,2,3} \geq (V_m)^{1/3}$, the molecules form primary effectons (coherent clusters), with properties of Bose condensates, described in Chapter 1.

The formula for external translational energy of de Broglie wave (Kaivarainen, 1989a, 1995) relates the frequency of de Broglie wave ($\omega_B$) to the product of its actual mass ($m_V^+$) and amplitude squared $A_B^2$:

$$\omega_B^{ext} = \frac{\hbar}{2m_V^+ (A_B^{ext})^2} \qquad 2.20$$

It is shown in our work (http://arxiv.org/pdf/physics/0207027), that the [corpuscle (C) ⇌ wave (W)] duality of fermions is a result of the modulation of quantum beats between the asymmetric 'actual' (torus) and 'complementary' (anti-torus) states of sub-elementary fermions and anti-fermions – by the empirical de Broglie wave frequency of these particles.

In the Hierarchic Theory of condensed matter we do not need the intrinsic mechanism of [C ⇌ W] pulsation and duality phenomena. The other book of this author with title: "Unified Theory of Bivacuum, Matter, Fields & Time Origination. New Approach to Normal & Paranormal Phenomena" submitted to NOVA Science Publishers (NY, USA) includes the duality mechanism and related problems.

There are several ways to express the total energy of the de Broglie wave (Kaivarainen, 1989a, 1995):

$$E_B^{ext} = \hbar \omega_B = \frac{\hbar^2}{2m_V^+ A_B^2} = 2m_V^+ A_B^2 \omega_B^2$$
$$E_B^{ext} = T_{kin} + V_B = m_V^+ \mathbf{v}_{gr} \mathbf{v}_{ph} = p \mathbf{v}_{ph} \qquad 2.21$$

Under the conditions of harmonization when $\mathbf{v}_{ph} = \mathbf{v}_{gr}$ and $T_{kin} = V$ the eq. (2.21) is simplified to:

$$E_B^{ext} = \frac{\hbar^2}{2m_V^+ (A_B^{ext})^2} \qquad 2.22$$

$$= 2T_{kin} = \frac{\hbar^2}{m_V^+ L_B^2} \qquad 2.23$$

One more condition of the de Broglie wave harmonization follows from eqs. 2.22 and 2.23:

$$2A_B^2 = L_B^2 \qquad 2.24$$

where $L_B = 1/k_B = \lambda_B/2\pi$, ($k_B$ is the wave number).

The constant value of the energy of the de Broglie wave allows the counterphase change of the instantaneous values $m^t$ and $(A_B^t)^2$ to be such that the product of these values remains constant:

$$m_V^{+t} (A_B^t)^2 = \text{const} \qquad 2.25$$

As $m_V^{+t} \to m_0$ decreases and $A_B^t \to A_0$ increases, the wave properties become apparent, while as $m_0 \to m_V^{+t}$ increases and $A_0 \to A_B^t$ decreases, the corpuscular properties of the de Broglie wave are manifested (Kaivarainen, 2006, http://arxiv.org/pdf/physics/0207027).

Differentiation of (2.25) gives:

$$\frac{\Delta m^t}{m^t} = -2\frac{\Delta A_B^t}{A_B^t} \qquad 2.26$$

Such types of dynamic phenomena – related to "pulsation" of de Broglie wave – are important for the realization of the wave-particle duality (Kaivarainen, 2006).

## 2.2. Parameters of de Broglie waves in condensed matter

The formulae given below allow one to calculate the frequencies of primary de Broglie waves in selected directions (1, 2, 3) in *a* and *b* states of primary effectons (translational and librational) (Kaivarainen, 1989a):

$$[v_{1,2,3}^a]_{tr,lb} = \left[\frac{v_p^{1,2,3}}{\exp(hv_p^{1,2,3}/kT) - 1}\right]_{tr,lb} \qquad 2.27$$

$$[v_{1,2,3}^b]_{tr,lb} = [v_{1,2,3}^a + v_p^{1,2,3}]_{tr,lb} \qquad 2.28$$

The most probable frequencies of photons $[v_p^{1,2,3}]_{tr,lb}$ are related to the wave numbers of the maxima of **their** corresponding bands (tr and lib) $[\tilde{v}_p^{1,2,3}]_{tr,lb}$ in oscillatory spectra:

$$[v_p^{1,2,3}]_{tr,lb} = c[\tilde{v}_p^{1,2,3}]_{tr,lb} \qquad 2.29$$

where (c) is the velocity of light. For water the most probable frequencies of IR photons, corresponding to $(a \Leftrightarrow b)_{tr}$ transitions of primary translational effectons are determined by maxima with the wave numbers: $\tilde{v}_p^{(1)} = 60 cm^{-1}$; $\tilde{v}_p^{(2)} = \tilde{v}_p^{(3)} = 190 cm^{-1}$.

The band $\tilde{v}_p^{(1)} = \tilde{v}_p^{(2)} = \tilde{v}_p^{(3)} = 700 cm^{-1}$ corresponds to the $(a \Leftrightarrow b)_{lb}$ transitions of primary librational effectons. The degeneracy of frequencies characterize the isotropy of given type of molecules dynamics.

The distribution (2.27) coincides with that of the Planck formula, for the case where:

$$v^a = \bar{n}_p v_p \qquad 2.30$$

where $\bar{n}_p = [\exp(hv_p/kT - 1)]^{-1}$ is the mean number of photons with the frequency $v_p$.

The transition $(a \to b)$ implies that $\bar{n}_p$ increases by one (1.0)

$$v^b = v^a + v_p = \bar{n}v_p + v_p = v_p(\bar{n} + 1) \qquad 2.31$$

The derivation of the formula (2.27) is based on the assumption that the $(a \Leftrightarrow b)_{1,2,3}$ transitions are analogous to the beats in a system of two weakly interacting quantum oscillators.

In such a case the frequency $(v_p^{1,2,3})$ of photons is equal to the difference between the frequencies of de Broglie waves forming primary effectons in *(b)* and *(a)* states (Grawford, 1973):

$$[v_p^{1,2,3} = v_{1,2,3}^b - v_{1,2,3}^a = \Delta v_B^{1,2,3}]_{tr,lb} \qquad 2.32$$

where $\Delta v_B^{1,2,3}$ is the most probable difference between frequencies of de Broglie waves in directions (1, 2, 3).

The ratio of concentration for de Broglie waves in a and b states $(n_B^a/n_B^b)$ is equal to the ratio of de Broglie wave periods $(T^a/T^b)$ or the inverse ratio of de Broglie wave frequencies $(v^b/v^a)$ in these states:

$$(T^a/T^b)_{1,2,3} = (v^b/v^a)_{1,2,3}.$$

On the other hand, the ratio of concentrations is determined by the Boltzmann distribution. So, we get the following formula:

$$\left(\frac{n_B^a}{n_B^b}\right)_{1,2,3} = \left(\frac{v^b}{v^a}\right)_{1,2,3} = \exp\left(\frac{hv_B^{1,2,3}}{kT}\right) = \exp\left(\frac{hv_p^{1,2,3}}{kT}\right) \qquad 2.33$$

Substituting the eq.(2.32) into (2.33) we derive the eq.(2.27), allows one to find $(v^a_{1,2,3})_{tr,lb}$ and $(v^b_{1,2,3})_{tr,lb}$ from the data of oscillation spectroscopy throughout the given temperature range.

The energies of the corresponding three de Broglie waves in (a) and (b) states ($E^a_{1,2,3}$ and $E^b_{1,2,3}$) and their sum, which determines the energy of primary effectons, as a 3D standing de Broglie waves with energies ($E^a_{ef}$ and $E^b_{ef}$) in a and b states are respectively equal to:

$$[E^a_{1,2,3} = hv^a_{1,2,3}]_{tr,lb}; \quad [E^a_{ef} = h(v^a_1 + v^a_2 + v^a_3)]_{tr,lb} \qquad 2.34$$

$$[E^b_{1,2,3} = hv^b_{1,2,3}]_{tr,lb}; \quad [E^b_{ef} = h(v^b_1 + v^b_2 + v^b_3)]_{tr,lb} \qquad 2.35$$

Consequently, the energies of quasi-particles in (a) and (b) states are determined only by the three selected coherent modes in the given directions (1,2,3). All the remaining degrees of freedom: $(3n - 3)$, where $n$ is the number of molecules forming effectons or deformons, are degenerate due to their coherence.

### 2.3. Hamiltonians of quasi-particles, introduced in Hierarchic Concept of Matter

In this approach, the quasi-particles of liquids and solids satisfying the condition (2.19) have the properties of a mesoscopic Bose condensate. The boiling temperature for liquid or melting temperature for solids are equal to the temperature of the initial state of the mesoscopic Bose condensations (mBC) of different kinds. These two phase transitions entails the assembly ⇌ disassembly of librational and translational primary effectons, respectively. This fact is confirmed by computer calculations, based on the Hierarchic Theory (see Figure 7a, and Figure 16 a and b).

Consequently, the phase transitions: [gas → liquid] and [liquid → solid] correspond to successive stages of these kinds of mBC. The first one entails the origination of the *librational primary effectons* only. The transition from the liquid to the solid phase is related to origination of translational primary effectons, as a coherent molecular clusters, accompanied also by the jump-way, abrupt increasing in the dimensions of librational effectons (Figure 16 $a, b$).

The Hamiltonian for a partially degenerate, weakly-nonideal Bose-gas can be expressed as:

$$H = E_0 + \sum_p n_p E(p) \qquad 2.36$$

where $E_0$ is the energy of the basic (primary) state for the quasi-particles in the Bose-condensate with momentum ($p = m\mathbf{v}_{gr}$) and packing number ($n_p = 0, 1, 2...$);

$$E(p) = \left[\frac{p^2 N_0}{2mV}v(p) + \frac{p^4}{2m^2}\right]^{1/2} \qquad 2.37$$

In (2.37), $E(p)$ is the energy of quasi-particles with corresponding packing numbers; $N_0 \leq N$ is the number of particles in the condensate ($N$ is the total number of particles in the system); $V$ is the volume occupied by $N$ particles of the Bose-gas; $v(p)$ is the Fourier component of the interaction potential $\Phi$ (x):

$$v(p) = \int \Phi(x) \exp(-ipx/h) dx \qquad 2.38$$

The formula (2.37) was obtained by Bogolyubov (1970).

*At small momentums* eq. (2.37) can be simplified to:

$$E(p) \approx p\mathbf{v}_s = m\mathbf{v}_{gr}\mathbf{v}_s \qquad 2.39$$

where $\mathbf{v}_s$ is the velocity of sound.

*At large momentums,* when the properties of a Bose-gas approach the ideal ones, the total energy is close to the kinetic energy:

$$E(p) \cong p^2/2m = T_{kin} \qquad 2.40$$

In the condition of *weak nonideality* of any given Bose-gas, the following holds:

$$\frac{a}{(V/n)^{1/3}} \ll 1 \qquad 2.41$$

where *(a)* is the so-called *length of scattering*, the Born approximation is valid:

$$v(p) \approx 4\pi h^2 a/m \qquad 2.42$$

The dispersion (or the "energy spectrum") of a low-density Bose-gas can be obtained with the help of the Green functions and by means of collective variables (Huang, 1964, Prokhorov, 1988).

In the general case, the dependence of quasiparticle energy on the wave vector $B(\vec{k}_B)$ is expressed through *the structure factor* $S(\vec{k})$ (Prokhorov, 1988):

$$E(\vec{k}) = \frac{\hbar^2 \vec{k}_B^2}{2m\, S(\vec{k})} \quad or: \qquad 2.43$$

$$E(\vec{k}) = \frac{\hbar^2}{2m\vec{L}_B^2} \frac{1}{S(\vec{k})} \qquad 2.43a$$

where

$$\vec{k} = \frac{1}{\vec{L}_B} = \frac{2\pi}{\vec{\lambda}_B} \quad and \quad S(\vec{k}) = \int g(x)\exp(i\vec{k}x)dx \qquad 2.44$$

$g(x)$ is the correlation function of the density.

*The structural factor* $S(\vec{k})$ can be determined from the neutron scattering data.

The formula (2.21) derived for external energy of the de Broglie wave is:

$$E_B = \frac{\hbar^2}{2mA_B^2} = m\mathbf{v}_{gr}\mathbf{v}_{ph} \qquad 2.45$$

Comparing (2.45) and (2.43), we find a new and useful relation between the most probable de Broglie wave amplitude ($A_B$), its length ($\lambda_B = 2\pi L_B$), the group and phase velocities ($\mathbf{v}_{gr}$ and $\mathbf{v}_{ph}$), the kinetic and total energies ($T_k$ and $E$) and *structural factor*. As far the amplitude of de Broglie wave squared can be presented like:

$$\vec{A}_B^2 = \vec{L}_B^2 S(\vec{k}) = (\vec{\lambda}_B/2\pi)^2\, S(\vec{k}) \qquad 2.46$$

we get for structural factor:

$$S(\vec{k}) = A_B^2/L_B^2 = T_k/E_B = \frac{\mathbf{v}_{gr}}{2\mathbf{v}_{ph}} \qquad 2.46a$$

where $T_k$ is kinetic energy of de Broglie wave:

$$T_k = \frac{p^2}{2m} = \frac{\hbar^2}{2mL_B^2}$$

The group and phase velocities ($\mathbf{v}_{gr}$ and $\mathbf{v}_{ph}$) are related to $T_k$ and $E_B$ in accordance with eq.(2.12).

The above description of the properties of primary and secondary effectons in *a* and *b* states, allows one to transform the Hamiltonian (2.36) to the following form, using the eq.(2.34, 2.35 and 2.22);

$$H_{1,2,3}^a = E_{1,2,3}^a + \sum_p n_p(E_{1,2,3}^a) \qquad 2.47$$

$$H^b_{1,2,3} = E^b_{1,2,3} + \sum_p n_p(E^b_{1,2,3}) \qquad 2.48$$

where $H^a_{1,2,3}$ and $H^b_{1,2,3}$ are the Hamiltonians of three most probable standing de Broglie waves forming the effectons in *(a)* and *(b)* states; the energies:

$$E^a_{1,2,3} = h\nu^a_{1,2,3} = m(\mathbf{v}^a_{gr}\mathbf{v}^a_{ph})_{1,2,3} \qquad 2.49$$

and

$$E^b_{1,2,3} = h\nu^b_{1,2,3} = m(\mathbf{v}^b_{gr}\mathbf{v}^b_{ph})_{1,2,3} \qquad 2.50$$

are the energies of the three de Broglie waves forming primary effectons in (a) and (b) states; $\nu^a_{1,2,3}$ and $\nu^b_{1,2,3}$ are the corresponding frequencies of de Broglie waves (eq. 2.34 and 2.35); $[P_a = m\mathbf{v}^a_{gr} = m\mathbf{v}^b_{gr} = P_b]_{1,2,3}$ are the most probable momentums of de Broglie waves, forming the effectons in a and b states, which are equal; $[\mathbf{v}^a_{ph}$ and $\mathbf{v}^b_{ph}]_{1,2,3}$ are the phase velocities of de Broglie waves for primary effectons in *(a)* and *(b)* states; $n_P = 0, 1, 2\ldots$ are the packing numbers.

The formulas (2.49) and (2.50) are similar to (2.39). For calculating the internal energy of matter it is necessary to know the mean values of the Hamiltonians ($\bar{H}^a_{1,2,3}$ and $\bar{H}^b_{1,2,3}$).

The mean de Broglie wave energy values of primary effectons are equal to their main (most probable) values ($\bar{E}^{a,b}_{1,2,3} = E^{a,b}_{1,2,3}$). Thus, the mean values of the Hamiltonians at $n_p = 0$ are equal to

$$\bar{H}^a_{1,2,3} = E^a_{1,2,3}; \qquad \bar{H}^b_{1,2,3} = E^b_{1,2,3} \qquad (n_p = 0) \qquad 2.51$$

*Now we must take into account a secondary effectons.* The effectons in $\bar{a}$ and $\bar{b}$ states as Bose-particles having the resulting momentum more than zero with the packing numbers $n^{a,b}_p \geq 1$ obey the Bose-Einstein distribution. It is known, that under the condition when $T < T_0$ ($T_0$ is the degeneration temperature), the chemical potential is close to zero ($\mu = 0$).

The mean packing numbers for $\bar{a}$ and $\bar{b}$ states are thereby expressed by the formula (1.27); the mean energies ($\bar{E}^a_{1,2,3} = h\bar{\nu}^a_{1,2,3}$ and $\bar{E}^b_{1,2,3} = h\bar{\nu}^b_{1,2,3}$) are derived using the Bose-Einstein distribution (1.21; 1.28), similar to the Planck formula with $\mu = 0$.

Finally, the averaged Hamiltonians of $(a, \bar{a})$ and $(b, \bar{b})$ states of the system containing primary and secondary effectons (translational and librational) which have the following formulae:

$$\left[\bar{H}^a_{1,2,3} = E^a_{1,2,3} + \bar{E}^a_{1,2,3} = h\nu^a_{1,2,3} + h\bar{\nu}^a_{1,2,3}\right]_{tr,lb} \qquad 2.52$$

$$\left[\bar{H}^b_{1,2,3} = E^b_{1,2,3} + \bar{E}^b_{1,2,3} = h\nu^b_{1,2,3} + h\bar{\nu}^b_{1,2,3}\right]_{tr,lb} \qquad 2.53$$

where

$$\left[\bar{\nu}^a_{1,2,3} = \frac{\nu^a_{1,2,3}}{\left[\exp(h\nu^a_{1,2,3})/kT - 1\right]} = \frac{\bar{\mathbf{v}}^a_{ph}}{\bar{\lambda}^{1,2,3}_a}\right]_{tr,lb} \qquad 2.54$$

$$\left[\bar{\nu}^b_{1,2,3} = \frac{\nu^b_{1,2,3}}{\left[\exp(h\nu^b_{1,2,3})/kT - 1\right]} = \frac{\bar{\mathbf{v}}^b_{ph}}{\bar{\lambda}^{1,2,3}_b}\right]_{tr,lb} \qquad 2.55$$

$\bar{\nu}^a_{1,2,3}$ and $\bar{\nu}^b_{1,2,3}$ are the mean frequency values of each of three types of coherent de Broglie waves forming effectons in $(\bar{a})$ and $(b)$ states; $\bar{\mathbf{v}}^a_{ph}$ and $\bar{\mathbf{v}}^b_{ph}$ are the corresponding phase velocities.

The resulting Hamiltonian for photons, which form the primary deformons and phonons which form secondary deformons, is obtained by term-wise subtraction of the formula (2.52) from the

formula (2.53):

$$|\Delta \bar{H}_{1,2,3}|_{tr,lb} = h\,|v^b_{1,2,3} - v^a_{1,2,3}|_{tr,lb} + h\,|\bar{v}^b_{1,2,3} - \bar{v}^a_{1,2,3}|_{tr,lb} =$$
$$= h(v^{1,2,3}_p)_{tr,lb} + h(v^{1,2,3}_{ph})_{tr,lb} \qquad 2.56$$

where the frequencies of the six IR photons, propagating in directions ($\pm 1, \pm 2, \pm 3$) and forming primary deformons within the volumes of their 3 standing waves superposition, are equal to:

$$(v^{1,2,3}_p)_{tr,lb} = |v^b_{1,2,3} - v^a_{1,2,3}|_{tr,lb} = (c/\lambda^{1,2,3}_p \times n)_{tr,lb} \qquad 2.57$$

where: $c$ and $n$ are the velocity of light and the refraction index of the sample; $\lambda^{1,2,3}_{ph}$ are the wavelengths of photons in directions (1, 2, 3); and the frequencies of phonons are:

$$(v^{1,2,3}_{ph})_{tr,lb} = |\bar{v}^b_{1,2,3} - \bar{v}^a_{1,2,3}|_{tr,lb} = (\mathbf{v}_s/\bar{\lambda}^{1,2,3}_{ph})_{tr,lb} \qquad 2.58$$

These are the frequencies of six phonons (translational and librational), propagating in the directions ($\pm 1, \pm 2, \pm 3$) and forming secondary acoustic deformons as a result of 3 acoustic standing waves superposition; $\mathbf{v}_s$ is a sound velocity; $\bar{\lambda}^{1,2,3}_{ph}$ are the wavelengths of phonons in three selected directions.

The corresponding energies of the photons and phonons are:

$$E^{1,2,3}_p = hv^{1,2,3}_p; \qquad \bar{E}^{1,2,3}_{ph} = h\bar{v}^{1,2,3}_{ph} \qquad 2.59$$

The formulae for the de Broglie wave lengths of primary and secondary effectons are derived from (2.27) and (2.54):

$$\lambda^{1,2,3}_a)_{tr,lb} = \lambda^{1,2,3}_b = \mathbf{v}^a_p / v^a_{1,2,3}$$
$$= (\mathbf{v}^a_p / v^{1,2,3}_p)\left[\exp(hv^{1,2,3}_p)/kT - 1\right]_{tr,lb} \qquad 2.60$$

$$\bar{\lambda}^{1,2,3}_a)_{tr,lb} = \bar{\lambda}^{1,2,3}_b = \bar{\mathbf{v}}^a_{ph}/\bar{v}^a_{1,2,3}$$
$$= (\bar{\mathbf{v}}^a_{ph}/\bar{v}^{1,2,3}_{ph})\left[\exp(h\bar{v}^{1,2,3}_{ph})/kT - 1\right]_{tr,lb} \qquad 2.61$$

The wavelengths of photons and phonons forming the *primary* and *secondary deformons* can be determined from (2.57) and (2.58) as follows

$$(\lambda^{1,2,3}_p)_{tr,lb} = (c/nv^{1,2,3}_p)_{tr,lb} = 1/(\tilde{v})^{1,2,3}_{tr,lb} \qquad 2.61a$$

where: $(\tilde{v})^{1,2,3}_{tr,lb}$ are wave numbers of the corresponding bands in the oscillatory spectra of condensed matter.

$$(\bar{\lambda}^{1,2,3}_{ph})_{tr,lb} = (\bar{\mathbf{v}}_s/\bar{v}^{1,2,3}_{ph})_{tr,lb}$$

For making calculations according to the formulae (2.59) and (2.60) it is necessary to find a way to calculate the resulting phase velocities of de Broglie waves forming primary and secondary effectons ($\mathbf{v}^a_{ph}$ and $\bar{\mathbf{v}}^a_{ph}$). This is the subject of the next section.

### 2.4. Phase velocities of de Broglie waves forming primary and secondary effectons

In crystals, three phonons with different phase velocities can propagate in three directions. In general case, two quasi-transversal waves: "fast"($\mathbf{v}^f_\perp$) and "slow" ($\mathbf{v}^s_\perp$) and one quasi-longitudinal ($\mathbf{v}_\parallel$) wave propagate in solids (Kittel, 1975, Ashkroft and Mermin, 1976).

The propagation of *transversal* acoustic waves is known to be accompanied by smaller deformations of the lattice, than that of *longitudinal* waves. The direction of latter coincide with vector of *external* momenta, exciting the acoustic waves in solid body. The thermal phonons, spontaneously originating and annihilating under conditions of heat equilibrium, may be accompanied by smaller perturbations of the structure and they can be considered as the transversal phonons.

Therefore, we assume, that in equilibrium conditions in isotropic solid body: $\mathbf{v}_\perp^f \approx \mathbf{v}_\perp^s = \mathbf{v}_{ph}^{1,2,3}$. The *resulting thermal phonons velocity* in **a** body with anisotropic properties is:

$$\mathbf{v}_s^{res} = (\mathbf{v}_\perp^{(1)} \mathbf{v}_\perp^{(2)} \mathbf{v}_\perp^{(3)})^{1/3} = \mathbf{v}_{ph} \qquad 2.62$$

In isotropic liquids, the resulting sound velocity, determined by the phase velocity of phonons, is isotropic:

$$\mathbf{v}_s^{liq} = \mathbf{v}_{ph}$$

According to our model, the resulting the velocity of sound in condensed matter is related to phase velocities of primary and secondary effectons in both - acoustic *(a)* and optic *(b)* states and phase velocity of deformons as:

$$\left[ \mathbf{v}_s = \mathbf{f}_a \mathbf{v}_{ph}^a + \mathbf{f}_b \mathbf{v}_{ph}^b + \mathbf{f}_d \mathbf{v}_{ph}^d \right]_{tr,lb} \qquad 2.63$$

$$\left[ \bar{\mathbf{v}}_s = \bar{\mathbf{f}}_a \bar{\mathbf{v}}_{ph}^a + \bar{\mathbf{f}}_b \bar{\mathbf{v}}_{ph}^b + \bar{\mathbf{f}}_d \bar{\mathbf{v}}_{ph}^d \right]_{tr,lb} \qquad 2.64$$

where: $\mathbf{v}_{ph}^a, \mathbf{v}_{ph}^b, \bar{\mathbf{v}}_{ph}^a, \bar{\mathbf{v}}_{ph}^b$ are phase velocities of the most probable (primary) and mean (secondary) effectons in the "acoustic" $(a, \bar{a})$ and "optic" $(b, \bar{b})$ states; and

$$\mathbf{v}_{ph}^d = \bar{\mathbf{v}}_{ph}^d = \mathbf{v}_s \qquad 2.65$$

are phase velocities of primary and secondary acoustic deformons, determined by velocity of sound.

Nonetheless, the $(a \to b)_{tr,lb}$ or $(b \to a)_{tr,lb}$ transitions of primary effectons are mainly related to absorption or emission of *photons,* the rate of such cycles is determined by the velocity of sound $(\mathbf{v}_s = \mathbf{v}_{ph})$. The absorption $\rightleftharpoons$ radiation of phonons with similar energy/frequency may occur during these transitions in the volume of macro- or super-deformons.

$$f_a = \frac{P_a}{P_a + P_b + P_d}; \quad f_b = \frac{P_b}{P_a + P_b + P_d}; \quad f_d = \frac{P_d}{P_a + P_b + P_d} \qquad 2.66$$

and

$$\bar{f}_a = \frac{\bar{P}_a}{\bar{P}_a + \bar{P}_b + \bar{P}_d}; \quad \bar{f}_b = \frac{\bar{P}_b}{\bar{P}_a + \bar{P}_b + \bar{P}_d}; \quad \bar{f}_d = \frac{\bar{P}_d}{\bar{P}_a + \bar{P}_b + \bar{P}_d} \qquad 2.67$$

are the probabilities of corresponding states of the primary (*f*) and secondary quasi-particles; $P_a, P_b, P_d$ and $\bar{P}_a, \bar{P}_b, \bar{P}_d$ - relative probabilities of excitation (thermal accessibilities) of the primary and secondary effectons and deformons (see eqs. 4.10, 4.11, 4.18, 4.19, 4.26 and 4.27).

Using eq. (2.12) it is possible to express the phase velocities in $b$ and $\bar{b}$ states of effectons ($\mathbf{v}_{ph}^b$ and $\bar{\mathbf{v}}_{ph}^b$) via ($\mathbf{v}_{ph}^a$ and $\bar{\mathbf{v}}_{ph}^a$) in the following way:

$$\left[ \frac{\mathbf{v}_{ph}^b}{\mathbf{v}_{gr}^b} \right]_{tr,lb} = \left[ \frac{E_{tot}^b}{2T_k^b} \right]_{tr,lb} = \left[ \frac{h\nu_{res}^b}{m(\mathbf{v}_{gr}^b)^2} \right]_{tr,lb} \qquad 2.68$$

From this equation, we obtain for the *most probable* phase velocity in *(b)* state:

$$(\mathbf{v}_{ph}^b)_{tr,lib} = \left[ \lambda_{ph}^{res} \nu_b^{res} \right]_{tr,lib} = \left[ (\mathbf{v}_{ph}^a) \frac{\nu_b^{res}}{\nu_a^{res}} \right]_{tr,lb} \qquad 2.69$$

Keep in mind that according to proposed model, $\mathbf{v}_{gr}^b = \mathbf{v}_{gr}^a$ and $\bar{\mathbf{v}}_{gr}^b = \bar{\mathbf{v}}_{gr}^a$, *i.e.* the group velocities of both states are equal, as their momentums and kinetic energies.

In (2.69) the resulting frequencies of the most probable (primary) effectons in $b$ and $a$ states are calculated using (2.27) and (2.28) as:

$$\begin{bmatrix} v^b_{\text{res}} = (v^b_1 v^b_2 v^b_3)^{1/3} \\ v^a_{\text{res}} = (v^a_1 v^a_2 v^a_3)^{1/3} \end{bmatrix}_{tr,lb} \qquad 2.70$$

Likewise (2.69) for the *mean* phase velocity, in the $\bar{b}$-state of effectons we have:

$$(\bar{\mathbf{v}}^b_{ph})_{tr,lb} = \left[ \left( \bar{\mathbf{v}}^a_{ph} \right) \frac{\bar{v}^{\text{res}}_b}{\bar{v}^{\text{res}}_a} \right]_{tr,lb} \qquad 2.71$$

The mean resulting frequencies of secondary effectons in $\bar{a}$ and $\bar{b}$ states:

$$\left[ \bar{v}^b_{\text{res}} = (\bar{v}^b_1 \bar{v}^b_2 \bar{v}^b_3)^{1/3} \right]_{tr,lb} \qquad 2.72$$

$$\left[ \bar{v}^a_{\text{res}} = (\bar{v}^a_1 \bar{v}^a_2 \bar{v}^a_3)^{1/3} \right]_{tr,lb} \qquad 2.73$$

can be evaluated according to eqs. (2.55 and 2.54).

Using eqs. (2.63 and 2.69), we find the formulas for the *resulting phase velocities* of the primary translational and librational effectons in the *(a)* state, involved in (eq. 2.69):

$$\left( \mathbf{v}^a_{ph} \right)_{tr,lb} = \left[ \frac{\mathbf{v}_s \frac{(1-f_d)}{f_a}}{1 + \frac{P_b}{P_a}\left(\frac{v^b_{\text{res}}}{v^a_{\text{res}}}\right)} \right]_{tr,lb} \qquad 2.74$$

Similarly, the resulting phase velocity of secondary effectons in the *($\bar{a}$)* state is obtained from eqs. (2.64) and (2.70), as:

$$\left( \bar{\mathbf{v}}^a_{ph} \right)_{tr,lb} = \left[ \frac{\mathbf{v}_s \frac{(1-\bar{f}_d)}{\bar{f}_a}}{1 + \frac{\bar{P}_b}{\bar{P}_a}\left(\frac{\bar{v}^b_{\text{res}}}{\bar{v}^a_{\text{res}}}\right)} \right]_{tr,lb} \qquad 2.75$$

As will be shown below, it is necessary to know these phase velocities: $\mathbf{v}^a_{ph}$ and $\bar{\mathbf{v}}^a_{ph}$ to determine the *concentration* of the primary and secondary effectons.

When the values of the resulting phase velocities in $a$ and $\bar{a}$ states of effectons: $\left(\mathbf{v}^a_{ph}\right)_{tr,lb}$ and $\left(\bar{\mathbf{v}}^a_{ph}\right)_{tr,lb}$ are known, then, from eqs. (2.69) and (2.70) it is easy to express the resulting phase velocities in $b$ and $\bar{b}$ states of translational and librational effectons.

The $(\bar{a} \Leftrightarrow \bar{b})_{tr,lb}$ frequency of transitions of secondary effectons is less than the frequency of $(a \Leftrightarrow b)_{tr,lb}$ the transitions of primary effectons – as revealed by the computer calculations (compare Figure 19, 20 and Figure 21):

$$\left( \bar{v}^{1,2,3}_{ph} = |\bar{v}_b - \bar{v}_a|^{1,2,3} < v^{1,2,3}_p = |v_b - v_a|^{1,2,3} \right)_{tr,lb} \qquad 2.76$$

The equations (2.63 - 2.65) are valid as long as the frequency of the velocity of sound, excited by *external momentums* ($v^{\text{ex}}_s$) is much less than the frequencies of $(\bar{a} \Leftrightarrow \bar{b})_{tr,lb}$ transitions of the secondary effectons $(\bar{v}_{ph})_{tr,lb}$. When the frequency of externally generated sound ($v^{\text{ex}}_s$), (*i.e.* "induced phonons") becomes comparable to the *resulting frequency* of secondary deformons, we obtain:

$$v^{\text{ex}}_s \to \bar{v}_d = (v^{(1)}_{ph} v^{(2)}_{ph} v^{(3)}_{ph})^{1/3}$$

This implies the attainment of a "resonance condition": $v^{\text{ex}}_s \cong (\bar{v}_d)_{tr,lb}$.

Under this condition $(\bar{a} \Leftrightarrow \bar{b})$ transitions are stimulated by high frequency external momentums $(v^{\text{ex}}_s)$ and maximum absorption of the external field should occur. It is shown by calculations (Figure

2) that the energy of the "acoustic" state of secondary effectons is higher than that of the "optic" state – in contrast to the situation with primary effectons:

$$\bar{E}_a - \bar{E}_b = h(\bar{\nu}_a - \bar{\nu}_b) \quad \text{where:} \qquad \qquad 2.77$$

$$\bar{E}_a - \bar{E}_b > 0 \qquad \qquad 2.77a$$

This implies means that the pumping of external acoustic field under the condition of resonance, the population of $(\bar{a})_{tr,lb}$ states of secondary effectons *increases* and that of $(\bar{b})_{tr,lb}$ states *decreases*. The equilibrium $(\bar{a} \Leftrightarrow \bar{b})_{tr,lb}$ should then shift to the left. Consequently the fraction $\bar{f}_a$ in (eq.2.64) increases. Because the phase velocity $(\bar{a})$ states are higher than that in *(b)* states for secondary effectons (as follows from this theory) the leftward equilibrium shift must lead to an increase of resulting (experimental) velocity of sound. This prediction of our theory can be vitrified experimentally.

The changes of $\bar{f}_a$ and $\bar{f}_b$ under the conditions of acoustic resonance may be calculated from the shifts of translational or librational bands in oscillatory spectra under the action of the high-frequency external acoustic field. Similar optoacoustic resonance effects can be induced by the presence of an external electromagnetic field, acting on the dynamic equilibrium $[a \rightleftharpoons b]$ of primary effectons.

# Chapter 3

# Concentrations and Properties of Quasi-Particles, Introduced in the Hierarchic Theory of Condensed Matter

It has been shown by Raleigh that the concentration of standing waves (of any type) with wave lengths within the range: $\lambda$ to $\lambda + d\lambda$ is equal to:

$$n_\lambda d\lambda = \frac{4\pi d\lambda}{\lambda^4} \qquad 3.1$$

Expressing the wave lengths via their frequencies and phase velocities $\lambda = \mathbf{v}_{ph}/\nu$ we get:

$$n_\nu d\nu = 4\pi \frac{\nu^2 d\nu}{\mathbf{v}_{ph}^2} \qquad 3.2$$

The concentration of standing waves in a frequency range from 0 up to any other frequency, for example, to the most probable ($\nu_a$) frequency of de Broglie wave, can be found as a result of integrating eq. (3.2):

$$n_a = \frac{4\pi}{\mathbf{v}_{ph}^3} \int_0^{\nu_a} \nu^2 d\nu = \frac{4}{3}\pi \left(\frac{\nu^a}{\mathbf{v}_{ph}}\right)^3 = \frac{4}{3}\pi \frac{1}{\lambda_a^3} \qquad 3.3$$

Jeans has shown that each standing wave formed by photons or phonons can be polarized twice. Taking into account this fact, the concentrations of standing photons and standing phonons in the all three directions (1,2,3) are equal, correspondingly, to:

$$n_p^{1,2,3} = \frac{8}{3}\pi \left(\frac{\nu_{1,2,3}^p}{c^{1,2,3}/n}\right)^3$$

$$\bar{n}_{ph}^{1,2,3} = \frac{8}{3}\pi \left(\frac{\bar{\nu}_{1,2,3}^{ph}}{\mathbf{v}_{ph}^{1,2,3}}\right)^3 \qquad 3.4$$

where: $c$ and $n$ are the speed of light in a vacuum and the refraction index of the sample; $\mathbf{v}_{ph} = \mathbf{v}_s$ is velocity of the sample's thermal phonons.

The standing de Broglie waves of atoms and molecules have only one linear polarization in the set of directions (1, 2, 3). Therefore, their concentrations are described by equation (3.3).

According to our model (See the Introduction and Chapter 1), superposition of three standing de Broglie waves of different orientation in space (1, 2, 3) forms the *effectons*. They are subdivided to most probable (primary) (with zero resulting momentum) and mean (secondary) effectons. The quasi-particles, formed by 3D superposition of standing photons and phonons, originating in the course of ($a \Leftrightarrow b$) and ($\bar{a} \Leftrightarrow \bar{b}$) transitions of the primary and secondary effectons, respectively, are termed primary and secondary *deformons* (See Table 1 in the Introduction).

Effectons and deformons (primary and secondary) are the result of thermal *translations (tr)* and *librations (lb)* of molecules in directions (1, 2, 3) and transitions between their acoustic ($a, \bar{a}$) and optic ($b, \bar{b}$) states. These quasi-particles may generally be approximated by a parallelepiped with symmetry axes (1, 2, 3).

*The three corresponding standing waves participate in the formation of each effecton and deformon.* This implies that the concentration of such quasi-particles must be three times lower, than the concentration of standing waves expressed by eqs. (3.3) and (3.4), respectively. Finally, we obtain the number of important parameters, presented below, which can be calculated.

### 3.1. **The concentration of primary effectons, primary transitons and convertons**

This concentration can be calculated as:

$$\left( n_{ef} \right)_{tr,lb} = \frac{4}{9}\pi \left( \frac{v^a_{res}}{\mathbf{v}^a_{ph}} \right)^3_{tr,lb} = n_t = n_c$$

where the resulting frequency of a-state of the primary effecton is

$$v^a_{res} = (v^a_1 v^a_2 v^a_3)^{1/3}_{tr,lb} \qquad 3.6$$

$v^a_1, v^a_2, v^a_3$ are the most probable frequencies of de Broglie waves in *a*-state in directions (1,2,3), which are calculated according to formula (2.27); $\mathbf{v}^a_{ph}$ - the resulting phase velocity of effectons in *a*-state, which corresponds to eq. (2.74).

### 3.2 The concentration of secondary (mean) effectons and secondary transitons

This concentration or density may be expressed in the same way as eq. (3.5):

$$(\bar{n}_{ef})_{tr,lb} = \frac{4}{9}\pi \left( \frac{\bar{v}^a_{res}}{\bar{\mathbf{v}}^a_{ph}} \right)^3_{tr,lb} = n_t \qquad 3.7$$

where *phase velocity* $\bar{\mathbf{v}}^a_{ph}$ corresponds to eq. (2.75); the resulting frequency of the mean de Broglie waves in $\bar{a}$-state of secondary (mean) effectons, eq. (2.73) is defined as:

$$\bar{v}^a_{res} = (\bar{v}^a_1 \bar{v}^a_2 \bar{v}^a_3)^{1/3} \qquad 3.8$$

The mean values $\bar{v}^a_{1,2,3}$ are found by applying the formula (2.54).

The *maximum concentrations* of the most probable and mean effectons ($n^{max}_{ef}$) and ($\bar{n}^{max}_{ef}$), as well as corresponding concentrations of transitons ($n^{max}_t$) and ($\bar{n}^{max}_t$) follow from the requirement that it should not be higher than the concentration of atoms.

If a molecule or *elementary cell* consists of [q] *atoms*, which have their own degrees of freedom and corresponding momentums, then

$$n^{max}_{ef} = n^{max}_t = \bar{n}^{max}_{ef} = \bar{n}^{max}_t = q\frac{N_0}{V_0} \qquad 3.9$$

For example for water molecule $q = 3$.

### 3.3. The concentration of the electromagnetic primary deformons

From eq. (3.4) we have:

$$\left( n_d \right)_{tr,lb} = \frac{8}{9}\pi \left( \frac{v^{res}_d}{c/n} \right)^3_{tr,lb} \qquad 3.10$$

where (*c*) and (*n*) are the speed of light and refraction index of matter;

$$\left( v^{res}_d \right)_{tr,lb} = \left( v^{(1)}_p v^{(2)}_p v^{(3)}_p \right)^{1/3}_{tr,lb} \qquad 3.11$$

the resulting frequency of primary deformons, where

$$\left( v^{1,2,3}_p \right)_{tr,lb} = c \left( \tilde{v}^{1,2,3}_p \right)_{tr,lb} \qquad 3.12$$

are the most probable frequencies of photons related to translations and librations; c - the speed of light; $\tilde{v}_p$ - the wave numbers, which may be found from oscillatory spectra of matter.

### 3.4. The concentration of acoustic secondary deformons

From eq. (3.4) we have:

$$\left( \bar{n}_d \right)_{tr,lb} = \frac{8}{9}\pi \left( \frac{\bar{v}_d^{res}}{\mathbf{v}_s} \right)^3_{tr,lb} \qquad 3.13$$

where $\mathbf{v}_s$ is the velocity of sound; and

$$\left( \bar{v}_d^{res} \right)_{tr,lb} = \left( \bar{v}_{ph}^{(1)} \bar{v}_{ph}^{(2)} \bar{v}_{ph}^{(3)} \right)^{1/3}_{tr,lb} \qquad 3.14$$

is the resulting frequency of *secondary* deformons (translational and librational); in this formula:

$$\left( \bar{v}_{ph}^{1,2,3} \right)_{tr,lb} = \left| \bar{v}^a - \bar{v}^b \right|^{1,2,3}_{tr,lb} \qquad 3.15$$

where $\left( \bar{v}_{ph}^{1,2,3} \right)_{tr,lb}$ are the frequencies of secondary phonons, radiated⇌ absorbed by secondary translational and librational effectons in directions (1,2,3), as a result of their $(\bar{a} \rightleftharpoons \bar{b})_{1,2,3}$ transitions, calculated from (2.54) and (2.55).

Since the primary and secondary deformons are the results of transitions ($a \Leftrightarrow b$ and $\bar{a} \Leftrightarrow \bar{b})_{tr,lb}$ of the primary and secondary effectons, respectively, the maximum concentrations of effectons, transitons and deformons must coincide (See eq. 3.9):

$$n_d^{max} = \bar{n}_d^{max} = n_{ef}^{max} = n_t^{max} = \bar{n}_{ef}^{max} = \bar{n}_t^{max} = q \frac{N_0}{V_0} \qquad 3.16$$

### 3.5. **The properties of macro-effectons (tr and lb)**

Macro-effectons, macro-transitons and macro-deformons are collective excitations representing the next higher level of the hierarchy in condensed matter. All these quasi-particles have the same macroscopic volume and concentration, determined by the volume $\left( V_d = 1/n_d \right)_{tr,lb}$ of primary electromagnetic deformons (tr or lb). Consequently, the concentrations of macro-effectons and macro-transitons are equal to (3.10):

$$\left[ n_M = n_M^t = n_d = \frac{8}{9}\pi \left( \frac{v_d}{c/n} \right)^3 \right]_{tr,lb} \qquad 3.17$$

In the A-state of macro-effectons the dynamics of *(a)* and (ā) states of primary and secondary effectons are *correlated.* The relative probability is given as:

$$\left[ P_M^A = P_a \bar{P}_a \right]_{tr,lb} \qquad 3.18$$

where: $P_a$ and $\bar{P}_a$ are thermo-accessibilities of *(a)* and *(ā)* states of the effectons (see eq. 4.10 and 4.18).

The relative probability of coherent (B) state of the macro-effectons can be expressed in a similar way:

$$\left[ P_M^B = P_b \bar{P}_b \right]_{tr,lb} \qquad 3.19$$

where $P_b$ and $\bar{P}_b$ can be calculated from eqs. (4.11) and (4.19). The energies of (A) and (B) states of macro-effectons (tr and lb) are introduced as:

$$\left[ E_M^A = -kT \ln P_M^A \right]_{tr,lb} \qquad 3.20$$

$$\left[ E_M^B = -kT \ln P_M^B \right]_{tr,lb} \qquad 3.21$$

These energies can be calculated from the data of oscillatory spectra of liquids or solids.

The dimensions of macro-effectons and Macrocon-deformon (macro-transitons) are much bigger than that of the primary effectons.

### 3.6. **The properties of macro-deformons and macro-transitons (tr and lb)**

Macro-deformons and macro-transitons represent fluctuations of concentration in the volume of macro-effectons, during their $(A \Leftrightarrow B)_{tr,lb}$ transitions corresponding to coherent $[a \Leftrightarrow b$ and $\bar{a} \Leftrightarrow \bar{b}]_{tr,lb}$ transitions of primary and secondary effectons. In the framework of the current model, the notions of macro-deformons and macro-transitons coincide; their volumes coincide as well. The volume and concentration of these particles is equal to those of the macro-effectons (3.17):

$$\left( n_M^D = n_M^T = n_M = n_d \right)_{tr,lb} \qquad 3.22$$

The relative excitation probability of macro-transitons (tr and lb) is determined from the product equations (3.18) and (3.19):

$$\left( P_M^T = P_M^D = P_M^A P_M^B \right)_{tr,lb} \qquad 3.23$$

The energy of these types of hierarchic quasi-particles: macro-deformons or macro-transitons (**tr** and **lb**) is equal to:

$$\left[ E_D \equiv E_T = -kT \ln P_M^D \right]_{tr,lb} \qquad 3.24$$

### 3.7. **The properties of super-effectons**

The *super-effectons* represent the highest and most complicated level in the hierarchical structure of condensed matter. This level can be characterized organized chaos.

*Super-effectons* correspond to such type of collective dynamic excitation, when correlation emerges between both: translational and librational *macro-effectons* in their $(A_{tr}, A_{lb})$ and $(B_{tr}, B_{lb})$ states.

The relative probabilities of such conditions can be calculated from (3.18) and (3.19) as:

$$P_S^{A^*} = \left( P_M^A \right)_{tr} \left( P_M^A \right)_{lb} \qquad 3.25$$

$$P_S^{B^*} = \left( P_M^B \right)_{tr} \left( P_M^B \right)_{lb} \qquad 3.26$$

Corresponding energies of excitations of $A^*$ and $B^*$ states of super-effectons are:

$$E_S^{A^*} = -kT \ln P_S^{A^*} \qquad 3.27$$

$$E_S^{B^*} = -kT \ln P_S^{B^*} \qquad 3.28$$

The concentration of super-effectons can be expressed by the concentrations of translational and librational macro-effectons, taking into account the their excitation probabilities as follows:

$$n_S = \frac{\left( P_M^D n_M \right)_{tr} + \left( P_M^D n_M \right)_{lb}}{\left( P_M \right)_{tr} + \left( P_M \right)_{lb}} = \frac{1}{V_s} \qquad 3.29$$

where: $V_S = 1/n_S$ is the averaged volume of super-effecton.

### 3.8. **The concentration of super-deformons (super-transitons)**

Super-transitons or super-deformons represent the large fluctuation of entropy and density in the volume of super-effectons during their $A^* \Leftrightarrow B^*$ transitions. Consequently, the concentration of super-deformons or super-transitons is equal to that of super-effectons (3.29):

$$n^S_{D^*} = n^S_{T^*} = n_s \qquad 3.30$$

The relative probability of excitation of super-deformons (super-transitons) is a product of (3.25) and (3.26):

$$P^{D^*}_S = P^{T^*}_S = P^{A^*}_S P^{B^*}_S = \left(P^D_M\right)_{tr} \left(P^D_M\right)_{lb} \qquad 3.31$$

The corresponding energy of super-deformon excitation can be calculated from (3.31) as:

$$E^{D^*}_S = 3h\nu^{D^*}_S = -kT \ln P^{D^*}_S \qquad 3.32$$

The maximum concentrations of the above presented quasi-particles are determined only by their volumes. For calculation their *real* concentration in condensed matter we should take into account the probabilities of their excitation.

### 3.9. The properties of convertons and related collective excitations

The *convertons* represent reversible transitions between different *a* and *b* states of *primary librational and translational effectons*. The volume of translational effectons, is much less than the volume of librational effectons; thus, the convertons could be considered as a [dissociation ⇌ association] processes of the coherent clusters: primary librational effectons. Transitions between their acoustic states $[lb \rightleftharpoons tr]^a$ are termed ($a_{lb}/a_{tr}$)-convertons and those between optic *b* – states $[lb \rightleftharpoons tr]^b$, are termed ($b_{lb}/b_{tr}$)-convertons. Simultaneous excitations of the *a* – convertons (*acon*) and *b* –convertons (*bcon*) in the same volume of primary librational effecton are named *Macro-convertons (cMt)*. The concentration of each of these quasi-particles ($n_{con}$) is assumed to be equal to concentration of primary librational effectons (eq.3.5):

$$n_{con} = n_{acon} = n_{bcon} = n_{cMt} = (n_{ef})_{lb} \qquad 3.33$$

### 3.10. The concentrations of acoustic deformons, excited by convertons

The concentrations of *con-deformons,* representing 3D standing phonons, excited by ($\mathbf{a}_{lb}/\mathbf{a}_{tr}$) - convertons and by ($\mathbf{b}_{lb}/\mathbf{b}_{tr}$)-convertons, named, correspondingly as (*cad*) and (*cbd*) are:

$$\left(n\right)_{cad,cbd} = \frac{8}{9}\pi \left(\frac{\nu_{ac,bc}}{\mathbf{v}_s}\right)^3 \qquad 3.34$$

where $\mathbf{v}_s$ is the velocity of sound and

$$\nu_{ac} = (\nu^a_{ef})_{lb} - (\nu^a_{ef})_{tr}, \qquad 3.35$$

$$\nu_{bc} = (\nu^b_{ef})_{lb} - (\nu^b_{ef})_{tr} \qquad 3.35a$$

are the characteristic frequencies of *a*- and *b*-convertons. These frequencies are equal to the difference between characteristic frequencies of primary librational and translational effectons (See eqs.4.8 and 4.9) in a and b- states, correspondingly.

The concentration of *[lb/tr] macrocon-deformons*, excited by *macro-convertons* is equal to:

$$n_{cMd} = \frac{8}{9}\pi \left(\frac{\nu_{cMd}}{\mathbf{v}_s}\right)^3 \qquad 3.36$$

where: $\nu_{cMd}$ is the characteristic frequency of [lb/tr] macrocon-deformons, equal to that of Macro-convertons.

*The maximum concentration of all convertons-related excitations* is also limited by concentration of molecules (eq.3.16).

### 3.11 Some quantitative characteristics of collective excitations

The volume of any type of collective excitations ($V_i$), described above, is simply the reciprocal

value of their concentration numbers, $(n_i)$ derived above:
$$\mathbf{V}_i = (n_i)^{-1} \qquad 3.37$$

The volume occupied by one molecule is:
$$\mathbf{V}_M = V_0/N_0, \qquad 3.38$$

where $V_0$ and $N_0$ are the molar volume and Avogadro's number.
Therefore, the number of molecules forming quasiparticle is:
$$n_M = \mathbf{V}_i/\mathbf{V}_M = N_0/(V_0 n_i) \qquad 3.39$$

The length of the edge of quasi-particles (3D standing waves) approximated by a cube is equal to:
$$l_i = (\mathbf{V}_i^{1/3}) = (n_i^{-1/3}) \qquad 3.40$$

In the general case quasi-particles are approximated by a parallelepiped with edges determined by the corresponding standing wave length. For effectons of all types, the de Broglie wave length is determined as a ratio of the resulting phase velocity $(\mathbf{v}_{ph})$ to the corresponding frequencies $(v_{1,2,3})$ in selected directions of de Broglie waves:
$$\lambda^{1,2,3} = \mathbf{v}_{ph}/v_{1,2,3} \qquad 3.41$$

For primary electromagnetic deformons in condensed matter:
$$\lambda_d^{1,2,3} = \frac{c/n}{(v_p)_{1,2,3}\ _{tr,lb}} \qquad 3.42$$

where: $c$ is the velocity of light, $n$ is the refraction index of the sample, and $(v_p = c\tilde{v})_{1,2,3}$ are the photon frequencies and corresponding wave numbers in the oscillatory spectra of condensed matter.

# Chapter 4
# Hierarchic Thermodynamics

## 4.1 The internal energy of matter as a hierarchic system of quasi-particles formed by 3D standing waves

The quantum theory of **a** crystal's heat capacity leads to the following equation for the density of thermal internal energy (Ashkroft and Mermin, 1976):

$$\epsilon = \frac{1}{V} \frac{\sum_i E_i \exp(-E_i/kT)}{\sum_i \exp(-E_i/kT)} \quad \quad 4.1$$

where $V$ is the crystal volume and $E_i$ is the energy of the $i-th$ stationary state.

In the current approach, the internal energy of matter is determined by several factors: by the concentration ($n_i$) of each type of quasi-particle, by the excitation probabilities of each of quasi-particles states ($P_i$), and by the energies of the corresponding states ($E_i$). The condensed matter may be considered as a complex dynamic system consisting of a hierarchic combination of 3D standing waves of different types:

- ● - de Broglie waves
  - phonons
  - IR photons.

The dynamic equilibrium between all of these excitations (especially in liquids) is very sensitive to many external physical factors: the electric, magnetic, acoustic and gravitational fields, mechanical perturbation, etc.

*The "total partition function"*, i.e. the sum of the relative excitation probabilities of each state of the quasi-particles (or, the resulting *thermo-accessibility* ) of any given state of condensed matter, is given by

$$Z = \sum_{tr,lb} \left\{ \left( P_{ef}^a + P_{ef}^b + P_d \right) + \left( \bar{P}_{ef}^a + \bar{P}_{ef}^b + \bar{P}_d \right) + (P_M^A + P_M^B + P_M^D) \right\}_{tr,lb}$$
$$+ \left( P_{ac} + P_{bc} + P_{cMd} \right) + (P_S^A + P_S^B + P_S^{D^*}) \quad \quad 4.2$$

Here we take into account the fact that the excitation probabilities of both primary and secondary transitons, as well as those of the deformons are the same, i.e., $(P_d = P_t \, ; \, \bar{P}_d = \bar{P}_t)$ – they are related to the same processes:

$(a \Leftrightarrow b)_{tr,lb}$ and $(\bar{a} \Leftrightarrow \bar{b})_{tr,lb}$ transitions, correspondingly.

An analogous situation occur regarding the probabilities of convertons and macro-convertons, as well as the equality of probabilities of corresponding acoustic deformon ($P_{cMd}$) and macrocon-deformon ($P_{cMt}$) excitations: $P_{cMd} = P_{cMt}$. This is the reason for taking these probabilities into account only one time in the partition function.

*The final formula for the total internal energy of* ($U_{tot}$) of one mole of matter following from the Hierarchic Model, is:

$$U^{tot} = V_0 \frac{1}{Z} \sum_{tr,lb} \left\{ \left[ n_{ef} \left( P_{ef}^a E_{ef}^a + P_{ef}^b E_{ef}^b + P_t E_t \right) + n_d P_d E_d \right] \right.$$
$$\left. + \left[ \bar{n}_{ef} \left( \bar{P}_{ef}^a \bar{E}_{ef}^a + \bar{P}_{ef}^b \bar{E}_{ef}^b + \bar{P}_t \bar{E}_t \right) + \bar{n}_d \bar{P}_d \bar{E}_d \right] \right.$$

$$+ \left[ n_M \left( P_M^A E_M^A + P_M^B E_M^B \right) + n_D P_M^D E_M^D \right]_{tr,lb}$$

$$+ V_0 \frac{1}{Z} \left[ n_{con} \left( P_{ac} E_{ac} + P_{bc} E_{bc} + P_{cMt} E_{cMt} \right) \right.$$

$$+ \left. \left( n_{cda} P_{ac} E_{ac} + n_{cdb} P_{bc} E_{bc} + n_{cMd} P_{cMd} E_{cMd} \right) \right]$$

$$+ V_0 \frac{1}{Z} n_s \left[ \left( P_S^{A^*} E_S^{A^*} + P_S^{B^*} E_S^{B^*} \right) + n_{D^*} P_S^{D^*} E_S^{D^*} \right] \qquad 4.3$$

where all of the effectons' contributions to the total internal energy are given by

$$U_{ef} = V_0 \frac{1}{Z} \sum_{tr,lb} \left[ n_{ef} \left( P_{ef}^a E_{ef}^a + P_{ef}^b E_{ef}^b \right) \right.$$

$$+ \bar{n}_{ef} \left( \bar{P}_{ef}^a \bar{E}_{ef}^a + \bar{P}_{ef}^b \bar{E}_{ef}^b \right) + n_M \left( P_M^A E_M^A + P_M^B E_M^B \right) \left. \right]_{tr,lb}$$

$$+ V_0 \frac{1}{Z} n_s \left( P_S^{A^*} E_S^{A^*} + P_S^{B^*} P_S^{B^*} \right) \qquad 4.4$$

and, as well, all of the deformons' contributions to $U^{tot}$ are calculated as follows:

$$U_d = V_0 \frac{1}{Z} \left\{ \sum_{tr,lb} \left( n_d P_d E_d + \bar{n}_d \bar{P}_d \bar{E}_d + n_M P_M^D E_M^D \right)_{tr,lb} + n_s P_S^{D^*} E_S^{D^*} \right\} \qquad 4.5$$

The contribution, related to *lb / tr* convertons, macro-convertons and related macrocon-deformons (cMd) is calculated as follows:

$$U_{con} = V_0 \frac{1}{Z} \left[ n_{con} \left( P_{ac} E_{ac} + P_{bc} E_{bc} + P_{cMt} E_{cMt} \right) \right.$$

$$+ \left. \left( n_{cda} P_{ac} E_{ac} + n_{cdb} P_{bc} E_{bc} + n_{cMd} P_{cMt} E_{cMd} \right) \right]$$

Contributions of all types of transitons ($U_t$) also can be easily calculated.

The intramolecular configurational dynamics of molecules are automatically taken into account in this approach. These dynamics have an influence on the intermolecular dimensions, on the concentration of quasi-particles, and, as well, on the excitation energies of their states. These dynamics affect the positions of the absorption bands in oscillatory spectra and the values of the velocity of sound, which we use for calculation of the internal energy.

The remaining small contribution of the intramolecular dynamics to $U^{tot}$ is related to the oscillation energy corresponding to the fundamental molecular modes ($v_p^i$). It may be estimated using the Planck distribution:

$$U_{in} = N_0 \sum_1^i h \bar{v}_p^i = N_0 \sum_1^i h v_p^i \left[ \exp\left( h v_p^i / kT \right) - 1 \right]^{-1}$$

where $i$ is the number of internal degrees of freedom, and is calculated as follows:
$$i = 3q - 6 \text{ for nonlinear molecules;}$$
$$i = 3q - 5 \text{ for linear molecules}$$

Also note: (q) is the number of atoms forming a molecule.

It has been shown by the computer simulations for the cases of water and ice that $U_{in} \ll U^{tot}$. This condition should, in general, hold for any type of condensed matter.

**The meaning of the variables in formulae (4.2 -4.5),**

**necessary for the internal energy calculations**

$V_0$ is the molar volume;

$n_{ef}$, $\bar{n}_{ef}$ are the concentrations of primary (eq. 3.5) and secondary (eq. 3.7) effectons;

$E^a_{ef}$, $E^b_{ef}$ are the energies of the primary effectons in $a$ and $b$ states, obtained as follows:

$$\left[ E^a_{ef} = 3h\nu^a_{ef} \right]_{tr,lb} \qquad 4.6$$

$$\left[ E^b_{ef} = 3h\nu^b_{ef} \right]_{tr,lb} \qquad 4.7$$

where

$$\left[ \nu^a_{ef} = \tfrac{1}{3}\left( \nu^a_1 + \nu^a_2 + \nu^a_3 \right) \right]_{tr,lb} \qquad 4.8$$

$$\left[ \nu^b_{ef} = \tfrac{1}{3}\left( \nu^b_1 + \nu^b_2 + \nu^b_3 \right) \right]_{tr,lb} \qquad 4.9$$

These two variables ($\nu^a_{ef}$, $\nu^b_{ef}$) are the characteristic frequencies of the primary effectons in the *(a)* and *(b)* states; $\nu^a_{1,2,3}$, $\nu^b_{1,2,3}$ are determined according to formulas (2.27 and 2.28);

$P^a_{ef}$, $P^b_{ef}$ are the relative excitation probabilities (thermo-accessibilities) of effectons in (a) and (b) states. Kaivarainen (1989a) introduced them as:

$$\left[ P^a_{ef} = \exp\left(-\frac{|E^a_{ef} - E_0|}{kT}\right) = \exp\left(-\frac{3h|\nu^a_{ef} - \nu_0|}{kT}\right) \right]_{tr,lb} \qquad 4.10$$

$$\left[ P^b_{ef} = \exp\left(-\frac{|E^a_{ef} - E_0|}{kT}\right) = \exp\left(-\frac{3h|\nu^b_{ef} - \nu_0|}{kT}\right) \right]_{tr,lb} \qquad 4.11$$

where

$$E_0 = 3kT = 3h\nu_0 \qquad 4.12$$

Here, $E_0$ is the equilibrium energy of any variety of quasi-particle determined by the temperature of the sample (T):

$$\nu_0 = \frac{kT}{h} \qquad 4.13$$

Here, $\nu_0$ is the equilibrium frequency, and,

$\bar{E}^a_{ef}$, $\bar{E}^b_{ef}$ are the characteristic energies of secondary effectons in $\bar{a}$ and $\bar{b}$ states respectively:

$$\left[ \bar{E}^a_{ef} = 3h\bar{\nu}^a_{ef} \right]_{tr,lb} \qquad 4.14$$

$$\left[ \bar{E}^b_{ef} = 3h\bar{\nu}^b_{ef} \right]_{tr,lb}, \qquad 4.15$$

$$\left[ \bar{\nu}^a_{ef} = \tfrac{1}{3}\left( \bar{\nu}^a_1 + \bar{\nu}^a_2 + \bar{\nu}^a_3 \right) \right]_{tr,lb} \qquad 4.16$$

$$\left[ \bar{\nu}^b_{ef} = \tfrac{1}{3}\left( \bar{\nu}^b_1 + \bar{\nu}^b_2 + \bar{\nu}^b_3 \right) \right]_{tr,lb} \qquad 4.17$$

Again, note that $\bar{\nu}^a_{ef}$ and $\bar{\nu}^b_{ef}$ are the characteristic frequencies of the mean effectons in $\bar{a}$ and $\bar{b}$ states; the selected frequencies: $\bar{\nu}^a_{1,2,3}$, $\bar{\nu}^b_{1,2,3}$ are determined according to formulae (2.54 and 2.55).

$\bar{P}^a_{ef}$, $\bar{P}^b_{ef}$ are the relative excitation probabilities (thermal-accessibilities) of the mean effectons in $\bar{a}$

and $\bar{b}$ states, introduced as:

$$\left[ P^a_{ef} = \exp\left( -\frac{|\bar{E}^a_{ef} - E_0|}{kT} \right) = \exp\left( -\frac{3h|\bar{v}^a_{ef} - v_0|}{kT} \right) \right]_{tr,lb} \qquad 4.18$$

$$\left[ \bar{P}^b_{ef} = \exp\left( -\frac{|\bar{E}^a_{ef} - E_0|}{kT} \right) = \exp\left( -\frac{3h|\bar{v}^b_{ef} - v_0|}{kT} \right) \right]_{tr,lb} \qquad 4.19$$

### 4.2 Parameters of deformons (**primary and secondary**):

$n_d$, $\bar{n}_d$ are the concentrations of primary (eq. 3.10) and secondary (eq. 3.13) deformons;
$E_d$, $\bar{E}_d$ are the characteristic energies of the *primary* and *secondary* deformons, equal to the energies of primary and secondary transitons:

$$\left[ E_d = 3hv^{res}_d = E_t \right]_{tr,lb} \qquad 4.20$$

$$\left[ \bar{E}_d = 3h\bar{v}^{res}_d = \bar{E}_t \right]_{tr,lb} \qquad 4.20$$

where these characteristic frequencies of the primary and secondary deformons are equal to:

$$\left[ v^{res}_d = \tfrac{1}{3}\left( v^{(1)}_p + v^{(2)}_p + v^{(3)}_p \right) \right]_{tr,lb} \qquad 4.22$$

$$\left[ \bar{v}^{res}_d = \tfrac{1}{3}\left( \bar{v}^{(1)}_{ph} + \bar{v}^{(2)}_{ph} + \bar{v}^{(3)}_{ph} \right) \right]_{tr,lb} \qquad 4.23$$

The frequencies of the primary photons are calculated from the experimental data of oscillatory spectra using (eq. 3.12).

The frequencies of *secondary phonons*, on the other hand, are calculated as:

$$\left( \overline{v^{1,2,3}_{ph}} \right)_{tr,lb} = |\bar{v}_a - \bar{v}_b|^{1,2,3}_{tr,lb} \qquad 4.24$$

where $\overline{v^{1,2,3}_a}$ and $\overline{v^{1,2,3}_b}$ are in accordance with (eq. 2.54) and (eq. 2.55).

$P_d$ and $\bar{P}_d$ are the relative probabilities of primary and secondary deformons excitation, introduced as the probabilities of intermediate transition states of primary and secondary effectons, correspondingly:

$$(a \Leftrightarrow b)_{tr,lb} \text{ and } (\bar{a} \Leftrightarrow \bar{b})_{tr,lb} :$$

$$\left( P_d = P^a_{ef} P^b_{ef} \right)_{tr,lb} \qquad 4.25$$

$$\left( \bar{P}_d = \bar{P}^a_{ef} \bar{P}^b_{ef} \right)_{tr,lb} \qquad 4.26$$

### 4.3 Parameters of transitons

The quantities $(n_t)_{tr,lb}$ and $(\bar{n}_t)_{tr,lb}$ are concentrations of the primary and secondary transitons, equal to concentration of primary (3.5) and secondary (3.7) effectons:

$$(n_t = n_{ef})_{tr,lb}; \quad (n_t = n_{ef})_{tr,lb} \qquad 4.27$$

Again, $(P_t$ and $\bar{P}_t)_{tr,lb}$ are the relative excitation probabilities of the primary and secondary transitons, equal to that of primary and secondary deformons:

$$(P_t = P_d)_{tr,lb}; \quad (\bar{P}_t = \bar{P}_d)_{tr,lb} \qquad 4.27a$$

Here, $(E_t \text{ and } \bar{E}_t)_{tr,lb}$ are the energies of the primary and secondary transitons:

$$\left[ E_t = E_d = h(v_p^{(1)} + v_p^{(2)} + v_p^{(3)}) \right]_{tr,lb} \qquad 4.28$$

$$\left\{ \bar{E}_t = \bar{E}_d = 3h \left[ \, | \bar{v}_{ef}^a - v_0 | + | \bar{v}_{ef}^b - v_0 | \, \right]^{1,2,3} \right\}_{tr,lb} \qquad 4.29$$

Primary and secondary deformons, in contrast to transitons, represent the quasi-elastic mechanism of the effectons interaction via the medium.

### 4.4 Parameters of macro-effectons:

$(n_M)_{tr,lb}$ is the concentration of macro-effectons from (eq. 3.17);

$(E_M^A \text{ and } E_M^B)_{tr,lb}$ are the energies of A and B states of macro-effectons, corresponding to (eq. 3.20) and (eq. 3.21);

$(P_M^A \text{ and } P_M^B)_{tr,lb}$ are the relative excitation probabilities of A and B states of macro-effectons corresponding to (eq. 3.18) and (eq. 3.19).

### 4.5 Parameters of macro-deformons:

$(n_M^D)_{tr,lb}$ is the concentration of macro-deformons equal to that of macro-effectons (macro-transitons) corresponding to eq.(eq. 3.22);

$(E_D^M)_{tr,lb}$ are the energies of macro-deformons (eq.3.24);

$(P_M^D)_{tr,lb}$ are the excitation probabilities of macro-deformons (eq.3.23).

### Parameters of convertons and related excitations

The frequency and energy of $(a_{lb} \rightleftharpoons a_{tr})$-convertons:

$$v_{ac} = | (v_{ef}^a)_{lb} - (v_{ef}^a)_{tr} |; \qquad E_{ac} = 3hv_{ac}$$

where $(v_{ef}^a)_{lb}$ and $(v_{ef}^a)_{tr}$ correspond to (eq. 4.8).

The frequency and energy of $(b_{lb} \rightleftharpoons b_{tr})$- convertons:

$$v_{bc} = | (v_{ef}^b)_{lb} - (v_{ef}^b)_{tr} |; \qquad E_{bc} = 3hv_{bc}$$

where $(v_{ef}^b)_{lb}$ and $(v_{ef}^b)_{tr}$ correspond to (eq. 4.9).

Probabilities of $(a_{lb} \rightleftharpoons a_{tr})$ and $(b_{lb} \rightleftharpoons b_{tr})$ convertons, equal to that of corresponding acoustic [lb/tr] con-deformons excitations:

$$\begin{pmatrix} P_{ac} = (P_{ef}^a)_{tr} (P_{ef}^a)_{lb} \\ P_{bc} = (P_{ef}^b)_{tr} (P_{ef}^b)_{lb} \end{pmatrix} \qquad 4.29a$$

*Probability and energy of Macro-convertons excitation [simultaneous excitation of $(a_{lb} \rightleftharpoons a_{tr})$ and $(b_{lb} \rightleftharpoons b_{tr})$ convertons, equal to that of [lb/tr] macrocon-deformons excitation is:*

$$P_{cMd} = P_{ac} P_{bc}; \qquad 4.29b$$

$$E_{cMt} = E_{cMd} = -kT P_{cMd} \qquad 4.29c$$

The characteristic frequency of macro-convertons and [lb/tr] macrocon-deformons is:

$$v_{cMt} = v_{cMd} = E_{cMd}/3h$$

The concentrations of convertons $(n_{con})$ and macro-convertons $(n_{cMd})$ are equal to those of primary librational effectons $(n_{ef})_{lb}$.

The concentration of acoustic $(a_{lb} \rightleftharpoons a_{tr})$ and $(b_{lb} \rightleftharpoons b_{tr})$, [lb/tr] con-deformons and macrocon-deformons, both originated and annihilated, as a result of convertons and macro-convertons excitation of acoustic waves (thermal phonons) in the surrounding medium, was given in Chapter 3

eqs.(3.34; 3.36).

### 4.6  Parameters of super-effectons:

$n_s$ is the concentration of super-effectons, corresponding to (eq. 3.29);
$E_s^{A^*}$ and $E_s^{B^*}$ are the energies of A and B states of super-effectons from (eq. 3.27) and (eq. 3.28);
$P_S^{A^*}$ and $P_S^{B^*}$ are the relative excitation probabilities of $A^*$ and $B^*$ states from (eq. 3.25) and (eq. 3.26).

### 4.7  Parameters of super-deformons:

$n_{D^*}$ is the concentration of super-deformons, corresponding to (eq. 3.30);
$E_S^{D^*}$ is the energy of super-deformons from (eq. 3.32);
$P_S^{D^*}$ is the relative probability of super-deformons from (eq. 3.31).

Substituting the parameters of quasi-particles, calculated in this way into eqs. (4.2 and 4.3), we obtain the total internal energy of one mole of matter in a solid or liquid phase. For water and ice the theoretical results coincide with experimental values fairly well (See Fig. 4).

The equations are the same for solid and liquid states (crystals, glasses, amorphous matter, dilute solutions and colloid systems). The difference in the primary experimental parameters, such as density, the velocity of sound, refraction index, positions of translational and librational bands determines the difference of more than 300 theoretical parameters of any given state of condensed matter. Most of them are hidden, *i.e.* inaccessible for current experimental evaluation.

### 4.8  Contributions of kinetic and potential energy to the total internal energy

The total internal energy of matter ($U^{\text{tot}}$) is equal to the sum of total kinetic ($T^{\text{tot}}$) and total potential ($V^{\text{tot}}$) energy:

$$U^{\text{tot}} = T^{\text{tot}} + V^{\text{tot}} \qquad 4.30$$

Proceeding from eq. (2.22), the kinetic energy of de Broglie waves $[T_B]$ of one molecule may be expressed using its total energy ($E_B$), mass ($m$), and phase velocity ($\mathbf{v}_{ph}$):

$$T_B = \frac{m\mathbf{v}_{gr}^2}{2} = \frac{E_B^2}{2m\mathbf{v}_{ph}^2} \qquad 4.31$$

The total mass ($M_i$) of 3D standing de Broglie waves forming effectons, transitons and deformons of different types are proportional to number of molecules in the volume of corresponding quasi-particle ($V_i = 1/n_i$):

$$M_i = \frac{1/n_i}{V_0/N_0} m \qquad 4.32$$

The limiting condition for minimum mass of quasi-particle is:

$$M_i^{\min} = m \qquad 4.33$$

Consequently the kinetic energy of each coherent effectons is equal to

$$\left[ T_{\text{kin}}^i = \frac{E_i^2}{2M_i \mathbf{v}_{ph}^2} \right] \qquad 4.34$$

where $E_i$ is a total energy of any given quasi-particle.

The kinetic energy of coherent primary and secondary deformons and transitons is expressed analogously to (eq. 4.34), but instead of the phase velocity of de Broglie waves we use the speed of light (for primary electromagnetic deformons) and the resulting the velocity of sound $\mathbf{v}^{\text{res}}$ (eq.2.62a) for secondary acoustic deformons, respectively:

$$\left[ T^i_{\text{kin}} = \frac{E_i^2}{2M_i c^2} \right]_d \quad \text{and} \quad \left[ \bar{T}^i_{\text{kin}} = \frac{E_i^2}{2M_i (\mathbf{v}_s^{\text{res}})^2} \right]_{\bar{d}} \qquad 4.35$$

The kinetic energies of $[tr/lb]$ convertons:

$$\left[ T^i_{\text{kin}} = \frac{(E_i/3)^2}{2M_i (\mathbf{v}_s^{\text{res}})^2} \right] = \left[ T^i_{\text{kin}} = \frac{E_i^2}{6M_i (\mathbf{v}_s^{\text{res}})^2} \right]_{\text{con}}$$

According to the model, the kinetic energies of the effectons in a and b and also in the $\bar{a}$ and $\bar{b}$ states are equal. Using (eq. 4.34) and (eq. 4.35) we obtain from (eq. 4.3) the total thermal kinetic energy for one mole of matter:

$$T^{\text{tot}} = V_0 \frac{1}{Z} \sum_{tr,lb} \left\{ \left[ n_{ef} \frac{\sum (E^a)^2_{1,2,3}}{2M_{ef}(\mathbf{v}_{ph}^a)^2} (P_{ef}^a + P_{ef}^b) + \bar{n}_{ef} \frac{\sum (\bar{E}^a)^2_{1,2,3}}{2M_{ef}(\bar{\mathbf{v}}_{ph}^a)^2} (\bar{P}_{ef}^a + \bar{P}_{ef}^b) \right] + \right.$$

$$+ \left[ n_t \frac{\sum (E_t)^2_{1,2,3}}{2M_t(\mathbf{v}_s^{\text{res}})^2} P_d + \bar{n}_t \frac{\sum (\bar{E}_t)^2_{1,2,3}}{2\bar{M}_t(\mathbf{v}_s^{\text{res}})^2} \bar{P}_d \right] +$$

$$+ \left[ n_d \frac{\sum (E_d)^2_{1,2,3}}{2M_d c^2} P_d + \bar{n}_d \frac{\sum (\bar{E}_d)^2_{1,2,3}}{2M_d(\mathbf{v}_s^{\text{res}})^2} \bar{P}_d \right] +$$

$$\left. + \left[ n_M \frac{\left(E_M^A\right)^2}{6M_M(\mathbf{v}_{ph}^A)^2} \left(P_M^A + P_M^B\right) + n_D \frac{\left(E_D\right)^2}{6M_D(\mathbf{v}_s^{\text{res}})^2} P_D^M \right] \right\}_{tr,lb} +$$

$$+ V_0 \frac{n_{\text{con}}}{Z} \left[ \frac{\left(E_{ac}\right)^2}{6M_c(\mathbf{v}_s^{\text{res}})^2} P_{ac} + \frac{\left(E_{bc}\right)^2}{6M_c(\mathbf{v}_s^{\text{res}})^2} P_{bc} + \frac{\left(E_{cMd}\right)^2}{6M_c(\mathbf{v}_s^{\text{res}})^2} P_{cMd} \right] +$$

$$V_0 \frac{1}{Z} n_{cda} \left[ \frac{\left(E_{ac}\right)^2}{6M_c(\mathbf{v}_s^{\text{res}})^2} P_{ac} + n_{cdb} \frac{\left(E_{bc}\right)^2}{6M_c(\mathbf{v}_s^{\text{res}})^2} P_{bc} + \frac{n_{cMd}\left(E_{cMd}\right)^2}{6M_c(\mathbf{v}_s^{\text{res}})^2} P_{cMd} \right] + \qquad 4.36$$

$$+ V_0 \frac{1}{Z} \left[ n_S \frac{(E_S^{A^*})^2}{6M_s\left(\mathbf{v}_{ph}^{A^*}\right)^2} (P_S^{A^*} + P_S^{B^*}) + n_S \frac{(E_{D^*})^2}{6M_S(\mathbf{v}_s^{\text{res}})^2} P_S^{D^*} \right]$$

where the effective phase velocity of A-state of macro-effectons is introduced as:

$$\left[ \frac{1}{\mathbf{v}_{ph}^A} = \frac{1}{\mathbf{v}_{ph}^a} + \frac{1}{\bar{\mathbf{v}}_{ph}^a} \right]_{tr,lb} \rightarrow \left[ \mathbf{v}_{ph}^A = \frac{\mathbf{v}_{ph}^a \bar{\mathbf{v}}_{ph}^a}{\mathbf{v}_{ph}^a + \bar{\mathbf{v}}_{ph}^a} \right]_{tr,lb} \qquad 4.37$$

and the effective phase velocity of super-effecton in $A^*$-state:

$$\mathbf{v}_{ph}^{A^*} = \frac{(\mathbf{v}_{ph}^A)_{tr} (\mathbf{v}_{ph}^A)_{lb}}{(\mathbf{v}_{ph}^A)_{tr} + (\mathbf{v}_{ph}^A)_{lb}} \qquad 4.38$$

Total potential energy is defined by the difference between total internal (eq. 4.3) and total kinetic energy (eq. 4.36):

$$V^{tot} = U^{tot} - T^{tot} \qquad 4.39$$

Consequently, we can separately calculate the kinetic and potential energy contributions to the total thermal internal energy of matter, using following experimental data:

1) *density or molar volume*
2) *the velocity of sound*
3) *refraction index*
4) *positions of translational and librational bands in oscillatory spectrum of condensed matter (liquid or solid).*

All these data should be obtained at the same temperature and pressure.

The contributions of all individual types of excitations to the kinetic and potential energy of matter, as well as many characteristics of these quasi-particles may also be calculated (See chapter 6).

### 4.9 Some useful parameters of condensed matter

Some of important parameters, characterizing the properties of individual collective excitations were introduced at the end of Chapter 3.

The total *Structural Factor of condensed matter* can be calculated as a ratio of the kinetic to the total energy of matter:

$$SF = T^{tot}/U^{tot} \qquad 4.40$$

The structural factors, related to the contributions of translations (*SFtr*) as well as the librations (*SFlb*) can be calculated separately as:

$$SFtr = T_{tr}/U^{tot} \quad \text{and} \quad SRlb = T_{lb}/U^{tot} \qquad 4.41$$

**The dynamic properties of quasi-particles, introduced in the Hierarchic Theory**

The frequency of c- macro-transitons or Macro-convertons excitation, representing [dissociation/association] of primary librational effectons - "flickering clusters" as a result of interconversions between primary [lb] and [tr] effectons is:

$$F_{cM} = \frac{1}{\tau_{Mc}} P_{Mc}/Z \qquad 4.42$$

where: $P_{Mc} = P_{ac} P_{bc}$ is a probability of macro-convertons excitation;
$Z$ is a total partition function (see eq.4.2) and the life-time of macro-convertons is:

$$\tau_{Mc} = (\tau_{ac} \tau_{bc})^{1/2} \qquad 4.43$$

In this formula, the cycle-periods of (ac) and (bc) convertons are determined by the sum of life-times of intermediate states of primary translational and librational effectons:

$$\tau_{ac} = (\tau_a)_{tr} + (\tau_a)_{lb};$$
$$\tau_{bc} = (\tau_b)_{tr} + (\tau_b)_{lb}; \qquad 4.44$$

The life-times of primary and secondary effectons (lb and tr) in *a*- and *b*-states are the reciprocal values of corresponding state frequencies:

$$[\tau_a = 1/\nu_a; \ \tau_{\bar{a}} = 1/\nu_{\bar{a}}]_{tr,lb}; \qquad [\tau_b = 1/\nu_b; \ \tau_{\bar{b}} = 1/\nu_{\bar{b}}]_{tr,lb} \qquad 4.45$$

$[(\nu_a) \text{ and } (\nu_b)]_{tr,lb}$ correspond to eqs. 4.8 and 4.9;
$[(\nu_{\bar{a}}) \text{ and } (\nu_{\bar{b}})]_{tr,lb}$ can be calculated using eqs.4.16; 4.17.

*4.9.1 The frequency of (ac) and (bc) convertons excitations [lb/tr]:*

$$F_{ac} = \frac{1}{\tau_{ac}} P_{ac}/Z \qquad 4.46$$

$$F_{bc} = \frac{1}{\tau_{bc}} P_{bc}/Z \qquad 4.47$$

where $P_{ac}$ and $P_{bc}$ are probabilities of corresponding convertons excitations (see eq.4.29a).

### 4.9.2 The frequency of excitations of Super-effectons and Super-transitons (super-deformons) cycle:

$$F_{SD} = \frac{1}{(\tau_{A^*} + \tau_{B^*} + \tau_{D^*})} P_S^{D^*}/Z \qquad 4.48$$

This frequency of the biggest fluctuations - cavitational ones in liquids and accompanied defects in solids is dependent on the round cycle-period of Super-effectons:

$$\tau_{SD} = \tau_{A^*} + \tau_{B^*} + \tau_{D^*} \qquad 4.48a$$

and the probability of super-deformons excitation ($P_S^{D^*}$).

The averaged life-times of super-effectons in $A^*$ and $B^*$ state in (4.48a) are dependent on similar states of both translational and librational macro-effectons:

$$\tau_{A^*} = [(\tau_A)_{tr} (\tau_A)_{lb}] = [(\tau_a \tau_{\bar{a}})_{tr} (\tau_a \tau_{\bar{a}})_{lb}]^{1/2} \qquad 4.49$$

and those in the B state:

$$\tau_{B^*} = [(\tau_B)_{tr} (\tau_B)_{lb}] = [(\tau_b \tau_{\bar{b}})_{tr} (\tau_b \tau_{\bar{b}})_{lb}]^{1/2} \qquad 4.50$$

*The life-time of Super-deformons excitation* is determined by the frequency of beats between $A^*$ and $B^*$ states of Super-effectons as:

$$\tau_{D^*} = 1/|(1/\tau_{A^*}) - (1/\tau_{B^*})| \qquad 4.51$$

### 4.9.3 The frequency of translational and librational macro-effectons and macro-transitons (deformons) cycle excitations:

It can be defined in a similar way as for Super-excitations:

$$\left[ F_M = \frac{1}{(\tau_A + \tau_B + \tau_D)} P_M^D/Z \right]_{tr,lb} \qquad 4.52$$

where

$$(\tau_A)_{tr,lb} = [(\tau_a \tau_{\bar{a}})_{tr,lb}]^{1/2} \qquad 4.53$$

and

$$(\tau_B)_{tr,lb} = [(\tau_b \tau_{\bar{b}})_{tr,lb}]^{1/2} \qquad 4.54$$

$$(\tau_D)_{tr,lb} = 1/|(1/\tau_A) - (1/\tau_B)|_{tr,lb} \qquad 4.55$$

### 4.9.4 The frequency of primary translational and librational effectons and transitons cycle excitations

This frequency of $(a \rightleftharpoons b)_{tr}$ cycles of translational effectons, involving corresponding transitons, can be expressed like:

$$F_{tr} = \frac{1/Z}{(\tau_a + \tau_b + \tau_t)_{tr}} (P_d)_{tr} \qquad 4.56$$

where $(P_d)_{tr}$ is a probability of primary translational deformons excitation; $(\tau_a, \tau_b$ and $\tau_t)_{tr}$ are the life-times of *(a)* and *(b)* states of primary *translational* effectons (eq.4.45) and the transition state between them.

The formula for frequency of primary librational effectons as $(a \rightleftharpoons b)_{lb}$ cycles excitations has

similar shape:

$$F_{lb} = \frac{1/Z}{(\tau_a + \tau_b + \tau_t)_{lb}} (P_d)_{lb} \qquad 4.57$$

where $(P_d)_{lb}$ is a probability of primary librational deformons excitation; $(\tau_a, \tau_b$ and $\tau_t)_{lb}$ are the life-times of (a) and (b) states of primary librational effectons defined as (eq. 4.45) and the transition state between them.

The life-time of primary transitons (tr and lb) as a result of quantum beats between (a) and (b) states of primary effectons can be introduced as:

$$\tau_t = \frac{1}{|1/\tau_a - 1/\tau_b|_{tr,lb}} \qquad 4.58$$

### 4.9.5 The fraction of molecules (Fr) in each type of independent collective excitations (quasi-particles):

$$Fr(i) = P(i)/Z \qquad 4.59$$

where $P(i)$ is the thermo-accessibility (relative probability) of given excitation and $Z$ is the total partition function (eq. 4.2).

### 4.9.6 The number of molecules in each types (i) of collective excitations:

$$N_m(i) = Fr(i) (N_A/V_0) = [P(i)/Z] (N_A/V_0) \qquad 4.60$$

where $N_A$ and $V_0$ are the Avogadro number and molar volume of matter.

### 4.9.7 The concentration of each type (i) of collective excitation:

$$N(i) = Fr(i) n(i) = [P(i)/Z] n(i) \qquad 4.61$$

where $n(i)$ is a concentration of given type *(i)* of quasi-particles; $Fr(i)$ is a fraction of corresponding type of quasi-particles.

### 4.9.8 The average distance between centers of i-type of collective excitations:

$$d(i) = 1/[N(i)]^{1/3} = 1/[(P(i)/Z) n(i)]^{1/3} \qquad 4.62$$

### 4.9.9 The ratio of the average distance between centers of excitations to their linear dimension $[l = 1/n(i)^{1/3}]$:

$$rat(i) = 1/[(P(i)/Z)]^{1/3} \qquad 4.63$$

It is also possible to calculate many other parameters, each characterizing different properties of condensed matter, using the Hierarchic Theory and the computer program based on this theory. These calculations are the topics of the following chapters.

# Chapter 5

# Heat capacity and radiation of matter in the Hierarchic Theory

In accordance with classical thermodynamics, enthalpy ($H$) is related to the total internal energy ($U$), pressure ($P$), and volume ($V$) of matter in the following manner:

$$H = U + PV$$

The heat capacity at constant pressure ($C_P$) is equal to:

$$C_P = \left(\frac{\partial H}{\partial T}\right)_P = \left(\frac{\partial U}{\partial T}\right)_P + \left(\frac{\partial V}{\partial T}\right)_P \qquad 5.2$$

The heat capacity at constant pressure and volume ($C_V$) is calculated as:

$$C_V = (\partial U/\partial T)_{P,V} \qquad 5.3$$

Substituting (eq. 4.3) into (eq. 5.3), we obtain the sum of the contributions to $C_V$ related to each type of quasi-particles:

$$C_V = \left(\frac{\partial U_{ef}}{\partial T} + \frac{\partial U_c}{\partial T} + \frac{\partial U_t}{\partial T} + \frac{\partial U_d}{\partial T}\right)_{tr,lb} + \left(\frac{\partial U_M}{\partial T}\right)_{tr,lb} + \frac{\partial U_S}{\partial T} \qquad 5.4$$

where $U_{ef}$, $U_c$, $U_t$, and $U_d$ are the contributions to the total energy of the effectons, convertons, transitons and deformons (primary and secondary); $U_M$ is the total contribution of macro-effectons and macro-deformons; $U_S$ is the resulting contribution of super-effectons and super-deformons.

In this simplified evaluation of heat capacity, we will ignore the contributions of convertons, transitons, macro- and super-effectons, as well as the corresponding macro- and super-deformons – as these contributions are much smaller than those of the primary and secondary effectons and deformons.

By substituting the corresponding elements of (eq. 4.3) into the first term of (eq. 5.4), one can specify the contributions of primary effectons and deformons to the resulting value for heat capacity:

$$(\partial U/\partial T)_{tr,lb} = V_0 \frac{1}{2}\left\{ \frac{\partial}{\partial T}\left[ n_{ef}\left( P^a_{ef}E^a_{ef} + P^b_{ef}E^b_{ef} \right) + n_d P_d E_d \right]_{tr,lb} \right\} +$$

$$+ V_0 \frac{\partial}{\partial T} \frac{1}{2}\left[ \bar{n}_{ef}\left( \bar{P}^a_{ef}\bar{E}^a_{ef} + \bar{P}^b_{ef}\bar{E}^b_{ef} \right) + \bar{n}_d \bar{P}_d \bar{E}_d \right]_{tr,lb} \qquad 5.5$$

The second term in (eq. 5.5) is similar to the first, where, instead of the most probable (primary) parameters of the effectons and deformons, the secondary (averaged) parameters are taken into account:

$$\left[ n_{ef};\ P^a_{ef};\ P^b_{ef};\ E^a_{ef};\ E^b_{ef};\ n_d;\ P_d;\ E_d \right]_{tr,lb} \rightarrow$$

$$\left[ \bar{n}_{ef};\ \bar{P}^a_{ef};\ \bar{P}^b_{ef};\ \bar{E}^a_{ef};\ \bar{E}^b_{ef};\ \bar{n}_d;\ \bar{P}_d;\ \bar{E}_d \right]_{tr,lb} \qquad 5.6$$

**1.** *Let us try at first to express in detail the derivative of (eq. 5.5), as it relates to primary effectons:*

$$\frac{\partial}{\partial T}\left[ n_{ef}\left( P^a_{ef}E^a_{ef} + P^b_{ef}E^b_{ef} \right) \right]_{tr,lb} = \left[ \frac{\partial n_{ef}}{\partial T}\left( P^a_{ef}E^a_{ef} + P^b_{ef}E^b_{ef} \right) + \right.$$

$$+ n_{ef} \frac{\partial \left( P^a_{ef} E^a_{ef} + P^b_{ef} E^b_{ef} \right)}{\partial T} \Bigg]_{tr,lb} \qquad 5.7$$

We subdivide the differentiation of (eq. 5.7) into number of stages (1.a - 1.c):

$(1.a)$:

$$\frac{\partial n_{ef}}{\partial T} = \frac{4\pi}{9\mathbf{v}^3_{res}} \frac{\partial}{\partial T}(v^a_1 v^a_2 v^a_3)_{tr,lb} =$$

$$= \frac{4\pi}{9\mathbf{v}^3_{res}} \left[ \frac{\partial v^a_1}{\partial T} v^a_2 v^a_3 + \frac{\partial v^a_2}{\partial T} v^a_1 v^a_3 + \frac{\partial v^a_3}{\partial T} v^a_2 v^a_3 \right]_{tr,lb} \qquad 5.8$$

Since $v^a_{1,2,3}$ is determined by (eq. 2.27), then

$$\left( \frac{\partial v^a_{1,2,3}}{\partial T} \right)_{tr,lb} = \frac{\partial}{\partial T} \left( \frac{v^{1,2,3}_{ph}}{\exp(hv^{1,2,3}_{ph}/kT) - 1} \right)_{tr,lb} \qquad 5.9$$

After some substitutions in the right-hand side, we obtain

$$\left( \frac{\partial v^a_{1,2,3}}{\partial T} \right)_{tr,lb} = \left( \frac{kc^2 a}{nb^2} \right)_{tr,lb} \qquad 5.10$$

where $k, h$ are the Boltzmann and Planck constants; $a, b, c$ are dimensionless parameters:

$$a = \exp\left( \frac{hv^{1,2,3}_{ph}}{kT} \right) = \exp(c); \quad c = \frac{hv^{1,2,3}_{ph}}{kT}; \quad b = a - 1 \qquad 5.11$$

$(\mathbf{1.b})$:

$$\frac{\partial}{\partial T} \left( P^a_{ef} E^a_{ef} + P^b_{ef} E^b_{ef} \right) = 3 \Bigg[ \frac{\partial P^a_{ef}}{\partial T} hv^a_{ef} +$$

$$+ P^a_{ef} \frac{\partial (hv^a_{ef})}{\partial T} + \frac{\partial P^b_{ef}}{\partial T} hv^b_{ef} + P^b_{ef} \frac{\partial (hv^b_{ef})}{\partial T} \Bigg] \qquad 5.12$$

where

$$E^a_{ef} = 3hv^a_{ef}; \quad v^a_{ef} = \frac{1}{3}\left( v^a_1 + v^a_2 + v^a_3 \right) \qquad 5.13$$

$$E^b_{ef} = 3hv^b_{ef}; \quad v^b_{ef} = \frac{1}{3}\left( v^b_1 + v^b_2 + v^b_3 \right) \qquad 5.14$$

where $v^b_{1,2,3} = v^a_{1,2,3} + v^{1,2,3}_{ef}$;

$$\frac{\partial v^b_{1,2,3}}{\partial T} = \frac{\partial v^a_{1,2,3}}{\partial T} = \frac{\partial v^{1,2,3}_{ef}}{\partial T} \qquad 5.15$$

$P^a_{ef}, P^b_{ef}$ are the thermo-accessibilities of $a$ and $b$ states of the effecton (see 4.10 and 4.11).

$(\mathbf{1.c})$:

$$\frac{\partial P^a}{\partial T} = \frac{\partial}{\partial T} \exp\left( -\frac{3h \mid v^a_{ef} - v_0 \mid}{kT} \right) = \frac{\partial}{\partial T} \exp\left( -\left| \frac{h\left( v^a_1 + v^a_2 + v^a_3 \right)}{kT} - 3 \right| \right) =$$

$$= P_{ef}^a \left[ \frac{h\left(v_1^a + v_2^a + v_3^a\right)}{kT} - \frac{h}{kT}\left(\frac{\partial v_1^a}{\partial T} + \frac{\partial v_2^a}{\partial T} + \frac{\partial v_3^a}{\partial T}\right) \right] \qquad 5.16$$

$$(d):$$

$$\frac{\partial P_{ef}^a}{\partial T} = P_{ef}^b \left[ \frac{h\left(v_1^b + v_2^b + v_3^b\right)}{kT} - \frac{h}{kT}\left(\frac{\partial v_1^b}{\partial T} + \frac{\partial v_2^b}{\partial T} + \frac{\partial v_3^b}{\partial T}\right) \right] \qquad 5.17$$

**2.** Then we derive, in a similar way, the contribution of secondary effectons in (eq. 5.5) to the heat capacity:

$$\frac{\partial}{\partial T}\left[ \bar{n}_{ef}\left(\bar{P}_{ef}^a \bar{E}_{ef}^a + \bar{P}_{ef}^b \bar{E}_{ef}^b\right)\right]_{tr,lb} =$$

$$= \left[ \frac{\partial \bar{n}_{ef}}{\partial T}\left(\bar{P}_{ef}^a \bar{E}_{ef}^a + \bar{P}_{ef}^b \bar{E}_{ef}^b\right) + \bar{n}_{ef}\frac{\partial\left(\bar{P}_{ef}^a \bar{E}_{ef}^a + \bar{P}_{ef}^b \bar{E}_{ef}^b\right)}{\partial T} \right]_{tr,lb} \qquad 5.18$$

where

(**2.a**) : $\frac{\partial \bar{n}_{ef}}{\partial T}$ has the form of (eq. 5.8), but after substitution of $v_{1,2,3}^a$ with:

$$v_{1,2,3}^a = \frac{v_{1,2,3}^a}{\exp\left(\frac{hv_{ph}^{1,2,3}}{kT}\right) - 1} \qquad 5.19$$

and the substitution of $\left(\frac{\partial v_{1,2,3}^a}{\partial T}\right)_{tr,lb}$ with:

$$\left(\frac{\partial \bar{v}_{1,2,3}^a}{\partial T}\right)_{tr,lb} = \frac{kc^2 a}{hdb^2}\left\{\frac{1}{d}\exp\left[\frac{hv_{ph}^{1,2,3}}{kT}\left(\frac{1}{b}-1\right)\left(1-\frac{ca}{b}\right)\right] - 1\right\}, \qquad 5.20$$

where $a_{1,2,3}, b_{1,2,3}, c_{1,2,3}$ are as in (eq. 5.11)

$$d_{1,2,3} = \exp\left(\frac{c_{1,2,3}}{b_{1,2,3}}\right) - 1 \qquad 5.21$$

(**2.b**) : $\frac{\partial}{\partial T}\left(\bar{P}_{ef}^a \bar{E}_{ef}^a + \bar{P}_{ef}^b \bar{E}_{ef}^b\right)$ has the form of (eq. 5.12), after replacing $v_{1,2,3}^b$ with

$$v_{1,2,3}^b = \frac{v_{1,2,3}^b}{\exp\left(\frac{hv_{1,2,3}^b}{kT}\right) - 1} \qquad 5.22$$

and after replacing $\left(\frac{\partial v_{1,2,3}^b}{\partial T}\right)$ with

$$\frac{\partial \bar{v}_{1,2,3}^b}{\partial T} = \frac{kc^2 a}{hS}\left\{\frac{1}{b^2} + \frac{1}{S}\left(\frac{1}{b}+1\right)\exp\left(\frac{hv_{ph}^{1,2,3}}{kT}\right)[g] + \frac{\Delta v_{ph}^{1,2,3}}{S\Delta T}\right\} \qquad 5.23$$

where the following parameters are added to the previous dimensionless parameters:

$$[g]_{1,2,3} = \left[1 + \frac{1}{b} - \frac{ca}{b^2} - \frac{T}{v_{ph}^{1,2,3}}\frac{\Delta v_{ph}^{1,2,3}}{\Delta T}\right]_{tr,lb}$$

$$S_{1,2,3} = \exp\left[\frac{h\nu_{ph}^{1,2,3}\left(\frac{1}{b}-1\right)}{kT}\right]_{1,2,3} - 1$$

(**2.c**) : $\frac{\partial \bar{P}_{ef}^a}{\partial T}$ has the form of (eq. 5.16), after replacing the most probable frequencies and their derivatives with the mean values from both (eq. 5.19) and (eq. 5.20).

(**2.d**) : $\frac{\partial \bar{P}_{ef}^b}{\partial T}$ has the form of (eq. 5.17) with the substitutions: (eq. 5.22) and (eq. 5.23).

Let us focus our attention on the second term in (eq. 5.5):

$$\frac{\partial}{\partial T}\left(\frac{1}{Z}\right) = -\frac{1}{Z^2}\left(\frac{\partial P_{ef}^a}{\partial T} + \frac{\partial P_{ef}^b}{\partial T} + \frac{\partial P_d}{\partial T} + \frac{\partial \bar{P}_{ef}^a}{\partial T} + \frac{\partial \bar{P}_{ef}^b}{\partial T} + \frac{\partial \bar{P}_d}{\partial T}\right)_{tr,lb} \qquad 5.24$$

where all parameters, except $\frac{\partial P_d}{\partial T}$ and $\frac{\partial \bar{P}_d}{\partial T}$, have already been discussed above. The derivatives of deformon probabilities are given below (eqs. 5.31 and 5.36).

*3. Let us further analyze the contribution of these primary deformons to $C_V$, in the first part of formula (eq. 5.5):*

$$\frac{\partial}{\partial T}\left(n_d P_d E_d\right) = \frac{\partial n_d}{\partial T}P_d E_d + \frac{\partial P_d}{\partial T}n_d E_d + 3\frac{\partial(h\nu_d)}{\partial T}n_d P_d, \qquad 5.25$$

where $n_d = \left(8\pi/9\mathbf{v}_{res}^3\right)\left(\nu_p^{(1)}\nu_p^{(2)}\nu_p^{(3)}\right)$ are concentrations of primary deformons, see (eq. 3.10):

(**3.a**) :

$$\frac{\partial n_d}{\partial T} = \left(8\pi/9\mathbf{v}_{res}^3\right)\frac{\partial}{\partial T}\left(\nu_p^{(1)}\nu_p^{(2)}\nu_p^{(3)}\right) =$$

$$\left(8\pi/9\mathbf{v}_{res}^3\right)\left[\frac{\partial \nu_p^{(1)}}{\partial T}\nu_p^{(2)}\nu_p^{(3)} + \frac{\partial \nu_{ph}^{(2)}}{\partial T}\nu_p^{(1)}\nu_p^{(3)} + \frac{\partial \nu_{ph}^{(3)}}{\partial T}\nu_p^{(1)}\nu_p^{(2)}\right] \qquad 5.26$$

Because

$$\left(\nu_p^{1,2,3}\right)_{tr,lb} = \left(\nu_{1,2,3}^b - \nu_{1,2,3}^a\right)_{tr,lb} = c\left(\tilde{\nu}_p^{1,2,3}\right)_{tr,lb}, \qquad 5.27$$

where: $c$ is the speed of light, and, $\left(\tilde{\nu}_p^{1,2,3}\right)_{tr,lb}$ are the wave numbers of the maxima associated with translational and librational bands in the oscillation spectra of a liquid or solid body, then, it follows that

$$\left(\frac{\partial \nu_p^{1,2,3}}{\partial T}\right)_{tr,lb} = c\frac{\left(\tilde{\nu}_p^{1,2,3}\right)_{tr,lb}}{\partial T} \qquad 5.28$$

Here, $\left(\frac{\partial \nu_p^{1,2,3}}{\partial T}\right)_{tr,lb}$ are temperature shifts of the band maxima in the IR-spectra, as found from the experimental data.

The energy of the primary deformons is:

$$E_d = 3h\nu_d; \qquad \nu_d = \frac{1}{3}\left(\nu_p^{(1)} + \nu_p^{(2)} + \nu_p^{(3)}\right) \qquad 5.29$$

The excitation probability of deformons (see eq. 4.26) is:

$$P_d = P_{ef}^a P_{ef}^b = \exp\left(-\frac{3k\nu_d}{kT}\right) \qquad 5.30$$

$$\frac{\partial P_d}{\partial T} = \frac{\partial P_{ef}^a}{\partial T}P_{ef}^b + \frac{\partial P_{ef}^b}{\partial T}P_{ef}^a, \qquad 5.31$$

where $\frac{\partial P_{ef}^a}{\partial T}$ corresponds to (eq. 5.16), $\frac{\partial P_{ef}^b}{\partial T}$ corresponds to (*eq.* 5.17); $P_{ef}^a$, $P_{ef}^b$ are obtained from (eq.

4.10) and (eq. 4.11).

*4. The contribution of secondary deformons to (**eq**. 5.5) has the form of (eq. 5.26), except for the replacement of primary parameters by secondary parameters:*

$$\overline{v_{ph}^{1,2,3}} = \left| \overline{v^a} - \overline{v^a} \right|_{tr,lb}^{1,2,3} \qquad 5.32$$

where $\bar{v}^a$ and $\bar{v}^b$ correspond to (eq. 5.19) and (eq. 5.22), and $\frac{\partial v_{ph}^{1,2,3}}{\partial T}$ is substituted for

$$\left( \frac{\partial \bar{v}_{1,2,3}^a}{\partial T} - \frac{\partial \bar{v}_{1,2,3}^b}{\partial T} \right)_{tr,lb}^{1,2,3} \qquad 5.33$$

where $\frac{\partial \bar{v}_{1,2,3}^a}{\partial T}$ corresponds to (eq. 5.20), and $\frac{\partial \bar{v}_{1,2,3}^b}{\partial T}$ corresponds to (eq. 5.23).

The probability of secondary deformon excitation is found as follows:

$$\left( \bar{P}_d \right)_{tr,lb} = P_a P_b = \exp\left(-\frac{2h\bar{v}_d}{kT}\right)_{tr,lb} = \exp\left(-\frac{E_d}{kT}\right)_{tr,lb} \qquad 5.34$$

where

$$v_d = \frac{1}{3}\left( \bar{v}_{ph}^{(1)} + \bar{v}_{ph}^{(2)} + \bar{v}_{ph}^{(3)} \right)_{tr,lb}; \qquad E_d = 3hv_d; \qquad 5.35$$

$$\frac{\partial \bar{P}_d}{\partial T} = \frac{\partial \bar{P}_{ef}^a}{\partial T} P_{ef}^b + \frac{\partial \bar{P}_{ef}^b}{\partial T} P_{ef}^a \qquad 5.36$$

$\frac{\partial \bar{P}_{ef}^a}{\partial T}$ and $\frac{\partial \bar{P}_{ef}^b}{\partial T}$ have the form of (eq. 5.16) and (eq. 5.17), respectively, after substituting the most probable (primary) frequencies for the mean (secondary) frequencies.

The concentration of secondary deformons, $\bar{n}_d$, corresponds to (*eq*. 3.13); $\partial \bar{n}_d / \partial T$; this has the form of (eq. 5.26) after substitution of the most probable frequencies with the derivatives of the mean frequencies: (eq. 5.32) and (eq. 5.33).

In the course of the computer calculations, $C_V$ values can be derived simply by numeric differentiation of the total internal energy ($U_{tot}$). To this end, it is enough to derive a set of $U_{tot}$-values in the temperature range of interest with increments of $\Delta T$. Then we have:

$$C_V = \left( U_{tot}^{(T+\Delta T)} - U_{tot}^{(T)} \right)/\Delta T = \Delta U/\Delta T \qquad 5.37$$

It will be demonstrated below that the Hierarchic Theory encompasses both the ideas of the Einstein and Debye models of condensed matter and can be reduced to them, after simplifications, as limiting cases.

### 5.1 The relation of hierarchic concept to Einstein and Debye theories of condensed matter

Our computer calculations proved that the main contribution to the heat capacity of the solid state of condensed matter (ice in our case) is determined mainly by secondary effectons (see Figure 4a). Let us analyze the corresponding formalism, based on the results of the previous section. The dependence of the partition function (Z) on the temperature can be neglected for the purpose of simplification (it is small indeed - see Figure 3). After making such an assumption, it follows from (eq. 5.4), (eq. 5.5) and (eq. 5.6), that the heat capacity can be given as:

$$C_V \approx \left( \bar{C}_V^{ef} \right)_{tr} + \left( \bar{C}_V^{ef} \right)_{lb} \approx \frac{2V_0}{2} \frac{\partial}{\partial T}\left[ \bar{n}_{ef}\left( \bar{P}_{ef}^a \bar{E}_{ef}^a + \bar{P}_{ef}^b \bar{E}_{ef}^b \right) \right] \qquad 5.38$$

where $\bar{n}_{ef}$ is the concentration of secondary effectons; $\bar{P}_{ef}^{a,b}$ and $\bar{E}_{ef}^{a,b}$ are their thermo-accessibilities and energy, correspondingly.

Coefficient 2 in the right-hand side appears as:

$$\left( \bar{C}_V \right)_{tr} \approx \left( \bar{C}_V \right)_{lb} \qquad 5.39$$

Eq. (5.38) can be simplified even more if, in accordance with our computer simulation calculations, we assume that:

$$Z \simeq 2 \quad \text{and} \quad P^a_{ef} E^a_{ef} > P^b_{ef} E^b_{ef} \qquad 5.40$$

Then we have from (eq. 5.38):

$$C_V \approx V_0 \frac{\partial}{\partial T} \left( \bar{n}_{ef} \bar{P}^a_{ef} \bar{E}^a_{ef} \right)_{tr} \qquad 5.41$$

Taking into account that the dependence of $(\bar{n}_{ef} \bar{P}^a_{ef})$ on temperature is much less, than that of $\bar{E}^a_{ef}$ and $\bar{P}^a_{ef} \to 1$ we obtain from (eq. 5.41):

$$C_V \simeq \frac{4}{9} \pi V_0 \left( \frac{\bar{v}^a_{res}}{\mathbf{v}_{res}} \right)^3 \left[ h\left( \frac{\partial \bar{v}^a}{\partial T} \right) \right]_{tr,lb} \qquad 5.42$$

where $\bar{v}^a_{res} = (\bar{v}^a_1 \bar{v}^a_2 \bar{v}^a_3)^{1/3}$ is the resulting frequency of the secondary effecton in the $(\bar{a})$ state; $\mathbf{v}^s_{res}$ is the resulting phase velocity of the de Broglie waves in a solid body, according to eq. (2.75).

The average frequency of the secondary effecton in (a) state according to (eq. 2.54) is:

$$\bar{v}^a = \frac{v^a}{\left( \frac{hv^a}{kT} \right) - 1} \qquad 5.43$$

then, the result of differentiating eq. (5.43) gives:

$$h \frac{\partial \bar{v}^a}{\partial T} = k \left( \frac{hv^a}{kT} \right) \exp\left( \frac{hv^a}{kT} \right) \times \left[ \exp\left( \frac{hv^a}{kT} \right) - 1 \right]^{-2} \qquad 5.44$$

eq. (5.42) with regard for (eq. 5.44) can be written in the following form:

$$C_V \simeq \frac{4}{3} \pi V_0 \left( \frac{\bar{v}^a}{\mathbf{v}_{res}} \right)^3 \times k \left( \frac{hv_a}{kT} \right)^2 \exp\left( \frac{hv_a}{kT} \right) \times \frac{1}{\left[ \exp\left( \frac{hv_a}{kT} \right) - 1 \right]^2} \qquad 5.45$$

In the low temperature range, when

$$hv_a \ll kT; \qquad \exp(hv_a/kT) \approx 1 + (hv_a/kT) \qquad 5.46$$

it follows from eq. (5.43) and eq. (5.42) that as $T \to 0$:

$$\bar{v}^a \to \frac{kT}{h} \qquad 5.47$$

Putting eq. (5.47) into eq. (5.45), we get, under these conditions:

$$C_V \simeq \frac{4}{3} \pi k \frac{V_0}{\mathbf{v}^3_{res}} \left( \frac{kT}{h} \right)^3 \frac{\left( \frac{hv_a}{kT} \right)^2 \exp\left( \frac{hv_a}{kT} \right)}{\left[ \exp\left( \frac{hv_a}{kT} \right) - 1 \right]^2} \qquad 5.48$$

From the limit conditions for secondary effectons (see eq.3.7 and eq. 3.9):

$$\bar{n}^{max}_{ef} = \frac{N_0}{V_0} = \frac{4}{9} \pi \left( \frac{\bar{v}^a_{res}}{\mathbf{v}_{res}} \right)^3_{max} \qquad 5.49$$

we derive the maximum resulting frequency of the secondary effecton $(\bar{v}^{max}_{res})$ and the corresponding characteristic temperature $(T_c)$, which is in agreement with the Debye temperature (eq. 5.55):

$$\bar{v}^{max}_{res} = \mathbf{v}_{res} \left( \frac{9}{4} \frac{N_0}{\pi V_0} \right)^{1/3} = \frac{kT_c}{h} \qquad 5.50$$

whence

$$\frac{1}{(T_c)^3} = \frac{4}{9}\pi \frac{V_0}{N_0} \frac{1}{\mathbf{v}_{res}^3} \left(\frac{k}{h}\right)^3 \qquad 5.51$$

Combining eq. (5.51) with eq. (5.48) and taking into account that $kN_0 = R$ we obtain the approximate formula for heat capacity, following from our theory:

$$C_V \simeq 3R\left(\frac{T}{T_c}\right)^3 \frac{\left(\frac{h\nu_a}{kT}\right)^2 \exp\left(\frac{h\nu_a}{kT}\right)}{\left[\exp\left(\frac{h\nu_a}{kT}\right) - 1\right]^2} \qquad 5.52$$

Let us compare this formula with the expressions for heat capacity, derived earlier by Einstein (1965) and Debye from much simpler models.

Einstein considered a solid body as a "gas of harmonic quantum oscillators." He obtained expression for heat capacity, using the Planck formula:

$$C_V = 3R\frac{\left(\frac{h\nu}{kT}\right)^2 \exp\left(\frac{h\nu}{kT}\right)}{\left[\exp\left(\frac{h\nu}{kT}\right) - 1\right]^2_{Einst.}} \qquad 5.53$$

where $(\nu)$ is the frequency of the solid body atoms treated as a system of independent quantum oscillators.

As $T \to T_c$, the formula eq. (5.52) coincides with that of Einstein, eq. (5.53).

Debye, on the other hand, considered a solid body as a continuous medium, with its internal energy determined solely by the energy of phonons.

The Debye formula for heat capacity at low temperature range is:

$$C_V = \frac{12}{5}\pi^4 \left(\frac{T}{\theta_D}\right)^3_{Debye} \qquad 5.54$$

where $\theta_D$ is the Debye temperature, defined by the maximum phonon frequency, when the phonons length becomes equal to separation between atoms. It is related to Debye frequency in the following way:

$$\theta_D = \frac{h\nu_D}{k}, \qquad 5.55$$

where the Debye frequency **is**:

$$\nu_D = \mathbf{v}_S \left(\frac{3}{4}\frac{N_0}{\pi V_0}\right)^{1/3} \qquad 5.56$$

The mean velocity of sound $\bar{\mathbf{v}}_S$ differs from the resulting velocity $\mathbf{v}_{res}$ that we used in (eq. 5.50). It was introduced by Debye as:

$$\frac{1}{\bar{\mathbf{v}}_S^3} = \frac{1}{3}\left(\frac{1}{\mathbf{v}_\parallel^3} + \frac{1}{\mathbf{v}_\perp^3}\right) \to \mathbf{v}_S = \mathbf{v}_\parallel \mathbf{v}_\perp \left(\frac{3}{\mathbf{v}_\parallel^3 + \mathbf{v}_\perp^3}\right)^{1/3} \qquad 5.57$$

We may see that our formula (eq. 5.52) qualitatively coincides with the Debye formula (eq. 5.54) at $T_c = \theta_D$ and at $h\nu_a \ll kT_c$ (eq. 5.46), when:

$$\frac{\left(\frac{h\nu_a}{kT}\right)^2 \exp\left(\frac{h\nu_a}{kT}\right)}{\left[\exp\left(\frac{h\nu_a}{kT}\right) - 1\right]^2} \to 1 \qquad 5.59$$

Consequently, the formula (eq. 5.52) encompasses the results obtained by both Einstein and Debye.

This is natural, as far the Hierarchic Theory takes into account both kind of condensed matter dynamics: the oscillations of molecules (Einstein model), represented by the effectons for one hand and molecular collective excitations - phonons (Debye model), represented by secondary (acoustic) deformons for other hand. The contribution of deformons to heat capacity can be driven to the form which is analogous to (eq. 5.52).

In the range of sufficiently high temperature, the condition (eq. 5.46) is correct and the factor in (eq. 5.45) becomes:

$$\left(\frac{h\nu_a}{kT}\right)^2 \exp\left(\frac{h\nu_a}{kT}\right) \frac{1}{\left[\exp\left(\frac{h\nu_a}{kT}\right) - 1\right]^2} \to 1$$

At the same time, the condition (eq. 5.49) is fulfilled. Consequently, the formula (eq. 5.45) reduces into the classical Dulong and Petit law:

$$C_V \approx 3kN_0 = 3R \qquad 5.60$$

We have undertaken this analysis to show that the Hierarchic Theory does not contradict conventional theories; also, our theory reduces to them as limiting cases. The current approach both unifies and develops the pioneering ideas of condensed matter, introduced by Einstein and Debye.

It will be demonstrated below, that the Stephan-Boltzmann law for radiation can also be derived from the current approach.

### 5.2 Relation of the Hierarchic Concept to the Stephan-Boltzmann law of radiation

According to Hierarchic theory, the density of phonon radiation in the volume of matter is determined by the contribution of secondary acoustic deformons in the internal energy (see eq. 4.5). Therefore, the density of the acoustic *phonon radiation* can be expressed as follows:

$$(\epsilon_{ph})_{tr,lb} \cong \frac{\bar{U}_d}{V_0} = \frac{1}{2} \sum_{tr,lb} [\bar{n}_d \bar{P}_d \bar{E}_d]_{tr,lb} \qquad 5.60$$

The density of the electromagnetic *photon radiation* in matter is equal to the primary deformons contribution **to** the internal energy of matter:

$$(\epsilon_p)_{tr,lb} \cong \frac{U_d}{V_0} = \frac{1}{2} \sum_{tr,lb} [n_d P_d E_d]_{tr,lb}, \qquad 5.61$$

where $\bar{n}_d$ and $n_d$ are the concentration of secondary and primary deformons (eq. 3.13):

$$\bar{n}_d = \frac{8}{9}\pi \left(\frac{\bar{\nu}_{res}^d}{\mathbf{v}_s}\right)^3; \qquad n_d = \frac{8}{9}\pi \left(\frac{\nu_{res}^d}{c/n}\right)^3, \qquad 5.62$$

and where $\mathbf{v}_s$ is the resulting the velocity of sound, which is equal to (eq. 2.62) for solids; $c$ is the velocity of light and:

$$\left.\begin{array}{l}\left(\bar{\nu}_{res}^d\right) = \left(\bar{\nu}_{ph}^{(1)} \bar{\nu}_{ph}^{(2)} \bar{\nu}_{ph}^{(3)}\right)^{1/3} \\ \left(\nu_{res}^d\right) = \left(\nu_p^{(1)} \nu_p^{(2)} \nu_p^{(3)}\right)^{1/3}\end{array}\right\} \qquad 5.63$$

are the resulting frequencies of secondary and primary deformons; and

$$\left(\bar{\nu}_{ph}^{1,2,3}\right) = |\bar{\nu}_a - \bar{\nu}_b| \qquad 5.64$$

are the frequencies of the mean phonons in directions (1, 2, 3) are found from the following formula:

$$\bar{E}_d = h\left(\bar{v}_{ph}^{(1)} + \bar{v}_{ph}^{(2)} + \bar{v}_{ph}^{(3)}\right) = 3h\bar{v}_{res}^d \qquad 5.65$$

Here, $\bar{E}_d$ is the energy of a secondary deformon, see (4.21) and (4.23). As it follows from our calculations that $\bar{v}_a \gg \bar{v}_b$, then from (eq. 5.64):

$$v_{ph}^{1,2,3} \approx v_a^{1,2,3} \qquad 5.66$$

and

$$\bar{E}_d \approx 3h\bar{v}_{res}^d \approx 3h\bar{v}_a \qquad 5.67$$

where $\bar{v}_a$ (eq. 2.54) is

$$\bar{v}_a = \frac{v_a}{\exp\left(\frac{hv_a}{kT}\right) - 1} \qquad 5.68$$

Putting (eq. 5.67) and (eq. 5.62) into (eq. 5.60) and taking the value of $(\bar{P}_d/Z) \sim 1/10$, based on the computer simulations, we obtain:

$$\epsilon_{ph} \sim h\frac{(\bar{v}_a)^4}{\mathbf{v}_{res}^3} \qquad 5.69$$

At sufficiently high temperatures, when $kT \gg hv_a$, the component of (eq. 5.68) becomes:

$$\exp\left(\frac{hv_a}{kT}\right) \approx 1 + \frac{hv_a}{kT} \qquad 5.70$$

and (eq. 5.68) becomes:

$$\bar{v}_a \approx kT/h \qquad 5.71$$

Putting (eq. 5.71) into (eq. 5.69), we derive an approximate formula for the density of phonons energy (acoustic field energy):

$$\bar{\epsilon}_{ph} \approx \frac{1}{\mathbf{v}_{res}^3}\frac{k^4 T^4}{h^3} \qquad 5.72$$

All the above arguments are true not only for phonons, but also for IR photons. The approximate formula for the density of photon radiation energy has the same form as, and differs from, (eq. 5.72) only by substitution of the resulting speed of sound ($\mathbf{v}_{res}$) for the speed of light ($c$):

$$\epsilon_p \approx \frac{k^4 T^4}{c^3 h^3} = \frac{k^4 T^4}{c^3 h^3} \qquad 5.73$$

On the other hand, the well-known *Stephan-Boltzmann law*, following from the Planck theory of radiation (Beizer, 1971), has the form:

$$\epsilon_p^{Pl} = \frac{8}{15}\pi^3 \frac{k^4 T^4}{c^3 h^3} = 16.5\frac{k^4 T^4}{c^3 h^3} \qquad 5.74$$

This formula differs from (eq. 5.73) only by a constant factor.

Taking into account these approximations and simplifications used when deriving (eq. 5.73), such an agreement can be considered as confirming the model once again.

Agreement between the new and conventional theories is known to be an important criteria of the correctness of the new theory. *However, the main evidence in support of the new theory is its ability to describe quantitatively the experimental results better than previous theories. This criterion is also fulfilled in our case, as it shown below.*

# Chapter 6

# Quantitative verification of The Hierarchic Theory for water and ice

### 6.1 General information about the physical properties of water and ice

The properties of water ($H_2O$), when compared to other substances composed of similar elements, such as $H_2S$, $NH_3$ etc., are highly unusual. Water is a very polar molecule and an excellent solvent. Water is capable of forming hydrogen bonds. Water can donate and simultaneously accept two of these H-bonds. It is the high H-bond capability of water that is responsible for its unique properties. Ice is less dense than water because most of the hydrogen bonds are saturated, and the H-O-H angles in ice are more linear than they are in water. This forces the water molecules further apart, thus lowering the density of the ice. In ice there are an average of 4 H-bonds per water molecule. In liquid water there are about 3.4 of H-bonds per molecule. Hydrogen bonds are examples of noncovalent interactions. Other types of noncovalent interactions in water are electrostatic, dipole-dipole interactions between molecules with nonzero permanent and induced dipole moments, van der Waals interactions, and quantum London dispersion forces.

Although hydrogen bonds are relatively weak compared to covalent bonds within the water molecule itself, the ability of H-bonds to provide cooperative interaction between many water molecules is responsible for a number of water's physical properties. One such property is its relatively high melting and boiling points. The chemically similar compound hydrogen sulfide ($H_2S$), which has much weaker hydrogen bonding, is a gas at room temperature even though it has twice the molecular weight of water. The extra bonding between water molecules also gives liquid water a large specific heat capacity.

The slight ability of water to ionize ($H_2O \rightleftharpoons H^+ + HO^-$) provides the formation of charged oxonium molecules $H_3^+O$, as well as high proton mobility in water.

For most substances, the solid phase is more dense than the liquid phase. However, ice is less dense than liquid water. The liquid water becomes denser with lower temperatures, as do other substances. But, at 4 °C water reaches its maximum density and, as water cools further toward its freezing point, the liquid water, under standard conditions, expands to become less dense. The physical reason for this is a competition between the thermal chaotization factor and tendency of water to cluster-formation. It is important to note, that our computer simulations based on the Hierarchic Theory confirm this anomaly – like many others. This will be demonstrated below.

### 6.2 The experimental data necessary for analysis of condensed matter, using pCAMP (computer program: Comprehensive Analyzer of Matter Properties)

All calculations, based on the Hierarchic Theory, were performed on personal computers. A special computer program: "Comprehensive Analyzer of Matter Properties" (pCAMP) was developed and verified on examples of water and ice. The program allows the user to evaluate more than three hundred parameters of condensed matter, if the following experimental input data are available – under the conditions of constant temperature and pressure:
  1. Sound velocity
  2. Density
  3. Refraction index
  4. Positions of translational and librational bands in oscillatory spectra

The demonstration version of pCAMP for water and ice can be downloaded from this author's homepage: web.petrsu.ru/ ~alexk.

### 6.2.1 The input parameters of ice as used in the computer calculations

The wave numbers ($\tilde{v}_{tr}$), corresponding to the positions of translational and librational bands in oscillatory IR spectra were taken from the work of Eisenberg and Kauzmann (1969). Wave numbers for ice at $0^0 C$ are:

$$\left(\tilde{v}_{ph}^{(1)}\right)_{tr} = 60 cm^{-1}; \quad \left(\tilde{v}_{ph}^{(2)}\right)_{tr} = 160 cm^{-1}; \quad \left(\tilde{v}_{ph}^{(3)}\right)_{tr} = 229 cm^{-1}$$

According to the model, the IR photons with corresponding frequencies are irradiated and absorbed as a result of $(a \Leftrightarrow b)$ primary translational deformons in ice. Temperature shifts of these band positions are close to zero:

$$\partial \left(\tilde{v}_{ph}^{1,2,3}\right)_{tr}/\partial T \approx 0$$

Wave numbers of the librational IR bands, corresponding to the absorption of photons, related to the $(a \Leftrightarrow b)_{lb}^{1,2,3}$ transitions of primary librational effectons of ice are:

$$\left(\tilde{v}_{ph}^{(1)}\right)_{lb} = \left(\tilde{v}_{ph}^{(2)}\right)_{lb} = \left(\tilde{v}_{ph}^{(3)}\right)_{lb} \approx 795 cm^{-1}.$$

The equality of wave numbers for all three directions (1, 2, 3) indicates the spatial isotropy of librations of $H_2O$ molecules. In this case deformons and effectons have a cubic geometry. In the general case, they have a shape of parallelepiped (similar to the quasi-particles of translational type) with each of three edges, corresponding to the most probable de Broglie wave length in the given direction.

The temperature shift of the position of the librational band maximum for ice is:

$$\partial \left(\tilde{v}_{ph}^{1,2,3}\right)_{lb}/\partial T \approx -0.2 cm^{-1}/C^0$$

The resulting thermal velocity of phonons in ice, responsible for secondary acoustic deformons, is taken to be equal to the transverse velocity of sound (Kikoin, 1976):

$$\mathbf{v}_s^{res} = 1.99 \times 10^5 cm/s$$

This velocity and the molar ice volume ($V_0$) are almost independent on temperature (Eisenberg and Kauzmann, 1969):

$$V_0 = 19.6 cm^3/M \simeq \text{const}$$

### 6.2.2 The input parameters of water for computer calculation

The wave numbers of translational bands in the IR spectrum, corresponding to the quantum transitions of primary translational effectons between acoustic (a) and optical (b) states with absorption or emission of photons, forming electromagnetic 3D translational deformons at $0^0 C$, are (Eisenberg and Kauzmann, 1969):

$$\left(\tilde{v}_{ph}^{(1)}\right)_{tr} = 60 cm^{-1}; \quad \left(\tilde{v}_{ph}^{(2)}\right)_{tr} \approx \left(\tilde{v}_{ph}^{(3)}\right)_{tr} \approx 199 cm^{-1}$$

with the temperature shifts calculated as follows:

$$\partial \left(\tilde{v}_{ph}^{(1)}\right)_{tr}/\partial T = 0; \quad \partial \left(\tilde{v}_{ph}^{(2,3)}\right)_{tr}/\partial T = -0.2 cm^{-1}/C^0$$

The primary librational deformons of water at $0^o C$ are characterized by the following degenerate wave numbers of librational bands in the IR spectrum:

$$\left(\tilde{v}_{ph}^{(1)}\right)_{lb} \approx \left(\tilde{v}_{ph}^{(2)}\right)_{lb} \approx \left(\tilde{v}_{ph}^{(3)}\right)_{lb} = 700 cm^{-1}$$

with the temperature shift calculated thus:

$$\partial \left(\tilde{v}_{ph}^{1,2,3}\right)_{lb}/\partial T = -0.7 cm^{-1}/C^0$$

Wave numbers are related to the frequencies ($v$) of corresponding transitions via the velocity of light as: $v = c\tilde{v}$.

The dependence of the velocity of sound ($v_s$) in water on temperature, within the temperature range $0 - 100^{\circ}C$, is expressed by the polynomial (Fine and Millero, 1973):

$$\mathbf{v}_s = 1402.385 + 5.03522t - 58.3087 \times 10^{-3}t^2 + 345.3 \times 10^{-6}t^3 -$$
$$- 1645.13 \times 10^{-9}t^4 + 3.9625 \times 10^{-9}t^5 \ (m/s).$$

The temperature dependence of a molar volume ($V_0$) of water within the same temperature range can be calculated using the following polynomial (Kell, 1975):

$$V_0 = 18000/[(999,83952 + 16.945176t - 7.98704 \times 10^{-3}t^2 -$$
$$- 4.6170461 \times 10^{-5}t^3 + 1.0556302 \times 10^{-7}t^4 - 2.8054253 \times 10^{-10}t^5)/$$
$$/(1 + 1.687985 \times 10^{-2}t)] \ (cm^3/M)$$

The refraction index for ice was taken as an independent of temperature ($n_{ice} = 1.35$) and that for water as a variable, depending on temperature, in accordance with experimental data, presented by Frontas'ev and Schreiber (1965).

The refraction index for water at $20^0 C$ is approximately:

$$n_{H_2O} \simeq 1.33$$

The temperature of different parameters for ice and water, computed using the formulas of the Hierarchic Theory, are presented in Figs.(2-25). This is only a part of the available information. It is possible to calculate more than 300 different parameters associated with the liquid and solid states, and their changes, in a course of first- and second-order phase transitions, using this computer program.

### 6.3 Discussion of theoretical temperature of different parameters of water and ice

It will be shown below, by use of examples of water and ice that the Hierarchic Theory makes it possible to calculate an unprecedented number of physical parameters for liquids and solids. Those parameters that where measured experimentally and published by different authors, are in excellent agreement with this theory.

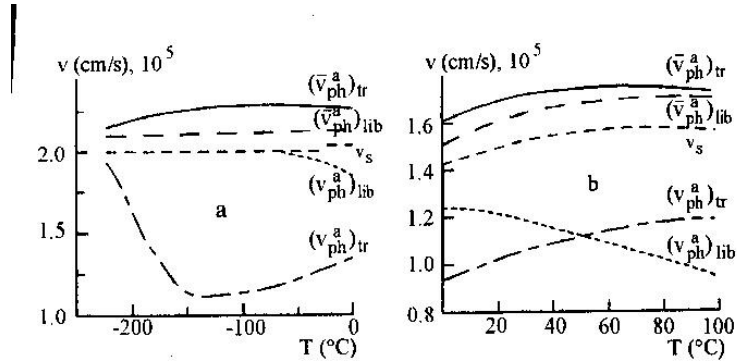

Figure 2. (**a**, **b**). Temperature dependencies of the resulting *phase velocities* of primary and secondary translational $(\mathbf{v}_{ph}^a, \bar{\mathbf{v}}_{ph}^a)_{tr}$ and librational $(\mathbf{v}_{ph}^a, \bar{\mathbf{v}}_{ph}^a)_{lib}$ effectons (eqs. 2.74 and 2.75) in the ground ($a, \bar{a}$) states and of the sound velocities [$\mathbf{v}_s$] for ice [**a**] and water [**b**].

It follows from Figures 2a, and 2b that the resulting phase velocities of secondary effectons in the ā-state in ice and water are all higher than the velocity of sound (phonons) ($\mathbf{v}_s = \mathbf{v}_{ph}$). On the other hand the phase velocities of primary effectons in (a)-state are lower than the velocity of sound. Phase velocities of the primary effectons in (b)-state are higher than that of (a) state, as follows from eq. 2.69)

(not shown in Figure 2).

*6.3.1. The temperature dependencies of the total partition function (Z) and some of its components*

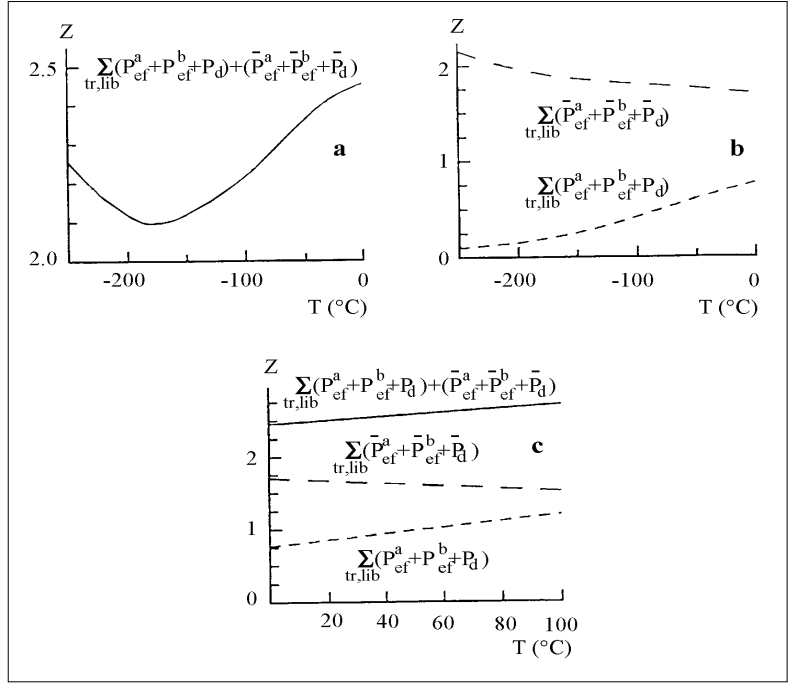

Figure 3. (*a, b, c*). Temperature dependencies of the total partition function (Z) (eq. 4.2) and contributions, related to primary and secondary effectons and deformons for ice (a, b) and water (c).

The resulting thermo-accessibility minimum (Figure 3 a) for ice (Z) corresponds to the temperature of $-170^0C$. The interval from $-198$ to $-173^0C$ is well-known for its anomalies, due to the fact that the heat equilibrium of ice "settles" very slowly in this range (Maeno, 1988, Haida *et al.*, 1972). This fact can be explained by the "less probable" and stable ice structure (minimum value of partition function Z ) near $-170^0C$.

On the other hand, the experimental anomaly, corresponding to maximum of heat capacity ($C_p$) of ice, also was observed near the same temperature. It can be explained, if we present the heat capacity (5.5) as:

$$C_p = \frac{\partial}{\partial T}(\frac{1}{Z}U^*) = -\frac{1}{Z^2}\frac{\partial Z}{\partial T}U^* + \frac{1}{Z}\frac{\partial U^*}{\partial T} \qquad 6.1$$

One can see that $C_p$ should be at a maximum when $(\partial Z/\partial T) = 0$, when the first negative member of $C_p$ in (1.1) is zero and Z as the denominator in the second member of (1.1) is minimal. This corresponds to minimum value of Z, following the theory as it is evaluated at $-170^0C$ (see Figure3 a). This important result is confirmed by the instability of the internal pressure of ice in the same temperature range (Figure 3 c). It is the first compelling evidence in proof of the Hierarchic Theory.

In liquid water the theoretical temperature dependencies of Z and its components are linear, similar to the situation with heat capacity. The thermo-accessibility of the mean secondary effectons in water decreases, while that of primary effectons increases, with temperature, similar to its action in ice (Figure 3 b, c).

On lowering the temperature, the total internal energy of ice (Figure 4 a) and its components decreases nonlinearly with temperature, coming closer to absolute zero. The same parameters for water decrease almost linearly within the interval $(100^0 - 0)^0C$ (Figure 4 b).

In computer calculations, the values of $C_p(t)$ can be determined by differentiating $U_{tot}$ numerically at any given temperature values.

*6.3.2. The temperature dependencies of the total internal energy and some of its contributions*

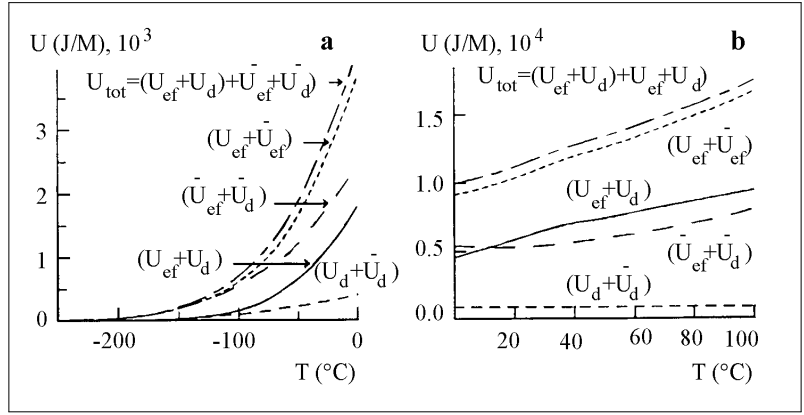

**Figure 4** (a, b). Temperature dependencies of the total internal energy ($U_{tot}$) and different contributions for ice *(a)* and water *(b)* (eqs. 4.3 - 4.5). The following contributions to $U_{tot}$ are presented: $(U_{ef} + \bar{U}_{ef})$ is the contribution of primary and secondary effectons; $(U_d + \bar{U}_d)$ is the contribution of primary and secondary deformons; $(U_{ef} + U_d)$ is the contribution of primary effectons and deformons; $(\bar{U}_{ef} + \bar{U}_d)$ is the contribution of secondary effectons and deformons.

The contributions of macro- and super-effectons to the total internal energy and corresponding macro- and super-deformons, as well as all types of convertons, are much smaller than those of the primary and secondary effectons and deformons.

It follows from Figure 4a, that the mean value of heat capacity for ice in the interval from -75 to $0^0 C$ is equal to:

$$\bar{C}_p^{ice} = \frac{\Delta U_{tot}}{\Delta T} = 39 J/MK = 9.3 \text{ cal}/MK$$

For water within the whole range $\Delta T = 100^0 C$, the change of the internal energy is (Figure 4b) is:

$$\Delta U = 17 - 9.7 = 7.3 kJ/M$$

This corresponds to the mean value of the heat capacity of water $C_p$:

$$C_p^{water} = \frac{\Delta U_{tot}}{\Delta T} = 73 J/M \times K \approx 17.5 cal/M \times K$$

These theoretical results agree well with the experimental mean values $C_p = 18$ *cal*/$M \times K$ for water and $C_p = 9$ *cal*/$M \times K$ for ice (Eisenberg and Kauzmann, 1969). This is the second piece of compelling evidence of a proof of the theory.

*6.3.3. The confirmation of quantum properties of the ice and water*

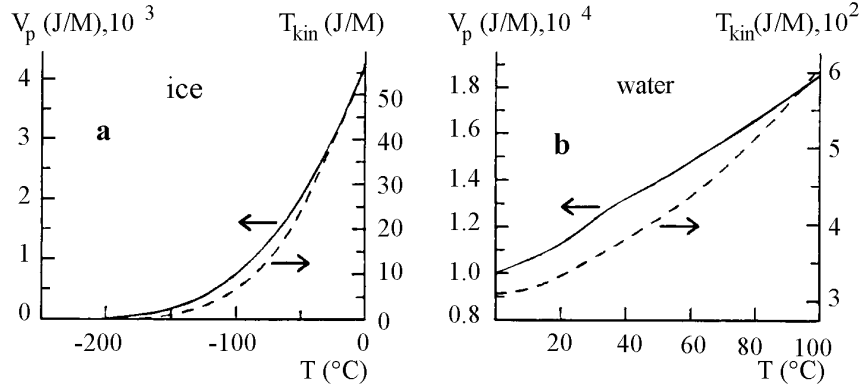

**Figure 5**. (a, b). Temperature dependencies of the kinetic ($T_{kin}$) and potential ($V_p$) energy of the ice *(a)* and water *(b)*. Note that $U_{tot} = T_{kin} + V_p$ (eqs. 4.31, 4.32).

It can be seen from Fig. 5 a, b, that the total kinetic energy of water ($T_{kin}$) is approximately 30 times less than the potential energy ($V_p$) at the same temperature. In the case of ice, these energies differ even more: ($T_{kin}/V$) < 1/100. The resulting $T_{kin}$ of water increases almost two times over the range (0 – 100°C) – from 313 to 585 J/M. However, the change of the total internal energy ($U_{tot} = T_{kin} + V_p$) is determined mainly by the change in the potential energy $V_p(t)$ of ice and water.

It is reasonable to analyze the above ratio between total kinetic and potential energies in terms of the *Viral theorem* (Clausius, 1870).

This theorem for a system of particles relates the averaged kinetic $\bar{T}_k(\vec{v}) = \sum_i \overline{m_i v_i^2/2}$ and potential $\bar{V}(r)$ energies as follows:

$$2\bar{T}_k(\vec{v}) = \sum_i \overline{m_i v_i^2} = \sum_i \overline{\vec{r}_i \partial V / \partial \vec{r}_i}$$

This equation is valid for both quantum-mechanical and classical systems.

If the potential energy $V(r)$ is a homogeneous n-order function such as:

$$V(r) \sim r^n \qquad\qquad 6.1a$$

then, the average kinetic and average potential energies are related as follows:

$$\overline{T_k(\vec{v})} = n \frac{\overline{V(r)}}{2} \qquad\qquad 6.2$$

For example, for a harmonic oscillator: $n = 2$ and $\bar{T}_k = \bar{V}$.
For Coulomb interaction: $n = -1$ and $\bar{T} = -\bar{V}/2$.

For water the calculation of $\bar{T}_k$ and $\overline{V(r)}$ gives: $n_w \sim 1/15$ and for ice: $n_{ice} \sim 1/50$. It follows from (6.1) that in water and ice the dependence of potential energy on distance (r) is very weak, namely:

$$V_{water}(r) \sim r^{(1/15)}; \quad V_{ice} \sim r^{(1/50)} \qquad\qquad 6.2a$$

This result can be considered as an indication of remote interaction between water molecules due to cooperative/orchestrated properties of water as an associative liquid. The role of distant van der Waals and London interactions, in stabilizing primary effectons, increases with increasing dimensions of these coherent clusters, i.e. with decreasing temperature.

This is strong evidence that water and ice can not be considered as classical systems. For classical equilibrium systems containing N-particles, the Virial theorem shows that average kinetic and potential energies related to each degree of freedom are the same, and are, in fact, equal to:

$$\bar{T}_k = \frac{1}{2}kT = \bar{V} \quad at \quad n = 2$$

This implies that, in the classical approximation, the particles of condensed matter should be considered as harmonic oscillators. For matter treated as a quantum system, however, this approximation is not successful in the general case.

*6.3.4. The temperature of properties of primary librational effectons of the ice and water*

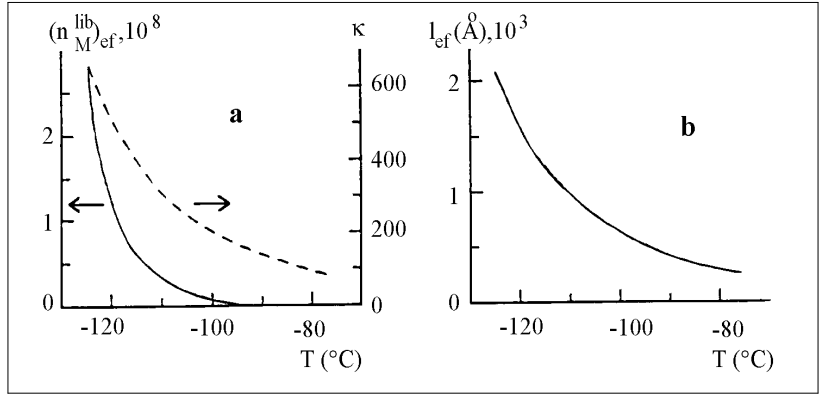

**Figure 6**. (a). Changes in the number of water molecules forming primary librational effectons of the ice $(n_M^{lib})_{ef} = V_{ef}/(V_0/N_0)$ within the $(-75 - 125°C)$ temperature range (left axis); the number of $H_2O$ molecules per length of the same effecton edge:
$\kappa = (n_M^{lib})_{ef}^{1/3}$ (right axis);
  (b) the temperature dependence of the length of the edge: $l = \kappa(V_0/N_0)^{1/3}$, of the same effecton with shape, approximated by a cube (b).

In accordance with this theory, the dimensions of all quasi-particle types increase upon lowering of the temperature due to a decrease in the molecular momentum. This leads in turn, to increasing of the thermal de Broglie wave length of molecules and an increase in the value of the fraction of mesoscopic Bose-condensation in liquids and solids.

The linear dimensions of the largest (primary) librational effectons in ice within the range from $-75$ to $-125°C$ are on the order of thousands of Angstroms, and increase strongly at temperature lowering $T \to 0$ $(K)$ (Figure 6). The primary librational effectons can determine the domain microcrystalline structure of solid bodies, glasses, liquid crystals and even polymers. Translational primary effectons are significantly smaller than librational effectons in both solid and liquid phase.

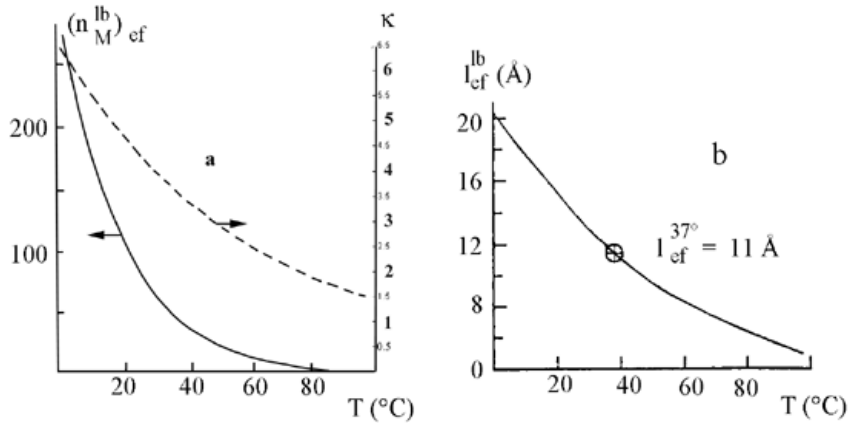

Figure **7**. (*a*) : The temperature dependencies of the number of $H_2O$ molecules in the volume of primary librational effectons ($n_M^{lb})_{ef}$, left axis) and the number of $H_2O$ per length of this effecton edge ($\kappa$, right axis);
 (*b*): the temperature dependence of the water – primary librational effecton's edge length $[l_{ef}^{lb} = \kappa(V_0/N_0)^{1/3}]$.

The number of $H_2O$ molecules within the *primary libration effectons* of water, which can be approximated by a cube, decreases from $n_M = 280$ at $0^0$ to $n_M \simeq 3$ at $100^0$ (Figure 7a). It should be noted that at physiological temperatures ($35 - 40^0$) such quasi-particles contain nearly 40 water molecules. This number is close to the number of of water molecules that can be enclosed in the open interdomain protein cavities – to judge from available X-ray data. The flickering of these clusters, i.e. their (*disassembly* ⇌ *assembly*) due to $[lb \Leftrightarrow tr]$ conversions is directly related to the large-scale dynamics of proteins (see chapter 15).

It is important to note, that the linear dimensions of such water clusters (11 Å) representing mesoscopic Bose condensate at physiological temperatures are close to that, common for protein domains, and clefts between domains and subunits in the 'open' to water state (Figure 7 b). Such spatial correlations indicate that the properties of water exert a strong influence on the biological evolution of macromolecules, namely, the spatial and dynamic parameters of such molecules.

### 6.3.5 Explanation of Drost-Hansen temperature anomalies

The Hierarchic Theory is the first theory which enables one to predict and give a clear explanation of deviations of temperature of some physical parameters of water from monotonic ones, discovered and studied by Drost-Hansen. The theory also clarifies the interrelation between these deviations (transitions) and the corresponding temperature anomalies in the properties of biosystems, such as dynamic equilibrium of the [assembly-disassembly] of microtubules and actin filaments, large-scale dynamics of proteins, enzyme activity, etc.

We assume here that integer and half-integer values of the number of water molecules per librational effecton's edge $[\kappa]$ (Figure 7a), approximated by the cube, correspondingly reflects the conditions of increased and decreased stabilities of the structure of water. This is apparently related to the stability of primary librational effectons as cooperative and coherent water clusters.

Nonmonotonic behavior of the properties of water with temperature is a widely-known and well-confirmed experimental fact (Drost-Hansen, 1976, 1992; Clegg and Drost-Hansen, 1991; Etzler, 1991; Roberts and Wang, 1993; Wang *et al.*, 1994). This phenomenon is interesting, as well as being important for the discipline of molecular biology. We can explain this phenomenon as a consequence of competition between two factors: *the quantum* and the *sterical factors* in the stability of primary librational effectons (coherent clusters in state of mesoscopic Bose condensation).

The *quantum factor*, such as the de Broglie wave length, determine the value of the effecton edge in this way:

$$\left[ l_{ef} = \kappa(V_0/N_0)^{1/3} \sim \lambda_B \right]_{lb}$$

This length decreases *monotonically* with increasing temperature.

The *sterical factor* is a discrete parameter depending on the water molecules' effective linear dimension: $l_{H_2O} = (V_0/N_0)^{1/3}$ and the number of these molecules $[\kappa]$ in the effecton's edge.

In accordance with the model, the shape of primary librational effectons in liquids and of primary translational effectons in solids can be approximated by a parallelepiped in the general case, or by cube, when corresponding thermal movements of molecules (lb and/or tr) are isotropic.

When ($l_{ef}$) corresponds to an *integer* number of $H_2O$ molecules, *i.e.*

$$[\kappa = (l_{ef}/l_{H_2O}) = 2, 3, 4, 5, 6 \dots]_{lb} \qquad \text{6.2a}$$

the competition between quantum and structural factors is minimal, and primary librational effectons are most stable.

On the other hand, when $(l_{ef}/l_{H_2O})_{lb}$ is a *half-integer*, the librational effectons are less stable (the *competition is maximum*). In the latter case $(a \Leftrightarrow b)_{lb}$ equilibrium of the effectons must be shifted rightward, to a less stable state of these coherent water clusters of higher potential energy. Consequently, the probability of dissociation of librational effectons to much smaller translational effectons increases.

Experimentally, the non-monotonic change of this probability with temperature can be registered by dielectric permittivity, refraction index measurements and by that of the average density of water. The refraction index and polarizability change, in turn, should lead to corresponding variations of surface tension, vapor pressure, viscosity, and lastly, self-diffusion (Kaivarainen, 1995; 2001).

The density of liquid water forming primary librational effectons (mesoscopic Bose condensate) is lower than the average density of bulk water. In primary effectons almost 100% of hydrogen bonds are saturated – just as in the case of "ideal" ice – in contrast to bulk water. Consequently, the conversion of primary librational effectons to translational effectons should be accompanied by an increase in average density of water.

We can see from Figure 7 a that the number of water molecules in a primary lb effecton edge ($\kappa$) is an integer value near the following temperatures:

$$6^0 (\kappa = 6); \quad 17^0 (\kappa = 5); \quad 32^0 (\kappa = 4); \quad 49^0 (\kappa = 3); \quad 77^0 (\kappa = 2) \qquad 6.2b$$

These temperatures agree very well with the maximums of relaxation time found in pure water, and with dielectric response anomalies (Roberts, et al., 1993; 1994; Wang, et al., 1994). The special temperature anomalies predicted by the current theory are also close to three classes of data:
- chemical kinetic data (Aksnes, Asaad, 1989; Aksnes, Libnau, 1991),
- refractometry data (Frontasev, Schreiber 1966), and
- IR-spectroscopy data (Prochorov, 1991).

Small discrepancies may result from the high sensitivity of water to any kind of perturbation, guest-effects, and the additional polarization of water molecules, induced by high frequency visible photons. Even such low concentrations of inorganic ions and NaOH, as used by Aksnes and Libnau (1991), may change the properties of the given sample of water. The increase of $H_2O$ polarizability under the effect of light also may lead to enhancement of the stability of water clusters, and to a corresponding high-temperature shift of non-monotonic changes of water properties.

The semi-integer numbers of $[\kappa]$ for librational effectons of pure water correspond to the following temperatures:

$$0^0 (\kappa = 6.5); \quad 12^0 (\kappa = 5.5); \quad 24^0 (\kappa = 4.5); \quad 40^0 (\kappa = 3.5); \qquad 6.2c$$
$$62^0 (\kappa = 2.5); \quad 99^0 (\kappa = 1.5)$$

These conditions (6.2c) characterize a water structure in the volume of primary librational effectons, which is less stable than that described by conditions (6.2b). The first-order phase transitions (freezing at $0^0$ and boiling at $100^0$) of water almost exactly correspond to semi-integer values $\kappa = 6.5$ and $\kappa = 1.5$. This fact is important for understanding the mechanism of first-order phase transitions, discussed in section 6.2.

The temperature anomalies of colloid water-containing systems, discovered by Drost-Hansen (1976) and studied by Etzler and coauthors (1987; 1991) occurred near $14\text{-}16^0$; $29\text{-}32^0$; $44\text{-}46^0$ and $59\text{-}62^0 C$. At these temperatures the extrema of viscosity, disjoining pressure and molar excess entropy of water between quartz plates – even with a separation (300 - 500) Å – have been observed. These temperatures are close to those predicted by the Hierarchic theory for bulk water anomalies, corresponding to integer values of $[\kappa]$ (see 5.7). Some deviations can be the result of interfacial water perturbations, induced by colloid particles and plates.

*We may conclude, then, that the Hierarchic Theory is the first theory which is able to predict and explain the existence of Drost-Hansen temperature anomalies in water.*

The dimensions, concentration and stability of water clusters (primary librational effectons) in the volume of vicinal water should all have larger values than do those same variables in bulk water – due to their lessened mobility and to longer de Broglie waves length (see section 13.3). Interesting ideas, concerning the role of water clusters in biosystems, were developed in the works of John Watterson (1988 a, b).

It was discovered (Kaivarainen, 1985; 1993) that non-monotonic changes of water around the Drost-Hansen temperatures are accompanied by in-phase changes of various types of large-scale protein dynamics, which are related to their functioning. Further investigation of similar phenomena is very important for understanding the molecular mechanisms of the function of biopolymers, as well as the thermal adaptation of living organisms.

### 6.3.6. The temperature dependencies of water density and some contributions to the total kinetic energy for water and ice

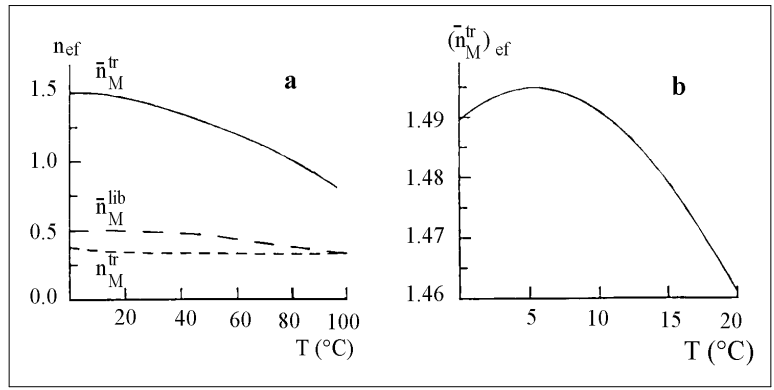

**Figure 8**. *(a)*: The changes of the number of $H_2O$ molecules composing the secondary translational effecton of water ($\bar{n}_M^{tr}$), the secondary librational effecton ($\bar{n}_M^{lb}$), and a primary translational effecton ($n_M^{tr}$) in the temperature range $(0 - 100^0 C)$; (b): the temperature dependence of $\bar{n}_M^{tr}$ in the range $(0 - 20^0 C)$ - to the increased scale. The number of molecules is calculated as: $n_M = V_{ef}/(V_0/N_0) = N_0/n_{ef}V_0$, where $V_{ef} = 1/n_{ef}$ is the volume of effectons with concentration $n_{ef}$ (eq. 3.5 and 3.7).

Inaccuracy in the experimental values of bands maxima wave numbers ($\tilde{v}_{tr}^{1,2,3}$ and $\tilde{v}_{lb}^{1,2,3}$) can lead to small temperature shifts ($\sim 2^0$) of the calculated extremum relative to experimental ones.

For example, the maximum number of water molecules in the volume of the secondary translational effecton $(\bar{n}_M^{tr})_{ef}$ (Figure 8 b), characterizing the conditions of maximum water stability and density, appears to correspond to $4^0$, rather than to $6^0 C$.

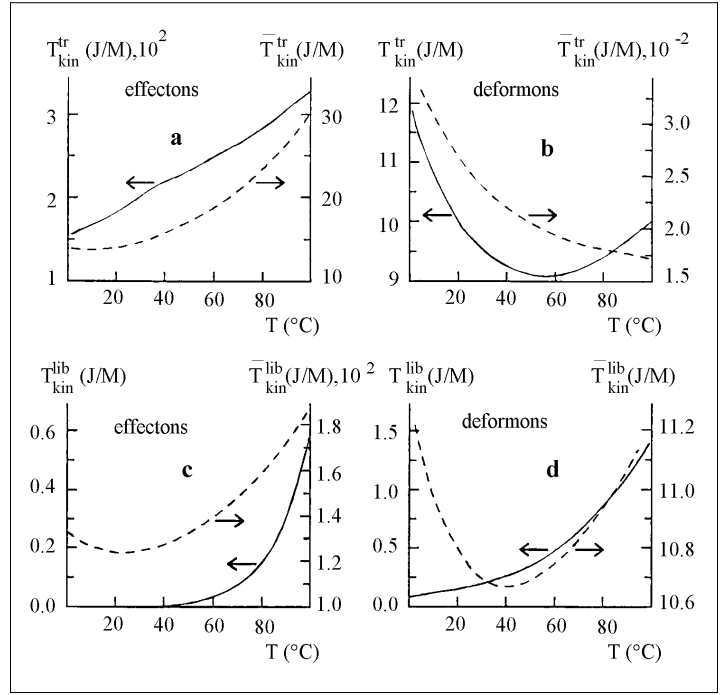

**Figure 9**. Temperature dependencies of contributions to the total kinetic energy of water $T^{tot}$ (eq. 4.31) of different quasi-particles: a) the kinetic energies of translational effectons: primary (left) and secondary (right) ones; b) the kinetic energies of translational deformons: primary (left) and secondary (right) ones; c) the kinetic energies of librational effectons: primary (left) and secondary (right) ones; d) the kinetic energies of librational deformons: primary (left) and secondary (right) ones.

Temperature dependencies for the contributions of different types of quasi-particles to the total kinetic energy of water (Figure 9) also are characterized by the extrema, which are close to the locations of the experimental temperature anomalies of water.

Thus, for example, the minimum of the kinetic energy $(T_{kin}^{tr})_d$ caused by primary translational deformons (Figure 9 b), is found at about $50^0C$. This is quite close to the minimum of the isothermal compressibility coefficient: $\gamma_T = -(1/V)(\partial V/\partial P)$, which is observed experimentally to occur at $46^0C$ (Eisenberg and Kauzmann, 1969).

The minimum kinetic energy $(\bar{T}_{kin}^{lb})_d$ of the secondary librational deformons (Figure 9 d) at $40^0C$ practically coincides with the region of minimum heat capacity for water $(37^0)$ and, as well, with the physiological temperature of warm-blooded animals.

Also, the deviation from the monotonic behavior, indicating stabilization of the water structure, is experimentally observed to occur at $37^0$.

The minimum $(\bar{T}_{kin}^{lb})_{ef}$ for the secondary librational effectons at $\simeq 20^0C$ is close to the location of the strong non-monotonic thermoinduced transition in water at $t \simeq 17^0$, a fact which has been experimentally established.

Judging from the data found in the literature, together with those data obtained in the laboratory by refractometry and IR spectroscopy methods, at the above-mentioned temperature, there occur changes in water which correspond to stabilization of its structure (Kaivarainen, 1989b).

The shift of the experimentally determined locations of these temperature anomalies in water relative to the theoretical values by $2 - 3^0$, as was mentioned earlier, may be the consequence of the inaccuracy in determining the locations of wide band maximums in the oscillation spectrum and certain uncontrolled conditions, such as pH, ionic traces, etc., which may affect the results of calculations.

Figure 10 *a, b* shows the resulting contributions to the total kinetic energy of water of two main subsystems: effectons and deformons. The minimum deformon contribution occurs at $43°C$, which is close to the physiological temperatures for warm-blooded animals, and thus, is critical for humans.

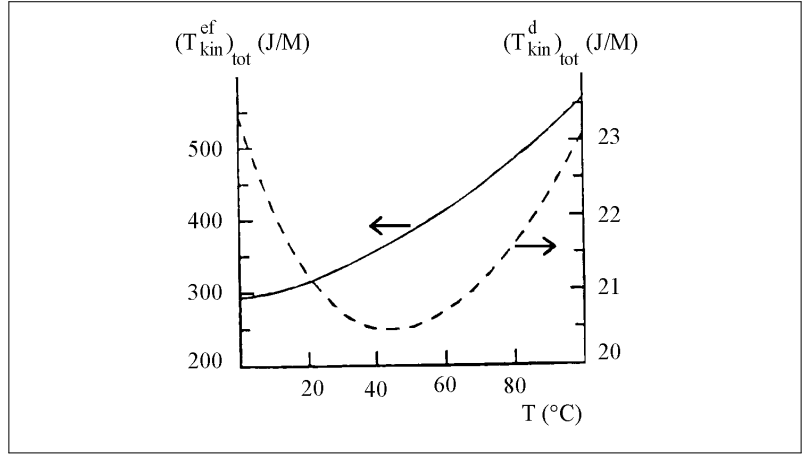

**Figure 10**. Temperature dependencies of two resulting contributions - effectons $(T_{kin}^{ef})$ and deformons $(T_{kin}^{d})$ of all types - to the total kinetic energy of water. The minima temperature dependencies of different contributions to the total kinetic energy of water in Figure 9 $(b, c, d)$ and Figure 10 correspond to the most probable/stable conditions in accordance with the *Principle of Least Action,* in the Maupertuis - Lagrange form.

In such a form, this principle is valid for the conservative holonomic systems, where limitations exist for the *displacements* of the particles of this system, rather than the magnitudes of their velocities. This Principle states that among all kinematically possible displacements of a system, from one configuration to another, without changing the total system energy, such displacements are most probable for which the action (eq. 6.5) is least: $\Delta W = 0$. Here $\Delta$ is the symbol of the total variation in coordinates, velocities and time.

Action is a fundamental physical parameter (which has the dimension of the product of energy and time) which characterizes the dynamics of a system.

According to Hamilton, the *action*:

$$S = \int_{t_0}^{t} L dt \qquad 6.3$$

is expressed through the Lagrange function:

$$L = T_{kin} - V \qquad 6.4$$

where $T_{kin}$ and $V$ are the kinetic and potential energies of the system or subsystem.

According to Lagrange, the action (W) can be expressed as:

$$W = \int_{t_0}^{t} 2T_{kin} dt \qquad 6.5$$

We can assume that, at the same integration limits, the minimum value of the action $\Delta W \simeq 0$ corresponds to the minimum value of $T_{kin}$. Then it can be said that at a temperature of about $45°$ the subsystem of deformons is the most stable (see Figure 10). This implies that the equilibrium between the acoustic and optic states of primary and secondary effectons should be most stable at this temperature.

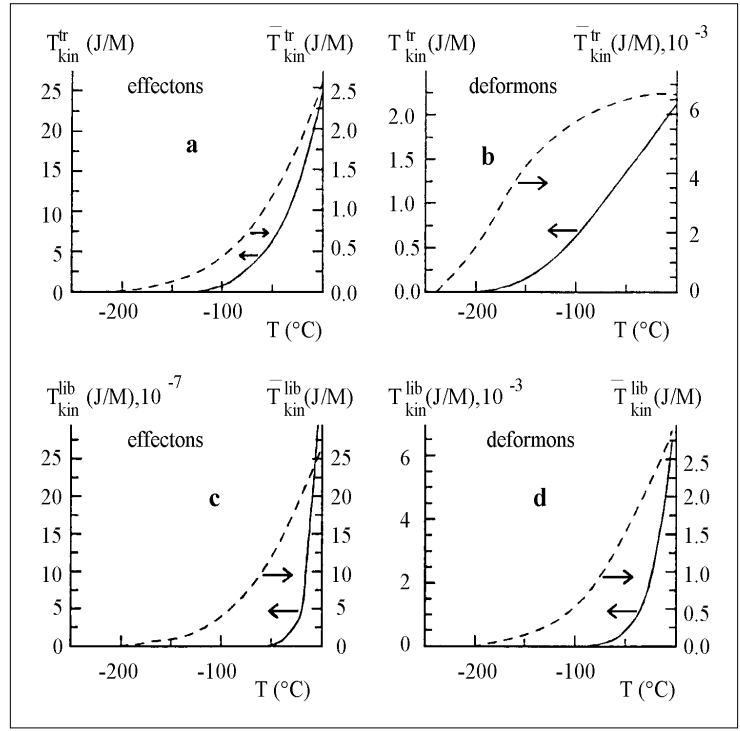

**Figure 11**. Temperature dependencies of contributions to the total kinetic energy of ice (eq. 4.36) for different types of quasi-particles:
a) the kinetic energies of translational effectons: primary (left) and secondary (right)
b) the kinetic energies of translational deformons: primary (left) and secondary (right)
c) the kinetic energies of librational effectons: primary (left) and secondary (right)
d) the kinetic energies of librational deformons: primary (left) and secondary (right)

The contributions of various quasi-particles to the kinetic energy of ice (Figure 11 a, b, c, d) decreases monotonically as the temperature lowers. The contribution of primary librational effectons approaches zero more rapidly than others. The largest contribution to the total kinetic energy of water and ice is represented by primary translational effectons $(T^{tr}_{kin})_{ef}$.

### 6.3.7. *The temperature dependencies of translational and librational velocities and corresponding de Broglie waves of $H_2O$ molecules in water and ice*

The current theory can also be used to calculate temperature dependencies for the most probable and mean thermal velocities of $H_2O$ molecules translations and librations forming primary and secondary effectons, respectively, for water (Fig. 12) and ice (Fig. 13).

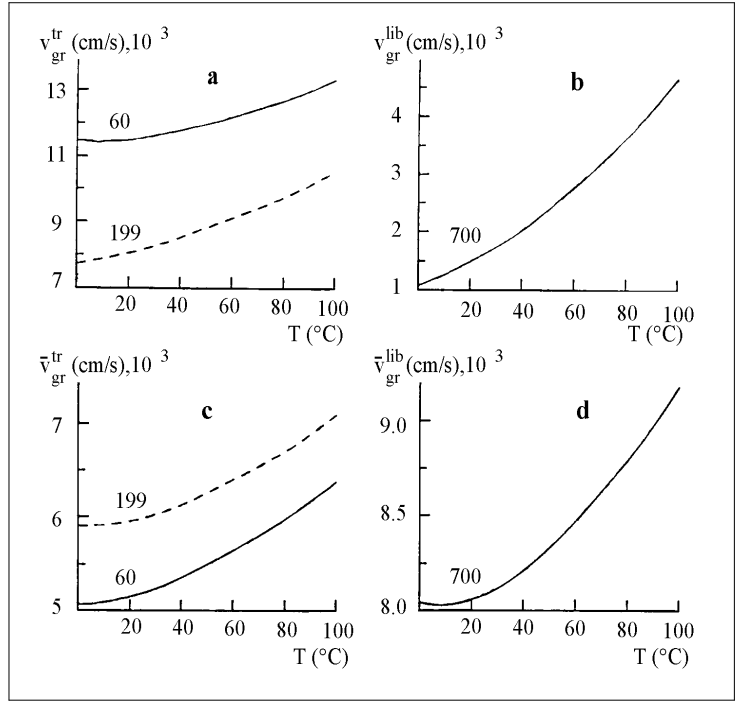

**Figure 12**. Theoretical temperature dependencies of the effective group velocities ($\mathbf{v}_{gr}$) of water molecules along main axes (1, 2, 3) of the effectons of the following types:
a) primary translational effectons
b) primary librational effectons
c) secondary translational effectons
d) secondary librational effectons

The ciphers denote the wave numbers ($\tilde{\nu}^{1,2,3}$) of the maxima of bands in the oscillatory spectra of water at $0^0 C$, according to which the calculations were made. The only maximum corresponding to librations ($\tilde{\nu}_{ph}^{(1)} = \tilde{\nu}_{ph}^{(2)} = \tilde{\nu}_{ph}^{(3)} = 700 cm^{-1}$) indicates the isotropy of this movement type ( *i.e.* the equality of the effective velocities in directions 1, 2, 3).

The effective group velocities were determined from the formula:
$(\mathbf{v}_{gr}^{1,2,3} = h/m\lambda_a^{1,2,3})_{tr,lb}$; $(\bar{\mathbf{v}}_{gr}^{1,2,3} = h/m\bar{\lambda}_a^{1,2,3})_{tr,lb}$, where $\lambda_a^{1,2,3}$ and $\bar{\lambda}_a^{1,2,3}$ were calculated from the (eq. 2.60 and 2.61); $m$ is the mass of the water molecule.

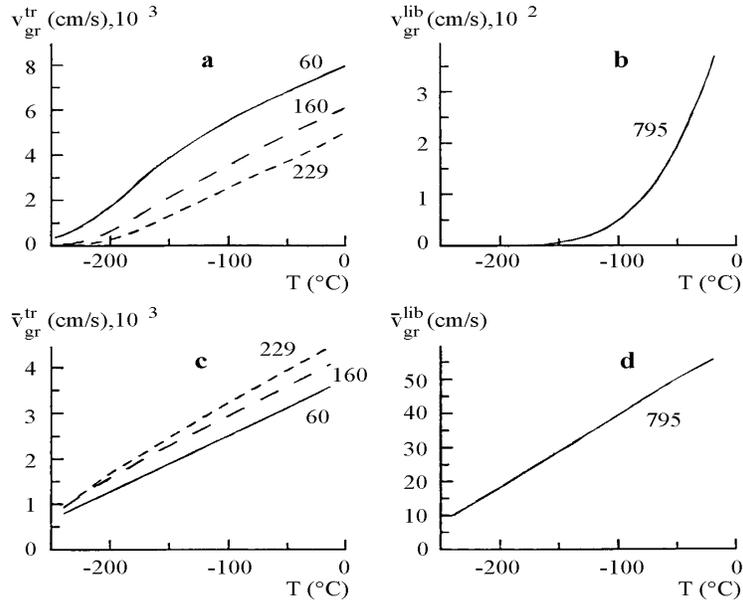

**Figure 13.** Theoretical temperature dependencies of the effective group velocities of water molecules along effectons symmetry directions (1, 2, 3) in composition of the following collective excitations in ice:
a) primary translational effectons
b) primary librational effectons
c) secondary translational deformons
d) secondary librational deformons
The ciphers denote the wave numbers of the band maxima in the oscillatory spectra of ice at $0^0 C$, according to which the calculations were made.

Thus, every de Broglie wave length $(\lambda_B = h/m\mathbf{v}_{gr})_{1,2,3}$ of water molecules, determining the length of three edges of an effecton, has its corresponding effective group velocity.

In the case of asymmetric translational effectons of ice (Figure 13 a, c), the de Broglie wave lengths and group velocities are different in selected directions (1, 2, 3). These quasi-particles are anisotropic and their shape is similar to a parallelepiped with unequal edges. As the temperature is lowered, the values of the most probable and mean velocities of $H_2O$ molecules approach zero.

The de Broglie waves forming the primary *translational* effectons have a group velocity $(\mathbf{v}_{gr})$ higher than that of librational effectons (Figure 13), however, this is *more than one order lower than* the speed of sound $(\mathbf{v}_{sound}^{H_2O} = 1.5 \times 10^5 cm/s)$. This fact confirms once again that the acoustic phonons and the de Broglie waves are different physical parameters and should be analyzed separately.

The temperature dependencies of de Broglie wave lengths for $H_2O$ molecules which determine the dimensions of primary and secondary effectons (tr and lib) in water and ice are presented in Figure 14.

The abrupt increase in the de Broglie wave length of primary librational effecton molecules $(\lambda_B^{lb})$ in ice in the region of temperatures lower than $-200^0$, as predicted by the theory, is an interesting fact. This abrupt increase implies that the size of this type of quasi-particle becomes macroscopic. In this range, substantial changes in the mechanical properties of ice should be expected.

The increasing of the contribution of quantum phenomena, such as the tunneling of protons, strongly increasing the effective velocity of their diffusion, also can also be predicted in ice as the temperature is lowered. *The length of tunneling (quantum leap) should correlate with the linear dimensions of primary effectons (tr and lb).*

These dimensions in ice can be hundreds and even thousands of times larger than in water, depending on both the temperature and the quality of the ice. This can explain the much higher rate of self-diffusion of protons in ice, than in liquid water (more than a hundred times).

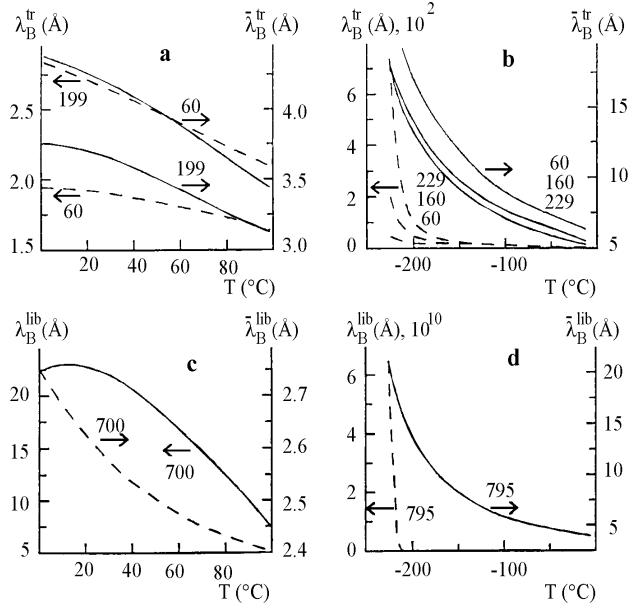

**Figure 14**. Calculated temperature dependencies of de Broglie wave lengths of vibrating $H_2O$ molecules along the directions (1, 2, 3) of the following types of effectons in water (a,c) and ice (b, d):
a) translational effectons of water: primary (left) and secondary (right)
b) translational effectons of ice: primary and secondary
c) librational effectons of water: primary (left) and secondary (right)
d) librational effectons of ice: primary (left) and secondary (right)
The ciphers denote the wave numbers of the band maxima for ice and water at $0^0 C$, according to which the calculations were carried out using formulae (2.60 and 2.6 1).

The de Broglie wave length of small guest molecules, incorporated in the volume of primary effectons, should be directly influence the length of the edges of the effectons.

### 6.4 Mechanism of the 1st-order phase transitions based on the Hierarchic Theory

The abrupt increase of the total internal energy (U) as a result of ice melting (Figure 15 a), equal to 6.27 *kJ/M*, calculated from the current theory, is close to the experimental value (6 *kJ/M*) (Eisenberg and Kauzmann, 1969). The resulting thermo-accessibility (Z) during the $[ice \to water]$ transition decreases abruptly, while the potential and kinetic energies increase (Figure 15b).

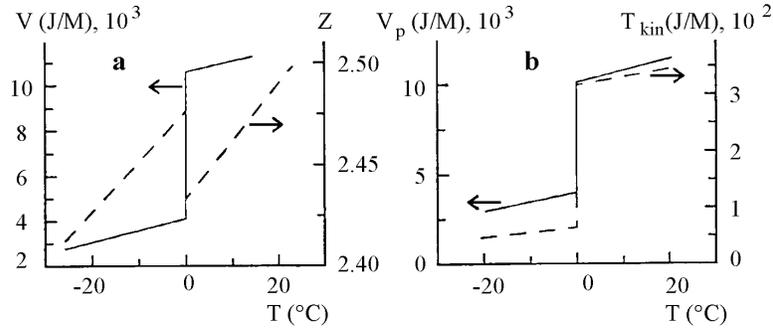

**Figure 15** *(a)*. The total internal energy ($U = T_{kin} + V_p$) change during an ice-water phase transition, and the corresponding change of the resulting thermo-accessibility (Z); **(b)** changes in kinetic ($T_{kin}$) and potential ($V_p$) energies as a result of the same transition.

It is important that at the melting point of $H_2O$, the number of molecules in a primary *translational* effecton $(n_M^{tr})_{ef}$ decreases from 1 to $\simeq 0.4$ (Figure 16 a). This fact implies that the volume of this type of quasi-particle becomes smaller than the volume occupied by one $H_2O$ molecule. According to the current model, under such conditions, the individual water molecules acquire independent translational mobility.

The number of water molecules forming a primary *librational* effecton decreases abruptly from about 3000 to 280, as a result of melting. The number of $H_2O$ molecules in the effective secondary librational effecton decreases correspondingly from $\sim 1.25$ to $0.5$ (Figure 16 b).

Figure 17 *a, b* contains more detailed information about changes in primary librational effecton parameters in the course of ice melting.

The theoretical dependencies obtained, allow us to give a clear interpretation of the first-order phase transitions. The condition of melting at $T = T_{cr}$ is realized in the course of heating when the number of molecules in the volume of primary translational effectons $n_M$ decreases and become less, than one:

$$n_M^{solid} \geq 1 \; (at \; T \leq T_{cr}) \xrightarrow{T_c} n_M^{liquid} \leq 1 \; (at \; T \geq T_{cr}) \qquad 6.6$$

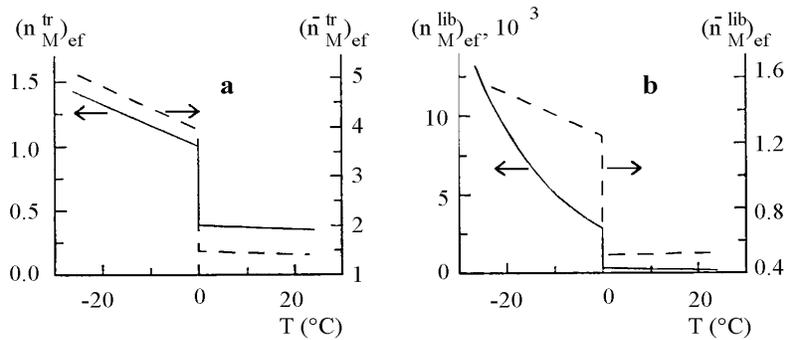

**Figure 16** *(a)*. Changes in the number of $H_2O$ molecules forming primary $(n_M^{tr})_{ef}$ and secondary $(\bar{n}_M^{tr})_{ef}$ translational effectons during an ice-water phase transition; **(b)** changes in the number of $H_2O$ molecules forming primary $(n_M^{lb})_{ef}$ and secondary $(\bar{n}_M^{lb})_{ef}$ librational effectons as a result of 1st order phase transitions.

The process of boiling, i.e. the [liquid → gas] transition is also determined by condition (6.6).

However, in this case the disassembly occurs not for primary translational effectons, but for primary librational effectons only. The former kind of coherent clusters are absent in liquid state and originate only after $[liquid \to solid]$ phase transition.

In other words, this means that the [gas → liquid] transition is a consequence of the assembly: mesoscopic Bose-condensation or primary librational effectons, as a cluster of molecules participating in coherent librations.

*In a liquid, as compared to a gas, the number of independent rotational degrees of freedom is decreased, due to librational primary effectons formation from number of coherently librating molecules (not independent already), but the number of translational degrees of freedom remains unchanged.*

*In turn, the number of independent translational degrees of freedom also decreases during the [liquid → solid] phase transition, as a result of primary translational effectons formation, containing number of coherent molecules with similar degrees of freedom.*

When the de Broglie wave length of molecules, determined by their translational vibrations, starts to exceed the mean distances between the centers of the molecules, this condition at corresponding temperature manifest the freezing process. The latter process of liquid →solid phase transition implies the initiation of translational mesoscopic Bose-condensation and the formation of primary translational effectons.

The dimensions of the librational effectons also increase strongly: from 20Å up to 50Å during a $[water \to ice]$ transition (Fig. 17b).

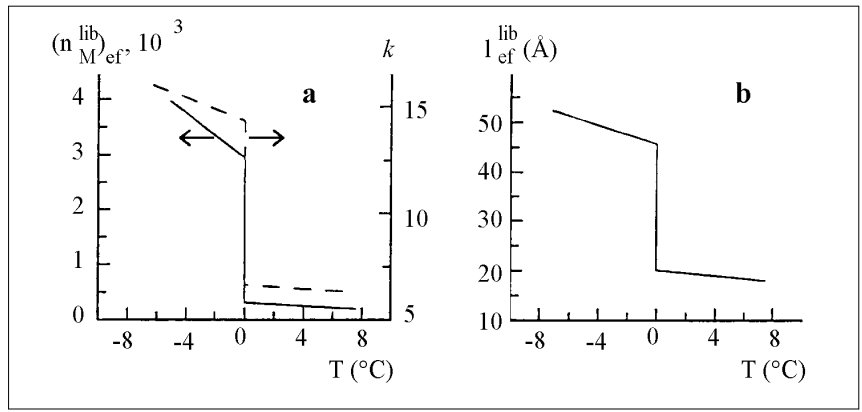

**Figure 17.** *(a)*: Changes of the number of $H_2O$ molecules forming a primary librational effecton $(n_M^{lb})_{ef}$, the number of $H_2O$ molecules ($\kappa$) along the edge of this effecton and *(b)*: the length of the effecton edge: $l_{ef}^{lb} = \kappa(V_0/N_0)^{1/3}$ during the ice-water phase transition.

### 6.5 Mechanism of the 2nd-order phase transitions following from Hierarchic theory

In contrast to first-order phase transitions, the second-order phase transitions are not related to the abrupt changes of primary effectons' volume, but only to their shape and symmetry changes. Such phenomena may be the result of a gradual [temperature/pressure] - dependent decrease in the difference between the dynamics and the energy of $a_i$ and $b_i$ states of one of the three standing de Broglie waves, forming these primary effectons, as follows:

$$\left\{ \begin{array}{l} \left[ h\nu_p = h\left( \nu_b - \nu_a \right) \right]_{tr,lb}^{i} \to 0 \\ \text{at } \left[ \lambda_b^{T_c} = \lambda_a^{T_c} \right]_{tr,lb}^{i} > \left( V_0/N_0 \right)^{1/3} \end{array} \right\} \qquad 6.7$$

As a result of the second-order phase transition, a new type of primary effecton, with a new geometry, appears. This implies the appearance of new values of energies of $(a)_{1,2,3}$ and $(b)_{1,2,3}$ - states, and a new constant of $(a \Leftrightarrow b)_{1,2,3}$ equilibrium. The polymerization of primary effectons to polyeffectons, stabilized by Josephson's junctions, tunneling effects and superposition of molecular de Broglie waves, accompanies this polymerization, turning the mesoscopic Bose condensation to macroscopic one. The examples of corresponding second-order phase transitions are superconductivity and superfluidity (sections 12.4 and 12.9).

The second-order phase transitions of ice, resulting in change of symmetry and dimensions of translational and librational effectons, can be induced by increasing the pressure at certain temperatures. This mechanism of 2nd order phase transition can be investigated in detail, using our optoacoustic system: Comprehensive Analyzer of Water Properties (CAMP). For description of CAMP see section 11.11 of this book.

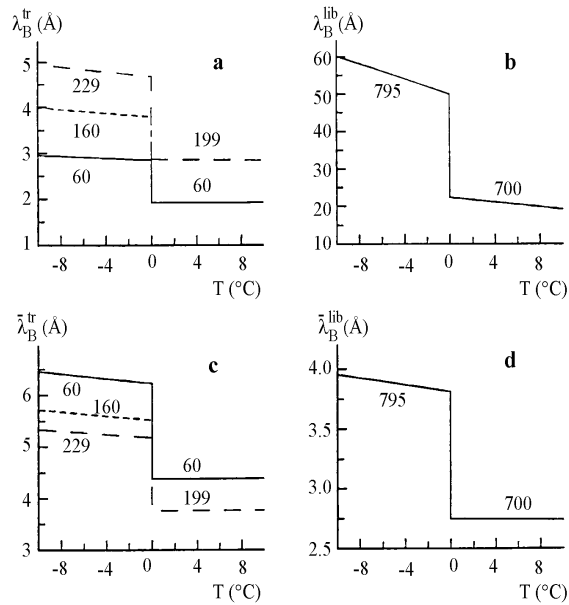

**Figure 18**. Calculated changes of the de Broglie wave lengths of $H_2O$ molecules forming primary and secondary effectons along directions (1, 2, 3) during ice-water phase transition for:
a) primary translational effectons
b) primary librational effectons
c) secondary translational effectons
d) secondary librational effectons

The ciphers denote the wave numbers of the band maxima in the oscillatory spectra of ice and water at $0^0 C$, according to which calculations were carried out using formulae (2.60 and 2.61).

*The second-order phase transition* is usually accompanied by non-monotonic changes of the velocity of sound and a low-frequency shift of translational and librational bands in oscillatory spectra (so-called "soft mode"). According to the current theory, these changes should be followed by an increase of heat capacity, compressibility, and in the coefficient of thermal expansion. The parameters of elementary cells, interrelated with the sizes and geometry of primary effectons, have to change also as a result of second-order phase transitions.

The Hierarchic theories of superfluidity and superconductivity are described in Chapter 12.

## 6.6 The energy of 3D quasi-particles discrete states in water and ice. Activation energy of macro- and super-effectons and corresponding deformons

Over the entire temperature range for water and ice, the energies of "acoustic" *a*-states of *primary effectons* (translational and librational) are lower than the energies of "optic" *b*-states (Figure 19). The energy of the 'ideal' effecton (3RT) with degenerate energies of *a*- and *b*-states has the following intermediate values.

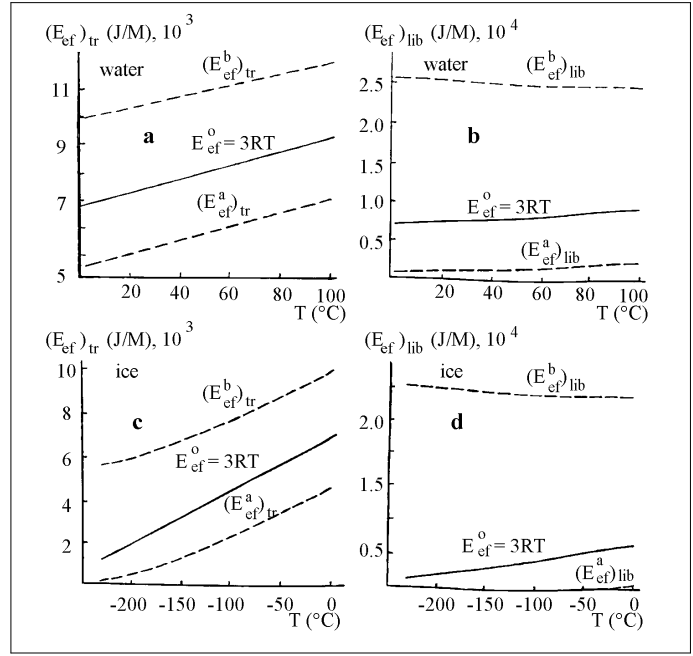

**Figure 19**. Temperature dependencies of the energy of primary effectons in "acoustic" *(a)* and "optical" *(b)* states and the energy of a harmonic 3D oscillator (the ideal effecton: $E_0 = 3RT$) in water and ice, calculated using formulas (4.6, 4.7 and 4.12):
a) for primary translational effectons of water in *a* and *b* states
b) for primary librational effectons of water in *a* and *b* states
c) for primary translational effectons of ice in *a* and *b* states
d) for primary librational effectons of ice in *a* and *b* states

According to (eq. 4.10 and 4.11) the *thermo-accessibilities* of *(a)* and *(b)* states are determined by the absolute values of the difference:

$$| E^a_{ef} - 3kT |_{tr,lb} \; ; \qquad | E^b_{ef} - 3kT |_{tr,lb} \; .$$

where $E_0 = 3kT = 3h\nu_0$ is the energy of the ideal effecton, pertinent for the gas-phase.

The $(a \Leftrightarrow b)$ transitions (quantum beats) can be considered as a jump over the thermal equilibrium state $(E_0)$, which is quantum-mechanically prohibited.

*The competition between the tendency of the primary effectons towards thermal equilibrium and their quantum properties can be considered as the driving force of quantum beats between (a) and (b) discrete states, IR photons radiation and absorption and other kinds of dynamics in condensed matter.*

The $(b \to a)$ transitions of primary effectons are accompanied by emission of photons, composing electromagnetic deformons, while the reverse transitions $(a \to b)$ are accompanied by photon absorption, i.e. origination and annihilation of primary IR deformons.

The nonequilibrium conditions in the subsystems of effectons and deformons, as was mentioned above, are a result of competition between the discrete quantum and continuous heat energy distributions, affecting the dynamics of these quasi-particles.

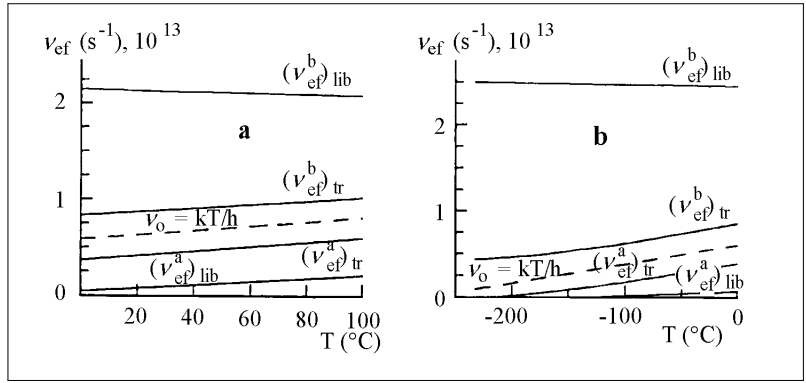

**Figure 20**. Temperature dependencies of the frequencies of (a) and (b) states of primary effectons - translational and librational for water (a) and ice (b), calculated from (Figure 19).

The relative distribution of frequencies in Figure 20 is the same as in Figure 19. The values of these frequencies reflect the minimal life-times of the corresponding states. The real life-time also depends on the probability of quantum "jumps" from one state to another.

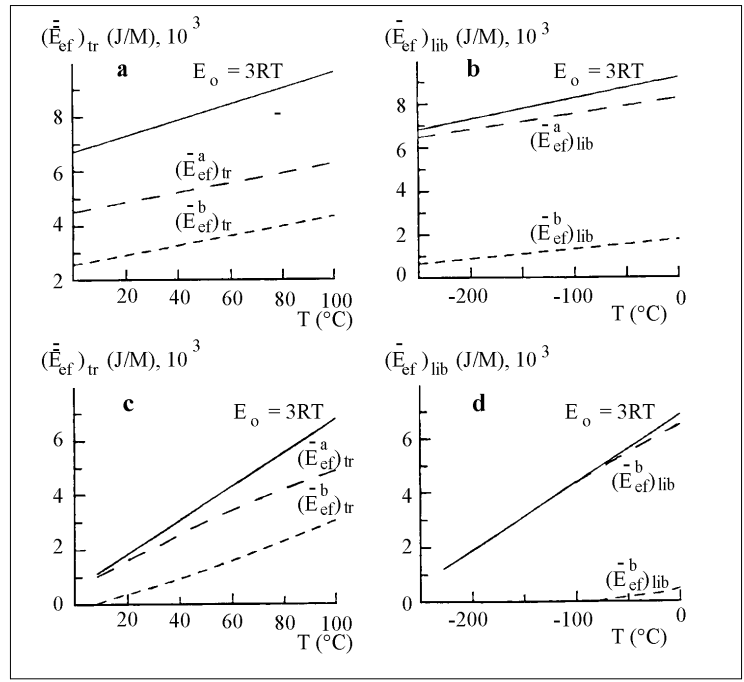

**Figure 21**. Temperature dependencies of the energy of secondary effectons in $\bar{a}$ and $\bar{b}$ states and energy of harmonic 3D oscillator (an ideal effecton: $E_0 = 3RT$) in water and ice, calculated using formulas (4.14, 4.15, and 4.12):
a) for secondary translational effectons of water
b) for secondary librational effectons of water
c) for secondary translational effectons of ice
d) for secondary librational effectons of ice

The energy of secondary effectons – the conventional quasi-particles in all cases – is lower than the energy of an ideal 'effecton' corresponding to the condition of thermal equilibrium: $E_0 = 3RT$ (Figure 21).

The thermo-accessibilities of $\bar{a}$ and $\bar{b}$ states are dependent on differences (see eq. 4.18; 4.19):

$$| \bar{E}_{ef}^{a} - 3kT |_{tr,lb} \; ; \qquad | \bar{E}_{ef}^{b} - 3kT |_{tr,lb}$$

where $E_0 = 3kT = 3h\nu_0$ is the energy of an 'ideal' effecton, pertinent for the gas phase.

$(\bar{a} \to \bar{b})$ transitions are related to the origination of phonons and acoustic deformons, while the reverse transitions $(\bar{b} \to \bar{a})$ are related to the absorption of them. Consequently these processes are accompanied by origination and annihilation of secondary acoustic deformons.

The secondary effectons are conventional, i.e. they are the result of averaging, using Bose-Einstein statistics; the parameters of transitions between their "acoustic" ($\bar{a}$) and "optic" ($\bar{b}$) states should also be considered as averaged values (Figure 22).

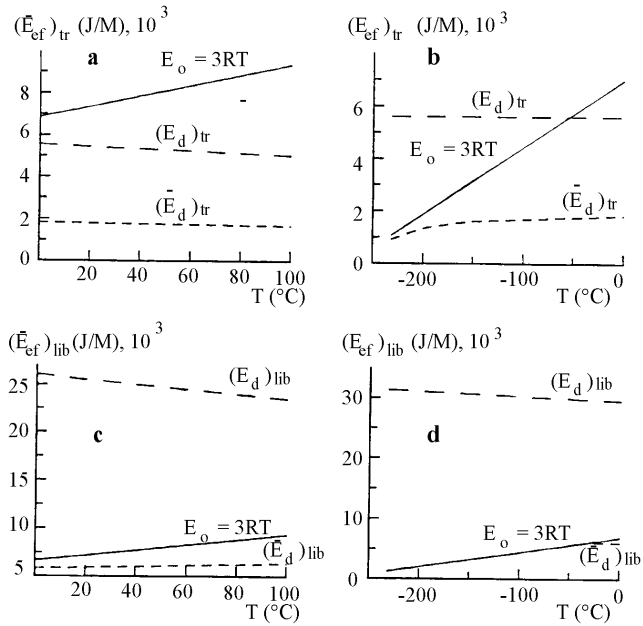

**Figure 22**. The temperature dependencies of the energies of primary and secondary deformons - translational (a, b) and librational (c, d) - and that of the energy of an 'ideal' deformon $E = 3RT$ for water (a, c) and ice (b, d). Calculations were carried out using formulae (4.20) and (4.21).

The energies of primary and secondary *deformons* of water and ice are placed asymmetrically - on the same side relative to 3RT (Figure 22) – in contrast to the situation with primary effectons. The energies of secondary deformons are usually lower than 3RT.

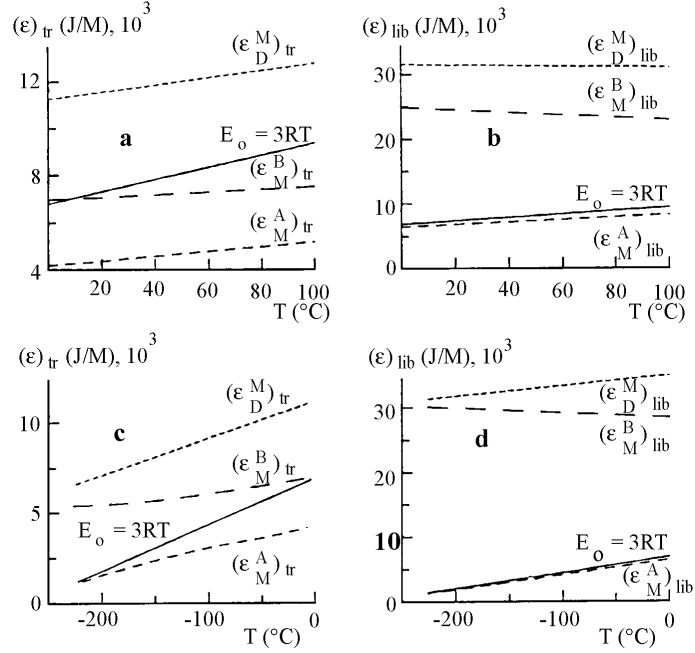

**Figure 23**. Temperature dependencies of the excitation energies: $(\epsilon_M^A)_{tr,lb}$ and $(\epsilon_M^B)_{tr,lb}$ of translational and librational macro-effectons in $A(a,\bar{a})$ and $B(b,\bar{b})$ states and the excitation energy of macro-deformons $(\epsilon_D^M)_{tr,lb}$ in water (a, b) and ice (c, d), calculated using formulas (6.8, 6.9 and 6.12). The $E_0 = 3RT$ is the energy of an ideal 3D excitation in thermal equilibrium.

The knowledge of excitation energies, presented on Figure 23, is very important for calculating of the viscosity and the coefficient of self-diffusion (see Chapter 11).

The A and B states of macro- and super-effectons represent large deviations from thermal equilibrium. The transitions between these states, causing the macro- and super-deformons, represent large fluctuations of polarizabilities, the refraction index, and dielectric permeability. It will be shown in Chapter 12 that the changes of the refraction index imply changes in the internal pressure and microviscosity.

The excitation energies of A and B states of macro-effectons (see eqs. 3.20 and 3.21) are determined in the following manner:

$$(\epsilon_M^A)_{tr,lb} = -RT\ln(P_{ef}^a \bar{P}_{ef}^a)_{tr,lb} = -RT\ln(P_M^A)_{tr,lb} \qquad 6.8$$

$$(\epsilon_M^B)_{tr,lb} = -RT\ln(P_{ef}^b \bar{P}_{ef}^b)_{tr,lb} = -RT\ln(P_M^B)_{tr,lb} \qquad 6.9$$

where $P_{ef}^a$ and $\bar{P}_{ef}^a$ are the thermo-accessibilities of the $(a) - (eq.\ 4.10)$ and $(\bar{a}) - (eq.\ 4.18)$ - states of the primary and secondary effectons, correspondingly; $P_{ef}^b$ and $\bar{P}_{ef}^b$ are the thermo-accessibilities of $(b) - (eq.\ 4.11)$ and $(\bar{b}) - (eq.\ 4.19)$ states of the same effectons.

The activation energy for super-deformons is:

$$\epsilon_{D^*}^s = -RT\ln(P_D^s) = -RT[\ln(P_D^M)_{tr} + \ln(P_D^M)_{lb}] = (\epsilon_D^M)_{tr} + (\epsilon_D^M)_l \qquad 6.10$$

The value $(\epsilon_D^M)_{tr} \approx 11.7 kJ/M \approx 2.8$ kcal/M characterizes the activation energy for *translational self-diffusion of water molecules*, and $(\epsilon_D^M)_{lb} \approx 31 kJ/M \approx 7.4$ kcal/M - the activation energy for librational self-diffusion of $H_2O$ (see Section 11.6). The latter value *is close to the energy of the*

*hydrogen bonds in water* (Eisenberg and Kauzmann, 1969).

It will be shown later (Section 11.5) that certain fluctuations related to translational and librational macro-deformons are responsible for the different contributions to viscosity. On the other hand, the largest fluctuations – super-deformons – are responsible for the process of cavitational fluctuations in liquids and the development of defects in solids. They determine both vapor pressure and sublimation (see Section 11.3).

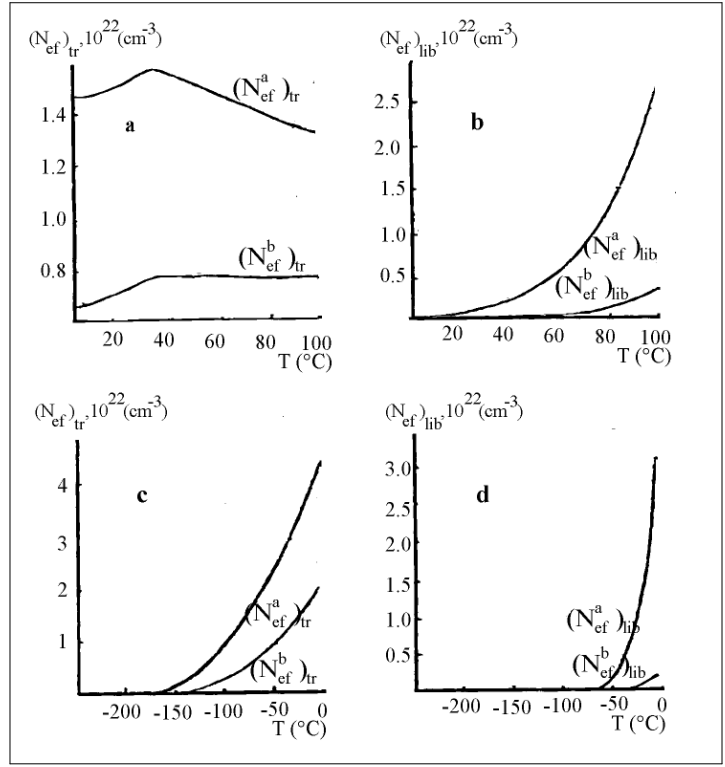

**Figure 24**. Temperature dependencies of the concentrations of primary effectons (translational and librational) in (a) and (b) states: $(N^a_{ef})_{tr,lb}, (N^b_{ef})_{tr,lb}$ and primary deformons $(N_d)_{tr,lb}$ for water (a and b); as well as similar dependencies for ice (c and d).
Concentrations of quasi-particles were calculated using eq. 4.3 as follows:

$$(N^a_{ef})_{tr,lb} = (n_{ef} P^a_{ef}/Z)_{tr,lb};$$

$$(N^b_{ef})_{tr,lb} = (n_{ef} P^b_{ef}/Z)_{tr,lb};$$

$$(N_d)_{tr,lb} = (n_d P_d/Z)_{tr,lb}$$

These parameters can be considered as the corresponding quasi-particles distribution functions.

To obtain such information, as presented at Fig.24, using conventional tools, i.e. by means of x-ray or neutron scattering methods, is impossible or very complicated task.

It is obvious then, from the scope of results presented in this Chapter, that the Hierarchic Theory can be useful for solving many problems of condensed matter physics in both liquid and solid phase on a quantitative level.

# Chapter 7

## Interaction of light with matter

The description of matter presented above, as a hierarchic system of collective excitations formed by 3D standing de Broglie waves of molecules, internal IR photons, phonons and their superpositions, allows one to go to a more advanced level of analysis of the interaction of such system with external electromagnetic fields.

### 7.1  Polarizabilities and induced dipole moments of atoms and molecules

The electron shells of atoms and molecules can be considered as a system of standing de Broglie waves of the electrons in the field of central forces. In a very simplified approach, every stationary electron's orbit has its own characteristic length ($\lambda_e$) and radius ($L_e = \lambda_e/2\pi$).

Following the terminology set forth, we may state that the atoms and molecules are coherent superposition of electronic and nuclear effectons. We assume that at the stationary electron orbits due to the conditions of intrinsic harmonic equilibrium in potential field, the eqs. (2.9, 2.10 and 2.24) must be satisfied:

$$\mathbf{v}_{ph}^{res} = \mathbf{v}_{gr}^{res}; \quad T_k = |V| = E_B/2; \quad 2A_e = L_e \qquad 7.1$$

where $\mathbf{v}_{ph}^{res}$ and $\mathbf{v}_{gr}^{res}$ are the resulting phase and group velocities of an electron, taking into account its orbital and spin dynamics composing the atom; $A_e$ and $L_e$ are the most probable amplitude and de Broglie wave radius of an electron; $T_k$, $V$ and $E_B$ are the kinetic, potential and total energies of the atom's electron.

The condition of the atom's stability, as a primary effecton, is the equality of the nuclear ($L_n$) and outer electron ($L_e$) de Broglie wave radii:

$$L_n = \hbar/m_n \mathbf{v}_{gr}^n = \hbar/m_e \mathbf{v}_{gr}^{res} = L_e \qquad 7.1a$$

The absorption and radiation of photons can be considered as a result of quantum beats between de Broglie waves of electrons in the main *(a)* and excited *(b)* states. If these states are characterized by frequencies $\omega_a$ and $\omega_b$, respectively (Grawford, 1973), then the beats frequency is:

$$\omega_p = \omega_a - \omega_b = \Delta\omega_a \qquad 7.2$$

where $\Delta\omega_a = \omega_p$ is the increment of the frequency of the electron's de Broglie wave resulting from the transition of the atom to excited state due to photon absorption.

If condition (7.1) is satisfied, the cyclic frequency of the electron in the atom, as a standing de Broglie wave, is equal to (see eq. 2.20):

$$\omega_e = \frac{\hbar}{2m_e A_e^2} = \frac{\hbar}{m_e L_e^2} \qquad 7.3$$

Differentiating this formula we have:

$$\Delta\omega_e = \frac{2\hbar}{m_e L_a^3} \mid \Delta L_e \mid = \omega_p \qquad 7.4$$

where $\omega_p$ is the incident photon frequency.

*Let us show that the most probable radius of the outer electrons' orbit in the atom or molecule cubed - has the meaning of polarizability:*

$$L^3 = \alpha = \frac{e^2}{m_e \omega_e} = \frac{1}{3}(\alpha_1 + \alpha_2 + \alpha_3) \qquad 7.5$$

where $\alpha_1$; $\alpha_2$ and $\alpha_3$ are the main components of the polarizability tensor.

Combining (7.4) and (7.5), it is possible to determine the de Broglie wave's outer electron radius

increment, induced by a photon:

$$\Delta L_e = \frac{\omega_p}{2\hbar} m_e \alpha \qquad 7.6$$

where $m_e$ is the electron mass and $\omega_p$ is the frequency of an incident photon.

Let us consider the atom under the effect of the external electric field E. Then the force affecting an electron and increasing its de Broglie wave's most probable radius by the quantity $\Delta l$ is equal to:

$$F = eE = D\Delta l = m_e \omega_e \Delta l \qquad 7.7$$

where $D = m_e \omega_e$ is the rigidity index as approximated by a harmonic oscillator.

From eq. (7.7)

$$E = \frac{m_e \omega_0^2 \Delta l}{e} \qquad 7.8$$

The induced dipole moment of the atom or molecule:

$$P_i = e\Delta l = \alpha E \qquad 7.9$$

Substituting eq. (7.8) into the right hand side of eq. (7.9) we get, for the polarizability:

$$\alpha = \frac{e^2}{m_e \omega_e^2} \qquad 7.10$$

This formula yields a relationship between the atom polarizability, the charge and the mass of an electron, and its de Broglie wave frequency at the outer orbit ($\omega_e$) (Kittel, 1978).

On the other hand, from a simple Bohr model, the condition of the stationary atom's outer orbit with radius $L_e$, is:

$$\frac{m_e v_{gr}^2}{L_e} = \frac{e^2}{L_e^2} \rightarrow m_e v_{gr}^2 = 2T_k = |V| = \frac{e^2}{L_e} \qquad 7.11$$

As far the potential energy of an outer electron in an atom as a harmonic oscillator is $|V| = m_e \omega_0 L_e$, then from eqs. (7.11) and (7.10) we obtain:

$$L_e^3 = \frac{e^2}{m_e \omega_e^2} = \alpha \qquad 7.12$$

The probability of the electron ($W_e$) to be located at a given point of space is proportional to its de Broglie wave amplitude squared ($A_e^2$):

$$W_e \sim A_e^2 = \frac{L_e^2}{2} \qquad 7.13$$

The change of this probability under the influence of outer [external?] field $\vec{E}$ may be expressed as:

$$\Delta W_e \sim 2A_e \Delta A_e = L_e \Delta L_e \qquad 7.14$$

Dividing (7.14) by (7.13), we derive the relative change of the probability to detect an electron at the given point of space:

$$\frac{\Delta W_e}{W_e} = 2\frac{\Delta A_e}{A_e} = 2\frac{\Delta L_e}{L_e} \qquad 7.15$$

We assume that the change of an electron's orbit radius under the effect of the outer EM field ($\vec{E} = \hbar\omega_p$) on a certain value ($\vec{\Delta l}$) is equal to the product of the relative probability for the location of the electron in the given direction ($\Delta W_e/W_e$) and the increment of its stationary orbit radius ($\vec{\Delta L_e}$):

$$\vec{\Delta l} = \frac{\Delta W_e}{W_e} \vec{\Delta L_e} = \frac{2\vec{\Delta L_e}^2}{L_e} \qquad 7.16$$

Putting (7.6) into (7.16), we obtain:

$$\vec{\Delta l} = \alpha \frac{\omega_p^2}{2} \left( \frac{\vec{L_e} m_e}{\hbar} \right)^2 \qquad 7.17$$

Putting (7.17) into (7.9), we derive a formula for the induced dipole moment:

$$\vec{P_i} = e\vec{\Delta l} = \alpha \frac{e\omega_p^2}{2} \left( \frac{\vec{L_e} m_e}{\hbar} \right)^2 = \alpha \vec{E} \qquad 7.18$$

From (7.18) we obtain the formula relating the *tension of the electric component of the electromagnetic field* at the atom with this field (photons) frequency ($\omega_p$) and the radius of the electron's outer orbit ($L_e$) in the atom:

$$\vec{E} = \vec{E_S} = \frac{e\omega_p^2}{2} \left( \frac{\vec{L_e} m_e}{\hbar} \right)^2 \qquad 7.19$$

where ($\vec{E_S}$) is the electric tension of the secondary EM wave just near an excited atom.
If, at a stationary orbit, the de Broglie wave frequency of an electron is:

$$\omega_e = \frac{\hbar}{m_e L_e^2}, \qquad 7.20$$

then the equation (7.19) can be represented as:

$$E_S = \frac{e}{2L_e^2} \frac{\omega_p^2}{\omega_e^2} = \frac{e}{2} \alpha^{-2/3} \left( \frac{\omega_p}{\omega_e} \right)^2 \qquad 7.21$$

Taking into account that the circular photon frequency is:

$$\omega_p = \frac{c}{L_p} = \frac{2\pi c}{\lambda_p} \qquad 7.22$$

the Compton wavelength of an electron is:

$$L_0 = \frac{\hbar}{m_e c}$$

and polarizability is: $\quad L_e^3 = \alpha \qquad 7.23$

for the case of spherical symmetry of the atom, (eq. 7.19) for the electric tension of the secondary EM wave, can be rewritten in the following form:

$$E_S = \frac{e}{L_0^2} L_e^2 \frac{2\pi^2}{\lambda_p} = \frac{e}{L_0^2} \frac{\alpha}{L_e} \frac{2\pi^2}{\lambda_p} \qquad 7.24$$

It will be shown below that by using (7.24), the Rayleigh formula for light scattering in gases can be derived.

### 7.2. The Rayleigh formula

It follows from (7.24) that under the influence of photons with the wavelength $\lambda_p$, an atom or a molecule radiate the *secondary* electromagnetic waves with the electric components $\vec{E_S}$, depending on the polarizability ($\alpha$).

The electric tension of the secondary/scattered wave ($\vec{E_S}$) decreases with the distance from the molecule as ($1/r$). This relationship also depends on the angle $\Phi$ between the primary $\vec{E}$ vector of the incident light (normal to the light direction) and the radius vector ($r$), directed from the radiating dipole to the point of observation.

The dependence of the secondary wave tension $\vec{E_r}$ on the distance between the induced dipole (oscillating with the frequency $\omega_p$) and the detector of the secondary wave can be presented as:

$$\vec{E}_r = \vec{E}_S \left| \frac{\Delta E(r)}{E_S} \right| \qquad 7.25$$

In the general case:
$$E(r) = \frac{q}{r^2} \quad \text{and} \quad |\Delta E(r)| = \frac{2q}{r^3} \Delta r$$

Therefore:
$$\frac{|\Delta E(r)|}{E_S} = \frac{|2\Delta r|}{r} \sin \Phi \qquad 7.26$$

If we assume that at $\Delta r_{max} = L_e$, we have $|E_{max}| = E_S$, then:
$$E_r = E_S \frac{2L_e}{r} \sin \Phi \qquad 7.27$$

Combining (7.24) and (7.27) we get:
$$E_r = \frac{e}{L_0^2} \frac{\alpha}{r} \frac{4\pi^2}{\lambda_p^2} \sin \Phi \qquad 7.28$$

Introducing the term:
$$E_0 = \frac{e}{L_0^2} \qquad 7.29$$

we can deduce from (7.28):
$$E_r = E_0 \frac{4\pi^2}{\lambda_p^2} \frac{\alpha}{r} \sin \Phi \qquad 7.30$$

This equation characterizes the *electric field tension* of the secondary wave on the distance ($r$) from the radiating molecule. The total tension $E_v$ from the volume element (**v**), containing on the average ($\bar{n}$) molecules, is equal to:

$$E_v = E_r \bar{n} = E_0 \bar{n} \frac{4\pi^2}{\lambda_p^2 r} \alpha \sin \Phi \qquad 7.31$$

As a result of the density fluctuations, the number of molecules in the different volume elements (**v**) can vary from the mean number by the magnitude (Vuks, 1977):

$$\Delta \bar{n} = n - \bar{n} \qquad 7.32$$

Due to those differences in the independent volume elements whose sizes are much less than the light wavelength, the secondary waves are not totally quenched as a result of interference. Consequently, a redundant (uncompensated) field appears, which is equal to:

$$\Delta \bar{E}_v = E_v - \bar{E}_v = E_0 \Delta \bar{n} \frac{4\pi^2}{\lambda_p^2 r} \alpha \sin \Phi. \qquad 7.33$$

The intensity of scattered light is determined by the uncompensated electric field mean value squared. Raising (7.33) to the second power and using the known formula from statistical physics:

$$(\Delta \bar{n})^2 = \bar{n} = n_M \mathbf{v}$$

we derive the intensity of light scattering:

$$I_v = (\Delta \bar{E}_v)^2 = I_0 n_M \mathbf{v} \frac{16\pi^4}{\lambda_p^4} \frac{\alpha^2}{r^2} \sin^2 \Phi \qquad 7.34$$

where $I_0 = E_0^2$ and $n_M$ is the equilibrium concentration of molecules.

If the volume elements (**v**) can be considered as independent of each other, similar to the case of a gas phase, then the resulting scattering of the gas with macroscopic volume $V$, is equal to the sum of the scattering of its elements (**v**).

Hence,

$$I_V = \Sigma I_v = I_0 n_M V \frac{16\pi^4}{\lambda_p^4} \frac{\alpha^2}{r^2} \sin^2\Phi \qquad 7.35$$

*We get the expression exactly coinciding with the Rayleigh formula for light scattering in gases.* This supports the correctness of our assumptions and the intermediate formulae used for deducing them.

One can introduce into (7.35) an angle $\theta$ between the polarized incident primary beam interacting with sample, and the scattered secondary beam. The direction of the light beam propagation is normal to $\vec{E}$, hence:

$$\Phi = \theta + \pi/2$$

or

$$\theta = \Phi - \pi/2$$

In this case,

$$\sin^2\theta = \cos^2\Phi$$

It is not difficult to show that the scattering intensity of the natural light (mixed in the $\vec{E}$ polarization) is expressed as:

$$I = I_0 n_M V \frac{8\pi^4}{\lambda^4} \frac{\alpha^2}{r^2}(1 + \cos^2\theta) \qquad 7.36$$

In common practice, the scattering coefficient $R = \frac{I}{I_0} \frac{r^2}{V}$ is usually measured at the right angle $\theta = 90^0$. Consequently, for such a condition we derive, from (7.36):

$$R = \frac{I}{I_0} \frac{r^2}{V} = \frac{8\pi^4}{\lambda^4} \alpha^2 n_M \quad (cm^{-1}) \qquad 7.37$$

where $n_M$ is the concentration of gas molecules.

*Let us try to use this approach for the description of light refraction and light scattering in transparent condensed media.*

# Chapter 8

# New approach to theory of light refraction

### 8.1. Refraction in gas

If the action of photons on the electrons of molecules is considered to be a force which activates a harmonic oscillator with decay, this concept leads to the known classical equations for a complex refraction index (Vuks, 1984).

The Lorentz-Lorenz formula obtained in such a way is convenient for practical needs. However, it does not describe the dependence of the refraction index on the incident light frequency; also, it does not take into account the intermolecular interactions. In the current theory of light refraction, we attempt to elucidate the relationship between these parameters.

The basic idea of our approach is that the dielectric permittivity of matter $\epsilon$, (equal in the optical interval of EM waves frequencies to the refraction index squared: $\epsilon = n^2$), is determined by the ratio of *partial volume energies of photon* in a vacuum to a similar partial volume energy of photon in matter:

$$\epsilon = n^2 = \frac{[E_p^0]}{[E_p^m]} = \frac{m_p c^2}{m_p c_m^2} = \frac{c^2}{c_m^2} \qquad 8.1$$

where $m_p = h\nu_p/c^2$ is the effective photon mass, $c$ is the velocity of light in a vacuum, and $c_m$ is the effective velocity of light in matter.

We introduce the notion of *partial volume energy of photon* in a vacuum $[E_p^0]$ and in matter $[E_p^m]$ as the product of the photon energy ($E_p = h\nu_p$) and the volume ($V_p$) occupied by the given 3D standing wave of photons in a vacuum and in matter, correspondingly:

$$[E_p^0] = E_p V_p^0 \qquad [E_p^m] = E_p V_p^m \qquad 8.2$$

The volume of primary electromagnetic deformon is determined by superposition of three different photons, propagating in space in directions, normal to each other in the space of these photons interception (see Introduction).

In a vacuum, the effect of excluded volume, provided by spatial incompatibility of the electronic shells of molecules with photons is absent; thus, the volume of a 3D photon standing wave (primary deformon) (see 3.4) is:

$$V_p^0 = \frac{1}{n_p} = \frac{3\lambda_p^2}{8\pi} \qquad 8.3$$

We will consider the interaction of light with matter *in this mesoscopic volume*, containing thousands of molecules of condensed matter. This is the reason why presented here theory of light refraction can be called a mesoscopic theory.

Putting (8.3) into (8.2), we obtain the formula for the partial volume energy of standing photon in vacuum:

$$[E_p^0] = E_p V_p^0 = h\nu_p \frac{9\lambda_p^2}{8\pi} = \frac{9}{4} hc\lambda_p^2 \qquad 8.4$$

Then we proceed from the assumption that de Broglie waves of photons *can not be spatially compatible with de Broglie waves of the electrons, which form the shells of atoms and molecules*. Hence, the effect of excluded volume appears during the propagation of an external electromagnetic wave through matter. This leads to the fact that, in matter, the volume occupied by a photon is equal to:

$$V_p^m = V_p^0 - V_p^{ex} = V_p^0 - n_M^p V_e^M \qquad 8.5$$

where $V_p^{ex} = n_M^p V_e^M$ is the excluded volume, which is equal to the product of the number of molecules in the volume of one photon standing wave ($n_M^p$) and the volume occupied by the electronic shell of one molecule ($V_e^M$).

$n_M^p$ is determined by the product of the volume of the photons 3D standing wave in the vacuum (8.3) and the concentration of molecules ($n_M = N_0/V_0$):

$$n_M^p = \frac{9\lambda_p^3}{8\pi} \left( \frac{N_0}{V_0} \right) \qquad 8.6$$

In the absence of the polarization by the external field and intermolecular interaction, the volume occupied by electrons of the molecule is:

$$V_e^M = \frac{4}{3} \pi L_e^3 \qquad 8.7$$

where $L_e$ is the radius of the most probable de Broglie wave ($L_e = \lambda_e/2\pi$) of the outer electron of a molecule. As it has been shown in (7.5), the mean molecule polarizability is:

$$\alpha = L_e^3 \qquad 8.8$$

Then taking (8.7) and (8.6) into account, the excluded volume of primary electromagnetic deformons in the matter is:

$$V_p^{ex} = \frac{9\lambda_p^3}{8\pi} n_M \frac{4}{3} \pi \alpha = \frac{3}{2} \lambda_p^3 n_M \alpha \qquad 8.9$$

Therefore, the partial volume energy of a photon in a vacuum is determined by (eq. 8.4), while the

partial volume energy in matter, according to (8.5) is:
$$[E_p^m] = E_p V_p^m = E_p[V_p^0 - V_p^{ex}] \qquad 8.10$$

Putting (8.4) and (8.10) into (8.1) we obtain:
$$\epsilon = n^2 = \frac{E_p V_p^0}{E_p(V_p^0 - V_p^{ex})} = \frac{V_p^0}{V_p^0 - V_p^{ex}} \qquad 8.11$$

or
$$\frac{1}{n^2} = 1 - \frac{V_p^{ex}}{V^0} \qquad 8.12$$

Then, putting (eq. 8.9) and (8.3) into (8.12) we derive:
$$\frac{1}{n^2} = 1 - \frac{4}{3}\pi n_M \alpha \qquad 8.13$$

or
$$\frac{n^2 - 1}{n^2} = \frac{4}{3}\pi n_M \alpha = \frac{4}{3}\pi \frac{N_0}{V_0} \alpha \qquad 8.14$$

where $n_M = N_0/V_0$ is a concentration of molecules.

In this equation $\alpha = L_e^3$ is the average static polarizability of molecules for the case when the external electromagnetic fields, as well as intermolecular interactions inducing the additional polarization, are absent or negligible. This situation is realized when $E_p = h\nu_p \to 0$ and $\lambda_p \to \infty$ in the gas phase. As it will be shown below, the resulting value of $\alpha^*$ in condensed matter is much larger.

### 8.2. Light refraction in liquids and solids

According to the Lorentz classical theory, the electric component of the outer electromagnetic field is amplified by an additional inner field ($E_{ad}$). This field is related to the interaction of induced dipole moments, which compose condensed matter, with one-another:
$$E_{ad} = \frac{n^2 - 1}{3} E \qquad 8.15$$

The mean Lorentz acting field $\bar{F}$ can be expressed as:
$$\bar{F} = E + E_{ad} = \frac{n^2 + 2}{3} E \quad (\text{at } n \to 1, \bar{F} \to E) \qquad 8.16$$

$\bar{F}$ has a dimension of electric field tension and, in the gas phase, tends to $E$ when $n \to 1$.

In accordance with the Hierarchic Model, if one excludes the Lorentz acting field, the total internal acting field includes two other contributions, which increase the polarizability ($\alpha$) of molecules in condensed matter:

1. The potential intermolecular field, including all the types of van der Waals interactions composing coherent collective excitations, even without an external electromagnetic field. Similar to the total potential energy of matter, this contribution must be dependent on temperature and pressure;

2. The primary internal field, related to primary electromagnetic deformons (tr and lb). This component of the total acting field also exists without external fields. Its frequencies corresponds to the IR range and its action is much weaker than the action of external visible light.

Let us try to estimate the energy of the total acting field, which we introduce as:
$$A_f = h\nu_f = \frac{hc}{\lambda_f} = A_L + A_V + A_D \qquad 8.17$$

and its effective frequency ($\nu_f$) and wavelength ($\lambda_f$).

$A_L$, $A_V$ and $A_D$ are contributions, related to the Lorentz field, the potential field, and the primary deformons field, respectively.

When the interaction energy of a molecule with a photon ($E_p = h\nu_p$) is less than the energy of resonance absorption, this leads to elastic polarization of the electron shell, and to origination of secondary photons, i.e. light scattering. We assume that the increment of the polarization of a molecule ($\alpha$) under the action of an external photon ($h\nu_p$) and the total active field ($A_f = h\nu_f$) can be expressed by way of the increase of the most probable radius of the electron's shell ($L_e = \alpha^{1/3}$). Thus, using (eq. 7.6):

$$\Delta L^* = \Delta L_e + \Delta L_f = \frac{(h\nu_p + A_f) m_e}{2\hbar^2} \alpha \qquad 8.18$$

where $\alpha = L_e^3$ is the average polarizability of molecules in a gas phase at $\nu_f \to 0$.

For a water molecule in a gas phase:

$$L_e = \alpha^{1/3} = 1.13 \times 10^{-10}\, m$$

is a known constant, determined experimentally (Eisenberg and Kauzmann, 1969).

The total increment of polarizability radius ($\Delta L^*$) can be found from the experimental refraction index ($n$) and the resulting polarizability of molecules composing condensed matter affected by the acting field: $\alpha^* = (L^*)^3$ from (8.14):

$$L^* = (\alpha^*)^{1/3} = \left[ \frac{3}{4\pi} \frac{V_0}{N_0} \frac{n^2 - 1}{n^2} \right]^{1/3} \qquad 8.19$$

as:

$$\Delta L^* = L^* - L_e \qquad 8.20$$

From (8.18) we obtain an expression for the increment of the radius of polarizability ($\Delta L_f$), induced by the total internal acting field:

$$\Delta L_f = \Delta L^* - \Delta L_e = \frac{A_f m_e}{2\hbar^2} \alpha \qquad 8.21$$

Similar to the total internal acting field energy (8.17), this total acting increment can be represented as a sum of contributions, related to the Lorentz field ($\Delta L_F$), the potential field ($\Delta L_V$) and the primary deformon field ($\Delta L_D$):

$$\Delta L_f = \Delta L_L + \Delta L_V + \Delta L_D = \Delta L^* - \Delta L_e \qquad 8.22$$

The increment $\Delta L_e$, induced by the external photon only, can be calculated from the known frequency ($\nu_p$) of the incident light (see 8.18):

$$\Delta L_e = \frac{h\nu_p m_e}{2\hbar^2} \alpha \qquad 8.23$$

This implies that $\Delta L_f$ can be found from (8.21) and (8.17), using (8.23). From (8.21) we can calculate the energy ($A_f = h\nu_f$), the effective frequency ($\nu_f$) and wave length ($\lambda_f$) of the total acting field as follows:

$$A_f = h\nu_f = hc/\lambda_f = 2 \frac{\Delta L_f \hbar^2}{m_e \alpha} \qquad 8.24$$

The computer calculations of $\alpha^*$; $L^* = L_e + \Delta L^* = (\alpha^*)^{1/3}$ and $A_f$ in the temperature range ($0 - 95^0$) are presented on Figure 25.

One must keep in mind that, in the general case, $\alpha$ and $L$ are tensors. This means that all the increments, calculated on the basis of (eq. 8.18) must be considered as the *effective* increments. Nevertheless, it is obvious that this new approach to the analysis of the acting field parameters can give useful additional information about the properties of transparent condensed matter.

The temperature dependencies of these parameters were computed using known from literature experimental data on the refraction index $n(t)$ for water, and are presented in Figure 25 a. The radius $L^*$ in the range $0 - 95^0 C$ increases less than 1% at a constant incident light wavelength

($\lambda = 546.1 nm$). The change of $\Delta L_f$ with temperature is determined by its potential field component change $\Delta L_V$.

The relative change of this component: $\Delta\Delta L_V/\Delta L_f$ ($t = 0^0 C$) is about 9%.

Corresponding to this increment of radius of polarizability ($\Delta L_f$) the increasing of the acting field energy $A_f$ (eq. 8.23) is about $8 kJ/M$ (see Fig 25 b) due to its potential field contribution.

It is important that the total potential energy of water in the same temperature range, according to our calculations, increases by the same magnitude (Figure 5 b). This fact points to a strong correlation between the potential intermolecular interaction in matter and the value of the acting field energy. In graphical calculations in Figure 25, the experimental temperature dependencies of the water refraction index were obtained by Frontas'ev and Schreiber (1965).

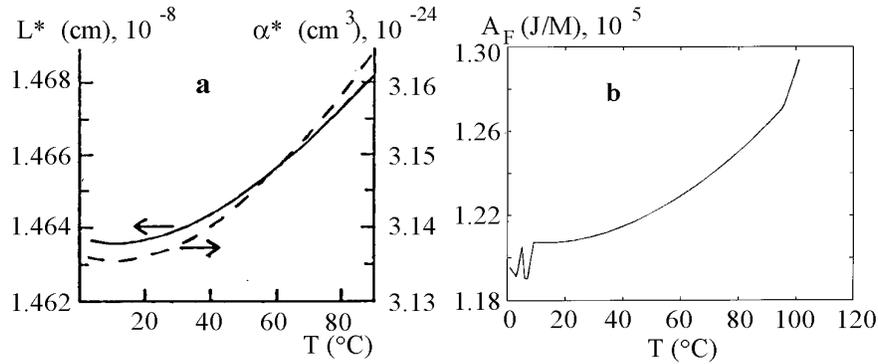

**Figure 25**. *(a)* - Temperature of the most probable outer electron shell radius of $H_2O$ ($L^*$) and the effective polarizability $\alpha^* = (L^*)^3$ in the total acting field;
*(b)* - Temperature dependence of the total acting field ($A_f$) energy in water at the wavelength of the incident light $\lambda_p = 5.461 \times 10^{-5} cm^{-1}$. The known experimental data for refraction index $n(t)$ were used in calculations. The initial electron shell radius is:
$L_e = \alpha_{H2O}^{1/3} = 1.13 \times 10^{-8} cm$ (Eisenberg and Kauzmann, 1969).

It was calculated, that at constant temperature ($20^0$) the energy of the acting field ($A_f$), (*eq*. 8.23) in water is practically independent of the wavelength of incident light ($\lambda_p$). At more than three time alterations of $\lambda_p$: from $12.56 \times 10^{-5} cm$ to $3.03 \times 10^{-5} cm$ and the water refraction index ($n$) from 1.320999 to 1.358100 (Kikoin, 1976), the value of $A_f$ changes less than by 1%.

Under the same conditions, the electron shell radius $L^*$ and the acting polarizability $\alpha^*$ thereby increase from $(1.45$ to $1.5) \times 10^{-10} m$ and from $(3.05$ to $3.274) \times 10^{-30} m^3$ respectively (Figure 26). These changes are due to the incident photons' action on the electronic shell only. For water molecules in the gas phase and $\lambda_p \to \infty$, the initial polarizability ($\alpha = L_e^3$) is equal to
$1.44 \times 10^{-24} cm^3$ (Eisenberg and Kauzmann, 1969), *i.e.* significantly less than in condensed matter under the action of external and internal fields.

Obviously, the temperature change of energy $A_f$ (Figure 25 b) is determined by the increasing internal pressure (section 11.2), related to an intermolecular interaction change, depending on the mean distances between molecules and, hence, on the concentration ($N_0/V_0$) of molecules in condensed matter.

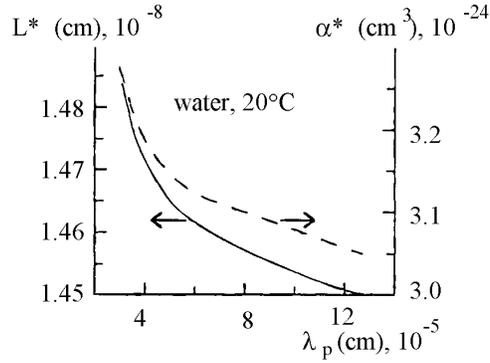

**Figure 26**. Diagram of the acting polarizability $\alpha^* = (L^*)^3$ and electron shell radius of water in the acting field $(L^*)$ on incident light wavelength $(\lambda_p)$, calculated from eq. (8.14) and experimental data $n(\lambda_p)$ (Kikoin, 1976). The initial polarizability of $H_2O$ in the gas phase as $\lambda_p \to \infty$ is equal to $\alpha = L_e^3 = 1.44 \times 10^{-24} cm^3$. The corresponding initial radius of the $H_2O$ electron shell is $L_e = 1.13 \times 10^{-8} cm$.

On the basis of the obtained data, we may conclude that the changes of $A_f$, calculated from (8.24), are caused mainly by the heat expansion of matter. The photon-induced increment of the polarizability $(\alpha \to \alpha^*)$ do not practically change the $A_f$.

The ability to obtain new and valuable information, pertaining to changes of molecular polarizability under the action of incident light and pertaining to temperature dependent molecular interaction in a condensed medium, markedly reinforces the usefulness of such a widely used method as refractometry.

*The above-defined relationship between the molecule polarizability and the wave length of the incident light allows one to make a new attempt to solve problems of light scattering.*

# Chapter 9

# The Hierarchic Theory of light scattering in condensed matter

## 9.1. The traditional approach

According to the conventional approach, light scattering in liquids, crystals, and in gases, takes place due to random heat fluctuations. In condensed media, fluctuations of density, temperature, and molecule orientation are possible, and determine the light scattering.

Density ($\rho$) fluctuations leading to dielectric permittivity ($\epsilon$) fluctuations are of major importance. This contribution is estimated by the Einstein formula for scattering coefficient of liquids:

$$R = \frac{Ir^2}{I_0 V} = \frac{\pi}{2\lambda^4} kT\beta_T \left( \rho \frac{\partial \epsilon}{\partial \rho} \right)_T \qquad 9.1$$

where $\beta_T$ is the isothermal compressibility.

Many authors have attempted to find a correct expression for the variable $(\rho \frac{\partial \epsilon}{\partial \rho})$.

The formula derived by Vuks (1977) is most consistent with the experimental data:

$$\rho \frac{\partial \epsilon}{\partial \rho} = (n^2 - 1) \frac{3n^2}{2n^2 - 1} \qquad 9.2$$

## 9.2. Fine structure of Brillouin light scattering

The fine structure spectrum for scattering in liquids is represented by two Brillouin components, with frequencies shifted relatively from the incident light frequency: $\nu_\pm = \nu_0 \pm \Delta\nu$, and one unshifted band, just as is found in gases ($\nu_0$).

The shift of the Brillouin components is caused by the Doppler effect resulting from a fraction of photons scattering on phonons moving at the speed of sound in opposite directions (Vuks. 1977).

This shift can be explained in a different way as well. If in the antinodes of the standing wave the density oscillation occurs at frequency ($\Omega$):

$$\rho = \rho_0 \cos \Omega t, \qquad 9.3$$

then the scattered wave amplitude will change at the same frequency. Such a wave can be represented as a superposition of two monochromatic waves having the frequencies: $(\omega + \Omega)$ and $(\omega - \Omega)$, where

$$\Omega = 2\pi f \qquad 9.4$$

is the elastic wave frequency at which scattering occurs when the Wolf-Bragg condition is satisfied:

$$2\Lambda \sin \varphi = 2\Lambda \sin \frac{\theta}{2} = \lambda' \qquad 9.5$$

or

$$\Lambda = \lambda'/(2 \sin \frac{\theta}{2}) = \frac{c}{n\nu}(2 \sin \frac{\theta}{2}) = \mathbf{v}_{ph}/f \qquad 9.6$$

where $\Lambda$ is the elastic wave length corresponding to the frequency $f$; $\lambda' = \lambda/n = c/n\nu$ ($\lambda'$ and $\lambda$ are the incident light wavelengths in matter and in vacuum, respectively); $\varphi$ is the angle of sliding; $\theta$ is the angle of scattering; $n$ is the refraction index of matter; $c$ is the speed of light.

The value of Brillouin splitting is represented as:

$$\pm \Delta\nu_{M-B} = f = \frac{\mathbf{v}_{ph}}{\Lambda} = 2\nu \frac{\mathbf{v}_{ph}}{c} n \sin \frac{\theta}{2} \qquad 9.7$$

where $\nu n/c = 1/\lambda$; $n$ is the refraction index of matter; $\nu$ is incident light frequency;

$$\mathbf{v}_{ph} = \mathbf{v}_S \qquad 9.8$$

is the phase velocity of a scattering wave, equal to hypersonic velocity.

The formula (9.7) is identical to that obtained from the analysis of the Doppler effect:

$$\frac{\Delta \nu}{\nu} = \pm 2 \frac{\mathbf{v}_S}{c} n \sin \frac{\theta}{2} \qquad 9.9$$

According to the classical theory, the central line, which is analogous to that observed in gases, is caused by entropy fluctuations in liquids, without any changes of pressure (Vuks, 1977). On the basis of the Frenkel theory of the liquid state, the central line can be explained by fluctuations of "hole" number - cavitational fluctuations (Theiner and White, 1969).

The thermodynamic approach of Landau and Plachek leads to the formula, which relates the intensities of the central ($I$) and two lateral ($I_{M-B}$) lines of the scattering spectrum to compressibilities and heat capacities:

$$\frac{I}{2I_{M-B}} = \frac{I_p}{I_{ad}} = \frac{\beta_T - \beta_S}{\beta_S} = \frac{C_p - C_\mathbf{v}}{C_\mathbf{v}} \qquad 9.10$$

where $\beta_T$ and $\beta_S$ are isothermal and adiabatic compressibilities; $C_p$ and $C_\mathbf{v}$ are isobaric and isohoric heat capacities.

In crystals, quartz for example, the central line in the fine structure of light scattering is usually absent or very small. However, instead of one pair of shifted components, observed in liquids, there appear *three* Brillouin components in crystals. One of them used to be explained by the scattering on the longitudinal phonons, and two by the scattering on the transversal phonons.

### 9.3. New hierarchic approach to Brillouin scattering

In the Hierarchic Theory of condensed matter the thermal "random" fluctuations are "organized" by different types of superimposed quantum excitations.

According to the Hierarchic Model, including microscopic, mesoscopic and macroscopic scales of matter (see *Introduction*), the most probable (primary) and mean (secondary) effectons, translational and librational are capable of quantum transitions between two discreet states: $(a \Leftrightarrow b)_{tr,lb}$ and $(\bar{a} \Leftrightarrow \bar{b})_{tr,lb}$ respectively. These transitions lead to origination/annihilation of photons and phonons, forming primary and secondary deformons.

The mean equilibrium heat energy of molecules and quasi-particles is determined by the value $3kT$, which, as the calculations show, has an intermediate value between the energies of a and b states of primary effectons (Figure 19), thus rendering the only allowed conditions in condensed matter as nonequilibrium ones.

The maximum deviations from thermal equilibrium, as well as from the average dielectric properties of matter, occur when the same states of primary and secondary quasi-particles, e.g. a,ā and b,b̄ occur simultaneously. Such a situation corresponds to the A and B states of macro-effectons.

The probability of the situation where two spatially non-correlated events coincide in time is equal to the product of their independent probabilities (Theiner and White, 1969). Thus, the probabilities of the coherent $(a,\bar{a})$ and $(b,\bar{b})$ states of primary and secondary effectons, corresponding to A and B states of the macro-effectons (tr and lib), independent on each other, are equal to:

$$\left( P_M^A \right)_{tr,lb}^{ind} = \left( P_{ef}^a \bar{P}_{ef}^a \right)_{tr,lb}^S \times \left( \frac{1}{Z^2} \right) = \left( \frac{P_M^A}{Z^2} \right)_{tr,lb} \qquad 9.11$$

$$\left( P_M^B \right)_{tr,lb}^{ind} = \left( P_{ef}^b \bar{P}_{ef}^b \right)_{tr,lb}^S \times \left( \frac{1}{Z^2} \right) = \left( \frac{P_M^B}{Z^2} \right)_{tr,lb} \qquad 9.12$$

where

$$\frac{1}{Z} \left( P_{ef}^a \right)_{tr,lb} \quad \text{and} \quad \frac{1}{Z} \left( \bar{P}_{ef}^a \right)_{tr,lb} \qquad 9.13$$

are the independent probabilities of $a$ and $\bar{a}$ states determined according to formulae (4.10 and 4.18), while probabilities $\left(P^b_{ef}/Z\right)_{tr,lb}$ and $\left(\bar{P}^b_{ef}/Z\right)_{tr,lb}$ are determined according to formulae (4.11 and 4.19);

$Z$ is the sum of probabilities of all types of quasi-particles states - (eq. 4.2).

The probabilities of molecules being involved in the spatially independent translational and librational macro-deformons are expressed as the products of (9.11) and (9.12):

$$\left(P^M_D\right)^{ind}_{tr,lb} = \left[\left(P^A_M\right)^{ind}\left(P^B_M\right)^{ind}\right]_{tr,lb} = \frac{P^M_D}{Z^4} \qquad 9.14$$

Formulae (9.11) and (9.12) may be considered as the probabilities of space-independent but coherent macro-effectons in A and B states, respectively.

For probabilities of independent $A^*$ and $B^*$ states of super-effectons in we obtain (see 3.25 and 3.26):

$$\left(P^{A^*}_S\right)^{ind} = \left(P^A_M\right)^{ind}_{tr}\left(P^A_M\right)^{ind}_{lb} = \frac{P^{A^*}_S}{Z^4} \qquad 9.15$$

$$\left(P^{B^*}_S\right)^{ind} = \left(P^B_M\right)^{ind}_{tr}\left(P^B_M\right)^{ind}_{tr} = \frac{P^{b^*}_S}{Z^4} \qquad 9.15$$

In a similar way using (9.14) we obtain the probabilities of super-deformons (see 3.31):

$$\left(P^{D^*}_S\right)^{ind} = \left(P^D_M\right)_{tr} \times \left(P^D_M\right)_{lb} = \frac{P^{D^*}_S}{Z^4} \qquad 9.16$$

The concentrations of molecules, the states of which markedly differ from the equilibrium, and which cause light scattering of macro-effectons and macro-deformons, are equal correspondingly to:

$$\left[N^A_M = \frac{N_0}{Z^2 V_0}\left(P^A_M\right)\right]_{tr,lb}; \quad \left[N^B_M = \frac{N_0}{Z^2 V_0}\left(P^B_M\right)\right]_{tr,lb} \qquad 9.17$$

$$\left[N^D_M = \frac{N_0}{Z^4 V_0}\left(P^D_M\right)\right]_{tr,lb}$$

The concentrations of molecules, involved in $(a_{tr} \rightleftharpoons a_{lb})$ - convertons, $(b_{tr} \rightleftharpoons b_{lb})$ - convertons and macro-convertons (see Introduction) are correspondingly:

$$N^{ac}_M = \frac{N_0}{Z^2 V_0} P_{ac}; \quad N^{bc}_M = \frac{N_0}{Z^2 V_0} P_{bc}; \quad N^C_M = \frac{N_0}{Z^4 V_0} P_{cMt} \qquad 9.18$$

The probabilities of convertons-related excitations are the same as those used in Chapter 4.

The concentration of molecules, participating in the independent super-effectons and super-deformons:

$$N^{A^*}_M = \frac{N_0}{Z^4 V_0} P^{A^*}_S; \quad N^{B^*}_M = \frac{N_0}{Z^4 V_0} P^{B^*}_S \qquad 9.19$$

$$N^{D^*}_M = \frac{N_0}{Z^8 V_0} P^{D^*}_S \qquad 9.20$$

where $N_0$ and $V_0$ are Avogadro's number and the molar volume of the matter in the sample.

Substituting (9.17 - 9.20) into the Rayleigh formula for the scattering coefficient (7.37), we obtain the values of the contributions from different states of quasi-particles to the resulting scattering coefficient:

$$\left(R^M_A\right)_{tr,lb} = \frac{8\pi^4}{\lambda^4}\frac{(\alpha^*)^2}{Z^2}\frac{N_0}{V_0}\left(P^A_M\right)_{tr,lb}; \quad R^S_A = \frac{8\pi^4}{\lambda^4}\frac{(\alpha^*)^2}{Z^4}\frac{N_0}{V_0}P^{A^*}_S \qquad 9.21$$

$$\left(R^M_B\right)_{tr,lb} = \frac{8\pi^4}{\lambda^4}\frac{(\alpha^*)^2}{Z^2}\frac{N_0}{V_0}\left(P^B_M\right)_{tr,lb}; \quad R^S_B = \frac{8\pi^4}{\lambda^4}\frac{(\alpha^*)^2}{Z^4}\frac{N_0}{V_0}P^{B^*}_S \qquad 9.22$$

$$(R_D^M)_{tr,lb} = \frac{8\pi^4}{\lambda^4} \frac{(\alpha^*)^2}{Z^2} \frac{N_0}{V_0} \left( P_M^D \right)_{tr,lb}; \quad R_D^S = \frac{8\pi^4}{\lambda^4} \frac{(\alpha^*)^2}{Z^4} \frac{N_0}{V_0} P_S^{D^*} \qquad 9.23$$

The contributions of the excitations, related to [tr/lb] convertons are:

$$R_{ac} = \frac{8\pi^4}{\lambda^4} \frac{(\alpha^*)^2}{Z^2} \frac{N_0}{V_0} R_{bc} = \frac{8\pi^4}{\lambda^4} \frac{(\alpha^*)^2}{Z^2} \frac{N_0}{V_0} P_{bc} \qquad 9.23a$$

$$R_{abc} = \frac{8\pi^4}{\lambda^4} \frac{(\alpha^*)^2}{Z^4} \frac{N_0}{V_0} P_{cMt} \qquad 9.23b$$

where $\alpha^*$ is the acting polarizability determined by (eq. 8.24) and (eq. 8.25).

The resulting coefficient of the isotropic scattering ($R_{iso}$) is defined as the sum of contributions (9.21-9.23b) and is subdivided into three kinds of scattering: that caused by translational quasi-particles, that caused by librational quasi-particles and that caused by mixed types of quasi-particles:

$$R_{iso} = [R_A^M + R_B^M + R_D^M]_{tr} + [R_A^M + R_B^M + R_D^M]_{lb} + [R_{ac} + R_{bc} + R_{abc}] + [R_A^S + R_B^S + R_D^S] \qquad 9.24$$

Total contributions, related to all types of convertons, super-effectons and super-deformons, are correspondingly:

$$R_C = R_{ac} + R_{bc} + R_{abc} \quad \text{and} \quad R_S = R_A^S + R_B^S + R_D^S$$

The polarizability of anisotropic molecules having no cubic symmetry is a tensor. In this case, total scattering (R) consists of scattering at density fluctuations ($R_{iso}$) and scattering at fluctuations of the anisotropy $\left(R_{an} = \frac{13\Delta}{6-7\Delta} R_{iso}\right)$:

$$R = R_{iso} + \frac{13\Delta}{6-7\Delta} R_{iso} = R_{iso} \frac{6+6\Delta}{6-7\Delta} = R_{iso} K \qquad 9.25$$

where $R_{iso}$ corresponds to (*eq. 9.24*); $\Delta$ is the depolarization coefficient.

The factor:

$$\left(\frac{6+6\Delta}{6-7\Delta}\right) = K \qquad 9.25a$$

was obtained by Cabanne and is called after him. In the case of isotropic molecules when $\Delta = 0$, the Cabanne factor is equal to 1.

The depolarization coefficient ($\Delta$) can be determined experimentally as the ratio:

$$\Delta = I_x/I_z, \qquad 9.26$$

where $I_x$ and $I_z$ are two polarized components of the beams scattered at right angle with respect to each other in which the electric vector is directed parallel and perpendicular to the incident beam, respectively. For example, in water $\Delta = 0.09$ (Vuks, 1977).

According to the proposed theory of light scattering in liquids, the central nonshifted (just as in gases) component of the Brillouin scattering spectrum is caused by fluctuations of concentration and self-diffusion of the molecules, participating in the convertons, macro-deformons (tr and lb) and super-deformons. The scattering coefficients of the central line ($R_{centr}$) and lateral lines ($2R_{side}$) in transparent condensed matter, as follows from (9.24) and (9.25), are equal correspondingly to:

$$R_{cent} = K\left[\left(R_D^M\right)_{tr} + \left(R_D^M\right)_{lb}\right] + K(R_C + R_S) \qquad 9.27$$

and

$$2R_{side} = \left(R_A^M + R_B^M\right)_{tr} + \left(R_A^M + R_A^M\right)_{lb} \qquad 9.27a$$

where $K$ is the Cabanne factor.

*The total coefficient of light scattering is:*

$$R = R_{\text{cent}} + 2R_{\text{side}} \qquad 9.28$$

In accordance with the current model, the fluctuations of anisotropy (Cabanne factor) should be taken into account for calculations of the central component only. The orientations of molecules composing A and B states of macro-effectons are correlated, and their coherent oscillations are not accompanied by fluctuations of anisotropy of polarizability (see Figure 27a).

The probability of the convertons, macro-deformons and super-deformons excitation (eqs.9.14, 4.16, 4.27) is much lower in crystals than in liquids, and hence, the central line in the Brillouin spectra of crystals is not usually observed.

The lateral lines in Brillouin spectra are caused by the scattering of the molecules forming (A) and (B) states of spatially independent macro-effectons, as was mentioned above.

The polarizabilities of the molecules forming the independent macro-effectons, synchronized in $(A)_{tr,lb}$ and $(B)_{tr,lb}$ states, and in the dielectric properties of these states, differ from each other and from that of transition states (macro-deformons). Such short-lived states should be considered as nonequilibrium states.

We must keep in mind, that the static polarizabilities in the more stable ground A state of the macro-effectons are higher than in the B state, because the energy of long-term van der Waals interaction between molecules of the A state is larger than that of the B-state.

If this difference is attributed mainly to the difference in the long-term dispersion interaction, then, from (1.33) we obtain:

$$E_B - E_A = V_B - V_A = -\frac{3}{2}\frac{E_0}{r^6}\left(\alpha_B^2 - \alpha_A^2\right) \qquad 9.29$$

where the polarizability of molecules in the A-state is higher than that in the B-state:

$$\alpha_A^2 > \left[\left(\alpha^*\right)^2 \simeq \alpha_D^2\right] > \alpha_B^2$$

The kinetic energy, and the dimensions of "acoustic" and "optic" states of macro-effectons are the same: $T_{\text{kin}}^A = T_{\text{kin}}^B$.

In the present calculations of light scattering we ignore this difference (9.29) between polarizabilities of molecules in the A and B states.

However, this difference can be taken into account if we assume that polarizabilities in (A) and (a), (B) and (b) states of primary effectons are such that:

$$\alpha_A \simeq \alpha_a \simeq \alpha^*; \quad \alpha_B \simeq \alpha_b$$

and the difference between the potential energy of (a) and (b) states is determined mainly by dispersion interaction (eq. 9.28).

The resulting polarizability $(\alpha^* \simeq \alpha_a)$ can be expressed as:

$$\alpha_a = f_a\alpha_a + f_b\alpha_b + f_t\alpha \qquad 9.29a$$

where $\alpha_t \simeq \alpha$ is the polarizability of molecules in the gas state (or transition state);

$$f_a = \frac{P_a}{P_a + P_b + P_t}; \quad f_b = \frac{P_b}{P_a + P_b + P_t};$$

$$\text{and} \quad f_t = f_d = \frac{P_t}{P_a + P_b + P_t}$$

are the fractions of *(a)*, *(b)* and transition (t) states (equal to 2.66) with $P_t = P_d = P_a \cdot P_b$.

On the other hand from (1.33), evaluated with $r = const$ we have:

$$\Delta V_{\text{dis}}^{b \to a} = -\frac{3}{4}\frac{(2\alpha\Delta\alpha)}{r^6}I_0 \quad (r_a = r_b; \ I_0^a \simeq I_0^b) \text{ and}$$

$$\frac{\Delta V_{\text{dis}}^{b \to a}}{V^b} = \frac{h\nu_p}{h\nu_b} = \frac{\Delta \alpha_a}{\alpha} \quad \text{or} \quad \Delta \alpha_a = \alpha_a \frac{\nu_p}{\nu_b} \qquad 9.29b$$

$$\alpha_b = \alpha_a - \Delta \alpha_a = \alpha_a \left(1 - \frac{\nu_p}{\nu_b}\right)$$

where $\Delta \alpha_a$ is a change of each molecule's polarizability as a result of the primary effecton's energy changing: $E_b \to E_a + h\nu_p$ with photon radiation; $\nu_b$ is a frequency of primary effecton in the (b)-state (eq. 2.28).

Combining (9.29) and (9.29b) we derive for $\alpha_a$ and $\alpha_b$ of the molecules composing primary translational or librational effectons:

$$\alpha_a = \frac{f_t \alpha}{1 - \left(f_a + f_b + f_b \frac{\nu_p}{\nu_b}\right)} \qquad 9.30$$

$$\alpha_b = \alpha_a \left(1 - \frac{\nu_p}{\nu_b}\right) \qquad 9.30a$$

The calculations by means of (9.30) are approximate in the framework of the assumptions mentioned above. However, they correctly reflect the tendencies of $\alpha_a$ and $\alpha_b$ changes with temperature.

The ratio of intensities or scattering coefficients of the central component to the lateral coefficients (see the previous reference) was described by the Landau-Plachek formula (9.10). According to the current theory this ratio can be calculated in another way, by considering formulae (9.27) and (9.28):

$$\frac{I_{\text{centr}}}{2I_{M-B}} = \frac{R_{\text{cent}}}{2R_{\text{side}}} \qquad 9.30b$$

Thus, by combining (9.30) and the Landau-Plachek formula (9.10) it is possible to calculate the ratio $(\beta_T/\beta_S)$ and $(C_P/C_V)$ using the Hierarchic Theory of light scattering.

### 9.4. Factors that determine the Brillouin line width

The equation for Brillouin shift is (see 9.7):

$$\Delta \nu_{M-B} = \nu_0 = 2 \frac{\mathbf{v}_s}{\lambda} n \sin(\theta/2) \qquad 9.31$$

where $\nu_s$ is the hypersonic velocity; $\lambda$ is the wavelength of incident light, n is the refraction index of matter, and $\theta$ is the scattering angle.

The deviation from $\nu_0$ that determines the Brillouin lateral line half width may be expressed as the result of fluctuations of $\nu_s$ and $n$, related to A and B states of tr and lib macro-effectons:

$$\frac{\Delta \nu_0}{\nu_0} = \left(\frac{\Delta \mathbf{v}_s}{\mathbf{v}_s} + \frac{\Delta n}{n}\right) \qquad 9.32$$

$\Delta \nu_0$ is the most probable line width, i.e. the true half width of the Brillouin line. It can be expressed as:

$$\Delta \nu_0 = \Delta \nu_{\text{exp}} - F \Delta \nu_{\text{inc}}$$

where $\Delta \nu_{\text{exp}}$ is the half width of the experimental line, $\Delta \nu_{\text{inc}}$ is the half width of the incident line, $F$ is the coefficient that takes into account the apparatus effects.

Let us analyze the first and the second terms in the right hand side of (9.32) separately.

The $\mathbf{v}_s$ squared is equal to the ratio of the compressibility modulus ($M$) and density ($\rho$):

$$\mathbf{v}_s^2 = M^2/\rho \qquad 9.33$$

Consequently, from (9.33) we have:

$$\frac{\Delta \mathbf{v}_s}{\mathbf{v}_s} = \frac{1}{2}\left(\frac{\Delta M}{M} - \frac{\Delta \rho}{\rho}\right) \qquad 9.34$$

In the case of independent fluctuations of $M$ and $\rho$:

$$\frac{\Delta v_s}{v_s} = \frac{1}{2}\left(\left|\frac{\Delta M}{M}\right| - \left|\frac{\Delta \rho}{\rho}\right|\right) \qquad 9.35$$

From equation (8.14) we obtain the following value for refraction index:

$$n^2 = \left(1 - \frac{4}{3}N\alpha^*\right)^{-1} \qquad 9.36$$

where $N = N_0/V_0$ is the molecular concentration.

From (9.36) we can derive:

$$\frac{\Delta n}{n} = \frac{1}{2}(n^2 - 1)\left(\frac{\Delta \alpha^*}{\alpha^*} + \frac{\Delta N}{N}\right) \qquad 9.37$$

where

$$(\Delta N/N) = (\Delta \rho/\rho) \qquad 9.38$$

and

$$\left(\frac{\Delta \alpha^*}{\alpha^*}\right) \simeq \left(\frac{\Delta M}{M}\right) \qquad 9.39$$

We can assume (eq. 9.39), for both parameters: polarizability ($\alpha^*$) and compressibility models (K) are related to the potential energy of intermolecular interaction.

On the other hand, one can assume that the following relation is true:

$$\frac{\Delta \alpha^*}{\alpha^*} \simeq \frac{|\bar{E}_{ef}^a - 3kT|}{3kT} = \frac{\Delta M}{M} \qquad 9.40$$

where $\bar{E}_{ef}^a$ is the energy of the secondary effectons in the (ā) state; $E_0 = 3kT$ is the energy of an "ideal" quasi-particle, considered as a superposition of 3D standing waves.

The density fluctuations can be estimated as a result of the free volume ($v_f$) fluctuations (see 11.45):

$$\left(\frac{\Delta v_f}{v_f}\right)_{tr,lb} = \frac{1}{Z}(P_D^M)_{tr,lb} \simeq (\Delta N/N)_{tr,lb} \qquad 9.41$$

Now, putting (9.40) and (9.41) into (9.37) and (9.34) and then into (9.32), we obtain the semiempirical formula for the Brillouin line half-width calculation:

$$\frac{\Delta v_f}{v_f} \simeq \frac{n^2}{2}\left[\frac{|\bar{E}_{ef}^a - 3kT|}{3kT} + \frac{1}{Z}\left(P_D^M\right)\right]_{tr,lb} \qquad 9.42$$

Brillouin line intensity depends on the half-width $\Delta v$ of the line in the following ways:
For a Gaussian line shape:

$$I(v) = I_0^{max} \exp\left[-0.693\left(\frac{v - v_0}{\frac{1}{2}\Delta v_0}\right)^2\right]; \qquad 9.43$$

For a Lorenzian line shape:

$$I(v) = \frac{I_0^{max}}{1 + \left[(v - v_0)/\frac{1}{2}\Delta v_0\right]^2} \qquad 9.44$$

*The traditional theory of Brillouin line shape* entails the possibility for calculation of $\Delta v_0$, taking into account the elastic (acoustic) wave dissipation.

The decay of the acoustic wave amplitude may be expressed as:

$$A = A_0 e^{-\alpha x} \quad \text{or} \quad A = A_0 e^{-\alpha v_s} \qquad 9.45$$

where $\alpha$ is the extinction coefficient; $x = v_s t$ is the distance from the source of waves; $v_s$ and $t$ are the velocity of sound, and time, correspondingly.

The hydrodynamic theory of sound propagation in liquids leads to the following expression for the extinction coefficient:

$$\alpha = \alpha_s + \alpha_b = \frac{\Omega^2}{2\rho v_s^3}\left(\frac{4}{3}\eta_s + \eta_b\right) \quad 9.46$$

where $\alpha_s$ and $\alpha_b$ are contributions to $\alpha$, related to share viscosity ($\eta_s$) and bulk viscosity ($\eta_b$), respectively; $\Omega = 2\pi f$ is the angular frequency of the acoustic waves.

When the lateral lines in Brillouin spectra broaden slightly, the following relation between their intensity (I) and shift $\Delta\omega = |\omega - \omega_0|$ from frequency $\omega_0$, corresponding to the maximum intensity ($I = I_0$) of lateral line is correct:

$$I = \frac{I_0}{1 + (\frac{\omega-\omega_0}{a})} \quad 9.47$$

where

$$a = \alpha v_s.$$

One can see from (9.46) that at $I(\omega) = I_0/2$, the half width may be calculated as follows:

$$\Delta\omega_{1/2} = 2\pi\Delta\nu_{1/2} = \alpha v_s \text{ and } \Delta\nu_{1/2} = \frac{1}{2}\pi\alpha v_s \quad 9.48$$

It will be shown in Chapter 12 how one can calculate the values of $\eta_s$ and consequently, also calculate $\alpha_s$ on the basis of the Hierarchic Theory of viscosity.

### 9.5. Quantitative verification of the Hierarchic Theory of Brillouin scattering

The calculations made according to the formulae (9.21 - 9.27) are presented in Figures 27-29. The proposed theory of scattering in liquids, based on the Hierarchic Model, has more descriptive and predictive power than the traditional Einstein, Mandelschtamm-Brillouin, Landau-Plachek theories based on classical thermodynamics. The Hierarchic theory describes experimental temperature and the $I_{centr}/2I_{M-B}$ ratio for water very well (Figure 27 c).

The calculations are made for the wavelength of incident light: $\lambda_{ph} = 546.1 nm = 5.461 \times 10^{-5} cm$. The experimental temperature dependence for the refraction index ($n$) at this wavelength was taken from the Frontas'ev and Schreiber paper (1965). The rest of the data for the calculating of various light scattering parameters of water (density the location of translational and librational bands in the oscillatory spectra) are identical to those used in Chapter 6 above.

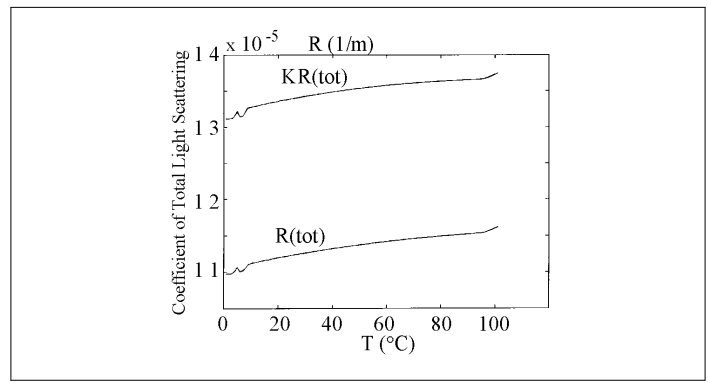

**Figure 27 a**. Theoretical temperature of the total scattering coefficient for water without taking into account the anisotropy of water molecules polarizability fluctuations in the volume of macro-effectons, responsible for lateral lines: [$R(tot)$] - (eq. 9.27a; 9.28) and taking them into account: $[KR(tot)]$, where $K$ is the Cabanne factor (eq. 9.25a).

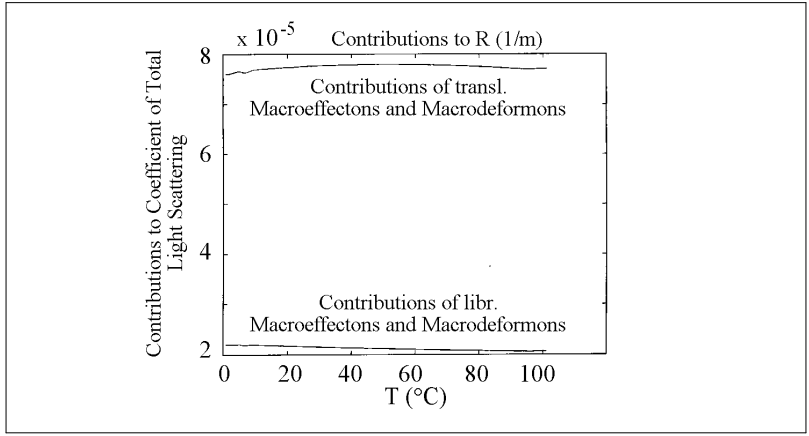

**Figure 27 b**. Theoretical temperature of contributions to the total coefficient of total light scattering (R) caused by translational and librational macro-effectons and macro-deformons (without taking into account fluctuations of anisotropy).

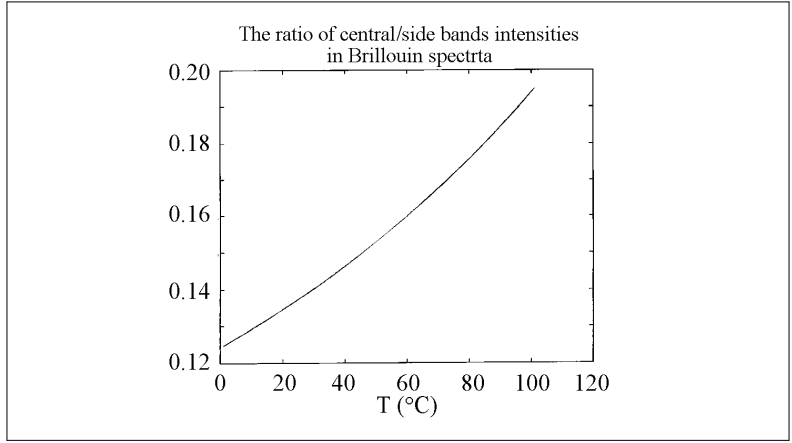

**Figure 27 c**. Theoretical temperature of central to side bands intensities ratio in Brillouin spectra (eq. 9.30).

The Hierarchic Theory of light scattering can be used to verify the correctness of our formula for the refraction index (eq. 8.14) in the case of condensed matter, when $\alpha = \alpha^*$:

$$\frac{n^2 - 1}{n^2} = \frac{4}{3}\pi \frac{N_0}{V_0} \alpha^* \qquad 9.48a$$

Now, compare this result with the Lorentz-Lorenz formula:

$$\frac{n^2 - 1}{n^2 + 1} = \frac{4}{3}\pi \frac{N_0}{V_0} \alpha \qquad 9.49$$

If the formula (9.48a) we have derived holds, then, in condensed matter the effective molecular polarizability ($\alpha^*$), used in (eq. 9.21-9.23) is:

$$\alpha^* = \frac{(n^2 - 1)/n^2}{(4/3)\pi(N_0/V_0)} \qquad 9.50$$

On the other hand, from the Lorentz-Lorenz formula (9.49) we have another value of polarizability:

$$\alpha = \frac{(n^2-1)/(n^2+2)}{(4/3)\pi(N_0/V_0)} \qquad 9.51$$

The light scattering coefficients (eq. 9.28), calculated for water, using (9.50) and (9.51) and taking the refraction index as: $n = 1.33$, differs by a factor of more than four:

$$\frac{R(\alpha^*)}{R(\alpha)} = \frac{(\alpha^*)^2}{(\alpha)^2} = \frac{(n^2-1)/n^2}{(n^2-1)/(n^2+2)} = \left(\frac{n^2+2}{n^2}\right)^2 = 4.56 \qquad 9.52$$

This implies that the correctness of 4.50 or 4.51 can be obtained from the experimental value of light scattering. At $25^0$ and $\lambda_{ph} = 546 nm$ the theoretical magnitude of the scattering coefficient for water, calculated from our formulas (9.27; 9.28) and (9.50) is equal (see Figure 27 a) to:

$$R = 11.2 \times 10^{-5} m^{-1} \qquad 9.53$$

*This result of our theory of Brillouin scattering agrees well with the most reliable experimental value* (Vuks, 1977):

$$R_{exp} = 10.8 \times 10^{-5} m^{-1} \qquad 9.53a$$

This implies that our formula (9.50) works for the calculation of molecular polarizability much better than the Lorentz-Lorenz formula (9.51), which yields a scattering coefficient a few times bigger than the experimental value.

Multiplication of the side Doppler bands contribution (9.27a) to the total Brillouin scattering ($2R_{side}$) by the Cabanne factor increases the calculated total scattering coefficient (9.53) by about 25%. This makes the correspondence with experiment worse. This fact confirms our assumption that fluctuations of anisotropy of polarizability composing A and B states of macro-effectons can be ignored in the light scattering evaluation, because of the spatial correlation of molecular dynamics in these states, in contrast to that of the macro-deformons, as a transition state of macro-effectons.

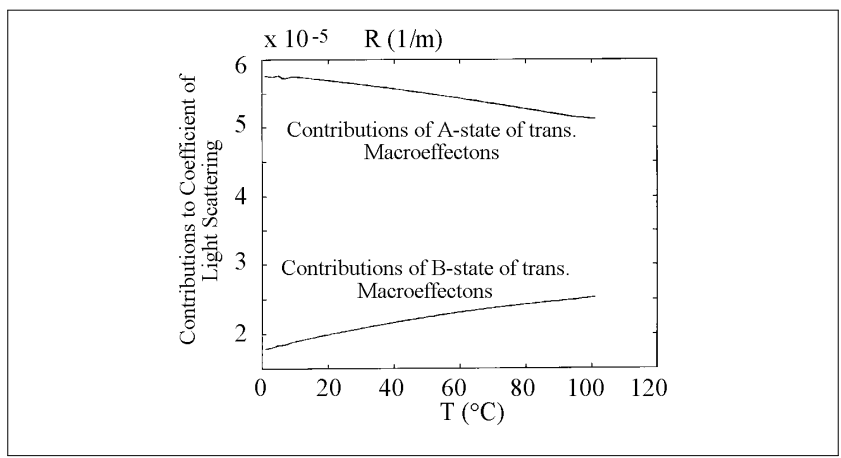

**Figure 28 a**. Theoretical temperature of the contributions of A and B states of translational *macro-effectons* to the total scattering coefficient of water (see also Figure 27 b).

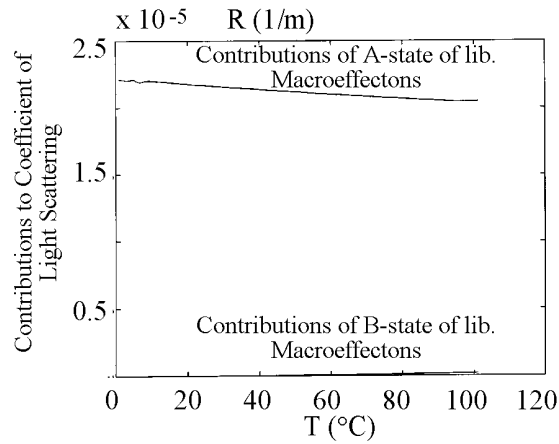

**Figure 28b**. Theoretical temperature of the contributions of the A and B states of librational macro-effectons to the coefficient of light scattering (R).

It follows from the Figure 28(a,b) that the light scattering depends on $(A \Leftrightarrow B)$ equilibrium of macro-effectons because $(R_A) > (R_B)$, *i.e.* scattering on $A$ states is bigger than that on $B$ states.

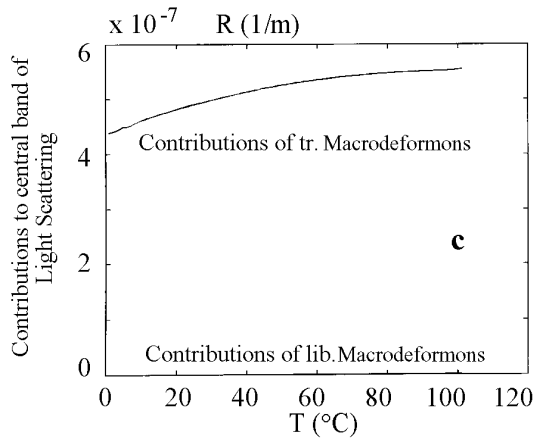

**Figure 28 c**. Theoretical temperature of the contributions to light scattering (central component), related to translational $(R_D)_{tr}$ and librational $(R_D)_{lb}$ macro-deformons.

Comparing Figs. 27 a, 27 c, and 28 c, one can see that the main contribution to the central component of light scattering is determined by $[lb/tr]$ convertons $R_c$ (see eq. 9.27).

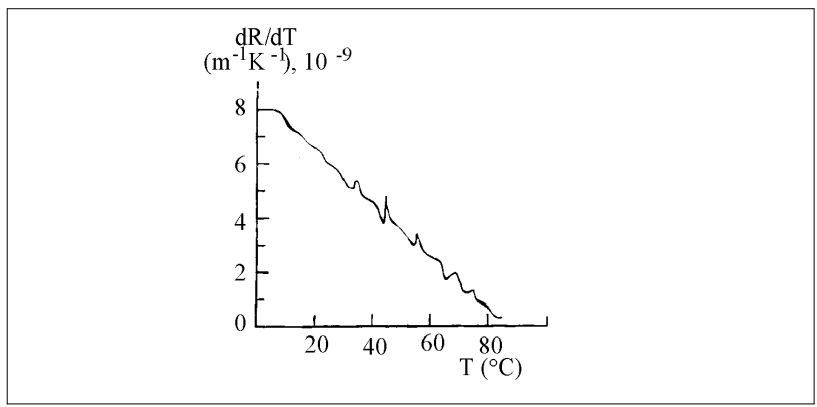

**Figure 29**. Theoretical temperature dependence for temperature derivative ($dR/dT$) of the total coefficient of light scattering of water.

Nonmonotonic deviations of the $dR/dT$ (Fig. 29) reflect the non-monotonic changes of the refraction index for water $n_{H_2O}(T)$, as indicated by available experimental data (Frontas'ev and Schreiber, 1965). The deviations of dependence $n_{H_2O}(t)$ from the monotonic path, in accordance with the Hierarchic Theory, are a consequence of the non-monotonic change in the stability of the water structure, i.e. a nonlinear change of $(A \Leftrightarrow B)_{tr,lb}$ equilibrium. Some possible reasons of such equilibrium change were discussed in Chapter 6.

It follows from (9.52) that the calculations, based on the Lorentz-Lorenz formula (9.51), give the scattering coefficient values of about 4.5 times smaller than experimental values. This means that the true $\alpha^*$ value should be calculated on the basis of our theory of light refraction (eq. 9.50).

The conventional Smolukhovsky-Einstein theory, valid for the integral light scattering only (eq. 9.1), yields values in the range of $R = 8.85 \times 10^{-5}\, m^{-1}$ to $R = 10.5 \times 10^{-5}\, m^{-1}$ (Eisenberg and Kauzmann, 1969, Vuks, 1977). *This means that the Hierarchic Theory of light scattering works better and is much more informative than the conventional theory of light scattering.*

### 9.6. Light scattering in solutions

If the guest molecules are dissolved in a liquid and their sizes are much less than the wavelength of the incident light, these molecules do not radically alter the solvent properties in small concentrations. For this case, the above-described mechanism of light scattering of pure liquids does not change qualitatively.

For such dilute solutions, the scattering due to the fluctuations of the concentration of dissolved molecules ($R_c$) is simply added to the scattering due to the density fluctuations of molecules of the host solvent (eq. 9.28). Taking into account the fluctuations of molecules' polarizability anisotropy (see 9.25), the total scattering coefficient of the solution ($R_S$) is:

$$R_S = R + R_c \qquad 9.54$$

Equations (9.21 - 9.28) can be used for calculating $R$ – until the critical concentration ($C_{cr}$) of the dissolved substance starts to destroy the solvent structure, so that the latter is no longer able to form primary librational effectons. The corresponding perturbations of the solvent structure will induce low-frequency shift of librational bands in the oscillatory spectrum of the solution.

If the experiment is made with a two-component solution of liquids, soluble in each other, e.g. water-alcohol, benzol-methanol etc., and the positions of translational and librational bands of solution components are different, then, at the concentration of the dissolved substance: $C > C_{cr}$, the dissolved substance and the solvent (the guest and host) can switch their roles. Then the translational and librational bands pertinent to the guest subsystem start to dominate. In this case, $R_t$ is to be calculated from the positions of the new bands corresponding to the "new" host-solvent. The total "melting" of the primary librational "host effectons" and the appearance of the dissolved substance "guest effectons" is similar to *the second-order phase transition* and should be accompanied by a heat capacity jump. Such experimental effects do indeed take place (Vuks, 1977).

According to Hierarchic concept, the coefficient $R_c$ in (eq. 9.54) is caused by the fluctuations of the concentration of dissolved molecules in the volume of translational and librational macro-deformons and super-deformons of the solvent. If the destabilization of the solvent is expressed in the low frequency shift of librational bands, then the coefficients $(R_A$ and $R_B)_{lb}$ increase (eq. 9.21 and 9.22) with the probability of macro-excitations. The probabilities of convertons and macro- and super-deformons and the central component of Brillouin spectra will increase also. Therefore, the intensity of the total light scattering increases correspondingly.

The fluctuations of concentration of the solute molecules, in accordance with the current model, occur in the volumes of macro-deformons and super-deformons. Consequently, the contribution of solute molecules in scattering ($R_c$ value in eq. 9.54) can be expressed by a formula, similar to (9.23), but containing the molecule polarizability of the dissolved substance ("guest"), equal to $(\alpha_g^*)^2$ instead of the molecule polarizability $(\alpha^*)$ of the solvent ("host"), and the molecular concentration of the "guest" substance in the solution $(n_g)$ instead of the solvent molecule concentration $(n_M = N_0/V_0)$. In this case $R_c$ can be presented as a sum of the following contributions:

$$(R_c)_{tr,lb} = \frac{8\pi^4}{\lambda^4}(\alpha_g^*)^2 n_g \left[(P_M^D)_{tr,lb} + P_S^{D^*}\right] \qquad 9.55$$

$$R_c^{D^*} = \frac{8\pi^4}{\lambda^4}(\alpha_g^*)^2 n_g (P_S^{D^*}) \qquad 9.55a$$

The resulting scattering coefficient $(R_e)$ on fluctuations of concentration in (9.54) is equal to:

$$R_c = (R_c)_{tr} + (R_c)_{lb} + R_c^{D^*} \qquad 9.56$$

If several substances are dissolved with concentrations lower than $(C_{cr})$, then their $R_c$ are summed additively.

Formulae (9.55) and (9.56) are valid also for the dilute solutions.

Eqs.(9.21-9.28) and (9.54-9.56) should, therefore, be used for calculating the resulting coefficient of light scattering in solutions $(R_S)$.

*The traditional theory represents* the scattering coefficient at fluctuations of concentration as (Vuks, 1977):

$$R_c = \frac{\pi^2}{2\lambda^4}\left(\frac{\partial \epsilon}{\partial x}\right)^2 \Delta x^2 \mathbf{v} \qquad 9.57$$

where $(\partial \epsilon/\partial x)$ is the dielectric penetrability derivative with respect to one of the components: $\Delta \bar{x}^2$ is the fluctuations of concentration of guest molecules squared in the volume element $v$.

The transformation of (9.57) on the basis of classical thermodynamics (Vuks, 1977) leads to the formula:

$$R_c = \frac{\pi^2}{2\lambda^4 N_0}\left(2n\frac{\partial n}{\partial x}\right)\left(\frac{9n^2}{(2n^2+1)(n^2+2)}\right)^2 x_1 x_2 V_{12} f, \qquad 9.58$$

where $N_0$ is Avogadro's number, $x_1$ and $x_2$ are the molar fractions of the first and second components in the solution, $V_{12}$ is the molar volume of the solution, $f$ is the function of fluctuations of concentration determined experimentally from the partial vapor pressures of the first $(P_1)$ and second $(P_2)$ solution components (Vuks, 1977):

$$\frac{1}{f} = \frac{x_1}{P_1}\frac{\partial P_1}{\partial x_1} = \frac{x_2}{P_2}\frac{\partial P_2}{\partial x_2} \qquad 9.59$$

In the case of ideal solutions

$$\frac{\partial P_1}{\partial x_1} = \frac{P_1}{x_1}; \quad \frac{\partial P_2}{\partial x_2} = \frac{P_2}{x_2}; \text{ and } f = 1.$$

*Application of our theory of light scattering to quantitative analysis of transparent liquids and solids yields much more information about properties of matter, its hierarchic dynamic structure, than the conventional theory.*

# Chapter 10

## The Hierarchic Theory of the Mössbauer effect

### 10.1. General background

When the atomic nucleus in the gas phase with mass (M) irradiates a $\gamma$-quantum with energy of

$$E_0 = h\nu_0 = m_p c^2 \qquad 10.1$$

where $m_p$ is the effective photon mass, then, according to the law of momentum conservation, the nucleus acquires additional velocity in the opposite direction:

$$\mathbf{v} = -\frac{E_0}{Mc} \qquad 10.2$$

The corresponding additional kinetic energy, calculated as follows:

$$E_R = \frac{M\mathbf{v}^2}{2} = \frac{E_0^2}{2Mc^2} \qquad 10.3$$

is termed *recoil energy*.

When an atom which irradiates a $\gamma$-quantum is a constituent of a solid body, three situations are possible:

1. The recoil energy of the atom is higher than the energy of atom - lattice interaction. In this case, the atom irradiating $\gamma$-quantum would be knocked out from its position in the lattice. That leads to the origination of defects ;

2. Recoil energy is insufficient for the appreciable displacement of an atom in the structure of the lattice, but is higher than the energy of the given phonon, which is equal to the energy of secondary transitons and phonons' excitations. In this case, recoil energy is spent in the process of heating the lattice;

3. Recoil energy is lower than the energy of primary transitons, related to the [emission / absorption] of IR translational and librational photons $(h\nu_p)_{tr,lb}$ and phonons $(h\nu_{ph})_{tr,lb}$. In this case, the probability (f) of $\gamma$-quantum irradiation without any loss of energy appears, termed the probability (fraction) of a recoilless process.

For example, when $E_R \ll h\nu_{ph}$ ($\nu_{ph}$ - the mean frequency of phonons), then the mean energy of recoil:

$$E_R = (1-f)h\nu_{ph} \qquad 10.4$$

Hence, the probability of recoilless effect is:

$$f = 1 - \frac{E_R}{h\nu_{ph}} \qquad 10.5$$

According to (eq. 10.3) the decrease of the recoil energy $E_R$ of an atom in the structure of a given lattice is related to an increase of its effective mass ($M$). In the current model $M$ corresponds to the mass of the effecton - that is, the given cluster of coherent molecules.

The effect of $\gamma$-quantum irradiation without recoil was discovered by Mössbauer in 1957 and is named after him. The value of the Mössbauer effect is determined by the value of $f$, when $f \leq 1$.

The large recoil energy may be transferred to the lattice by portions that are resonant with the frequency of IR photons (tr and lb) and phonons. The possibility of super-radiation of any given IR quanta stimulation as a result of such a recoil process is one consequence of the Hierarchic Model.

The scattering of $\gamma$-quanta without lattice excitation, i.e., when $E_R \ll h\nu_{ph}$, is termed the elastic scattering. The general expression (Wertheim, 1964, Shpinel, 1969) for the probability of such phonon-less elastic $\gamma$-quantum radiation is equal to:

$$f = \exp\left(-\frac{4\pi <x^2>}{\lambda_0^2}\right) \qquad 10.6$$

where $\lambda_0 = c/\nu_0$ is the real wavelength of the $\gamma$-quantum; $<x^2>$ is the nucleus oscillation's mean amplitude squared in the direction of $\gamma$-quantum irradiation.

The $\gamma$-quanta wavelength parameter may be introduced in the following way:

$$L_0 = \lambda_0/2\pi, \qquad 10.7$$

where $L_0 = 1.37 \times 10^{-5} cm$ for $Fe^{57}$, then (eq. 10.6) can be written as follows:

$$f = \exp\left(-\frac{<x^2>}{L_0^2}\right) \qquad 10.8$$

It may be shown (Shpinel, 1969, Singvi and Sielander, 1962), proceeding from the model of a crystal as a system of 3N identical quantum oscillators, that when the given temperature (T) is much lower than the Debye associated temperature, ($\theta_D$) then:

$$<x^2> = \frac{9\hbar^2}{4Mk\theta_D}\left\{1 + \frac{2\hbar^2 T^2}{3\theta_D^2}\right\}, \qquad 10.9$$

where $\theta_D = h\nu_D/k$ and $\nu_D$ is the Debye frequency.

From (10.1), (10.3) and (10.7) we have:

$$\frac{1}{L} = \frac{E_0}{\hbar c} \qquad 10.10$$

where $E_0 = h\nu = c(2ME_R)^{1/2}$ is the energy of $\gamma$-quantum.

Substituting eqs.(10.9 and 10.10) into (eq. 10.8), we obtain the Debye-Valler formula:

$$f = \exp\left[-\frac{E_R}{k\theta_D}\left\{\frac{3}{2} + \frac{\pi^2 T^2}{\theta_D}\right\}\right] \qquad 10.11$$

when $T \to 0$, then

$$f \to \exp\left(-\frac{3E_R}{2k\theta_D}\right) \qquad 10.12$$

### 10.2. Probability of elastic effects

The displacement squared $<x^2>$ in the elastic (recoilless) effect of $\gamma$ − quanta scattering is determined by mobility of the atoms and molecules, forming primary and secondary effectons (translational and librational).

We will ignore in our approach the contributions of macro- and super-effectons, related to the Mössbauer effect, as very small. Then, the resulting probability of elastic effects associated with $\gamma$-quantum radiation is determined by the sum of the following contributions:

$$f = \frac{1}{Z}\sum_{tr,lb}\left[\left(P_{ef}^a f_{ef}^a + P_{ef}^b f_{ef}^b\right) + \left(\bar{P}_{ef}^a \bar{f}_{ef}^a + \bar{P}_{ef}^b \bar{f}_{ef}^b\right)\right]_{tr,lb} \qquad 10.13$$

where $P_{ef}^a$, $P_{ef}^b$, $\bar{P}_{ef}^a$, $\bar{P}_{ef}^b$ are the relative probabilities of the acoustic and optic states for primary and secondary effectons (see eqs. 4.10 - 4.19); $Z$ is the total partition function (4.2).

These parameters are calculated as described in Chapter 4. Each of the contributions to the resulting probability of the elastic effect can be calculated separately as:

$$\left(f_{ef}^a\right)_{tr,lb} = \exp\left[-\frac{<\left(x^a\right)^2>_{tr,lb}}{L_0^2}\right] \qquad 10.14$$

$\left( f_{ef}^{a} \right)_{tr,lb}$ is the probability of elastic effect, related to the dynamics of primary translational and librational effectons in the a-state.

Mean square displacements $< x^2 >$ of either atoms or molecules in condensed matter (eq. 10.8) are not related to excitation of thermal photons or phonons ( i.e. primary or secondary transitons). According to the current concept, $< x^2 >$ type displacements are caused by the mobility of the atoms forming effectons; they differ for primary and secondary translational and librational effectons in $(a,\bar{a})_{tr,lb}$ and $(b,\bar{b})_{tr,lb}$ states.

$$\left( f_{ef}^{b} \right)_{tr,lb} = \exp\left[ -\frac{< \left( x^b \right)^2 >_{tr,lb}}{L_0^2} \right] \qquad 10.15$$

$\left( f_{ef}^{b} \right)_{tr,lb}$ is the probability of elastic effect in primary translational and librational effectons in the $b$-state;

$$\left( \bar{f}_{ef}^{a} \right)_{tr,lb} = \exp\left[ -\frac{< \left( \bar{x}^a \right)^2 >_{tr,lb}}{L_0^2} \right] \qquad 10.16$$

$\left( \bar{f}_{ef}^{a} \right)_{tr,lb}$ is the probability for secondary effectons in the $\bar{a}$ -state;

$$\left( \bar{f}_{ef}^{b} \right)_{tr,lb} = \exp\left[ -\frac{< \left( \bar{x}^b \right)^2 >_{tr,lb}}{L_0^2} \right] \qquad 10.17$$

$\left( \bar{f}_{ef}^{b} \right)_{tr,lb}$ is the probability of elastic effect, related to secondary effectons in the $\bar{b}$-state.

The mean square displacements within different types of effectons in eqs.(10.14-10.17) are related to their phase and group velocities. First we express the displacements using group velocities of the de Broglie waves ($\mathbf{v}_{gr}$) and periods of corresponding oscillations ($T$) as:

$$< \left( x^a \right)^2 >_{tr,lb} = \frac{< (\mathbf{v}_{gr}^a)^2_{tr,lb} >}{< v_a^2 >_{tr,lb}} = < \left( \mathbf{v}_{gr}^a T^a \right)^2 >_{tr,lb} \qquad 10.18$$

where $(T^a)_{tr,lb} = (1/v_a)_{tr,lb}$ is a relation between the period and the frequency of primary translational and librational effectons in the a-state;

$(\mathbf{v}_{gr}^a = \mathbf{v}_{gr}^b)_{tr,lb}$ are the group velocities of atoms forming these effectons equal in both (a) and (b) states.

In a similar way, we can express the displacements of atoms forming the *(b)* state of primary effectons (tr and lib):

$$< \left( x^b \right)^2 >_{tr,lb} = \frac{< (\mathbf{v}_{gr}^b)^2_{tr,lb} >}{< v_b^2 >_{tr,lb}} \qquad 10.19$$

where $v_b$ is the frequency of *primary* translational and librational effectons in the *b*-state.

The mean square displacements of atoms forming *secondary* translational and librational effectons in $\bar{a}$ and $\bar{b}$ states is:

$$< \left( \bar{x}^a \right)^2 >_{tr,lb} = \frac{< (\bar{\mathbf{v}}_{gr}^a)^2_{tr,lb} >}{< \bar{v}_a^2 >_{tr,lb}} \qquad 10.20$$

$$\left\langle \left( \bar{x}^b \right)^2 \right\rangle_{tr,lb} = \frac{\left\langle (\bar{\mathbf{v}}_{gr}^b)^2_{tr,lb} \right\rangle}{\left\langle \bar{v}_b^2 \right\rangle_{tr,lb}} \qquad 10.21$$

where $(\bar{\mathbf{v}}_{gr}^a = \bar{\mathbf{v}}_{gr}^b)_{tr,lb}$

Group velocities of atoms in primary and secondary effectons may be expressed using the corresponding phase velocities ($\mathbf{v}_{ph}$) and formulae for de Broglie wave lengths as follows:

$$\left( \lambda_a \right)_{tr,lb} = \frac{h}{m \left\langle \mathbf{v}_{gr} \right\rangle_{tr,lb}} = \left( \frac{\mathbf{v}_{ph}^a}{v_a} \right)_{tr,lb} = \qquad 10.22$$

$$= \left( \lambda_b \right)_{tr,lb} = \left( \frac{\mathbf{v}_{ph}^b}{v_b} \right)_{tr,lb}$$

hence, for the square of the value of the group velocities of the atoms or molecules forming primary effectons (**tr** and **lb**) we have:

$$\left( \mathbf{v}_{gr}^{a,b} \right)^2_{tr,lb} = \frac{h^2}{m^2} \left( \frac{v_{a,b}}{\mathbf{v}_{ph}^{a,b}} \right)^2_{tr,lb} \qquad 10.23$$

In accordance with the Hierarchic Theory, the de Broglie wave lengths, the momenta and group velocities in *a* and *b* states of the effectons are equal. Similarly to (10.23), we obtain the group velocities of particles, composing secondary effectons:

$$\left( \bar{\mathbf{v}}_{gr}^{a,b} \right)^2_{tr,lb} = \frac{h^2}{m^2} \left( \frac{\bar{v}_{a,b}}{\bar{\mathbf{v}}_{ph}^{a,b}} \right)^2_{tr,lb} \qquad 10.24$$

Substituting eqs.(10.23) and (10.24) into (10.18-10.21), we find the important expressions for the *average coherent displacement* of particles squared as a result of their oscillations in the volume of the effectons (*tr, lib*) in both discreet states (acoustic and optic):

$$\langle (x^a)^2_{tr,lb} \rangle = (h/m\mathbf{v}_{ph}^a)^2_{tr,lb} \qquad 10.25$$

$$\langle (x^b)^2_{tr,lb} \rangle = (h/m\mathbf{v}_{ph}^b)^2_{tr,lb} \qquad 10.26$$

$$\langle (\bar{x}^a)^2_{tr,lb} \rangle = (h/m\bar{\mathbf{v}}_{ph}^a)^2_{tr,lb} \qquad 10.27$$

$$\langle (\bar{x}^b)^2_{tr,lb} \rangle = (h/m\bar{\mathbf{v}}_{ph}^b)^2_{tr,lb} \qquad 10.28$$

Then, substituting these values into eqs.(10.14-10.17) we obtain a set of different contributions to the resulting probability of effects without recoil:

$$\left. \begin{array}{l} \left( f_f^a \right)_{tr,lb} = \exp\left[ -\left( \frac{h}{mL_0 \mathbf{v}_{ph}^a} \right)^2 \right]_{tr,lb} ; \\[2mm] \left( f_f^b \right)_{tr,lb} = \exp\left[ -\left( \frac{h}{mL_0 \mathbf{v}_{ph}^b} \right)^2 \right]_{tr,lb} ; \end{array} \right\} \qquad 10.29$$

$$\left. \begin{array}{l} \left( \bar{f}_f^a \right)_{tr,lb} = \exp\left[ -\left( \frac{h}{mL_0 \bar{\mathbf{v}}_{ph}^a} \right)^2 \right]_{tr,lb} ; \\[2mm] \left( \bar{f}_f^b \right)_{tr,lb} = \exp\left[ -\left( \frac{h}{mL_0 \bar{\mathbf{v}}_{ph}^b} \right)^2 \right]_{tr,lb} ; \end{array} \right\} \qquad 10.30$$

where the phase velocities $(\mathbf{v}_{ph}^a, \mathbf{v}_{ph}^b, \bar{\mathbf{v}}_{ph}^a, \bar{\mathbf{v}}_{ph}^b)_{tr,lb}$ are calculated from the resulting velocity of sound and the positions of translational and librational bands in the oscillatory spectra of matter at a given

temperature using eqs.2.69-2.75. The wavelength parameter:
$$L_0 = \frac{c}{2\pi v_0} = \frac{hc}{2\pi E_0} = 1.375 \times 10^{-11}\,m$$
for gamma-quanta, radiated by the nucleus of $Fe^{57}$, with energy, is:
$$E_0 = 14.4125\,\text{keV} = 2.30167 \times 10^{-8}\,\text{erg}$$

Substituting eqs.(10.29) and (10.30) into (10.13), we find the total probability of recoil*less* effects ($f_{tot}$) in the given substance. The corresponding computer calculations for ice and water are presented on Figures 30 and 31.

As for the second-order phase transitions in the general case, they are accompanied by alterations of the velocity of sound, and the positions of translational and librational bands; they should also be accompanied by alterations of $f_{tot}$ and its components.

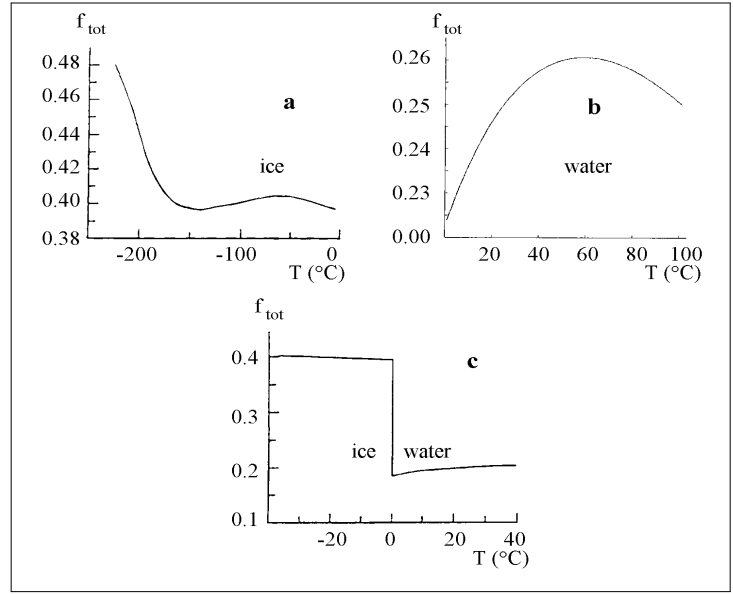

Figure 30. Temperature dependencies of the total probability (f) for the elastic effect without recoil and phonon excitation: (a) in ice; (b) in water; (c) during a phase transition. The calculations were performed using (eq. 10.13).

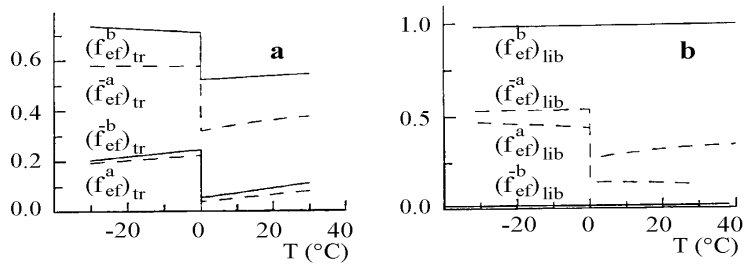

**Figure 31**. *(a)* - The contributions to probability of elastic effect (f), presented in Figure 30, related to primary $(f_{ef}^{a,b})_{tr}$ and secondary $(\bar{f})_{tr}$ translational effectons; (b) the same kind of contributions, related to librational effectons $(f_{ef}^{a,b})_{lb}$ and $(\bar{f})_{lb}$ near the temperature of the [ice ⇔ water] phase transition.

Both the total probability (*f*), and its components, caused by primary and secondary quasi-particles,

were calculated according to formula (10.13). The value of (*f*) determines the magnitude of the Mössbauer effect registered by $\gamma$ –resonances spectroscopy.

The band width, caused by recoilless effects, is determined by the Uncertainty Principle; it is expressed as follows:

$$\Gamma = \frac{h}{\tau} \approx \frac{10^{-27}}{1.4 \times 10^{-7}} = 7.14 \times 10^{-21} \text{ erg} = 4.4 \times 10^{-9} eV \qquad 10.31$$

where $\tau$ is the lifetime of a given nucleus in the excited state (for $Fe^{57}$ $\tau = 1.4 \times 10^{-7} s$).

The position of the band depends on the mean square velocity of the associated atoms, i.e. on the second-order Doppler effect. In experiments, such an effect is compensated for by the velocity of $\gamma$-quanta source motion relative to absorbent. In the framework of the current model, this velocity can be interrelated with the mean velocity of the secondary effectons' diffusion in condensed matter.

### 10.3. Doppler broadening in spectra of nuclear gamma-resonance (NGR)

The Mössbauer effect is characterized by the non-broadened component of the NGR spectra only, with the probability of the elastic effect determined by (eq. 10.13).

When the energy of an absorbed $\gamma$-quanta exceeds the energy of thermal IR photons (tr,lib) or phonons excitation, the absorption band broadens as a result of the Doppler effect. Within the framework of the hierarchic concept, the Doppler broadening is caused by thermal displacements of the particles during $[a \Leftrightarrow b$ and $\bar{a} \Leftrightarrow \bar{b}]_{tr,lb}$ transitions of primary and secondary effectons, leading to the origination/annihilation of the corresponding type of deformons (electromagnetic and acoustic).

The flickering clusters: [*lb/tr*] convertons (a and b), can also contribute to the NGR line broadening.

In this case, the value of Doppler broadening ($\Delta\Gamma$) of the band in the NGR spectrum can be estimated from the corresponding kinetic energies of these excitations, as related to their group velocities (see eq. 4.31). We take into account the "reduced to one molecule" kinetic energies of primary and secondary translational and librational transitons, a-convertons and b-convertons. The contributions of macro-convertons, macro- and super-deformons are much smaller, due to their small probability and concentration:

$$\Delta\Gamma = \frac{V_0}{N_0 Z} \sum_{tr,lb} \left( n_t P_t T_t + \bar{n}_t \bar{P}_t \bar{T}_t \right)_{tr,lb} + \qquad 10.32$$

$$+ \frac{V_0}{N_0 Z} (n_{ef})_{lb} [P_{ac} T_{ac} + P_{bc} T_{bc}]$$

where $N_0$ and $V_0$ are the Avogadro number and molar volume; Z is the total partition function (*eq*. 4.2); $n_t$ and $\bar{n}_t$ are the concentrations of primary and secondary transitons, equal to that of corresponding effectons (eqs.3.5 and 3.7);

$(n_{ef})_{lb} = n_{con}$ is a concentration of primary librational effectons, equal to that of (lb/tr) convertons; $P_t$ and $\bar{P}_t$ are the relative probabilities of primary and secondary transitons (eqs. 4.27a); $P_{ac}$ and $P_{bc}$ are relative probabilities of (*a* and *b*) convertons (4.29a);

$T_t$ and $\bar{T}_t$ are the kinetic energies of primary and secondary transitons, related to the corresponding total energies of these excitations ($E_t$ and $\bar{E}_t$), their masses ($M_t$ and $\bar{M}_t$), and the resulting velocity of sound ($\mathbf{v}_s$, see eq.2.40) in the following form:

$$(T_t)_{tr,lb} = \frac{\sum_1^3 (E_t^{1,2,3})_{tr,lb}}{2M_t (\mathbf{v}_s^{res})^2} \qquad 10.33$$

$$(T_t)_{tr,lb} = \frac{\sum_1^3 (\bar{E}_t^{1,2,3})_{tr,lb}}{2\bar{M}_t (\mathbf{v}_s^{res})^2} \qquad 10.34$$

The kinetic energies of (a and b) convertons (**ac** and **bc**) are expressed in a similar way:

$$(T_{ac}) = \frac{\sum_1^3 (E_{ac}^{1,2,3})_{tr,lb}}{2M_c(\mathbf{v}_s^{res})^2}$$

$$(T_{bc}) = \frac{\sum_1^3 (E_{bc}^{1,2,3})_{tr,lb}}{2M_c(\mathbf{v}_s^{res})^2}$$

where $E_{ac}^{1,2,3}$ and $E_{bc}^{1,2,3}$ are the energies of selected states of corresponding convertons; $M_c$ is the mass of the convertons, equal to that of primary librational effectons.

The broadening of NGR spectral lines by the Doppler effect in liquids is generally expressed using the diffusion coefficient (D) – with the assumption that the motion of any given Mössbauer atom has the character of unlimited diffusion (Singvi and Sielander, 1962):

$$\Delta\Gamma = \frac{2E_0^2}{\hbar c^2} D \qquad 10.35$$

where $E_0 = h\nu_0$ is the energy of gamma quanta; $c$ is the velocity of light and

$$D = \frac{kT}{6\pi\eta a} \qquad 10.36$$

where $\eta$ is the viscosity, (a) is the effective Stokes radius of the atom of $Fe^{57}$.

The probability of recoilless $\gamma$-quantum absorption by the matter containing, for example, $Fe^{57}$, decreases due to diffusion and a corresponding Doppler broadening of the band ($\Delta\Gamma$):

$$f_D = \frac{\Gamma}{\Gamma + \Delta\Gamma} \qquad 10.37$$

where $\Delta\Gamma$ corresponds to (eq. 10.32).

The formulae obtained here make it possible to experimentally verify a set of consequences of the Hierarchic Theory using the gamma-resonance method. A more detailed interpretation of the data obtained by this method also becomes possible.

The magnitude of ($\Delta\Gamma$) was calculated according to formula (10.32). This magnitude corresponds well to experimentally determined Doppler widening in the nuclear gamma resonance (NGR) spectra of ice.

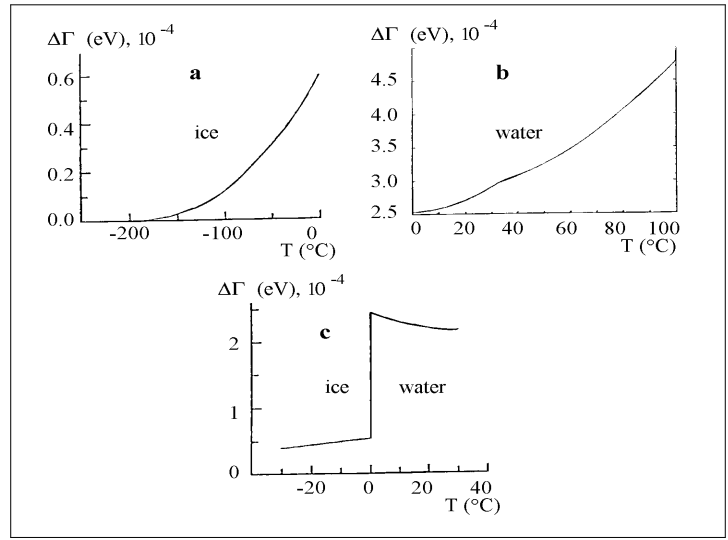

**Figure 32**. The temperature dependencies of the parameter $\Delta\Gamma$, characterizing the nonelastic effects and related to the excitation of thermal phonons and IR photons: **a)** in ice;

**b)** in water; **c)** near the phase transition.

### 10.4. Acceleration and forces, related to thermal dynamics of molecules and ions. Hypothesis of vibro-gravitational interaction

During the period of any particles' thermal oscillations (tr and lb), their instantaneous velocity, acceleration and corresponding forces alternatively and strongly change.

The change of de Broglie wave instantaneous group velocity, averaged during the molecular oscillation period composing the (a) and (b) states of the effectons, determines the average acceleration:

$$\left[ a_{gr}^{a,b} = \frac{d\mathbf{v}_{gr}^{a,b}}{dt} = \frac{\mathbf{v}_{gr}^{a,b}}{T} = \mathbf{v}_{gr} v^{a,b} \right]_{tr,lb}^{1,2,3} \qquad 10.38$$

Keep in mind that group velocities, momenta, and de Broglie wave lengths in (a) and (b) states of the effectons are equal, in accordance with this model.

Corresponding to (10.38) forces:

$$[F^{a,b} = m a_{gr}^{a,b}]_{tr,lb}^{1,2,3} \qquad 10.39$$

The energies of the molecules in (a) and (b) states of the effectons also can be expressed via the corresponding accelerations:

$$\left[ E^{a,b} = h v^{a,b} = F^{a,b} \lambda = m a^{a,b} \times \lambda = m a^{a,b} (\mathbf{v}_{ph}^{a,b}/v^{a,b}) \right]_{tr,lb}^{1,2,3} \qquad 10.40$$

From (10.40) one can express these accelerations of particles in the primary effectons of condensed matter, using their phase velocities as a de Broglie waves:

$$\left[ a_{gr}^{a,b} = \frac{h(v^{a,b})^2}{m \mathbf{v}_{ph}^{a,b}} \right]_{tr,lb}^{1,2,3} \qquad 10.41$$

The accelerations of particles composing secondary effectons have a similar form:

$$\left[ \bar{a}_{gr}^{a,b} = \frac{h(\bar{v}^{a,b})^2}{m \bar{\mathbf{v}}_{ph}^{a,b}} \right]_{tr,lb}^{1,2,3} \qquad 10.42$$

These parameters are important for understanding the dynamic properties of condensed systems. The accelerations of the atoms, forming primary and secondary effectons, can be calculated, using eqs.(2.74-2.75), to determine phase velocities, and eqs. (2.27, 2.28, 2.54, 2.55) - for determination of frequencies.

Multiplying (10.41) and (10.42) by the atomic mass *m*, we derive the most probable and mean forces acting upon the particles in both states of primary and secondary effectons in condensed matter:

$$\left[ F_{gr}^{a,b} = \frac{h(v^{a,b})^2}{\mathbf{v}_{ph}^{a,b}} \right]_{tr,lb}^{1,2,3} \qquad \left[ \bar{F}_{gr}^{a,b} = \frac{h(\bar{v}^{a,b})^2}{\bar{\mathbf{v}}_{ph}^{a,b}} \right] \qquad 10.43$$

The comparison of calculated accelerations with empirical data of the Mössbauer effect supports the correctness of new approach, based on Hierarchic theory.

According to (eq. 2.54), in the low temperature range, when $h v_a \ll kT$, the frequency of *secondary tr and lb* effectons in the *(a)* state can be estimated as:

$$v^a = \frac{v_a}{\exp\left(\frac{h v_a}{kT}\right) - 1} \approx \frac{kT}{h} \qquad 10.44$$

For example, at $T = 200K$ we have $\bar{v}^a \approx 4 \times 10^{12} s^{-1}$.

If the phase speed in (eq. 10.42) is taken equal to $\bar{v}_{ph}^a = 2.1 \times 10^5 cm/s$ (see Figure2a) and the

mass of water molecule:
$$m = 18 \times 1.66 \cdot 10^{-24} g = 2.98 \times 10^{-23} g,$$
then from (10.42) we get the acceleration of molecules composing secondary effectons of ice in *(a)* state:
$$\bar{a}_{gr}^a = \frac{h(\bar{v}^a)^2}{m \bar{\mathbf{v}}_{ph}^a} = 1.6 \times 10^{16} cm/s^2 \qquad 10.45$$

This value is about $10^{13}$ times more than that of free fall acceleration ($g = 9.8 \times 10^2 cm/s^2$). This agrees well with experimental data, obtained for solid bodies (Wertheim, 1964; Sherwin, 1960).

Accelerations of $H_2O$ molecules composing *primary librational* effectons ($a_{gr}^a$) in ice at 200K and in water at 300K are equal to: $0.6 \times 10^{13} cm/s^2$ and $2 \times 10^{15} cm/s^2$, correspondingly. *They also exceed by many orders of magnitude the free fall acceleration.*

It was shown experimentally (Sherwin, 1960), that the heating of a solid body leads to a decreasing of gamma-quanta frequency (red Doppler shift) i.e. and increase of the corresponding quantum transitions period. This can be explained as the relative time-pace decreasing due to an elevation of the average thermal velocity of atoms (Wertheim, 1964).

The thermal vibrations of particles (atoms, molecules) composing primary effectons as a partial Bose-condensate are coherent. The increase of the dimensions of such clusters, determined by the most probable de Broglie wave length, as a result of cooling, pressure elevation, or magnetic field action (see section 14.6), leads to an enhancement of these coherent regions.

*Each cluster of coherently vibrating particles with big alternating accelerations, like librational and translational effectons, is a source of coherent vibro-gravitational waves with amplitude, proportional to the number of coherent particles.*

The frequency of these vibro-gravitational waves (VGW) is equal to the frequency of particle vibrations ( i.e. frequency of the effectons in a or b states). The amplitude of VGW ($A_G$) is proportional to the number of vibrating coherent particles ($N_G$) of these primary effectons:
$$A_G \sim N_G \sim V_{ef}/(V_0/N_0) = (1/n_{ef})/(V_0/N_0)$$

The resonant remote vibro-gravitational interaction between coherent clusters of the same macroscopic object or between different objects is possible. Such vibro-gravitational interaction (VGI) could prove to be a background of Bivacuum Virtual Guides – mediating nonlocal, remote interaction (Kaivarainen, 2006).

The important role of distant resonant VGI in biosystems and in interactions between them can be quite adequately provided by the presence and functioning of primary librational water effectons in microtubules associated with nerve cells, thus providing remote interaction between 'tuned nerve cells', firing simultaneously (see "Hierarchic Model of elementary act consciousness" described in section 17.5).

# Chapter 11

# Interrelation between mesoscopic and macroscopic parameters of matter, resulting in number of new theories, verified quantitatively

### 11.1. A state equation for a real gas

The Clapeyrone-Mendeleyev equation defines the relationship between pressure ($P$), volume ($V$) and temperature ($T$) values for the ideal gas containing $N_0$ molecules (one mole):

$$PV = N_0 kT = RT \qquad 11.1$$

In a real gases and condensed matter the interaction between the molecules and their sizes should be taken into account. It can be achieved by entering the corresponding amendments into the left part, to the right or to the both parts of eq. (11.1).

It was van der Waals who chose the first way many years ago (1873) and derived the equation:

$$\left(P + \frac{a}{V^2}\right)\left(V - b\right) = RT \qquad 11.2$$

where the attraction forces between molecules are accounted for by the amending term ($a/V^2$), while the repulsion forces and the effects of the excluded volume are taken into account by term (b).

Equation (11.2) correctly describes a changes in $P, V$ and $T$ related to liquid-gas transitions on the qualitative level. However, the quantitative analysis by means of (11.2) is approximate and needs the adjustments. The parameters ($a$) and ($b$) in (11.2) are not constant forgiven substance and depend on temperature. Hence, the van der Waals equation is only some approximation describing the state of a real gas or condensed matter.

*In our hierarchical approach* we modified the right part of (eq. 11.1), substituting it for the part of the kinetic energy ($T_{kin}$) of 1 mole of the substance (eq. 4.36) in real gas, formed only by secondary effectons and deformons with nonzero momentum, affecting the pressure:

$$PV = \frac{2}{3}\bar{T}_{kin} = \frac{2}{3}V_0 \frac{1}{Z}\sum_{tr,lb}\left[\bar{n}_{ef}\frac{\sum_1^3(\bar{E}^a_{1,2,3})^2}{2m(\bar{\mathbf{v}}^a_{ph})^2}\left(\bar{P}^a_{ef} + \bar{P}^b_{ef}\right) + \right.$$

$$\left. + \bar{n}_d \frac{\sum_1^3(\bar{E}^{1,2,3}_d)^2}{2m(\mathbf{v}_s)^2}\bar{P}_d\right]_{tr,lb} \qquad 11.3$$

The contribution to pressure, caused by primary effectons with the zero resulting momentum, is close to zero, as well as probability of their origination in the gas phase.

It is assumed in such approach that for real gases the model of a system of weakly interacted *oscillators* is working. The validity of such approach for water vapor is confirmed by the available experimental data indicating the presence of dimers, trimers and larger $H_2O$ clusters in the vapor (Eisenberg and Kauzmann, 1969).

The water vapor has an intensive band in oscillatory spectra at $\tilde{\nu} = 200 cm^{-1}$. Possibly, it is this band that characterizes the frequencies of quantum beats between "acoustic" *(a)* and "optic" *(b)* translational oscillations in pairs of molecules and small clusters. The frequencies of librational collective modes in vapor are absent, pointing to the absence of primary librational effectons.

The energies of primary gas micro-effectons ($h\nu_a$ and $h\nu_b$) can be calculated on the basis of formulas (4.6-4.9).

However, to calculate the energies of secondary quasi-particles in (ā) and *(b)* states the Bose-Einstein distribution must be used for the case, when the *temperature is higher than the Bose-condensation temperature* ($T > T_0$) and the chemical potential is not equal to zero ($\mu < 0$).

According to this distribution:

$$\begin{cases} \bar{E}^a = h\bar{\nu}^a = \dfrac{h\nu^a}{\exp\left(\dfrac{h\nu^a-\mu}{kT}\right)-1} \end{cases}_{tr,lb}$$

$$\begin{cases} \bar{E}^b = h\bar{\nu}^b = \dfrac{h\nu^b}{\exp\left(\dfrac{h\nu^b-\mu}{kT}\right)-1} \end{cases}_{tr,lb}$$

11.4

The kinetic energies of effectons $(\bar{a})_{tr,lb}$ and $(b)_{tr,lb}$ states are equal, only the potential energies differ, like in the case of condensed matter. All other parameters in basic equation (11.3) can be calculated as previously described.

### 11.2. New general state equation for condensed matter

Using our (eq. 4.3) for the total internal energy of condensed matter ($U_{tot}$), we can present state equation (11.1) in a more general form than (11.3).

For this end we introduce the notions of *internal pressure* ($P_{in}$), including *all type of interactions* between molecules/atoms in condensed matter and the excluded molar volume ($V_{exc}$):

$$V_{exc} = \frac{4}{3}\pi a^* N_0 = V_0\left(\frac{n^2-1}{n^2}\right) \qquad 11.5$$

where $a^*$ is the acting polarizability of molecules in condensed matter (**eq. 9.50**); $N_0$ is Avogadro number, and $V_0$ is molar volume.

*The state equation for condensed matter, general for liquids and solids,* following from our approach, taking into account structural factor of condensed matter: $S = T_{kin}/U_{tot}$, can be expressed, introducing the effective internal energy $U_{ef} = U_{tot}^2/T_{kin} = U_{tot}/S$ as:

$$P_{tot}V_{fr} = (P_{ext}+P_{in})V_{fr} = \frac{U_{tot}}{S} = U_{ef} \qquad 11.6$$

where $P_{tot} = P_{ext}+P_{in}$ is total pressure, $P_{ext}$ and $P_{in}$ are external and internal pressures;

$$V_{fr} = V_0 - V_{exc} = V_0/n^2 \qquad 11.6a$$

is a free molar volume, following from (11.5); $U_{tot} = V + T_{kin}$ is the total internal energy, $V$ and $T_{kin}$ are the potential and kinetic energies of one mole of matter.

$U_{ef} = U_{tot}(1 + V/T_{kin}^t) = U_{tot}^2/T_{kin} = U_{tot}/S$ is the effective internal energy and:

$$(1 + V/T_{kin}) = U_{tot}/T_{kin} = S^{-1} \qquad 11.6b$$

is the reciprocal value of the total structural factor (*eq. 2.46a*);

For the limit case of ideal gas, when $P_{in} = 0$; $V_{exc} = 0$ and the potential energy $V = 0$, we get from (11.6) the Clapeyrone - Mendeleyev equation (see 11.1):

$$P_{ext}V_0 = T_{kin} = RT$$

#### 11.2.1 The analysis of new general state equation of condensed matter. The internal pressure calculation for ice and water

One can use the state equation (11.6) for estimation of sum of *all types of internal matter interactions*, which determines the internal pressure $P_{in}$:

$$P_{in} = \frac{U_{ef}}{V_{fr}} - P_{ext} = \frac{n^2 U_{tot}^2}{V_0 T_{kin}} - P_{ext} \qquad 11.7$$

where the molar free volume: $V_{fr} = V_0 - V_{exc} = V_0/n^2$;

and the effective total energy: $U_{ef} = U_{tot}^2/T_{kin} = U_{tot}/S$, where $U_{tot} = T_{kin} + V_{pot}$ and the total structural factor is:

$$S = T_{kin}/U_{tot} \quad \text{and} \quad \frac{1}{S} = \frac{U_{tot}}{T_{kin}} = 1 + \frac{V_{pot}}{T_{kin}} \qquad 11.7a$$

For solids and most of the liquids with good approximation $P_{in} \gg [P_{ext} \sim 1\ atm = 10^5\ Pa]$ and, consequently, (11.7) turns to:

$$P_{in} \cong \frac{n^2 U_{tot}^2}{V_0 T_{kin}} = \frac{n^2 U_{tot}}{V_0 S} = \frac{n^2}{V_0} U_{tot}\left(1 + \frac{V_{pot}}{T_{kin}}\right) \cong P_{tot} \qquad 11.8$$

where $T_{kin}$ and $V_{pot}$ are the total kinetic and potential energies of one mole of condensed matter, respectively.

For example, for 1 mole of water under standard conditions we obtain:

$$V_{exc} = 8.4 cm^3;\quad V_{fr} = 9.6\ cm^3;\quad V_0 = V_{exc} + V_{fr} = 18\ cm^3;$$

$$P_{in} \cong 380000\ atm = 3.8 \times 10^{10}\ Pa\ (1 atm = 10^5 Pa)$$

The parameters, such as the velocity of sound, molar volume and positions of translational and librational bands in oscillatory spectra, which determine the effective total energy $U_{ef} = U_{tot}^2/T_{kin}$ depend on external pressure and temperature.

The results of computer calculations of the internal pressure $P_{in}$ (eq. 11.7) for ice and water at standard external atmospheric pressure are presented on Figure 33 a,b.

Polarizability and, consequently, free volume ($V_{fr}$) and $P_{in}$ in (11.6) depend on energy of external electromagnetic fields (Figure26).

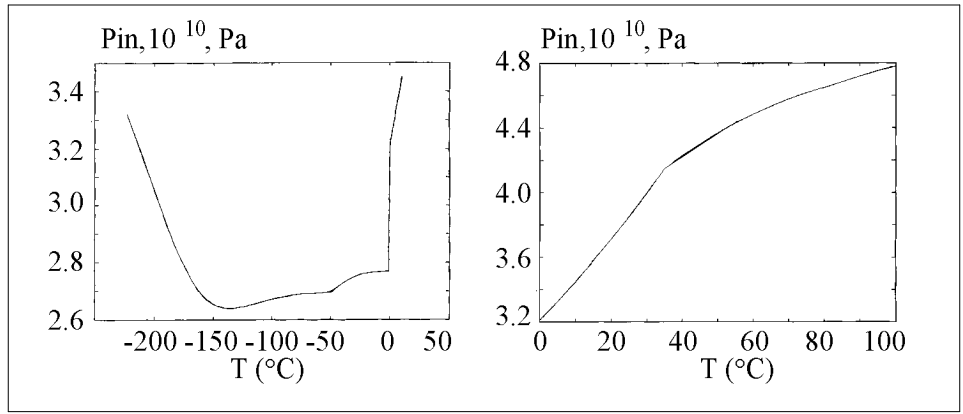

**Figure 33**. *(a)* Theoretical temperature dependence of internal pressure ($P_{in}$) in ice including the point of [ice ⇔ water] phase transition; *(b)* Theoretical temperature dependence of internal pressure ($P_{in}$) in water. Computer calculations were performed using eq. (11.7).

The minima of $P_{in}(T)$ for ice at $-140^0 C$ and $-50^0 C$ in accordance with (11.9) correspond to the ice most stable structure, related to temperature transition. The high T- transition in $P_{in}$ occur near melting and looks quite natural. However, the low T - transition of $P_{in}$ around $-170^0 C$ is not evident. For the hand, this transition coincides with calculated minimum of total partition function (Figure3a), pointing to the less stable state of the ice structure at this temperature. Experimentally this phenomena is confirmed in calorimetric experiments as a maximum of the ice heat capacity at this temperature and slowest relaxation time to heat equilibrium at $-170^0 C$ (Maeno, 1988).

In water the extremum in $\left(\partial P_{in}/\partial T\right)$ dependence appears around $35^0 C$, *i.e.* near physiological temperature.

There may exist conditions when the derivatives of internal pressure $P_{in}$ are equal to zero:

$$(a): \left(\frac{\partial P_{in}}{\partial P_{ext}}\right)_T = 0 \text{ and } (b): \left(\frac{\partial P_{in}}{\partial T}\right)_{P_{ext}} = 0 \qquad 11.9$$

This condition corresponds to the *minima of potential energy, i.e. to the most stable structure of given matter*. In a general case there may be a few metastable states when conditions (11.9) are fulfilled.

Equation of state (11.7) may be useful for the study of mechanical properties of condensed matter and their change under different influences.

Differentiation of (11.6) by external pressure gives us at $T = const$:

$$V_{fr} + V_{fr}\frac{\partial P_{in}}{\partial P_{ext}} + \frac{\partial V_{fr}}{\partial P_{ext}}(P_{ex} + P_{in}) = \frac{\partial P_{ef}}{\partial P_{ext}} \qquad 11.10$$

Dividing the left and right part of (11.10) by free volume $V_{fr}$ we obtain:

$$\left(\frac{\partial P_{in}}{\partial P_{ext}}\right)_T = \left(\frac{\partial P_{ef}}{\partial P_{ext}}\right)_T - \left[1 + \beta_T(P_{ext} + P_{in})\right]_T \qquad 11.11$$

where

$$\beta_T = -(\partial V_{fr}/\partial P_{ext})/V_{fr} \qquad 11.12$$

is isothermal compressibility.

Using (11.9a) and (11.11) we derive condition for isothermal *maximum stability* of matter structure at $\left(\frac{\partial P_{in}}{\partial P_{ext}}\right)_T = 0$:

$$\left(\frac{\partial P_{ef}}{\partial P_{ext}}\right)_T = 1 + \beta_T^0 P_{tot}^{opt} \qquad 11.13$$

where $P_{tot}^{opt} = P_{ext} + P_{in}^{opt}$ is the "optimum" total pressure, providing maximum stability of condensed matter at $T = const$.

The derivative of (11.6) by temperature gives us at isobaric condition $P_{ext} = const$:

$$(P_{ext} + P_{in})\left(\frac{\partial V_{fr}}{\partial T}\right)_{P_{ext}} + V_{fr}\left(\frac{\partial P_{in}}{\partial T}\right)_{P_{ext}} = \left(\frac{\partial U_{ef}}{\partial T}\right)_{P_{ext}} = C_V \qquad 11.14$$

When condition (11.9b) is fulfilled, we obtain for optimum internal pressure $(P_{in}^{opt})$ from (11.14):

$$P_{in}^{opt} = C_V/\left(\frac{\partial V_{fr}}{\partial T}\right)_{P_{ext}} - P_{ext} \qquad 11.16$$

or

$$P_{in}^{opt} = \frac{C_V}{V_{fr}\gamma} - P_{ext} \qquad 11.17$$

where

$$\gamma = (\partial V_{fr}/\partial T)/V_{fr} \qquad 11.18$$

is the thermal expansion coefficient;

$V_{fr}$ is the total free volume in 1 mole of condensed matter (see 11.6a):

$$V_{fr} = V_0 - V_{exc} = V_0/n^2 \qquad 11.19$$

as far the excluded volume, occupied by molecules in 1 mole of matter (11.5):
$$V_{exc} = \frac{4}{3}\pi a^* N_0 = V_0\left(\frac{n^2-1}{n^2}\right)$$

It is taken into account in (11.14) and (11.19) that

$$(\partial V_{exc}/\partial T) \cong 0 \qquad 11.20$$

because, as has been shown earlier (Figure 25a),

$$\partial \alpha^*/\partial T \cong 0$$

Dividing the left and right parts of (11.14) by $P_{tot}V_{fr} = U_{ef}$, we obtain for the heat expansion coefficient:

$$\gamma = \frac{C_V}{U_{ef}} - \frac{1}{P_{tot}}\left(\frac{\partial P_{in}}{\partial T}\right)_{P_{ext}} \qquad 11.21$$

Under metastable states, when condition (11.9,b) is fulfilled,

$$\gamma^0 = C_V/U_{ef} \qquad 11.22$$

Putting (11.8) into (11.13), we obtain for isothermal compressibility of metastable states corresponding to (11.9a) following formula:

$$\beta_T^0 = \frac{V_0 T_{kin}}{n^2 U_{tot}^2}\left(\frac{\partial U_{ef}}{\partial P_{ext}} - 1\right) \qquad 11.23$$

Our unified state equation (11.7), based on the Hierarchic Theory, can be applied for study of different types of external influences (pressure, temperature, electromagnetic radiation, deformation, etc.) on the thermodynamic and mechanic properties of solids and liquids.

### 11.3. New theory of vapor pressure and its computerized confirmation

When a liquid is incubated long enough in a closed vessel at constant temperature, then an equilibrium between the liquid and vapor is attained. At this moment, the number of molecules evaporated and condensed back to liquid are equal. The same is true of the process of sublimation. There is no satisfactory quantitative theory for *vapor pressure* calculation yet.

We suggested such a theory using introduced in the Hierarchic Theory notion of *super-deformons,* representing the biggest thermal fluctuations (see Table 1 and Introduction). The basic idea is that the external equilibrium vapor pressure is related to internal one ($P_{in}^S$) with the factor determined by the probability of cavitational fluctuations (super-deformons) in the *surface layer* of liquids or solids.

In other words due to excitation of super-deformons with probability ($P_D^S$), the internal pressure ($P_{in}^S$) in surface layers, determined by the total contributions of all intramolecular interactions turns to external one, responsible for vapor pressure ($P_V$). This process resembles the compressed spring energy releasing after the trigger switching off.

For taking into account the difference between the surface and bulk internal pressure ($P_{in}$) we introduce the semiempirical surface pressure factor ($q^S$) as:

$$P_{in}^S = q^S P_{in}^{bulc} - P_{ext} = q^S \frac{n^2 U_{tot}}{V_0 S} - P_{ext} \qquad 11.24$$

where $P_{in}$ corresponds to (eq. 11.7); $S = T_{kin}/U_{tot}$ is a total structure factor.

The value of surface factor ($q^S$) for liquid and solid states is not the same:

$$q_{liq}^S = \left(\frac{P_{in}^{surf}}{P_{in}^{bulc}}\right)_{liq} < q_{solid}^S = \left(\frac{P_{in}^{surf}}{P_{in}^{bulc}}\right)_{solid} \qquad 11.25$$

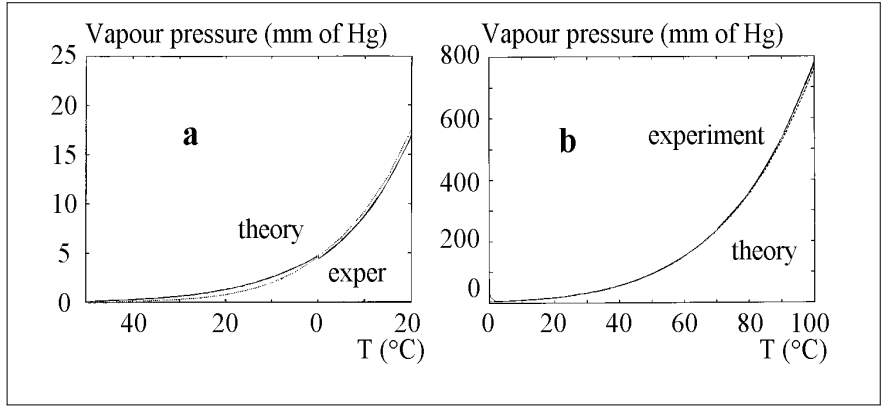

**Figure 34**. a) Theoretical (−) and experimental (· ·) temperature dependencies of vapor pressure ($P_{vap}$) for ice *(a)* and water *(b)* including phase transition region. Computer calculations were performed using eq. (11.26).
The experimental data were taken from Handbook of Chem. & Phys., 67 ed., CRC press, 1986-1987.

Multiplying (11.24) to probability of super-deformons excitation we obtain for vapor pressure, resulting from evaporation or sublimation, the following formulae:

$$P_V = P_{in}^S P_D^S = \left( q^S \frac{n^2 U_{tot}^2}{V_0 T_{kin}} - P_{ext} \right) \exp\left( -\frac{E_D^S}{kT} \right) \qquad 11.26$$

where

$$P_D^S = \exp\left( -\frac{E_D^S}{kT} \right) \qquad 11.27$$

is a probability of super-deformons excitation (see eqs. 3.31 and 3.32).

Theoretical calculated temperature dependencies of vapor pressure, described by (11.26) coincide very well with experimental ones at surface pressure factor $q_{liq}^S = 3.1$ for water and $q_{solid}^S = 18$ for ice (Figure 34).

The almost five-times difference between $q_{solid}^S$ and $q_{liq}^S$ means that the *surface* properties of ice differ from the internal *bulk* ones much more than that of liquid water.

The surface pressure factors $q_{liq}^S$ and $q_{sol}^S$ should be considered as a fit parameters. They are the only fit parameters, that was used in the Hierarchic Theory of condensed matter. The calculation of surface pressure factors from the known vapor pressure or surface tension can give an important information itself.

### 11.4. New theory of surface tension and its quantitative confirmation

The resulting surface tension is introduced in the Hierarchic Model as a sum:

$$\sigma = (\sigma_{tr} + \sigma_{lb}) \qquad 11.28$$

where $\sigma_{tr}$ and $\sigma_{lb}$ are translational and librational contributions to surface tension. Each of these components can be expressed using our mesoscopic state equation (11.6 and 11.7), taking into account the difference between surface and bulk total energies ($q^S$), introduced in previous section:

$$\sigma_{tr,lb} = \frac{1}{\frac{1}{\pi}(V_{ef})_{tr,lb}^{2/3}} \left[ \frac{q^S P_{tot}(P_{ef} V_{ef})_{tr,lb} - P_{tot}(P_{ef} V_{ef})_{tr,lb}}{(P_{ef} + P_t)_{tr} + (P_{ef} + P_t)_{lb} + (P_{con} + P_{cMt})} \right] \qquad 11.29$$

where $(V_{ef})_{tr,lb}$ are volumes of primary translational and librational effectons, related to their concentration $(n_{ef})_{tr,lb}$ as:

$$(V_{ef})_{tr,lb} = (1/n_{ef})_{tr,lb};$$

$$r_{tr,lb} = \frac{1}{\pi}(V_{ef})_{tr,lb}^{2/3}$$

is an effective radius of the primary translational and librational effectons, localized on the surface of condensed matter; $q^S$ *is the surface factor, equal* to that used in (eq. 11.24-11.26); $\left[P_{tot} = P_{in} + P_{ext}\right]$ is a total pressure, corresponding to (eq. 11.6); $(P_{ef})_{tr,lb}$ is a total probability of primary effecton excitations in the *(a)* and *(b)* states:

$$(P_{ef})_{tr} = (P_{ef}^a + P_{ef}^b)_{tr}$$

$$(P_{ef})_{lb} = (P_{ef}^a + P_{ef}^b)_{lb}$$

$(P_t)_{tr}$ and $(P_t)_{lb}$ in (11.29) are the probabilities of corresponding transitons excitation; $P_{con} = P_{ac} + P_{bc}$ is the sum of probabilities of $[a_{tr}/a_{lb}]$ and $[b_{tr}/b_{lb}]$ *convertons;* $P_{cMt} = P_{ac} P_{bc}$ is the probability of Macroconverton (see Introduction and Chapter 4).

The eq. (11.29) contains the ratio:

$$(V_{ef}/V_{ef}^{2/3})_{tr,lb} = l_{tr,lb} \qquad 11.30$$

where $l_{tr} = (1/n_{ef})_{tr}^{1/3}$ and $l_{lb} = (1/n_{ef})_{lib}^{1/3}$ are the length of the edges of the primary translational and librational effectons, approximated by cube.

Using (11.30) and (11.29) the resulting surface tension (11.28) can be presented as:

$$\sigma = \sigma_{tr} + \sigma_{lb} = \pi \frac{P_{tot}(q^S - 1) \times \left[(P_{ef})_{tr}l_{tr} + (P_{ef})l_{lb}\right]}{(P_{ef} + P_t)_{tr} + (P_{ef} + P_t)_{lb} + (P_{con} + P_{cMt})} \qquad 11.31$$

where translational component of surface tension is:

$$\sigma_{tr} = \pi \frac{P_{tot}(q^s - 1)(P_{ef})_{tr}l_{tr}}{(P_{ef} + P_t)_{tr} + (P_{ef} + P_t)_{lb} + (P_{con} + P_{cMt})} \qquad 11.32$$

and librational component of $\sigma$ is:

$$\sigma_{lb} = \pi \frac{P_{tot}(q^S - 1)(P_{ef})_{lb}l_{lb}}{(P_{ef} + P_t)_{lb} + (P_{ef} + P_t)_{lb} + (P_{con} + P_{cMt})} \qquad 11.33$$

Under the boiling condition when $q^S \to 1$ as a result of $(U_{tot}^S \to U_{tot})$, then $\sigma_{tr}$, $\sigma_{lb}$ and $\sigma$ tends to zero. The maximum depth of the surface layer, which determines the $\sigma_{lb}$ is equal to the length of edge of cube $(l_{lb})$, that approximates the shape of primary *librational* effectons. It decreases from about 20 Å at $0^0C$ till about 2.5 Å at $100^0C$ (see Figure 7b). Monotonic decrease of $(l_{lb})$ with temperature could be accompanied by nonmonotonic change of probabilities of [lb/tr] convertons and macroconvertons excitations (see comments to Figure 7a from Chapter 6). Consequently, the temperature dependence of surface tension on temperature can display anomalies at definite temperatures. This consequence of our theory is confirmed experimentally (Adamson, 1982; Drost-Hansen and Lin Singleton, 1992).

The thickness of layer $(l_{tr})$, responsible for contribution of *translational* effectons in surface tension $(\sigma_{tr})$ has the dimension of one molecule in all temperature interval for liquid water.

The results of computer calculations of $\sigma$ (eq. 11.31) for water and experimental data are presented at Figure 35.

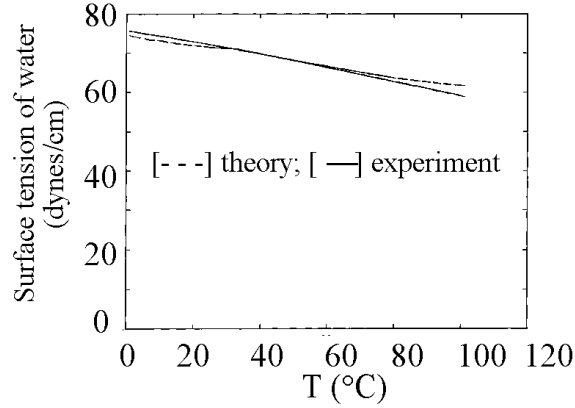

**Figure 35**. Experimental (—) and theoretical (- - -) temperature dependencies of the surface tension for water, calculated from (eq. 11.31). The experimental data were taken from Handbook of Chem. & Phys., 67 ed., CRC press, 1986-1987.

It is obvious, that the correspondence between theory and experiment is very good, confirming in such a way the correctness of our model and Hierarchic concept in general.

### 11.5. Hierarchic Theory of thermal conductivity

Thermal conductivity may be related to phonons, photons, free electrons, holes and [electron-hole] pairs movement. We will discuss here only the main type of thermal conductivity in condensed matter, related to phonons.

The analogy with the known formula for thermal conductivity ($\kappa$) in the framework of the kinetic theory for gas is used (Blackmore, 1985):

$$\kappa = \frac{1}{3} C_v \mathbf{v}_s \Lambda \qquad 11.34$$

where $C_v$ is the heat capacity of condensed matter, $\mathbf{v}_s$ is the velocity of sound, characterizing the speed of phonon propagation in matter, and $\Lambda$ is the average *length of free run* of phonons.

The value of $\Lambda$ depends on the scattering and dissipation of phonons on the other phonons and different types of defects. Usually decreasing temperature increases $\Lambda$.

Different factors influencing a thermal equilibrium in the system of phonons are discussed. Among them are the so called U- and N- processes describing the types of phonon-phonon interaction. However, the traditional theories are unable to calculate $\Lambda$ directly.

The Hierarchic Theory introduce two contributions to thermal conductivity:

- related to phonons, irradiated by secondary effectons and forming secondary translational and librational deformons ($\kappa_{sd})_{tr,lb}$ and
- related to phonons, irradiated by ($a_{tr}/a_{lb}$) and ($b_{tr}/b_{lb}$) convertons, exciting the convertons-induced deformons ($\kappa_{cd})_{ac,bc}$:

$$\kappa = (\kappa_{sd})_{tr,lb} + (\kappa_{cd})_{ac,bc} = \frac{1}{3} C_v \mathbf{v}_s [(\Lambda_{sd})_{tr,lb} + (\Lambda_{cd})_{ac,bc}] \qquad 11.35$$

where the *free runs* of secondary phonons (translational and librational) are represented as:

$$1/(\Lambda_{sd})_{tr,lb} = 1/(\Lambda_{tr}) + 1/(\Lambda_{lb}) = (\bar{v}_d)_{tr}/\mathbf{v}_s + (\bar{v}_d)_{lb}/\mathbf{v}_s$$

consequently:

$$1/(\Lambda_{sd})_{tr,lb} = \frac{\mathbf{v}_s}{(\bar{v}_d)_{tr} + (\bar{v}_d)_{lb}} \qquad 11.36$$

and free runs of convertons-induced phonons:

$$1/(\Lambda_{cd})_{ac,bc} = 1/(\Lambda_{ac}) + 1/(\Lambda_{bc}) = (v_{ac})/\mathbf{v}_s + (v_{bc})/\mathbf{v}_s$$

$$\text{consequently: } (\Lambda_{sd})_{tr,lb} = \frac{\mathbf{v}_s}{(v_d)_{tr} + (v_d)_{lb}} \qquad 11.37$$

The heat capacity: $C_V = \partial U_{tot}/\partial T$ can be calculated also from our theory (See chapters 4 and 5).

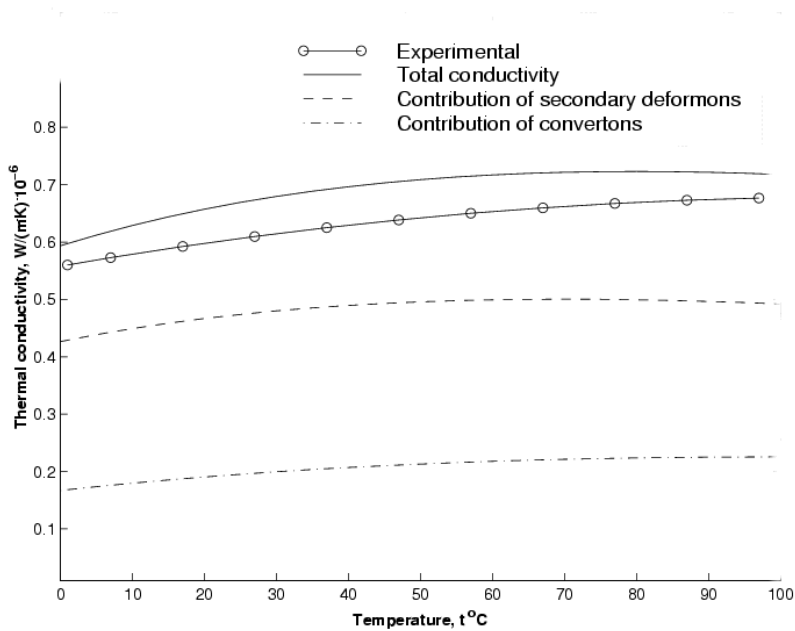

**Figure 36**. Temperature dependencies of total thermal conductivity for water and contributions, related to acoustic deformons and [lb/tr]convertons. The dependencies were calculated, using eq. (11.35).
The experimental data were taken from Handbook of Chem. & Phys., 67 ed., CRC press, 1986-1987.

Quantitative calculations show that formula (11.35), based on the Hierarchic Model, works well for water (Figure 36). It could be used for any other condensed matter also if positions of translational and librational bands, the velocity of sound and molar volume for this matter at the same temperature interval are known.

The small difference between experimental and theoretical data can reflect the contributions of non-phonon process in thermal conductivity, related to macro-deformons, super-deformons and macroconvertons, *i.e.* big fluctuations.

### 11.6. Hierarchic Theory of viscosity

The viscosity is determined by the energy dissipation as a result of medium (liquid or solid) structure deformation. Viscosity corresponding to the shift deformation is named *shear viscosity*. So-called *bulk viscosity* is related to deformation of volume parameters and corresponding dissipation. These types of viscosity have not the same values and nature.

The statistical theory of irreversible process leads to the following expression for shear viscosity (Prokhorov, 1988):

$$\eta = nkT\tau_p + (\mu_\infty - nkT)\tau_q \qquad 11.38$$

where n is the concentration of particles, $\mu_\infty$ is the modulus of instant shift characterizing the instant elastic reaction of medium, $\tau_p$ and $\tau_q$ are the relaxation times of momentums and coordinates, respectively.

However, (eq. 11.38) is inconvenient for practical purposes due to difficulties in determination of $\tau_p, \tau_q$ and $\mu_\infty$.

Sometimes in a narrow temperature interval the empirical Ondrade equation is working:

$$\eta = A(T) \times \exp(\beta/T) \qquad 11.39$$

A(T) is a function poorly dependent on temperature.

A good results in study the microviscosity problem were obtained by combining the model of molecular rotational relaxation (Dote *et al.*, 1981) and the Kramers equation (Åkesson *et al.*, 1991). However, the using of the fit parameters was necessarily in this case also.

*We present here the Hierarchic Theory of viscosity*. To this end the dissipation processes, related to $(A \rightleftharpoons B)_{tr,lb}$ cycles of translational and librational macro-effectons, $(a_{tr}/a_{lb})$ and $(b_{tr}/b_{lb})$-*convertons* excitations were used. The same approach was employed for elaboration of the Hierarchic Theory of diffusion in condensed matter (see next section).

*In contrast to liquid state, the viscosity of solids* is determined by the biggest fluctuations: *super-effectons* and *super-deformons,* resulting from simultaneous excitations of translational and librational macro-effectons and macro-deformons in the same volume.

The dissipation phenomena and ability of particles or molecules to diffusion are related to the local fluctuations of the free volume $(\Delta \mathbf{v}_f)_{tr,lb}$. According to the Hierarchic Theory, the fluctuations of free volume and that of density occur in the almost macroscopic volumes of translational and librational macro-deformons and in mesoscopic volumes of *macroconvertons*, equal to volume of primary librational effecton at the given conditions. Translational and librational types of macro-effectons determine two types of viscosity, *i.e.* translational $(\eta_{tr})$ and librational $(\eta_{lb})$ ones. They can be attributed to the bulk viscosity. The contribution to viscosity, determined by *(a and b)- convertons is much more local and may be responsible for microviscosity and mesoviscosity.*

Let us start from calculation of the additional free volumes $(\Delta \mathbf{v}_f)$ originating from fluctuations of density, accompanied the translational and librational macro-deformons (macro-transitons).

For one mole of condensed matter the following ratio between free volume and concentration fluctuations is true:

$$\left(\frac{\Delta \mathbf{v}_f}{\mathbf{v}_f}\right)_{tr,lb} = \left(\frac{\Delta N_0}{N_0}\right)_{tr,lb} \qquad 11.40$$

where $N_0$ is the average number of molecules in 1 mole of matter

$$\text{and} \qquad (\Delta N_0)_{tr,lb} = N_0 \left(\frac{P_D^M}{Z}\right)_{tr,lb} \qquad 11.41$$

is the number of molecules changing their concentration as a result of translational and librational macro-deformons excitation.

The probability of translational and librational macro-effectons excitation (see eqs. 3.23; 3.24):

$$\left(\frac{P_D^M}{Z}\right)_{tr,lb} = \frac{1}{Z} \exp\left(-\frac{\epsilon_D^M}{kT}\right)_{tr,lb} \qquad 11.42$$

where Z is the total partition function of the system (Chapter 4).

Putting (11.41) to (11.40) and dividing to Avogadro number ($N_0$), we obtain the fluctuating free volume, reduced to 1 molecule of matter:

$$\Delta \mathbf{v}_f^0 = \frac{\Delta \mathbf{v}_f}{N_0} = \left[ \frac{\mathbf{v}_f}{N_0} \left( \frac{P_D^M}{Z} \right) \right]_{tr,lb} \qquad 11.43$$

It has been shown above (eq. 11.19) that the average value of free volume in 1 mole of matter is:

$$\mathbf{v}_f = V_0/n^2$$

Consequently, for reduced fluctuating (additional) volume we have:

$$(\Delta \mathbf{v}_f^0)_{tr,lb} = \frac{V_0}{N_0 n^2} \frac{1}{Z} \exp\left(-\frac{\epsilon_D^M}{kT}\right)_{tr,lb} \qquad 11.44$$

Taking into account the dimensions of viscosity and its physical sense, it should be proportional to the work (activation energy) of fluctuation-dissipation, necessary for creating the unit of additional free volume: $(E_D^M/\Delta \mathbf{v}_f^0)$, and the period of $(A \rightleftharpoons B)_{tr,lb}$ cycles of translational and librational macro-effectons $\tau_{A \rightleftharpoons B}$, determined by the life-times of all intermediate states (eq. 11.46).

In turn, the energy of dissipation should be strongly dependent on the *structural factor* (S): the ratio of kinetic energy of matter to its *total* internal energy. We assume here that this dependence for viscosity calculation is cubical: $(T_k/U_{tot})^3 = S^3$.

Consequently, the contributions of translational and librational macro-deformons to resulting viscosity we present in the following way:

$$\eta_{tr,lb}^M = \left[ \frac{E_D^M}{\Delta \mathbf{v}_f^0} \tau^M \left( \frac{T_k}{U_{tot}} \right)^3 \right]_{tr,lb} \qquad 11.45$$

where reduced fluctuating volume $(\Delta \mathbf{v}_f^0)$ corresponds to (11.44); the energy of macro-deformons: $[E_D^M = -kT (\ln P_D^M)]_{tr,lb}$.

The cycle-periods of the *tr* and *lib* macro-effectons has been introduced as:

$$[\tau^M = \tau_A + \tau_B + \tau_D]_{tr,lb} \qquad 11.46$$

where characteristic life-times of macro-effectons in A, B-states and that of transition state in the volume of primary electromagnetic deformons can be presented, correspondingly, as follows (see eqs.4.49-4.51):

$$\left[ \tau_A = (\tau_a \tau_{\bar{a}})^{1/2} \right]_{tr,lb} \quad \text{and} \quad \left[ \tau_A = (\tau_a \tau_{\bar{a}})^{1/2} \right]_{tr,lb} \qquad 11.47$$

$$\left[ \tau_D = |(1/\tau_A) - (1/\tau_B)|^{-1} \right]_{tr,lb}$$

Using (11.47, 11.46 and 11.44) it is possible to calculate the contributions of $(A \rightleftharpoons B)$ cycles of translational and librational macro-effectons to viscosity separately, using (11.45).

The averaged contribution of macro-excitations (**tr** and **lb**) in viscosity is:

$$\eta^M = \left[ (\eta)_{tr}^M (\eta)_{lb}^M \right]^{1/2} \qquad 11.48$$

The contribution of $a_{tr}/a_{lb}$ and $b_{tr}/b_{lb}$ *convertons* to viscosity of liquids could be presented in a similar to (11.44-11.48) manner after substituting the parameters of *tr* and *lb* macro-effectons with parameters of convertons:

$$\eta_{ac,bc} = \left[ \frac{E_c}{\Delta \mathbf{v}_f^0} \tau_c \left( \frac{T_k}{U_{tot}} \right)^3 \right]_{ac,bc} \qquad 11.49$$

where reduced fluctuating volume of convertons $(\Delta v_f^0)_{ac,bc}$ corresponds to:

$$(\Delta \mathbf{v}_f^0)_{ac,bc} = \frac{V_0}{N_0 n^2} \frac{1}{Z} P_{ac,bc} \qquad 11.50$$

where $P_{ac}$ and $P_{bc}$ are the relative probabilities of *tr/lb* interconversions between $(a_{tr}/a_{lb})$ and

($b_{tr}/b_{lb}$) states of translational and librational primary effectons (see Introduction and Chapter 4); $E_{ac}$ and $E_{bc}$ are the excitation energies of ($a_{tr}/a_{lb}$) and ($b_{tr}/b_{lb}$) convertons (See chapter 4);

Characteristic life-times of corresponding convertons [tr/lb] in the volume of primary librational effectons ("flickering clusters") can be presented as:

$$\tau_{ac} = (\tau_a)_{tr} + (\tau_a)_{lb} = (1/\nu_a)_{tr} + (1/\nu_a)_{lb}$$
$$\tau_{bc} = (\tau_b)_{tr} + (\tau_b)_{lb} = (1/\nu_b)_{tr} + (1/\nu_b)_{lb}$$
11.51

The averaged contribution of the both types of convertons in viscosity is:

$$\eta_c = (\eta_{ac}\, \eta_{bc})^{1/2}$$
11.52

This contribution could be responsible for microviscosity or better term: mesoviscosity, related to volumes, equal to that of primary librational effectons.

The resulting viscosity (Figure 37) is a sum of the averaged contributions of macro-deformons and convertons:

$$\eta = \eta^M + \eta_c$$
11.53

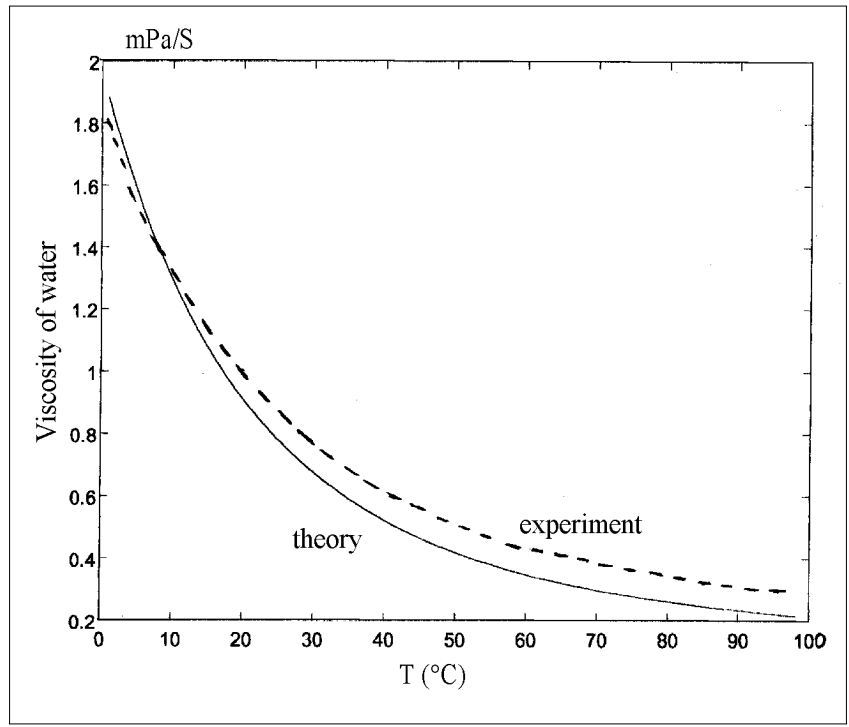

**Figure 37**. Theoretical and experimental temperature dependencies of viscosities for water. Computer calculations were performed using eqs. (11.44 - 11.53) and (4.3; 4.36). The experimental data were taken from Handbook of Chem. & Phys., 67 ed., CRC press, 1986-1987.

The best correlation between theoretical and experimental data was achieved after assuming that only ($\pi/2 = 2\pi/4$) part of the period of above described fluctuation cycles is important for dissipation and viscosity. Introducing this factor to equations for viscosity calculations gives up very good correspondence between theory and experiment in all temperature interval (0-100°C) for water (Figure37).

As will be shown below the same factor, introducing the effective time of fluctuations [$\frac{\tau}{\pi/2}$], leads to best results for self-diffusion coefficient calculation.

In the classical hydrodynamic theory the sound absorption coefficient ($\alpha$) obtained by Stokes includes share ($\eta$) and bulk ($\eta_b$) averaged macroviscosities:

$$\alpha = \frac{\Omega}{2\rho \mathbf{v}_s^3}\left(\frac{4}{3}\eta + \eta_b\right), \qquad 11.54$$

where $\Omega$ is the angular frequency of sound waves; $\rho$ is the density of liquid.

Bulk viscosity ($\eta_b$) is usually calculated from the experimental $\eta$ and $\alpha$. It is known that for water:

$$(\eta_b/\eta) \sim 3.$$

### 11.6a. The viscosity of solids

In accordance with our model, the biggest fluctuations: super-effectons and super-deformons (see Introduction) are responsible for viscosity and diffusion phenomena in solid state. super-deformons are accompanied by the emergency of cavitational fluctuations in liquids and the defects in solids. The presentation of viscosity formula in solids ($\eta_s$) is similar to that for liquids:

$$\eta_s = \frac{E_S}{(\Delta \mathbf{v}_f^0)_S}\,\tau_S\left[\frac{T_k}{U_{tot}}\right]^3 \qquad 11.55$$

where reduced fluctuating volume, related to super-deformons excitation $(\Delta \mathbf{v}_f^0)_s$ is:

$$(\Delta \mathbf{v}_f^0)_S = \frac{V_0}{N_0 n^2}\frac{1}{Z}P_S \qquad 11.56$$

where $P_s = (P_D^M)_{tr}\,(P_D^M)_{lb}$ is the relative probability of super-deformons, equal to product of probabilities of tr and lb macro-deformons excitation (see 11.42); $E_s = -kT \ln P_s$ is the energy of super-deformons (See chapter 4);

Characteristic cycle-period of $(A^* \rightleftharpoons B^*)$ transition of super-effecton is related to its life-times in $A^*$, $B^*$ and transition $D^*$ states (see eq.11.46) as was shown in section 4.3:

$$\tau_S = \tau_{A^*} + \tau_{B^*} + \tau_{D^*} \qquad 11.56a$$

The viscosity of ice, calculated from (eq. 11.55) is bigger than that of water (eq. 11.53) to about $10^5$ times. This result is in accordance with available experimental data.

### 11.7. Brownian diffusion

The important formula obtained by Einstein in his theory of Brownian motion is for translational motion of particle:

$$r^2 = 6Dt = \frac{kT}{\pi\eta a}t \qquad 11.57$$

and that for rotational Brownian motion:

$$\varphi^2 = \frac{kT}{4\pi\eta a^3}t \qquad 11.58$$

where a - radius of spherical particle, much larger than dimension of molecules of liquid. The coefficient of diffusion D for Brownian motion is equal to:

$$D = \frac{kT}{6\pi\eta a} \qquad 11.59$$

If we take the angle $\bar{\varphi}^2 = 1/3$ in (11.59), then the corresponding rotational correlation time comes to the form of the known Stokes- Einstein equation:

$$\tau = \frac{4}{3}\pi a^3 \frac{1}{k}\left(\frac{\eta}{T}\right) \qquad 11.60$$

All these formulas (11.57 - 11.60) include macroscopic share viscosity ($\eta$) corresponding to our (11.53).

### 11.8. Self-diffusion in liquids and solids

Molecular theory of self-diffusion, as well as general concept of *transfer phenomena* in condensed matter is extremely important, but still unresolved problem.

Simple semiempirical approach developed by Frenkel leads to following expression for diffusion coefficient in liquid and solid:

$$D = \frac{a^2}{\tau_0} \exp(-W/kT) \qquad 11.61$$

where [a] is the distance of fluctuation jump; $\tau_0 \sim (10^{-12} \div 10^{-13})\,s$ is the average period of molecule oscillations between jumps; $W$ - activation energy of jump.

The parameters: $a$, $\tau_0$ and $W$ should be considered as a fit parameters.

In accordance with the Hierarchic Theory, the process of self-diffusion in liquids, like that of viscosity, described above, is determined by two contributions:

a) the *collective contribution,* related to translational and librational macro-deformons ($D_{tr,lb}$);

b) the *local contribution,* related to coherent clusters flickering: [dissociation/association] of primary librational effectons (a and b)- convertons ($D_{ac,bc}$).

Each component of the resulting coefficient of self-diffusion (D) in liquid could be presented as the ratio of fluctuation volume cross-section surface: $[\Delta v_f^0]^{2/3}$ to the period of macrofluctuation ($\tau$). The first contribution to coefficient **D**, produced by translational and librational macro-deformons is:

$$D_{tr,lb} = \left[ (\Delta \mathbf{v}_f^0)^{2/3} \frac{1}{\tau^M} \right]_{tr,lb} \qquad 11.62$$

where the surface cross-sections of reduced fluctuating free volumes (see eq.11.43) fluctuations composing macro-deformons (*tr* and *lb*) are:

$$(\Delta \mathbf{v}_f^0)^{2/3}_{tr,lb} = \left[ \frac{V_0}{N_0 n^2} \frac{1}{Z} \exp\left(-\frac{\epsilon_D^M}{kT}\right)_{tr,lb} \right]^{2/3} \qquad 11.63$$

$(\tau^M)_{tr,lb}$ are the characteristic $(A \Leftrightarrow B)$ cycle-periods of translational and librational macro-effectons (see eqs. 11.46 and 11.47).

The averaged component of self-diffusion coefficient, which takes into account both types of nonlocal fluctuations, related to translational and librational macro-effectons and macro-deformons, can be find as:

$$D^M = [(D)_{tr}^M (D)_{lb}^M]^{1/2} \qquad 11.64$$

*The formulae for the second, local contribution to self-diffusion* in liquids, related to ($a_{tr}/a_{lb}$ and $b_{tr}/b_{lb}$) convertons ($D_{ac,bc}$) are symmetrical by form to that, presented above for nonlocal/collective processes:

$$D_{ac,bc} = \left[ (\Delta \mathbf{v}_f^0)^{2/3} \frac{1}{\tau_S} \right]_{ac,bc} \qquad 11.65$$

where reduced fluctuating free volume of (a and b) convertons $(\Delta v_f^0)_{ac,bc}$ is the same as was used above in the Hierarchic Theory of viscosity (eq. 11.50):

$$(\Delta \mathbf{v}_f^0)_{ac,bc} = \frac{V_0}{N_0 n^2} \frac{1}{Z} P_{ac,bc} \qquad 11.66$$

where $P_{ac}$ and $P_{bc}$ are the relative probabilities of *tr/lib* interconversions between $a_{tr} \rightleftharpoons a_{lb}$ and $b_{tr} \rightleftharpoons b_{lb}$ states of translational and librational primary effectons (see Introduction and Chapter 4)

The averaged local component of self-diffusion coefficient, which takes into account both types of convertons is:

$$D_C = [(D)_{ac} (D)_{bc}]^{1/2} \qquad 11.67$$

In similar way we should take into account the contribution of macroconvertons ($D_{Mc}$):

$$D_{Mc} = \left( \frac{V_0}{N_0 n^2} \frac{1}{Z} P_{Mc} \right)^{2/3} \frac{1}{\tau_{Mc}} \qquad 11.67a$$

where $P_{Mc} = P_{ac} P_{bc}$ is a probability of macroconvertons excitation;
the life-time of macroconvertons is:

$$\tau_{Mc} = (\tau_{ac} \tau_{bc})^{1/2} \qquad 11.67b$$

The cycle-period of (*ac*) and (*bc*) convertons are determined by the sum of life-times of intermediate states of primary translational and librational effectons:

$$\tau_{ac} = (\tau_a)_{tr} + (\tau_a)_{lb}; \quad \text{and} \quad \tau_{bc} = (\tau_b)_{tr} + (\tau_b)_{lb} \qquad 11.67c$$

The life-times of primary and secondary effectons (lb and tr) in *a*- and *b*-states are the reciprocal values of corresponding state frequencies:

$$[\tau_a = 1/\nu_a; \quad \tau_{\bar{a}} = 1/\nu_{\bar{a}}; \quad \text{and} \quad \tau_b = 1/\nu_b; \quad \tau_{\bar{b}} = 1/\nu_{\bar{b}}]_{tr,lb} \qquad 11.67d$$

$[\nu_a \text{ and } \nu_b]_{tr,lb}$ correspond to eqs. 4.8 and 4.9; $[\nu_{\bar{a}} \text{ and } \nu_{\bar{b}}]_{tr,lb}$ could be calculated using eqs. 2.54 and 2.55.

The resulting coefficient of self-diffusion in liquids (D) is a sum of ($D^M$) and local ($D_c$, $D_{Mc}$) contributions (see eqs.11.64 and 11.67):

$$D = D^M + D_c + D_{Mc} \qquad 11.68$$

The effective fluctuation times were taken the same as in previous section for viscosity calculation, using the correction factor $[(\pi/2) \times \tau]$.

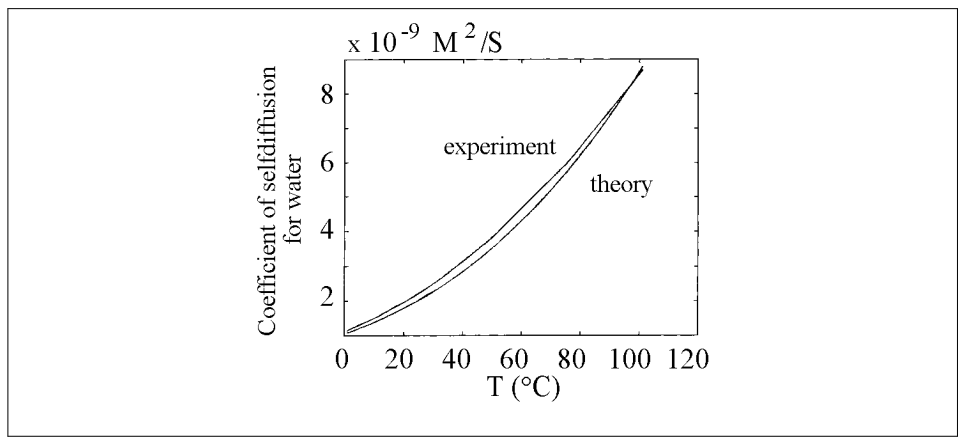

**Figure 38**. Theoretical and experimental temperature dependencies of self-diffusion coefficients in water. Theoretical coefficient was calculated using eq. 11.68. The experimental data were taken from Handbook of Chem. & Phys., 67 ed., CRC press, 1986-1987.

Like in the cases of thermal conductivity, viscosity and vapor pressure, the results of theoretical calculations of self-diffusion coefficient coincide well with experimental data for water (Figure 38) in temperature interval $(0 - 100^0 C)$.

### The self-diffusion in solids

In solid state only the biggest fluctuations: super-deformons, representing simultaneous excitation of translational and librational macro-deformons in the same volumes of matter are responsible for diffusion and the viscosity phenomena. They are related to origination and migration of the defects in solids. The formal presentation of super-deformons contribution to self-diffusion in solids ($D_s$) is similar to that of macro-deformons for liquids:

$$D_S = (\Delta \mathbf{v}_f^0)_S^{2/3} \frac{1}{\tau_S} \qquad 11.69$$

where reduced fluctuating free volume composing super-deformons $(\Delta v_f^0)_S$ is the same as was used above in the Hierarchic Theory of viscosity (eq. 11.56):

$$(\Delta \mathbf{v}_f^0)_S = \frac{V_0}{N_0 n^2} \frac{1}{Z} P_S \qquad 11.70$$

where $P_S = (P_D^M)_{tr} (P_D^M)_{lb}$ is the relative probability of super-deformons, equal to product of probabilities of tr and lb macro-deformons excitation (see 11.42).

Characteristic cycle-period of super-effectons is related to that of tr and lb macro-effectons like it was presented in (eq. 11.56a):

$$\tau_S = \tau_{A^*} + \tau_{B^*} + \tau_{D^*} \qquad 11.71$$

The self-diffusion coefficient for ice, calculated from eq.11.69 is less than that of water (eq. 11.53) to about $10^5$ times. This result is in accordance with available experimental data.

Strong decreasing of D in a course of phase transition: [water → ice] predicted by the Hierarchic Theory also is in accordance with experiment (Figure 39).

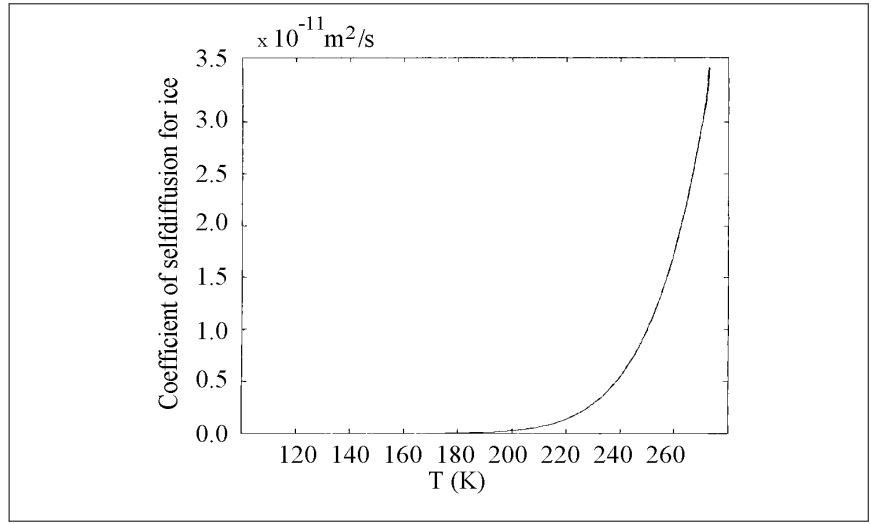

**Figure 39**. Theoretical temperature dependencies of self-diffusion coefficients in ice.

All these results allow to consider the Hierarchic Theory of transfer phenomena as the quantitatively confirmed one. They point that the "mesoscopic bridge" between Micro- and Macro Worlds is wide and reliable indeed. It gives new possibilities for understanding and detailed description of very different phenomena in solids and liquids.

### The consequences of the Hierarchic Theory of viscosity and self-diffusion

One of the important consequences of our theory of viscosity and diffusion is the possibility of explaining numerous nonmonotonic temperature changes, registered by a number of physicochemical methods in various aqueous systems during the study of temperature dependencies (Drost-Hansen, 1976, Drost-Hansen and Singleton, 1992; Johri and Roberts, 1990; Aksnes and Asaad, 1989; Aksnes and Libnau, 1991; Kaivarainen, 1985, 1989b).

Most of them are related to diffusion or viscosity processes and may be explained by nonmonotonic changes of the refraction index, included in our eqs: 11.44, 11.45, 11.50 for viscosity and 11.69, 11.70 for self-diffusion. For water these temperature anomalies of refraction index were revealed experimentally, using few wave lengths in the temperature interval $3 - 95^0$ (Frontas'ev and Schreiber, 1966). They are close to Drost-Hansen's temperatures. The explanation of these effects, related to periodic variation of primary librational effectons stability with monotonic temperature change was presented as a comments to Figure 7a (Chapter 6).

*Another consequence* of our theory is the elucidation of a big difference between librational $\eta_{lb}$ (11.48), translational $\eta_{tr}$ (11.45) viscosities and mesoviscosity, determined by [lb/tr] convertons (11.49 and 11.52). The effect of mesoviscosity can be checked as long as the volume of a Brownian particle does not exceed much the volume of primary librational effectons (eq. 11.15). If we take a Brownian particle, much bigger than the librational primary effecton, then its motion will reflect only averaged share viscosity (eq. 11.53).

*The third consequence* of the Hierarchic Theory of viscosity is the prediction of nonmonotonic temperature behavior of the sound absorption coefficient $\alpha$ (11.51). Its temperature dependence must have anomalies in the same regions, where the refraction index has.

The experimentally revealed temperature anomalies of refraction index (n) also follow from our theory as a result of nonmonotonic $(a \Leftrightarrow b)_{lb}$ equilibrium behavior, stability of primary librational effectons and probability of [lb/tr] convertons excitation (See chapter 6: Discussion to Figure7a).

*Our model predicts also that in the course of transition from the laminar type of flow to the turbulent one the share viscosity ($\eta$) will increases due to increasing of structural factor $(T_k/U_{\text{tot}})$ in eq. 11.45.*

The superfluidity ($\eta \to 0$) in the liquid helium could be a result of inability of this liquid at the very low temperature for translational and librational macro-effectons excitations, *i.e.* $\tau^M \to 0$ and dissipation is absent. In turn, it is a consequence of tending to zero the concentration and life-times of secondary effectons and deformons in eqs.(11.45), responsible for dissipation processes, due to their Bose-condensation, increasing the fraction of primary effectons. (see also Chapter 12). *The polyeffectons, stabilized by Josephson's junctions between primary effectons form the superfluid component of liquid helium only.*

### 11.9 Hierarchic approach to proton conductivity in water, ice and other systems, containing hydrogen bonds

The numerous models of proton transitions in water and ice are usually related to migration of two types of defects in the ideal Bernal-Fouler structure (Antonchenko *et al.*, 1991):

1. Ionic defects originated as a result of $2H_2O$ dissociation to hydroxonium and hydroxyl ions:

$$2H_2O \Leftrightarrow H_3O^+ + OH^-$$

2. Orientational Bjerrum defects are subdivided to D (dopplet) and L (leer) ones.

*D-defect* (positive) corresponds to situation, when *2 protons* are placed between two oxygen atoms, instead of the normal structure of hydrogen bond: $O\ldots H - O$ containing 1 proton.

*L-defect* (negative) corresponds to opposite anomaly, when even 1 proton between two oxygen atoms is absent. Reorientation of dipole moment of $H_2O$ in the case of D- and L-defects leads to origination of charges:

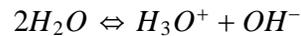

$$q_B = q_D^+ = |\ q_L^-\ | = 0.38e \qquad 11.72$$

The interrelation between the charge of electron (e), Bjerrum charge ($q_B$) and ionic charge ($e_I$) (Onsager, Dupius, 1962) is:

$$e = e_I + q_B \qquad 11.73$$

The general approach to problem of proton transition takes into account both types of defects: ionic and orientational. It was assumed that orientational defects originate and annihilate in the process of continuous migration of ions $H^+$ and $OH^-$ through the water medium. Krjachko (1987) considers DL-pairs as a cooperative water cluster with linear dimensions of about 15Å and with "kink". The Bjerrum's DL-pair is a limit case of such model.

The protons conductivity in water must decrease with temperature increasing due to decreasing and disordering of water clusters and chains.

The *kink-soliton* model of orientational defects migration along the $H_2O$ chain was developed by Sergienko (1986). Mobility of ionic defects exceeds the orientational ones about 10 times.

But it is important to point out that the strong experimental evidence confirming the existence of just Bjerrum type orientational defects are still absent.

The Hierarchic approach proton diffusion in ice, water and other hydrogen bonds containing systems includes following stages:

1. Ionization of water molecules composing super-deformons and *ionic* defects origination;
2. Bordering by $H_3^+O$ and $HO^-$ the opposite surface-sides of primary librational effectons;
3. Tunneling of proton through the volume of primary librational effecton as mesoscopic Bose condensate (mBC);
4. Diffusion of ions $H_3^+O$ and $HO^-$ in the less ordered medium between primary librational effectons can be realized via fluctuation mechanism described in Section 11.8 above . The velocity of this stage is less than quantum tunneling.

Transitions of protons and hydroxyl groups can occur also due to *exchange processes* (Antonchenko, 1991) like:

$$H_3^+O + H_2O \Leftrightarrow H_2O + H_3^+O \qquad 11.74$$

$$H_2O + HO^- \Leftrightarrow HO^- + H_2O \qquad 11.75$$

The rate of ions transferring due to exchange is about 10 times more, than diffusion velocity, but slower than that, determined by tunneling jumps.

5. The orientational defects can originate as a result of $H_2O$ molecules rearrangements and conversions between translational and librational effectons composing super-deformons. Activation energy of super-deformons and macroconvertons in water is 10.2 kcal/M and about 12 kcal/M in ice (see 6.12; 6.13). The additional activation energy about 2-3 kcal/M is necessary for subsequent reorientation of surrounding molecules (Bjerrum, 1951).

Like the ionic defects, positive (D) and negative (L) defects can form a separated pairs on the opposite sides of primary effectons, approximated by parallelepiped. Such pairs means the effectons polarization.

Probability of $H^+$ or $HO^-$ tunneling through the coherent cluster - primary effecton in the *(a)* -state is higher than that in the *(b)* -state as far (see 1.30-1.32):

$$[E_a = T_{kin}^a + V_a] < [E_b = T_{kin}^b + V_b] \qquad 11.76$$

where the kinetic energies $T_{kin}^a = T_{kin}^b$ and the difference between total and potential energies: $E_b - E_a = V_b - V_a$.

In accordance with known theory of *tunneling*, the probability of passing the particle with mass *(m)* through the barrier with wideness *(a)* and height *($\epsilon$)* has a following dependence on these parameters:

$$|\psi_a| \sim \exp\left(-\frac{a}{b}\right) = \exp\left(-\frac{a(2m\epsilon)^{1/2}}{\hbar}\right) \qquad 11.77$$

$$or: \quad |\psi_a| \sim \exp\left(-\frac{a - \lambda_B}{A_B}\right) \qquad 11.77a$$

where

$$b = \hbar/(2m\epsilon)^{1/2} \qquad 11.78$$

is the effective wave function fading length.

Parameter *(b)* is similar to de Broglie wave most probable amplitude $(A_B)$ with total energy $E_B = \epsilon$ (see eq. 2.21):

$$b = A_B = \hbar/(2mE_B)^{1/2} \qquad 11.79$$

From our theory of corpuscle-wave duality (Kaivarainen, 2006), it follows that the tunneling becomes possible when the de Broglie wave length of tunneling particle become equal or bigger that of the barrier wideness $(\lambda_B \geq a)$. As a result, we get (11.77a).

With temperature decreasing the $(a \Leftrightarrow b)_{tr,lb}$ equilibrium of primary effectons shifts to the left:

$$K_{a \Leftrightarrow b} = (P_a/P_b) \to \infty \qquad 11.80$$

where $P_a \to 1$ and $P_b \to 0$ are the thermo-accessibilities of *(a)* and *(b)* states of primary effectons (see eqs. 4.10-4.12). The linear dimensions of primary effectons also tend to infinity at $T \to 0$. This means that mesoscopic Bose condensation in solids turns to macroscopic one.

In water the tunneling stage of proton conductivity can be related to primary *librational* effectons *only* and its contribution increases with temperature decreasing. Dimensions of translational effectons in water does not exceed the dimensions of one molecule (see eq. 6.6).

Increasing of protons conductivity in ice with respect to water, in accordance with our model, is a consequence of participation of translational primary effectons in tunneling of $[H^+]$ besides librational ones, as well as significant elevation of primary librational effectons dimensions. Increasing of the total contribution of tunneling process in protons migration in ice rise up their resulting transferring velocity comparing to water.

*The external electric field induce:*

a) redistribution of positive and negative charges on the surface of primary librational effectons determined by ionic defects and corresponding orientational defects;

b) orientation of polarized primary lb-effectons in electric field, stimulating the growth of quasi-continuous polyeffectons chains and superclusters.

These effects create the conditions for $H^+$ and $H_3^+O$ propagation in direction of electric field and $[HO^-]$ in the opposite one.

*In accordance with the Hierarchic Model, the $H^+$ transition mechanism includes the combination of the tunneling, the exchange process and usual diffusion.*

### 11.10. Regulation of pH and shining of water by electromagnetic and acoustic fields

In accordance with our theory, water dissociation reaction:

$$H_2O \Leftrightarrow H^+ + HO^-$$

leading to *increase of protons concentration* is dependent on probability of $[A_S^* \to B_S^*]$ transitions of super-effectons (Table 1). This means that stimulation of $[A_S^* \to B_S^*]$ transitions (super-deformons) by ultrasound with resonant frequencies, corresponding to frequency of these transitions, should lead to decreasing of pH, i.e. to increasing the concentration of protons $[H^+]$.

The $[A_S^* \to B_S^*]$ transitions of super-effectons can be accompanied by origination of cavitational fluctuations (cavitational microbubbles). The opposite $[A_S^* \to B_S^*]$ transitions are related to the collapse of these microbubbles. As a result of this adiabatic process, water vapor in the bubbles is heated up to $4000 - 6000\ ^0K$. The usual energy of super-deformons in water (Section 6.3):

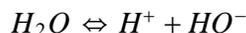

$$\epsilon_D^S = 10.2\ kcal/M \simeq RT^* \qquad 11.81$$

correspond to local temperature $T^* \simeq 5000\ ^0K$. For the other hand it is known, that even $2000\ ^0K$ is enough already for partial dissociation of water molecules inside bubbles (about 0.01% of total amount of bubble water).

The variable pressure (P), generated by ultrasound in liquid is dependent on its intensity ($I$, $wt/cm^2$) like:

$$P = (\rho \mathbf{v}_s I)^{1/2} \times 4.6 \cdot 10^{-3} (\text{atm})\qquad 11.82$$

where $\rho$ is density of liquid; $\mathbf{v}_s$ - the velocity of sound ($m/s$).

$[A_S^* \to B_S^*]$ transitions and cavitational bubbles origination can be stimulated also by IR radiation with frequency, corresponding to the activation energy of corresponding big fluctuations, described in the Hierarchic Theory by *super-deformons* and *macroconvertons.*

In such a way, using IR radiation and ultrasound it is possible to regulate a lot of different processes in aqueous systems, depending on pH and water activity.

The increasing of ultrasound intensity leads to increased cavitational bubble concentration. The dependence of the resonance cavity radius ($R_{res}$) on ultrasound frequency (f) can be approximately expressed as:

$$R_{res} = 3000/f \qquad 11.83$$

At certain conditions the water placed in the ultrasound field, *begins to shine* in the visible region: $300 - 600 nm$ (Guravlev and Akopyan, 1977). This shining, named *sonoluminiscence,* is a consequence of electronic excitation of water ions and molecules in the volume of overheated cavitational bubbles.

When the conditions of ultrasound *standing wave* exist, the number of bubbles and intensity of shining is maximal.

The intensity of shining is nonmonotonicly dependent on temperature with maxima around $15, 30, 45$ and $65^0$ (Drost-Hansen, 1976). This temperature corresponds to extremes of stability of primary librational effectons, related to the number of $H_2O$ per effecton's edge ($\kappa$) (see Figure 7a). An increase in inorganic ion concentration, destabilizing *(a)* -state of these effectons, elevate the probability of super-deformons and consequently, shining intensity.

The most probable reason of photon radiation is recombination of water molecules in an exited state:

$$^-OH + H^+ \rightleftharpoons H_2O \to H_2O + h\nu_p \qquad 11.84$$

Very different chemical reactions can be stimulated in the volume of cavitational fluctuation by the external fields. The optimal resonant parameters of these fields could be calculated using the Hierarchic Theory.

*We propose here that the water molecules recombination (11.84), accompanied cavitational fluctuation, and excitation of molecular dipoles oscillation in proteins and DNA, could be responsible for coherent "biophotons" radiation by cell's and microbes cultures and living organisms in wide spectral range.* The biophotons research has been strongly advanced by Popp and his team (1992-2006).

In accordance to our model, the cell's body filaments - microtubules "catastrophe" (cooperative reversible disassembly) is a result of the internal water cavitational fluctuations due to super-deformons excitation. Such collective process should be accompanied by dissociation and recombination (11.84) of part of water molecules, localized in the hollow core of microtubules, leading to high-frequency electromagnetic radiation (Kaivarainen, 1994). The coherent photons super-radiated in infrared (IR) range, are the consequence of $(a \rightleftharpoons b)_{tr,lb}$ transitions of the water primary librational effectons in microtubules.

The brief description of the optoacoustic device - CAMP for quantitative evaluation of lot of parameters of the ice, water and other condensed matter, following from our approach, is presented below.

### 11.11. A New optoacoustic device: Comprehensive Analyzer of

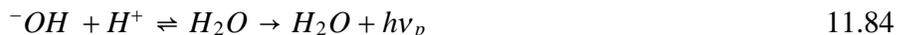

## Matter Properties (CAMP), based on the Hierarchic Theory

The Hierarchic Theory based computer program was copyrighted by this author in 1997, USA. It allows to calculate more than 300 output physical parameters of liquids and solids if only four input parameters are available. The theory and program were quantitatively verified on examples of water and ice in wide temperature range, as demonstrated in the previous part of book.

Among the calculated output parameters are: total internal energy, heat capacity, thermal conductivity, surface tension, vapor pressure, viscosity, self-diffusion and a lot of intermediate parameters, inaccessible for direct evaluation.

Most of intermediate parameters of calculations, using CAMP computer program (pCAMP) are hidden for direct experimental measurements, but carry an important information about physical properties of matter. This makes the CAMP system very valuable for investigations of complex perturbations in condensed matter. The computerized evaluation of hundreds of output parameters is possible in a second, if the following input experimental parameters are available:

1. Sound velocity;
2. Density;
3. Refraction index;
4. Positions of translational or librational bands in far and middle IR range: 30-1000) 1/cm.

These data should be obtained at the same temperature and pressure from the same sample (liquid or solid). The Hierarchic Theory and computer program pCAMP have been verified from comparisons of calculated and experimental (founded in literature) properties of water and ice, using listed above input parameters for water and ice in wide temperature range: 20 – 373 K, including liquid$\rightleftharpoons$solid phase transition (Kaivarainen, 2001; http://arxiv.org/abs/physics/0102086). The coincidence between theory and experiment is very good (see chapters 6-11 of this book).

Such possibilities of pCAMP computer program led this author to idea of new optoacoustic device: Comprehensive Analyzer of Matter Properties (CAMP), providing a simultaneous measurements of the above listed input parameters of the same sample at the same conditions (T, P). The interface of such device with computer, supplied with pCAMP program almost immediately provides us with huge amount of information of water and ice or any other pure condensed matter.

The table-top system for measurement of the velocity of sound, density and refraction index of the same liquid, almost at the same time is available already on the market. It includes the equipment, manufacturing by Anton-Paar Corporation (Graz, Austria): DSA 5000 (the velocity of sound and density) and RXA 156 or RXA 170 (refraction index). The FT-IR spectrometer can be used as a part of CAMP system for registration of translational and librational absorption bands positions in far and middle IR region.

This author investigated already the perturbation of water properties and their relaxation time (memory) after permanent magnetic field treatment, using CAMP system (Kaivarainen, 2004: http://arxiv.org/abs/physics/0207114).

Other possible configuration of CAMP system, proposed by this author, may include the Brillouin light scattering spectrometer, based on Fabry-Perrot interferometer. This configuration makes it possible a simultaneous measurement of the velocity of sound (from the Doppler shift of side bands of Brillouin spectra) and positions of intermolecular bands (translational and librational) from the Stokes/antiStokes satellite components on the central (Raman) peak of Brillouin spectra.

The CAMP system could be applied for monitoring of perturbation of very different physical properties of water, ice and other condensed matter (natural or artificial materials) under the influence of guest molecules, temperature, pressure or external electromagnetic or acoustic fields.

Our CAMP could be an ideal instrument for monitoring and improving of drinking water quality, using 'fingerprints' containing more than 300 physical parameters. The CAMP could be used in environmental research (effects of different kind of contamination, including EM pollution, using water as a test system) and in the control of water purification technology in the enterprises of each town.

The demonstrational (DEMO) version of pCAMP computer program for evaluation of pure water and ice properties in the range: 20-373K can be downloaded from the front page of this author site: web.petrsu.ru/~alexk [download pCAMP].

### 11.11.1. Possible projects with applications of CAMP system for aqueous systems and corresponding partners

- Project: "Monitoring of physical properties of dilute aqueous solutions of regular drugs and homeopathic drugs solutions. The correlation of the water structure perturbations with biological activity of these solutions". This research could be a scientific background of homeopathy and unexplored problems of water perturbation biological effects, induced by drug molecules. *Possible partners: Pharmaceutical companies.*

- Monitoring of electromagnetic pollution of environment, including mobile phones, using different water physical parameters. *Possible partners: WHO and organizations, studying the ecology problems.*

- Project: "Modulation of physical properties of drinking water by special filters and its monitoring by CAMP system. The correlation of water structure perturbation with lowering blood pressure, concentration of cholesterol, liver and kidney diseases healing, etc." *Possible partners: Companies, producing drinking water and water purification enterprises.*

- Project: "Influence of EM and acoustic fields on water clusters and hydration shell of colloid particles for regulation of their [coagulation - peptization] equilibrium, improving the quality of paper." *Possible partners: Companies, producing paper.*

- Project: "Monitoring of EM and acoustic fields influence on physical properties of water by CAMP. Correlation with vegetables growth - stimulating effect". *Possible partners: Companies, related to green house products (tomato, onions, etc.).*

- Project: "Correlation between physical parameters of beer and vine and their taste". *Possible partners: Alcohol beverages producing companies.*

- Project: "Influence of specifically treated water on the hardness and density of concrete and bricks". *Possible partners: Companies producing concrete.*

# Chapter 12

# Hierarchic background of the turbulence, superfluidity and superconductivity

The new approach to turbulence, superfluidity and superconductivity, based on the Hierarchic Model, is proposed (Kaivarainen, 1995; 2001; 2007). The modified theories are in good agreement with experimental data. The common features between superfluidity and superconductivity are demonstrated. The conditions, when mesoscopic Bose condensation (mBC), revealed in our work, turns to macroscopic one, are discussed.

## 12.1. Turbulence. General description

The type of flow when particles move along the straight trajectory without mixing with adjoining layers is termed the *laminar* flow.

If the layers of the liquid of the laminar flow are moving relative to each other at different velocities, then the forces of internal friction ($F_{fr}$) or viscosity forces originate between them:

$$F_{fr} = \eta \left| \frac{\Delta \mathbf{v}}{\Delta d} \right| S, \qquad 12.1$$

where $\Delta \mathbf{v}$ is relative liquid layer velocity; $S$ is the contact surface; $\eta$ is dynamic viscosity; $\left| \frac{\Delta \mathbf{v}}{\Delta d} \right|$ is the module of velocity gradient directed to the surface of layers.

Near the walls of a straight tube the velocity of laminar flow is equal to zero and in the center of the tube it is maximum.

The relation between the layer velocity and its distance from central axes of the tube (r) is parabolic:

$$\mathbf{v}(r) = \mathbf{v}_0 \left( 1 - \frac{r^2}{a_t^2} \right) \qquad 12.2$$

where $a_t$ is the tube radius; $\mathbf{v}_0$ is the velocity of the liquid on the central axis, depending on the difference of pressure at the ends of the tube:

$$\Delta P = P_1 - P_2 \qquad 12.3$$

as follows:

$$\mathbf{v}_0 = \frac{P_1 - P_2}{4\eta l} a_t^2 \qquad 12.4$$

where $l$ is tube length and $\eta$ is *dynamic viscosity*.

The *flux of liquid*, *i.e. the volume* of liquid flowing over the cross-section of the tube during a time unit is determined by the Poiseuille formula:

$$Q = \frac{(P_1 - P_2)\pi a_t^2}{8\eta l} \qquad 12.5$$

This formula has been used frequently for estimation of dynamic viscosity $\eta$.

The corresponding mass of flowing liquid is:

$$m = \rho Q \qquad 12.6$$

and corresponding kinetic energy:

$$T_k = \frac{\rho}{4} Q \mathbf{v}_0^2 \qquad 12.7$$

where $\rho$ is the density of liquid.

The work of internal friction force is:

$$A = -4\eta \mathbf{v}_0 lQ/\rho R^2 \qquad 12.7a$$

In the case of the laminar movement of a spherical body relative to liquid the force of internal friction (viscosity force) is determined by the *Stokes law*:

$$F_{fr} = 6\pi a \mathbf{v} \eta, \qquad 12.8$$

where *(a)* is the radius of sphere and (**v**) is its relative velocity.

As a result of liquid velocity (**v**) and/or the characteristic dimension *(a)* increasing, the *laminar* type of liquid flow could change to the *turbulent* one.

This begins at certain values of the dimensionless Reynolds number:

$$R = \rho \mathbf{v}_c a/\eta = \mathbf{v}_c a/\nu, \qquad 12.9$$

where $\rho$ is liquid density; $\mathbf{v}_c$ is characteristic (average) flow velocity; $\nu = \eta/\rho$ is kinematic viscosity of liquid.

For the round tube with radius *(a)* the critical value of R is about 1000. A turbulent type of flow is accompanied by rapid irregular pulsations of liquid velocity and pressure, representing a kind of self-organization.

In the case of noncontinuous movement, the flow can be characterized by 2 additional dimensionless parameters.

1. *Strouhal number:*

$$S = \mathbf{v}_c \tau /a \qquad 12.10$$

where $\tau$ is the characteristic time of velocity ($\mathbf{v}_c$) pulsations;

2. *March number*:

$$M = \mathbf{v}_c/\mathbf{v}_s \qquad 12.11$$

where $\mathbf{v}_s$ is the velocity of sound in liquid.

### 12.2 The Hierarchic mechanism of turbulence

The physical scenario of transition from a laminar flow to a turbulent one is still unclear. It is possible to propose the mechanism of this transition based on hierarchic concept of matter.

Let us start from the assumption that in the case of laminar flow, the thickness of parallel layers is determined by the thickness of primary librational electromagnetic deformons ($\sim 10^5 \text{Å}$), equal to linear dimensions of corresponding type of macro-effectons and macro-deformons. This provides certain polarization and ordering of water molecules in layer.

The total internal energy and the internal pressure of neighboring layers are not equal. The surface between such layers can be characterized by corresponding surface tension $[\sigma(r)]$. Surface tension prevents mixing between layers with different laminar flow velocities.

According to our theory of surface tension (Section 11.4), the thickness of the two outer borders of each layer (skin-surface) is determined by the effective linear dimensions of primary (tr and lib) effectons $[l_{tr,lb} \sim (3-15)\text{Å}]$ related to the corresponding most probable de Broglie wave length ($\lambda_{tr,lb}$) of molecules in liquid (see eq. 11.30):

$$l_{tr,lb} = (V_{ef}/V_{ef}^{2/3}) \qquad 12.12$$

Decreasing of $l_{tr,lb}$, depending on most probable momentum of liquid molecules, as a result of increased flow velocity $[\mathbf{v}_1(r)]$ see eq. (12.13) and/or temperature elevation in accordance with our theory of surface tension (see eq. 11.33), leads to reducing of $\sigma(r)$. In turn, this effect strongly decreases the work of cavitational fluctuations, *i.e.* the bubbles formation (12.27) and increases their concentration (12.29). *These cavitational fluctuations induce mixing of laminar layers, the instability of laminar flow and its changing to the turbulent type of flow.*

The critical flow velocity: $\mathbf{v}_c = \mathbf{v}^1(r)$ is determined by the critical librational de Broglie wave

length:

$$\lambda_{lb}^{1,2,3} = (h/m)\left[\left(\mathbf{v}_{gr}^{1,2,3}\right)_{lb} + \mathbf{v}^1(r)\right] \qquad 12.13$$

corresponding to the condition (6.6) of liquid-gas first-order phase transition:

$$\left[(V_{ef})_{lb}/(V_0/N_0)\right] = \left[\frac{9}{4\pi}(\lambda^{(1)}\lambda^{(2)}\lambda^{(3)})_{lb}/(V_0/N_0)\right] \leq 1 \qquad 12.14$$

where $m$ is molecular mass; $(\mathbf{v}_{gr}^{1,2,3})_{lb}$ is the most probable librational group velocity of molecules of liquid in selected directions $(1, 2, 3)$; $\mathbf{v}^1(r)$ is the flow velocity of a liquid layer in the tube at the distance $(r)$ from the central axes of the tube (12.2).

Increasing of $\mathbf{v}^1(r)$ at $r \to 0$ decreases $\lambda_{lb}^1$ and $(V_{ef})_{lb}$ in accordance with (12.13) and (12.14). The value of $\lambda_{lb}^1$ is also related to phase velocity $(\mathbf{v}_{ph}^a)$ and frequency $(v_1^a)$ of the primary librational effecton in (a) state (see 2.60):

$$\lambda_{lib}^1 = h/m\left[(\mathbf{v}_{gr}^1)_{lb} + \mathbf{v}^1(r)\right] = \left(\frac{\mathbf{v}_{ph}^a}{v_1^a}\right)_{lb} =$$

$$= (\mathbf{v}_{ph}^a/v_p^1)\left[\exp\left(h(v_p^1)_{lb}/kT\right) - 1\right] \qquad 12.15$$

where

$$(v_p^1)_{lb} = (v_1^b - v_1^a)_{lb} = c(\tilde{v}_p^1)_{lb} \qquad 12.16$$

is the frequency of $(a \Leftrightarrow b)_{lb}$ transitions of the primary librational effecton of *flowing* liquid determined by librational band wave number $(\tilde{v}_p^1)_{lb}$ in the oscillatory spectra.

It was calculated earlier for water under stationary conditions that the elevation of temperature from $0^0$ to $100^0 C$ till the phase transition condition (12.14) is accompanied by the increase in $(\mathbf{v}_{gr})_{lb}$ from $1.1 \times 10^3 cm/s$ to $4.6 \times 10^3 cm/s$ (Figure 12b).

This means that at $30^0 C$, when $(\mathbf{v}_{gr})_{lb} \simeq 2 \times 10^3 cm/s$, the critical flow velocity $\mathbf{v}^1(r)$, necessary for *mechanical boiling* of water (conditions 12.13 and 12.14) should be about $2.6 \times 10^3 cm/s = 26\ m/s$.

The reduced number of primary librational effectons $(N_{ef})$ in the volume $(V_D^M)$ of primary electromagnetic deformons (**tr** and **lb**) also increases with temperature and/or flow velocity:

$$\left(N_{ef}\right)_{tr,lb}^D = \left[\frac{P_a + P_b}{Z} n_{ef} V_D^M\right]_{tr,lb} \qquad 12.17$$

The reduced number of primary transitons $(N_t)$ has a similar dependence on $T$ and $\mathbf{v}^1(r)$, due to increasing of $n_{ef}$, and $V_D^M$ as far:

$$\left(N_t\right)_{tr,lb}^D = \left(N_{ef}\right)_{tr,lb}^D \qquad 12.18$$

The analysis of (eq. 12.15) *predicts* that at T = *const* an increase in $\mathbf{v}^1(r)$ must be accompanied by the low-frequency shift of the librational band: $\tilde{v}_{lib}^{(1)} \simeq 700 cm^{-1}$ and/or by the decrease in the velocity of sound $(\mathbf{v}_{ph}^a)_{lb}$ (eq. 2.74) in the direction (1) of flow.

It follows also from the Hierarchic Theory that these changes should be accompanied by a rise in dynamic viscosity (11.45) due to increased structural factor $(T_{kin}/U_{tot})$.

Turbulent pulsations of flow velocity $(\Delta \mathbf{v})$ originate under developed turbulence conditions:

$$\Delta \mathbf{v} = \mathbf{v}_{tur} = \mathbf{v} - \bar{\mathbf{v}} \qquad 12.19$$

where $\bar{\mathbf{v}}$ is average flow velocity and $(v)$ is instant flow velocity.

The frequencies of large-scale pulsations have the order of:

$$v = \bar{\mathbf{v}}/\lambda, \qquad 12.20$$

where $\lambda$ is the main scale of pulsations.

The $\lambda$ can correlate with the dimensions of librational electromagnetic deformons and can be determined by the transverse convection rate, depending on the bubbles dimensions.

The pulsations of flow velocity ($\Delta \mathbf{v}$) can result from the following factors:

a) mixing of parallel layers with different flow velocity and

b) fluctuation of viscosity force (eq.12.8) due to fluctuations in bubbles radius and concentration, as well as density, viscosity and thermal conductivity;

c) movements and emergence of the bubbles as a result of the Archimedes force.

The bubbles have two opposite types of influence on *instant* velocity. The layer mixing effect induced by them can increase flow velocity. On the other hand, the bubbles can simultaneously decrease flow velocity due to enhanced internal friction.

In the case of developed turbulence with different scales of pulsations it is reasonable to introduce the characteristic *Reynolds* number:

$$R_\lambda = \mathbf{v}_{tur}\lambda/\nu_{tur} \qquad 12.21$$

where $\lambda$ is a scale of pulsations; $\mathbf{v}_{tur}$ is the velocity of pulsation and $\nu_{tur} = (\eta/\rho)_{tur}$ characteristic kinematic viscosity.

The ratio between turbulent kinematic viscosity ($\nu_{tur}$) and a laminar one ($\nu$) is related to the corresponding Reynolds numbers like (Landau, Lifshits, 1988):

$$\frac{\nu_{tur}}{\nu} \sim \frac{R}{R_{tur}} \qquad 12.22$$

One can see from (12.21) that it means:

$$\mathbf{v}_{tur}\lambda \sim const \qquad 12.23$$

Based on dimensional analysis, the turbulent kinematic viscosity can be expressed as follows:

$$\nu_{tur} \sim \Delta \mathbf{v}\, l \sim \mathbf{v}_\lambda\, \lambda \qquad 12.24$$

and the dissipation energy as:

$$\epsilon_{dis} \sim \nu_{tur}(\mathbf{v}_{tur}/\lambda)^2 \sim \frac{\mathbf{v}_{tur}^3}{\lambda} \qquad 12.25$$

This expression leads to the Kholmogorov-Obuchov law:

$$\mathbf{v}_{tur} \sim (\epsilon_{dis}\lambda)^{1/3} \qquad 12.26$$

Large-scale pulsations correspond to high $\lambda$ values and low $\nu_{tur}$ values. *i.e.* to high characteristic turbulence Reynolds numbers (see 12.21).

According to proposed model, the maximum energy dissipation occurs in the volume of super-deformons (or super-transitons). The mechanically induced boiling under conditions of turbulence (eqs. 12.13 and 12.14) is accompanied by the emergence of gas bubbles related to the increased super-deformons probability and decreased surface tension between layers.

The critical bubble origination work ($W$) is strongly dependent on surface tension between layers ($\sigma$). A general classical theory (Nesis, 1973) gives:

$$W = \frac{4}{3}\pi a^2 \sigma = \frac{16\pi\sigma^3}{3(P - P_{ext})^2} \qquad 12.27$$

where

$$P = P_{ext} + \frac{2\sigma}{a} = P_{a=\infty}\exp\left(-\frac{2\sigma V_e}{akT}\right) \qquad 12.28$$

$P$ is the internal gas pressure in a bubble with radius ($a$); $V_e$ is the volume of liquid occupied by one

molecule.

The bubbles quantity ($N_b$) has an exponential dependence on $W$:

$$N_b = \exp\left(-\frac{W}{kT}\right) \qquad 12.29$$

One can see from the Hierarchic Theory of surface tension (eqs. 11.31 - 11.33 and 12.12) that under *mechanical boiling* conditions the skin-surface thickness (12.12): $l \to (V_0/N_0)^{1/3}$ and $q^s \to 1$, the surface tension between layers ($\sigma$) tends to zero, $W$ decreases and $N_b$ increases.

We can see, that the hierarchic scenario of mechanical boiling presented here is base for quantitative physical theory of turbulence and other hydrodynamic instabilities like Taylor's and Benar's ones.

### 12.3. Superfluidity: a general description

The macroscopic superfluidity has been revealed for two helium isotopes: $^4$He with boson's properties ($S = 0$) and $^3$He with fermion properties ($S = 1/2$). The interactions between the atoms of these liquids is very weak. It will be shown below, that the values of normal the velocity of sound at temperatures higher than those of second-order phase transition ($\lambda$-point) are lower than the most probable thermal velocities of the atoms of these liquids.

The first theories of superfluidity were proposed by Landau (1941) and Feynman (1953).

first-order phase transition [gas $\to$ liquid] occurs at $4.22K$. second-order phase transition, when superfluidity originates, $^4$He $\to$ He II takes place at $T_\lambda = 2.17K$ ($P_{\text{ext}} = 1$ atm.). This transition is accompanied by:

a) heat capacity jump to higher values;
b) abruptly increased thermal conductivity;
c) markedly decreased cavitational fluctuations and bubbles in liquid helium.

For explanation of experimental data Landau supposed that at $T < T_\lambda$ the He II consists of two components:

- the *superfluidity component* with relative fraction of density $\rho_S/\rho$, increasing from zero at $T = T_\lambda$ to 1 at T = 0 K. The properties of this component are close to those of an ideal liquid with a potential type of flow. The entropy of this component is zero and it does not manifest the viscous friction on flowing through narrow capillaries;

- the *normal component* with density

$$\rho_n = \rho - \rho_s \qquad 12.30$$

decreasing from 1 at $T = T_\lambda$ to zero at T = 0 K. This component behaves as a usual viscous liquid which exhibits dumping of the oscillating disk in He II. Landau considered this component to be a gas of two types of excitations: *phonons* and *rotons*.

The hydrodynamics of normal and superfluid components of He II are characterized by *two velocities*: normal ($\mathbf{v}_n$) and superfluid one:

$$\mathbf{v}_{sf} = (\hbar/m)\nabla\varphi \qquad 12.31$$

where $\nabla\varphi \sim k_{sf} = 1/L_{sf}$ is a phase of Bose-condensate wave function - see eq. 12.36.

As a result of two types of hydrodynamic velocities and densities, the corresponding 2 types of sound waves propagate in the volume of He II.

The *first sound* ($U_1$) is determined by the usual formula valid for normal condensed matter:

$$U_1^2 = (\partial P/\partial \rho)_S \qquad 12.32$$

In this case density oscillations spread in the form of phonons.

The *second sound* ($U_2$) is related to oscillations of temperature and entropy (S):

$$U_2^2 = \rho_S TS^2/c\rho_n \qquad 12.33$$

In normal condensed media the temperature oscillation fade at the distance of the order of wave length.

Landau considered the second sound as density waves in the gas of quasi-particles: rotons and phonons.

*The third sound* ($U_3$) propagates in thin surface films of He II in the form of "ripplons", *i.e.* quantum capillary waves related to the isothermal oscillations of the superfluid component.

$$U_3 = (\rho_S/\rho_S)\, d\frac{\partial E}{\partial d}(1 + TS/L), \qquad 12.34$$

where $(\bar{\rho}_S/\rho_S)$ is the relative density of superfluid component averaged in the thickness of the film ($d$); E is the potential of Van- der-Waals interactions of $^4$He atoms with the bottom surface; $L$ is evaporation heat.

The *fourth sound* ($U_4$) propagates in He II, located in very narrow capillaries, when the length of quasi-particles (phonons and rotons) free run is compatible or bigger than the diameter of these capillaries or pores.

The hydrodynamic velocity ($\mathbf{v}_n$) of the normal component under such conditions is zero and $\rho_n/\rho \ll \rho_{sf}/\rho$:

$$U_4^2 = (\rho_S/\rho)U_1^2 + (\rho_n/\rho)U_2^2 \simeq (\rho_S/\rho)U_1^2 \qquad 12.35$$

In accordance with Bose-Einstein statistics, a decrease in temperature, when $T \to T_\lambda$, leads to condensation of bosons in a minimum energy state.

This process results in the origination of a superfluid component of He II with the coherent thermal and hydrodynamic movement of atoms.

Coherence means that this movement of atoms can be described by the single wave function:

$$\psi = \rho_S^{1/2}\, e^{i\varphi} \qquad 12.36$$

The movement of the superfluid component is *potential* as far its velocity ($\vec{\mathbf{v}}_{sf}$) is determined by eq.12.31 and:

$$\operatorname{rot} \mathbf{v}_{sf} = 0 \qquad 12.37$$

### 12.3.1. Vortex filaments in He II

When the rotation velocity of a cylindrical vessel containing He II is high enough, then the emergency of so-called vortex filaments becomes thermodynamically favorable. The filaments are formed by the superfluid component of He II in such a way, that their kinetic energy of circulation decreases the total energy of He II in a rotating vessel.

The shape of filaments in this case is like a straight rodes with *thickness* of the order of atom's dimensions, increasing with lowering the temperature at $T < T_\lambda$.

Vortex filaments are continuous. They are closed or limited within the boundaries of a liquid. For each surface, surrounding a vortex filament the condition (12.37) is valid.

The values of velocity of circulation around the axis of filaments are determined (Landau, 1941) as follows:

$$\oint \mathbf{v}_{sf} dl = 2\pi r \mathbf{v}_{sf} = 2\pi\kappa \qquad 12.38$$

and

$$\mathbf{v}_{sf} = \kappa/r \qquad 12.39$$

Increasing the radius of circulation ($r$) leads to decreased circulation velocity ($\mathbf{v}_{sf}$). Substituting $\mathbf{v}_{sf}$ in eq.12.31, we obtain:

$$\oint \mathbf{v}_{sf} dl = \frac{\hbar}{m}\Delta\Phi, \qquad 12.40$$

where $\Delta\Phi = n2\pi$ is a phase change as a result of circulation, $n = 1, 2, 3\ldots$ is the integer number.

Comparing (12.40) and (12.38) gives:

$$\kappa = n\frac{\hbar}{m} \qquad 12.41$$

It has been shown, that only curls with $n = 1$ are thermodynamically stable.

Taking this into account, we have from (12.39) and (12.41):

$$r = \frac{\hbar}{m\mathbf{v}_{sf}} = L \qquad 12.42$$

*This means that the condition of stability of superfluid rotating filaments corresponds to condition of de Broglie standing wave of atoms of He II.*

The increasing of the angle frequency of rotation of the cylinder containing He II results in the increased density of superfluid vortex filaments distribution in the cross-section of the cylinder.

As a result of interaction between these filaments and the normal component of He II, the filaments move in the rotating cylinder with normal liquid.

The flow of He II through the capillaries also can be accompanied by emergence of vortex filaments.

*In ring-shaped vessels the circulation of closed vortex filaments is stable, like electric current in closed superconductors.* This stability is related to the condition of He II de Broglie standing waves (12.42), like stability of the electrons on their stable orbits in atoms.

Let us consider now the phenomena of superfluidity in He II in the framework of the Hierarchic Theory.

### 12.4 Hierarchic scenario of superfluidity

It will be shown below how the Hierarchic Model (Table 1) can be used to explain He II properties, its excitation spectrum (Figure 40), increased heat capacity at $\lambda$-point and the vortex filaments formation.

We assume here, that the formulae obtained earlier for internal energy ($U_{tot}$ –eq. 4.3), viscosity (eqs. 11.48, 11.49 and 11.55), thermal conductivity (eq. 11.37), vapor pressure (eq. 11.26) remain valid for both components of He II.

The theory proposed by Landau (Lifshits, Pitaevsky, 1978) qualitatively explains only the lower branch *(a)* in the spectrum (Figure 40) as a result of phonons and *rotons* excitation.

But the upper branch *(b)* points that the real process is more complicated and needs introduction of other quasi-particles and excited states for its explanation.

The Hierarchic Model of superfluidity interrelates the *lower branch* with the ground acoustic *(a)* state of *primary effectons* in liquid $^4$He and the upper branch with their excited optical *(b)* state. In accordance with our model, the dissipation and viscosity friction (see section 11.6) arise in the *normal component of He II* due to thermal phonons radiated and absorbed in the course of the $\bar{b} \to \bar{a}$ and $\bar{a} \to \bar{b}$ transitions of secondary effectons correspondingly, in-phase to macro-deformons excitation.

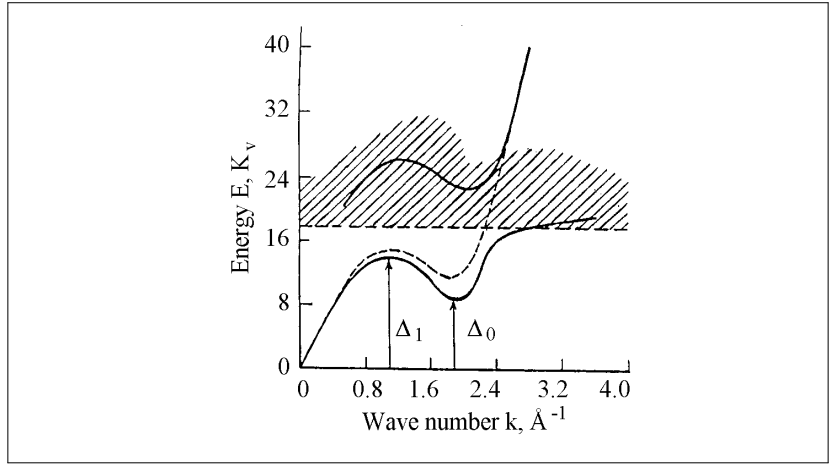

**Figure 40**. Excitation spectrum of liquid $^4He$ from neutron scattering measurements (March and Parrinello, 1982). Spectrum is characterized by two branches, corresponding to *(a)* and *(b)* states of the primary librational/rotational effectons (rotons) in accordance to the Hierarchic Model.

Landau described the minimum in the region of $\lambda$-point using the expression:

$$E = \Delta_0 + \frac{(P - P_0)^2}{2m^*}, \qquad 12.43$$

where $\Delta_0$ and $P_0$ are the energy and momentum of liquid $^4He$ at $\lambda$-point (Figure 40) and $m^* = 0.16m$ is the effective mass of the $^4He$ atom ($m_{He} = 4 \times 1.44 \cdot 10^{-24} g = 5.76 \cdot 10^{-24} g$). The effective mass $m^*$ can be determined experimentally.

Feynman (1953) explained the same part of the excitation spectra by the nonmonotonic behavior of the structure factor $S(k)$ and the formula:

$$E = \hbar\omega = \frac{\hbar^2 k^2}{2mS} = \frac{\hbar^2}{2mL^2 S} \qquad 12.44$$

where

$$k = 1/L = 2\pi/\lambda \qquad 12.45$$

is the wave number of neutron interacting with liquid $^4He$.

*The Hierarchic Model allows to unify the Landau's and Feynman's approaches.*

The total energy of de Broglie wave either free or as a part of condensed matter can be expressed through its amplitude squared ($A^2$), or effective mass ($m^*$) in the following manner (see 2.45 and 2.46):

$$E_{tot} = T_k + V = m\mathbf{v}_{gr}\mathbf{v}_{ph} = \frac{\hbar^2}{2mA^2} = \frac{\hbar^2}{2m^*L^2} \qquad 12.46$$

where $\mathbf{v}_{gr}$ and $\mathbf{v}_{ph}$ are the most probable group and phase velocities.

In accordance with our model (eq. 2.46a), the structural factor $S(k)$ is equal to the ratio of kinetic ($T_k$) energy to the total one ($E_{tot}$):

$$S = T_k/E_{tot} = A^2/L^2 = m^*/m \qquad 12.47$$

where

$$T_k = P^2/2m = \frac{\hbar}{2mL} \qquad 12.48$$

Combining (12.46), (12.47) and (12.48), we obtain the following set of equations for the energy of $^4He$ at transition $\lambda$-point:

$$\left.\begin{array}{c} \Delta_0 = E_0 = \dfrac{\hbar^2}{2mA_0^2} = \dfrac{\hbar^2}{2m^*L_0^2} \\ \Delta_0 = \dfrac{\hbar^2}{2mL_0^2 S} = \dfrac{T_k^0}{S} \end{array}\right\} \quad 12.49$$

These approximate formulae for the total energy of liquid $^4He$ made it possible to estimate the most probable de Broglie wave length, forming the primary librational/rotational effectons at $\lambda$-point:

$$\lambda_0 = \frac{h}{m\mathbf{v}_{gr}^0} = 2\pi L_0 = 2\pi A_0 (m/m^*)^{1/2} \quad 12.50$$

where the critical amplitude of de Broglie wave:

$$A_0 = \hbar \left(\frac{1}{2mE_0}\right)^{1/2} \quad 12.51$$

can be calculated from the experimental $E_0$ values (Figure 40). Putting into (12.51) and (12.50) the available data:

$$\Delta_0 = E_0 = k_B \times 8.7K = 1.2 \times 10^{-15} \, erg$$

the mass of atom: of ($^4He$) $m = 5.76 \times 10^{-24} g$ and ratio $(m^*/m) = 0.16$, we obtain:

$$\lambda_0 \cong 14 \times 10^{-8} cm = 14 \, \text{Å} \quad 12.52$$

the corresponding most probable group velocity of $^4He$ atoms is: $\mathbf{v}_{gr}^0 = 8.16 \times 10^3 cm/s$.

It is known from the experiment that the volume occupied by *one atom of liquid* $^4He$ is equal: $\mathbf{v}_{(^4He)} = 46 \, \text{Å}^3$/atom. The edge length of the corresponding cubic volume is:

$$l = (\mathbf{v}_{^4He})^{1/3} = 3.58 \, \text{Å} \quad 12.53$$

From (12.52) and (12.53) we can calculate the number of $^4He$ atoms in the volume of primary librational (rotational) effecton at $\lambda$-point:

$$n_V^0 = \frac{V_{ef}}{\mathbf{v}_{^4He}} = \frac{(9/4\pi)\lambda_0^3}{l^3} = 43 \text{ atoms} \quad 12.54$$

One edge of such an effecton contains $(43)^{1/3} \cong 3.5$ atoms of liquid $^4He$.

We must take into account, that these parameters can be *lower than the real* ones as in above simple calculations we did not consider the contribution of secondary effectons, transitons and deformons to total internal energy (see eq. 4.3).

On the other hand, in accordance with our model, the conditions of the maximum stability of primary effectons correspond to the *integer* number of particles in the edge of these effectons (see chapter 6 and Figure 6a).

Consequently, we have to assume that the true number of $^4He$ atoms forming a primary effecton at $\lambda$-point is equal to $n_V^0 = 64$. It means that the edge of cube as the effecton shape approximation contains $q^0 = 4$ atoms of $^4He$:

$$n_e^0 = (n_V^0)^{1/3} = 64^{1/3} = 4 \quad 12.55$$

The primary librational/rotational effectons of such a type may correspond to rotons introduced by Landau to explain the high heat capacity of He II.

The thermal momentums of $^4He$ atoms in these coherent clusters can totally compensate each other and the resulting momentum of primary effectons is equal to zero. Further decline in temperature gives rise to dimensions of primary effectons. The most stable of them contain in their edges the integer number of helium atoms:

$$q = q^0 + k \quad 12.56$$

where $k = 0, 1, 2, 3...$

$\lambda_0$, $n_V^0$ and $q^0$ can be calculated more accurately using eqs. (2.60) and (3.5), if the required experimental data on oscillatory spectroscopy and sound velocimetry are available.

### 12.5. Superfluidity as a hierarchic self-organization process

Let us analyze the phenomena observed in $^4$He in the course of temperature decline to explain Figure 40 in the framework of Hierarchic Model:

*1. In accordance to our model, the gas-liquid first-order phase transition in $^4$He occurs under condition (6.6).* This condition means that the most probable de Broglie wave length of atoms related to their librations/rotations starts to exceed the average distance between $^4He$ atoms in a liquid phase:

$$\lambda = h/m\mathbf{v}_{gr} \geq \left(\frac{V_0}{N_0}\right)^{1/3} \geq 3.58 \text{Å} \quad\quad 12.57$$

The corresponding value of the most probable group velocity is

$$\mathbf{v}_{gr} \leq 3.2 \times 10^4 cm/s.$$

The *translational* thermal momentums of particles are usually bigger than those related to *librations*. In accordance with our model of first-order phase transitions (Section 6.4), this fact determines the difference in the temperatures of [gas → liquid] and [liquid → solid] transitions.

The *freezing* of liquid $^4He$ occurs at a sufficiently high pressure of ∼ 25 atm only. This means the emergency of *primary translational* effectons at these conditions. The pressure increase as well as the drop in temperature declines the momentums of particles and stimulates the distant van der Waals interaction between them, responsible for coherent clusters formation, representing the translational mesoscopic Bose condensate.

In normal component of liquid $^4$He II, like in regular liquid, the existence of primary and secondary effectons, convertons, transitons and deformons is possible. The contributions of each of these quasi-particles determine the total internal energy (eq. 4.3), kinetic and potential energies (eqs. 4.33 and 4.36), viscosity (11.45), thermal conductivity (11.35), vapor pressure (11.26) and many other parameters of normal component of $^4$He II.

We assume that the lower branch in the excitation spectrum of Figure 40 reflects the *(a)* state and the upper branch the *(b)* state of primary (lb and tr) effectons.

*2. Decreasing the temperature to $\lambda$-point: $T_\lambda = 2.17K$ and below is accompanied by the condition (12.55), which means Bose-condensation of atoms of $^4$He, the dimensions of primary librational or rotational effectons increasing and starting of Bose-condensation of secondary effectons.*

This leads to emergency of primary librational/rotational polyeffectons, representing the vortex filaments of superfluid subsystem, stabilized by distant van der Waals interactions and Josephson junctions between neighboring effectons at the de Broglie wave standing wave condition of vortex-filaments (12.42).

The polymerization of librational effectons is accompanied by the $(a)$ - states probability jump-way increasing ($P_a \to 1$, eq. 4.10) and that of $(b)$ -states decreasing ($P_b \to 0$, eq. 4.11). Probability of primary and secondary deformons ($P_d = P_a P_b$;  $\bar{P}_d = \bar{P}_a \bar{P}_b$) decreases correspondingly. In the excitation spectrum (Figure 40) these processes are displayed as the shift of $(b)$ -branch toward the $(a)$ - branch due to degeneration of $(b)$ -branch at very low temperature. In the IR spectra this process displays itself as a 'soft' mode.

Like in theory of 2nd order phase transitions proposed by Landau (see Landau and Lifshits, 1976), we can introduce the *parameter of order* as:

$$\eta = 1 - \kappa = 1 - \frac{P_a - P_b}{P_a + P_b} \quad\quad 12.58$$

where  $\kappa = \frac{P_a - P_b}{P_a + P_b}$  is the *equilibrium parameter*.

One can see that at $P_a = P_b$, the equilibrium parameter $\kappa = 0$ and $\eta = 1$ (the system is far from 2nd order phase transition). On the other hand, at conditions of 2nd order phase transition: $T \to T_\lambda$ when $P_b \to 0$, $\kappa \to 1$ and parameter of order ($\eta$) tends to zero.

According to Landau's theory, the equality of his specific parameter of order to zero is also a criterion of 2nd order phase transition. As usual, this transition is followed by a decrease in structural symmetry with a decline in temperature.

The important point of our scenario of superfluidity is the statement that the leftward shift of ($a \Leftrightarrow b$) equilibrium of the primary effectons (tr and lb) becomes stable starting from $T_\lambda$ due to their "head to tail" polymerization and the *vortex filaments formation*. This process of Bose-condensation, including conversion of secondary effectons to primary ones, differs from condensation of an *ideal* Bose-gas described by eq. (1.26). Such kind of Bose-condensation means the enhancement of the concentration of *(a)*-state of primary effectons with lower energy, related to degeneration of the all other states. The polymerization of primary effectons in He II gives rise to macroscopic filament-like (or chain-like) *polyeffectons*. This process can be considered as self-organization in helium on macroscopic scale. These vortical filament- polyeffectons, representing superfluid component of liquid helium, can form a closed circles or three-dimensional (3D) isotropic networks in a vessel with *HeII*. *The remnant fraction of liquid represent normal fraction of He II.*

Polyeffectons are characterized by the dynamic equilibrium: $\left[ assembly \Leftrightarrow disassembly \right]$. Temperature decreasing and pressure increasing shift this equilibrium to the left, increasing the dimensions of primary effectons, their side-by-side interaction and number of Josephson junctions.

The probability of tunneling of *HeII* atoms between coherent clusters increases also, correspondingly.

The relative movement (sliding) of flexible "snake-like" polyeffectons occurs without phonons excitation in the volumes of macro-deformons in superfluid liquid helium. Just macro-deformons excitation is responsible for dissipation and viscosity in normal liquids (see section 11.6). The absence of macro-deformons excitation in the volume, occupied by vortical filaments - librational polyeffectons explains the superfluidity phenomenon, according to our model.

Breaking of symmetry in a three-dimensional polyeffectons network can be induced by the external fields, gravitational gradient, mechanical perturbation and surface effects. It is possible as far the coherent polyeffectons system is highly cooperative.

In rotating cylindrical vessel the filament-like polyeffectons originate from 3D isotropic net and they tend to be oriented along the cylinder axis. The rotation of these vortical filaments around own main axis occur in the direction opposite to that of cylinder rotation, minimizing the kinetic energy of liquid helium. The radius of the filaments (12.42) is determined by the tangential group velocity of the coherent $^4$He atoms, which form part of the primary effectons ($\mathbf{v}_{gr} = \mathbf{v}_{sf}$). The numerical value of $\mathbf{v}_{gr}$ must be equal to or less than $6 \times 10^3 cm/s$, this corresponding to conditions (12.55 and 12.56). At $T \to 0$, $\mathbf{v}_{gr}$ decreases and the filament radius (12.42) increases to reach the values corresponding to $\mathbf{v}_{gr}^{min} = \mathbf{v}^0$ determined by the zero-point oscillations of $^4$He atoms. Under these conditions the polymerization of translational primary effectons in *(a)*-state can occur, leading to liquid-solid phase transition in $^4$He.

The self-organization of highly cooperative coherent polyeffectons in the $\lambda$ – point and ($a \rightleftharpoons b$) equilibrium leftward shift should be accompanied by heat capacity jump.

*The successive mechanisms of polymerization of primary effectons into vortical filaments could be responsible for second-order phase transitions, leading to emergency of superfluidity and superconductivity.*

*The second sound* in such a model can be attributed to phase velocity in a system of polyeffectons. The propagation of the second sound through fraction of filaments - polyeffectons should be accompanied by their elastic deformation and [assembly ⇔ disassembly] equilibrium oscillations.

*The third sound* can be also related to the elastic deformation of polyeffectons and equilibrium constant oscillations, but in the *surface layer* with properties different from those in bulk volume. Such a difference between the surface and volume properties is responsible (see eqs.11.31-11.33) for surface tension ($\sigma$) in He II and its jump at $\lambda$-point. The increasing of surface tension explains the disappearance of cavitational bubbles at $T < T_\lambda$ in superfluids.

*The fourth sound* can be a consequence of the primary effectons and polyeffectons dimensions increasing and change in their phase velocity, as a result of interaction of liquid helium with narrow capillary's walls, following by immobilization atoms of He II.

The *normal* component of He II is related to the fraction of He II atoms not involved in polyeffectons - vortical filaments formation. This fraction involves primary and secondary effectons, keeping the ability for $(a \Leftrightarrow b)$ and $(\bar{a} \Leftrightarrow \bar{b})$ transitions in contrast to superfluid fraction of liquid helium. In accordance with the Hierarchic Model, these transitions composing macro-effectons and macro-deformons are accompanied by the emission and absorption of thermal phonons and dissipation.

The manifestation of viscous properties in normal liquid and normal component of He II is related to fluctuations of concentration in the volume of macro-deformons (11.45-11.48).

On the other hand, the macro-deformons and convertons are absent in the superfluid component of He, as far in primary polyeffectons at $T < T_\lambda$: the probability of B-state of macro-effectons tends to zero: $P_B = P_b \bar{P}_b \to 0$; the probability of A-state of the effectons: $P_A = P_a \bar{P}_a \to 1$ and, consequently, the probability of macro-deformons also tends to zero: $P_D^M = P_B P_A \to 0$.

The decreasing of probability of super-deformons $P_D^S = (P_D^M)_{tr}(P_D^M)_{lb} \to 0$ means the decreased concentration of cavitational bubbles and vapor pressure in superfluid state.

We can explain the minimum in $E(k)$ at Figure 40 around $T_\lambda$ by reducing the contributions related to *(b)* state of primary effectons, degeneration of secondary effectons due to their Bose-condensation and concomitant elimination of the contribution of secondary acoustic deformons (*i.e.* phonons) to the total energy of liquid $^4He$.

One can see from eqs. (11.45 - 11.53) that under conditions of superfluidity at the absence of secondary effectons, when the life-time of secondary effectons and cycle-period of macro-effectons $(\tau^M)_{tr,lb}$ tends to zero, the viscosity also tends to zero: $\eta \to 0$.

In accordance with the Hierarchic Theory of thermal conductivity (see eqs. 11.35 - 11.37), the elimination of secondary acoustic deformons at $T \leq T_\lambda$ must lead also to enhanced thermal conductivity. This effect was also registered experimentally.

The increasing in $E(k)$ at Figure 40 in the process of further temperature decreasing at $T < T_\lambda$ can be induced by the enhanced contribution of primary polyeffectons (vortical filaments) to the total energy of He II and the factor: $U_{tot}/T_k = S^{-1}$ in (eq. 12.44).

The maximum in Figure 40 at $0 < T < T_\lambda$, corresponding to $\Delta_1$, is the result of competition of two opposite factors:

1) a rise in the total energy of He II due to progress of primary effectons polymerization and

2) its reduction due to the decline in the most probable group velocity ($\mathbf{v}_{gr}$), accompanied by secondary effectons and deformons degeneration (see eq.12.46).

The latter process dominates at $T \to 0$, providing tending of $E(k)$ to zero. The development of polyeffectons superfluid subsystem is accompanied by a corresponding decreasing of the normal component in He II ($\rho_S \to 1$ and $\rho \to 0$). The normal component has a bigger internal energy than superfluid one.

The dimensions of primary translational and librational effectons, composing polyeffectons, increases at $T \to 0$.

### Verification of the inaccessibility of the b-state

### of primary effectons at $T \leq T_\lambda$

Let us analyze our formula (2.74) for the phase velocity of primary effectons in the *(a)* -state for the condition $T \leq T_\lambda$, when filament - like polyeffectons originate:

$$\mathbf{v}_{ph}^a = \frac{\mathbf{v}_S \frac{1-f_d}{f_a}}{1 + \frac{P_b}{P_a}\left(\frac{v_{res}^b}{v_{res}^b}\right)} \qquad 12.59$$

where $\mathbf{v}_S$ is the velocity of sound; $P_b$ and $P_a$ are the thermo-accessibility of the *(b)* and *(a)* states of primary effectons; $f_d$ and $f_a$ are the probabilities of primary deformons and primary effectons in *(a)* state excitations (see eq. 2.66).

One can see from (12.59, 2.66 and 2.75) that if:

$$P_b \to 0, \text{ then } P_d = P_b P_a \to 0 \text{ and } f_d \to 0 \text{ at } T \leq T_\lambda$$

then phase velocity of the effecton in *(a)* state tends to the velocity of sound:

$$\mathbf{v}_{ph}^a \to \mathbf{v}_S \qquad 12.60$$

For these $\lambda -$ point conditions, the total energy of $^4$He atoms forming polyeffectons due to Bose-condensation of secondary effectons (see 12.46) can be presented as:

$$E_{tot} \sim E_a = m\mathbf{v}_{gr}\mathbf{v}_{ph}^a \to m\mathbf{v}_{gr}\mathbf{v}_S \qquad 12.61$$

where the empirical the velocity of sound in He II is $\mathbf{v}_S = 2.4 \times 10^4 cm/s$.

The kinetic energy of de Broglie wave at the same conditions is $T_k = m\mathbf{v}_{gr}^2/2$. Dividing $E_{tot}$ by $T_k$ we have, using (12.47):

$$\frac{\mathbf{v}_S}{\mathbf{v}_{gr}} = \frac{E_{tot}}{2T_k} = \frac{1}{2S} = \frac{1}{2(m^*/m)} \qquad 12.62$$

and

$$\mathbf{v}_{gr}^0 = \mathbf{v}_s \times 2S^0 = 2.4 \cdot 10^4 \times 0.32 = 7.6 \times 10^3 cm/s. \qquad 12.63$$

$m^* = 0.16m$ is the semiempirical effective mass at $T = T_\lambda$.

The most probable de Broglie wave length corresponding to (12.63) at $\lambda$-point:

$$\lambda^0 = h/m\mathbf{v}_{gr}^0 = 15.1 \text{ Å} \qquad 12.64$$

The number of $^4He$ atoms in the volume of the same effecton calculated in accordance with (12.54) is equal: $q^0 = (n_V^0)^{1/3} = 3.8$.

This result is even closer to one predicted by our model (see 12.55) than (12.53). It confirms that at $T \leq T_\lambda$ the probability of b-state $P_b \to 0$ and conditions (12.60) and (12.61) take place indeed.

In such a way Hierarchic Model of superfluidity explains the available experimental data on liquid $^4$He as a limit case of the Hierarchic Theory of viscosity for normal liquids (see section 11.6).

### 12.6. Superfluidity of $^3$He

The scenario of superfluidity, described above for Bose-liquid of $^4$He ($S = 0$) in principle is valid for Fermi-liquid of $^3$He ($S = \pm 1/2$) as well. A basic difference is determined by an additional preliminary stage related to the formation of Cooper pairs of $^3$He atoms with total spins, equal to 1, *i.e.* with boson's properties. The bosons only can form effectons as coherent clusters containing particles with *equal* energies.

We assume in our model that Cooper's pairs can be formed between neighboring $^3$He atoms. It means that the minimum number of $^3$He atoms forming part of the primary effecton's edge at $\lambda$-point must be 8, *i.e.* two times more than that in $^4$He (condition 12.55). Correspondingly, the number of $^3$He atoms in the volume of an effecton is $(n_V^0)_{3_{He}} = 8^3 = 312$. These conditions explains the fact that superfluidity in $^3$He arises at temperature $T = 2.6 \times 10^{-3} K$, *i.e.* lower than that in $^4$He. The formation of flexible filament-like polyeffectons, representing macroscopic Bose-condensate in liquid $^3$He responsible for superfluidity, is a process, similar to that in $^4$He described above.

### 12.7. Superconductivity

#### 12.7.1 General properties of metals and semiconductors

The dynamics of conductance electrons in metals and semiconductors is determined by three main factors (Kittel, 1978, Ashkroft and Mermin, 1976, Blakemore, 1985):
1. The electric field influencing the energy of electrons.
2. The magnetic fields changing the direction of electrons motions.
3. Scattering on the other electrons, ions, phonons, defects.

The latter factor determines the values of the electron conductance and resistance.

In spite of the small mean distances between electrons in metals (2-3)Å their mean free run length at room temperatures exceeds $10^4$Å and grows by several orders at $T \to 0$. It is related to the fact that *only electrons having energy higher than Fermi energy ($\epsilon_F$) may be involved in collisions.* The fraction of these electrons in the total number of electrons is very small and decreases on lowering the temperature as $(kT/\epsilon_F)^2$. At room temperatures the scattering of electrons in metals occurs mainly on phonons.

The mean free run length of electrons in indium at 2K is about 30 cm.

The analysis of electric and magnetic fields influence on an electron needs the notion of its effective mass ($m^*$). It is introduced as a proportionality coefficient between the force acting on the electron and the acceleration *(a)* in the electric field (E):

$$F = -eE = m^*a; \qquad a = d\mathbf{v}_{gr}/dt \qquad 12.65$$

In a simple case of an isotropic solid body the effective mass of an electron is a scalar (Kittel, 1978):

$$m^* = \frac{h^2}{d^2\epsilon/dk^2} \qquad 12.66$$

where $\epsilon$ is the kinetic energy of an electron, having a quadratic dependence on the wave number ($k = 1/L_B$):

$$\epsilon = \frac{\hbar^2 k^2}{m^*} = \frac{\hbar^2}{2mL^2} \qquad 12.67$$

In a general case, for electrons in solid bodies with a complex periodic structure, the effective mass is a tensor:

$$[m_{ij}^*] = \hbar^2/[\partial^2 \epsilon/\partial k_i \partial k_j] \qquad 12.68$$

The effective mass tensor can have positive components for some directions and negative for others.

#### 12.7.2 Plasma oscillations

At every displacement of the electron gas relative to the subsystem of ions in a solid body, a returning electric field appears. As a consequence of that, the subsystem of electrons will oscillate relative to the subsystem of ions with the characteristic plasma frequency (Ashkroft and Mermin, 1976):

$$\omega_{pl} = 2\pi\nu_{pl} = \left(\frac{4\pi n e^2}{m^*}\right)^{1/2} \qquad 12.69$$

where *(n)* is the number of electrons in $1 cm^3$, *(e)* is the charge and *(m*)* is the effective mass of an electron.

The quantified collective oscillations of electron plasma are termed *plasmons*. With decreasing *(n)* from $10^{22}$ to $10^{10} cm^{-3}$ the frequencies $\omega_{pl}$ decrease from $6 \times 10^{15} s^{-1}$ to $6 \times 10^3 s^{-1}$. For the metals $\omega_{pl}$ corresponds to an ultraviolet frequency range, and for semiconductors - to an IR frequency range.

For longitudinal plasma oscillations at small wave vectors the dependence of frequency on the wave number ($k = 1/L = 2\pi/\lambda$) can be approximately represented (Kittel, 1978) as:

$$\omega \approx \omega_{pl}\left(1 + \frac{3k^2 \mathbf{v}_F^2}{10\omega_{pl}^2} + ...\right) \quad 12.70$$

where $\mathbf{v}_F$ is the Fermi velocity of an electron (see eq.12.77).

The screening length ($l$), characterizing the electron-electron interaction in plasmons of metals when Fermi-gas is degenerated, is equal to:

$$l = \mathbf{v}_F/v_p \sim 1\text{Å} \quad 12.71$$

For the cases of non-degenerated Fermi-gas, when the concentrations of free electrons are sufficiently low (in semiconductors) or at high temperatures $T \sim 10^4 K$, the screening length ($l_d$) is dependent on thermal electron velocity:

$$\mathbf{v}_{th} = (3k_B T/m^*)^{1/2} \quad 12.72$$

and

$$l_D = \mathbf{v}_{th}/v_p \cong \left(\frac{\epsilon k_b T}{4\pi n e^2}\right)^{1/2} \quad 12.73$$

where $v_{pl}$ corresponds to (12.69), $\epsilon$ is the dielectric constant.

For example, if in a semiconductor $n = 5 \times 10^{17} cm^{-3}$ and $\epsilon = 12$, then $l_D = 60\text{Å}$ (March, Parrinello, 1982).

### 12.7.3 Fermi energy

The notion of Fermi energy ($\epsilon_F$) can be derived from the Pauli principle, forbidding the fermions to be in the same energetic states.

The formula for Fermi energy for the case of ideal electron gas includes the electron mass (m), the Planck constant ($h = 2\pi\hbar$) and the concentration of free (nonvalent) electrons ($n_e = N_e/V$):

$$\epsilon_F = \frac{h^2}{2m}\left(\frac{3}{8\pi}n_e\right)^{3/2} = \frac{2\pi^2\hbar^2}{m}\left(\frac{3}{8\pi}n_e\right)^{3/2} \quad 12.73a$$

where $N_e$ is the number of free electrons in selected volume ($V$). For real electron gas, $m$ must be substituted by its effective mass: $m \to m^*$.

The formula (12.73a) can also be derived using the idea of standing de Broglie waves of the nonvalent electrons of matter. The condition under which the concentration of twice polarized standing de Broglie waves of the electrons is equal to the concentration of electrons themselves:

$$n_B^F = \frac{N_e}{V} = \frac{8\pi}{3(\lambda_B^F)^3} = n_e \quad 12.74$$

The de Broglie wave length of an electron corresponding to this condition is:

$$\lambda_B^F = \left(\frac{8\pi}{3n_e}\right)^{1/3} = \frac{h}{m\mathbf{v}_{gr}^f} \quad 12.75$$

The kinetic energy of the nonvalent electrons de Broglie waves ($T_k$) could be expressed through their length and mass. It appears that the kinetic energy of the electrons standing de Broglie waves, limited by their concentration/density ($n_e$), is equal to Fermi energy:

$$T_k^F = \frac{h^2}{2m\lambda_F^2} = \frac{h^2}{2m}\left(\frac{3n_e}{8\pi}\right)^{2/3} = \frac{P_F^2}{2m} = \epsilon_F, \quad 12.76$$

where Fermi momentum:

$$P_F = m\mathbf{v}_F = h\left(\frac{3n_e}{8\pi}\right)^{1/3} = \hbar(3\pi^2 n_e)^{1/3} \quad 12.77$$

The Fermi energy corresponds to Fermi temperature ($T_F$):

$$\epsilon_F = kT_F = h\nu_F \qquad 12.78$$

At $T < T_F$ electron gas is in a strongly "compressed" state. The more the relation $(T/T_F) = kT/\epsilon_F$, the more the probability of the appearance of "free volume" in a dense electron gas and more 'free' electrons appear. On lowering the temperature, when the momentum of electrons decreases and the heat de Broglie wave length increases, the "effective pressure" of the electron gas grows up, following by its Bose-condensation.

### 12.7.4 Cyclotronic resonance

The magnetic field $B_z$ in the direction ($z$) influencing the electron by the Lorentz force, changes the direction of its motion without changing the energy. If an electron's energy does not dissipate, then the electrons rotate in the plane $xy$, around $z$-axis. Such an electron with effective mass $m^*$ has a circulation orbit of the radius $r$, with rotation frequency $\omega_c$. From the condition of equality between the Lorentz force ($r\omega_c e B_z$) and the centrifugal force ($m^* \omega_c^2 r$) the formula is derived for angular cyclotron frequency (Kittel, 1978):

$$\omega_c = eB_z/m^* \qquad 12.79$$

The kinetic energy, corresponding to rotation is equal to:

$$T_k = \frac{1}{2} m^* (\omega_c)^2 r^2 \qquad 12.80$$

In the range of radio-frequencies ($\omega$) such a value of the magnetic induction $B_z$ can be selected that at this value the resonance energy absorption occurs, when $\omega = \omega_c$.

Such experiments on the cyclotron resonance can be done to determine $m^*$ in selected directions.

In a simple case, an electron revolves around the Fermi sphere with the *zero momentum component in z-direction.*

The radius of this sphere is determined by the Fermi momentum $P_F$ (see eq. 12.77). In the real space:

$$r_F \sim \hbar/P_F \qquad 12.81$$

The energy of *free* particles near the Fermi surface:

$$\epsilon(P_F) = \mathbf{v}_F(P - P_F) \qquad 12.82$$

where $\mathbf{v}_F$ and $P_F = m^* \mathbf{v}_F$ are the Fermi velocity and momentum: $P > P_F$ is the momentum of thermal electron at $T > 0$ near the Fermi surface.

The solution of the Schrödinger equation, modified by Landau for electrons in a magnetic field in real space leads to the following total energy eigenvalues (Blakemore, 1985):

$$\epsilon = \frac{\hbar^2 k_z^2}{m^*} + \left(l + \frac{1}{2}\right) \hbar \omega_c, \qquad 12.83$$

where the first term of the right part represent the energy of *translational* motion of electrons, which does not depend on magnetic field magnitude; $k_z = 1/L_z$ is the wave number of this motion; the second term is responsible for rotational energy, $l = 0, 1, 2 \ldots$ is the integer quantum number for rotational motion in magnetic field $B_z$. Every value of $l$ means a corresponding Landau level.

Thus, free electrons in a magnetic field move along the helical trajectory of the radius:

$$r_l = [(2l+1)\hbar/m^* \omega_c]^{1/2} \qquad 12.84$$

At the transition from real space to the wave number space, the radius of the orbit ($k_p$) and its area is quantified as:

$$S_l = \pi k_p^2 = \frac{2\pi eB}{\hbar c}\left(l + \frac{1}{2}\right) \qquad 12.85$$

This formula is valid not only for the free electron model, but also for real metals.

The magnitude $2\pi(\hbar c/e)$ termed a *flux quantum*.

In a strong magnetic field the quantization of electrons energy leads to the periodic dependence of the metal magnetic moment on the magnetic field (B): the de Haaz - van Alfen effect (Kittel, 1978, Ashkroft and Mermin, 1976).

### 12.7.5. Electroconductivity

According to the Sommerfeld theory (Blakemore, 1985), electroconductivity ($\sigma$) depends on the free run time of an electron ($\tau$) between collisions:

$$\sigma = ne^2\tau/m, \qquad 12.86$$

where $n$ is the concentration of electrons, (e) and (m) are electron charge and mass.

The free run time is equal to the ratio of the average free run distance ($\lambda$) of electrons to the Fermi speed ($\mathbf{v}_F$):

$$\tau = \lambda/\mathbf{v}_F \qquad 12.87$$

The free run distance is determined by scattering at defects ($\lambda_D$) and scattering at phonons ($\lambda_{ph}$):

$$1/\lambda = 1/\lambda_D + 1/\lambda_{ph} \qquad 12.88$$

The resistance ($R = 1/\sigma$) could be expressed as:

$$R = 1/\sigma_D + 1/\sigma_{ph} = R_D + R_{ph} \qquad 12.89$$

the contribution $R_D$ depends mainly on the concentration of the conductors defects, and the phonon contribution $R_{ph}$ depends on temperature.

Formula (12.89) expresses the Mattisen rule.

A transition to a superconducting state means that the free run time and distance tend to infinity: $\tau \to \infty$; $\lambda \simeq \lambda_{ef} = h/P_{ef} \to \infty$, while the *resulting* group velocity of the electrons ($\mathbf{v}_{gr}^{res}$) and momentums tends to zero:

$$P_{ef} = m\mathbf{v}_{gr}^{res} \to 0$$

The emergency of macroscopic Bose-condensation of secondary ionic effectons and Cooper pairs corresponds to this condition.

*We assume in our hierarchic approach, that the absence of the non-elastic scattering and dissipation of electrons energy is observed as superconductivity, when the probability of secondary ionic effectons and deformons tends to zero, leading to emergency of primary polyeffectons from electronic Cooper pairs, like vortical filaments, formed by the helium atoms in conditions of superfluidity.*

Let us consider first a conventional microscopic approach to the problem of superconductivity.

### 12.8. Microscopic theory of superconductivity (BCS)

This theory (BCS) was created by Bardin, Cooper and Schriffer in 1957. The basic, experimentally proven assumption of this theory, is that electrons at sufficiently low temperatures are grouped into Cooper pairs with oppositely directed spins, composing Bose-particles with a zero spin. The charge of the pair is equal to $e^* = 2e$ and mass $m^* = 2m_e$.

Such electron pairs obey the Bose-Einstein statistic. The Bose- condensation of this system at the temperature below the Bose-gas condensation temperature ($T < T_k$) leads to the *superfluidity of the electron liquid*. This superfluidity (analogous to the superfluidity of liquid helium) is manifested as superconductivity.

According to BCS's theory, the Cooper electron pair formation mechanism is the consequence of virtual phonon exchange through the lattice.

The energy of binding between the electrons in a pairs is very low: $2\Delta \sim 3kT_c$. It determines a minimum energetic gap ($\Delta$) separating a state of superconductivity from a state of usual conductivity.

Notwithstanding that the kinetic energy of electrons in a superconducting state is greater than $\epsilon_F$, the contribution of the potential energy of attraction between electron pairs is such that the total energy of the superconducting state $(E_a^e)$ is smaller than the Fermi energy $(\epsilon_F)$ (Kittel, 1978). The presence of the energetic gap $(\Delta)$ makes a superconducting state stable after switching-off the external voltage. *The middle of the gap coincides with Fermi level.*

The rupture of a pair can happen due to photon absorption by superconductor with the energy: $\Delta = h\nu_p \approx 3kT_c$. Superconductivity usually disappears in the frequency range $10^9 < \nu_p < 10^{14} s^{-1}$.

In the BCS theory, the magnitude of gap $\Delta$ is proportional to the number of Cooper pairs and grows on lowering the temperature.

The excitation energy of quasi-particles in a superconducting state, which is characterized by the wave number $(k = 1/L)$, is:

$$E_k = (\epsilon_k^2 + \Delta^2)^{1/2} \qquad 12.90$$

where

$$\epsilon_k = \frac{\hbar^2}{m}(k^2 - k_F^2) \approx \frac{\hbar^2}{m} k_F(k - k_F) \qquad 12.91$$

and

$$\delta k_F = (k - k_F) \ll k_F = 1/L_F$$

The critical speed of the electron gas $(\mathbf{v}_c)$, for exciting a transition from a superconducting state to a normal one is determined from the condition:

$$E_k = \hbar k \mathbf{v}_c \text{ and } \mathbf{v}_c = \frac{E_k}{\hbar k} \qquad 12.92$$

The wave function $\Phi(r)$, which describes the properties of electron pairs in the BCS theory, is the superposition of one-electron functions with energies in a range of about $2\Delta$ near $\epsilon_F$. Therefore, the dispersion of momentums for the electrons, involved in formation of pairs can be expressed as:

$$\Delta = \delta\epsilon_F = \delta\left(\frac{P_F^2}{2m}\right) = \left(\frac{P_F}{m}\right)\delta P_F \approx \mathbf{v}_F \delta P_F \qquad 12.93$$

where $\mathbf{v}_F$ is the Fermi velocity of the electrons; $P_F$ - Fermi momentum of the electrons.

The characteristic *coherence length* $(\xi_c)$ of the pair function $\Phi(r)$ has the value (Ashkroft and Mermin, 1976, Lifshits and Pitaevsky, 1978):

$$\xi_c \sim \hbar/\delta P_F \simeq \frac{\hbar \mathbf{v}_F}{\Delta} \simeq \frac{1}{k_F} \frac{\epsilon_F}{\Delta} \qquad 12.94$$

The magnitude $(\epsilon_F/\Delta)$ is usually $10^3 - 10^4$, and $k_F = 1/L_F \sim 10^8 cm^{-1}$. Thus, from (12.94):

$$\xi_c \sim (10^3 - 10^4) \text{Å} \qquad 12.95$$

Inside the region of coherence length $(\xi_c)$ there are millions of pairs. The momentums of Cooper pairs in such regions are correlated in such a way that their resulting momentum is equal to zero, like in our primary effectons and the result of their head-to-tail polymerization (filaments).

At $T > 0$ some of the pairs turn to a dissociated state and the concentration of superconducting electrons $(n_s)$ decreases. The coherence length $(\xi_c)$ also tends to zero with increase in temperature.

The important parameter, characterizing the properties of a superconductor, is the value of the critical magnetic field $(H_c)$, above which the superconductor switches to a normal state.

With a rise in the temperature of the superconductor when $T \to T_c$, the critical field tends at zero: $H_c \to 0$. And vice versa, at lowering of temperature, when

$$T < T_c, \text{ the } H_c \text{ grows up as:}$$

$$H_c = H_0[1 - (T/T_c)^2]$$

where $H_0$ corresponds to $T = 0$.

The *Meisner effect* - representing the "forcing out" of the external magnetic field from superconductor, is also an important feature of superconductivity.

The depth of magnetic field penetration into the superconductor ($\lambda$) (Kittel, 1978, Ashkroft and Mermin, 1976) is:

$$\lambda = (mc^2\varepsilon_0/e^2 n_s)^{1/2} \simeq (10^{-6} - 10^{-5}) cm,$$

where $n_s$ the density of electrons in a superfluid state; $\varepsilon_0$ - dielectric constant.

On temperature raising from $0\ K$ to $T_c$, the $\lambda$ grows as:

$$\lambda = \frac{\lambda_0}{\left[1 - \left(\frac{T}{T_c}\right)^4\right]^{1/2}}, \qquad 12.96$$

where $\lambda_0$ corresponds to $\lambda$ at $T = 0$.

The superconductors with magnetic field penetration depth ($\lambda$) less than coherence length $\xi$ :

$$\lambda \ll \xi \qquad 12.97$$

are termed the *first-order superconductors* and those with

$$\lambda \gg \xi \qquad 12.98$$

are *second-order superconductors*.

*Nowadays, in connection with the discovery of high temperature superconductivity (Bednorz, Miller, 1986, Nelson, 1987) the mechanism of stabilizing electron pairs by means of virtual phonons in the BCS theory evokes a doubts.*

## 12.9 The Hierarchic scenario of superconductivity

A new mechanism of electron pair formation and their subsequent Bose-condensation, without virtual phonons as a mediators is proposed. Such a process includes the formation of primary effectons (mesoscopic Bose condensate) from Cooper pairs, as a 1st stage, and the system of filaments of polyeffectons (macroscopic Bose condensate), as a 2nd stage, like in superfluidity phenomena (Kaivarainen, 2007).

Two basic questions have been answered in this approach to superconductivity:

I. Why does energy dissipation in the system [conductivity electrons + lattice] disappear at $T \leq T_c$ ?

II. How does the coherence in this system, accompanied by the electrons Cooper pair formation and their Bose condensation, originate ?

The correctness of the idea about the electrons 3D de Broglie standing waves formation in condensed matter is confirmed by the increasing experimental data of high-density filamentation or clusterization of the free electrons. These data point to existing of strong attractive force between the electrons (Cooper pairs), dominating over Coulomb repulsion at certain condition (Zhitenev, 1999 and Shoulders, 1991).

It will be shown below how these problems can be solved in the framework of the Hierarchic Theory.

*The following factors can affect electron's dynamics, scattering and dissipation near Fermi energy:*

1. Interaction of electrons with *primary and secondary ionic/atomic effectons* in acoustic *(a)* and ($\bar{a}$) states, stimulating formation of the electronic Cooper pairs;

2. Interaction with primary and secondary *effectons* of lattice in optic *(b)* and ($\bar{b}$) states, accompanied by origination of *polarons* and Cooper pairs dissociation;

3. Interaction with *transitons* in the course of $(a \Leftrightarrow b)$ and $(\bar{a} \Leftrightarrow \bar{b})$ transitions of primary and secondary effectons;

4. Interaction with [$tr/lb$] *convertons* (interconversions between primary translational and

librational effectons;

    5. Interaction with primary electromagnetic *deformons* - the result of ($a \Leftrightarrow b$) transitions of primary effectons;

    6. Interaction with secondary acoustic *deformons* (possibility of polaron formation);

    7. Interaction with macro-effectons in A- and B-states;

    8. Interaction with macro- and super-deformons, stimulating the defectons origination.

It follows from our model that the oscillations of all types of quasi-particles in the conductors and semiconductors are accompanied not only by electron-phonon scattering and corresponding dissipation, but also by electromagnetic interaction of primary deformons with nonvalent-free electrons.

At $T > T_c$ the fluctuations of nonvalent electrons with energy higher than Fermi one under the influence of factors (1 - 8) are random (noise-like) and no selected order of fluctuations in normal conductors exists. It means that the ideal Fermi-gas approximation for such electrons is sufficiently good. In this case, the effective electron mass can be close to that of a free electron ($m^* \simeq m$).

The electric current in normal conductors at external voltage should dissipate due to fluctuations and energy exchange of the electrons with lattice determined by factors (1 - 8).

Coherent in-phase acoustic oscillation of *the ionic primary* effectons in *(a)* -state *is* the "ordering factor" simulating electron gas coherence due to electromagnetic interactions. But its contribution in normal conductors at $T > T_c$ is very small. For the *ideal* electron gas, the total energy ($E_{tot}$) of each electron as de Broglie **wave** is equal to its kinetic energy ($T_k$), as far potential energy ($V = 0$):

$$E_{tot} = \hbar\omega = \frac{\hbar^2}{2mA^2} = T_k + V = \frac{\hbar^2}{2mL^2} \qquad 12.99$$

where

$$T_k = \frac{\hbar^2}{2mL^2} \quad \text{and} \quad V = 0 \qquad 12.100$$

and

$$L = \frac{\hbar}{m\mathbf{v}_{gr}} = \frac{1}{k} \qquad 12.101$$

One can see from (12.100) that for ideal gas, when $m = m^*$, the most probable amplitude (A) and de Broglie wave length (L) are equal:

$$A = L \quad \text{if} \quad E_{tot} = T_k \quad \text{and} \quad V = 0 \qquad 12.102$$

*Like in the case of liquid helium at conditions of superfluidity, the Bose- condensation in metals and semiconductors is related to increasing of the concentration of the (a) -state of primary effectons with the lowest energy and corresponding decrease in the concentrations of (b) - state and all other excitations. The Bose-condensation and degeneration of secondary ionic effectons and deformons, followed by formation of electronic polyeffectons from Cooper pairs, is responsible for second-order phase transition like transition from the regular conductivity to superconductivity.*

The cooperative character of 2nd order phase transition [conductor → superconductor] is determined by the feedback reaction between the lattice and electron subsystems. It means that the collective Bose-condensation in both subsystems is starting at the same temperature: $T = T_c$.

Under such conditions the probabilities of the *(a)* -states of ionic ($P_a^i$) and electronic ($P_a^e$) effectons tend to 1 at $T \leq T_c$:

$$\left.\begin{array}{l} P_a^i \to 1; \; P_a^e \to 1 \\ P_b^i \to 0; \; P_b^e \to 0 \end{array}\right\} \qquad 12.103$$

The equilibrium parameter for both subsystems:

$$\kappa^{i,e} = \left(\frac{P_a - P_b}{P_a + P_b}\right)^{i,e} \to 1 \qquad 12.104$$

and the *order parameter*:

$$\eta^{i,e} = (1 - \kappa^{i,e}) \to 0 \qquad 12.105$$

is like in description of 2-nd order phase transition for liquid helium (see eq. 12.58).

In our model the coherent Cooper pairs are formed as Bose particles with resulting spin equal to 0 and 1 from the *neighboring electrons*, in contrast to BCS theory, when they are separated.

The experimentally proved formation of clusters even in pure electron gas, nonetheless of Coulomb repulsion, confirms the possibility of formation of Bose condensate from non-remote electronic Cooper pairs.

Such pairs can compose a primary electron's effectons (*e-effectons*) as a coherent clusters with a resulting momentum *equal to zero*. Formation of secondary *e-effectons* with nonzero resulting momentum is possible also. The interaction of this secondary e-effectons with lattice is responsible for electric resistance in normal conductors. Degeneration of such type of excitations in the process of their Bose-condensation and their conversion to primary e-effectons means the transition to superconducting state.

The in-phase coherent oscillations of the integer number of the electrons pairs, forming primary *e-effectons*, correspond to their acoustic *(a)* - state, and the counterphase oscillations to their optic *(b)* state, like in ionic or molecular effectons.

We assume that superconductivity can originate only when the fraction of nonvalent coherent electrons forming *primary e-effectons in definite regions of conductor* strongly prevails over the fraction of noncoherent *secondary* e-effectons. Due to feedback reaction between subsystems of lattice and electrons this fraction should be equal to ratio of wave length of primary and secondary ionic effectons. This condition can be introduced as:

$$\left(\frac{\lambda_a^e}{\bar{\lambda}_a^e}\right)_{T_c} = \left(\frac{\lambda_a^i}{\bar{\lambda}_a^i}\right)_{T_c} = \left(\frac{\bar{v}_a^i}{v_a^i}\right)_{T_c} \geq 10 \qquad 12.106$$

where $\lambda_a^e / \bar{\lambda}_a^e$ is the ratio of wave lengths of primary and secondary *e- effectons,* equal to that of the ionic primary and secondary effectons of lattice $\lambda_a^i / \bar{\lambda}_a^i = \frac{\mathbf{v}_s/v_a^i}{\mathbf{v}_s/\bar{v}_a^i} = \bar{v}_a^i/v_a^i$ where $\mathbf{v}_s = \mathbf{v}_{ph}^i \simeq \bar{\mathbf{v}}_{ph}^i$.

Otherwise the condition (12.106) can be expressed as:

$$\lambda_a^e = \lambda_a^i \simeq 10 \times \bar{\lambda}_a^e \simeq 20 n_e^{1/3} \qquad 12.107$$

As far the oscillations of e-pairs in the Bose-condensate (a-state of e- effectons) are modulated by the electromagnetic field, excited by oscillating ions in the *(a)* -state of primary ionic effectons they must have the same frequency.

Condition (12.107) means that the *number of electrons* in the volume ($V_e \sim \lambda^3$) of primary e-effectons is about $10^3$ times more than that in secondary e-effectons. The lattice and electronic effectons subsystems as bosonic ones are spatially compatible.

As far the effective mass ($m^*$) of the electrons in the coherent macroscopic Bose-condensate formed by e-polyeffectons at conditions of superconductivity tends to infinity:

$$\begin{aligned} T &< T_c \\ m^* &\mapsto \infty \\ T &\to 0 \end{aligned} \qquad 12.108$$

the plasma frequency (eq. 12.69) tends to zero:

$$T < T_c$$
$$2\pi \nu_{pl} = \omega_{pl} \mapsto 0$$
$$at \ T \to 0$$

and, consequently, the screening length (eq. 12.71) tends to infinity: $l \to \infty$. *This condition corresponds to that of macroscopic Bose-condensation emergency.*

Under these conditions the momentum (see eq.12.93) originates in addition to Fermi's one:

$$\delta P_F = \hbar \delta k_F > 0 \qquad 12.109$$

but a decrease in the potential energy of both electron's and ion's subsystems due to leftward $(a \Leftrightarrow b)$ equilibrium shift *leads to the emergence of the gap near the Fermi surface* $(2\Delta)$ depending on the difference of energy between *(a)* and *(b)* states of primary effectons.

The linear dimension of coherent *primary electronic effectons* (e-effectons), which is equal to coherence length in the BCS theory (see eq. 12.94) is determined by additional momentum $\delta P_F$:

$$\xi = \lambda_a^e = \frac{\mathbf{v}_s}{\mathbf{v}_a} = h/\Delta p_F = h/m_e \Delta \mathbf{v}_F \qquad 12.110$$

In turn, the primary e-effecton in *(a)*-state can form e- polyeffectons as a result of their polymerization, like in superfluid liquids. The starting point of this collective process represents the macroscopic Bose - condensation, accompanied by second-order phase transition in accordance with our model.

### 12.9.1 Interpretation of the experimental data, confirming our superconductivity model

The energy gap between normal and superconductive states can be calculated directly from the Hierarchic Theory, as the difference between the total energy of matter before $\left[ U_{tot}^{T>T_c} \right]$ and after $\left[ U_{tot}^{T<T_c} \right]$ second-order phase transition:

$$2\Delta = U^{T>T_c} - U^{T<T_c} \qquad 12.111$$

However, such experimental parameters as the velocity of sound, density and the positions of bands in a far IR region should be available around the transition temperature $(T_c)$ for calculation of (12.111).

This gap must be close to the energy of $(a \to b)^i$ transitions of the *ionic* primary effectons, related to the energy of $(a \to b)^e$ transitions of the electronic *e-effectons* (see 12.103 - 12.105).

This statement of our model of superconductivity coincides well with switching off the superconductivity state by IR-radiation with minimum photon frequency $(\nu_p)$, corresponding to the energy gap $(2\Delta)$ at given temperature:

$$h\nu_p = 2\Delta \sim (E_b - E_a) \qquad 12.112$$
$$or: \quad \nu_p = 2\Delta/h \qquad 12.112a$$

Another general feature of superconductivity for low- and high-temperature superconductors is the almost constant ratio:

$$\frac{2\Delta_0}{kT_c} \simeq 3.5 \qquad 12.113$$

where the gap: $2\Delta = 2\Delta_0 = h\nu_p$ at $T = T_c$ and $\Delta = 0$ at $T > T_c$.

It will be shown below that the experimental result (12.113) is follows from the condition (12.106) of the Hierarchic Model of superconductivity.

Considering (12.112), (12.113) and (2.27), the frequency of a primary ionic effectons in a-state near transition temperature can be expressed as:

$$v_a^i = \frac{v_p^0}{\exp\left(\frac{2\Delta}{kT_c}\right) - 1} \simeq v_p^0/32.1 = 0.03(2\Delta/h) \qquad 12.114$$

consequently:

$$hv_a^i = 0.03 \times 2\Delta \qquad 12.115$$

The frequency of secondary lattice effectons in $(\bar{a})$-state in accordance with (2.54) is:

$$\overline{v_a^i} = \frac{v_a^i}{\exp\left(\frac{hv_a^i}{kT_c}\right) - 1} \qquad 12.115a$$

$$\text{or :} \quad \frac{\overline{v_a^i}}{v_a^i} = \frac{1}{\exp\left(\frac{hv_a^i}{kT_c}\right) - 1} \qquad 12.115b$$

as far:

$$\frac{hv_a}{kT_c} = 0.03\frac{2\Delta}{kT_c} = 0.1 \ll 1 \qquad 12.116$$

we have:

$$\frac{1}{\exp\left(\frac{hv_a^i}{kT_c}\right) - 1} \sim \frac{kT_c}{hv_a} \sim 10 \qquad 12.116a$$

consequently from 12.116a, we get for transition conditions:

$$hv_a^i \simeq 0.1 \times kT_c \qquad 12.117$$

Now, using (12.17), (12.116a) and (12.115b), we confirm the correctness of superconductivity condition (12.106):

$$\left(\frac{\lambda_a^e}{\overline{\lambda}_a^e}\right)_{T_c} = \left(\frac{\lambda_a^i}{\overline{\lambda}_a^i}\right)_{T_c} = \left(\frac{\overline{v}_a^i}{v_a^i}\right)_{T_c} = \frac{1}{\exp\left(\frac{hv_a^i}{kT_c}\right) - 1} \sim 10 \qquad 12.118$$

where

$$\lambda_a^e = h/2m_e(\mathbf{v}_{gr}^a)^e = (\mathbf{v}_s/v_a)^i = \lambda_a^i \qquad 12.119$$

is the most probable de Broglie wave length of coherent electron pairs composing a primary e-effecton; $(v_{gr}^a)^e$ is a group velocity of electron pairs in a-state of primary e-effectons, stimulated by ionic lattice oscillations:

$$\overline{\lambda_a^e} = h/2m^*\overline{\mathbf{v}_{gr}^a} = (\mathbf{v}_s/\overline{v_a})^i = \overline{\lambda_a^i} \qquad 12.120$$

is the mean de Broglie wave length of electron pair forming the effective secondary e-effecton.

*Our theory predicts also the another condition of coherency between ionic and electronic subsystems, leading to superconductivity, when the linear dimension of primary translational ionic effectons grows up to the value of coherence length (see eq.12.94):*

$$(\lambda_a^i)_{tr} = \frac{\mathbf{v}_s}{(v_a^i)_{tr}} \geq_{T_c} \zeta = \frac{h\mathbf{v}_F}{\Delta} \qquad 12.121$$

In simple metals a relation between the velocity of sound $(v_s)$ and Fermi velocity $(v_F)$ is determined by the electron to ion mass ratio $(m_e/M)^{1/2}$ (March and Parinello, 1982):

$$\mathbf{v}_s = \left(\frac{zm_e}{2M}\right)^{1/2}\mathbf{v}_F \qquad 12.122$$

where $(z)$ is the valence of the ions in metal.

Putting $(v_a^i)_{tr}$ from (eq. 12.115) and (eq. 12.122) in condition (12.121), and introducing instead electron mass its effective mass $(m_e \to m^*)$ composing e-effecton, we get at $T = T_c$:

$$m^* \stackrel{T_c}{\simeq} 2 \cdot 10^{-4} \times \frac{M}{z} \qquad 12.123$$

As far we assume in our approach, that at the transition temperature $(T_c)$ the volumes of primary lattice (ionic) effectons and primary e-effectons coincide, then the number of electrons in the volume of primary ionic effectons is:

$$N_e = n_e V^i_{ef} = n_e \frac{9}{4\pi} \left(\frac{\mathbf{v}_s}{v^i_a}\right)^3_{T_c} \qquad 12.124$$

where $n_e$ is a concentration of the electrons; $V^i_{ef} = V^e_{ef}$ is the volume of primary translational ionic effectons.

Using the relation between primary and secondary ionic de Broglie waves (12.106) at transition temperature $(T_c)$ as $\lambda_a = 10\bar{\lambda}_a$, taking into account (12.117) and (12.121) we get:

$$\left(\frac{zm^*_e}{2}\right)^{1/2} \frac{1}{T_c M^{1/2}} = \frac{k}{10 \times 2\pi\Delta} \qquad 12.125$$

or:

$$T_c(M/z)^{1/2} = \frac{10\pi\Delta}{k}(2m^*_e)^{1/2} \qquad 12.126$$

As far for different isotopes the energetic gap is almost constant $(2\Delta \simeq const)$, then the left part of (12.126) is constant also. Consequently for isotopes we have:

$$T_c(M/z)^{1/2} \simeq const \qquad 12.127$$

Such an important correlation between transition temperature $(T_c)$ and isotope mass $(M)$ is experimentally confirmed for many metals:

It follows from (12.121) that the more rigid is lattice and the bigger is the velocity of sound $(\mathbf{v}_s)$, the higher is transition temperature of superconductivity. The anisotropy of $(\mathbf{v}_s)_{x,y,z}$ means the anisotropy of superconductor properties and can be affected by external factors such as pressure.

*It was shown in this section, that all most important phenomena, related to superconductivity can be explained in the framework of the Hierarchic Theory.*

# Chapter 13
# The Hierarchic Theory of complex systems

### 13.1. **The independent experimental and theoretical results**, **confirming mesoscopic Bose condensation (mBC) in condensed matter**

The existence of domains and granules, composed from the atoms/molecules are very common in solids, liquid crystals, and polymers of artificial and biological origin.

In liquids, as is seen from the X-ray data and neutron scattering, the local order is also kept like in solid bodies. Just like in the case of solids, local order in liquids can be caused by the primary librational effectons, but smaller in size.

In water, the relatively stable clusters of molecules where revealed by the quasi-elastic neutron scattering method. The diameter of these clusters are (20-30) Å and the lifetime is of the order of $10^{-10} s$ (Gordeyev and Khaidarov, 1983). These parameters are close to those we have calculated for librational water effectons (Figure 7).

A coherent-inelastic neutron scattering, performed on heavy water - $D_2O$ at room temperature revealed collective high-frequency sound mode. Observed collective excitation has a solid-like character with dimension around 20 Å, resembling water clusters with saturated hydrogen bonds. The observed the velocity of sound in these clusters is about 3300 m/s, *i.e.* close to velocity of sound in ice and much bigger than that in liquid water: 1500 m/s (Teixeira *et al.*, 1985). Such data like the previous ones (Gordeyev and Khaidarov, 1983), confirm the existence of primary librational effectons as a molecular mesoscopic Bose condensate (mBC), following from the Hierarchic Theory (Figure7).

Among the earlier theoretical models of water the model of "flickering clusters" proposed by Frank and Wen (1957) is closer to our model than others. The "flickering" of a cluster consisting of water molecules is expressed by the fact that it dissociates and associates with a short period $(10^{-10} - 10^{-11})$ s. Near the non-polar molecules this period grows up (Frank and Wen, 1957, Frank and Evans, 1945) and "icebergs" appear. The formation of hydrogen bonds in water is treated as a cooperative process. Our [*lb/tr*] convertons, *i.e.* interconversions between primary lb and tr effectons, reflect the properties of flickering clusters better than other quasi-particles of Hierarchic Model.

Proceeding from the flickering cluster model, Nemethy and Scheraga (1962), using the methods of statistical thermodynamics, calculated a number of parameters for water (free energy, internal energy, entropy) and their temperature dependencies, which agree with the experimental data in the limits of 3%. However, calculations of heat capacity were less successful. The quantity of water molecules decreases from 91 at $0^0 C$ to 25 at $70^0 C$ (Nemethy and Scheraga, 1962). It is in rather good agreement with our results (Figure 7a) on the change of the number of water molecules in a primary librational effecton with temperature.

The stability of primary effectons (clusters, domains), forming the condensed media, is determined by the coherence of heat motions of molecules, atoms and atomic groups and superposition of their de Broglie waves - increases the distant van der Waals interaction in the volume of the effectons. The mesoscopic Bose condensation (mBC) or coherent clusters formation may start, when the value of *most probable* de Broglie waves length of vibrating molecules of condensed matter exceeds the average distance between molecules:

$$\lambda_B = \frac{h}{m\mathbf{v}} \geq \left(\frac{V_0}{N_0}\right)^{1/3} \qquad 13.1$$

This condition can be achieved not only in homogenic by molecular/atomic composition condensed matter, like pure water. The equality of the most probable 3D standing de Broglie waves of *different* molecules, atoms or ions is possible, if their most probable momentums and kinetic energies are equal:

$$P_i = m_1 \mathbf{v}_1 = m_2 \mathbf{v}_2 = \ldots = m_i \mathbf{v}_i \qquad 13.2$$

then:

$$\lambda_1 = \frac{h}{m_1 \mathbf{v}_1} = \lambda_2 = \frac{h}{m_2 \mathbf{v}_2} = \ldots = \lambda_i = \frac{h}{m_i \mathbf{v}_i} \qquad 13.2a$$

In this case the differences in masses are compensated by differences in the most probable group velocities of these oscillating particles of condensed matter, so that their most probable momentums are equal. Corresponding different particles may compose the same primary effecton - mesoscopic Bose condensate, if their de Broglie waves length $\lambda_i$ satisfy the condition 13.1.

The domains or the crystallites in heterogenic by atomic composition solid bodies *can be considered as a primary effectons,* containing a big number of elementary cells of nonidentical atoms.

The transitions between different types of elementary cells (second-order phase transitions) means cooperative redistribution of spatial symmetry of momentums of coherent atoms, leading to origination of primary effectons of different shape and volume.

In accordance to our theory of superfluidity and superconductivity (see chapter 12), the second-order phase transitions can be related also to primary effectons polymerization and vortical filaments formation, accompanied by strong shift of $(a \Leftrightarrow b)$ equilibrium of primary effectons to the left.

The presented below material gives the additional evidence, that the Hierarchic Theory can be used to describe a wide range of physical-chemical and biological phenomena.

### 13.2. Protein domain mesoscopic organization

It follows from numerous data of X-ray analysis, that the dimensions of protein cavities are close to that of water librational effectons (Figure7). The evolution of biological macromolecules could have gone in such a way that they "learned" to use the cooperative properties of water clusters and their 'flickering' for regulation of biopolymers large-scale dynamics and signal transmission.

*For example, the calculated, using our theory based computer program (pCAMP), the frequency of [lb/tr] convertons $(10^6 - 10^7) c^{-1}$ coincides with frequency of protein cavities large-scale pulsations and protein domains relative fluctuations.*

All sufficiently large globular proteins consist of domains whose dimensions under normal conditions vary in the narrow limits: (10-20)Å (Kaivarainen, 1985; 1995).

The anticipated correlation between dimensions of water clusters and size of domains and interdomain cavities means, that the lower is the physiological temperature of given warm blood animals or birds or normal temperature of surrounding medium of cold-blooded organisms, like fish, insects, ants, etc., the larger are the interdomain cavities and the domains of proteins with similar function. The author did not find the corresponding comparative analysis of correlations in literature.

It is known that the water in pours or cavities with diameter less than 50 Å freeze out at very low temperatures (about $-60^0 C$) and water viscosity is high (Martini *et al.*, 1983). For the other hand, our calculations shows (Figure17b), that freezing in normal conditions should be accompanied by increasing the linear size of primary librational effectons at least till 50 Å.

As far the condition for *lb* effectons growth in narrow pours is absent, the formation of primary *tr* effectons, necessary for freezing is also violated, as far these two process are interrelated. As a result of that, the condition (6.6) for [liquid → solid] phase transition occurs in small cavities at much lower temperature than in bulk water and this ice should differ from the regular one. So these results confirms our theory of 1st order phase transition.

If we get experimentally the velocity of sound in protein and the position of band in oscillatory spectra, which characterize the librations of coherent atomic groups, then the most probable de Broglie wave length of aminoacids groups and atoms, forming protein nodes or domains $(\lambda_1, \lambda_2, \lambda_3)$ can be calculated using our computer program - pCAMP.

If the volume of the primary effecton can be approximated by sphere:

$$V_{ef} = \frac{9(\lambda_1\lambda_2\lambda_3)}{4\pi} = \frac{4}{3}\pi r^3, \qquad 13.3$$

then its radius:

$$r = \left[\frac{27(\lambda_1\lambda_2\lambda_3)}{16\pi^2}\right]^{1/3} = 0.555\lambda_{res}, \qquad 13.4$$

where $\lambda_{res} = (\lambda_1\lambda_2\lambda_3)^{1/3}$, and the diameter of the effecton: $d = 2r = 1.11\lambda_{res}$.

The collective properties of protein's primary effectons presented by $\alpha$-structures, $\beta$-sheets and whole domains can determine the cooperative properties of biopolymers and their folding $\rightleftharpoons$ unfolding mechanism.

*Heat oscillations of atoms and atom groups, forming the protein effectons (domains and knots) must be coherent, like in the primary effectons of any other condensed matter.*

The notion of "knots" in proteins was introduced by Lumry and Gregory (1986). The knots are regions, containing very slow $H \Leftrightarrow D$ exchangeable protons composing compact cooperative structures.

The dimensions of knots are less than dimensions of domains. It looks that knots could represent a translational primary effectons, in contrast to librational effectons, possibly representing whole domains.

The molecular dynamic computer simulations of proteins reveal, indeed, a highly correlated collective motion of groups of atoms, inhomogeneously distributed in proteins structure (Swaminathan *et al.*, 1982).

It looks that the traditional theory of protein tertiary structure self-organization/folding from primary one (Cantor and Shimmel, 1980) is not totally successful as far it does not take into account the quantum effects, like mesoscopic Bose condensation in proteins.

The change of interdomain interactions, the stabilization of its small-scale dynamics of proteins by ligands (decreasing of momentum) leads to the increase of $\lambda_{res} = h/p$ (see eq.13.4). This factor of protein folding can provide the long- distance signal transmission in macromolecules and allosteric effects in oligomeric proteins. The $[lb/tr]$ convertons, *i.e.* dissociation of interdomain water clusters in the protein cavities can be a trigger in the above mentioned processes.

The described events interrelate a small-scale dynamics of atoms and the large-scale dynamics of domains and subunits with dynamics of water clusters in protein interdomain cavities, dependent in turn on the properties of bulk water (Kaivarainen, 1985,1989b, Kaivarainen *et al.*, 1993).

Our hierarchic water clusters - mediated mechanism of signal transmission in biopolymers is alternative to *Davidov's soliton* mechanism of signal transmission in biosystems. The latter is good only for highly ordered systems with small dissipation.

### 13.3. Quantum background of lipid domain organization in biomembranes

The importance of lipid domains in membranes for cells functioning is known, but the physical background for domains origination remains unclear.

Our approach was used for computer simulations of lipid domain dimensions in model membranes. The known positions of IR bands, corresponding to asymmetric $[N-(CH_3)_3]$ stretching between *trance* ($920 cm^{-1}$) and *gauche* conformations (900 and 860 $cm^{-1}$) where taken for calculations. The known change of the velocity of sound, as a result of phase transitions in these membranes: $(1.97 \times 10^5 cm/s)^{38°} \to (1.82 \times 10^5 cm/s)^{42°}$ also was used for calculations of lipid domains dimensions, using (eq. 2.59). The results of calculations are presented on Figure 41. In our model the lipid domains in the membranes are considered as a primary librational - *effectons*, formed by 3D superposition of the most probable de Broglie waves $(\lambda = h/p)_{1,2,3}$, determined by coherent thermal torsion oscillations of $CH_3$ groups of lipid molecules. The lesser the value of the most probable momentum $(p = m\mathbf{v})_{1,2,3}$ of lipid, the bigger is corresponding de Broglie wave length $\lambda_{1,2,3} = h/p_{1,2,3}$ and the effecton volume (eq. 13.3).

According to our calculations, a rise in temperature from 0 to $70^0$ leads to decrease of the most probable $\lambda$ from 88 to 25 Å. In the phase transition region ($38 - 42^0$) de Broglie wave length $\lambda$ decreases from 46 to 37 Å (Figure 41). The former process corresponds to decreasing of the lipid domains volume from (50 to 2) × $10^4$ $Å^3$ and the latter one from (7.5 to 5) × $10^4$ $Å^3$, respectively (Figure 42). The values of these changes coincide with available experimental data.

Like the calculations for ice and water, these results provide further support to our theory.

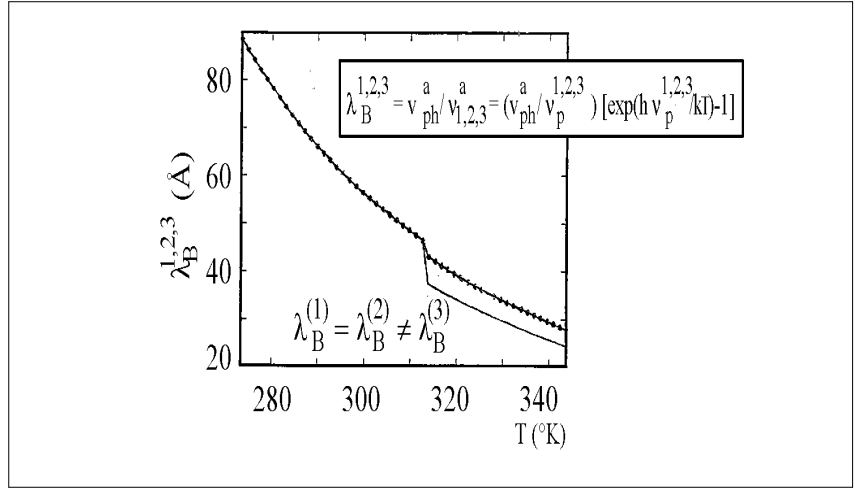

**Figure 41**. Temperature of the most probable de Broglie wave length of groups of lipids ($\lambda_1, \lambda_2, \lambda_3$), related to their stretching, including *trance - gauche* change of lipids conformation.

The values of $\lambda_1, \lambda_2, \lambda_3$ determine spatial dimensions of lipid domains. Domains are considered as a coherent clusters - primary effectons (mBC), formed by 3D superposition of the most probable de Broglie waves of $CH_3$ groups of lipids.

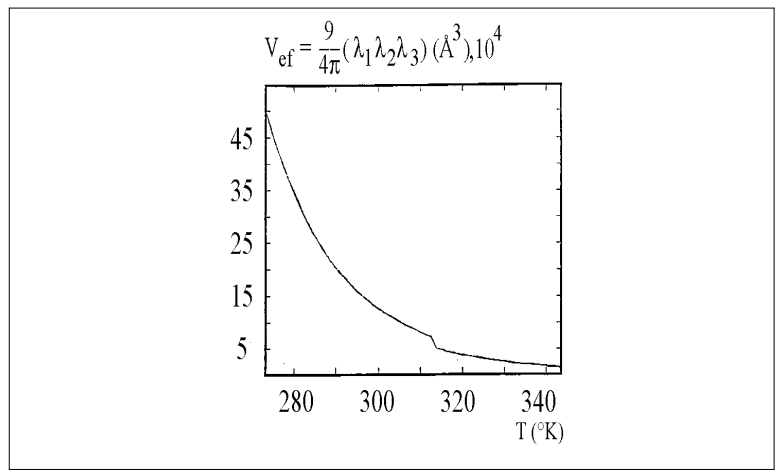

**Figure 42**. Temperature dependence of the volume of lipid domains: $V = \frac{9}{4\pi}\lambda_1\lambda_2\lambda_3$. The volume is determined by 3D superposition of the most probable de Broglie waves of lipid molecules or their fragments. Phase transition occurs in the region around $312K$ ($39^0$) and is accompanied by *trance - gauche* change of lipids conformation.

### 13.4 Hierarchic approach to theory of solutions and colloid systems

The action of the dissolved molecules can lead to the shift of ($a \Leftrightarrow b$) equilibrium of librational effectons of solvent to the right or to the left. In the former case the lifetime of unstable *b*- state for primary effectons increases, and in the latter case the stabilization of *a* - state of molecular associates (clusters) takes place.

The same is true for convertons equilibrium: $[lb \Leftrightarrow tr]$, reflecting [association $\Leftrightarrow$ dissociation] of water clusters - primary librational effectons and relatively independent water molecules.

The effects of stabilization of clusters can be *reinforced* at such concentration of dissolved molecules (guests), when the mean distances between them (r) coincide with primary librational effectons edges $(\lambda_1, \lambda_2, \lambda_3)$ or integer number of the edge length:

$$r = \frac{11.8}{(C_M)^{1/3}} \; (\text{Å}) = nl \qquad 13.5$$

where $C_M$ is the molar concentration; $l$ is the edge of primary *lb* effecton and $n = 1, 2, 3...$ (the integer number). So, the concentration dependence of the stabilizing action brought by the dissolved substance upon the solvent can be non-monotonic and periodic. The experimental data, confirming our statement have been published indeed (Tereshkevitch *et al.*, 1974).

If the de Broglie wave length of the dissolved particles exceeds the dimensions of primary effectons, then it should increase the degree of mesoscopic Bose condensation. In the opposite case the ordering of liquid structure decreases. To prove this consequence of our approach, it should be noted that the *structure-forming* ions, with a positive hydration ($Li^+$, $Na^+$, $F^-$), as a rule, have a lesser mass, lesser thermal momentum $p = m\mathbf{v}_{gr}$ and consequently, a larger value of $\lambda_B = h/m\mathbf{v}_{gr}$, than those with a negative hydration: $Rb^+$, $Cs^+$, $Br^-$, $I^-$ of the same charges.

As anticipated, the nonpolar atoms, minimally distorting the hydrogen bonds of pure liquid, like *He*, *Ne* ones, have a maximum structuring action on water, stabilizing primary librational effectons.

*In host-guest systems a following situations are possible:*

1) guest molecules stabilize *(a)* -states of host effectons and increases their dimensions. The $(a \Leftrightarrow b)$ equilibrium of the primary librational effectons and $[lb \Leftrightarrow tr]$ equilibrium of convertons becomes shifted leftward decreasing potential energy of a system, corresponding to stabilization effect of guest (solute) particle;

2) guest molecules destabilize *(a)* -states of host effectons. The $(a \Leftrightarrow b)$ and $[lb \Leftrightarrow tr]$ equilibriums of the primary effectons and convertons correspondingly are shifted rightward, inducing general destabilization effect of the system;

3) guest and host molecules form separate individual effectons (mesophase) without separation in two macrophases.

With the increasing concentration of guest in solution of two molecular liquids (for example, water - ethanol) the roles of guest and host may change.

It was shown that conductivity of aqueous solutions of NaCl, containing ions: $Na^+$, $Cl^-$, $H_3O^+$ and $OH^-$ show to vary in a different linear fashion over two ranges of temperature: $273 \leq T \leq 323K$ and $323 \leq T \leq 360 \; K$. The change in slope of the plot shows transition in the character of water-ions interaction near 323 K (Roberts *et al.*1994). In the same work was revealed, that the above aqueous systems exhibited some "memory" of the temperature effects after changing the temperature from low to high and then from high to low values. This phenomena displays itself in a form of hysteresis. Such memory could mean a slow relaxation process, accompanied by redistribution between populations of different excitations in water and their equilibrium constants shifts.

The Hierarchic Model may be useful for elaboration a general theory of solutions. In accordance with our theory, any reorganizations of liquid structure, induced by solute molecules, must be accompanied by correlated changes of following parameters:

1) the velocity of sound;
2) positions of translational and librational bands in oscillatory spectra;

3) density;
4) refraction index;
5) share and bulk viscosities;
6) coefficient of self-diffusion
7) light scattering;
8) heat capacity and thermal conductivity;
9) vapor pressure and surface tension.

There are also a lot of other parameters and properties that may change in solutions as a result of solute-solvent interaction. It could be revealed by computer simulations, using our computer program: "Comprehensive Analyzer of Matter Properties (pCAMP)".

The first four of the above listed parameters are the input ones for pCAMP computer program and must be determined under similar temperature and pressure.

The existing experimental data confirm the consequences of our theory, pointing to interrelation between the above listed parameters. The changes of the velocity of sound in different water - ethanol mixtures as well as that of light and neutron scattering were studied in detail by D'Arrigo and Paparelli (1988a, 1988b, 1989), Benassi *et al.* (1988), D'Arrigo and Teixeira (1990).

Correlations between density, viscosity, the refractive index, and the dielectric constant of mixed molecular liquids at different temperatures were investigated by D'Aprano's group in Rome (D'Aprano *et al.*, 1989, 1990a, 1990b).

The interaction of a solute (guest) molecule with *librational* solvent effectons can be subdivided into *two cases*: when the rotational correlation time of a guest molecule ($\tau_M^{rot}$) is *less (a)* and *more (b)* than the rotational correlation time of *librational effectons* ($\tau_{ef}^{lb}$):

$$a) \quad \tau_M^{rot} < \tau_{ef}^{lb} \qquad \qquad 13.6$$

and

$$b) \quad \tau_M^{rot} > \tau_{ef}^{lb} \qquad \qquad 13.7$$

In accordance with the Stokes-Einstein formula the corresponding rotational correlation times of guest molecule and librational effecton are:

$$\tau_M^{rot} = \frac{V_M}{k} \frac{\eta}{T} \quad \text{and} \quad \tau_{ef}^{lb} = \frac{V_{ef}^{lb}}{k} \frac{\eta}{T} \qquad \qquad 13.8$$

where $\eta$ and $T$ are the share viscosity and absolute temperature of the solvent;

$\tau_M^{rot}$ and $\tau_{ef}^{lb}$ are dependent on the effective volumes of a guest molecule ($V_M$) and the volume of primary librational effecton ($V_{ef}^{lb}$):

$$V_{ef}^{lb} = n_M^{lb}(V_0/N_0) = \frac{9}{4}\pi(\lambda_1\lambda_2\lambda_3) \qquad \qquad 13.9$$

$\lambda_{1,2,3}$ are most probable de Broglie wave length in 3 selected directions; $n_M^{lb}$ - number of molecules in a librational primary effecton, depending on temperature: in water it decreases from 280 till to 3 in the temperature interval $0 - 100^0 C$ (Figure 7a).

When the condition (13.6) is realized, a small and neutral guest molecules affect presumably only the translational effectons. In the second case (13.7) guest macromolecules can change the properties of both types of effectons: translational and librational and shift the equilibrium [$lb \Leftrightarrow tr$] of convertons to the left, stimulating the cluster-formation.

### 13.4.1. Definitions of hydrophilic, hydrophobic and the newly-introduced clusterphilic interactions, based on the Hierarchic Theory

In accordance with our model, the *hydrophilic interaction* is related to shift of the ($a \Leftrightarrow b$) equilibrium of *translational* effectons to the *left*. As far the potential energy of the *(a)* state ($V_a$) is less than that in the *(b)* state ($V_b$), it means that such solvent-solute (host - guest) interaction will decrease the potential and free energy of the solution. Hydrophilic interaction does not need the realization of condition (13.7).

*Hydrophobic interaction* is a consequence of shift of the $(a \Leftrightarrow b)_{lb}$ equilibrium of *librational effectons* to the *right*. Such a shift increases the potential energy of the system and destabilize it. The dimensions of coherent water clusters representing librational effectons under condition (13.7) may even increase. However, the decrease in of entropy ($\Delta S$) in this case is more than that in enthalpy ($\Delta H$) and, consequently, free energy will increase: $\Delta G = \Delta H - T\Delta S > 0$. This is a source of hydrophobic interaction, leading to aggregation of hydrophobic particles in water.

*Clusterphilic interaction* was introduced by this author in 1980 (Kaivarainen, 1980, 1985; 1995; 2001) to describe the cooperative water cluster interaction with nonpolar protein cavities. This idea has got support in the framework of hierarchic concept. The clusterphilic interaction is related to:

a) the leftward shift of $(a \Leftrightarrow b)_{lb}$ equilibrium of primary librational effectons under condition (13.7),

b) the leftward shift of the $[lb \Leftrightarrow tr]$ equilibrium of convertons and

c) increasing of *lb* primary effectons dimensions due to water molecules immobilization at condition 13.1.

The latter effect is a result of decreasing of the rotational correlation time of librational effectons and decreasing of the most probable momentums of water molecules (13.2), related to librations under the effect of the guest particles.

*The clusterphilic interactions can be subdivided to:*

**1**. *Intramolecular interaction* - when water cluster is formed in the "open" states of big interdomain or inter-subunit cavities and

**2**. *Intermolecular clusterphilic interaction*. Intermolecular clusterphilic interactions in associated liquids can be induced by any sufficiently big macromolecules.

*Clusterphilic interactions* can play an important role in the self-organization of biosystems, especially multiglobular allosteric proteins, microtubules and the actin filaments. Cooperative properties of the cytoplasm, formation of thixotropic structures, signal transmission in biopolymers, membranes and distant interactions between macromolecules can be mediated by both types of clusterphilic interactions.

From (4.4) the contributions of primary translational and librational effectons to the total internal energy are:

$$U_{ef}^{tr,lb} = \frac{V_0}{Z} \left[ n_{ef} \left( P_{ef}^a E_{ef}^a + P_{ef}^b E_{ef}^b \right) \right]_{tr,lb} \qquad 13.10$$

The contributions of this type of effectons to total kinetic energy (see 4.36) are:

$$T_{ef}^{tr,lb} = \frac{V_0}{Z} \left[ n_{ef} \frac{\sum_1^3 (E^a)_{1,2,3}^2}{2m(V_{ph}^a)^2} \left( P_{ef}^a + P_{ef}^b \right) \right]_{tr,lb} \qquad 13.10a$$

Subtracting (13.10a) from (13.10), we get the potential energy of primary effectons:

$$V_{ef} = V_{tr} + V_{lb} = (U_{ef} - T_{ef})_{tr} + (U_{ef} - T_{ef})_{lb} \qquad 13.11$$

The driving force of *clusterphilic interaction* and self-organization in colloid systems is mainly the *decreasing* of librational contribution to potential energy $V_{lb}$ in the presence of macromolecules.

*Hydrophilic* interaction, in accordance with our model, is a result of translational contribution $V_{tr}$. decreasing.

On the other hand, *hydrophobic interaction* is a consequence of both contributions: $V_{lb}$ and $V_{tr}$ increasing in the presence of guest molecules as a result of $a \rightleftharpoons b$ equilibrium of primary effectons to the right.

The important for self-organization of aqueous colloid systems *clusterphilic interactions* has been revealed, for example, in of freezing temperature ($T_f$) for buffer solutions of polyethyleneglycol (PEG) on its molecular mass and concentration (Figure 82 in book: Kaivarainen, 1985). The anomalous increasing of $T_f$ in the presence of PEG with molecular mass more than 2000 D and at concentration less than 30 mg/ml, pointing to increasing of water activity, were registered by the cryoscopy method. It may be explained as a result of clusterphilic intermolecular interaction, when the fraction of ice-like water clusters (librational effectons) increases. A big macromolecules and small ions may have the opposite: positive and negative effects on the stability and volume of primary librational effectons.

Macromolecules or polymers with molecular mass less than 2000 do not satisfy the condition (13.7) and can not stimulate the growth of librational effectons. On the other hand, a considerable increase in the concentrations of even big polymers, when the average distance between them (eq. 13.5) comes to be less than the dimensions of a librational water effecton, perturbs clusterphilic interactions and decreases freezing temperature, reducing water activity (Kaivarainen, 1985, Figure82).

*In general case each guest macromolecule has two opposite effects on clusterphilic interactions.* The equilibrium between these tendencies depends on the temperature, viscosity, concentration of a guest macromolecule, its dynamics and water activity.

When solute particles are sufficiently small, they can associate and compose mesoscopic Bose condensate (mBC) if [solute-solute] interaction is more preferable than [solute-solvent] and the conditions (13.1 and 13.2) are fulfilled.

The important confirmation of this consequence of the Hierarchic Theory of complex systems is the observation of compact clusters of ions even in dilute salt solutions. For example, the extended x-ray absorption fine structure data showed that the average distance between $Zn^{2+}$ and $Br^-$ is $2.37 Å$ in 0.089M $ZnBr_2$ aqueous solutions and $2.30 Å$ in 0.05 M solutions (Lagarde *et al.*, 1980). These values are very close to inter ionic separation, observed in the crystalline state of $ZnBr_2$ ($2.40 Å$).

The same results where obtained for $NiBr_2$ ethyl acetate solutions (Sadoc *et al.*, 1981) and aqueous $CuBr_2$ solutions (Fontaine *et al.*,1978). The calculated average distance between ions in the case of the absence of association (see eq.13.5) is much bigger than experimental one, pointing to ionic clusters formation.

Our theory of solutions consider the formation of crystallites (inorganic ionic clusters) even in dilute solutions, as a solute coherent primary effectons ( *i.e.* the solute molecules/ions mesoscopic Bose condensation). The ion-ion interaction in these cases is more favorable process, than ion-water interactions.

The Combined Analyzer of Matter Properties (CAMP), proposed by this author (section 11.11), could be the main tool for such a measurements and study of complicated properties of dilute solutions and pure liquids and solids.

### 13.5. **The multi-fractional nature and properties of interfacial water**, **based on the Hierarchic Theory**

We can present here a classification and description of four interfacial water fraction properties, based on the Hierarchic Theory:

1. The first hydration fraction - primary hydration shell with maximum interaction energy with surface of macromolecule/colloid particle/surface. The structure and dynamics of the 1st fraction can differ strongly from those of bulk water. The thickness of the 1st fraction: (3-10 Å) corresponds to 1-3 solvent molecule;

2. The second hydration fraction - *vicinal water (VW)* is formed by elongated primary librational *(lb)* effectons with saturated hydrogen bonds. It is a result of th bulk water primary *lb* effecton adsorption on the primary hydration shell of the surface of colloid particle or rigid surface. The thickness of this fraction of interfacial water: (30-75 Å) corresponds to 10-25 molecules and is dependent on temperature, dimensions of colloid particles and their surface mobility. The VW can exist in rigid pores of corresponding dimensions;

3. The third fraction of interfacial water - the *surface-stimulated Bose-condensate (SSBC)*, represented by 3D network of filaments, formed by polymerized primary *lb* effectons has a thickness of (50-300 Å). It is a next hierarchical level of interfacial water self-organization on the surface of second vicinal fraction (VW). The time of gradual polymerization of primary librational effectons by head-to-tail principle and formation of 3D net of librational filaments - polyeffectons, is much longer than that of VW and it is more sensitive to temperature, pressure and any other external perturbations. The second and third fractions of interfacial water can play an important role in biological cells activity;

4. The biggest and most fragile forth fraction of interfacial water, named *"super-radiation orchestrated water (SOW)"*, is a result of slow orchestration of bulk primary effectons in the volume of primary (electromagnetic) lb deformons. The primary deformons appears as a result of three standing IR photons (lb) interference. Corresponding IR photons are super-radiated (see Introduction, item 5) by the enlarged lb effectons of vicinal water and 3D hydration fraction: *surface-stimulated Bose-condensate (SSBC)*. The linear dimension of primary IR deformons is about half of librational IR photons, *i.e.* 5 microns ($5 \times 10^4 Å$). This *"super-radiation orchestrated water (SOW)"* fraction easily can be destroyed not only by temperature increasing, ultrasound and Brownian movement, but even by mechanical shaking. The time of spontaneous reassembly of this fraction after destruction has an order of hours and is dependent strongly on temperature, viscosity and dimensions of colloid particles. The processes of self-organization of third (SSBC) and forth (SOW) fractions are interrelated.

*13.5.1. Consequences and predictions of new model of interfacial solvent structure*

In accordance to generally accepted and experimentally proved models of hydration of macromolecules and colloid particles, we assume that the first 2-3 layers of strongly bound water molecules of the *1st hydration fraction*, serves like the intermediate shell, neutralizing or "buffering" the specific chemical properties of surface (charged, polar, nonpolar, etc.). Such strongly bound water can remain partially untouched even after strong dehydration of samples in vacuum.

This first fraction of interfacial water serves like a matrix for *2nd fraction - the vicinal water (VW)* shell formation. The therm *'paradoxical effect'*, introduced by Drost-Hansen (1985) means that the properties of vicinal water are independent on specific chemical structure of the surface like quartz plates, mineral grains, membranes or large macromolecules (Clegg and Drost-Hansen, 1991). This can be a result of "buffering" effect of the 1st fraction of hydration shell.

The *vicinal water (VW)* can be defined as a water with structure, modified in the volume of pores, by curved and plate interfaces and interaction with strongly 'bound' water of 1st fraction.

For discussion of Vicinal water (VW) properties we proceed from the statement that if the rotational correlation time of hydrated macromolecule is bigger, than that of primary librational effecton of bulk water; $\tau_M > \tau_{eff}^{lb}$ corresponding to condition (13.7 -13.8), these effectons should have a tendency to "condensate" on 1st fraction of hydration shell of large colloid particles. It is a condition of their life-time and stability increasing, because the resulting momentum of the primary effectons is close to zero, in accordance to our model.

The decreasing of most probable librational thermal momentums of water molecules, especially in direction, normal to the surface of macromolecule or colloid particle, should lead to increasing of corresponding edges of adsorbed primary librational effectons (vicinal water) as compared to the bulk ones:

$$[\lambda = h/m\mathbf{v}_{gr}^{\perp}]_{lb}^{vic} > [\lambda = h/m\mathbf{v}_{gr}^{\perp}]_{lb}^{bulk} \qquad 13.12$$

This turns the cube-like shape of effectons, pertinent to the bulk water to shape of elongated parallelepiped..

Consequently, the volume of these *vicinal librational (lb) primary* effectons is bigger, than that of bulk water effectons. As far the stability and life-time of these vicinal *lb* effectons are increased, it means the increasing of their concentration as well.

As far we assume in our model of interfacial water, that VW is a result of "condensation" of librational effectons on primary hydration shell and their elongation in direction, normal to surface, we can make some predictions, related to properties of this 2nd water fraction:

1. The thickness of VW can be about 30-75 Å, depending on properties of surface (geometry, polarity), temperature, pressure and presence the perturbing solvent structure agents;

2. This water should differ from the bulk water by number of physical parameters, such as:

a) lower density;
b) bigger heat capacity;
c) bigger sound velocity
d) bigger viscosity;
e) smaller dielectric relaxation frequency, etc.

These differences should be enhanced in a course of third fraction of interfacial water formation: surface-stimulated Bose-condensate (SSBC) as far the concentration of primary librational effectons in this fraction is also bigger than that in bulk water. The time, necessary for SSBC three-dimensional structure self-organization can have an order of minutes or hours, depending on temperature, geometry of surface and average distance between adjacent surfaces.

From Figure 4b we can see that the linear dimension of primary librational effecton of bulk water at $25^0 C$ is about $[\lambda]_{lb}^{bu} \sim 15 Å$. The lower mobility of water molecules of vicinal water is confirmed directly by almost 10-times difference of dielectric relaxation frequency ($2 \times 10^9$ Hz) as respect to bulk one ($19 \times 10^9 Hz$) (see Clegg and Drost-Hansen, 1991). The consequence of less mobility and most probable momentum of water molecules should be the increasing of most probable de Broglie wave length and dimensions of primary effectons. The enhancement of *lb* primary effecton edge should be more pronounced in the direction, normal to the interface surface. In turn, such elongation of coherent cluster can be resulted in increasing the intensity of librational IR photons super-radiation.

The increasing of temperature should lead to decreasing the vicinal librational effectons dimensions and their destabilization.

The dimensions of primary translational effectons of water is much less than of librational ones (see Figure 4a). The contribution of translational effectons in vicinal effects is correspondingly much smaller, than that of librational one.

Our model predicts that not only dimensions but also the concentrations of primary librational effectons should increase near the rigid surfaces.

In accordance to our model, the Drost-Hansen thermal anomalies of vicinal water behavior near $15^0$, $30^0$, $45^0$ and $60^0$ has the same explanation as presented in comments to Figure 4a for bulk water. Because the dimensions, stability and concentration of vicinal librational effectons are bigger than that of bulk water effectons the temperature of anomalies for these two water fraction can differ also.

As far the positions of disjoining pressure sharp maxima of water between quartz plates did not shift markedly when their separation change from 100 Å to 500 Å it points that the temperatures of D-H anomalies are related to bulk librational effectons and fragile surface-stimulated Bose-condensate (SSBC - 3D fraction) and super-radiation-orchestrated water (SOW - 4th fraction). The same is true for viscosity measurements of water between plates with separation: 300-900 Å.

The possible explanation of Drost-Hansen temperature values stability, is that the vicinal water layer (50 -70Å) has the bigger dimensions of primary librational effectons edge, than that of bulk water, presumably only in the direction, normal to the surface of the interface.

The elongated structure of primary librational effectons, composing SSBC, should increase all the effects, related to coherent super-radiation of lb IR photons in directions of the longest edge of the effectons (see Introduction). These effects include the formation of most fragile *"super-radiation-orchestrated water (SOW)"*.

The 3D structure of SSBC and SOW could be easily destroyed by mechanical perturbation of colloid system or heating. The relaxation time of "regeneration" or self-organization of these water fractions can take an hours.

The sharp conditions of maximum stability of librational primary effectons at certain temperatures, corresponding to integer number of the molecules in the edge of librational effectons (see comments to Figure 4a):

$$\kappa = 6, 5, 4, 3, 2 \qquad 13.13$$

These conditions are responsible for periodic deviations of temperature for many parameters of water from monotonic ones.

The vicinal water (VW) and surface-stimulated Bose-condensation of water near biological membranes and filaments (microtubules, actin polymers) can play an important regulation role in cells and their compartments dynamics and function. Its highly cooperative properties and thermal sensitivity near Drost-Hansen temperatures can be used effectively in complicated processes, related to cells proliferation, differentiation and migration.

The increasing of water activity, as a result of VW and SSBC disassembly and decreasing in a course of it 3D structure self-organization, can affect strongly the dynamic equilibrium [association ⇌ dissociation] of oligomeric proteins, their allosteric properties and passive osmotic processes in cells, their swelling or shrinking.

### *13.5.2 Comparison of experimental data with theoretical predictions of the interfacial water model*

It will be shown below, that the Hierarchic Model of interfacial water explains the comprehensive and convincing experimental data, available on subject.

The following experiments, illustrating the difference between the second fraction of hydration shell- vicinal water (VW) and third, surface-stimulated Bose-condensate (SSBC) fraction of interfacial and bulk water will be discussed:

1. The lower density of vicinal water near plates and in pores (~0.970g/cm$^3$) as compared to bulk one (~1.000 g/cm) (Etzler and Fagundus, 1983; Low, 1979; Clegg 1985);
2. Different selectivity of vicinal water in pores to structure-breaking and structure-making inorganic ions;
3. Volume contraction on sedimentation of particles dispersed in water (Braun *et al.* 1987);
4. Higher heat capacity of vicinal water as compared to bulk water (Etzler 1988);
5. Higher velocity of sound in the interfacial water;
6. Higher ultrasound absorption in the interfacial water;
7. Higher viscosity of interfacial water (Peshel and Adlfinger 1971) and its dependence on shearing rate;
8. Sharp decreasing of the effective radius of dilute solution of polysterone spheres, in a course of temperature increasing.

In accordance to our model, the water molecules composing primary librational effectons (see Figure 4a) are four-coordinated like in ideal ice structure with lowest density. In contrast to that, the water in the volumes of translational effectons, [lb/tr] convertons and lb and tr macro-effectons and super-effectons has the nonsaturated hydrogen bonds and a higher density. The compressibility of primary [lb] effectons should be lower and the velocity of sound - higher than that of bulk water. It is confirmed by results of Teixeira et. al. (1985), obtained by coherent- inelastic- neutron scattering.

They point on existence in heavy water at 25 $^0$C the solid-like collective excitations with bigger sound velocity than in bulk water. These experimental data can be considered as a direct confirmation of our primary librational effectons existence.

As far the fraction of water involved in primary librational effectons in vicinal water is much higher than in the bulk one, this explains the result [1] at the list above. The biggest decreasing of density occur in pores, containing enlarged primary librational effectons, due to stronger water molecules immobilization.

Different selectivity of vicinal water in pores of silica gel (result [2]) to structure-breaking and structure-making inorganic ions (Wiggins 1971, 1973), leading to higher concentration of the former (like $K^+$) as respect to latter (like $Na^+$) ones in pores was revealed. It is in total accordance with our model as far for penetration into the pore the ion have at first to break the ordered structure of enlarged librational effectons in the volume of pore. Such kind of Na/K selectivity can be of great importance in the passive transport of ions throw the pores of biological membranes.

Result [3] of volume contraction of suspension of 5-$\mu m$ silica particles in a course of their sedimentation - is a consequence of mechanical perturbation of cooperative and unstable 3D fraction: surface - stimulated Bose condensation (SSBC), its partial 'melting' and increasing of water fraction with nonsaturated hydrogen bonds and higher density.

The available experimental data of the vicinal water thickness evaluate it as about 50-70 Å (Drost-Hansen, 1985). In totally or partly closed volumes like in silica pores the vicinal effect must be bigger than near the plain surface. This explains the maximum heat capacity of water at $25^0$ (result [4] from the list above) in silica pores with radii near 70 Å (Etzler and White, 1988).

As far the cooperative properties of 2nd and 3d fractions of interfacial water, corresponding to vicinal water and surface - stimulated Bose condensation are higher than that of bulk water, it explains the bigger heat capacity of both of these fractions.

The higher the velocity of sound in the VW and SSBC fractions as compared to bulk water (result [5]) is a consequence of higher concentration of primary librational effectons with low compressibility due to saturated H-bonds.

Higher absorption of ultrasound by interfacial water (result [6]) can be a consequence of dissipation processes, accompanied the destruction of water fraction, corresponding to SSBC by ultrasound.

The higher viscosity of vicinal water (result [7]) is a result of higher activation energy of librational macro-deformons excitation $(\varepsilon_D^M)_{lb}$ in a more stable system of vicinal polyclustrons (see the Hierarchic Theory of viscosity in section 11.6 of this book).

Sharp decreasing of the effective dimensions of dilute solution of polysterone spheres (0.1%), in a course of temperature increasing (result [8]) - is a consequence of cooperative destruction (melting) of 3d fraction of interfacial water (SSBC).

The corresponding transition occur at 30-34 $^0$C, as registered by the photon correlation spectroscopy and is accompanied by the effective Stocks radius of PS spheres decreasing onto 300 Å.

*The "regeneration time" of this process is about 20 hours. It may include both: time of SAPS self-organization and self-organization of the less stable forth interfacial fraction: "Water, orchestrated in the volume of IR primary deformons".*

The frequent non-reproducibility of results, related to properties of interfacial water, including Drost-Hansen temperature anomalies, can be resulted from different methods of samples preparation and experimental conditions.

For example, if samples where boiling or strongly heating just before the experiment, the 3d and 4th fractions of interfacial water can not be observable. The same negative result is anticipated if the colloid system in the process of measurement is under stirring or intensive ultrasound radiation influence.

### 13.5.3. *The predictions, related to both the third fraction (SSBC) and the forth fraction: super-radiation-orchestrated water (SOW) of new interfacial model*

1. Gradual increasing of pH of distilled water in a course of these fractions formation near nonpolar surface - due to enhancement of probability of super-deformons excitation. Corresponding increasing of concentration of cavitational fluctuations are accompanied by dissociation of water molecules:

$$H_2O \rightleftharpoons H^+ + HO^-$$

and the protons concentration elevation;

2. Increasing the UV and visible photons spontaneous emission near nonpolar surface as a result of increasing of the frequency of water molecules recombination:

$$H^+ + HO^- \rightarrow H_2O + photons$$

These experiments should be performed in dark box, using sensitive photon counter or photo-film.

The both predicted effects should be enhanced in system, containing parallel nonpolar multi-layers, with distance ($l$) between them, corresponding to conditions of librational IR photons standing waves formation: $l = 5, 10, 15, 20$ microns.

3. More fast and ordered spatial self-organization of macromolecules, like described in the next section is anticipated also in the volume of forth fraction of interfacial water.
The dynamics of such process can be registered by optical confocal or tunneling microscope.

4. Our model predicts also that the external weak coherent electromagnetic field, generated by IR laser, like the internal one, also can stimulate a process of self-organization in colloid systems.

*The vicinal water in combination with osmotic processes could be responsible for coordinated intra-cell spatial and dynamic reorganizations (see section 8 of this Chapter).*

### 13.6. **Distant solvent-mediated interaction between macromolecules**

The most of macromolecules, including proteins, can exist in dynamic equilibrium between two conformers (A and B) with different hydration ($n_{H_2O}$) and flexibility:

$$A + n_{H_2O} \Leftrightarrow B \qquad \text{13.11 a}$$

Usually the correlation time of more hydrated B - conformer ($\tau_B$) and its *effective* volume are less than that of more rigid A-conformer ($\tau_A$):

$$\tau_{A,B} = \frac{V_{A,B}}{k}(\eta/T) \qquad \text{13.11b}$$

$$\tau_A > \tau_B \quad \text{and} \quad V_A > V_B$$

This means that flexibility of more hydrated B-conformer, determined by large-scale dynamics, is higher, than that of A-conformer.

For such case the change of the bulk water activity ($a_{H_2O}$) in solution by addition of other macromolecules or inorganic ions induce the change of the equilibrium constant: $K_{A \Leftrightarrow B} = (K_{B \Leftrightarrow A})^{-1}$ and the dynamic behavior of macromolecules (Kaivarainen, 1985, 1995):

$$\Delta \ln K_{B \Leftrightarrow A} = n_{H_2O} \Delta \ln a_{H_2O} \qquad \text{13.12}$$

The proposed by this author method of separate calculation of two correlation times of spin-label (see section 3.6 in book by Kaivarainen, 1985) characterize:

a) the frequency of spin-label rotation as respect to protein surface ($v_R \sim \tau_R$), depending on microviscosity of macromolecule matrix and

b) the resulting frequency of rotation of macromolecule as a whole, *i.e.* the averaged ($v_{A,B} \sim \tau_{A,B}$), where $\tau_{A,B}$ is the resulting correlation time of A and B - conformers, depending on $A \Leftrightarrow B$ equilibrium shift.

In mixed systems: [PEG + spin-labeled antibody] the dependence of large-scale (LS) dynamics of antibody on the molecular mass of polyethyleneglycol (PEG) is similar to dependence of water activity and freezing temperature ($T_f$) of PEG solution, discussed above (Kaivarainen, 1985, Figure 82).

The presence of PEG with mass and concentration increasing ($a_{H_2O}$) and ($T_f$) stimulate the LS-dynamics of proteins decreasing their effective volume $V$ and correlation time $\tau_M$ in accordance to eqs.(13.12 and 13.11b).

If $\Delta T_f = T_f^0 - T_f$ is the difference between the freezing point of a solvent ($T_f^0$) and solution ($T_f$), then the relation between water activity in solution and $\Delta T_f$ is given by known relation:

$$\ln a_{H_2O} = -\left[\frac{\Delta H}{R(T_f^0)^2}\right]\Delta T_f \qquad 13.13$$

where $\Delta H$ is the enthalpy of solvent melting; R is the gas constant.

In our experiments with polymer solutions the 0.1 M phosphate buffer $pH\ 7.3 + 0.3M$ NaCl was used as a solvent (Kaivarainen, 1985, Figure 82).

One can see from (13.13) that the negative values of $\Delta T_f$ in the presence of certain polymers means the increasing of water activity in three component [water - ions - polymer] system ($\Delta \ln a_{H_2O} > 0$). In turn, it follows from (13.12) that the increasing of $a_{H_2O}$ shifts the $[A \Leftrightarrow B]$ equilibrium of proteins in solutions to the right. Consequently, the flexibility of the proteins will increase as far $\tau_A > \tau_B$.

The correlation between freezing temperature $T_f$, the water activity ($a_{H_2O}$) and immunoglobulin flexibility ($\tau_M$), corresponding to (13.11 - 13.13) was confirmed in our experiments (Kaivarainen, 1985, Table 13).

It was shown that *protein-protein* distant interaction depends on their LS dynamics and conformational changes induced by ligand binding or temperature (Kaivarainen, 1985).

The temperature of correlation time of spin labeled human serum albumin (HSA-SL), characterizing the rigidity of macromolecule in regular solvent and in presence of 3% $D_2O$ is presented on Figure 42a.

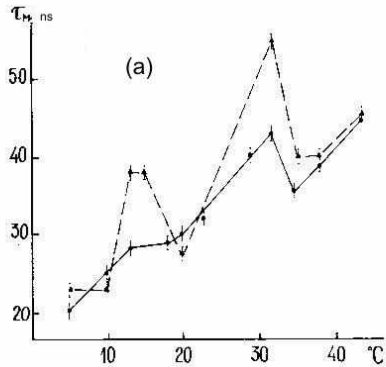

**Figure 42a.** The temperature dependence of resulting correlation time ($\tau_M$) of spin-labeled human serum albumin (HSA-SL) and similar dependence in presence of 3% $D_2O$ (dotted line). Concentration of HSA was 25 mg/cm$^3$ in 0.01 M phosphate buffer (pH 7.3) + 0.15 M NaCl.

Figure 42b demonstrates a distant, solvent-mediated interaction between human serum albumin (HSA) and spin labeled hemoglobin (Hb-SL). We can see, that in presence of HSA its temperature - induced changes of flexibility/rigidity (correlation time, Figure 42a) influence the flexibility of Hb-SL. The peak of correlation time of HSA around 15 $^0$C, stimulated by presence of 3% $D_2O$ also induce a corresponding transition in large-scale dynamics of Hb-SL.

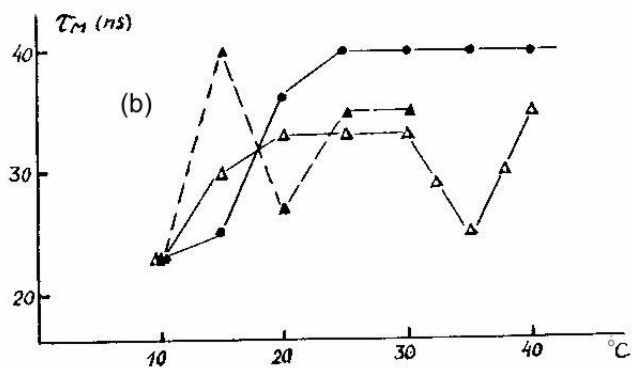

**Figure 42b**. The temperature dependence of resulting correlation time ($\tau_M$) of:
a) spin-labeled oxyhemoglobin (Hb-SL) - black dots;
b) similar dependence in presence of human serum albumin (HSA);
c) similar dependence in presence of human serum albumin (HSA) and 3% $D_2O$ (dotted line).
The concentration of oxyhemoglobin was 20 mg/cm$^3$, the concentration of HSA was 10 mg/cm$^3$ in solvent: 0.01 M phosphate buffer (pH 7.3) + 0.15 M NaCl.

Between the clusterphilic interaction, discussed in previous section, and protein flexibility or rigidity, characterized by $\tau_B$, the positive correlation exists. It means that the increasing of dimensions *of librational bulk water effectons is accompanied by enhancement of the water clusters dimensions in protein cavities, following, in turn, by shift of A ⇔ B* equilibrium of the cavities to the right - to the more flexible conformer ($\tau_B < \tau_A$).

Our interpretation is confirmed by fact that lowering the temperature, increasing the dimensions of bulk librational effectons has increases the flexibility of protein.

At *low concentration of macromolecules* ($C_M$), when the average distance (r) between them (eq. 13.5) is much bigger than dimensions of primary librational water effecton ($r \gg \lambda_{lb}$) the large-scale [$A \Leftrightarrow B$] pulsations of proteins, accompanied by exiting of acoustic waves in solvent, can enhance the *activity of water*.

Such influence of pulsing proteins on solvent can be responsible for the distant solvent- mediated interaction between macromolecules at low concentration (Kaivarainen, 1985, 1995).

Acoustic momentums in protein solutions are the result of the jump-way $B \to A$ transition of interdomain or inter-subunit cavities with characteristic time about $10^{-10}$ sec. This rapid transition follows the cavitational fluctuation of a water cluster formed by 30 - 50 water molecules in space between domains and subunits of oligomeric proteins. This fluctuation is a result of conversion of librational primary effecton to translational one.

In very concentrated solutions of macromolecules, when the distance between macromolecules starts to be less than linear dimension of primary librational effecton of water: $r \leq \lambda_{lb}$, the trivial aggregation process begins to dominate. It is a consequence of decreasing of water activity in solutions.

### 13.6.1. *Possible mechanism of water activity increasing in solutions of macromolecules*

Let us analyze in more detail the new effect, discovered in our work, - the increasing of water activity ($a_{H_2O}$) in presence of macromolecules in three component [water - salt - macromolecules] system. The Gibbs-Duhem law for this case can be presented as (Kaivarainen, 1988):

$$X_{H_2O}\Delta \ln a_{H_2O} + X_M \frac{\Delta \mu_M}{RT} + X_i \Delta \ln a_i = 0 \qquad 13.14$$

where $X_{H_2O}$, $X_M$, $X_i$ are the molar fractions of water, macromolecules and ions in the system;

$$a_j = y_j X_j = \exp\left(-\frac{\mu_0 - \mu_j}{RT}\right) = \exp\left(-\frac{\Delta \mu_j}{RT}\right) \qquad 13.15$$

is the activity of each component related to its molar fraction ($X_j$) and coefficients of activity ($y_j$);

$$\mu_M = f_B G_B + f_A G_A \simeq f_B(G_B - G_A) + G_A \qquad 13.16$$

$$\Delta \mu_M \simeq (G_B - G_A)\Delta f_B \qquad 13.17$$

is the mean chemical potential ($\mu_M$) of a macromolecule (protein), pulsing between A and B conformers with corresponding partial free energies $G_A$ and $G_B$ and its change ($\Delta \mu_M$), as a result of $A \rightleftharpoons B$ equilibrium shift, taking into account that $f_B + f_A \cong 1$ and $\Delta f_B = -\Delta f_A$ and

$$\Delta \ln a_i = (\Delta a_i/a_i) \simeq -\Delta \kappa_i \qquad 13.18$$

where the fraction of thermodynamically excluded ions (for example, due to ionic pair formation):

$$\kappa_i = 1 - y_i \qquad 13.19$$

One can see from (13.15) that when $a_{H_2O} < 1$, it means that

$$\mu_{H_2O}^0 > \mu_{H_2O}^S = H_{H_2O}^S - TS_{H_2O}^S \qquad 13.20$$

It follows from (13.20) that the decreasing of water entropy ($\bar{S}$) in solution related to hydrophobic and clusterphilic interactions may lead to increasing of $\mu_{H_2O}^S$ and water activity.

It is easy to see from (13.14) that the elevation the concentration and $X_M$ of macromolecules in a system at constant temperature and $\Delta \bar{\mu}_M$ may induce a rise in water activity ($a_{H_2O}$) only if the activity of ions ($a_i = y_i X_i$) is decreased. The latter could happen due to increasing of fraction of thermodynamically excluded ions ($\kappa$) (eqs. 13.18 and 13.19).

There are *two processes* which may lead to increasing the probability of ionic pair formation and fraction $\kappa$ elevation.

The *first* one is the forcing out of the ions from the ice-like structure of enlarged librational effectons, stimulated by the presence relatively high concentration of macromolecules and strong interfacial effects (2d and 3d fractions of hydration shell (section 13.5). This effect of excluded volume increases the effective concentration of inorganic ions and probability of association, accompanied by their dehydration.

The *second* process dominates at the low concentration of $[A \Leftrightarrow B]$ pulsing macromolecules, when the thixotropic structure fail to form ($r = 11.8/C_M^{1/3} \gg \lambda_{lb}$). The acoustic waves in solvent, generated by pulsing proteins stimulate the fluctuation of ion concentration (Kaivarainen, 1988) also increasing the probability of neutral ionic pairs formation and their dehydration.

In accordance to Gibbs-Duhem law (13.14), the decreasing of ionic activity: $\Delta \ln a_i < 0$, should be accompanied by increasing of water activity: $\Delta \ln a_{H_2O} > 0$ and increasing of $\Delta \mu_M$, meaning shift of $[A \Leftrightarrow B]$ equilibrium to the right - toward more hydrated and flexible B-conformer of proteins, when $\Delta f_B > 0$ (see eq.13.17).

### 13.7. **Spatial self-organization (thixotropic structure formation) in water-macromolecular systems**

The new type of self-organization in aqueous solutions of biopolymers was revealed in Italy

(Giordano *et al.*, 1981). The results obtained from viscosity, acoustic and light-scattering measurements showed the existence of long-range structures that exhibit a *thixotropic behavior*. This was shown for solutions of lysozyme, bovine serum albumin (BSA), hemoglobin and DNA. Ordered structure builds up gradually in the course of time to become fully developed after more than 10-15 hours.

When a sample is mechanically shaken this type of self-organization is destroyed. The "preferred distance" between macromolecules in such an ordered system was about $L \simeq 50 \mathring{A}$ as revealed by small angle neutron scattering (Giordano *et al.*, 1991). It is important that this distance can be much less than the average statistical distance between proteins at low molar concentrations ($C_\mathbf{M}$) (see eq.13.5).

*This fact points to attraction force between macromolecules.* In accordance with our described in section (13.4.1), this attraction is a result of intermolecular clusterphilic interaction. The water clusters serves as a 'glue' between macromolecules, stabilizing their interaction at certain separation.

It has been shown experimentally that hypersonic velocity in the ordered thixotropic structures of 10% lysozyme solutions is about 2500 *m/s*, *i.e.* 60% higher than that in pure water (1500 *m/s*). This result is in-line with data of quasi-elastic and elastic neutron scattering in 10% lysozyme aqueous system. These data show, that at $20^0 C$ the dynamics of the "bound" water in thixotropic system is similar to dynamics of pure water at $16^0 C$, *i.e.* $(3-4)^0 C$ below the actual temperature (Giordano *et al.*, 1991).

The experimental evidence of heat capacity increasing in lysozyme solutions during 10-15 hours of self-organization was obtained by adiabatic microcalorimetry (Bertolini *et al.*, 1992). The character of this process is practically independent on pH and disappears only at very low concentration of protein ($< 0.2\%$), when the average distance between macromolecules becomes too big for realization of inter-macromolecule clusterphilic interaction. Increasing the temperature above $40^0$ also inhibits thixotropic-type of self - organization in water - macromolecular systems. Our theory explains the latter as a consequence of decreasing of water clusters dimensions and stability with temperature (Figure7).

In series of experiments using artificial polymers - polyethyleneglycol (PEG) with different molecular mass it was shown, that self-organization (SO) in 0.1 M phosphate buffer ($pH 7.3, \ 25^0$) exists in the presence of PEG with molecular mass (MM) of 20.000, 10.000 and 8.000 daltons, but disappears at MM of 2.000 and lower (Salvetti *et al.*, 1992).

These data are in good agreement with the influence of PEG on the freezing temperature and intramolecular interaction discussed in section 13.4.

Independently of Italian group, similar ordering process were observed in Japan (Ise and Okubo, 1980) for aqueous solutions of macroions. The process of "compactization" or "catheterization" of solute (guest) molecules in one volumes of solution or colloid system, should lead to emergency of voids in another ones. Such a phenomena were revealed by means of confocal laser scanning microscope, ultramicroscope and video image analyzer (Ito *et al.*,1994; Yoshida *et al.*,1991; Ise *et al.*,1990). Inhomogeneity of guest particles distribution were revealed in different ionic systems, containing ionic polymers or macroions, like sodium polyacrilate, the colloid particles, like polystyrene latex ($N 300, \ 1.3 \mu C/cm^2$) and Langmuir-Blodgett films. The time evolution of the numbers of different clusters from such particles were followed during few hours, using above listed technics.

The colloid crystal growth at 25 $^0$C in $H_2O$ and $(H_2O - D_2O)$ systems where subdivided on four stages (Yoshida et all.,1991):
 - in the first stage the particles were diffusing freely.
 - in second stage the local concentration of the particles took place.
 - in third stage clusters from 3-10 particles were formed.
 - in the last fourth stage the smaller clusters turns to the bigger ones and the macroscopically well ordered structures were formed. Simultaneously the huge voids as large as $50 - 150 \mu m$ were observed.

Such an ordering can be immediately destroyed by mechanical shaking or by adding of inorganic salts (NaCl), even in such relatively small concentrations as $10^{-4}$ mol/dm$^3$.

Authors conclude that the repulsion between macroions, as only one assumption is not enough for explanation of the phenomena observed. Their model, considers a *short-range repulsive* interaction and *long-range attraction* between particles of similar charge. The authors try to explain the attraction between similar charged macroions by presence of small inorganic counter-ions. Authors believe that short range repulsion can be overwhelmed by attraction. However, such a simple model does not explain "clusterization" of electrically neutral guest macromolecules and acceleration of this process due to decreasing the concentration of inorganic salts in presence of ionic sorbents (macroions).

*Despite a large amount of different experimental data, the mechanism of spontaneous type of self-organization in colloid systems remains unknown.* It is evident, however, that it can not be attributed to trivial aggregation.

Our explanation of distant attraction between macromolecules, mediated by *clusterphilic interaction* and polymerization of *clustrons* (see Table 2) in the volume of primary electromagnetic deformons, is more adequate description of self-organization phenomena in aqueous colloid systems, than the counter-ion hypothesis.

Just *clusterphilic interactions* determines the attraction (glue effect) between macromolecules or colloid particles even at low concentration. They are responsible for number of vicinal water effects described below.

Our computer calculations of water properties show that the volume of primary librational effecton of water at $25^0$ includes about 100 water molecules (Figure 7a).

In accordance with condition (13.7), the macromolecules with volume:
$V_M > 100\mathbf{v}_{H_2O} = 100\ (V_0/N_0)$, (where $\mathbf{v}_{H_2O} = V_0(t)/N_0$ is a volume, occupied by one water molecule) can decrease the librational mobility of $H_2O$ and their momentums. Consequently, such macromolecules increases the dimensions of the librational effectons of water (see eqs. 13.2 and 13.9). Obviously, the direct correlation between the effective Stokes radius of a macromolecule - colloid particle (its molecular mass) and the process of self-organization in colloid systems must exist, if our approach is correct.

The mass of a lysozyme (Lys) is about 13.000 D, and available experimental data (Giordano *et al.*, 1991) show that the mobility of water in a hydration shell composing thixotropic system at $25^0$ is about 3 times less than that of pure water. This mobility is directly related to the value of most probable group velocity and momentum of librating $H_2O$ ($p = m\mathbf{v}_{gr}$). It means a three-fold increase in the dimensions of librational effectons in the presence of lysozyme (see eq. 13.1):

$$[\lambda_{H_2O} \approx 15\text{Å}] \rightarrow [L_{(H_2O)} \approx 45\text{Å}] \qquad 13.20a$$

This value is quite close to the experimental *preferred distance* (50 $\overset{o}{A}$) between proteins after self-organization (SO). Because the shape of the effectons can be approximated by parallelepiped or cube, we suggest that each of its 6 sides can be bordered and stabilized by one macromolecule (Table 2). We termed the corresponding type of quasi-particle, representing: *librational effecton, approximated by cube + 6 interfaced macromolecules, as "clustron"*.

The enlarged librational effectons serve as a "glue", promoting interaction between bordering of this effecton/cluster macromolecules or colloid particles. The probability of librational effectons disassembly to translational ones, *i.e.* [*lb/tr*] convertons excitation (see Introduction) - decreases composing *clustrons*.

The cooperative water clusters in the volume of clustrons are very sensitive to perturbation by temperature or by ion-dipole interactions. When the colloid particles have their own charged groups on the surface, very small addition of inorganic salts can influence the clustrons formation.

Our model of *clustrons* can answer the following questions:

1. Why the polymers with a molecular mass less than 2000 $D$ do not stimulate self-organization in aqueous systems?

2. Why the self-organization process (thixotropic structure formation) in colloid systems is absent at sufficiently high temperature ($> 40^0$) ?

3. What causes experimentally confirmed repulsive hydration force between colloid particles ?

The answer to the 1st question is clear from condition (13.7) and eqs.(13.8). Small polymers can not stimulate the growth of librational effectons and clustrons formation due to their own high mobility.

The response to the 2nd question is that a decrease in the dimensions of librational effectons with temperature (Figure 7) leads to deterioration of their cooperative properties and stability. The probability of clustrons formation drops and thixotropic structures can not develop.

The reason of the repulsion (3d question), like the attraction between colloid particles can be explained by introduced in our work *clusterphilic* interaction. The increasing of external pressure should shift the $(a \Leftrightarrow b)_{lb}$ equilibrium of librational water effectons to the left, like temperature decreasing. This means the enhancement of librational water effecton (and clustron) stability and dimensions. For this reason a *repulsive or disjoining hydration force* arises in different colloid systems. The swelling of clays against imposed pressure also is a consequence of clusterphilic interactions enhancement, *i.e.* enlargement of water librational effectons due to $H_2O$ immobilization on the surface of big enough colloid particles.

In accordance with our model, the effectons and their derivatives, *clustrons*, are responsible for self-organization in colloid systems on *mesoscopic* level (Stage II in Table 2).

The *macroscopic* level of self-organization (polyclustrons), responsible for the increase of viscosity and long-distance interaction between clustrons, originates due to interaction between the electric and magnetic dipoles of clustrons (Stage III, Table 2). Due to coherence of thermal vibrations of $H_2O$ molecules in the volume of clustrons, the *dipole moment of clustron* is proportional to the number of $H_2O$ in its volume.

It means, that the dipoles moments of individual water molecules are the additive parameters in clustron volume.

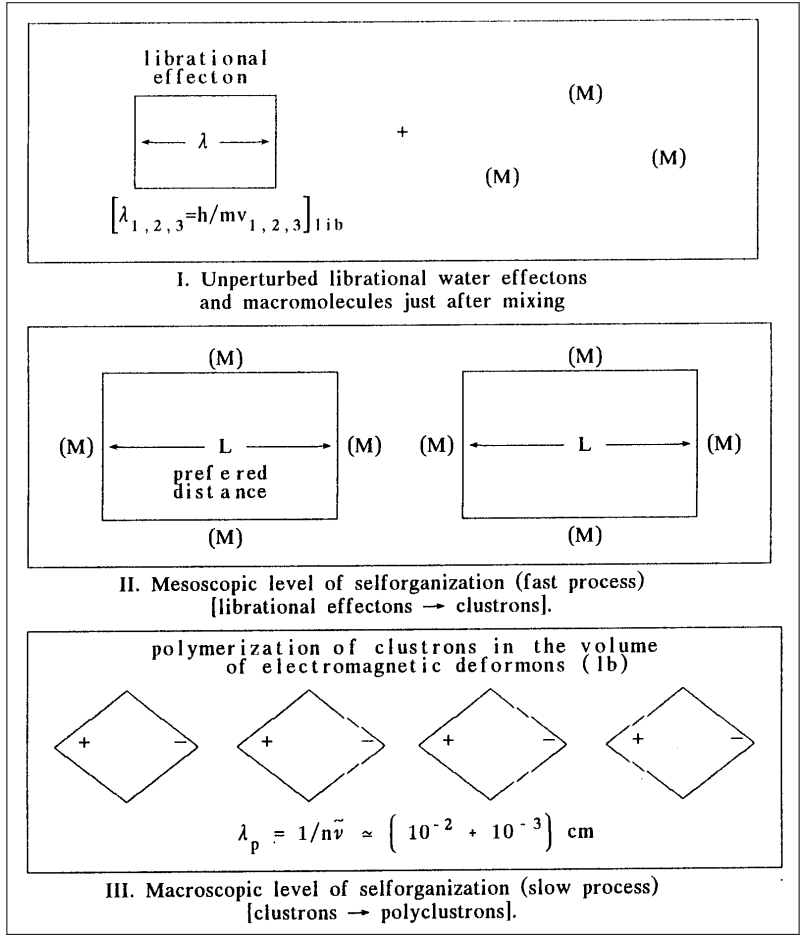

**Table 2.** Schematic representation of three stages ($I \to II \to III$) of gradual spatial self-organization in aqueous solutions of macromolecules ($M$), bordering each of 6 sides of primary librational effectons of water (clustrons - stage II). It is supposed, that the formation of *polyclustrons* (Stage III), representing space-time correlated large group of clustrons, as a result of their polymerization, can be stimulated in the volume of primary *tr and lb* electromagnetic deformons (see Introduction*).*

The *clustrons* (stage II) are complexes of enlarged primary librational effectons, bordered by macromolecules from each of their 6 side. The increasing of librational effecton dimensions ($\lambda \to L$) is related to decreasing of water molecule mobility (most probable librational momentum) due to interfacial effects.

The *polyclustrons* (stage III) are space and time correlated systems of a large group of clustrons in the volume of electromagnetic deformons. Their linear dimensions: $\lambda_{tr,lb} = n\tilde{v}_{tr,lb} = (35\mu - 10\mu)$ correspond to translational and librational wave numbers:

$$\tilde{v}_{tr} \simeq 200 cm^{-1} \quad \text{and} \quad \tilde{v}_{lb} \simeq 700 cm^{-1}$$

and the refraction index of water $n \simeq 1.33$.

We suppose that for optimal process of self-organization, the molar concentration of macromolecules ($C_M$) and corresponding most probable separation between macromolecules ($r$) must satisfy the condition:

$$10L > [r = \frac{11.8}{C_M^{-1/3}} \; (\text{Å})\,] \geq L,$$

where $L \cong q\lambda_{lb} = h/m(\mathbf{v}_{lb}/q)$ is the dimension of a clustron ($\lambda_{lb}^{1,2,3} = h/m\mathbf{v}_{lb}^{1,2,3}$) is the dimension of primary librational effecton in pure water; $r$ - statistically most probable distance between macromolecules (eq. 13.5); $q$ is the *lib* effecton magnification factor, reflecting the effect of water molecules immobilization (decreasing of their group velocity) on the surface of macromolecules or colloid particles.

Under condition $r > 10L$, the formation of *clustrons and their aggregates* can be accompanied by big voids origination in colloid system. At the opposite limit condition $r < L$, the trivial aggregation will dominate, accompanied by "melting" of vicinal librational effectons. It can be accompanied by conversion of clusterphilic interaction to hydrophobic or hydrophilic ones.

Our model predicts that the external electromagnetic field, like internal one, also can stimulate a process of self-organization in colloid systems. On the other hand, the agents perturbing the structure of librational effectons (temperature, ions, organic solvents) should have the opposite effect. The slowest stage of self-organization, polyclustrons and polyclustrons net formation (thixotropic - like structure), is also sensitive to mechanical shaking. Due to the cumulative effect, the resulting energy of interaction between clustrons in polyclustrons *can be higher than kT* at sufficiently low temperatures. The formation of clustrons and polyclustrons are the processes, directly interrelated with described in section 13.5 interfacial water fractions: *vicinal water (VW), surface-stimulated Bose-condensate (SSBC) and super-radiation orchestrated water (SOW)*.

The described model of self-organization in colloid systems explains the increasing of viscosity, heat capacity and the velocity of sound by the enhancement of cooperative units with ice-like water structure - *clustrons* and *polyclustrons* (Table 2).

The density of ice-like structures in clustrons and polyclustrons is lower than in pure water. Consequently, one can predict that the *free volume of water will increase as the result of self-organization* in aqueous systems of macromolecules. Such a type of effect was revealed, indeed, in our microcalorimetry study of large-scale and small-scale protein dynamics contributions in the resulting heat capacity of solvent (Kaivarainen *et al.*, 1993).

The additional free volume ($\mathbf{v}_f$) in $[0.2 - 0.3\%]$ (**w/v**) concentration of different proteins is quite close to the volume occupied by macromolecules in the solution itself. The $\mathbf{v}_f$ is dependent on the active site state (ligand-dependent) and large-scale dynamics of proteins. The more intensive the large-scale pulsations of proteins and their flexibility and the less its effective volume, the smaller the additional free volume solvent. This correlation could be resulted from acoustic impulses generated by pulsing protein. This additional acoustic noise can stimulate dissociation (melting) of water clusters - librational effectons with saturated hydrogen bonds in a system (Kaivarainen *et al.*, 1993). In terms of Hierarchic Model it means increasing the probability of [*lb/tr*] convertons (flickering clusters) excitation.

### 13.8. The antifreeze and ice-nucleation proteins
*http://arxiv.org/abs/physics/0105067*

The antifreeze proteins (AFP), discovered in arctic fish by P. Scholander et al., in 1957 are widely spread in the nature. The AFP are used for surviving at low temperatures not only by Arctic and Antarctic fish, but also by frogs, insects, plants, crops, microbes, etc.

The interest to AFP is substantial first of all because of undergoing large research programs on the development of genetically manipulated crop plants adapted to cold climate. Quite recently, the importance of AFP has been even intensified by more straightforward technical applications. For example, AFP can be used for:

- preservation of blood substances;

- protection of human organs for transplantation at low temperatures;
- preservation of texture of frozen food after melting.

All these applications exploit AFP´ability to inhibit formation of big ice crystals, damaging the cytoplasmic supramolecular structure of cells, cell membranes and tissues. However, the mechanism of specific influence of AFP on primary hydrated shell, vicinal water and bulk water is still obscure. On the other hand, this understanding is very important for the full exploitation of AFP.

The interesting observation is that AFP accumulates at the interface between ice and water. A. Haymet (University of Houston) hypothesized that a hydrophobic reaction between the protein and the neighboring water prevents water from forming ice crystals. To test this idea Harding (University of Sydney) synthesized several mutant flounder proteins and compared them with wild type proteins. Two mutants lacking all threonine residues - *the hydroxyl groups necessary for hydrogen bonding* behaved exactly like the wild type proteins. These results mean that hydrogen [protein-water] bonding does not play the main role in antifreeze activity (Haymet et al., 1998).

In a series of experiments, Davies and colleagues compared antifreeze proteins found in fish with much stronger ones produced in some moths and beetles. The insect proteins inhibited ice growth up to $-6^0$ degrees Celsius. This is about four times the efficacy of known fish proteins. The results point out, in turn, that threonine with its hydroxyl groups plays an important role in AFP activity, since this amino acid is more abundant in insect proteins than in fish proteins.

The fact that all antifreeze proteins studied are rich in hydrogen-bonding amino acid groups cannot be ignored, especially since many other structural and sequential aspects of antifreeze proteins differ drastically from type to type.

The understanding of general principles of how the different types of antifreeze mechanisms work could help to build up effective synthetic molecules for different practical purposes. The magnitude of the thermal hysteresis (difference between freezing and melting of water in presence of AFP) depends upon the specific activity and concentration of AFP.

For understanding the mechanisms of AFP, it is important to compare them with *ice nucleation proteins* (INP) with the opposite effect on water structure. The INP are responsible for another strategy to survive freezing temperature - to make the animals and plants freeze tolerant. Such organisms secrete INP to the extracellular environment. INP effects that ice-formation outside the cells decreases since it minimizes the extent of supercooling, so that freezing becomes less destructive to cells, making a time for cells to the cold acclimation. Such adaptation should be accompanied by dehydration of cells due to decreasing of water activity in the space out of cells and corresponding passive osmotic diffusion of water from cells.

The tertiary structure of the ice-nucleation protein is likely to be regular, consistent with the expectation of its forming a template for ice-matching. INP stimulates the formation of "ice embryo" - a region with some characteristics of macroscopic ice.

INP catalyze ice formation at temperatures much higher than most organic or inorganic substances (Gurian-Sherman and Lindow, 1993). Models of INPs predict that they form a planar array of hydrogen bonding groups that are closely complementary to the ice crystal. The circular dichroism spectra of the INP indicate the presence of large beta-sheet folds. This may explain the tendency of INPs to aggregate.

Bacterial ice-nucleation proteins are among the most active nucleants. The three-dimensional theoretical model of such proteins, based on molecular modeling, predicts a largely planar extended molecule, with one side serving as a template for ice-like spatial orientation of water molecules and the other side interacting with the membrane (Kajava and Lindow, 1993). This model predicts also that INPs can form big aggregates.

In contrast to INP, the AFP do not display any tendency to aggregation at physiological concentrations. Moreover, the hydroxyl groups on the surface of AFPs, combined with hydrophobic properties looks to be important for their inhibition of the process of ice formation.

### *13.8.1 Theoretical model of the antifreeze proteins (AFP) action*

We will proceed from our model of 1st order phase transition, described in section 6.2, and experimental data available. The model of AFP action, based on our Hierarchic theory of condensed matter, is able to explain the following experimental results:

a) decreasing of freezing point of water at relatively small concentrations of purified AFP of about (0.5-1) mg/ml at weak buffers at physiological pH (7.0 - 7.5). At these concentrations the water activity and osmotic pressure practically do not change thus explaining the hysteresis between freezing and melting;

b) changing of shape of ice crystals from hexagonal, pertinent for pure water, to elongated bipyramidal one;

c) accumulation of AFPs at the interface between ice and water;

d) "controversy" of AFPs´surface properties, i.e. the presence of regular hydroxyl groups necessary for hydrogen bonding, like threonine residues and their ability to inhibit the ice microcrystal growth;

e) significant increase of viscosity of water in presence of AFPs, directly correlated with their antifreezing activity.

We suppose, that the mechanism of AFP action, is related to specific influence of these proteins on the properties of primary librational effectons and translational mobility of water molecules.

Our model includes the following stages:

1. Tendency of AFP to surround the primary [lb] effectons with their four-coordinated ideal ice structure, resulted from certain stereo-chemical complementary between distribution of side group on the surface of AFP and ice;

2. Increasing the dimensions of this coherent clusters due to water molecules librations immobilization near the surface and vicinal water, corresponding decreasing of the most probable [lb] impulse (momentum) and increasing of de Broglie wave (wave B) length. Such enlarged water clusters (primary [lb] effecton) surrounded by macromolecules we named "clustron" (see Table 2);

3. Changing the symmetry of librational water molecules, thermal oscillations, from almost isotropic to anisotropic ones. Thus it means redistribution of thermal energy between three librational degrees of freedom again due to special spatial/dynamic properties of AFP surface. The consequence of this effect can be the change of ice-nucleation centers shape from hexagonal to bipyramidal, making the process of freezing less favorable. It should result in corresponding change in ice microcrystal form;

4. Bordering of the enlarged primary [lb] effectons by AFP provides the "insulation effect", preventing the association of these effectons to big enough nucleation centers, which normally accompany the water-ice transition, judging from our computer simulations;

5. Increasing the number of defects in the process of ice lattice formation in presence of AFP, as far the freezing temperature of water inside clustron can be much lower than in free bulk water. It is known experimental fact that the freezing point of water in pores or cavities with dimensions of about nanometers is much lower, than in bulk water (Kaivarainen, 1985). Such the interface effect may explain accumulation of AFP between solid and liquid phase;

6. Shifting the equilibrium between the ground - acoustic (a) state of enlarged primary [lb] effectons and optic (b) state to the latter one. This shift increases the potential energy of the affected [lb] effecton because the potential energy of (b) state is higher that of (a) and the kinetic energy of both states are the same in accordance to our model;

7. Increasing the probability and frequency of excitation of [lb-tr] convertons, i.e. [association - dissociation] of water molecules in the volume of clustron, accompanied by water density fluctuations and corresponding pulsation of the clustron's volume;

8. Radiation of the ultrasound (US) waves by clustrons, pulsing with frequency of convertons $(10^6$-$10^7)$ s$^{-1}$. These US waves destroy the ice-nucleation centers and activate the translational mobility (impulse) of water molecules, decreasing the translational wave B length ($\lambda_{tr} = h/mv_{tr}$). In accordance to condition of the 1st order phase transition (see section 6.3) this effect decreases the freezing point of water. The melting point remains unchanged, explaining the T-hysteresis provided by AFP, because the clustron pulsation in solid phase of system are inhibited;

9. Tendency to association of pulsing clustrons to form thixotropic-like structures in liquid phase, could be a consequence of Bjerknes hydrodynamic attractive forces, existing between pulsing particles, which radiate the acoustic density waves and dipole-dipole interaction between clustrons. The formation of thixotropic structures, resulting from clustron association, explains the increasing of viscosity in AFP solutions. This model of AFP action (Kaivarainen, 2001), includes the mechanism of interfacial water formation and theory of solutions, described in previous sections.

*13.8.2 Consequences and predictions of proposed model of AFP action*

1. The radius of clustrons enlarged primary [lb] effectons, surrounded by AFP can be about 30-50 °A, depending on properties of surface (geometry, polarity), temperature, pressure and presence the perturbing solvent structural agents.

2. Water, involved in clustron formation should differ by number of physical
parameters from the bulk water. It should be characterized by:
a) lower density;
b) bigger heat capacity;
c) bigger sound velocity
d) bigger viscosity;
e) smaller dielectric relaxation frequency, etc.

The formation of thixotropic structure in AFP-water systems should be accompanied by non-monotonic spatial distribution of AFP in the volume due to interaction between clustrons. The compressibility of primary [lb] effectons should be lower and sound velocity higher than that of bulk water. It is confirmed by results of Teixeira et. al. (1985), obtained by coherent- inelastic-neutron scattering. This was proved in heavy water at 25 C with solid-like collective excitations with bigger sound velocity than in bulk water. These experimental data can be considered as a direct confirmation of primary librational effectons existence.

The largest decrease of water density occurs in pores, containing enlarged primary librational effectons, due to stronger immobilization of water molecules. The existing experimental data point, that the freezing of water inside pores occurs much below $0^0$C.

We can conclude, that most of consequences of proposed model of AFP action is in a good accordance with available experimental data and the verification of the rest ones is a matter of future work.

### 13.9. **Properties of [bisolvent - polymer system]**

Let us consider the situation when the ability of nonaqueous cosolvent (for example, alcohol) in [(water-cosolvent) + polymer] system to form *clustrons* is less, than that in water.

The resulting refraction index ($n_{res}$) in [bisolvent + polymer] total system can be expressed as:

$$n_{res} = X_{H_2O}n_{H_2O} + X_{cs}n_{cs} + X_p n_p \qquad 13.21$$

$$X_{H_2O} = C_{H_2O}/(C_{H_2O} + C_{cs} + C_p)$$

where
$$X_{cs} = C_{cs}/(C_{H_2O} + C_{cs} + C_p)$$

$$X_p = C_p/(C_{H_2O} + C_{cs} + C_p)$$

are the molar fractions of water, cosolvent and polymer correspondingly; $n_{H_2O}$ the average molar refraction index of water in system; $n_{cs}$ is molar refraction index of cosolvent; $n_p$ is the average molar refraction index of polymer.

The average molar refraction index of solvent component of the total system only is:

$$\bar{n}_s = \bar{n}_{H_2O} + \bar{n}_{cs} = X_{H_2O} f_{cl}(n_{H_2O}^{cl}) + X_{H_2O}(1 - f_{cl}) n_{H_2O}^{bulk} + X_{cs} n_{cs}^{bulk} \qquad 13.22$$

$X_{H_2O} f_{cl}$ is a fraction of water composing clustrons and polyclustrons; $X_{H_2O}(1 - f_{cl})$ is a remnant bulk water fraction; $n_{H_2O}^{cl}$, $n_{H_2O}^{bulk}$ and $n_{cs}^{bulk}$ are the refraction indexes of corresponding fractions of water clusters, bulk water and bulk cosolvent correspondingly, related to their polarizabilities ($\alpha^{cl} \simeq \alpha^{bulk} \neq \alpha^{cs}$) and concentrations of molecules ($N_M^{cl}$ and $N_r^{bulk}$) in accordance with our eq. (8.13):

$$\left(1/n_{H_2O}^{cl}\right)^2 = 1 - (4/3)\alpha^{cl} N_M^{cl}$$

$$\left(1/n_{H_2O}^{bulk}\right)^2 = 1 - (4/3)\alpha^{bulk} N_M^{bulk}$$

As far the ice-like water structure composing clustrons and polyclustrons, discussed in previous section is more loose than the structure of normal bulk water, it means that:

$$N_M^{cl} < N_M^{bulk} \quad \text{and} \quad n_{H_2O}^{cl} < n_{H_2O} \qquad 13.23$$

Let us consider the factors, affecting the resulting refraction index $n_{res}$ in total [(water - cosolvent) + polymer] system, assuming that the molar refraction index of polymer ($n_p$) is much more stable parameter, than that of solvent ($\bar{n}_s$):

1. Increasing the fraction of water, composing clusters ($f_{cl}$) with smaller dense structure and smaller refraction index ($n_{H_2O}^{clust} < n_{H_2O}^{bulk}$), than has the bulk fraction $(1 - f_{cl}) \equiv f^{bulk}$.

This phenomenon in accordance with our model develops in time and space in the process of [(water - cosolvent) + polymer] system self-organization and should display itself in *decreasing* of $\bar{n}_{H_2O}$, $\bar{n}_s$ and resulting refraction index $n_{res}$;

2. After saturation of the processes of self-organization process and formation of thixotropic-like structures, the solvent refraction indexes $\bar{n}_s$ and resulting index $n_{res}$ can be affected by all types of perturbants: temperature, inorganic ions, urea, etc. Clustrons and polyclustrons stabilized by librational water effectons must be very sensitive to perturbations, due to their high cooperativeness. Under the action of perturbants the fraction of water cluster ($f_{cl}$) can decrease, accompanied by the elevation of resulting refraction index of the system $n_{res}$ symbatically with water density increasing.

The molecules of cosolvent do not participate in the ice-like water clusters with saturated hydrogen bonds composing clustrons. *Consequently, the concentration of cosolvent in bulk water should be higher than the averaged one.*

If the refraction index of cosolvent molecules is bigger than $n_{H_2O}^{bulk}$, then the difference between effective bulk solvent refraction index and that of water in clustrons ($n_{H_2O}^{cl}$) : $\Delta\bar{n} = \bar{n}^{bulk} - \bar{n}_{H_2O}^{cl}$

$$\bar{n}^{bulk} = \left[ X_{H_2O}(1 - f_{cl}) n_{H_2O}^{bulk} + X_{cs} n_{cs}^{bulk} \right] / \left[ X_{H_2O}(1 - f_{cl}) + X_{cs} \right]$$

will increase as compared to [water-polymer] system without cosolvent. Such inhomogeneity in dielectric properties of [(water - cosolvent) + polymer] system may develop even more, if the cosolvent violate the structure of bulk water, making it more dense, *i.e.* increasing $N_M^{bulk}$ and $n_{H_2O}^{bulk}$. The dimensions of clustrons, depending on molecular mass of polymers also can strongly affect optical properties, such as diffraction ability of the system.

Elevation of temperature in such a microphase system can strongly decrease the fraction of clustered water ($f_{cl}$) destroying the polyclustron subsystem. On the other hand, the radiation of such a system with laser beam enhancing the polarizability of water molecules (see chapter 8) can induce the opposite effect.

The more is the difference between $n^{bulk}$ and $n^{cl}_{H_2O}$, the more are $\bar{n}_s$ and $n_{res}$ sensitive to change of $f_{cl}$.

*Cosolvent can not compete with water in clustron formation, if the dimensions and stability of cosolvent primary effectons are much less than that of water at the same conditions.*

The dimensions of primary librational effectons of cosolvent can be easily calculated using our eqs. (2.60) and (2.74) and experimental values of the velocity of sound, positions of translational and librational bands in oscillatory spectra of pure cosolvent.

Our model predicts a big difference in ability of cosolvents to *clustron* formation in [(water - ethanol) + polymer] system, depending on difference of primary librational effectons dimensions, formed by water and ethanol molecules.

In such a case, if the cosolvent (ethanol), non-forming the clustrons, dominates in three - component system:

$$X_{cs} \gg X_{H_2O} > X_p$$

and the most of the water fraction is involved in clustrons:

$$X_{H_2O} f_{cl} \gg X_{H_2O}(1 - f_{cl})$$

if $f_{cl} \simeq 1$, then for the averaged solvent refraction index ($\bar{n}_s$) we have from (13.22):

$$\bar{n}_s \cong X^{cl}_{H_2O}(n^{cl}_{H_2O}) + X_{cs} n^{bulk}_{cs}$$

Obviously the temperature-induced change of resulting refraction index (13.21) in this conditions can be expressed as:

$$\Delta n_{res} \simeq \Delta X^{cl}_{H_2} n^{cl}_{H_2O} + X^{cl}_{H_2O} \Delta n^{cl}_{H_2O}$$

as far

$$\Delta(X_{cs} n^{bulk}_{cs}) \simeq 0 \quad \text{and} \quad \Delta(X_p n_p) \simeq 0$$

where $\Delta n_{H_2O} = n^{bulk}_{H_2O} - n^{cl}_{H_2O}$ is a difference between refraction index of bulk water and water composing clustrons; $\Delta X^{cl}_{H_2O}$ is the change of water fraction composing clustrons.

*The limited case of such three - component system is a system of so-called "reversed micelles", widely used in biochemistry and biotechnology.* It forms spontaneously after mixing of nonpolar oil, water and bipolar macromolecules like detergents or lipids.

As a result of self-organization of such a system, where the oil component is a dominating one, a small droplets of water, bordered with bipolar macromolecules in nonpolar oil (*reversed micelles*), originate.

### 13.9.1 The possibility of some nontrivial effects, following from structural and optical properties of [bisolvent - polymer] systems.

The laser beam, penetrating through described above [water - cosolvent - polymer] system, can change the fraction of water involved in clustrons ($X^{cl}_{H_2}$) and the resulting refraction index of system ($n_{res}$). As far the heating is dependent on the intensity of laser beam, this can produce the nonlinear optical effect, such as beam's phase modulation. For example, let us consider the propagation of He-Ne laser beam with photons wave length $\lambda_0 = 0.6328 \mu m$ through the cuvette with optical path length $l = 63.28 \mu m$ and resulting refraction index $n_{res} = 1.35$.

The number of the electromagnetic standing waves in the working optical path in this case will be:

$\kappa = n_{res}(l/\lambda_0) = n_{res}100 = 135$. This means that the changing of resulting refraction index $(n_{res})$ necessary for optical beam's phase change as $\varphi = 2\pi$ is equal to $\Delta n_{res} = 0.01$. The relative increasing of $n_{res}$ with laser induced temperature elevation $(\Delta n/n)_{res} \sim 1/135$ is less than 1%.

High - frequency shift of laser beam photons should lead to increasing effect of phase modulation, until the total destruction of the clustron's subsystem, as a consequence of its thermal librational effectons melting. The experimental evidence for such type of nonlinear optical phenomena has been obtained (Kodono *et al.*, 1993, Peiponen *et al.*, 1993) on the example of system [(water - ethanol) + extract of Chinese tea].

The elevation of water fraction till 15% or more destroy the nonlinear optical effects. Our model explains this phenomenon by reducing $\Delta X^{cl}_{H_2O}$ and $\Delta n^{cl}_{H_2O}$, resulting from averaging the properties of solvent inside and outside the clustrons's and polyclustron's structure, inducing their destabilization and dissociation.

The mechanical shaking or strong acoustic impulses also can destroy a fragile polyclustrons, responsible for thixotropic structure formation (see previous section 13.7). The polyclustron system is characterized by dynamic equilibrium of [assembly $\rightleftharpoons$ disassembly], very sensitive to any type of external influences and with long relaxation time.

The interesting phenomena can be anticipated, as a consequence of suggested mechanism of self - organization in described three-component system. If the linear dimensions of particles (clustron or polyclustrons), scattering the photons of laser beam, are bigger than the wave length of coherent photons, *i.e.* more than $6330 Å$ in the case of He-Ne laser, the reflection of photons from the surface of such particles is possible. In such conditions, a system of corresponding 3D standing waves can originate in colloid system.

*In accordance with our model, these 3D standing photons can influence the dynamic properties of [bisolvent - polymer] system and stimulate its spatial self-organization. The sufficiently large macromolecules or colloid particles are necessary for this nontrivial light-induced self-organization.*

### 13.10. Osmosis and solvent activity

*13.10.1. The traditional approach*

It was shown by Van't Hoff in 1887 that the osmotic pressure ($\Pi$) in dilute molar concentration of solute molecules (c) follows a simple expression:

$$\Pi = RTc \qquad 13.24$$

where T is the absolute temperature.

Since the concentration of solute $c = n_2/V$, where $(n_2)$ is the number of moles of the solute in the volume of solution (V), the above relation can also be written as

$$\Pi V = RTn_2 \qquad 13.25$$

which looks very much like the ideal gas law.

This formula can be obtained from the *equilibrium condition* between solvent and ideal solution after saturation of diffusion - osmotic process of the solvent through a semipermeable membrane, separating it from solution:

$$\mu_1^0(P) = \mu_1(P, X_i) = const \qquad 13.26$$

where $\mu_1^0$ and $\mu_1$ are the chemical potentials of a pure solvent and a solvent in presence of solute molecules; $P$ - external pressure; $X_1$ is the solvent fraction in solution.

At equilibrium conditions of osmosis: $d\mu_1^0 = d\mu_1 = 0$ and from (13.24a) we get

$$d\mu_1 = \left[\frac{\partial \mu_1}{\partial P_1}\right]_{X_1} dP_1 + \left[\frac{\partial \mu_1}{\partial X_1}\right]_{P_1} dX_1 = 0 \qquad 13.27$$

Because
$$\mu_1 = \left(\partial G/\partial n_1\right)_{P,T} = \mu_1^0 + RT\ln X_1 \qquad 13.27a$$

where $(n_2)$ is the number of moles of the solvent in the volume of solution $(V)$, then

$$\left(\frac{\partial \mu_1}{\partial P_1}\right)_{X_1} = \left(\frac{\partial^2 G}{\partial P\, \partial n_1}\right)_{P,T,X} = \left(\frac{\partial V}{\partial n_1}\right) = \bar{V}_1 \qquad 13.27b$$

where $\bar{V}_1$ is the partial molar volume of the solvent. For dilute solution: $\bar{V}_1 \simeq V_1^0$ (molar volume of pure solvent).

From (13.27a) we have:

$$\frac{\partial \mu_1}{\partial X_1} = RT\left(\frac{\partial \ln X_1}{\partial X_1}\right)_{P,T} \qquad 13.28$$

Putting (13.27b) and (13.28) into (13.27) we get

$$dP_1 = -\frac{RT}{V_1^0 X_1} dX_1 \qquad 13.28a$$

The integration:

$$\int_P^{p+\pi} dP_1 = -\frac{RT}{V_1^0}\int_1^{x_1} d\ln X_1 \qquad 13.29$$

gives:

$$\Pi = -\frac{RT}{V_1^0}\ln X_1 = -\frac{RT}{V_1^0}\ln(1 - X_2) \qquad 13.29a$$

and for the dilute solution $(X_2 \ll 1)$ we finally obtain Van't Hoff equation:

$$\Pi = \frac{RT}{V_1^0} X_2 \cong RTc \qquad 13.30$$

where the fraction of solute molecules in solution is:

$$X_2 = n_2/(n_1 + n_2) \cong n_2/n_1 \qquad 13.31$$

and

$$\frac{n_2/n_1}{V_1^0} = c \qquad 13.32$$

Considering a real solution, we have to replace a solvent fraction $X_1$ in (13.29) by solvent activity: $[X_1 \to a_1]$. Then taking into account (13.30), we can express osmotic pressure as follows:

$$\Pi = -\frac{RT}{\bar{V}_1}\ln a_1 = \frac{\Delta \mu_1}{\bar{V}_1} \qquad 13.33$$

where $\Delta \mu_1 = \mu_1^0 - \mu_1$ is the difference between the chemical potentials of a pure solvent and the one perturbed by solute at the starting moment of osmotic process, *i.e.* the driving force of osmosis; $\bar{V}_1 \cong V_1$ is the molar volume of solvent at dilute solutions.

Although the osmotic effects are widespread in Nature and are very important, especially in biology, the physical mechanism of osmosis remains unclear (Watterson, 1992).

The explanation following from Van't Hoff equation (13.30), means that the osmotic pressure is equal to that induced by solute molecules, if they behave as an ideal gas molecules in the same volume at a given temperature. It can't be considered as satisfactory.

*13.10.2. A new approach to osmosis, based on the Hierarchic Theory of condensed matter*

The phenomenon of osmosis can be explained quantitatively on the basis of the Hierarchic Theory and state equation (11.7). To this end, we have to introduce the rules of conservation of the main internal parameters of solvent in the presence of guest (solute) molecules or particles in conditions of equilibrium:

$$\left.\begin{array}{l} 1.\ \text{Internal pressure of solvent: } P_{in} = \text{const} \\ 2.\ \text{The total energy of solvent: } U_{tot} = \text{const} \end{array}\right\} \quad 13.34$$

These two conservation rules can be considered as the consequence of Le Chatelier principle in general form.

Using (11.7), we have for internal pressure of the pure solvent ($P_{in}$) and solvent perturbed by a solute ($P_{in}^1$), a following two equations, respectively:

$$P_{in} = \frac{U_{tot}}{V_{fr}^0}\left(1 + \frac{V}{T_k}\right) - P_{ext} \qquad 13.35$$

$$P_{in}^1 = \frac{U_{tot}^1}{V_{fr}^1}\left(1 + \frac{V_1}{T_k^1}\right) - P_{ext}^1, \qquad 13.36$$

where from (11.19):

$$V_{fr}^0 = \frac{V_0}{n^2} \quad \text{and} \quad V_{fr}^1 = \frac{V_0}{n_1^2} \qquad 13.37$$

are the free volumes of pure solvent and solvent in presence of solute (guest) molecules, as a ratio of molar volume of solvent ($V_0$) to correspondent values of refraction index squared ($n^2$ and $n_1^2$).

The equilibrium conditions *after osmotic process saturation*, leading from our conservation rules (13.34) are

$$P_{in} = P_{in}^1 \quad \text{when} \quad P_{ext}^1 = P_{ext} + \Pi \qquad 13.38$$

$$U_{tot} = V + T_k = V_1 + T_k^1 = U_{tot}^1 \qquad 13.39$$

From (13.39) we have:

$$\text{Dif} = T_k - T_k^1 = V_1 - V \qquad 13.40$$

The index (1) denote perturbed by solute molecules solvent parameters.

Comparing (13.35) and (13.36) and taking into account (13.37 - 13.39), we derive a new formula for osmotic pressure, based on new state equation (11.7):

$$\Pi = \frac{n^2}{V_0} U_{tot}\left[\frac{n_1^2 T_k - n^2 T_k^1}{T_k T_k^1}\right] \qquad 13.41$$

where $n$, $V_0$, $U_{tot}$ and $T_k$ are the refraction index, molar volume, total energy and total kinetic energy of a pure solvent, respectively; $T_k^1$ and $n_1$ are the total kinetic energy and refraction index of the solvent in the presence of guest (solute) molecules, that can be calculated from the Hierarchic Theory (eq. 4.33), using our computer program pCAMP.

For the case of dilute solutions, when $T_k T_k^1 \cong T_k^2$ and $n \cong n_1$, the (eq. 13.41) can be simplified:

$$\Pi = \frac{n^2}{V_0}\left(\frac{U_{tot}}{T_k}\right)^2 (T_k - T_k^1) = \frac{n^2}{V_0} \frac{1}{S^2}(T_k - T_k^1) \qquad 13.42$$

or using (13.40):

$$\Pi = \frac{n^2}{V_0}\left(\frac{U_{tot}}{T_k}\right)^2 (V_1 - V) \qquad 13.43$$

The ratio:

$$S = T_k/U_{tot} \qquad 13.44$$

is generally known as a structural factor (see eq. 2.46):

We can see from (13.42) and (13.43) that osmotic pressure is proportional to the difference between total kinetic energy of a free solvent ($T_k$) and that of the solvent perturbed by guest molecules:

$$\Delta T_k = T_k - T_k^1$$

or corresponding difference between the total potential energy of perturbed and pure solvent:
$\Delta V = V_1 - V$, where $\Delta T_k = \Delta V \equiv \text{Dif}$ (see Figure 43).

As far $\Delta T_k > 0$ and $\Delta V > 0$, it means that:

$$T_k > T_k^1$$
$$\text{or} \qquad 13.45$$
$$V_1 > V$$

Theoretical temperature dependence of the difference

$$\text{Dif} = \Delta T_k = \Delta V$$

calculated from (13.42) or (13.43) at constant osmotic pressure: $\Pi \equiv Pos = 8$ atm, pertinent to blood is presented on Figure 43.

The next Figure 44 illustrate theoretical temperature dependence of osmotic pressure (13.43) in blood at the constant value of

$\text{Dif} = 6.7 \times 10^{-3} (J/M)$, corresponding on Figure 43 to physiological temperature ($37^0 C$).

The ratios of this Dif value to total potential ($V$) and total kinetic energy ($T_k$) of pure water at $37^0$ (see Figure 5b) are equal to:

$$(\text{Dif}/V) \simeq \frac{6.7 \cdot 10^{-3}}{1.3 \cdot 10^4} \cong 5 \times 10^{-7} \text{ and}$$

$$(\text{Dif}/T_k) \simeq \frac{6.7 \cdot 10^{-3}}{3.5 \cdot 10^2} \cong 2 \times 10^{-5}$$

*i.e.* the relative changes of the solvent potential and kinetic energies are very small.

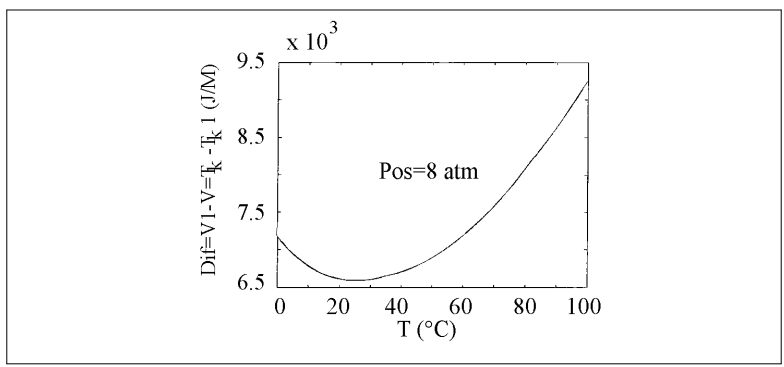

**Figure 43**. Theoretical temperature dependence of the difference: $\text{Dif} = V_1 - V = T_k - T_k^1$ at constant osmotic pressure: $\Pi \equiv Pos = 8$ atm, characteristic for blood. The computer calculations were performed using eqs. (13.42) or (13.43).

For each type of *concentrated macromolecular solutions the optimum amount of water is needed to minimize the potential energy of the system*. The realization of conservation rules (13.34) in solutions of macromolecules may be responsible for *driving force of osmosis* in different compartments of biological cells.

Comparing (13.43) and (13.33) and assuming equality of the molar volumes $V_0 = \bar{V}_1$, we find a relation between the difference in potential energies and chemical potentials ($\Delta\mu$) of unperturbed solvent and that perturbed by the solute:

$$\Delta\mu = \mu_1^0 - \mu_1 = n^2 \left(\frac{U_{tot}}{T_k}\right)^2 (V_1 - V) \qquad 13.46$$

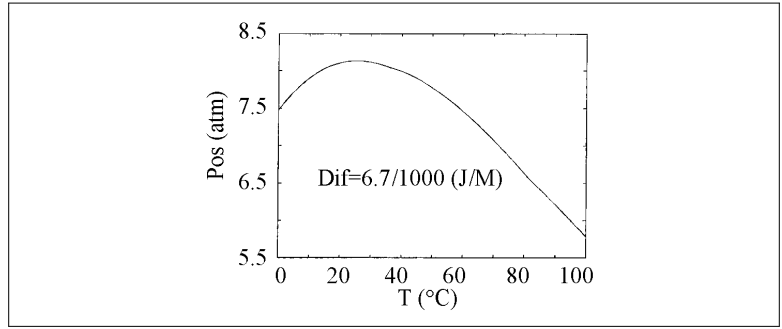

**Figure 44**. Theoretical temperature dependence of osmotic pressure (eq. 13.43) in blood at constant value of difference: $Dif = \Delta T = \Delta V = 6.7 \times 10^{-3} J/M$. This value in accordance with Figure 43 corresponds to physiological temperature ($37^0$).

*The results obtained above mean that solvent activity ($a_1$) and a lot of other thermodynamic parameters of solutions can be calculated on the basis of the Hierarchic Theory:*

$$a_1 = \exp\left(-\frac{\Delta\mu}{RT}\right) = \exp\left[-\left(\frac{n}{S}\right)^2 \frac{V_1 - V}{RT}\right] \qquad 13.47$$

where $S = T_k/U_{tot}$ is a structural factor for the solvent.

The molar coefficient of activity is:

$$y_i = a_i/c_i, \qquad 13.48$$

where

$$c_i = n_i/V \qquad 13.49$$

is the molar quantity of $i$-component ($n_i$) in of solution; $V$ is solution volume in liters.

The molar activity of the solvent in solution is related to its vapor pressure ($P_i$) as:

$$a_i = P_i/P_i^0 \qquad 13.50$$

where $P_i^0$ is the vapor pressure of the pure solvent. Theoretical temperature dependence of water activity ($a_1$) in blood at constant difference: $Dif = \Delta T = \Delta V = 6.7 \cdot 10^{-3} J/M$ is presented on Figure 45.

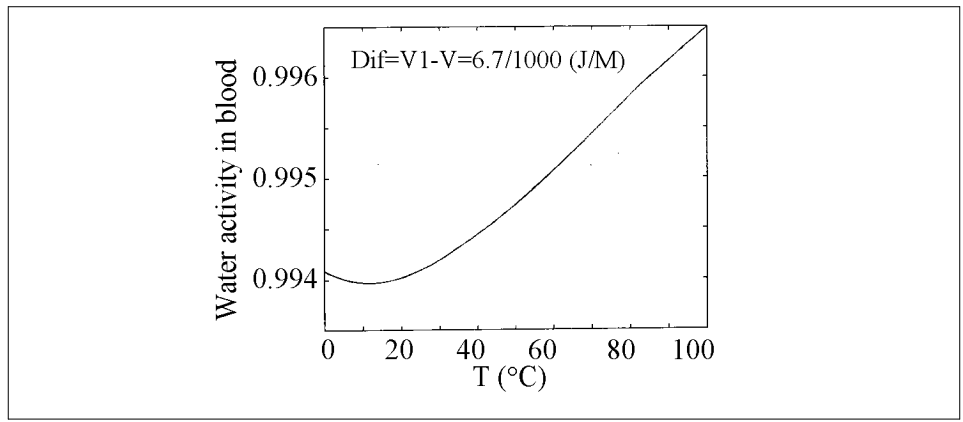

**Figure 45**. Theoretical temperature dependence of water activity ($a_1$) (*eq.* 13.47) in blood at constant difference: $Dif = \Delta T = \Delta V = 6.7 \times 10^{-3} J/M$.

Another *colligative parameter* such as low temperature shift of freezing temperature of the solvent ($\Delta T_f$) in the presence of guest molecules also can be calculated from (13.47) and (13.13):

$$\Delta T_f = -\frac{R(T_f^0)^2}{\Delta H}\ln a_1 \frac{(T_f^0)^2}{\Delta H\, T}\left(\frac{n}{S}\right)^2 (V_1 - V) \qquad 13.51$$

where $T_f^0$ is the freezing temperature of the pure solvent; $T$ is the temperature corresponding to the conditions of calculations of $V_1(T)$ and $V(T)$ from (eqs. 4.33 and 4.36).

The partial molar enthalpy ($\bar{H}_1$) of solvent in solution are related to solvent activity like:

$$\bar{H}_1 = \bar{H}_1^0 - RT^2 \frac{\partial \ln a_1}{\partial T} = \bar{H}_1^0 + \bar{L}_1^0 \qquad 13.52$$

where $\bar{H}_1^0$ is the partial enthalpy of the solvent at infinitive dilution;

$$\bar{L}_1 = -RT^2 \frac{\partial \ln a_1}{\partial T} = T^2 \frac{\partial}{\partial T}\left[\left(\frac{n}{S}\right)^2 \frac{V_1 - V}{T}\right] \qquad 12.53$$

is the relative partial molar enthalpy of solvent in a given solution.

From (13.52) we obtain partial molar heat capacity as:

$$C_p^1 = \frac{\partial}{\partial T}(H_1) = C_p^0 - R\left(T^2 \frac{\partial^2 \ln a_1}{\partial T^2} + 2T\frac{\partial \ln a_1}{\partial T}\right) \qquad 13.54$$

An analogous equation exists for the *solute* of this solution as well as for partial *molar volume* and other important parameters of the solvent, *including solvent activity* (Godnev et al., 1982).

*It is obvious, the application of the Hierarchic Theory to solvent activity determination might be of great practical importance for different processes in chemical and colloid technology.*

### 13.11. The external and internal water activity, as a regulative factor in cellular processes

*Four most important factors can be responsible for spatial-time processes in living cells:*

1. Self-organization of supramolecular systems in the form of membranes, oligomeric proteins and filaments. Such processes can be mediated by water-dependent hydrophilic, hydrophobic and introduced by us clusterphilic interactions;

2. Compartmentalization of cell volume by semipermeable lipid- bilayer membranes and due to different cell's organelles formation;

3. Changes in the volumes of different cell compartments by osmotic process (active and passive). These correlated changes are dependent on water activity regulated by intra-cell [assembly ⇌ disassembly] equilibrium shift of microtubules and actin's filaments.

4. The changing of the fraction of *vicinal water*, representing a developed system of enlarged primary librational effectons, stabilized by surface of proteins (see section 13.5). The disassembly of filaments leads to water activity decreasing due to increasing of 'free surface' of proteins and vicinal water fraction. Their assembly induces the opposite effect. In turn, the [*assembly* ⇌ *disassembly*] equilibrium could be modulated by $Ca^{2+}$, $Mg^{2+}$ and other ions concentration, polarization of neuron's membranes in a course of nerve excitation.

The thickness of vicinal water layer is about 30-50 Å, depending on temperature and mobility of intra-cell components. The vicinal water with more ordered and cooperative structure than the bulk water represents the dominant fraction of intra-cell water. A lot of experimental evidences, pointing to important role of vicinal water in cell physiology were presented in reviews of Drost-Hansen and Singleton (1992) and Clegg and Drost-Hansen (1991).

The third fraction of interfacial water, introduced in our model: surface-stimulated Bose-condensation have an order of hundreds angstroms (see section 13.5). It is too unstable to exist at physiological temperatures of warm-blood animals - around $37^0 C$. However it can play an important role in the internal spatial organization of cytoplasm and internal properties of plenty of water containing plants: tomatoes, cucumbers, cold blood and marine organisms, like jellyfish.

The dynamic equilibrium: { I ⇔ II ⇔ III } between three stages of macromolecular self-organization, discussed in Section 13.7 (Table 2), may play an important role in biosystems: blood, lymph as well as inter- and intra-cell media. This equilibrium is dependent on the water activity (inorganic ions, pH), temperature, concentration and surface properties of macromolecules. Large-scale protein dynamics, decreasing the fraction of vicinal and surface-stabilized water (Kaivarainen, 1985, Kaivarainen *et al.*, 1995) is modulated by the protein's active site ligand state. These factors may play a regulative role in [*coagulation* ⇔ *peptization*] and [*gel* ⇔ *sol*] transitions in the cytoplasm of mobile cells ( *i.e.* macrophages), necessary for their migration - active movement.

A lot of spatial cellular processes such as the increase or decrease in the length of microtubules or actin filaments are dependent on their *self-assembly* from corresponding subunits ($\alpha, \beta$ tubulins and actin).

The self-assembly of such superpolymers is dependent on the [association $(A)$ ⇔ dissociation $(B)$] equilibrium constant ($K_{A \Leftrightarrow B} = K_{B \Leftrightarrow A}^{-1}$). In turn, this constant is dependent on water activity ($a_{H_2O}$), as was shown earlier (eqs.13.11a; 13.12).

*The double helix of actin filaments*, responsible for the spatial organization and cell's shape, is composed of the monomers of globular protein - actin (MM 42.000). The rate of actin filament polymerization or depolymerization, responsible for cells shape adaptation to environment, is very high and strongly depends on ionic strength (concentration of $Na^+$, $Cl^-$, $Ca^{++}$, $Mg^{++}$). For example, the increasing of $Na^+$, $Cl^-$ concentration and corresponding decreasing of $a_{H_2O}$ stimulate the actin polymerization. The same is true of $\alpha$ and $\beta$ tubulin polymerization in the form of microtubules.

The activity of water in cells and cell compartments can be regulated by $[Na^+ - K^+]\ ATP$ –dependent pumps. Even the equal concentrations of $Na^+$ and $K^+$ decrease water activity $[a_{H_2O}]$ differently due to their different interaction with bulk and, especially with ordered vicinal water (Wiggins 1971, 1973).

Regulation of pH by proton pumps, incorporated in membranes, also can be of great importance for intra-cell $a_{H_2O}$ changing.

Cell division is strongly correlated with dynamic equilibrium: [assembly ⇔ disassembly] of microtubules of centrioles. Inhibition of tubulin subunits dissociation (disassembly) by *addition of $D_2O$*, or *stimulation* this process by *decreasing temperature or increasing hydrostatic pressure* stops cell mitosis - division (Albert *et al.*, 1983). These data confirms our hypothesis, that microtubules (MT) assembly are mediated by clusterphilic interaction (see section 17.5). The clusterphilic interaction regulates the $A \Leftrightarrow B$ equilibrium of cavities between $\alpha$ *and* $\beta$ tubulins of ($\alpha\beta$) dimers, composing microtubules.

The decrease of temperature is accompanied by the enhancement of dimensions of librational water effectons (mesoscopic Bose condensate). This leads to increasing of the internal disjoining force of water in MT's and their disassembly.

Microtubules are responsible for coordination of intra-cell space organization and movements, including chromosome movement at the mitotic cycle, coordinated by centrioles.

The communication between different cells by means of inter-cell channels can regulate the ionic concentration and correspondent $a_{H_2O}$ gradients in the embryo.

*In accordance with our hypothesis, the gradient of water activity, regulated by change of vicinal, surface-modulated water fraction in different compartments of the same cell and in different cells can play a role of so-called morphogenic field, responsible for differentiation of embryo cells. The coherent biophotons in IR range, emitted ⇌ absorbed by coherent clusters (mBC) in microtubules, providing remote interaction between systems [centrioles + chromosomes] of remote cells realize the organism morphogenesis on quantum level (see also section 17.5).*

### 13.12. Distant solvent-mediated interaction between proteins and the cells

The unknown earlier phenomena, like noncontact, solvent - mediated interaction between different proteins (Fig. 42b) and proteins and cells (Fig. 46) was revealed in our experiments (Kaivarainen, 1985, section 8.5). The latter kind of interaction, modulated by temperature and protein ligand state, was discovered on examples of mixed systems: erythrocyte suspension + human serum albumin (HSA), using turbidimetry (light scattering) method (Fig. 46).

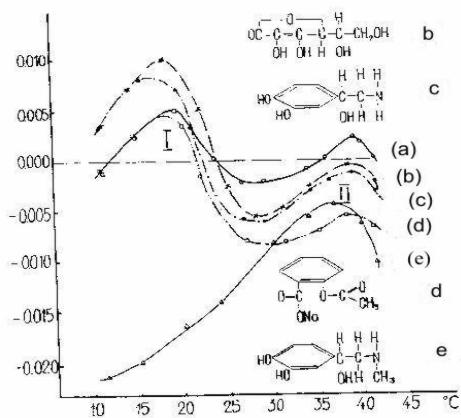

**Figure 46**. The differential temperature dependence of turbidity of human erythrocytes suspension at the light wave length $\lambda = 600$ nm in the presence and absence of human serum albumin (HSA) in different ligand state:

$$\Delta D^* = D^*_{ER+HSA} - D^*_{ER}$$

a) the interaction with intact HSA; b) HSA + ascorbic acid; c) HSA + noradrenaline; d) HSA + sodium acetyl-salicylate; e) HSA + adrenaline. The experiments where performed in modified physiological Henx solution (pH7.3), where the glucose was excluded to avoid possible artefacts on osmotic processes (Kaivarainen, 1985, Figure84).

The concentration of erythrocytes was $2 \times 10^5$ cm$^{-1}$ and the concentration of HSA was 15mg/cm$^3$. All ligands had fivefold molar excess over HSA. The erythrocyte suspension was incubated at least 24 hours before measurements for exhaustion of cellular ATF and elimination of the active osmosis. A number of special experimental controls where performed to prove the absence of absorption of HSA on the membranes of cells and direct effect of ligands on turbidity of erythrocyte suspension

(Kaivarainen, 1985, section 8.5).

The *swelling* and *shrinking* of erythrocytes, reflected by increasing and decreasing of their suspension turbidity, is a consequence of increasing and decreasing of the external water activity (Figure46). The latter factors are responsible for the direction of water diffusion across the membranes of cells ('in' or 'out') as a result of passive osmosis. They are in-phase with corresponding changes of HSA large-scale dynamics (flexibility). The latter was revealed in special experiments with spin-labeled HSA-SL (see Figure 42a). The same kind of temperature dependence of HSA - SL flexibility was confirmed in modified Henx solution.

We got the modified Boyle van Hoff equation, pointing to direct correlation between the increments of cell's volume and the external water activity (Kaivarainen, 1985): $\Delta \mathbf{V} \sim \Delta \mathbf{a}_{H_2O}$.

Both thermoinduced conformational transitions of serum albumin occur in physiological region of surface tissues of body. Consequently, presented data point to possible role of albumin (the biggest protein fraction in blood) in thermal adaptation of many animals, including human, using the discovered *solvent - mediated mechanism of protein - protein and protein - cell interaction.*

### 13.13 The cancer cells selective destructor

We put forward a hypothesis that one of the reasons of unlimited cancer cell division is related to partial disassembly of cytoskeleton's actin-like filaments and microtubules due to some genetically controlled mistakes in biosynthesis and increasing the osmotic diffusion of water into transformed cell.

Decreasing of the intra-cell concentration of any types of ions ($Na^+$, $K^+$, $H^+$, $Mg^{2+}$ etc.), as the result of corresponding ionic pump destruction, incorporated in biomembranes, also may lead to disassembly of filaments.

The shift of equilibrium: [assembly ⇔ disassembly] of microtubules (MTs) and actin filaments to the right increases the amount of intra-cell water, involved in hydration shells of protein and decreases water activity. As a consequence of concomitant osmotic process, cells tend to swell and acquire a ball-like shape. The number of direct contacts between transformed cells decrease and the water activity in the intercell space increases also.

Certain decline in the external inter-cell water activity could be a triggering signal for inhibition of normal cell division. The shape of normal cells under control of cell's filament is a specific one, providing good dense intercell contacts with limited amount of water, in contrast to transformed cells.

In accordance to mechanism proposed, the absence of contact inhibition in the case of cancer cells is a result of insufficient decreasing of intercell water activity due to loose [cell-cell] contacts.

If our model of cancer emergency is correct, then the problem of tumor inhibition problem can be related to the problem of inter - and intra-cell water activity regulation by means of chemical and physical factors.

Another approach to cancer healing, based on mechanism described above, is the IR laser treatment of transformed cells with IR photons frequencies (about $3 \times 10^{13}$ s$^{-1}$), stimulating excitation of cavitational fluctuations in water of cytoplasm, disassembly of actin filaments and gel → sol transition (see chapter 18). The corresponding centrioles destruction will prevent cells division and should give a good therapeutic effect. The IR Laser based selective cancer cells destructor, proposed by this author, can be combined with preliminary ultrasound (US) treatment of tissues and blood. Such US treatment with frequency of about 40 kHz in accordance to the Hierarchic Theory of water should make cancer cells more "tuned" to IR laser beam.

The method of cancer cells elimination, proposed here, is based on assumption that stability of MTs in transformed cells is weaker than that of normal cells. This difference should provide the selectivity of destructive action of the ultrasound and laser beam on the cancer cells, remaining the normal cells undamaged.

Our computer calculations predicts except the pointed above carrying frequency of IR radiation,

the modulation frequencies in radio frequency range (about $10^7$ Hz), which should increase the yield of cancer cells destruction effect.

# Chapter 14

# Macroscopic oscillations and slow relaxation (memory) in condensed matter.
# The effects of a magnetic field

## 14.1. Theoretical background

One of the consequences of our concept is of special interest. It is the possibility for oscillation processes in solids and liquids. The law of energy conservation is not violated thereupon because the energies of two quasi-particle subsystems related to effectons and deformons, can change in opposite phases. The total internal energy of matter keeps constant.

The energy redistribution between primary and secondary effecton and deformon subsystems may have a periodical character, coupled with the oscillation of the ($a \Leftrightarrow b$) equilibrium constant of primary effectons ($K_{a \Leftrightarrow b}$) and correlated oscillations of primary electromagnetic deformons concentration. According to our model (Table 1) the ($a \to b$) transition of primary effecton is related to photon absorption, i.e. a decrease in primary electromagnetic deformon concentration, while the ($b \to a$) transition on the contrary, radiate photons. If, therefore, the $[a \Leftrightarrow b]$ and $[\bar{a} \Leftrightarrow \bar{b}]$ equilibriums are shifted rightward and equilibrium constants $K_{a \Leftrightarrow b}$ and $\bar{K}_{a \Leftrightarrow b}$ decreases, then concentrations of primary and secondary deformons ($n_d$ and $\bar{n}_d$) also decreases. If $K_{a \Leftrightarrow b}$ grows up, i.e. the concentration of primary effectons in a-states increases, then $n_d$ increases. We remind that a and b states of the primary effectons correspond to the more and less stable molecular clusters, correspondingly (see Introduction). In accordance with our model, between dynamic equilibrium of primary and secondary effectons the strong interrelation exists. Equilibrium of primary effectons is more sensitive to any perturbations. However, the equilibrium shift of secondary effectons affect the total internal energy, the entropy change and possible mass defect (see below) much stronger than that of primary effectons.

As we have shown (Figure 28 a,b), the scattering ability of A-states is more than two times as high as that of B-states. Their polarizability, refraction index and dielectric permeability are also higher. It makes possible to register the oscillations in the condensed matter in different ways.

In accordance with our theory the oscillation of refraction index must induce the corresponding changes of viscosity and self-diffusion in condensed matter (See chapter 11). The diffusion variations are possible, for example, in solutions of macromolecules or other Brownian particles. In such a way self-organization in space and time gradually may originate in appropriate solvents, solutions, colloid systems and even in solid bodies.

The period and amplitude of these oscillations depend on the times of relaxation processes which are related to the activation energy of equilibrium shifts in the effectons, polyeffectons or coherent superclusters of primary effectons subsystems.

The reorganizations in the subsystems of translational and librational effectons, macro- and super-effectons, as well as chain-like polyeffectons, whose stabilities and sizes differ from each other, must go on at different rates. It should, therefore, be expected that in the experiment the presence of several oscillation processes would be revealed. These processes are interrelated but going on with different periods and amplitudes. Concomitant oscillations of self-diffusion rate also must be taken into account. In such a way Prigogine's dissipative structures could be developed (Prigogine, 1984). The properties of liquids and solids as an active mediums are determined by two states of the effectons and feedback link between their equilibrium constant and primary and secondary effectons concentrations. Instability in the degree of ordering in time and space is accompanied by the slow oscillation of entropy of the whole macroscopic system.

### 14.2. The entropy - information content of matter as a hierarchic system

The well known Boltzmann formula, interrelate the entropy of macrosystem with its statistical weigh (P):

$$S = k \ln P = k \ln W \qquad 14.1$$

The statistical weigh for macrosystem (P), equal to number of microstates (W), corresponding to given macrostate, necessary for entropy calculation using (14.2) could be presented as:

$$P = W = \frac{N!}{N_1! \times N_2! \times \ldots N_q!} \qquad 14.1a$$

where:

$$N = N_1 + N_2 + \ldots N_q \qquad 14.2$$

is the total number of molecules in macrosystem;

$N_i$ is the number of molecules in the i-th state;

$q$ is the number of independent states of all quasiparticles (collective excitations) in macrosystem.

We can subdivide macroscopic volume of $1 cm^3$ into 24 types of quasiparticles in accordance with our hierarchic model (see Table 1 in Introduction).

In turn, each type of the effectons (primary, secondary, macro- and super-effectons) is subdivided on two states: ground $(a, \bar{a}, A)_{tr,lb}$ and excited $(b, \bar{b}, B)_{tr,lb}$ states. Taking into account two ways of the effectons origination - due to thermal translations ($tr$) and librations ($lb$), excitations, [$lb/tr$] convertons, macro- and super deformons, the total number of states is 24 also. It is equal to number of relative probabilities of excitations, composing partition function Z (see eq.4.2). Consequently, in eqs.(14.1a and 14.2) we have:

$$q = 24$$

The number of molecules, in the unit of volume of condensed matter ($1 cm^3$), participating in each of 24 excitation states ($i$) can be calculated as:

$$N_i = \frac{(\mathbf{v})_i}{V_0/N_0} n_i \frac{P_i}{Z} = \frac{N_0}{V_0} \frac{P_i}{Z} \qquad 14.3$$

where: $(\mathbf{v})_i = 1/n_i$ is the volume of (i) quasiparticle, equal to reciprocal value of its concentration ($n_i$); $N_0$ and $V_0$ are Avogadro number and molar volume, correspondingly; Z is partition function and $P_i$ are relative probabilities of independent excitations in composition of Z $(eq. 4.2)$.

The total number of molecules of (i)-type of excitation in any macroscopic volume of matter ($V_{mac}$) is equal to

$$N^i_{mac} = N_i V_{mac} = V_{mac} \frac{N_0}{V_0} \frac{P_i}{Z} \qquad 14.4$$

Putting (14.3) into (14.1a) and (14.1), we can calculate the statistical weight and entropy.

For the large values of $N_i$ it is convenient to use a Stirling formula:

$$N_i = (2\pi N)^{1/2} (N/e)^N \exp(\Theta/12N) \sim (2\pi N)^{1/2} (N/\Theta)^N \qquad 14.5$$

Using this formula and (14.1), one can obtain the following expression for entropy:

$$S = k \ln W = -k \sum_{i}^{q} (N_i + \frac{1}{2}) \ln N_i + \text{const} = S_1 + S_2 + \ldots S_i \qquad 14.6$$

From this equation we can see that the temperature increasing or [solid → liquid] phase transition will lead to the entropy elevation:

$$\Delta S = S_L - S_S = k \ln(W_L/W_S) > 0 \qquad 14.7$$

where: $S_L$ and $S_S$ are the entropy of liquid and solid state, corresponding to $W_L$ and $W_S$.

It follows from (14.6 and 14.3), that under conditions when $P_i$ and $N_i$ undergoes oscillations this lead to oscillations of contributions of different types of quasiparticles to the entropy of system and to oscillations of total entropy of system. The coherent oscillations of $P_i$ and $N_i$ can be induced by different external fields: acoustic, electromagnetic and gravitational. The autooscillations may arise spontaneously also in the sensitive and highly cooperative active mediums.

The experimental evidence for such phenomena will be discussed in the next section.

The notions of probability of given microstate ($p_i = 1/W$), entropy ($S_i$) and information ($I_i$) are strongly interrelated. The smaller the probability the greater is information (Nicolis 1986):

$$I_i = \lg_2 \frac{1}{p_i} = -\lg_2 p_i = \lg_2 W_i \qquad 14.8$$

where $p_i$ is defined from the Boltzmann distribution as:

$$p_i = \frac{\exp(-E_i/kT)}{\sum_{m=0}^{\infty} \exp(-n_m h\nu_i/kT)} \qquad 14.9$$

where $n_m$ is quantum number; $h$ is the Planck constant; $E_i = h\nu_i$ is the energy of ($i$)-state.

There is strict relation between the entropy and information, leading from comparison of (14.8) and (14.1):

$$S_i = (k_B \ln 2) I_i = 2.3 \times 10^{-24} I_i \qquad 14.10$$

The informational entropy is given as expectation of the information in the system (Nicolis,1986; Haken, 1988).

$$<I> = \Sigma P_i \lg_2(1/p_i) = -\Sigma p_i \lg_2(p_i) \qquad 14.11$$

From (14.10) and (14.11) we can see that variation of probability $p_i$ and/or $N_i$ in (14.3) will lead to changes of entropy and information, characterizing the matter as a hierarchical system.

The reduced information (entropy), characterizing the information quality, related to selected collective excitation of any type of condensed matter, we introduce here as a product of corresponding component of information [$I_i$] to the number of molecules (atoms) with similar dynamic properties in composition of this excitation:

$$q_i = (\mathbf{v}_i/\mathbf{v}_m) = N_0/(V_0 n_i) \qquad 14.12$$

where: $\mathbf{v}_i = 1/n_i$ is the volume of quasiparticle, reversible to its concentration ($n_i$); $\mathbf{v}_m = V_0/N_0$ is the volume, occupied by one molecule.

The product of (14.11) and (14.12), i.e. the *reduced information* gives the quantitative characteristic not only about quantity but also about the quality of the information:

$$(Iq)_i = p_i \lg_2(1/p_i) N_0/(V_0 n_i) \qquad 14.13$$

*This new formula could be considered as a useful modification of known Shannon equation.*

### 14.3. Experimentally revealed macroscopic oscillations

A series of experiments was conducted in our laboratory to study oscillations in the buffer (pH 7.3) containing 0.15 M NaCl as a control system and immunoglobulin G (IgG) solutions in this buffer at the following concentrations: $3 \times 10^{-3} mg/cm^3$; $6 \times 10^{-3} mg/cm^3$; $1.2 \times 10^{-2}$ and $2.4 \times 10^{-2} mg/cm^3$.

The turbidity ($D^*$) of water and the solutions were measured every 10 seconds with the spectrophotometer at $\lambda = 350 nm$. Data were obtained automatically with the time constant 5 s during 40 minutes. The number of $D^*$ values in every series was usually equal to 256. The total number of the fulfilled series was 30.

The time series of $D^*$ were processed using computer program for time series analysis. The time trend was thus subtracted and the auto-covariance function and the spectral density were calculated.

The empty quartz cuvette with the optical path about 1 cm were used as a basic control.

Only the optical density of water and water dissolved substances, which really exceeded background optical density in the control series were taken into account. It is shown that the noise of the photoelectronic multiplier does not contribute markedly to dispersion of $D^*$.

The measurements were made at temperatures of $17^0 C$, $28^0$, $32^0$, and $37^0$. The period of the trustworthily registered oscillation processes related to changes in $D^*$, had 2 to 4 discrete values over the range of $(30 - 600) s$ under our conditions. It does not exclude the fact that the autooscillations of longer or shorter periods exist. For example, in distilled water at $32^0 C$ the oscillations of the scattering ability are characterized by periods of 30, 120 and 600 s and the spectral density amplitudes 14, 38 and 78 (in relative units), respectively. With an increase in the oscillation period their amplitude also increases. At $28^0 C$ the periods of the values 30s, 41s and 92s see have the corresponding normalized amplitudes 14.7, 10.6 and 12.0.

Autooscillations in the buffer solution at $28^0 C$ in a 1 cm wide cuvette with the optical way length 1 cm (i.e. square section) are characterized with periods: 34*s*, 52*s*, 110*s* and 240 s and the amplitudes: 24, 33, 27 and 33 relative units. In the cuvette with a smaller (0.5 cm) or larger (5 cm) optical wavelength at the same width (1 cm) the periods of oscillations in the buffer change insignificantly. However, amplitudes decreased by 50% in the 5 cm cuvette and by 10-20% in the 0.5 cm-cuvette. This points to the role of geometry of space where oscillations occur, and to the existence of the finite correlation radius of the synchronous processes in the volume. The radius of coherency is macroscopic and comparable with the size of the cuvette.

The dependence of the oscillations amplitude on the concentration of the protein - immunoglobulin G has a sharp maximum at the concentration of $1.2 \times 10^{-2} \, mg/cm^3$. Oscillations in water and water solutions with nearly the same periods have been registered by the light-scattering method by Chernikov (1985).

Chernikov (1990d) has studied the dependence of light scattering fluctuations on temperature, mechanic perturbation and magnetic field in water and water hemoglobin and DNA solution. It has been shown that an increase in temperature results in the decline of long-term oscillation amplitude and in the increase of short-time fluctuation amplitude. Mechanical mixing removes long- term fluctuations and over 10 hours are spent for their recovery. Regular fluctuations (oscillations) appear when the constant magnetic field above 240 $A/m$ is applied; the fluctuations are retained for many hours after removing the field. The period of long-term oscillations has the order of 10 minutes. It has been assumed that the maintenance of long-range correlation of molecular rotation-translation fluctuation underlies the mechanism of long-term light scattering fluctuations.

It has been shown (Chernikov, 1990b) that a pulsed magnetic field (MF), like constant magnetic field, gives rise to light scattering oscillations in water and other liquids containing H atoms: glycerin, xylol, ethanol, a mixture of unsaturated lipids. All this liquids also have a distinct response to the constant MF. "Spontaneous" and MF-induced fluctuations are shown to be associated with the isotropic component of scattering. These phenomena do not occur in the nonproton liquid (carbon tetrachloride) and are present to a certain extent in chloroform (containing one hydrogen atom in its molecule). The facts obtained indicate an important role of hydrogen atoms and cooperative system of hydrogen bonds in "spontaneous" and induced by external perturbations macroscopic oscillations.

The understanding of such phenomena can provide a physical basis for of self-organization (Prigogine, 1980, 1984, Babloyantz, 1986), the biological system evolution and chemical processes oscillations (Field and Burger, 1988).

*It is quite probable that macroscopic oscillation processes in biological liquids, e.g. blood and liquor, caused by the properties of water, are involved in animal and human physiological processes.*

We have registered the oscillations of water activity in the protein-cell system by means of light microscopy using the apparatus "Morphoquant", through the change of the erythrocyte sizes, the erythrocytes being ATP-exhausted and fulfilling a role of the passive osmotic units. The revealed oscillations have a few minute-order periods.

Preliminary data obtained from the analysis of oscillation processes in the human cerebrospinal liquor indicate their dependence on some pathology. Perhaps, the autooscillations spectrum of the liquor can serve as a sensitive test for the physiological status of the organism. The liquor is an electrolyte and its autooscillations can be modulated with the electromagnetic activity of the brain.

We suggested that the activity of the central nervous system and the biological rhythms of the organism are dependent with the oscillation processes in the liquor. If it is the case, then the directed influence on these autooscillations processes, for example, by means of magnetic field makes it possible to regulate the state of the organism and its separate organs. Some of reflexo-therapeutic effects can be caused by correction of biorhythms.

During my three month joint work with G.Salvetty in the Institute of Atomic and Molecular Physics in Pisa (Italy) in 1992, the oscillations of heat capacity $[C_p]$ in 0.1 M phosphate buffer (pH7) and in 1% solution of lysozyme in the same buffer at $20^0 C$ were revealed. The sensitive adiabatic differential microcalorimeter was used for this aim. The biggest relative amplitude changing: $[\Delta C_p]/[C_p] \sim (0.5 \pm 0.02)\%$ occurs with period of about 24 hours, i.e. corresponds to circadian rhythm.

Such oscillations can be stimulated by the variation of magnetic and gravitational conditions for the each point of the Earth during cycle of its rotation as respect to Sun.

*14.3.1 The coherent radio-frequency oscillations in water, revealed by C.Smith*

It was shown experimentally by C. Smith (1994) that the water display a coherent properties under the action of biofield radiated by palm. He shows also, that water is capable of retaining the frequency of an alternating magnetic field. For a tube of water placed inside a solenoid coil, the threshold for the alternating magnetic field, potentiating electromagnetic frequencies into water, is 7.6 $\mu T$ (rms). He comes to conclusion that the frequency information is carried on the magnetic vector potential.

He revealed also that in a course of yeast cells culture synchronously dividing, the radio-frequency emission around 1 MHz ($10^6$ 1/$s$), 7-9 MHz (7-9×$10^6$ 1/$s$) and 50-80 MHz (5-9×$10^7$ 1/$s$) with very narrow bandwidth (~50 Hz) might be observed for a few minutes.

These frequencies could correspond to frequencies of different water collective excitations, introduced in our Hierarchic theory, like [lb/tr] macroconvertons, the $[a \rightleftharpoons b]_{lb}$ transitons, etc. (see Fig. 48), taking into account the deviation of water properties in the colloid and biological systems as respect to pure one.

Cyril Smith has proposed that the increasing of coherence radius in water could be a consequence of coherent water clusters association due to Josephson effect (Josephson, 1965). As far primary librational effectons are resulted from partial Bose-condensation of molecules, this idea looks quite probable in the framework of our Hierarchic theory. The similar idea was used in our Hierarchic model of superfluidity for explanation of chain-like polyeffectons formation (see sections 12.3-12.6). The mechanism of slow macroscopic oscillations, proposed in section 14.2 could be valid for faster oscillations with radio-frequencies as well.

The coherent oscillations in water, revealed by C. Smith could be induced by coherent electromagnetic radiation of microtubules (MT) of cells of palm's skin, produced by correlated intra-MT water excitations in neuronal ensembles (see Section 17.5 and Fig. 48).

**14.4. Phenomena in water and aqueous systems, induced by magnetic field**

The changes of refraction index of bidistilled water during 30 min. in rotating test-tube under magnetic field treatment with tension up to 80 kA/m has been demonstrated already (Semikhina, 1981). For each magnetic field tension the optimal rotation speed, corresponding to maximum effect, has been revealed. For example, for geomagnetic field tension (46.5 A/m) the optimal number of rotation per minute (rpm) was 790 rpm.

In the work of (Semikhina and Kiselev, 1988, Kiselev et al., 1988, Berezin et al., 1988) the influence of the artificial weak magnetic field on the dielectric losses, the changes of dissociation constant, density, refraction index, light scattering and electroconductivity, the coefficient of heat transition, the depth of super-cooling for distilled water and for ice was studied. This weak alternating field was used as a modulator a geomagnetic field action.

The absorption spectra and the fluorescence of the dye (rhodamine 6G) and proteins in solutions also changed under the action of weak magnetic fields on water.

The influence of permanent geomagnetic, modulated by weak alternating magnetic fields on water and ice in the frequency range $(10^4 - 10^8) Hz$ was studied. The maximum sensitivity to field action was observed at the frequency $v_{max} = 10^5 Hz$. In accordance with our calculations, this frequency corresponds to frequency of super-deformons excitations in water (see Fig. 48d).

A few of physical parameters changed after the long (nearly 6 hour) influence of the variable fields ($\tilde{H}$), modulating the geomagnetic field of the tension $[\mathbf{H} \sim \mathbf{H}_{geo}]$ with the frequency in the range of $f = (1 \div 10) \times 10^2 Hz$ (Semikhina and Kiselev, 1988, Kiselev et al., 1988)[18,19]:

$$\tilde{\mathbf{H}} = \mathbf{H} \cos 2\pi ft \qquad 14.14$$

In the range of modulating magnetic field (**H**) tension from $0.08\,A/m$ to $212\,A/m$ the *eight maxima of dielectric losses tangent* in the above mentioned (*f*) range were observed. Dissociation constant decreases more than other parameters (by 6 times) after the incubation of ice and water in magnetic field. The relaxation time ("memory") of the changes, induced in water by fields was in the interval from 0.5 to 8 hours.

The authors explain the experimental data obtained, because of influence of magnetic field on the probability of proton transfer along the net of hydrogen bonds in water and ice, which lead to the deformation of this net.

However, this explanation is doubtful, as far the *equilibrium constant* for the reaction of dissociation:

$$H_2O \Leftrightarrow OH^- + H^+ \qquad 14.15$$

in ice is less by almost six orders ($\simeq 10^6$) than that for water. On the other hand the values of the *field-induced effects in ice are several times more than in water*, and the time for reaching them in ice is less.

In the framework of our Hierarchic theory all the aforementioned phenomena could be explained by the shift of the ($a \Leftrightarrow b$) equilibrium of primary translational and librational effectons to the left and decreasing the ions hydration. In turn, this shift stimulates polyeffectons or coherent superclusters growth, under the influence of magnetic fields. Therefore, parameters such as light scattering have to enhance, while the $H_2O$ dissociation constant depending on the probability of super-deformons must decrease. The latter correlate with declined electric conductance, revealed in experiments.

As far, the magnetic moments of molecules within the coherent superclusters or polyeffectons, formed by primary librational effectons are the additive parameters, then the values of changes induced by magnetic field must be proportional to polyeffectons sizes. These sizes are markedly higher in ice than in water and decrease with increasing temperature.

Inasmuch the effectons and polyeffectons interact with each other by means of phonons (i.e. the subsystem of secondary deformons), and the velocity of phonons is higher in ice than in water, then the saturation of all concomitant effects and achievement of new equilibrium state in ice is faster than in water.

The frequencies of geomagnetic field modulation, at which changes in the properties of water and ice have maxima can correspond to the eigen-frequencies of the $[a \Leftrightarrow b]$ equilibrium constant of primary effectons oscillations, determined by [assembly $\Leftrightarrow$ disassembly] equilibrium oscillations for coherent super clusters or polyeffectons.

The presence of dissolved molecules (ions, proteins) in water or ice can influence on the initial $[a \Leftrightarrow b]$ equilibrium dimensions of polyeffectons and, consequently the interaction of solution with outer field.

Narrowing of $^1$H-NMR lines in a salt-containing water and calcium bicarbonate solution was observed after magnetic field action. This indicates that the degree of ion hydration is decreased by magnetic treatment. On the other hand, the width of the resonance line in *distilled water* remains unchanged after 30 minute treatment in the field (135 $kA/m$) at water flow rate of 60 $cm/s$ (Klassen, 1982).

The hydration of diamagnetic ions ($Li^+$, $Mg^{2+}$, $Ca^{2+}$) decreases, while the hydration of paramagnetic ions ($Fe^{3+}$, $Ni^{2+}$, $Cu^{2+}$) increases. It leads from corresponding changes in ultrasound velocity in ion solutions (Duhanin and Kluchnikov, 1975).

There are numerous data which pointing to increasing of coagulation of different particles and their sedimentation velocity after magnetic field treatment. These phenomena point to dehydration of particles in treated by magnetic field aqueous solutions, stimulating particles association. In turn, this provide a reducing the scale formation in heating systems, widely used in practice. Crystallization and polymerization also increase in magnetic field. It points to decrease of water activity, deterioration of its properties as a solvent.

### *14.4.1. The quantitative analysis of magnetic field influence on water using program Comprehensive Analyzer of Matter Properties (pCAMP)*

For better understanding the character of perturbations, induced by magnetic fields on water, and the mechanism of water 'memory' we perform a systematic quantitative investigation, using our Hierarchic theory based computer program *Comprehensive Analyzer of Matter Properties (pCAMP)* (copyright, 1997, Kaivarainen).

In order to keep pH stable, the 1 mM phosphate buffer, pH 7.0 has been used as a test system. The influence of permanent magnetic field with tension:

$$\mathbf{H} = 200\,mT = 2000\,G = 160\,kA/m \qquad 14.16$$

with space between the two pairs of North and South poles of permanent magnets, equal to 17.5 mm on weak buffer, has been studied.

The volume of liquid of about 6 ml in standard glass test-tubes with diameter 10 mm was rotated during 21 hours and sometimes 70 hours at room temperature: $t = (23 \pm 0.5)^0 C$.

In our work for study the non-monotonic, nonlinear character of **H**- field interaction with water, we used few rotational frequencies (rotations per minute, rpm) of test-tube in magnetic field (2.1) with weak buffer: [50] rpm; [125] rpm and [200] rpm. For the control measurements, the same volume of buffer in similar test-tube, similar temperature and light conditions were used in the absence of external magnetic field (2.1) and static conditions. The possible influence of test-tube rotation on water properties inside without magnetic field action has been controlled also.

Two degassing procedures of buffer samples in test-tubes in vacuum chamber were performed: the 1st - before starting the magnetic field treatment of test-tubes and 2nd - after treatment just before measurements. The relaxation time (memory) of the induced by magnetic field changes in water at $23^0$ C was at least 48 hours. This time decreases with temperature of treated buffer increasing.

Three parameters has been checked for evaluation of magnetic field influence on water structure in buffer, using the same samples during 1.5-2 hours after finishing of H -treatment, much less than relaxation time at $25^0$C:
1. Density;
2. Sound velocity;
3. Refraction index;
4. Position of translational (tr) and librational (lb) bands in IR spectra.

The density and sound velocity were measured simultaneously in the same cuvette at $(25.000 \pm 0.001)^0 C$, using device from Anton - Paar company (Austria): DSA- 5000. The molar volume of substance is equal to ratio of its molecular mass (gr./M) to density (gr/m$^3$):

$$\mathbf{V}_M (m^3/M) = \frac{Mol.\ mass\ (gr/M)}{density\ (gr/m^3)} \qquad 14.17$$

It means that the bigger is density of substance, the less is its molar volume ($\mathbf{V}_M$).

The refraction index was measured, using differential refractometer (Knouffer, Germany) of high sensitivity at temperature 25$^0$C. The IR spectra of attenuated total reflection (ATR) in the middle IR range of the control and **H**-treated samples were registered by FR-IR spectrometer (Perkin -Elmer).

The results of density, sound velocity, refraction index and positions of translational and librational bands in FT- IR spectra are presented at the different frequencies of test-tube rotation during 21 hours are presented in Table 1.

**TABLE 3.**
**The basic experimental parameters of 1 mM phosphate buffer, pH 7.0 ($25^0 C$), treated by H-field at different frequencies of test-tube rotation and the control - untreated buffer**

| Frequency of test-tube rotation (rpm) | 50 | 125 | 200 | Untreated by H-field buffer |
|---|---|---|---|---|
| Molar volume (m$^3$/M) | $1,80541 \times 10^{-5}$ | $1,80473 \times 10^{-5}$ | $1,80540 \times 10^{-5}$ | $1,80538 \times 10^{-5}$ |
| Sound velocity (m/sec.) | 1496,714 | 1495,014 | 1496,874 | 1496,614 |
| Refraction index | 1,333968 | 1,333928 | 1,333970 | 1,333978 |
| Positions of tr. bands (cm$^{-1}$) | 194 | 194 | 194 | 194 |
| Positions of lb. bands (cm$^{-1}$) | 682,5 | 682,5 | 682,5 | 682,5 |

All the experiments were repeated 3-5 times with close results. The described experiments was repeated 5 times with qualitatively same results.

The strongest effects of H-field treatment of buffer were obtained at test-tube rotating velocity 125 rpm during 22 hours at 23 C. The control buffer was incubated in similar test-tube without H -field and rotation.

The molar volume and sound velocity in water after magnetic treatment at 50 rpm and 200 rpm increase as respect to control and are smaller and opposite to these values, obtained at 125 rpm. The date obtained point to complicated nonlinear polyextremal dependence of treated buffer effects on rotation frequency of test-tube in permanent magnetic field.

The difference between treated by magnetic field 1mM phosphate buffer, pH 7.0 (sample) and control buffer (reference) was measured on differential refractometer just after evaluation of density and sound velocity of the same liquids. In all test tube rotation frequencies, the refraction index decreases. Like for density and sound velocity, the strongest change in refraction index were obtained at rotating velocity 125 rpm during 22 hours at 23 C.

The reversibility and relaxation time (memory) after incubation of the treated at 125 rpm buffer at 23 C during 22 h and degassing was studied. The starting values of differences with control of density, sound velocity and refraction index decreases about 5 times. However, the relaxation even over 70 h at 23 C of the effects in water, induced by magnetic field was not total. After incubation of the magnetized and reference/control buffer in water bath at 100$^0$C during 5 minutes, the differences between them decreases about 10 times and sound velocity - about 4 times.

The change of refraction index of buffer at 50 rpm and 200 rpm in H-field is few times smaller, than at (125) rpm, however, has the same sign, in contrast to changes of density and sound velocity. The control experiment on the influence of rotation of tube with buffer in magnetic field in conditions, when rotation is absent, shows, that the changes in density and sound velocity are negligible, however, the change of refraction index has almost the same value as at 50 and 200 rpm. These results points that different physical parameters of water have different sensibility to magnetic field and rotation effects and could be related to different levels of water structure changes (microscopic / mesoscopic / macroscopic) in magnetic field. Pure rotation of the test-tube at 125 rpm in the absence of H-field do not influence water properties.

The opposite changes of primary input parameters of buffer at different test-tube rotation frequencies, presented at Table 3, are accompanied by the opposite deviations of big number of simulated by our computer program output parameters of quantum excitations of buffer (see Table 2).

The clear correlation has been revealed between the sign of changes of primary (input) parameters and most of calculated (output) ones.

We can see from Table 4, that the molar volume and sound velocity of buffer has been increased after treatment at 50 rpm and 200 rpm, in contrast to decreasing of these parameters at 125 rpm. The refraction index decreases in all three cases.

The full description of parameters, presented in tables could be found in chapter 4 and in my paper: "New Hierarchic Theory of Condensed Matter and its Computerized Application to Water and Ice" on line: http://arxiv.org/abs/physics/0102086.

**TABLE 4.**

**The primary experimental and a part of calculated parameters for control buffer ($p_c$) and their relative deviations**: $\Delta p/p = (p_c - p_H)/p_c$ **in treated by magnetic field buffer ($p_H$) at different frequencies of test-tube rotation ($25^0 C$).**

| Experimental parameters of buffer | Control: untreated by H-field buffer | $\Delta p/p$ 50 rpm | $\Delta p/p$ 125 rpm | 2( |
|---|---|---|---|---|
| Molar volume (m$^3$/M) | $1,805384 \times 10^{-5}$ | $-1.60077 \times 10^{-5}$ | $3.60976 \times 10^{-4}$ | $-4.98$ |
| Sound velocity (m/sec) | $1496,614$ | $-6.68175 \times 10^{-5}$ | $1.06908 \times 10^{-3}$ | $-1.73$ |
| Refraction index | $1,3339779$ | $1 \times 10^{-5}$ | $3.74819 \times 10^{-5}$ | $5.99$ |
| Positions of translational bands (cm$^{-1}$) | 194 | 0 | 0 | |
| Positions of librational bands (cm$^{-1}$) | 682,5 | 0 | 0 | |
| **Calculated parameters of buffer** | | | | |
| 1. Contribution of translations to the total internal energy, J/M | 6456.831 | $1.84378 \times 10^{-4}$ | $-2.852 \times 10^{-3}$ | 5.159 |
| 2. Contribution of librations to the total internal energy, J/M | 4743.6564 | $1.69679 \times 10^{-4}$ | $-2.624 \times 10^{-3}$ | 4.748 |
| 3. The total internal energy, J/M | 11256.701 | $1.78027 \times 10^{-4}$ | $-2.753 \times 10^{-3}$ | 4.981 |

| | Untreated by H-field buffer | Δp/p 50 rpm | Δp/p 125 rpm | Δp/p 200 rpm |
|---|---|---|---|---|
| 4. Total kinetic energy, J/M | 320.643 | $3.17206 \times 10^{-4}$ | $-4.988 \times 10^{-3}$ | $8.6102 \times 1$ |
| 5. Total potential energy, J/M | 10936.058 | $1.7392 \times 10^{-4}$ | $-2.688 \times 10^{-3}$ | $4.87561 \times$ |
| 6. The length of edges of libr. primary effectons, Å | 13.636054 | $-6.6808 \times 10^{-5}$ | $1.069 \times 10^{-3}$ | $-1.73731 \times$ |
| 7. Number of molecules in the edge of primary libr. effectons | 4.270908 | $-6.68242 \times 10^{-5}$ | $1.06909 \times 10^{-3}$ | $-1.737 \times 1$ |
| 8. The ratio of total potential energy to total kinetic energy | 35.1066 | $-1.3923 \times 10^{-4}$ | $2.223 \times 10^{-3}$ | $-3.63151 \times$ |
| 9. Number of mol. in primary librational effectons | 84.5759 | $-1.8441 \times 10^{-4}$ | $2.844 \times 10^{-3}$ | $-5.16258 \times$ |
| 10. Aver. dist. between primary librational effectons, Å | 46.057 | $-6.6808 \times 10^{-5}$ | $1.069 \times 10^{-3}$ | $-1.7372 \times$ |
| 11. The internal pressure, Pa | $3.8951763 \times 10^{10}$ | $6.983 \times 10^{-5}$ | $-8.1 \times 10^{-4}$ | $1.52188 \times$ |
| 12. Total thermal conductivity, $10^{-6}$ W/(mK) | 0.75014597 | $2.8661 \times 10^{-6}$ | $6.1415 \times 10^{-5}$ | $4.5058 \times 1$ |
| 13. Vapor pressure, Pa | 3255.4087 | $6.98223 \times 10^{-5}$ | $-8.101 \times 10^{-4}$ | 1.5218 |
| 14. Viscosity of liquids, Pa·s | $7.570872 \times 10^{-4}$ | $4.48641 \times 10^{-4}$ | $-6.988 \times 10^{-3}$ | 1.1057 |
| 15. Total coeff. of self-diffusion in liquids, $m^2/s$ | $1.9779126 \times 10^{-9}$ | $-2.06784 \times 10^{-5}$ | $1.907 \times 10^{-4}$ | $-1.1325$ |
| 16 Surface tension, dynes/cm | 71.2318 | $3.01831 \times 10^{-6}$ | $2.598 \times 10^{-4}$ | $-2.1507$ |
| 17. Acting polarizability of molecules, $cm^3$ | $3.13507 \times 10^{-30}$ | $3.15782 \times 10^{-6}$ | $4.57 \times 10^{-4}$ | 1.0366 |
| 18. The acting field energy, J | $1.76833 \times 10^{-19}$ | $1.459 \times 10^{-5}$ | $2.0917 \times 10^{-3}$ | 4.745 |
| 19. Coeff. of total light scattering, $m^{-1}$ | $1.3686059 \times 10^{-4}$ | $2.24316 \times 10^{-5}$ | $5.532 \times 10^{-4}$ | 2.5793 |
| 20. Contrib of primary libr. effect in (a) state to the total reduced inform, bit | 11.483 | $-1.844 \times 10^{-4}$ | $2.844 \times 10^{-3}$ | $-5.1632$ |
| 21. Contrib. of primary libr. effect in (b) state to the total reduced inform, bit | 0.25 | $-1.844 \times 10^{-4}$ | $2.844 \times 10^{-3}$ | $-5.1625$ |

*14.4.2. Possible mechanism of water properties perturbations under magnetic field treatment*

We can propose few possible targets and corresponding mechanisms of interaction of external permanent magnetic field with pure water and weak buffer, moving in this field:

1) Influence on the coherent molecular clusters (molecular Bose condensate), oriented by laminar flow of water, stimulated their polymerization via Josephson junctions; 2) Influence of Lorentz force on trajectory (orbit) of ions, like $H_3^+O$ and hydroxyl ions $HO^-$. These ions originate as a result of dissociation of water molecules to proton and hydroxyl in the process of cavitational fluctuations, accompanied by high temperature fluctuations:

$$H_2O \rightleftharpoons H^+ + HO^- \rightleftharpoons H_3^+O + HO^- \qquad 14.18$$

In buffers, containing inorganic ions, i.e. phosphate ions $PO_4^{2-}$ and others, these ions are also under the Lorentz force influence.

The stability of closed orbits of ions and their influence on stability of polyeffectons and other properties of water is related to conditions of standing de Broglie waves of corresponding ions in flowing liquid. Other interesting form of self-organization of water systems in permanent magnetic field is a possibility of interaction of internal EM fields with radio-frequency, corresponding to frequency of certain fluctuations and $H_2O$ dipoles reorientations (collective excitations) in water at conditions, close to cyclotronic resonance:

$$\omega_c = 2\pi \nu_c = \frac{ZeB_z}{m^*c} \qquad 14.19$$

The permanent magnetic field, inducing cyclotronic orbits of ions and the alternating EM field, accompanied cyclotronic resonance at certain conditions, could be external as respect to liquid or internal, induced by resulting magnetic field of polyeffectons and correlated in volume density/symmetry of water dipoles fluctuations.

*Let us discuss at first the interaction of magnetic field with macroscopically diamagnetic matter*, like water, as an example. The magnetic susceptibility ($\chi$) of water is a sum of two opposite contributions (Eizenberg and Kauzmann, 1969):

1) average negative diamagnetic part, induced by external magnetic field:

$$\bar{\chi}^d = \frac{1}{2}(\chi_{xx} + \chi_{yy} + \chi_{zz}) \cong -14.6(\pm 1.9) \times 10^{-6} \qquad 14.20$$

2) positive paramagnetism related to the polarization of water molecule due to asymmetry of electron density distribution, existing without external magnetic field. Paramagnetic susceptibility ($\chi^p$) of $H_2O$ is a tensor with the following components:

$$\chi^p_{xx} = 2.46 \times 10^{-6}; \quad \chi^p_{yy} = 0.77 \times 10^{-6}; \quad \chi^p_{zz} = 1.42 \times 10^{-6} \qquad 14.21$$

The resulting susceptibility:

$$\chi_{H_2} = \bar{\chi}^d + \bar{\chi}^p \cong -13 \times 10^{-6} \qquad 14.22$$

The second contribution in the magnetic susceptibility of water is about 10 times lesser than the first one. But the first contribution to the magnetic moment of water depends on external magnetic field and must disappear when it is switched out in contrast to second one.

The coherent primary librational effectons of water even in liquid state contain about 100 molecules $\left[ (n_M^{ef})_{lb} \simeq 100 \right]$ at room temperature (Fig. 7). In ice this number is $(n_M^{ef})_{lb} \geq 10^4$ (fig.6). In (a)-state the vibrations of all these molecules are synchronized in the same phase, and in (b)-state - in counterphase. Correlation of $H_2O$ forming effectons means that the energies of interaction of water molecules with external magnetic field are additive: $\epsilon^{ef} = n_M^{ef} \mu_p H$

In such a case this total energy of effecton interaction with field may exceed thermal energy: $\epsilon^{ef} > kT$

In the case of polyeffectons formation, as a result of association of primary librational effectons via Josephson junctions, this inequality becomes much stronger. Corresponding chains, formed by mesoscopic magnets may self-organize to long-life metastable 3D structures, influencing macroscopic properties of water, like viscosity, self-diffusion, surface tension, thermal conductivity, light scattering. These macroscopic changes are interrelated with changes on microscopic molecular level, like polarizabilities and related parameters, like refraction index, acting field energy and light scattering (Kaivarainen,.2001)

The life-time of closed ionic orbits, corresponding to standing waves conditions and 3D polyeffectons determines the relaxation time of changes in water properties (memory) after magnetic treatment.

The energy of interaction of primary librational effectons and other collective excitations in water with external magnetic field is dependent on few factors:

1. Dimensions of primary librational effectons;
2. Stability of primary librational effectons;
3) Interaction with $H_3^+O$ and hydroxyl ions (see eq. 2.3) and the guest molecules (inorganic ions or organic molecules, polymers, including proteins, lipids, etc.).

The effectons dimensions are determined by the most probable momentums of water molecules in selected directions: $\lambda_{1,2,3} = h/p_{1,2,3}$

The stability of the effectons is dependent on number of molecules $[n_M(L_{lb})]$ in the edge of primary librational effectons, approximated by cube.

The closer this number is to integer one, the more stable are effectons and more favorable condition for their polymerization are existing. For the other hand, the closer this number is to semi-integer values, the less stable are primary effectons and smaller is probability of their 'head' to 'tail' polymerization and closed cycles/rings of polyeffectons formation.

If water is flowing in a tube, it increases the relative orientations of all effectons in volume and stimulate the coherent superclusters formation. All the above discussed effects must increase. Similar ordering phenomena happen in a rotating test-tube with liquid. After switching off the external magnetic field, the relaxation of *induced ferromagnetism* in water begins. It may be accompanied by the oscillatory behavior of water properties.

Remnant ferromagnetism in water was experimentally established using a SQUID superconducting magnetometer by Kaivarainen et al. in 1992 (unpublished data). Water was treated in permanent magnetic field $500\,G$ for two hours. Then it was frozen and after switching off the external magnetic field the remnant ferromagnetism was registered at helium temperature. Even at this low temperature a slow relaxation in form of decreasing of ferromagnetic signal was revealed.

*14.4.3 Cyclotronic frequency of ions in rotating tube with water and possible mechanism of water structure stabilization/destabilization under magnetic field treatment*

The frequency of rotation of charged particle with charge (Ze) and effective mass ($\mathbf{m}^*$), moving in permanent magnetic field $\mathbf{H}$ with velocity $\mathbf{v}$ along circle orbit with radius ($\mathbf{r}$), is determined by the equality of the Lorentz force ($\mathbf{F}$) and centripetal force:

$$\mathbf{F} = e\mathbf{E} + \frac{|Ze|}{c}[\mathbf{v}\,\mathbf{B}] = \frac{\mathbf{m}^*\mathbf{v}^2}{\mathbf{r}} \qquad 14.23$$

In the absence of electric field ($\mathbf{E} = 0$), we get from (2.11) the known formula for cyclotronic frequency:

$$\omega_c = 2\pi\nu_c = \frac{1}{\mathbf{r}/\mathbf{v}} = \frac{|Ze|\mathbf{H}}{\mathbf{m}^*\mathbf{c}} \qquad 14.24$$

The radius of cyclotronic orbit is dependent on velocity of charged particle and its de Broglie wave length radius ($\mathbf{L}_B = \hbar/\mathbf{m}^*\mathbf{v}$) as:

$$\mathbf{r} = \frac{\mathbf{c}}{|Ze|\mathbf{H}}\mathbf{m}^*\mathbf{v} = \frac{\hbar\mathbf{c}}{|Ze|\mathbf{H}}\frac{1}{\mathbf{L}_B} = n\mathbf{L}_B \qquad 14.25$$

The condition of orbit stability, corresponds to that of de Broglie standing waves condition: $\mathbf{r} = 1\mathbf{L}_B;\ 2\mathbf{L}_B;\ 3\mathbf{L}_B,\ldots n\mathbf{L}_B$

From (14.25) we get these conditions of stability in form:

$$n = 1,\,2,\,3\ldots = \frac{\hbar\mathbf{c}}{|Ze|\mathbf{H}}\frac{1}{\mathbf{L}_B^2} = \frac{\mathbf{c}}{|Ze|\mathbf{H}}\frac{(\mathbf{m}^*\mathbf{v})^2}{\hbar} \qquad 14.26$$

The velocity of particle, moving in the volume of liquid in rotating test-tube with internal radius 0.5 cm on distance $\mathbf{R} = 0.3$ cm from the center is:

$$\mathbf{v} = \frac{2\pi\mathbf{R}}{T_{rot}} = 2\pi\nu_{rot}\,\mathbf{R} = \omega_{rot}\mathbf{R} \qquad 14.27$$

Putting (14.27) into (14.26), we get the formula for rotation frequency of test-tube in permanent magnetic field, corresponding to stable cyclotronic orbits of charge (Ze) with effective mass ($\mathbf{m}^*$) at given distance ($\mathbf{R}$) from center of rotation:

$$\omega_{rot} = \frac{n}{\mathbf{R}\mathbf{m}^*}\left(\frac{\hbar\,|Ze|\mathbf{H}}{c}\right)^{1/2} \qquad 14.28$$

or in dimensionless mode:

$$\mathbf{n} = \omega_{rot}\frac{\mathbf{R}\mathbf{m}^*}{(\hbar\,|Ze|\mathbf{H}/c)^{1/2}} = \text{integer number} \qquad 14.29$$

Taking into account formula for cyclotronic resonance (14.24), we get from (14.28) the relation between the test-tube rotation angle frequency and cyclotron frequencies:

$$\omega_{rot} = \frac{n}{\mathbf{R}}\left(\frac{\hbar}{\mathbf{m}^*}\omega_c\right)^{1/2} \qquad 14.30$$

One of the consequence of our formulas is that the stability of cyclotronic orbit and related stability of water structure in test-tube should have the polyextremal dependence on rotation frequency ($\nu_{rot}$) of the test tube. A stable water structure would corresponds to integer values of (**n**) and unstable water structure to its half-integer values. We have to keep in mind that formulas (14.27 - 14.30) are approximate, because the component of Lorentz force: $\frac{Ze}{c}[\mathbf{v}\,\mathbf{B}]$ is true only for particles, *moving in the plane, normal* to direction of **H** vector. Obviously, in the case of rotating test-tube with water and ions this condition do not take a place. However, for qualitative explanation of nonlinear, polyextremal dependence of the water parameters changes on rotation frequency of test-tube, the formulas obtained are good enough. They confirm the proposed mechanism of water structure stabilization or destabilization after magnetic treatment, as a consequence of stable or unstable cyclotronic orbits of ions origination in water and aqueous solutions.

The cyclotronic resonance represents the absorption of alternating electromagnetic field energy with frequencies ($\omega_{EM}$), equal or multiple to cyclotron frequency:

$$\hbar\omega_{EM} = n\omega_c, \quad where\ n = 1,2,3,\ldots \qquad 14.31$$

The integer **n** correspond to condition of maximum of EM energy absorption and half-integer value - to minimum absorption. If the frequency $\omega_{EM}$ is close to one of the radio-frequencies of correlated fluctuations in water, presented on Fig.3.2, then the auto-cyclotron resonance effect in liquid is possible. Such new phenomena may represent a new kind of macroscopic self-organization in water and other, ions-containing liquids at selected conditions.

### 14.5 Influence of weak magnetic field on the properties of solid bodies

*It has been established that as a result of magnetic field action on solids with interaction energy ($\mu_B H$) much less than kT, many properties of matter such as hardness, parameters of crystal cells and others change significantly.*

The short-time action of magnetic field on silicon semiconductors is followed by a very long (many days) relaxation process. The action of magnetic field was in the form of about 10 impulses with a length of 0.2 ms and an amplitude of about $10^5 A$/m. The most interesting fact was that this relaxation had an oscillatory character with periods of about several days (Maslovsky and Postnikov, 1989).

Such a type of long period oscillation effects has been found in magnetic and nonmagnetic materials.

This points to the general nature of the macroscopic oscillation phenomena in solids and liquids.

The period of oscillations in solids is much longer than in liquids. This may be due to bigger energy gap between (*a*) and (*b*) states of primary effectons and polyeffectons and much lesser probabilities of ($a \rightarrow b$) transitons and deformons excitation. Consequently, the relaxation time of $(a \Leftrightarrow b)_{tr,lb}$ equilibrium shift in solids is much longer than in liquids. The oscillations originate due to instability of dynamic equilibrium between the subsystems of effectons and deformons and external perturbations.

The attempt to make a theory of magnetic field influence on water based on other model were made earlier (Yashkichev, 1980). However, this theory does not take into account the quantum properties of water and cannot be considered as satisfactory one.

The huge material obtained by Shnoll and his team (1965-2006) when studying the macroscopic oscillations of very various processes, reveals their fundamental character and dependence on gravitation and certain cosmic factors. The interpretation of these unusual results is given by our Unified theory of Bivacuum, particles duality, fields and time (Kaivarainen, 2006a, 2006b).

The next Chapter is devoted to water-protein interaction on mesoscopic level and its role in proteins function.

# Chapter 15

# Role of inter-domain water clusters in large-scale dynamics of proteins

The functioning of proteins, like antibodies, numerous enzymes, hemoglobin, etc. is caused by the physico-chemical properties, geometry and dynamics of their active sites. The mobility of an active site is related to dynamics of the residual part of a protein molecule, its hydration shell and properties of the bulk solvent.

The dynamic model of a proteins proposed by this author in 1975 and confirmed later by numerous data (Kaivarainen, 1985, 1989b, 1995), is based on the following statements:

1. A protein molecule contains one or more inter-domain and inter-subunit cavities or clefts capable to large - scale (LS) pulsations between two states: "closed" (A) and "open" (B) with lesser and bigger accessibility to water, correspondingly.

The frequency of pulsations ($\nu_{A \Leftrightarrow B}$):

$$10^4 s^{-1} \leq \nu_{A \Leftrightarrow B} \leq 10^7 s^{-1}$$

depends on the structure of protein, its ligand state, temperature and solvent viscosity. Transitions between A and B states are the result of the relative rotational-translational displacements of protein domains and subunits forming the cavities;

2. The water, interacting with protein, consists of two main fractions:
   - the 1st major fraction solvates the outer surface of protein and has less cooperative properties than
   - the 2nd, minor fraction of water is confined in protein cavities.

The water molecules, interacting with cavity in the "open" (B)-state form a cooperative cluster, with the lifetime ($\geq 10^{-10} s^{-1}$). The properties of clusters are determined by the volume, geometry, stability and polarity of the cavities and by temperature and pressure.

It is seen from X-ray structural data that the protein cavities, like the active sites (AS) and space between subunits of oligomeric proteins, usually are highly nonpolar.

In contrast to small intra-domain holes, isolated from the outer medium, containing only few $H_2O$ molecules as a structural elements of proteins, the big interdomain and inter-subunit cavities in open state can contain several dozens of molecules (Fig. 47). This water fraction easily exchange with bulk water in the process of proteins large-scale pulsation.

The development of this dynamic model has led us to the following classification of the native globular proteins dynamics:

*1. Small-scale (SS) dynamics:*

the low amplitude ($\leq 1$ Å) thermal fluctuations of atoms, aminoacids residues, and displacements of $\alpha$-helixes and $\beta$-structures within domains and subunits. The SS dynamics do not accompanied by the change of the effective Stokes radius of proteins. However, the SS dynamics of domains and subunits can differ in the content of A and B conformers (Fig. 47, dashed line). The range of characteristic times at SS dynamics is $(10^{-4} - 10^{-11})s$, determined by activation energies of corresponding transitions.

*2. Large-scale (LS) dynamics:*

is subdivided into LS-*pulsations* and LS-*librations,* represented limited translations and rotation of domains and subunits in proteins:

a) *pulsations* are the relative translational-rotational displacements of domains and subunits at distances ≽ 3Å. The cavities, formed by domains and subunits, fluctuate between less (A) and more (B) accessible to water states. The life-times of these states depending on stability of cavities, water clusters and temperature are in the range $(10^{-4} - 10^{-7})$ s.

One of this life-time factors is a frequency of excitations of $[lb \rightleftharpoons tr]$ macroconvertons, i.e. frequency of water cluster flickering. The frequency of macroconvertons excitation at normal conditions is about $10^7$ (1/s).

The pulsation frequency of big central cavity in oligomeric proteins is about $10^4$ (1/s). It could be defined by stronger fluctuations of water cluster in this cavity, like macro-deformons or super-deformons (Fig.49 c,d).

The life-times of (A) and (B) conformer markedly exceeds the transition-time between them: $(10^{-9} - 10^{-11})$ s.

The $(A \Leftrightarrow B)$ pulsations of various cavities in proteins are correlated. The corresponding A and B conformers have different Stokes radii and effective volumes.

The geometrical deformation of the inter-subunits large central cavity of oligomeric proteins and the destabilization of the water cluster located in it, lead to corresponding relaxation of $(A \Leftrightarrow B)$ equilibrium constant:

$$K_{A \Leftrightarrow B} = \exp\left(-\frac{G_A - G_B}{RT}\right).$$

The dashed line at Fig. 49c means that the stability and the small-scale dynamics of domains and subunits in the content of A and B conformers can differ from each other. The $[A \Leftrightarrow B]$ pulsations are accompanied by reversible sorption$\rightleftharpoons$desorption of $(20 - 70) H_2O$ molecules from the cavities.

Structural protein domains are space-separated formations with a mass of $(1 - 2) \times 10^3$ D.

The protein subunits ($MM \geq 2 \times 10^3 D$), as a rule, consist of 2 or more domains. The domains can consist only of $\alpha$ or only of $\beta$-structure or may have no secondary structure at all (Schulz, Schirmer, 1979).

The shift of $A \Leftrightarrow B$ equilibrium of central cavity of oligomeric proteins determines their cooperative properties during consecutive ligand binding in the active sites. Signal transmission from the active sites to the remote regions of macromolecules is also dependent on $(A \Leftrightarrow B)$ equilibrium of big central cavity in multi-subunit proteins.

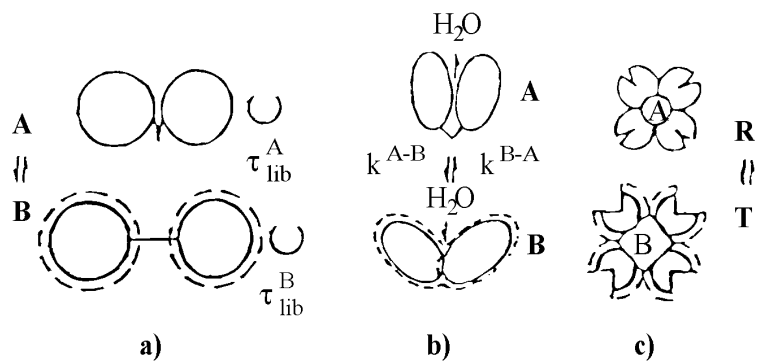

**Figure 47**. Examples of large-scale (LS) protein dynamics: $A \Leftrightarrow B$ pulsations and librations with correlation times $(\tau_{lb}^B < \tau_{lb}^A)$ (Kaivarainen, 1985, 1989).

 a) mobility of domains connected by flexible hinge or contact region, like in the light chains of immunoglobulins;
 b) mobility of domains that form the active sites of proteins, like in hexokinase, papain, pepsin, lysozyme etc. due to flexibility of contacts;
 c) mobility of subunits forming the oligomeric proteins like hemoglobin. Besides transitions of the active sites of each subunit, the $(A \Leftrightarrow B)$ pulsations of big central with frequencies of $(10^4 - 10^6)$ $s^{-1}$ stand for allosteric properties of oligomeric proteins;

The *librations* represent the relative rotational - librational motions of domains and subunits in the content of both: A and B conformers with different correlation times $\tau_M \simeq (1-5) \times 10^{-8}$s.

The process of "flickering" of water cluster in the open cavity between domains or subunits, necessary for librations, is close to the reversible first-order phase transition, when:

$$\Delta G_{H_2O} = \Delta H_{H_2O} - T\Delta S_{H_2O} \approx 0$$

Such type of transitions in water-macromolecular systems could be responsible for so called "enthalpy-entropy compensation effects" (Lumry and Biltonen, 1969).

The "flickering clusters" means excitation of [*lb/tr*] conversions between librational and translational primary water effectons, accompanied by reversible dissociation of coherent water cluster (primary librational effectons) to number of much smaller translational effectons (see Fig. 18*a,b*).

The water cluster (primary *lb* effecton) association and dissociation in protein cavities in terms of hierarchic model represent the $(ac)$ - convertons or $(bc)$ - convertons. These excitations stimulate the LS- librations of domains in composition of B-conformer. The frequencies of $(ac)$ and $(bc)$ convertons, has the order of about $10^8 c^{-1}$. This value coincides well with experimental characteristic times for protein domains librations within open B-conformer.

The (ac) and (bc) convertons represent transitions between similar states of primary librational and translational effectons: $[a_{lb} \rightleftharpoons a_{tr}]$ and $[b_{lb} \rightleftharpoons b_{tr}]$ (see Introduction).

For the other hand, the Macroconvertons occur at simultaneous excitation of $(ac + bc)$ convertons and accompanied by big fluctuation of water density and exchange with bulk water. These fluctuations are responsible for $[B \rightleftharpoons A]$ large-scale pulsations of proteins.

The librational mobility of domains and subunits is revealed by the fact that the experimental value of $\tau_M$ is less than the theoretical one ($\tau_M^t$) calculated on the Stokes-Einstein formula:

$$\tau_M^t = \frac{V}{k} \eta/T$$

where *k* is a Boltzmann constant; $\eta$ and *T* are the viscosity and temperature of solution.

This formula is based on the assumption that the whole protein can be approximated by a rigid sphere with volume *V*. It means, that the large-scale dynamics can be characterized by the "flexibility factor", in the absence of aggregation equal to ratio:

$$fl = (\tau_M/\tau_M^t) \leq 1$$

Antonchenko (1986) has demonstrated, using the Monte-Carlo method for simulations, that the *disjoining pressure* of a liquid in the pores onto the walls changes periodically depending on the distance (**L**) between the limiting surfaces. If the water molecules are approximated by rigid globes, then the maxima of the wedging pressure lie on the values of distance **L**: 9.8; 7; and 3.3Å. It points, that small changes in the geometry of cavities can lead to significant changes in their $A \Leftrightarrow B$ equilibrium constant ($K_{A \Leftrightarrow B}$).

After our model, the large-scale transition of the protein cavity from the "open" B-state to the "closed" A-state consists of the following stages:

1. Small reorientation (libration) of domains or subunits, which form an "open" cavity (B-state). This process is induced by $(ac)$ or $(bc)$ convertons of water librational effecton, localized in cavity (flickering of water cluster);

2. Cavitational fluctuation of water cluster, containing $(20-50)$ $H_2O$ molecules and the destabilization of the B-state of cavity as a result of $[lb \rightleftharpoons tr]$ *macroconvertons* excitation;

3. Collapsing of a cavity during the time about $10^{-10}$ *s*, dependent on previous stage and concomitant rapid structural change in the hinge region of interdomain and inter-subunit contacts: $[B \rightarrow A]$ transition.

The $b \to a$ transition of one of the protein cavities can be followed by similar or the opposite $A \to B$ transition of the other cavity in the same macromolecule/protein.

It should be noted that the collapsing time of a cavitation bubble with the radius: $r \simeq (10 - 15)Å$ in *bulk* water and collapsing time of interdomain cavity are of the same order: $\Delta t \sim 10^{-10} s$ (Shutilov, 1980).

If configurational changes of macromolecules at $B \to A$ and $A \to B$ transitions are sufficiently quick and occur as a jumps of the effective volume, they accompanied by appearance of the shock acoustic waves in the bulk medium.

When the cavitational fluctuation of water in the "open" cavity does not occur, then $(b \to a)$ or $(B \to A)$ transitions are slower processes, determined by continuous diffusion of domains and subunits. This happens when $[lb \to tr]$ macroconvertons are not excited.

In their review, Karplus and McCammon (1986) analyzed data on alcoholdehydrogenase, myoglobin and ribonuclease, which have been obtained using molecular dynamics approach. It has been shown that large-scale reorientation of domains occur together with their deformation and motions of $\alpha$ and $\beta$ structures.

It has been shown also (Karplus and McCammon, 1986) that activation free energies, necessary for $[A \Leftrightarrow B]$ transitions and the reorganization of hinge region between domains, do not exceed (3-4) kcal/mole. Such low values were obtained for proteins with even rather dense interdomain region, as seen from X-ray data. The authors explain such low values of activation energy by the fact that the displacement of atoms, necessary for such transition, does not exceed 0.5 Å, i.e. they are comparable with the usual amplitudes of atomic oscillation at temperatures $20 - 30^0 C$. It means that they occur very quickly within times of $10^{-12} s$, i.e. much less than the times of $[A \to B]$ or $[B \to A]$ domain displacements $(10^{-9} - 10^{-10} s)$. Therefore, the high frequency small-scale dynamics of hinge is responsible for the quick adaptation of hinge geometry to the changing distance between the domains and for decreasing the total activation energy of $[A \Leftrightarrow B]$ pulsations of proteins,.

Recent calculations by means of molecular dynamics reveal that the oscillations in proteins are harmonic at the low temperature (T< 220K) only. At the physiological temperatures the oscillations are strongly anharmonic, collective, global and their amplitude increases with hydration (Steinback et al., 1996). Water is a "catalyzer" of protein anharmonic dynamics.

It is obvious, that both small-scale (SS) and large-scale (LS) dynamics, introduced in our model, are necessary for protein function. For quantitative description of LS dynamics, this author have proposed the unified Stokes-Einstein and Eyring-Polany equation.

### 15.1. Description of large-scale dynamics of proteins, based on generalized Stokes-Einstein and Eyring-Polany equation

In the case of the continuous Brownian diffusion of a particle, the rate constant of diffusion is determined by the Stokes-Einstein law:

$$k = \frac{1}{\tau} = \frac{k_B T}{V \eta} \qquad 15.1$$

where: $\tau$ is correlation time, i.e. the time, necessary for rotation of a particle by the mean angle determined as $\bar{\varphi} \approx 0.5$ of the full turn, or the characteristic time for the translational movement of a particle with the radius (a) on the distance $(\bar{\Delta}_x)^{1/2} \simeq 0.6a$ (Einstein, 1965); $V = 4\pi a^3/3$ is the volume of the spherical particle; $k_B$ is the Boltzmann constant, T and $\eta$ are the absolute temperature and bulk viscosity of the solvent.

On the other hand, the rate constant of $[A \to B]$ reaction for a molecule in the gas phase, which is related to passing over the activation barrier $G^{A \to B}$ in the moment of collision, is described by the Eyring-Polany equation:

$$k^{A \to B} = \frac{kT_B}{h} \exp\left(-\frac{G^{A \to B}}{RT}\right) \qquad 15.2$$

To describe the large-scale dynamics of macromolecules in solution related to fluctuations of domains and subunits (librations and pulsations), this author proposed the equation, taking into account the diffusion and activation processes simultaneously.

The rate constant ($\mathbf{k}_c$) for the rotational- translational diffusion of the particle (domain or subunit) in composition of protein or other macromolecule can be expressed by *generalized Stokes-Einstein and Eyring-Polany formula* (Kaivarainen, 1995, 2000):

$$\mathbf{k}_c = \frac{k_B T}{\eta V} \exp\left(-\frac{G_{st}}{RT}\right) = \tau_c^{-1} \qquad 15.3$$

where: $V$ is the effective volume of domain or subunit, which are capable to the Brownian mobility independently from the rest part of the macromolecule, with the probability:

$$P_{lb} = \exp\left(-\frac{G_{st}}{RT}\right), \qquad 15.4$$

where: $G_{st}$ is the activation energy of structural change in the contact (hinge) region of a macromolecule, necessary for independent mobility of domain or subunit; $\tau_c$ is the effective correlation time for the continuous diffusion of this relatively independent particle. The effective volume $V$ can change under the action of temperature, perturbants and ligands.

The generalized Stokes-Einstein and Eyring-Polany equation (15.3) is applicable also to diffusion of the Brownian particle, dependent on the surrounding medium fluctuations with activation energy ($G_a$). For example, the ligand diffusion throw the active site cavity of proteins is a such type of process.

To describe the *noncontinuous process*, the formula for rate constant ($k_{\text{jump}}$) of the jump-like translations of particle, related to emergency of cavitational fluctuations (holes) near the particle was proposed (Kaivarainen 1995):

$$k_{\text{jump}} = \frac{1}{\tau_{\text{jump}}^{\min}} \exp\left(-\frac{W}{RT}\right) = \frac{1}{\tau_{\text{jump}}}, \qquad 15.5$$

where:

$$W = \sigma S + n_s(\mu_{\text{out}} - \mu_{\text{in}}) \qquad 15.6$$

is the work of *cavitation fluctuation*, when $n_s$ molecules of the solvent (water) change its effective chemical potential from $\mu_{\text{in}}$ to $\mu_{\text{out}}$ and the cavity with surface $S$ is formed.

The dimensions of fluctuating cavities near particle must be comparable with particles size. In a homogeneous phase (i.e. pure water) under equilibrium conditions we have: $\mu_{\text{in}} = \mu_{\text{out}}$ and $W = \sigma S$.

With an increase of particle sizes and the surface (S) and work of cavitational fluctuation (W), the corresponding probability of cavitation fluctuations falls down as:

$$P_{\text{jump}} = \exp(-W/RT)$$

The notion of the surface energy ($\sigma$) retains its meaning even at very small "holes" because of its molecular nature (see Section 11.4).

The $\tau_{\text{jump}}^{\min}$ in eq. (15.5) is a minimal possible jump-time of the particle with mass (m) over the distance $\lambda$ with mean velocity:

$$\mathbf{v}_{\max} = (2kT/m)^{1/2} \qquad 15.7$$

Hence, we derive for the maximal jump-rate at $W = 0$

$$k_{\text{jump}}^{\max} = \frac{1}{\tau_{\text{jump}}^{\min}} = \frac{V_{\max}}{\lambda} = \frac{1}{\lambda}\left(\frac{2kT}{m}\right)^{1/2} \qquad 15.8$$

In the case of hinged domains, forming macromolecules their relative $A \rightleftharpoons B$ displacements (pulsations) are related not only to possible holes forming in the interdomain (inter-subunit) cavities or near their outer surfaces, but to the structural change of hinge regions as well.

If the activation energy of the necessary hinge structure changes is equal to $G_{st}^{A \rightleftharpoons B}$, then eq. (15.5), with regard for (15.8), is transformed into

$$k_{\text{jump}}^{A \Leftrightarrow B} = \frac{1}{\lambda} \left( \frac{2kT}{m} \right)^{1/2} \exp\left( -\frac{W_{A,B} + G_{st}^{A \Leftrightarrow B}}{RT} \right) \qquad 15.9$$

where: $W_{A,B}$ is the work of cavitational fluctuations of water; this work can be different in two directions: $(W_B)$ is accompanied the $B \to A$ transition of protein cavity; $(W_A)$ is accompanied $A \to B$ transition of cavity.

Under certain conditions $A \rightleftharpoons B$ transitions between protein conformers (LS- pulsations) can include both, the *jump-way and continuous* types of relative diffusion of domains or subunits, representing two stage reaction. In this case, the resulting rate constant of these transitions can be expressed through (15.9) and (15.3) as:

$$k_{\text{res}}^{A \Leftrightarrow B} = k_{\text{jump}}^{A \Leftrightarrow B} = k_c^{A \Leftrightarrow B} = \frac{1}{\lambda} \left( \frac{2kT}{m} \right)^{1/2} \exp\left( -\frac{W_{A,B} + G_{st}^{A \Leftrightarrow B}}{RT} \right) + $$

$$+ \frac{kT}{\lambda V} \exp\left( -\frac{G_{st}^{A \Leftrightarrow B}}{RT} \right) \qquad 15.10$$

The interaction between two domains in *A-conformer* can be described using microscopic Hamaker - de Bohr theory. One of the contributions into $G_{st}^{A \Leftrightarrow B}$ is the energy of dispersion interactions between domains of the *radius (a)* (Kaivarainen, 1989b, 1995, 2000):

$$[G_{st}^{dis} \sim U_H \approx -A^* a/12H]_{A,B} \qquad 15.11$$

where

$$A^* \approx (A_s^{1/2} - A_c^{1/2})^2 \approx \frac{3}{2} \pi h v_0^s [\alpha_s N_s - \alpha_c N_c]^2 \qquad 15.12$$

is the complex Hamaker constant; $H$ is the slit thickness between domains in the closed *A-state*; $A_c$ and $A_s$ are the simple Hamaker constants, characterizing the properties of water molecules in the *A-state* of the cavity and in the bulk solvent, correspondingly. They depend on the concentration of corresponding water molecules ($N_c$ and $N_s$) and their polarizability ($\alpha_c$ and $\alpha_s$):

$$A_c = \frac{3}{2} \pi h v_0^c \alpha_c^2 N_c^2 \quad \text{and} \quad A_s = \frac{3}{2} \pi h v_0^s \alpha_s^2 N_s^2; \qquad 15.12a$$

where: $h v_0^c \approx h v_0^s$ are the ionization potentials of $H_2O$ molecules in the cavity and in the bulk solvent.

In the "closed" A-state of protein cavity the water between domains has more compact packing (density) as compared to water cluster in the B-state of cavity. As far $H_A < H_B$ the dispersion interaction (15.11) between domains in A- state of cavity is stronger, than that in B-state: $U_H^A > U_H^B$.

The disjoining pressure of water in the cavities can be described as

$$\Pi = -A^*/6\pi H^3 \qquad 15.12b$$

It decreases with the increase of the complex Hamaker constant ($A^*$) that corresponds to the increase of the attraction energy ($U_H$) between domains.

We can evaluate the changes of $\alpha_s N_s$, measuring the solvent refraction index, as far from (8.20a):

$$(n_s^2 - 1)/n_s^2 = \frac{4}{3} \pi \alpha_s N_s \quad \text{or:} \quad \alpha_s N_s = \frac{3}{4} \pi \frac{n_s^2 - 1}{n_s^2}$$

The water properties in closed cavities are more stable than that in bulk water. As a consequence,

the thermoinduced nonmonotonic transition in the bulk water refraction index ($n_s$) must be accompanied by the in-phase nonmonotonic changes of the complex Hamaker constant (15.12) and the $[A \Leftrightarrow B]$ equilibrium constant ($K_{A \Leftrightarrow B}$). As far the (A) and (B) conformers usually have different flexibility, the changes of $K_{A \Leftrightarrow B}$ will be manifested in the changes of protein large-scale dynamics. The water viscosity itself should have the nonmonotonic temperature dependence due to the nonmonotonic dependence of $n^2(t)$ as it follows from eqs.11.44, 11.45 and 11.48.

Thus, the thermoinduced transitions of macromolecules can be caused by nonmonotonic changes of solvent properties, like water refraction index.

*The influence of $D_2O$ and other perturbants on protein dynamics can be explained in a similar way.*

The effect of deuterium oxide ($D_2O$) is a result of substitution of $H_2O$ from protein cavities and corresponding change of complex Hamaker constant (15.12).

Generalized equation (15.3) is applicable not only for evaluating the frequency of macromolecules transition between A and B conformers, but also for the frequency of the dumped librations of domains and subunits within A and B conformers. Judging by various data (Kaivarainen, 1985, 1989b, 1995), the interval of $A \rightleftharpoons B$ pulsation frequency is:

$$v_{A \Leftrightarrow B} = \frac{1}{t_A + t_B} \approx k^{A \rightarrow B} = (10^4 - 10^7) \, s^{-1} \qquad 15.13$$

where: $t_A$ and $t_B$ are the lifetimes of A and B conformers.

The corresponding interval of the total activation energy of $A \Leftrightarrow B$ pulsations can be evaluated from the eq. (15.9). We assume for this end that the pre-exponential multiplier is about $10^{10} s^{-1}$, as a frequency of cavitational fluctuations in water with the radius of cavities about (10-15) Å.

Taking a logarithm of (15.9) we derive:

$$G_{res}^{A \Leftrightarrow B} = (W_{A,B} + G_{st}^{A \Leftrightarrow B}) \approx RT \times (\ln 10^{10} - \ln v_{A \Leftrightarrow B}) \qquad 15.14$$

We get from this formula, that at physiological temperatures the following energy range corresponds to the frequency range of $A \Leftrightarrow B$ pulsations (15.13)

$$G_{res}^{A \Leftrightarrow B} \approx (4 - 8) \text{ kcal/mole} \qquad 15.15$$

These energies are pertinent to a wide range of biochemical processes.

The jump-way pulsations of proteins can excite the acoustic shock - waves in the solvent, stimulating the inorganic ions pairs formations. The corresponding increasing of water activity provides a distant interaction between different proteins and proteins and cells, described in previous sections.

The *Kramers equation* (1940), widely used for describing diffusion processes, has a shape:

$$k = \frac{A}{\eta} \exp\left(-\frac{H^*}{RT}\right) \qquad 15.16$$

where A is the empirical constant and $\eta$ is a solvent viscosity.

In contrast to Kramers equation, the pre-exponential factor in our *generalized Stokes-Einstein and Eyring-Polany formula* (15.3) contains not only the viscosity, but also the temperature and the effective volume of Brownian particle. It was proved in our experiments that formula (15.3) describes the dynamic processes in solutions of macromolecules/proteins much better than the Kramers equation (15.16).

### 15.2. Dynamic model of protein-ligand complexes formation

According to our model of specific complexes formation, it is accompanied by the following order of stages (Fig. 48):

1. Ligand (L) collides the active site (AS), formed usually by two domains, in its open (b) state: the structure of water cluster in the AS is being perturbed and water is forced out of AS cavity totally or partially;

    2. Transition of AS from the open (b) to the closed (a) state occurs due to strong shift of $[a \Leftrightarrow b]$ equilibrium to the left;

    3. The dynamic adaptation of complex of ligand with active site [L+AS], accompanied by directed ligand diffusion in AS cavity and 'tuning' to its geometry, supported by domains small-scale dynamics;

    4. For the case of oligomeric proteins with few AS, the complex formation with each subunit cause nonlinear perturbation of the geometry of big central cavity between subunits in the open (B) state and destabilization of the large central water cluster, following by nonlinear shift of the $A \rightleftharpoons B$ equilibrium, corresponding to $R \rightleftharpoons T$ equilibrium of quaternary structure leftward. Due to the feedback mechanism such shift of central cavity can influence the $[a \Leftrightarrow b]$ equilibrium of the remaining free from the ligand active sites, promoting their reaction with the next ligand. Every new ligand stimulates this process, promoting the positive cooperativeness. So, the cooperative properties of central water cluster and its 'melting' provides the cooperative character of ligands binding by oligomeric proteins, like in case of hemoglobin;

    5. The final stage of $[protein - ligand]$ complex formation is the relaxation process, representing the deformation of the domains and subunits tertiary structure after relatively fast quaternary structure change. This process could be much slower than the initial ones (1-3), as far the resulting activation barrier is much higher. The stability of specific complex grows up in the process of this relaxation.

*Dissociation of specific complex* is a set of reverse processes to that described above which starts from the $[a^* \rightarrow b]$ fluctuation of the active site cavity.

In multi-domain proteins, like antibodies, which consist of 12 domains, and in oligomeric proteins, like hemoglobin, *the cooperative properties of $H_2O$ clusters in the cavities can determine the mechanism of signal transmission* from the active sites to remote effector regions and interaction between different active sites, providing the allosteric properties.

The stability of librational water effecton, as a coherent cluster, strongly depends on its sizes and geometry. This means that very small deformations of protein cavity, which violate the [cavity - cluster] complementary condition and *clusterphilic* interaction, induce a cooperative shift of $[A \Leftrightarrow B]$ equilibrium leftward. The clusterphilic interaction, introduced in section 13.4 turns to hydrophobic one.

This process can be developed step by step. For example, the reorientation of variable domains, which form the antibodies active sites after reaction with the antigen determinant or hapten deforms the next cavity between pairs of variable and constant domains forming $F_{ab}$ subunits (Fig.48). The leftward shift of $[A \Leftrightarrow B]$ equilibrium of this cavity, in turn, changes the geometry of the big central cavity between $F_{ab}$ and $F_c$ subunits. Therefore, the signal transmission from the AS to $F_c$ subunits occurs due to conversion of clusterphilic interactions and hydrophobic interactions. This mechanism of signal transmission from the active sites to remote regions of multi-domain proteins can be quite general.

The leftward shift of $[A \Leftrightarrow B]$ equilibrium in a number of cavities in the elongated multi-domain proteins, induced by complex-formation of ligands with active sites should lead to the significant decrease of proteins linear size and dehydration. This consequence of our model has a lot of experimental confirmations.

The mechanism of muscular contraction, described in chapter 16, is probably based on described phenomena, as a result of interplay between clusterphilic and hydrophobic interactions.

The $\left[ protein\ cavities\ +\ water\ clusters \right]$ system can be considered, as a nonlinear system, like a "house of cards". Energy is necessary for reorientation of the first couple of domains only. The process then goes on spontaneously with decreasing the resulting protein chemical potential.

The chemical potential of the *A* conformer is usually lower than that of *B* ($\bar{G}_A < \bar{G}_B$) and the relaxation is accompanied by the leftward $A \Leftrightarrow B$ equilibrium shift of cavities. The hydration of proteins must, therefore, decrease, when clusterphilic [cluster-cavity] interaction turns to hydrophobic one.

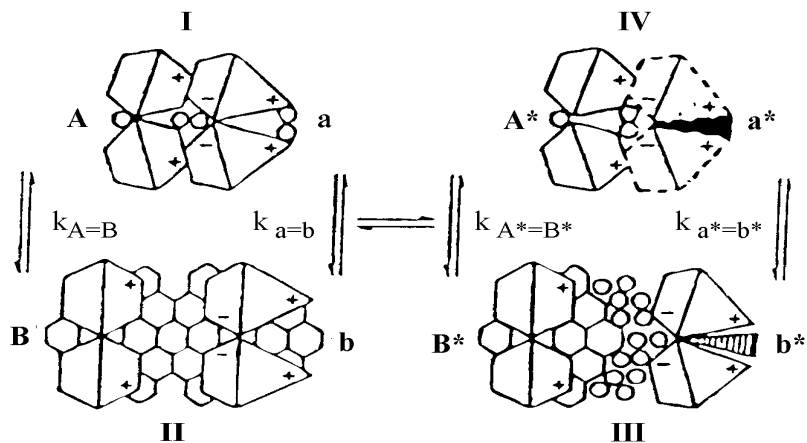

**Fig**. **48**. The schematic picture of the protein - ligand complex formation (the Fab subunit of antibody with hapten, for example), which is accompanied by the destabilization of water clusters in cavities, according to the dynamic model (Kaivarainen, 1985; 1995). The dotted line denotes the perturbation of the tertiary structure of the domains forming the active site. Antibodies of IgG, IgA and IgE types with general Y-like structure, contain usually two such Fab subunit and one Fc subunit, conjugated by flexible hinge.

The dynamic, water mediated model of signal propagation in proteins, described above, is the alternative one to solitonic mechanism of non-dissipative signal transmission (Davidov, 1973). The propagation of solitonic wave is well known nonlinear process in ordered *homogeneous* mediums. The solitons can originate, when the nonlinear effects are compensated by the wave dispersion effects. Dispersion is reflected in fact that the longer waves spreads in medium with higher velocity than the shorter ones.

For the other hand, the biological systems are inhomogeneous and fluctuating ones. They have not a proper features for solitons excitation and propagation. Our dynamic model takes into account the real multidomain and multiglobular dynamic structure of the proteins and their hydration shell fractions being in dynamic exchange with the bulk water. In contrast to solitonic model, the dissipation processes, like reversible melting of water clusters, accompanied the large-scale dynamics of proteins, are the crucial stages in the described above mechanism of signal transmission in biosystems, involving dynamic equilibrium between *hydrophobic and clusterphilic interactions*.

The evolution of the ideas of the ligand - active site complex formation proceeded in the following sequence:

1. The idea of "key - lock" or the *rigid conformity* between the geometry of the active site and that of a ligand (Fisher, 1894);

2. "Hand - glove" or the so-called principle of *induced conformity* (Koshland, 1960);

3. At the current stage, when the role of protein dynamics gets clearer, the *"principle of stabilized conformity"* instead of *"induced conformity"* in protein-ligand reaction was suggested by this author (Kaivarainen, 1985; 1995).

The principle of *stabilized conformity* means that the geometry of the active site, optimal for reaction from energetic and stereochemical point of view, is already existing before reaction with ligand. The optimal geometry of the active site is just selected from the number of possible and stabilized by ligand. However, it is not induced "de nova".

For example, the [a ⇔ b] large-scale pulsations of the active sites due to domain relative thermal fluctuations and stabilization of their closed (a) state by ligand are necessary for the initial stages of reaction. These pulsations of the active site decreases the total activation energy, necessary for the final complex formation.

*15.2.1 Possible mechanism of distant specific attraction between ligands and proteins*

The kind of "memory of water", discovered by Jacques Benveniste team in 1988 (Davenas, et.al. 1988), includes the ability of water to carry information about biologically active guest molecules and possibility to record, transmit and amplify this information. This phenomenon involves the successive diluting and shaking of [water + guest] system to a degree where the final solution contains no guest molecules more at all. However, using hypersensitive biological cells-containing test systems, he observed that this highly diluted solution initiated a reaction in similar way, as if the active guest molecules were still present in water. From the first high dilution experiments in 1984 to the present, thousands of experiments have been made in DigiBio company (Paris), enriching and considerably consolidating the initial knowledge of such kind of water memory. It was demonstrated also by DigiBio team, that low frequency ( 20 kHz = $2 \times 10^4$ s$^{-1}$) electromagnetic waves are able to activate biological cells. These results where interpreted that the molecular signal is composed of waveforms in the $(10-44)$ kHz range, which are specific to each molecular entity. This prompted Benveniste to hypothesize that the molecular signal is composed of such low frequency waves and that the ligand co-resonates with the cell receptor on these frequencies, stimulating specific attraction between them.

*However, we have to note, that just in this frequency range the resonant cavitational fluctuations in water can be excited by EM or acoustic fields (Kaivarainen, 1995, 2001). This effects, following by excitation of acoustic waves, also can be responsible for the membranes perturbations and cells activation.*

The perturbation of water properties, induced by haptens and antibodies, concurrent inhibitors and enzymes, viruses and cells in separate and mixed solutions can be studied, using our Hierarchic theory of condensed matter and this theory based Comprehensive Analyzer of Matter Properties (CAMP). The ways for specific water treatment by EM fields, corresponding to activation or inhibition of concrete biological processes, can be found out. The limiting stages of relaxation of water perturbations, induced by 'guest' molecules after successive dilution and shaking, responsible for 'memory' of water can be investigated also. Such investigations in case of success could turn the homeopathy from the art to quantitative science.

*It can be calculated, that the trivial Brownian collisions between interacting molecules in aqueous medium can not explain a high rate of specific complex formation.* One of possible explanation of specific distant interaction/attraction between the ligand and its receptor, is the exchange resonant EM interaction, proposed by DigiBio team.

This author's explanation of the specific attraction between sterically and dynamically complementary macromolecules and molecules in water, like in system: [antibody + hapten] or [enzyme + substrate] is based on Unified theory of Bivacuum, particles, duality and theory of Virtual Replicas of material objects (Kaivarainen, 2006; http://arxiv.org/abs/physics/0207027).

This new kind of Bivacuum mediated interaction between Sender [S] (ligand) and Receiver [R] (protein) can be a result of superposition of two their *virtual replicas*: [**VR**$^S$ ⇌ **VR**$^R$]. The carrying frequency of Virtual Pressure Waves (VPW$^\pm$) is equal to basic frequency of [S] and [R] elementary particles [Corpuscle ⇌ Wave] pulsation ($\omega_0 = m_0 c^2/\hbar)^i$, different for electrons and protons and responsible for their rest mass and charge origination. These basic frequencies are modulated by thermal vibrations of atoms and molecules. Just this modulation phenomena explains the nature of de Broglie waves of particles and provide a possibility of resonant distant interaction between molecules with close internal (quantum) and external (thermal) dynamics/frequencies via Virtual guides of spin, momentum and energy (**VirG**$_{S,M,E}$). A three-dimensional superposition of modulated by elementary particles of the object standing $\mathbf{VPW}_m^\pm$ compose the internal and external Virtual Replicas (VR$_{in}$ and VR$_{ext}$). The Virtual Replicas (VR) has a properties of 3-dimensinal Virtual Quantum Holograms (Kaivarainen, 2006). The $VR = VR_{in} + VR_{ext}$ of atoms, molecules and macroscopic objects reflect:

a) the internal properties of the object ($VR_{in}$), including its dynamics, inhomogeneity, asymmetry, etc.;

b) the external surface properties and shape of macroscopic objects ($VR_{ext}$).

*The virtual replicas (VR) and their superpositions may exist in Bivacuum and in any gas or condensed matter. It latter case the properties of matter can be slightly changed.*

Our Unified theory, including VR, is in-line with Bohm and Pribram holographic paradigm and make it more detailed and concrete.

The proposed kind of Bivacuum mediated interaction (Kaivarainen, 2006) should be accompanied by the increase of dielectric permittivity between interacting molecules, decreasing the Van-der-Waals interactions between water molecules and enhancing the coefficient of diffusion in selected space between active site of protein and specific to this site ligand. The probability of cavitational fluctuations in water with average frequency of about $10^4$ Hz (like revealed by DigiBio group), also should increase in the volume of $\mathbf{VR}^S$ and $\mathbf{VR}^R$ superposition, i.e. in the space between active site and the ligand.

The following three mechanisms of specific complex formation can be provided by:

a) the thermal fluctuations and diffusion in solution of protein and specific ligand;

b) the electromagnetic resonance exchange interaction between oscillating dipoles of protein active site and ligand (Benveniste hypothesis) and

c) the Bivacuum-mediated remote attraction between the ligand and active site, as a result of their Virtual replicas superposition (Kaivarainen, 2003 - 2006).

It is possible, that all three listed mechanisms of specific complex - formation are interrelated and enhance each other. The elucidation of the role of each of them in specific distant interaction/attraction between ligands and protein's active sites *in vitro* and *in vivo* - is a intriguing subject of future research.

*15.2.2. Virtual replica of drugs in water and possible mechanism of 'homeopathic memory'*

The memory of water, as a long relaxation time from nonequilibrium to equilibrium state, may have two explanations. One of them, for the case of magnetically treated water was suggested by this author in paper: "New Hierarchic Theory of Water & its Application to Analysis of Water Perturbations by Magnetic Field. Role of Water in Biosystems", placed to the arXiv: http://arxiv.org/abs/physics/0207114 and described in previous section.

The another one, described below, may explain the memory of guest molecules properties in solvent even after multiple dilution and shaking, as common in homeopathy. The hypothesis of 'homeopathic memory' is based on two consequences of our Unified theory (UT), including theory of Virtual Replicas of the actual objects (Kaivarainen, 2006; http://arxiv.org/abs/physics/0207027):

1. It follows from UT, that between any actual object (**AO**), like guest molecule in water, and its virtual replica (**VR**), - the *direct* and *back* reaction is existing: (**AO** ⇌ **VR**). For example, when the drug molecule interact with binding site of receptor or substrate with enzyme both of reagents are already 'tuned' by their virtual replicas superposition. We remind, that **VR** may exist in Bivacuum and as well in any matter in gas, liquid or solid state;

2. It follows from UT, that a stable virtual replica of the guest molecule, as a system of 3D standing virtual pressure waves ($\mathbf{VPW}_m^\pm$): $VR = VR_{in} + VR_{ext}$ and its ability to infinitive spatial multiplication [**VRM**(**r**, **t**)], may exist even after transferring the object from the primary location to very remote place or even its total destruction/disintegration.

If we consider a solution of any biologically active guest molecule in water (or in other liquid, in general case), then it follows from the above consequences of Unified theory, that the guest **VR**$_{guest}$ may retain its ability to affect the target via its superposition with virtual replica of the target **VR**$_{t\arg et}$ (i.e. the active site of cell's receptor, antibody, or enzyme). This can be a result of mentioned above back reaction of modulated virtual replica of target on the actual target (object):
[**VR**$_{guest}$ ⋈ **VR**$_{t\arg et}$] → **AO** even after super-dilution, when no one guest molecule (ligand) is no longer present in solution.

The liquid shaking (potentiation) after each step of homeopathic drug dilution is important in 'memorizing' of peculiar information about drug properties. However, the role of this procedure was obscure.

Our approach suggest the following explanation of such 'imprinting' phenomena, accompanied the shaking or vigorous stirring, involving two stages:

**1**. Each *act of shaking* of water after dilution is accompanied by the collective motion of huge number of water in the test vessel and activation of the collective de Broglie wave of water molecules (modulation wave), participating in all 24 quantum excitations of water ($i = 24$) in accordance to our Hierarchic theory. The most probable de Broglie wave length of big number of water molecules, created in the moment of shaking or stirring can be expressed as:

$$\vec{\lambda}_s = \frac{h}{m_{H_2O} \vec{v}_{col}} \qquad 15.17$$

where: $m_{H_2O}$ is a mass of one water molecule; $\vec{v}$ is a velocity of collective motion of water molecules in stirred volume, determined by velocity and direction of stirring or shaking.

Similar formula is valid for de Broglie wave length of the solute/guest molecules in the process of shaking. However instead of mass of water molecules, the mass of guest molecule ($m_{guest}$) should be used.

As a result of shaking or stirring and creation of collective de Broglie wave of water, the degree of entanglement between coherent water clusters in state of mesoscopic Bose condensation (mBC) can be increased by two different mechanisms, depending on velocity of shaking or stirring ($\vec{v}_{col}$), in accordance to our theory of turbulence (section 12.2).

Theory predicts that at $30^0 C$, when the most probable group velocity, related to water librations is $(v_{gr})_{lb} \simeq 2 \times 10^3 cm/s$, the critical flow velocity $v^1(r)$, necessary for *mechanical boiling* of water, corresponding to transition from the laminar flow to turbulent one (conditions 12.13 and 12.14) should be about $2.6 \times 10^3 cm/s = 26$ m/s.

At relatively low $\vec{v}_{col} < 10$ m/s, when the collective $\vec{\lambda}_s$ is bigger, than the average separation between primary librational effectons: about 55 Å at 25 $^0$C (see Fig. 51), this makes possible their unification on this base, representing partial transition from mesoscopic to macroscopic Bose condensation.

At higher velocity of shaking/stirring, when $\vec{v}_{col} \gtrsim 25$ m/s and the short-time turbulence originates in the whole volume of water under treatment, this means synhronization of conversions between primary librational and translational effectons or synhronized macro-convertons excitation in this volume. This process is accompanied by the enhancement of the exchange interaction between remote clusters in state of mBC by means of phonons and librational photons. This is followed by unification of mBC to fragile nonuniform macroscopic Bose condensation of all water volume in the vessel.

**2**. The water properties modulation by shaking or stirring, increasing the correlation between water clusters in state of mesoscopic BC. It makes the superposition of virtual replica of drag/guest molecule (**VR**$_{guest}$ - also related to its de Broglie wave length, atomic and elementary particles composition) with virtual replicas of coherent water molecules $[\sum \mathbf{VR}^i_{H_2O}]$ more effective. This superposition represents the act of **VR**$_{guest}$ imprinting in the ordered fraction of water in the treated volume.

The process of imprinting is accompanied the guest molecule Virtual Replica spatial multiplication [$\mathbf{VRM(r,t)}$] in new aqueous environment (Kaivarainen, 2006, http://arxiv.org/abs/physics/0207027, sections 13.4-13.5). We assume in our explanation of homeopathic memory, that the $\mathbf{VR}_{guest}$ of guest/drug molecule retains some physical properties of this molecule itself. The proposed mechanism of homeopathic memorization of VR in space and time [$\mathbf{VRM(r,t)}$] may explain the homeopathic drugs action in super-dilute solutions, below Avogadro number, prepared by successive shaking or stirring.

The mechanism proposed has already some experimental confirmation. Louis Rey (2003) in Switzerland, has published a paper in the mainstream journal, Physica A, describing the experiments that suggest water does have a memory of molecules that have been diluted away. With good reproducibility this fact was demonstrated by thermoluminescence method.

In this technique, the material is 'activated' by irradiation at low temperature, with UV, X-rays, electron beams, or other high-energy elementary particles. This causes electrons to come loose from the atoms and molecules, creating 'electron-hole pairs' that become separated and trapped at different energy levels. When the irradiated material is warmed up, it releases the absorbed energy and the trapped electrons and holes come together and recombine. This is accompanied by release of a characteristic glow of light, peaking at different temperatures depending on the magnitude of the separation between electron and hole.

As a general rule, the phenomenon is observed in crystals with an ordered arrangement of atoms and molecules, but it is also seen in disordered materials such as glasses. In this mechanism, imperfections in the atomic/molecular lattice are considered to be the sites at which luminescence appears.

Rey used this technique to investigate water, starting with heavy water or deuterium oxide that's been frozen into ice at a temperature of 77K.

As the ice warms up, a first peak (I) of luminescence appears near 120K, and a second peak (II) near 166 K. Heavy water gives a much stronger signal than regular water. In both cases, samples that were not irradiated gave no signals at all.

*It was shown, that the peak II comes from the hydrogen-bonded network within ice, whereas peak I comes from the individual molecules.* Rey then investigated what would happen when he dissolved some chemicals in the water and diluted it in steps of one hundred fold with vigorous stirring (as in the preparation of homeopathic remedies), until he reached a concentration of $10^{-30}$ g/cm$^3$ and compare that to the control that has not had any chemical dissolved in it *and treated in the process of imitated dilution in the same way*. The samples were frozen and activated with irradiation as usual.

When lithium chloride (LiCl) was added, and then diluted away, the thermoluminescent glow became reduced, but the reduction of peak II was greater relative to peak I. Sodium chloride (NaCl) had the same effect albeit to a lesser degree.

It appears, therefore, that substances like LiCl and NaCl can modify the hydrogen-bonded network of water, and that this modification remains even when the molecules have been diluted away. The fact that this 'memory' remains because of vigorous stirring or shaking at successive dilutions, indicates that the 'memory' depends on a dynamic process, like introduced in our explanation collective de Broglie wave of water molecules or instant turbulence, increasing the interaction between water clusters in state of mesoscopic Bose condensate.

### 15.3. **The life-time of water quasiparticles and frequencies of their excitation**.
### **The correlation with protein dynamics**

The set of formulas, describing dynamic properties of quasiparticles of condensed matter, introduced in our hierarchic theory, was presented in Chapter 4 of this book.

For the case of $(a \Leftrightarrow b)^{1,2,3}$ transitions of primary and secondary effectons *(translational and*

*librational),* their life-times in (a) and (b) states are the reciprocal value of corresponding frequencies: $[\tau_a = 1/\nu_a$ and $\tau_b = 1/\nu_b]_{tr,lb}^{1,2,3}$. These parameters and the resulting ones could be calculated from eqs.(2.27; 2.28) for primary effectons and (2.54; 2.55) for secondary ones.

The results of calculations, using eq.(4.56 and 4.57) for frequency of excitations of both states of primary *tr and lb water* effectons and frequency of $(a \Leftrightarrow b)^{1,2,3}$ transitions state, taking into account their probabilities, are plotted on Fig. 49 a,b.

The frequencies of Macroconvertons and Super-deformon excitations were calculated using eqs.(4.42 and 4.48).

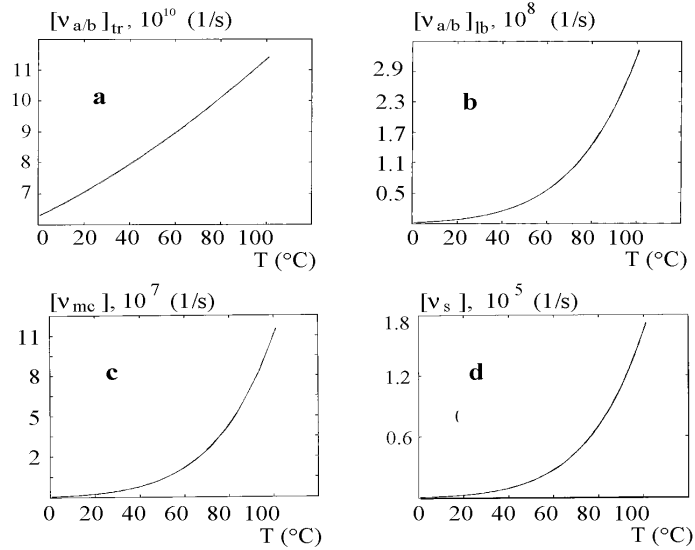

**Fig**. 49. (a) - Frequency of primary [tr] effectons excitations, calculated from eq.(4.56);
(b) - Frequency of primary [lb] effectons excitations, calculated from eq.(4.57);
(c) - Frequency of [*lb/tr*] macroconvertons (flickering clusters) excitations, calculated from eq.(4.42);
(d) - Frequency of Super-deformon excitations, calculated from eq.(4.48).

At the temperature interval $(0 - 100)^0 C$ the frequencies of translational and librational macro-deformons (tr and lb) are in the interval of $(1.3 - 2.8) \times 10^9 s^{-1}$ and $(0.2 - 13) \times 10^6 s^{-1}$ correspondingly. The frequencies of (ac) and (bc) convertons could be defined also using our software and formulae, presented at the end of Chapter 4.

The frequency of primary translational effectons $[a \Leftrightarrow b]$ transitions at $20^0 C$, calculated from eq.(4.56) is $\nu \sim 7 \times 10^{10} (1/s)$. It corresponds to electromagnetic wave length in water with refraction index $(n = 1.33)$ of:

$$\lambda = (cn)/\nu \sim 6mm$$

For the other hand, there are a lot of evidence, that the irradiation of very different biological systems just with such coherent electromagnetic waves has a big influences on their properties (Grundler and Keilman, 1983).

These data point again, that between the dynamics and function of proteins, membranes, etc. and dynamics of their aqueous environment, calculated using our theory, the strong correlation is existing.

The frequency of macroconvertons, representing big density fluctuation in the volume of primary librational effectons of water at $37^0 C$ is about $10^7 (1/s)$ (Fig. 49c).

The frequency of librational macro-deformons at the same temperature is about $10^6 s^{-1}$, i.e. coincides well with frequency of large-scale protein cavities pulsations between the open and closed to water states (see Figs.47 and 48). These quantitative results confirm our hypothesis that the clusterphilic interaction is responsible for stabilization of the protein cavities open states and that the transitions from their open states to the closed ones are induced by coherent water cluster dissociation (mesoscopic 1st order phase transition).

The frequency of Super-deformons excitation (Fig.49d) is much lower:

$$\nu_s \sim (10^4 - 10^5) \, s^{-1} \qquad 15.18$$

Super-deformons are responsible for cavitational fluctuations in liquids and origination of defects in solids. The dissociation of oligomeric proteins, like hemoglobin or disassembly (peptization) of actin and microtubules could be also stimulated by such big fluctuations of water (see Section 17.5).

# Chapter 16

# Hierarchic mechanism of enzymatic catalysis.
# Conditions of quantum beats between substrate and catalytic groups

The mechanism of enzymatic catalysis is one of the most intriguing and unresolved yet problems of molecular biophysics. It becomes clear, that it is interrelated not only with a spatial complementarity of the substrate and active site, but also with complicated dynamics of proteins (see book: "The Fluctuating Enzyme", Ed. by G.R.Welch, 1986).

The [proteins + solvent] system should be considered as cooperative one with feedback links (Kaivarainen, 1980, 1985, 1992, 1995). Somogyi and Damjanovich (1986) proposed a similar idea that collective excitations of protein structure are interrelated with dynamics of vicinal water.

The enzymatic reaction can be described in accordance with our model as a consequence of the following number of stages (Kaivarainen, 1995, 2000).

*The first stage:*

$$(I) \quad E^b + S \rightleftharpoons E^b S \qquad 16.1$$

- the collision of the substrate (S) with the open (b) state of the active site [AS] cavity of enzyme (E).

The frequency of collisions between the enzyme and the substrate, whose concentrations are $[C_E]$ and $[C_S]$, respectively, is expressed with the known formula (Cantor and Schimmel, 1980):

$$v_{col} = 4\pi r_0 (D_E + D_S) \times N_0 [C_E][C_S] \qquad 16.2$$

where: $r_0 = a_E + a_S$ is the sum of the enzyme's and substrate's molecular radii; $N_0$ is the Avogadro number;

$$D_E = \frac{kT}{6\pi\eta a_E} \quad \text{and} \quad D_S = \frac{kT}{6\pi\eta a_S} \qquad 16.3$$

- are the diffusion coefficients of the enzyme and substrate; $k$ is the Boltzmann constant; $T$ is absolute temperature; $\eta$ is a solvent viscosity.

The probability of collision of the open ($b$) state of the active site with substrate is proportional to the ratio of the b-state cross section area to the whole enzyme surface area:

$$P_b = \frac{\O^b}{\O_E} F_b \qquad 16.4$$

where:

$$F_b = \frac{f_b}{f_b + f_a} \qquad 16.5$$

is a fraction of time, when the active site [AS] is in open (b) - state.

So, the frequency of collision between the substrate and the ($b$) state of the active site (AS) as the first stage of enzymatic reaction is:

$$v^b_{col} = v_{col} P_b = k_I \qquad 16.6$$

*The second stage* of reaction is a formation of the primary enzyme-substrate complex:

$$(II) \quad E^b S \Leftrightarrow [E^{a^*} S]^{(1)} \qquad 16.7$$

It corresponds to transition of the active site cavity from the open ($b$) state to the closed ($a$) one and stabilization of the latter by the ligand.

The rate constant of the $[b \rightarrow a]$ transitions is described by Stokes-Einstein and Eyring - Polany generalized equation (15.3):

$$k_{\text{II}}^{b \rightarrow a^*} = \frac{kT}{\eta V} \exp\left(-\frac{G_{st}^{b \rightarrow a}}{RT}\right) \qquad 16.8$$

where: $\eta$ is the solvent viscosity; $V$ is the effective volume of the enzyme domain, whose diffusional reorientation accompanies the $(b \rightarrow a)$ transition of the active site [AS].

The leftward shift of the $[a \Leftrightarrow b]$ equilibrium between two states of the active site corresponds to this stage of the reaction. It reflects the introduced *principle of stabilized conformity,* provided by AS domains relative dynamics (see Chapter 15).

*The third stage of enzymatic reaction:*

$$(III) \qquad [E^a S] \Leftrightarrow [E^{a^*} S^*] \qquad 16.9$$

represents the formation of the secondary specific complex. This process is related to directed ligand diffusion into the active site cavity and the dynamic adaptation of ligand geometry to the geometry of the active site.

Here the *principle of stabilized microscopic conformity* is realized, when the AS change its geometry and dynamics from $(a)$ to $(a^*)$ without additional domains reorientation. The rate constant of this stage is determined by the rate constant of substrate diffusion in the closed $(a)$ state of the active site cavity. It is also described by *generalized kinetic equation* (16.8), but with other values of variables:

$$k_S^{1 \rightarrow 2^*} = \frac{1}{\tau_s^*} \exp\left(-\frac{G_S^a}{RT}\right) = k_{\text{III}} \qquad 16.10$$

where: $\tau_s = (\mathbf{v}_s/k)\eta^{\text{in}}/T$ is the correlation time of substrate of volume $(\mathbf{v}_s)$ in the $(a)$ state of the active site; $\eta^{\text{in}}$ is the internal effective viscosity; $G_S^a$ is the activation energy of thermal fluctuations of groups, representing small-scale dynamics (SS), which determine the directed diffusion of a substrate in the active site [AS] closed cavity.

The vectorized character of ligand diffusion in the AS cavity can be determined by minor changes of the AS structure by the ligand. These changes were observed in many reactions of specific protein-ligand complexes formation.

We assume that at this important stage, the waves B of the attacking catalytic atoms $(\lambda_B^c)$ and the attacked substrate atoms $(\lambda_B^S)$ start to superimpose and interfere, providing the conditions for quantum beats between corresponding de Broglie waves of the AS and substrate.

According to classical statistics, every degree of freedom gets the energy, which is equal to $kT/2$. This corresponds to harmonic approximation when the mean potential and kinetic energy of particles are equal:

$$V = T_k = m\mathbf{v}^2/2 \approx kT/2 \qquad 16.10a$$

The corresponding to such classical conditions wave B length is equal to:

$$\lambda_B = \frac{h}{m\mathbf{v}} = \frac{h}{(mkT)^{1/2}} \qquad 16.11$$

For example, the de Broglie wave length of a particle with mass of proton at room temperature is nearly 2.5Å, for a carbon atom it is about three times smaller and for oxygen - four times as small.

So, in classical conditions, the thermal wave B length of atoms is comparable or even less than the size of the atom itself. In such conditions their waves can not overlap and quantum beats between them are not possible. However, in *real condensed systems,* including the active sites of enzymes, the harmonic approximation of atomic oscillations is not valid because $(T_k/V) \ll 1$ (see results of our computer calculations at Fig. 5). Consequently, the kinetic energy of atoms of the active site:

$$T_k \ll (1/2)kT \quad \text{and} \quad \lambda_B \gg h/(mkT)^{1/2}$$

It must be taken into account that librations, in a general case are presented by rotational-translational motions of atoms and molecules, but not by their pure rotations (Coffey et al., 1984). The length of waves B of atoms caused by a small translational component of librations is greater than that caused by pure translations:

$$[\lambda_{lb} = (h/P_{lb})] > [\lambda_{tr} = (h/P_{tr})]$$

Even in pure water the linear sizes of primary librational effectons are several times bigger than that of translational effectons and linear space, occupied by one $H_2O$ molecule (see Fig. 7).

In composition of the active site (AS) rigid core the librational waves B of atoms also can significantly exceed the sizes of the atoms themselves and superimpose with waves B of substrate atoms. In this case the quantum beats between waves B of catalytic atoms of the active site and the attacked atoms of substrate complex in [ES] complex is quite possible. The chemical reaction [$S^* \to P$], accelerated by these quantum beats, is the *next 4th stage of enzymatic process. The result of this stage is chemical transformation of substrate to product*:

$$(IV) \qquad [E^{a^*}S^*] \to [E^{a^*}P] \qquad 16.12$$

The angular wave B frequency of the *attacked atoms of substrate* with mass $m_S$ and the amplitude $A_S$ can be expressed by eq.(2.20):

$$\omega^S = \hbar/2m_S A_S^2 \qquad 16.13$$

The wave B frequency of the attacking catalytic atom (or a group of atoms) of the active site is equal to:

$$\omega^{cat} = \frac{\hbar}{2m_c A_c^2} \qquad 16.14$$

*The frequency of quantum beats* which appear between waves B of catalytic and substrate atoms is:

$$\omega^* = \omega^{cat} - \omega^S = \hbar\left(\frac{1}{2m_c A_c^2} - \frac{1}{2m_S A_S^2}\right) \qquad 16.15$$

The corresponding energy of beats:

$$E^* = E^{cat} - E^S = \hbar\omega^* \qquad 16.16$$

It is seen from these formulae that the smaller is the mass of the catalytic atom ($m_c$) and its wave B amplitude ($A_c$), the more frequently beats occur at the constant parameters of substrate ($m_S$ and $A_S$). *The energy of quantum beats is transmitted to the waves B of the attacked substrate atom from the catalytic atoms, making chemical reaction possible.*

It is known from the theory of oscillations (Grawford, 1973) that the effect of beats is maximal, if the amplitudes of the interacting oscillators are equal:

$$A_c \approx A_S \qquad 16.17$$

Consequently the difference $\left(\frac{1}{2m_c A_c} - \frac{1}{2m_S A_S}\right)$ in (16.15) can be provided in these conditions only by difference between $m_c$ and $m_S$.

The [substrate $\to$ product] conversion is a result of the substrate wave B transition from the main [S] state to excited [P] state. The rate constant of such a reaction in the absence of the catalyst ($k^{S \to P}$) can be presented by the modified Eyring-Polany formulae, following from our eq.(2.27) at condition: $\exp(h\nu_P/kT) \gg 1$

$$\nu_A^S = k^{S \to P} = \frac{E^P}{h}\exp\left(-\frac{E^P - E^S}{kT}\right) = \nu_B^P \exp\left[-\frac{h(\nu_B^P - \nu_B^S)}{kT}\right] \qquad 16.18$$

where: $E^S = h\nu^S$ and $E^P = h\nu^P$ are the main and excited (transitional to product) states of substrate, correspondingly; $v^S$ and $v^P$ are the substrate wave B frequencies in the main and excited states, respectively.

If catalyst is present, which acts by the above described mechanism, then the energy of the substrate $E^S$ is increased by the magnitude $E^{cat}$ with the quantum beats frequency $\omega^*$ (16.15) and gets equal to:

$$E^{Sc} = E^S + E^{cat} \qquad 16.19$$

Substituting $E^P = hv_B^P$ and $E^{Sc} = h(v_B^S + v_B^{cat})$ in (16.18), we derive the rate constants for the catalytic reaction in the moment of quantum beats between de Broglie waves of the active site and substrate ($k^{Sc \to P}$).

This process describes the 4th stage of enzymatic reaction:

$$(IV): \quad k^{Sc \to P} = v^P \exp\left[-\frac{h(v^P - v^S - v^{cat})}{kT}\right] = k_{IV} \qquad 16.20$$

where: $v^P, v^S$ and $v^c$ are the most probable B wave frequencies of the transition $[S \to P]$ state, of the substrate and of the catalyst atoms, correspondingly.

Hence, in the presence of the catalyst the coefficient of acceleration ($q_{cat}$) is equal to:

$$q_{cat} = \frac{k^{Sc \to P}}{k^{S \to P}} = \exp\left(\frac{hv^{cat}}{kT}\right) \qquad 16.21$$

For example, at

$$hv^c/kT \approx 10; \qquad q_{cat} = 2.2 \times 10^4 \qquad 16.22$$

At the room temperatures this condition corresponds to

$$E^{cat} = hv^{cat} \simeq 6 \text{ kcal/mole}.$$

In accordance to proposed model of enzymatic reaction, the acts of quantum beats, in-phase with the transitions of a substrate molecule to its excited state ($S \to Sc$), can be accompanied by the absorption of phonons or photons with the resonant frequency $\omega^*$ (16.15). In this case, the irradiation of [substrate + catalyst] system with ultrasound or electromagnetic waves with frequency $\omega^*$ should strongly accelerate the reaction.

This consequence of our model can be easily verified experimentally.

It is possible that the resonance effects of this type can account for the experimentally revealed response of various biological systems to electromagnetic field radiation with the frequency about $6 \times 10^{10} Hz$ (Deviatkov et al., 1973).

Changes in structure and electronic properties of substrate molecule after transition to the product, change its interaction energy with the active site by the magnitude:

$$\Delta E_{SP}^{a^*} = (E_S^{a^*} - E_P^{a^*}) \qquad 16.23$$

This must destabilize the closed state of the active site and increase the probability of its reverse $[a^* \to b^*]$ transition. Such transition promotes the last *5th stage* of the catalytic cycle - the dissociation of the enzyme-product complex:

$$(V) \qquad [E^{a^*}P] \Leftrightarrow [E_P^b P] \Leftrightarrow E^b + P \qquad 16.24$$

The resulting rate constant of this stage, like stage (II), is described by the generalized Stokes-Einstein and Eyring-Polany equation (16.8), but with different activation energy $G_{st}^{a^* \to b}$ valid for the $[a^* \to b]$ transition:

$$k_P^{a^* \to b} = \frac{kT}{\eta V} \exp\left(-\frac{G_{st}^{a^* \to b}}{kT}\right) = 1/\tau_P^{a^* \to b} \qquad 16.25$$

If the lifetime of the ($a^*$) state is sufficiently long, then the desorption of the product can occur irrespective of $[a^* \to b]$ transition, but with a longer characteristic time as a consequence of its diffusion out of the active site's "closed state". The rate constant of this process ($k_P^*$) is determined by small-scale dynamics. It practically does not depend on solvent viscosity, but can grows up with rising temperature, like at the 3d stage, described by (16.10):

$$k_P^* = \frac{kT}{\eta_{a^*}^{in} \mathbf{v}_P} \exp\left(-\frac{G_P^{a^*}}{kT}\right) = 1/\tau_P^* \qquad 16.26$$

where $\eta_{a^*}^{in}$ is active site interior viscosity in the $(a^*)$ state; $\mathbf{v}_P$ is the effective volume of product molecules; $P_{a^*} = \exp(-G_P^a/RT)$ is the probability of small-scale, *functionally important motions* necessary for the desorption of the product from the $(a^*)$ state of the active site; $G_P^{a^*}$ is the free energy of activation of such motions.

Because the processes, described by the eq.(16.25) and (16.26), are independent, the resulting product desorption rate constant is equal to:

$$k_V = k_P^{a^* \to b} + k_P^* = 1/\tau_P^{a^* \to b} + 1/\tau_P^* \qquad 16.27$$

The characteristic time of this final stage of enzymatic reaction is:

$$\tau_V = 1/k_V = \frac{\tau_P^{a^* \to b} \tau_P^*}{\tau_P^{a^* \to b} + \tau_P^*} \qquad 16.28$$

*This stage is accompanied by the relaxation of the deformed active site domains and overall protein structure to the initial state.*

After the whole reaction cycle is completed the enzyme gets ready for the next cycle. The number of cycles (catalytic acts) in the majority of enzymes is within the limits of $(10^2 - 10^4)\,s^{-1}$. It means that the $[a \rightleftharpoons b]$ pulsations of the active site cavities must occur with higher frequency as far it is only one of the five stages of enzymatic reaction cycle.

In experiments, where various sucrose concentrations were used at constant temperature, the dependence of enzymatic catalysis rate on solvent viscosity $(T/\eta)$ was demonstrated (Gavish and Weber, 1979). The amendment for changing the dielectric penetrability of the solvent by sucrose was taken into account. There are reasons to consider stages (II) and/or (V) in the model described above as the limiting ones of enzyme catalysis. According to eqs.(16.8) and (16.25), these stages depend on $(T/\eta)$, indeed.

The resulting rate constant of the enzyme reaction could be expressed as the reciprocal sum of life times of all its separate stages (I-V):

$$k_{res} = 1/\tau_{res} = 1/(\tau_I + \tau_{II} + \tau_{III} + \tau_{IV} + \tau_V) \qquad 16.29$$

where

$$\tau_I = 1/k_I = \frac{1}{v_{col} P_b}; \quad \tau_{II} = 1/k_{II} = \frac{\eta V}{kT} \exp\left(\frac{G_{st}^{b \to a}}{RT}\right);$$

$$\tau_{III} = 1/k_{III} = \frac{\eta^{in} \mathbf{v}_S}{kT} \exp\left(\frac{G_{SS}^a}{RT}\right);$$

$$\tau_{IV} = 1/k_{IV} = \frac{1}{(v^p)} \exp\left[\frac{h(v^s - v^p - v^c)}{kT}\right]$$

$\tau_V$ corresponds to eq.(16.28).

The slowest stages of the reaction seem to be stages (II), (V), and stage (III). The latter is dependent only on the small-scale dynamics in the region of the active site.

Sometimes *product desorption* goes on much more slowly than other stages of the enzymatic process, i.e.
$$k_V \ll k_{III} \ll k_{II}$$

Then the resulting rate of the process ($k_{res}$) is represented by its limiting stage (eq. 16.27):

$$k_{res} \approx \frac{kT}{\eta V}\exp\left(-\frac{G_{st}^{a^* \to b}}{RT}\right) + k_P^* = 1/\tau_{res} \qquad 16.30$$

The corresponding period of enzyme turnover: $\tau_{res} \approx \tau_V$ (eq.16.28). The internal medium viscosity ($\eta^{in}$) in the protein regions, which are far from the periphery, is 2-3 orders higher than the viscosity of a water-saline solvent ($\eta \sim 0.001$ P) under standard conditions:

$$(\eta^{in}/\eta) \geq 10^3$$

Therefore, the changes of sucrose concentration in the limits of 0-40% at constant temperature can not influence markedly internal small- scale dynamics in proteins, its activation energy ($G^{a^*}$) and internal microviscosity (Kaivarainen, 1989b). This fact was revealed using the spin-label method. It is in accordance with viscosity dependencies of tryptophan fluorescence quenching in proteins and model systems related to acrylamide diffusion in protein matrix (Eftink and Hagaman, 1986). In the examples of parvalbumin and ribonuclease $T_1$ it has been shown that the dynamics of internal residues is practically insensitive to changing solvent viscosity by glycerol over the range of 0.01 to 1 P.

It follows from the above data that the moderate changes in solvent viscosity ($\eta$) at constant temperature do not influence markedly the $k_P^*$ value in eq.(16.30).

Therefore, the isothermal dependencies of $k_{res}$ on $(T/\eta)_T$ with changing sucrose or glycerol concentration must represent straight lines with the slope:

$$tg\alpha = \frac{\Delta k_{res}}{\Delta(T/\eta)_T} = \frac{k}{V}\exp\left(-\frac{G_{st}^{a^* \to b}}{RT}\right)_T \qquad 16.31$$

The interception of isotherms at extrapolation to ($T/\eta \to 0$) yields (16.26):

$$(k_{res})_{(T/\eta) \to 0} = k_P^* = \frac{kT}{\eta^{in}\,\mathbf{v}_P}\exp\left(-\frac{G^{a^*}}{RT}\right) \qquad 16.32$$

The volume of the Brownian particle (V) in eq.(16.31) corresponds to the effective volume of one of the domains, which reorientation is responsible for ($a \rightleftharpoons b$) transitions of the enzyme active site.

Under conditions when $G_{st}^{a^* \to b}$ weakly depends on temperature, it is possible to investigate the temperature dependence of the effective volume V, using eq.(16.31), analyzing a slopes of set of isotherms (16.31).

Our model predicts the increasing of V with temperature rising. This reflects the dumping of the large-scale dynamics of proteins due to water clusters melting and enhancement the Van der Waals interactions between protein domains and subunits (Kaivarainen, 1985; 1989b, Kaivarainen et al., 1993). The contribution of the small-scale dynamics ($k^*$) to $k_{res}$ must grow due to its thermoactivation and the decrease in $\eta^{in}$ and $G^{a^*}$ ($eq.\,16.32$).

*The diffusion trajectory* of ligands, substrates and products of enzyme reactions in "closed" (*a*) states of active sites is probably *determined by the spatial gradient of minimum wave B length (maximum momentums) values of atoms, forming the active site cavity.* Apparently, *functionally important motions (FIM)*, introduced as a term by H.Frauenfelder et al., (1985, 1988), are determined by specific geometry of the momentum space characterizing the distribution of small-scale dynamics of domains in the region of protein's active site.

*The analysis of the momentum distribution in the active site area and energy of quantum beats between de Broglie waves of the atoms of substrate and active sites, modulated by solvent-dependent large-scale dynamics, should lead to complete understanding of the physical background of enzyme catalysis.*

## 16.1 The mechanism of ATP hydrolysis energy utilization in muscle contraction and proteins filaments polymerization

A great number of biochemical reactions are endothermic, i.e. they need additional thermal energy in contrast to exothermic ones. The most universal and common source of this additional energy is a reaction of adenosine-triphosphate (ATP) hydrolysis:

$$\text{ATP} \underset{k_{-1}}{\overset{k_1}{\Leftrightarrow}} \text{ADP} + \text{P} \qquad 16.33$$

The reaction products are adenosine-diphosphate (ADP) and inorganic phosphate (P).

The equilibrium constant of the reaction depends on the concentration of the substrate [ATP] and products [ADP] and [P] like:

$$K = \frac{k_1}{k_{-1}} = \frac{[\text{ADP}][\text{P}]}{[\text{ATP}]} \qquad 16.34$$

The equilibrium constant and temperature determine the reaction free energy change:

$$\Delta G = -RT \ln K = \Delta H - T\Delta S \qquad 16.35$$

where: $\Delta H$ and $\Delta S$ are changes in enthalpy and entropy, respectively.

Under the real conditions in cell the reaction of *ATP* hydrolysis is highly favorable energetically as is accompanied by strong free energy decrease: $\Delta G = -(11 \div 13)$ kcal/M. It follows from (16.35) that $\Delta G < 0$, when

$$T\Delta S > \Delta H \qquad 16.36$$

and the entropy and enthalpy changes are positive ($\Delta S > 0$ and $\Delta H > 0$). However, the specific molecular mechanism of these changes in different biochemical reactions, including muscle contraction, remains unclear.

Acceleration of actin polymerization and tubulin self-assembly to the microtubules as a result of the ATP and nucleotide GTP splitting, respectively, is still obscure as well.

Using our model of water-macromolecule interaction (Sections 13.3 - 13.5), we can explain these processes by the "melting" of the water clusters - librational effectons (mesoscopic Bose condensate) in cavities between neighboring domains and subunits of proteins. This mesoscopic melting is induced by absorption of energy of ATP or GTP hydrolysis and represents [*lb/tr*] conversion of primary librational effectons to translational ones. It leads to the partial dehydration of domains and subunits cavities. The concomitant transition of interdomain/subunit cavities from the "open" B-state to the "closed" A-state should be accompanied by decreasing of linear dimensions of a macromolecule. This process is usually reversible and responsible for the large-scale dynamics and proteins contraction.

In the case when *disjoining clusterphilic interactions* that shift the $[A \Leftrightarrow B]$ equilibrium to the right are stronger than Van der Waals interactions stabilizing A-state, the expansion of the macromolecule dominates on the contraction and can induce a mechanical "pushing" force.

In accordance to proposed by this author model, this *"swelling driving force"* is responsible for shifting of myosin "heads" with respect to the actin filaments and muscle contraction. This 1st relaxation "swelling working step" is accompanied by dissociation of products of ATP hydrolysis from the active sites of myosin heads (heavy meromyosin).

*The 2-nd stage of reaction*, the dissociation of the complex: [myosin "head" + actin], is related to the absorption of ATP at the myosin active site. At this stage the $A \Leftrightarrow B$ equilibrium between the heavy meromyosin conformers is strongly shifted to the right, i.e. to an expanded form of the protein.

The next *3-d stage* represents the ATP hydrolysis, (ATP → ADP + P) and expelling of (P) from the active site. The concomitant local enthalpy and entropy jump leads to the melting of the water clusters in the cavities, $B \to A$ transitions and the contraction of free meromyosin heads.

*The energy of the clusterphilic interaction at this stage is accumulated in myosin like in a squeezed spring.* After this 3d stage is over, the complex [myosin head + actin] forms again.

We assume here that the interaction between myosin head and actin induces the releasing of the product (ADP) from myosin active site. It is important to stress that the driving force of "swelling working stage": $[A \to B]$ transition of myosin cavities is provided by new *clusterphilic interactionns* which was not considered earlier et all (see Section 13.4).

A repetition of such a cycle results in the relative shift of myosin filaments with respect to actin ones and finally in muscle contraction. The mechanism proposed does not need the hypothesis of Davydov's soliton propagation (Davydov, 1982) along a myosin macromolecule. It seems that this nondissipative process scarcely takes place in strongly fluctuating biological systems. Soliton model does not take into account the real mesoscopic structure of macromolecules and their dynamic interaction with water.

The polymerization of actin, tubulin and other globular proteins, composing cytoplasmic and extracell filaments due to hydrophobic interaction can be accelerated as a result of their selected dehydration, turning the clusterphilic interactions to hydrophobic ones. This happens due to local temperature jumps in mesoscopic volumes where the ATP and GTP hydrolysis takes place.

The [assembly ⇔ disassembly] equilibrium of subunits of filaments is shifted to the left, when the [*protein − protein*] interfacial Van-der-Waals interactions becomes stronger than the clusterphilic ones. The latter is mediated by librational water effecton stabilization in interdomain or intersubunit cavities.

*It looks that the clusterphilic interactions, introduced in our work, play an extremely important role in the self-organization and dynamics of very different biological systems on mesoscale and even macroscopic scale..*

# Chapter 17

# Some new applications of hierarchic theory to biology

## 17.1 Possible nature of biological field

One of the conclusion of our Hierarchic concept of matter and field is that every type of condensed matter, including every biological organism is a source of three types of fields: *electromagnetic, acoustic and vibro-gravitational* ones (Kaivarainen 1992, 1995, 1996, 2001, 2006).

These fields are resulted from coherent vibrations (translational and librational) of atoms and molecules in composition of the effectons. The effectons represents a clusters with properties of mesoscopic Bose-condensate (mBC), with dimensions, determined by 3D superposition of de Broglie standing waves of molecules, related to their librations and translations. The quantum transitions of the effectons between two kinds of coherent anharmonic oscillations: in-phase (acoustic-like) and counterphase (optic-like) are accompanied by emission/absorption of IR photons (the case of primary effectons) or phonons (the case of secondary effectons).

It is known that any type of vibrations should be followed by excitation of vibro - gravitational waves with corresponding frequencies, as related with coherent accelerations change.

The 3D superposition due to interception of photons, phonons and vibro-gravitational waves, in turn, lead, correspondingly, to emergency of:
  1) electromagnetic deformons;
  2) acoustic deformons and

3) vibro-gravitational effectons, deformons and convertons.

These kinds of physical fields irradiated by living organisms and their superposition could be considered as a different components of "biological field".

In our later works, devoted to theory of Bivacuum and its interaction with matter, it was demonstrated that each physical object, including biological ones has also specific *virtual replica* (or virtual hologram), reflecting not only the external properties (shape, volume), like the regular hologram, but also the internal spatial and dynamic properties of the objects (Kaivarainen, 2006).

### 17.2. **The electromagnetic and acoustic waves**, radiated by bodies

According to our hierarchic theory, the $(a \rightleftharpoons b)_{tr,lb}^{1,2,3}$ transitions of primary translational and librational effectons are related to emission and absorption of number of coherent IR photons in six selected directions (normal to each plane of primary effectons, approximated by parallelepiped), with different frequencies in general case. Such spontaneous collective type of dipole emission have a features of *superradiance* (Dicke, 1954).

The primary electromagnetic deformons represent a pulsing 3D quasiparticles formed by superposition of three normal to each other standing IR photons. As far the photons move at light speed in selected directions, they can give rise a huge number of electromagnetic deformons on their way, depending on density of primary effectons in the volume of condensed matter.

In the ordered systems with high degree of coherence, the number of IR photons, emitted and absorbed by the primary effectons can be equal. But usually such ideal equilibrium is unstable and as a result of energy exchange of system with external medium, some amount of coherent electromagnetic IR photons, originated in solids and liquids, not participating in primary deformons formation, can be irradiated in accordance with Stephan-Boltzmann law (see eqs.5.73 and 5.74). The decoherence factors, provided by thermal fluctuations, make it difficult the experimental registration of coherent radiation by liquids.

The rigid crystals and liquid crystals with an ordered structure should have more discrete and sharp distribution of the emitted photons frequencies as compared to amorphous matter.

Biological membranes have a properties of liquid crystals. The membranes, cell's organelles, cytoskeleton, filaments, microtubules (MT) and intra-microtubules water (chapter 18) are a source of electromagnetic, acoustic and vibro-gravitational deformons also. The coherent biophotons (Popp, 1985; 1992) can be explained in such a way.

One end of microtubule (MT) is often fixed on the cell's membrane and the other end - starts from two centrioles. The ends of MTs are the most powerful source of coherent librational IR photons, in accordance to theory of superradiation (Dicke, 1954).

### 17.3. **Coherent IR superradiation and distant Van der Waals interaction**

The effect of coherent IR photons, emitted by macroscopic solid bodies due to superradiation, following from our Hierarchic theory of condensed matter (see Introduction and Kaivarainen, 1992; 1995; 2001), can explain the Van der Waals interactions between bodies at distances exceeding 1000 Å. The synchronized $(b \Leftrightarrow a)_{tr,lb}^{I}$ transitions of primary *tr* and *lb* effectons in one of interacting bodies with frequency $[\nu = 1/T]_{tr,lb}$ leads to superradiation of coherent IR standing photons with volume:

$$[V_{el.d} = (3/4\pi)\lambda^3 = (3/4\pi)(c/n\nu)^3]_{tr,lb} \qquad 17.1$$

existing outside these bodies in medium with dielectric constant: $\epsilon = n^2$.

If the distance between the interacting bodies have the same order as the IR photons (tr, lb) wave length: $[l \approx \lambda]_{tr,lb} \geq 10^5 \text{Å}$, then the energy, radiated by one of the bodies can be spent for the excitation of $(a \Leftrightarrow b)_{tr,lb}^{II}$ transitions in the next body.

The synchronization of the effecton's and deformon's dynamics of the neighboring bodies, induced by such exchange processes, means the coherence emergency between fluctuations of polarizabilities and dipole moments of the molecules, composing them. This coherence leads to the resonant electromagnetic attraction of bodies. In this case, the sum of potential energies of isolated bodies is bigger than that of interacting bodies.

The mechanism described, explains the physical nature of long-wave electromagnetic fluctuations introduced by E. Lifshitz in 1955 in his theory of macroscopic Van der Waals forces. This theory was later extended for two bodies of different nature, separated with the medium of the certain dielectric penetrability (Dzeloshinsky et al., 1961). The corresponding Van der Waals force is inversely proportional to the distance between bodies ($l$) in third power:

$$F = h\bar{\omega}/8\pi l^3 \qquad 17.1a$$

where $\bar{\omega}$ is the generalized frequency characteristic for the absorption in all three media.

If the mechanism of synchronized pulsations of effectons in interacting bodies proposed here is correct, then the dependence of $F$ on distance $l$ must be nonmonotonic.

The maxima at the decreasing function $F(l)$ must correspond to the conditions:

$$l = i\,\lambda_S = i\,(c/v_S), \qquad 17.2$$

where $i = 1, 2, 3...$ is an integer number.

*The proposed mechanism of long-range Van der Waals interaction as well as Resonant Vibro-Gravitational Interaction, discussed in the next Section may play an important role in distant informational communication between objects, containing similar coherent clusters, like primary effectons.*

## 17.4. Gravitational effectons, transitons and deformons

Vibro-Gravitational waves (VGW) are caused by the coherent oscillations of any type of particles with alterable accelerations. As far the molecules (i.e. water, lipids of biomembranes, etc.) in composition of translational and librational effectons are synchronized, the gravitational waves of the same frequency, radiated by them are coherent also. The amplitude of VGW, irradiated by the object is proportional to the number of coherent particles (molecules) in its volume. Three-dimensional superpositions of gravitational waves are reflecting the properties of effectons, exciting them, like frequency of collective oscillations of molecules and their volume. The $a$ and $b$ states of molecular effectons excite the corresponding $a$ and $b$ states of *gravitational effectons* with similar frequencies. Because librational effectons are much larger than the translational ones, the former are sources of vibro-gravitational waves with bigger amplitude and intensity.

According to eq.(10.41), the most probable accelerations in ($a$) and ($b$) states of the primary effectons in three selected directions, normal to each other, are:

$$\left[a_a = \frac{h(\nu_a)^2}{m\mathbf{v}_{ph}^a} = \frac{h}{m\lambda_a}\nu_a;\quad a_b = \frac{h(\nu_b)^2}{m\mathbf{v}_{ph}^b} = \frac{h}{m\lambda_b}\nu_b\right]_{tr,lb}^{1,2,3} \qquad 17.3$$

where $\nu_a$ and $\nu_b$ are the frequencies of the effectons in corresponding states; $\mathbf{v}_{ph}^a$ and $\mathbf{v}_{ph}^b$ are phase velocities of waves B forming effectons in $a$ and $b$ states; $[m]$ is the mass of molecules forming effectons.

The wave B lengths of the effectons, which determine their sizes in $a$ and $b$ states, are equal in both states:

$$[\lambda_a = \lambda_b = h/m\mathbf{v}_{gr}^a = h/m\mathbf{v}_{gr}^b]_{tr,lb}^{1,2,3} \qquad 17.4$$

where: $\mathbf{v}_{gr}^a = \mathbf{v}_{gr}^b$ are the group velocities of molecules forming primary effectons in (a) and (b) states.

The accelerations, related to (ac) and (bc) convertons could be calculated as:

$$a_{ac} = \frac{h}{m[(\lambda_{a,b})_{lb} - (\lambda_{a,b})_{tr}]}\nu_{ac};\quad a_{bc} = \frac{h}{m[(\lambda_{a,b})_{lb} - (\lambda_{a,b})_{tr}]}\nu_{bc}$$

$\nu_{ac} = |(\nu_a)_{tr} - (\nu_a)_{lb}|$ and $\nu_{bc} = |(\nu_b)_{tr} - (\nu_b)_{lb}|$ are frequencies of quantum beats between corresponding states.

Acceleration of molecules, accompanied the macroconvertons excitation, is a sum of two previous ones, like the corresponding forces with same directions:

$$a_{cMt} = a_{ac} + a_{bc} \qquad 17.4a$$

The 3D interceptions/superpositions of three gravitational waves with different orientations, induced by convertons form corresponding *gravitational convertons*.

Taking into account (17.3) and (17.4), the coherent changes of wave B accelerations as a result of $[a \rightleftharpoons b]$ transitions are determined as:

$$[a_t = a_b - a_a = \frac{h}{m\lambda_a}(\nu_b - \nu_a) = \frac{h}{m\lambda_a}\nu_t]_{tr,lb}^{1,2,3} \qquad 17.5$$

where: $[\nu_t = \nu_p = \nu_b - \nu_a]_{tr,lb}^{1,2,3}$ are the frequencies of transitons of primary effectons, equal to that of IR photons, forming primary electromagnetic deformons.

The 3D superposition of gravitational waves, induced by such processes, form quasiparticles which can be termed *gravitational primary transitons.*

The *secondary* gravitational effectons, transitons and deformons can be stimulated by dynamics of corresponding types of quasiparticles: secondary effectons, secondary transitons and secondary acoustic deformons. Macro- and super-effectons and macro- and super-deformons of condensed matter also should have their gravitational analogs.

The hierarchic system of 3D gravitational standing waves radiated by the all possible states of 24 quasiparticles of any objects contains much more information as compared to a *3D* system of acoustic or electromagnetic waves, excited by the transition states of primary, secondary effectons and convertons only.

Other advantages of a gravitational 3D standing waves, as a carrier and storage of information are follows:

- the gravitational effectons and deformons (primary and secondary) have practically no obstacles, as far they penetrate throw any types of screens;

- the principle of their organization as 3D standing waves, makes it possible to conserve the information they contain for a long time in the form of stable vacuum excitations.

The increasing of the correlation of big group of primary effectons dynamics in composition of macro- and super-effectons or due to another orchestrating factors will lead to enhancement of the gravitational waves amplitude and energy, excited by coherent oscillations.

Superposition of all types of gravitational effectons, convertons, transitons, deformons and their combinations represents hierarchic system of gravitational 3D standing waves, like holograms. This superposition contains all information about dynamics and spatial dimensions of any coherent mesoscopic regions of any condensed matter, including living organisms.

The stable characteristic modes of particles collective vibrations, generating standing waves, could produce stable gravitational excitations in vacuum. By analogy with distant Van der Waals interaction, the distant Vibro-Gravitational Interaction could be introduced.

### 17.5 Resonant remote vibro-gravitational interaction (VGI)

In accordance with our model, Vibro-Gravitational Interaction (VGI) is based on the ability of coherent thermal oscillations of molecules (atoms) of one of few distant macroscopic bodies to induce the similar type of in-phase oscillations in these bodies by means of coherent gravitational waves. The feedback reaction affect the dynamic properties of body-inductor as well.

*All types of described in previous section gravitational excitations can contribute in VGI.* The role of (*a*) and (*b*) states of primary *lb* effectons in such interaction is bigger than that of primary *tr* effectons because the number of coherent oscillators in former quasiparticle is more than in latter one.

If our hypothesis is correct, then the decreasing of temperature have to lead to enhancement of (VGI), based on principle of resonant excitation. The total potential energy of a system bodies, tuned by means of distant Van der Waals and VGI is less than a sum of potential energies of isolated bodies. Thermal vibrations in interacting system of bodies are more coherent than in isolated bodies. This phenomena could be considered as a kind of macroscopic self-organization of system in time.

*A change in emotional state or a process of braining* should be accompanied by a corresponding change in the combination of effectons and deformons in the cytoskeleton, microtubules and membranes of nerve cells and their ensembles. Then, according to our hypothesis, it must be followed by the perturbation of the properties of the electromagnetic and gravitational excitations, surrounding the organism.

According to the aforesaid, the state of the central nerve system, the character and degree of its excitation must affect not only the properties and organization of neuronal nets, but also the mesoscopic-domain structure of individual nerve cell membranes and their ability to form synaptic contacts. It can be expected that during the training and emergency of the long-term memory, the properties of membranes and a system of microtubules will change in a definite way. The orchestration *of nerve cells* via systems of microtubules as a consequence of meditation, increasing the amplitude of 3D standing waves is an important factor for biofield radiation and perception.

The investigation of results of acupuncture points stimulation, correcting the biorhythms, on cytoskeleton of the nerve cells and neurons net organization looks to be very interesting.

### 17.6. Possible explanation of Kirlian effect and its dependence on the emotional state

One of the sensitive physical methods for control of the emotional state of a human being is based upon the Kirlian effect or its technically advanced version, named Gas Discharge Visualization (GDV) (Korotkov, 2001).

The effect is that around an object placed into a high-impulse electromagnetic field, there appears a corona discharge which can be registered with a photo film or photomultiplier. It is a consequence of radiation, excited by collisions between electrons, emitted from the object surface in strong gradient of external electric field and the air molecules. Numerous experiments have shown that the shining of the skin in the various parts of the human body surface and palm depends on kind and tension of psychic activities (intellectual efforts, concentration and relaxation, etc.).

The dependence of these phenomena on the state of biofield in we can explain by the change in dynamics of membranes and coherent radiation of microtubules (MT) of skin cells. One of the consequences of nerve excitation is the definite change of the amplitude-frequency properties of UV, visible and IR deformons, radiated by skin. The high-frequency radiation of microfilament system (250-600 nm) could be a result of its collective disassembly of the actin system, accompanied by *gel-sol* transition, accompanied by cavitational fluctuations of cytoplasmic water, dissociation and recombination of water molecules (see sections 11.10 and 17.5). The coherent IR photons radiation and absorption is due to $[a \rightleftharpoons b]$ transitions of primary translational and librational effectons of intra-cellular water. The fraction of this electromagnetic radiation of body, not involved in formation of standing waves, may affect the kinetic energy of the emitted electrons and probability of excitation of the air molecules, resulting from the electrons collisions with molecules.

The electrostatic interface - skin polarization effect, related to polarization $\rightleftharpoons$ depolarization equilibrium shift of the nerve cells, close to surface of body, may also influence the velocity and kinetic energy of the emitted by surface of the object electrons.

Consequently, the shining (radiation) frequency of the air molecules in Kirlian effect is dependent on properties of electromagnetic "aura" around the body, emitted by cells and skin electric polarization.

For example, the decreasing of the kinetic energy of electrons in the restraining electric field of skin must induce the *red shift* of shining spectrum of Kirlian effect. The decreasing of triggering the electrons velocity skin radiation/polarization follows by the opposite - *blue shift* of Kirlian effect spectrum. The electromagnetic field, radiated by primary effectons, around inorganic objects and their interface static electric polarization can be more stable and strong than around the living ones. However it also could be dependent on temperature, pressure and phase transitions, induced by them.

The components of biofield, like *vibro-gravitational interaction,* described above and *Virtual Replica* (Kaivarainen, 2006 a,b) are able penetrate through any types of screens and on the big distance. The Virtual Replica (VR) of the object may affect the electric permittivity of surrounding space and influence on threshold of air molecules ionization by electrons. In some condition the life-time of VR as a system of modulated standing virtual pressure waves (VPW$^\pm$) can exceed the life time of the object itself (Kaivarainen, 2006 a,b). This phenomena may be responsible for morphological 'memory' of Kirlian picture of the actual object even after its partial destruction. This effect of such virtual 3D memory was confirmed experimentally on example of the plants leaves and their parts.

## Chapter 18

## Elementary act of consciousness or "Cycle of Mind", involving distant and nonlocal interaction

Each macroscopic process can be subdivided on the quantum and classical stages.
*Particularly, the quantum stages of Cycle of Mind involves:*
1) the stimulation of dynamic correlation between water clusters in the same and remote microtubules (MTs) in state of mesoscopic Bose condensation (mBC) by phonons (acoustic waves) and by librational IR photons (electromagnetic-EM waves) distant exchange;
2) the transition from distant EM interaction between remote MTs to nonlocal quantum interaction, induced by IR photons exchange between clusters, the clusters Virtual Replicas multiplication (VRM) and virtual guides (VirG) formation between elementary particles of remote coherent water molecules. This process represents transition from mesoscopic Bose condensation to macroscopic nonuniform semivirtual Bose condensation (VirBC) (Kaivarainen, 2006, http://arxiv.org/abs/physics/0207027);
3) the collapsing of corresponding macroscopic wave function, as a result of the optical bistability of the entangled water clusters and their disassembly due to librational photons pumping, shifting clusters to less stable state;
4) turning the clusterphilic interaction between water clusters in the open state of cavities between alpha and beta tubulins to hydrophobic one and the in-phase shift of these cavities to the closed state due to clusters disassembly (Kaivarainen, http://arxiv.org/abs/physics/0102086).

*The classical stages of our model of elementary act of consciousness or "Cycle of Mind" are following:*
a) the nerve cells membrane depolarization;
b) the gel → sol transition induced by disconnection of microtubules (MTs) with membranes and disassembly of the actin filaments;
c) the shape /volume pulsation of dendrites of big number of coherently interacting nerve cells, accompanied by jump-way reorganization of synaptic contacts on the dendrites surface;
d) the back sol → gel transition in corresponding cells ensembles, stabilizing (memorizing) of new state by formation of new system of microtubules and MTs associated proteins in the dendrites (MAPs).

The 'Period of Cycle of Mind' is determined by the sum of the life-time of quantum phase - semivirtual macroscopic Bose condensation, providing coherence and macroscopic entanglement and the life-time of collapsed mesoscopic Bose condensation, induced by decoherence factors. The life-time of entangled coherent phase can be many times shorter than classical phase. Consequently, the $[coherence \rightleftharpoons decoherence]$ dynamic equilibrium in macroscopic system of neurons is strongly shifted to the right. However, even extremely short time of macroscopic entanglement is enough for nonlocal remote interaction. This is true not only for systems under consideration, but for any kind of oscillating macroscopic entanglement.

Our approach to elementary act of consciousness has some common features with well-known Penrose - Hameroff model, interrelated act of consciousness with the wave function of microtubules collapsing. So we start from description of this Orchestrated objective reduction (Orch OR) model.

### 18.1 The basis of Orchestrated objective reduction (Orch OR) model

## of Penrose and Hameroff and comparison with Cycle of Mind model

Non-computable self-collapse of a quantum coherent wave function within the brain may fulfill the role of non-deterministic free will after Penrose and Hameroff (1995).

For biological qubits Penrose and Hameroff chose the open and closed clefts between the pair of tubulin subunits in microtubules. Tubulin qubits would interact and compute by entanglement with other tubulin qubits in microtubules in the same and different neurons.

It was known that the pair of alpha and beta tubulin subunits flexes 30 degrees, giving two different conformational shapes as superpositions.

The authors considered three possible types of tubulin superpositions: separation at the level of the entire protein, separation at the level of the atomic nuclei of the individual atoms within the proteins, and separation at the level of the protons and neutrons (nucleons) within the protein.

The calculated gravitational energy (E) at the level of atomic nuclei of tubulins had the highest energy, and would be the dominant factor in the wave function collapsing.

The best electrophysiological correlate of consciousness is gamma EEG, synchronized oscillations in the range of 30 to 90 Hz (also known as "coherent 40 Hz") mediated by dendritic membrane depolarizations (not axonal action potentials). This means that roughly 40 times per second (every 25 milliseconds) neuronal dendrites depolarize synchronously throughout wide regions of brain.

Using the indeterminacy principle $E = \hbar/t$ for OR, the authors take $t = 25$ ms, and calculated (E) in terms of number of tubulins (since $E$ was already evaluated for one tubulin).

The number of tubulins to be required in isolated superposition to reach OR threshold in $t = 25$ ms turned out to be $2 \times 10^{11}$ tubulins.

Each brain neuron is estimated to contain about $10^7$ tubulins (Yu and Bass, 1994). If 10% of each neuron's tubulins became coherent, then Orch OR of tubulins within roughly 20,000 (gap-junction connected) neurons would be required for a 25 ms conscious event, 5,000 neurons for a 100 ms event, or 1,000 neurons for a 500 ms event, etc.

These estimates (20,000 to 200,000 neurons) fit very well with others from more conventional approaches suggesting tens to hundreds of thousands of neurons are involved in consciousness at any one time.

The environmental decoherence can be avoided in quasi-solid (gelatinous: "gel") phases due to polymerization states of the actin. In the actin-polymerized gel phase, cell water and ions are ordered on actin surfaces, so microtubules are embedded in a highly structured (i.e. non-random) medium. Tubulins are also known to have C termini "tails", negatively charged peptide sequences extending string-like from the tubulin body into the cytoplasm, attracting positive ions and forming a plasma-like Debye layer which can also shield microtubule quantum states. Finally, tubulins in microtubules were suggested to be coherently pumped laser-like into quantum states by biochemical energy (as proposed by H. Fröhlich).

Actin state dependent gel⇌sol cycling occur with frequency 40 Hz. Thus during classical, liquid (sol) phases of actin depolymerization, inputs from membrane/synaptic inputs could "orchestrate" microtubule states. When actin gelation occurs, quantum isolation and computation ensues until OR threshold is reached, and actin depolymerizes.

The result of each OR event (in terms of patterns of tubulin states) would proceed to organize neuronal activities including axonal firing and synaptic modulation/learning. Each OR event (e.g. 40 per second) is proposed to be a conscious event.

One implication of the Orch OR model is that consciousness is a sequence of discrete events, related to collapsing of general for these states wave function.

However, the following problems are not clear in the Hamroff-Penrose approach:
a) what is the mechanism responsible for coherence of big number of remote tubulins and their entanglement;

b) how the microtubules can be unified by single wave function, like in the case of macroscopic Bose condensation, e.g. superfluidity or superconductivity?

*Our model of elementary act of consciousness or cycle of mind has the answers to this crucial questions.*

In our approach we explain the selection of certain configurational space of huge number of 'tuned' neurons, not by structural changes of tubulins like in Hameroff-Penrose model, but by increasing of mass of water in state of macroscopic BC in brain in the process of condensation of spatially separated mesoscopic BC - mBC (coherent water clusters in MTs), stimulated by IR photons exchange.

However, the macroscopic BC resulting from unification of mesoscopic BC is initiated by correlated shift of dynamic equilibrium of nonpolar clefts formed by tubulins, between the open (b) and closed (a) states to the open one, increasing the fracture of water clusters and their resulting mass. The corresponding structural rearrangements of tubulins pairs in the process of shift of [open $\rightleftharpoons$ closed] clefts to the right or left itself, do not change their mass. So, they can not be a source of wave function collapsing "under its own weight" in contrast to increasing of mass of water in state of entangled nonuniform semivirtual BC, proposed in our approach.

The dynamics of increasing $\rightleftharpoons$ decreasing of the entangled water mass in state of macroscopic BC is a result of correlated shift of dynamic equilibrium between primary librational (lb) effectons (coherent water clusters - mesoscopic BC), stabilized by open inter-tubulins cavities and primary translational (tr) effectons (independent water molecules), corresponding to closed cavities.

The correlated conversions between librational (lb) and translational (tr) effectons of water in remote MTs, representing the [*association* $\rightleftharpoons$ *dissociation*] of the entangled water clusters in state of mBC reflect, in fact, the reversible cycles of [coherence $\rightleftharpoons$ decoherence] corresponding to cycles of mesoscopic wave function of these clusters collapsing.

The relatively slow oscillations of dynamic equilibrium of $lb \rightleftharpoons tr$ conversions with period about $1/40 = 25$ ms are responsible for alternating contribution of macroscopic quantum entanglement and macroscopic wave function collapsing in human consciousness. *Each such reversible dynamic process represents the "Cycle of Mind".*

## 18.2. The basis of Cycle of Mind model, including the distant and nonlocal interactions

In accordance to our model of Cycle of Mind, each specific kind of neuronal ensembles excitation, accompanied by jump-way reorganization of big number of *dendrites and synaptic contacts* - corresponds to certain change of hierarchical system of three - dimensional (3D) standing waves of following kinds:
- thermal de Broglie waves, produced by anharmonic translations and librations of molecules;
- electromagnetic (IR) waves;
- acoustic waves;
- virtual pressure waves (VPW$^{\pm}$);
- inter-space waves (ISW), in-phase with VPW$^{\pm}$;
- virtual spin waves and (VirSW).

Scattering of all-penetrating virtual waves of Bivacuum, representing the *reference waves,* on the de Broglie waves of atoms and molecules of any mesoscopic or macroscopic object - generates the *object waves* by analogy with optical holography.

The interference of Bivacuum *reference waves* with the *object waves* forms the primary Virtual Replica (VR) of any systems, including: water clusters, microtubules (MTs) and whole neurons. The notion of primary VR of any material object was introduced earlier in theory of Bivacuum (Kaivarainen, 2006; 2007).

The *primary* **VR** of any macroscopic object represents 3D interference pattern of Bivacuum *reference waves*:
- virtual pressure waves (**VPW**$^{\pm}$),
- virtual spin waves (**VirSW**$^{\pm 1/2}$) and
- inter-space waves (**ISW**)

with the *object waves* of similar nature.

The resulting *primary* **VR** can be subdivided on the *surface* and *volume* virtual replicas: **VR**$_S$ and **VR**$_V$, correspondingly. The **VR**$_S$ contains the information about shape of the object, like regular optical hologram. The **VR**$_V$ contains the information about internal spatial and dynamic properties of the object.

The multiplication/iteration of the *primary* **VR** with properties of 3D standing waves in space and time was named Virtual Replica Multiplication: **VRM(r, t)**.

The **VRM(r,t)** can be named *Holoiteration* by analogy with **hologram**. In Greece '*holo*' means the 'whole' or 'total'.

The factors, responsible for conversion of mesoscopic Bose condensation (mBC) to macroscopic BC are following:

a) the IR photons exchange interaction between coherent water clusters in the same and remote microtubules (MTs),

b) the spatial Virtual Replica multiplication: **VRM(r)** and

c) formation of Virtual Guides of spin, momentum and energy (**VirG**$_{SME}$), connecting the entangled protons and neutrons of similar (like H$_2$O) remote coherent molecules and atoms (see Kaivarainen 2006; 2007).

*The nonuniform macroscopic BC can be considered as a unified sub-systems of connected by bundles of* **VirG**$_{SME}$ *nucleons of water clusters (actual mesoscopic BC), abridged by* **VRM(r,t)** *and* **VirG**$_{SME}$ *(virtual BC).*

The dynamics of [increasing ⇌ decreasing] of the entangled water mass in state of macroscopic BC in the process of elementary act of consciousness is a result of correlated shift of dynamic equilibrium between open and closed cavities formed by alpha and beta tubulins. The closed or open configuration of cavities, the relative orientation of microtubules and their stability as respect to disassembly can be regulated by variable system of microtubule - associated - proteins (MAPs).

The **Reason and Mechanism** of macroscopic Wave Function collapsing, induced by decoherence of VirBC can be following:

*The Reason* of collapsing is destabilization of macroscopic Bose condensation of coherent water clusters in state of mesoscopic BC in MTs. *The Mechanism* is a consequence of such known quantum optic phenomena, as *bistability*. The bistability represents the $a \rightleftharpoons b$ equilibrium shift of librational primary effectons between the acoustic (a) and optic (b) state to the less stable (b) state, as a consequence of librational IR photons pumping and absorption by water clusters, saturating b-state. This saturation can be considered as the decoherence factor, triggering the macroscopic Wave Function collapsing.

The result of *bistability* is the collective dissociation of water clusters in MTs between tubulins and shift the nonpolar cavities states equilibrium to the closed state. This process is accompanied by shrinkage of MTs. *It is a transition from quantum to classical stages of "elementary act of consciousness".*

The shrinkage of MTs induces the disjoining of the MTs ends from the internal surfaces of membranes of nerve cell bodies. The consequence of disjoining is the [gel ⇌ sol] transition in cytoplasm, accompanied by disassembly of actin filaments.

Strong increasing of the actin monomers free surface and the fraction of water, involved in hydration shells of these proteins, decreases the internal water activity and initiate the passive osmosis of water into the nerve cell from the external space. The cell swallows and its volume increases. Corresponding change of cell's body volume and shape of dendrites is followed by synaptic contacts reorganization. *This is a final classical stage of "Cycle of Mind"*.

### 18.3. Mesoscopic Bose condensation (mBC) at physiological temperature. Why it is possible?

The possibility of mesoscopic (intermediate between microscopic and macroscopic) Bose condensation in form of coherent molecular and atomic clusters in condensed matter (liquid and solid) at the ambient temperature was rejected for a long time. The reason of such shortcoming was a ***wrong starting assumption***, that the thermal oscillations of atoms and molecules in condensed matter are harmonic ones (see for example: Beck and Eccles, 1992). The condition of harmonic oscillations means that the averaged kinetic ($T_k$) and potential ($V$) energy of molecules are equal to each other and linearly dependent on temperature ($T$):

$$T_k = V = \frac{1}{2}kT \qquad 18.1$$

where: $k$ is a Boltzmann constant

The averaged kinetic energy of the oscillating particle may be expressed via its averaged momentum ($p$) and mass ($m$):

$$T_k = p^2/2m \qquad 18.2$$

The most probable wave B length ($\lambda_B$) of such particle, based on wrong assumption (1), is:

$$\lambda_B = h/p = \frac{h}{(mkT)^{1/2}} \qquad 18.3$$

It is easy to calculate from this formula, that around the melting point of water $T = 273\,K$ the most probable wave B length of water molecule is about 1Å, i.e. much less than the distance between centers of $H_2O$ ($l \sim 3$ Å). This result leads to shortcoming that no mesoscopic Bose condensation (BC) is possible at this and higher temperature and water and ice are classical systems.

It is known from theory of Bose condensation that mesoscopic BC is possible only at conditions, when the length of waves B of particles exceeds the average distance between their centers ($l$):

$$l = (V_0/N_0)^{1/3} \qquad 18.4$$

$$L > \lambda_B > l \qquad 18.5$$

where: $L$ is a macroscopic parameter, determined by dimensions of the whole sample.

Condition (18.5) is a condition of partial or mesoscopic Bose condensation in form of coherent molecular clusters - named primary effectons, which confirmed to be correct in our Hierarchic theory of condensed matter and related computer calculations (see new book online: http://arxiv.org/abs/physics/0102086).

Consequently, the incorrect assumption (18.1) leads to formula (18.3) and incorrect result:

$$\lambda_B < l = (V_0/N_0)^{1/3} \qquad 18.6$$

meaning the absence of mesoscopic Bose condensation (mBC) in condensed matter at room and higher temperature.

The right way to proceed is to evaluate correctly the ratio between internal kinetic and potential energy of condensed matter and after this apply to Virial theorem (Clausius,1870).

It is shown (Kaivarainen, 1995; 2007), that the structural factor (S) of collective excitation, which can be calculated using our pCAMP computer program reflects the ratio of kinetic energy ($T_k$) to the total energy ($E = V + T_k$) of quasiparticle.

If $S = T_k/(V + T_k) < 1/2\,(V > T_k)$, this points to anharmonic oscillations of particles and nonclassical properties of corresponding matter in accordance to Virial theorem.

The Virial theorem in general form is correct not only for classical, but as well for quantum systems. It relates the averaged kinetic $\bar{T}_k(\vec{v}) = \sum_i \overline{m_i \mathbf{v}_i^2/2}$ and potential $\bar{V}(r)$ energies of particles, composing these systems:

$$2\bar{T}_k(\vec{r}) = \sum_i \overline{m_i \mathbf{v}_i^2} = \sum_i \overline{\vec{r}_i \partial V/\partial \vec{r}_i} \qquad 18.7$$

It follows from Virial theorem, that if the potential energy $V(r)$ is a homogeneous $n - order$ function:

$$V(r) \sim r^n \qquad 18.8$$

then the average kinetic and the average potential energies are related as:

$$n = \frac{2\overline{T_k}}{\overline{V(r)}} \qquad 18.9$$

For example, for a harmonic oscillator, when $\bar{T}_k = \bar{V}$, we have $n = 2$ and condition (1).
For Coulomb interaction: $n = -1$ and $\bar{T} = -\bar{V}/2$.

For water our hierarchic theory based computer calculations of $\bar{T}_k$ and $\bar{V}$ gives: $n_w \sim 1/15$ ($\bar{V}/\bar{T}_k \sim 30)_w$ and for ice: $n_{ice} \sim 1/50$ ($\bar{V}/\bar{T}_k \sim 100)_{ice}$ (see http://arxiv.org/abs/physics/0102086).

It follows from (18.8) and our results, that in water and ice the dependence of potential energy on distance (r) is very weak:

$$V_w(r) \sim r^{(1/15)}; \qquad V_{ice} \sim r^{(1/50)} \qquad 18.10$$

Such weak dependence of potential energy on the distance can be considered as indication of long-range interaction due to the expressed cooperative properties of water as associative liquid and the ability of its molecules for mesoscopic Bose condensation (mBC).

The difference between water and ice (18.10) and our computer simulations prove, that the role of distant Van der Waals interactions, stabilizing primary effectons (mesoscopic molecular Bose condensate), is increasing with temperature decreasing and [liquid→solid] phase transition. This correlates with strong jump of dimensions of $H_2O$ clusters in state of mBC just after freezing, evaluated in our work.

The conditions (18.10) are good evidence that oscillations of molecules in water and ice are strongly anharmonic and the condensed matter in both phase can not be considered as a classical system. For real condensed matter we have:

$$\overline{T_k} \ll \overline{V} \quad and \quad \lambda_B > l = (V_0/N_0)^{1/3} \qquad 18.11$$

It is important to note, that when the average momentum ($\bar{p}$) is tending to zero for example with temperature decreasing or pressure increasing, the kinetic energy is also tending to zero:

$$\overline{T_k} = \bar{p}^2/2m \to 0 \qquad 18.12$$

then the ratio: $n = \frac{2\overline{T_k}}{\overline{V(r)}} \to 0$ and the interaction between particles becomes independent on the distance between them:

$$V_w(r) \sim r^{(n \to 0)} = 1 = const \qquad 18.13$$

*Consequently the interaction turns from the regular distant interaction to the nonlocal one.*

This transition is in-phase with turning of the particles system from state of mesoscopic Bose condensation to macroscopic one. We came to important conclusion, that the conditions of nonlocality (18.12 and 18.13) become valid at conditions of macroscopic BC at $\bar{p} \to 0$:

$$\bar{\lambda}_B = (h/\bar{p}) \to \infty \qquad 18.14$$

This macroscopic BC can be *actual*, like in the case superconductivity and superfluidity, it can be virtual BC, composed from Bivacuum dipoles (see http://arxiv.org/abs/physics/0207027) and it can be *semiactual or semivirtual BC*. The latter will be discussed in the next section.

### 18.4. The transition from distant electromagnetic interaction between remote water clusters to nonlocal interaction

The role of Virtual Replica of clusters in state of mBC spatial multiplication VRM(r) (see http://arxiv.org/abs/physics/0207027) is to create the virtual connections between remote actual clusters. The subsequent formation of Virtual Channels between big number of pairs of coherent elementary particles (electrons, protons and neutrons) of opposite spins, - turns the mesoscopic BC to macroscopic one. This is a result of unification of remote clusters wave function to integer linear superposition of its eigenvalues, resulting in macroscopic wave function. Part of these eigenvalues of the integer wave function corresponds to virtual replica (3D virtual standing waves) of the cluster and other to clusters themselves. The corresponding nonuniform macroscopic Bose condensate, consequently, is partly virtual and partly actual.

The transition from distant EM interaction between remote MTs to nonlocal quantum interaction is a result of entanglement between clusters - mBC, stimulated by IR photons exchange and Virtual Replicas multiplication (VRM) of the clusters. This transition is accompanied by nonuniform macroscopic virtual Bose condensation (VirBC). *The nonuniform VirBC become possible only at certain spatial separation and orientation of coherent water clusters as 3D standing de Broglie waves of water molecules in the entangled microtubules.*

The mechanism of this transition is based on our theories of Virtual Replica multiplication in space - VRM(r), virtual Bose condensation (VirBC) and nonlocal virtual guides (VirG$_{S,M,E}$) of spin, momentum and energy (Kaivarainen, 2006; 2007; http://arxiv.org/abs/physics/0207027).

The Virtual Guides have a shape of virtual microtubules with properties of quasi-1-dimensional virtual Bose condensate (Fig.50). The VirG are constructed from 'head-to-tail' polymerized Bivacuum bosons $BVB^{\pm}$ or Cooper pairs of Bivacuum fermions $[BVF^{\uparrow} \bowtie BVF^{\downarrow}]$. The bundles of VirG$_{S,M,E}$, connecting coherent atoms of Sender (S) and Receiver (S), named the *Entanglement Channels* (see formula 18.5), are responsible for macroscopic entanglement, providing nonlocal interaction, telepathy, remote healing and telekinesis. The poltergeist can be considered as a private case of telekinesis, realized via Entanglement Channels, connecting coherent elementary particles of psychic and the object.

The changes of de Broglie waves of atoms and molecules, participating in elementary act of consciousness, modulate virtual pressure waves of Bivacuum (**VPW**$^{\pm}$). These modulated standing virtual waves form the *Virtual Replica (VR)* of 'tuned' neuronal ensembles and of microtubules (MTs) systems.

*The notion of Virtual Replica of any material object* was introduced by this author in Unified theory of Bivacuum, duality of particles, time and fields (Kaivarainen, 2006; 2007). The Virtual Replica (VR) or *Virtual Hologram* of any material object - from elementary particles up to molecules and complex hierarchical systems, like biological one, planetary systems, galactic, etc.

In general case complex VR is a result superposition of big number of elementary VR.

The VR of the object can be subdivided to the

*a) surface VR* and

*b) volume VR*

The surface VR has a similarity with regular optical hologram, but the role of coherent *reference*

*waves* (like the laser ones) play Virtual Pressure Waves (VPW+/-) of Bivacuum, exciting as a result of symmetric quantum transitions of Bivacuum dipoles (see Appendix). The role of the scattered *object waves* play a secondary $VPW^{\pm}$, scattered and modulated by de Broglie waves of atoms and molecules of the object.

The *volume VR* is a new notion as respect to regular hologram. It is a consequence of ability of $VPW^{\pm}$ to penetrate throw any objects (in contrast to photons) and as a result to be modulated/scattered on the internal de Broglie waves of the object.

Consequently, the total $VR = VR_{sur} + VR_{vol}$ is much more informative than the regular optical hologram. In the conditions of macroscopic entanglement between remote coherent systems, the VR exchange can be responsible for teleportation.

Like the regular hologram, the VR or Virtual Hologram is a result of interference of the *reference waves* with the surface and volume *object waves*. The VR can multiply in space and evolve in time (the latter process is like self-organization to number of metastable states). The VR multiplication: VRM(r,t) is nonlocal process in conditions of virtual macroscopic Bose condensation of Bivacuum dipoles (Kaivarainen, 2006).

The nonlocal superposition of VRM(r,t) of similar by shape and internal properties objects, may influence on the spatial and dynamic organization of these objects. This feedback interaction between VR and the number of objects with similar shape and internal quantum dynamics, producing such type of VR (Virtual Hologram) can be responsible for *morphic resonance,* proposed by Rupert Sheldrake in 1983.

The iteration of the same ideas in humans or animal brains/nerve systems is accompanied by iteration of the same configurations of neuron' ensembles, dendrites, microtubules, etc. This process is also accompanied by the activation and multiplication of corresponding VR. These Virtual Replicas again may have the feedback reaction on the nerve systems and brains of morphogenetically and dynamically close organisms. This phenomena explain the number of experimental facts, confirming the existence of the active Informational Field or NOOSPHERE around the Earth.

*The classical stages Cycle of Mind* represents the gel-sol transition in cytoplasm of the neuron bodies and the tuned neurons ensembles in-phase pulsation and axonal firing, accompanied by redistribution of synaptic contacts between the starting and final states of dendrites.

*The quantum stages of Cycle Mind* involves the periodical collapsing of unified wave function of big number of the entangled coherent water clusters (mBC) in remote microtubules. This means a transition from the state of non-uniform semivirtual macroscopic Bose condensation (BC) to mesoscopic BC. It represents the reversible dissipative process.

The Hameroff - Penrose model considers only the coherent conformational transition of cavities between big number of pairs of tubulins between open to closed states in remote entangled MTs as the act of wave function collapsing.

In our model this transition is only the triggering act, stimulating quantum transition of the big number of water clusters in state of coherent macroscopic Bose condensation (entangled state) to decoherence state of mesoscopic Bose condensation (mBC).

*Consequently, the Cycles of Mind can be considered as a reversible transitions of the certain parts of brain between coherent and noncoherent state, involving quantum and classical stages.* The non-uniform coherent state of semivirtual macroscopic Bose Condensation of water in microtubules system in time and space, e.g. periodically entangled, but spatially separated flickering clusters of mBC is different of continuous macroscopic BC, pertinent for superfluidity and superconductivity.

The Penrose-Hameroff model considers only the correlated conformational transition of cavities between pairs of tubulins between open to closed states of big number of MTs as the act of wave function collapsing. In our model this transition is only the triggering act, stimulating quantum transition of the big number of entangled water clusters in state of coherent macroscopic Bose condensation to state of non-entangled mesoscopic Bose condensation (mBC). Consequently, the cycles of consciousness can be considered as a reversible transitions of certain part of brain between *coherence* and *decoherence.*

The most important collective excitations providing the entanglement and quantum background of consciousness and can be the quantum integrity of the whole organism are coherent water clusters (primary librational effectons), representing mBC of water in microtubules.

Due to rigid core of MTs and stabilization of water librations (decreasing of most probable librational velocity $\mathbf{v}_{lb}$) the dimensions of mBC inside the MTs ($\lambda_B^{lb} = h/m\mathbf{v}_{lb}$) are bigger, than in bulk water and cytoplasm. The dimensions and stability of mBC is dependent on relative position of nonpolar cavities between $\alpha$ and $\beta$ tubulins, forming MTs and cavities dynamics.

In the open state of cavities the water clusters (mBC) are assembled and stable, making possible the macroscopic BC via entanglement in a big number of neurons MTs and in closed state of protein cavities the clusters are disassembled and macroscopic entanglement is destroyed.

The quantum beats between the ground - acoustic (a) and excited - optic (b) states of primary librational effectons (mBC) of water are accompanied by super-radiation of coherent librational IR photons and their absorption (see Introduction and Kaivarainen, 1992). The similar idea for water in microtubules was proposed later by Jibu at al. (1994, p.199).

The process of coherent IR photons radiation ⇌ absorption is interrelated with dynamic equilibrium between open (B) and closed (A) states of nonpolar clefts between $\alpha$ and $\beta$ tubulins. These IR photons exchange interaction between 'tuned' systems of MTs stands for *distant* interaction between neurons in contrast to nonlocal interaction provided by conversion of mesoscopic BC to macroscopic BC.

The collective shift in geometry of nonpolar clefts/pockets equilibrium from the open to closed state is accompanied by the shrinkage of MTs is a result of turning of clusterphilic interaction to hydrophobic ones and *dissociation of water clusters*. This process induce the disjoining of the MTs ends from the membranes of nerve cell bodies and gel → sol transition in cytoplasm, accompanied by disassembly of actin filaments.

Strong abrupt increasing of the actin monomers free surface and the fraction of water, involved in hydration shells of these proteins, decreases the internal water activity and initiate the water passive osmosis into the nerve cell from the external space. The cell swallows and its volume increases. Corresponding change of cell's body volume and shape of dendrites is followed by synaptic contacts reorganization. This is a final stage of multistage act of consciousness.

The *bistability* represents the water clusters polarization change as a result of $a \rightleftharpoons b$ equilibrium shift in librational primary effectons to the right. In turn, this shift is a consequence of librational IR photons pumping and the excited b-state of librational effectons saturation.

The related to above phenomena: the *self-induced transparency* is due to light absorption saturation by primary librational effectons (Andreev, et al., 1988). This saturation can be followed by the pike regime (light emission pulsation, after $b$ −state saturation of librational effectons and subsequent super-radiation of big number of entangled water clusters in state of mBC in the process of their correlated $\sum(\mathbf{b} \to \mathbf{a})$ transitions.

The entanglement between coherent nucleons of opposite spins of H and O of remote water clusters in a big number of MTs, in accordance to our theory of nonlocality, can be realized via bundles of Bivacuum virtual guides $\mathbf{VirG}_{SME}$ of spin, momentum and energy (Kaivarainen, 2006, 2007).

<div style="text-align:center">

*18.4.1 The mechanism of the Entanglement channels formation between remote coherent de Broglie waves of the nucleons*

</div>

The bundles of $\mathbf{VirG}_{SME}$, connecting pairs of protons and neutrons of opposite spins of remote coherent molecules in state of mesoscopic Bose condensation (mBC) were named the *Entanglement channels* (Kaivarainen, 2006, 2007):

$$\textbf{Entanglement channel} = \left[ \mathbf{N(t,r)} \times \sum^{n} \mathbf{VirG}_{SME} (\mathbf{S \iff R}) \right]^{i}_{x,y,z} \quad\quad 18.15$$

where: (**n**) is a number of pairs of similar tuned elementary particles (protons, neutrons and electrons) of opposite spins of the remote entangled atoms and molecules; $\mathbf{N(t,r)}$ is a number of coherent atoms/molecules in the entangled molecular (e.g. water) clusters in state of mBC.

The Virtual Guides (microtubules), connecting the remote elementary particles have a properties of quasi- 1- dimensional virtual Bose condensate.

A *double* Virtual Guides are composed from Cooper pairs of Bivacuum fermions of opposite spins ($\mathbf{BVF^\uparrow \bowtie BVF^\downarrow}$):

$$\textit{Double } \mathbf{VirG}^{\mathbf{BVF^\uparrow \bowtie BVF^\downarrow}}_{SME} = \mathbf{D(r,t)} \times [\mathbf{BVF^\uparrow_+ \bowtie BVF^\downarrow_-}]^{s}_{S=0} \quad\quad 18.17$$

where: $\mathbf{D(r,t)}$ is a number of Bivacuum dipoles in Virtual guides, dependent on the distance (**r**) between remote but tuned de Broglie waves of elementary particles of opposite spins. The diameter of these dipoles and spatial gap between their torus and antitorus are pulsing in-phase.

Just the *Entanglement channels* are responsible for nonlocal Bivacuum mediated interaction between the mesoscopic BC, turning them to macroscopic BC. For the entanglement channels activation the interacting mBC systems should be in non-equilibrium state.

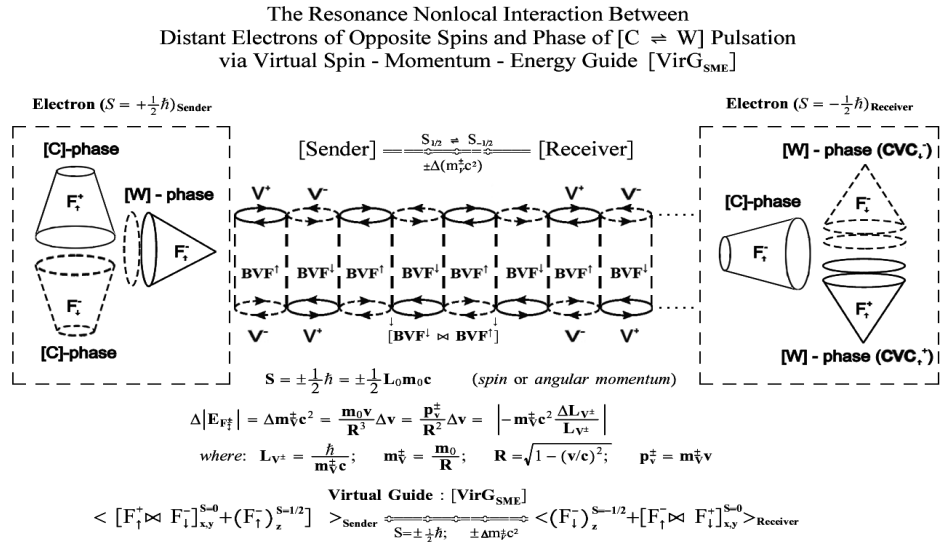

**Figure 50**. The mechanism of nonlocal Bivacuum mediated interaction (entanglement) between two distant unpaired sub-elementary fermions of 'tuned' elementary triplets (particles) of the opposite spins $< [\mathbf{F^+_\uparrow \bowtie F^-_\downarrow}] + \mathbf{F^-_\uparrow} >^{i}_{\text{Sender}}$ and $< [\mathbf{F^+_\downarrow \bowtie F^-_\uparrow}] + \mathbf{F^-_\downarrow} >^{i}_{\text{Receiver}}$, with close frequency of $[\mathbf{C \rightleftharpoons W}]$ pulsation and close de Broglie wave length ($\lambda_B = h/m^+_V v$) of particles. The tunnelling of momentum and energy increments: $\Delta|m^\pm_V c^2| \sim \Delta|\text{VirP}^+| \pm \Delta|\text{VirP}^-|$ from Sender to Receiver and vice-verse via Virtual spin-momentum-energy Guide $[\mathbf{VirG}^i_{SME}]$ is accompanied by instantaneous pulsation of the inter-space gap between torus and antitorus of virtual Cooper pairs of Bivacuum fermions: $[\mathbf{BVF^\uparrow \bowtie BVF^\downarrow}]$ composing virtual guide, and their diameter. The nonlocal spin state exchange between [S] and [R] can be induced by the change of polarization of Cooper pairs: $[\mathbf{BVF^\uparrow \bowtie BVF^\downarrow}] \rightleftharpoons [\mathbf{BVF^\downarrow \bowtie BVF^\uparrow}]$ and Bivacuum bosons, composing $\mathbf{VirG}_{SME}(\mathbf{S \iff R})^i$.

The assembly of huge number of bundles of virtual microtubules of Bivacuum, like Virtual Channels side-by-side can compose virtual multilayer membranes. Each of this layer, pulsing in counterphase with the next one between the excited and ground states are interacting with each other via dynamic exchange by pairs of virtual pressure waves $[\mathbf{VPW}^+ \bowtie \mathbf{VPW}^-]$. This process occur without violation of the energy conservation law and is accompanied by nonlocal Bivacuum gap oscillation over the space of virtual BC of Bivacuum dipoles. The value of spatial gap between the actual and complementary torus and antitorus of Bivacuum fermions is dependent on their excitation state quantum number ($\mathbf{n = 0, 1, 2, 3\ldots}$):

$$[\mathbf{d}_{\mathbf{v}^+ \Updownarrow \mathbf{v}^-}]_n = \frac{h}{\mathbf{m}_0 \mathbf{c}(1 + 2\mathbf{n})} \quad\quad 18.18$$

The Bivacuum gap oscillations and corresponding inter-space waves (ISW), correlated with $VPW^\pm$, can be responsible for the *lateral or transversal nonlocality of Bivacuum in contrast to longitudinal one,* connecting the nucleons with opposite spin, realized via $\mathbf{VirG}_{SME}$ (Kaivarainen, 2006 a, b).

The gel-sol transition in the number of entangled neurons is accompanied by decreasing of viscosity of cytoplasm. The tuned - parallel orientation of MTs in tuned remote cells change as a result of Brownian motion, accompanied by decoherence and loosing the entanglement between water clusters in MTs.

This is followed by relaxation of the internal water + microtubulins to normal dynamics and grows of (+) ends of MTs up to new contacts formation with cells membranes, stabilizing cells dendrites new geometry and synaptic contacts distribution. *This new configuration and state of the nerve system and brain, represents the transition of the Virtual Replica of the brain from the former state to the new one.*

We assume in our model the existence of back reaction between the properties of Virtual Replica of the nerve system of living organisms, with individual properties, generated by systems:

[*microtubules of neurons + DNA of chromosomes*]

and the actual object - the organism itself. The corresponding subsystems can be entangled with each other by the described above Virtual Channels.

The interference of such individual (self) virtual replica VR{self} with virtual replicas of other organisms and inorganic macroscopic system may modulate the properties of VR{self}.

Because of back reaction of VR{self} on corresponding organism, the interaction of this organism with resulting/Global virtual replica of the external macroscopic world may be realized.

The twisting of centrioles in cells to parallel orientation, corresponding to maximum energy of the MTs interaction of remote neurons, is a first stage of the *next* elementary act of consciousness.

The superradiated photons from enlarged in MTs water clusters have a higher frequency than $(\mathbf{a} \rightleftharpoons \mathbf{b})_{lb}$ transitions of water primary librational effectons of cytoplasmic and inter-cell water. This feature provides the regular *transparency* of medium between 'tuned' microtubules of remote cells for librational photons.

The [gel→sol] transitions in cells is interrelated with tuned nerve cells (ensembles) coherent excitation, their membranes depolarization and the *axonal firing*.

### 18.5 Two triggering mechanisms of Cycle of Mind

It is possible in some cases, that the excitation/depolarization of the nerve cells by the external factors (sound, vision, smell, tactical feeling) are the triggering - *primary* events and [gel→sol] transitions in nerve cells are the *secondary* events.

However, the opposite mechanism, when the tuning of remote cells and [gel → sol] transitions are the *primary* events, for example, as a result of thinking/meditation and the nerve cells depolarization of cells are *secondary* events, is possible also.

The 1st mechanism, describing the case, when depolarization of nerve membranes due to external factors is a *primary* event and *gel → sol* transition a *secondary* one, includes the following stages of elementary act of consciousness:

a) simultaneous depolarization of big enough number of neurons, forming ensemble, accompanied by opening the potential-dependent channels and increasing the concentration of $Ca^{2+}$ in cytoplasm of neurons body;

b) collective disassembly of actin filaments, accompanied by [gel → sol] transition of big group of depolarized neurons stimulated by $Ca^{2+}$ − activated proteins like gelsolin and villin. Before depolarization the concentration of $Ca^{2+}$ outside of cell is about $10^{-3}M$ and inside about $10^{-7}M$. Such strong gradient provide fast increasing of these ions concentration in cell till $10^{-5}M$ after depolarization.

c) strong decreasing of cytoplasm viscosity and disjoining of the (+) ends of MTs from membranes, making possible the spatial fluctuations of MTs orientations inducing decoherence and switching off the entanglement between mBC;

d) volume/shape pulsation of neuron's body and dendrites, inducing reorganization of ionic channels activity and synaptic contacts in the excited neuron ensembles. These volume/shape pulsations occur due to reversible decrease of the intra-cell water activity and corresponding swallow of cell as a result of increasing of passive osmotic diffusion of water from the external space into the cell.

In the opposite case, accompanied the process of braining, the depolarization of nerve membranes, the axonal firing is a *secondary* event and *gel → sol* transition a *primary* one, stimulated in turn by simultaneous dissociation of big number of water clusters to independent molecules. The latter process represents the conversion of primary librational effectons to translational ones, following from our theory (see 'convertons' in the Introduction of book: Kaivarainen, 2007, http://arxiv.org/abs/physics/0102086).

The frequency of electromagnetic field, related to change of ionic flux in excitable tissues usually does not exceed $10^3$ Hz (Kneppo and Titomir, 1989).

The electrical recording of human brain activity demonstrate a coherent (40 to 70 Hz) firing among widely distributed and distant brain neurons (Singer, 1993). Such synchronization in a big population of groups of cells points to possibility of not only the regular axon-mediated interaction, but also to fields-mediated interaction and quantum entanglement between remote neurons bodies.

The dynamic virtual replicas (VR) of all hierarchical sub-systems of brain and its space-time multiplication VRM(r,t) contain information about all kind of processes in condensed matter on the level of coherent elementary particles (Kaivarainen, 2006 a,b). Consequently, our model agrees in some points with ideas of Karl Pribram (1977), David Bohm and Basil Hiley (1993) of holographic mind, incorporated in the hologram of the Universe.

### 18.6. The comparison of Cycle of Mind and Quantum brain dynamics models

Our approach to Quantum Mind problem has some common features with model of Quantum Brain Dynamics (QBD), proposed by L.Riccardi and H.Umezawa in 1967 and developed by C.I.Stuart, Y.Takahashi, H.Umezava (1978, 1979), M.Jibu and K.Yasue (1992, 1995).

In addition to traditional electrical and chemical factors in the nerve tissue function, this group introduced two new types of *quantum* excitations (ingredients), responsible for the overall control of electrical and chemical signal transfer: *corticons and exchange bosons* (dipolar phonons).

The *corticons* has a definite spatial localization and can be described by Pauli spin matrices. The

*exchange bosons*, like phonons are delocalized and follow Bose-Einstein statistics. "By absorbing and emitting bosons coherently, corticons manifest global collective dynamics..., providing systematized brain functioning" (Jibu and Yasue,1993). In other paper (1992) these authors gave more concrete definitions:

"*Corticons* are nothing but quanta of the molecular vibrational field of biomolecules (quanta of electric polarization, confined in protein filaments). *Exchange bosons* are nothing but quanta of the vibrational field of water molecules...".

It is easy to find analogy between spatially localized "corticons" and our primary effectons as well as between "exchange bosons" and our secondary (acoustic) deformons. It is evident also, that our Hierarchic theory is more developed as far as it is based on detailed description of all collective excitations in any condensed matter (including water and biosystems) and their quantitative analysis.

Jibu, Yasue, Hagan and others (1994) discussed a possible role of quantum optical coherence in microtubules for brain function. They considered MTs as a *wave guides* for coherent EM superradiation. They also supposed that coherent photons, penetrating in MTs, lead to *"self-induced transparency"*. Both of these phenomena are well known in fiber and quantum optics. We also use these phenomena for explanation of transition from mesoscopic entanglement of water clusters in MTs to macroscopic one, as a result of IR photons exchange between coherent clusters (mBC). However, we have to note, that the transition of mBC to macroscopic BC in 'tuned' MTs is possible without self-induced transparency also.

It follows also from our approach, that the mechanism of macroscopic BC of water clusters do not need the hypothesis of Frölich that the proteins (tubulins of MTs) can be coherently pumped into macroscopic quantum states by biochemical energy.

We also do not use the idea of Jibu et al. that the MTs works like the photons wave - guides without possibility of side radiation throw the walls of MTs. The latter in our approach increases the probability of macroscopic entanglement between remote MTs and cells of the organism.

### 18.7. The Properties of the Actin Filaments, Microtubules (MTs) and Internal Water in MTs

There are six main kind of actin existing. Most general F-actin is a polymer, constructed from globular protein G-actin with molecular mass 41800. Each G-actin subunit is stabilized by one ion $Ca^{2+}$ and is in noncovalent complex with one ATP molecule. Polymerization of G-actin is accompanied by splitting of the last phosphate group. The velocity of F-actin polymerization is enhanced strongly by hydrolysis of ATP. However, polymerization itself do not needs energy. Simple increasing of salt concentration (decreasing of water activity), approximately till to physiological one - induce polymerization and strong increasing of viscosity.

The actin filaments are composed from two chains of G-actin with diameter of 40 Å and forming double helix. The actin filaments are the polar structure with different properties of two ends.

*Let us consider the properties of microtubules (MT) as one of the most important component of cytoskeleton, responsible for spatial organization and dynamic behavior of the cells.*

The microtubules (MTs) are the nanostructures of cells, interconnecting the quantum and classical stages of the Cycle of Mind.

The [assembly ⇔ disassembly] equilibrium of microtubules composed of $\alpha$ and $\beta$ tubulins is strongly dependent on internal and external water activity ($a$), concentration of $Ca^{2+}$ and on the electric field gradient change due to MTs piezoelectric properties.

The $\alpha$ and $\beta$ tubulins are globular proteins with equal molecular mass ($MM = 55.000$), usually forming $\alpha\beta$ dimers with linear dimension 8 nm. Polymerization of microtubules can be stimulated by NaCl, $Mg^{2+}$ and GTP (1:1 tubulin monomer) (Alberts *et al.*, 1983). The presence of heavy water (deuterium oxide) also stimulates polymerization of MTs.

In contrast to that the presence of ions of $Ca^{2+}$ even in micromolar concentrations, action of colhicine and lowering the temperature till $4^0C$ induce disassembly of MT.

Due to multigenic composition, *α* and *β* tubulins have a number of isoforms. For example, two-dimensional gel electrophoresis revealed 17 varieties of *β* tubulin in mammalian brain (Lee *et al.*, 1986). Tubulin structure may also be altered by enzymatic modification: addition or removal of amino acids, glycosylation, etc.

*Microtubules* are hollow cylinders, filled with water. Their internal diameter about $d_{in} = 140$Å and external diameter $d_{ext} = 280$Å (Figure 51). These data, including the dimensions of *αβ* dimers were obtained from x-ray crystallography (Amos and Klug, 1974). However we must keep in mind that under the conditions of crystallization the multiglobular proteins and their assemblies tends to more compact structure than in solutions due to lower water activity.

This means that in natural conditions the above dimensions could be a bit bigger.

The length of microtubules (MT) can vary in the interval:

$$l_t = (1 - 20) \times 10^5 \text{Å} \qquad \qquad 18.19$$

The spacing between the tubulin monomers in MT is about 40 Å and that between *αβ* dimers: 80 Å are the same in longitudinal and transversal directions of MT.

Microtubules sometimes can be as long as axons of nerve cells, *i.e.* tenth of centimeters long. Microtubules (MT) in axons are usually parallel and are arranged in bundles. Microtubules associated proteins (MAP) form a "bridges", linking MT and are responsible for their interaction and cooperative system formation. Brain contains a big amount of microtubules. *Their most probable length is about $10^5$Å, i.e. close to librational photon wave length.*

The viscosity of ordered water in such narrow microtubules seems to be too high for transport of ions or metabolites at normal conditions.

All 24 types of quasi-particles, introduced in the Hierarchic Theory of matter (Table 1), also can be pertinent for ordered water in the microtubules (MT). However, the dynamic equilibrium between populations of different quasi-particles of water in MT must be shifted towards primary librational effectons, comparing to bulk water due to increased clusterphilic interactions (Kaivarainen, 1985, 2000, 2007). The dimensions of internal primary librational effectons have to be bigger than in bulk water as a consequence of stabilization of MT walls the mobility of water molecules, increasing their most probable de Broglie wave length.

The interrelation must exist between properties of internal water in MT and structure and dynamics of their walls, depending on [*α* − *β*] tubulins interaction. Especially important can be a quantum transitions like convertons [$tr \Leftrightarrow lb$]. The convertons in are accompanied by [dissociation/association] of primary librational effectons, i.e. flickering of coherent water clusters, followed by the change of angle between *α* and *β* subunits in tubulin dimers.

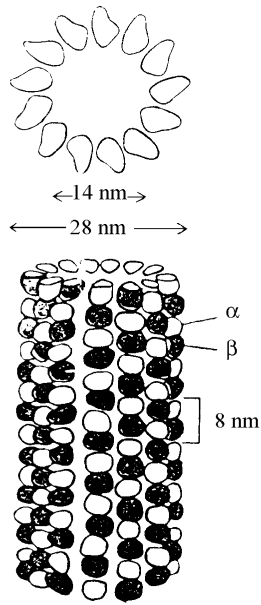

**Figure 51**. Construction of microtubule from $\alpha$ and $\beta$ tubulins, globular proteins with molecular mass 55 kD, existing in form of dimers ($\alpha\beta$). Each $\alpha\beta$ dimer is a dipole with negative charges, shifted towards $\alpha$ subunit (De Brabander, 1982). Consequently, microtubules, as an oriented elongated structure of dipoles system, have the piezoelectric properties (Athestaedt, 1974; Mascarennas, 1974).

Intra-microtubular *clusterphilic interactions* stimulate the growth of tubules from $\alpha\beta$ tubulin dimers. The structural physical-chemical asymmetry of $\alpha\beta$ dimers composing microtubules determines their different rates of growth from the opposite ends ([+] and [-]).

The equilibrium of "closed" (A) and "open"(B) states of nonpolar cavities between $\alpha$ and $\beta$ tubulins in ($\alpha\beta$) dimers can be shifted to the (B) one under the change of external electric field in a course of membrane depolarization. It is a consequence of piezoelectric properties of MTs and stimulate the formation of coherent water clusters in the open cavities of ($\alpha\beta$) dimers. The open cavities serve as a centers of water cluster formation and molecular Bose condensation.

The parallel orientation of MT in different cells, optimal for maximum [MT-MT] resonance interaction could be achieved due to twisting of centrioles, changing spatial orientation of MT. However, it looks that the normal orientation of MT as respect to each other corresponds to the most stable condition, *i.e.* minimum of potential energy of interaction (see Albreht-Buehner, 1990).

It is important to stress here that the orientation of two centrioles as a source of MT bundles in each cell are always normal to each other.

The linear dimensions of the primary librational effectons edge ($l_{ef}^{lb}$) in pure water at physiological temperature ($36^0C$) is about 11 Å and in the ice at $0^0C$ it is equal to 45 Å.

We assume that in the rigid internal core of MT, the linear dimension (edge length) of librational effecton, approximated by cube is between 11Å and 45 Å *i.e.* about $l_{ef}^{lb} \sim 23$Å. It will be shown below, that this assumption fits the spatial and symmetry properties of MT very well.

The most probable group velocity of water molecules forming primary *lb* effectons is:

$$\mathbf{v}_{gr}^{lb} \sim h/(m_{H_2O} \times l_{ef}^{lb})$$  18.20.

The librational mobility of internal water molecules in MT, which determines ($\mathbf{v}_{gr}^{lb}$) should be about 2

times less than in bulk water at $37^0 C$, if we assume for water in microtubules: $l^{lb}_{ef} \sim 23 Å$.

Results of our computer simulations for pure bulk water shows, that the distance between centers of primary [lb] effectons, approximated by cube exceed their linear dimension to about 3.5 times (Fig 52 b). For our case it means that the average distance between the effectons centers is about:

$$d = l^{lb}_{ef} \times 3.5 = 23 \times 3.5 \sim 80 Å \qquad 18.21$$

This result of our theory points to the equidistant (80 Å) localization of the primary *lb* effectons in clefts between α and β tubulins of each (αβ) dimer in the internal core of MTs.

In the case, if the dimensions of librational effectons in MTs are quite the same as in bulk water, i.e. 11 Å, the separation between them should be: $d = l^{lb}_{ef} \times 3.5 = 11 \times 3.5 \sim 40$ Å.

This result points that the coherent water clusters can naturally exist not only between α and β subunits of each pair, but also between pairs of (αβ) dimers.

In the both cases the spatial distribution symmetry of the internal flickering clusters in MT (Fig 51; 52) may serve as an important factor for realization of the signal propagation along the MT (conformational wave), accompanied by alternating process of closing and opening the clefts between neighboring α and β tubulins pairs.

This large-scale protein dynamics is regulated by dissociation ⇌ association of water clusters in the clefts between (αβ) dimers of MT (Fig.52) due to [*lb/tr*] convertons excitation and librational photons and phonons exchange between primary and secondary effectons, correspondingly.

The dynamic equilibrium between *tr* and *lb* types of the intra MT water effectons must to be very sensitive to α − β tubulins interactions, dependent on nerve cells excitation and their membranes polarization.

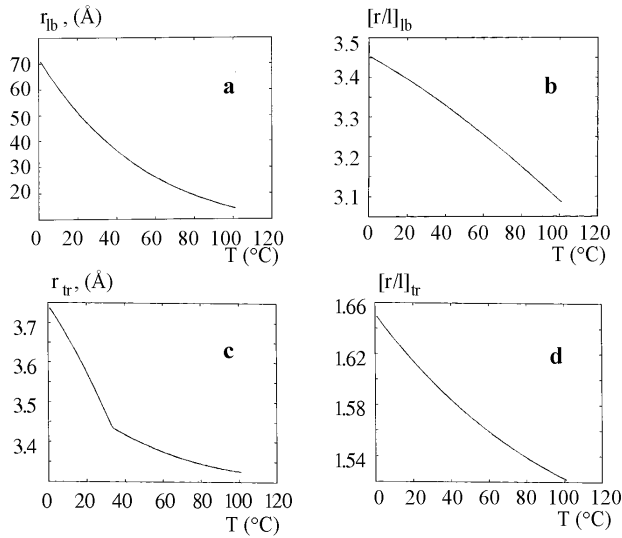

**Figure 52**. Theoretical temperature dependencies of:
(a) - the space between centers of primary [lb] effectons, calculated in accordance to eq.4.62;
(b) - the ratio of space between primary [lb] effectons to their length, calculated, using eq.4.63 ;
(c) - the space between centers of primary [tr] effectons in accordance to eq.4.62;
(d) - the ratio of space between primary [tr] effectons to their length from eq.4.63.

**Two statements of our Cycle of Mind model are important**:
1. The ability of intra-MT primary water effectons (tr and lb) for superradiation of six coherent IR

photons from each of the effectons side, approximated by parallelepiped:

two identical - "longitudinal" IR photons, penetrating along the core of microtubule, forming the longitudinal standing waves inside it and two pairs of identical - "transverse" IR photons, also responsible for the distant, nonlocal interaction between microtubules. *In accordance to superradiation mechanism the intensity of longitudinal radiation of MTs is much bigger than that of transverse one;*

2. The parameters of the water clusters radiation (frequency of librational photons, coherency, intensity) are regulated by the interaction of the internal water with MT walls, dependent on the [open ⇔ closed] states dynamic equilibrium of cavities between $\alpha$ and $\beta$ tubulins.

### 18.8 The role of librational and translational IR photons in the microtubules

We found out that the average length of microtubules ($l$) correlates with length of standing electromagnetic waves of librational and translational IR photons, radiated by corresponding primary effectons:

$$l_{lb} = \kappa \frac{\lambda_p^{lb}}{2} = \frac{\kappa}{2n\tilde{\nu}_p^{lb}} \qquad 18.22$$

and

$$l_{tr} = \kappa \frac{\lambda_p^{tr}}{2} = \frac{\kappa}{2n\tilde{\nu}_p^{tr}} \qquad 18.23$$

here $\kappa$ is the integer number; $\lambda_p^{lb,tr}$ is a librational or translational IR photon wave length equal to:

$$\lambda_p^{lb} = (n\tilde{\nu}_p^{lb})^{-1} \simeq 10^5 \text{Å} = 10\mu \qquad 18.24$$

$$\lambda_p^{tr} = (n\tilde{\nu}_p^{tr})^{-1} \simeq 3.5 \times 10^5 \text{Å} = 30\mu \qquad 18.25$$

where: $n \simeq 1.33$ is an approximate refraction index of water in the microtubule; $\tilde{\nu}_p^{lb} \simeq (700 - 750) \, cm^{-1}$ is wave number of librational photons and $\tilde{\nu}_p^{tr} \simeq (200 - 180) \, cm^{-1}$ is wave number of translational photons.

It is important that the most probable length of MTs in normal cells is about $10\mu$ indeed. So, just the librational photons and the corresponding primary effectons play the crucial role in the entanglement inside the microtubules and between MTs. The necessary for this quantum phenomena tuning of molecular dynamics of water is provided by the electromagnetic interaction between separated coherent water clusters in state of mesoscopic Bose condensate.

In the normal animal-cells, microtubules grow from pair of centriole in center to the cell's periphery. In the center of plant-cells the centrioles are absent. Two centrioles in cells of animals are always oriented at the right angle with respect to each other. The centrioles represent a construction of 9 triplets of microtubules (Fig. 52), i.e. two centriole are a source of: $(2 \times 27 = 54)$ microtubules. The centriole length is about 3000 Å and its diameter is 1000 Å.

These dimensions mean that all 27 microtubules of each centrioles can be orchestrated in the volume ($\mathbf{v}_d$) of one translational or librational electromagnetic deformon:

$$[\mathbf{v}_d = \frac{9}{4\pi} \lambda_p^3]_{tr,lb} \qquad 18.26$$

where: $(\lambda_p)_{lb} \sim 10^5 \text{Å}$ and $(\lambda_p)_{tr} \sim 3.5 \times 10^5 \text{Å}$

Two centrioles with normal orientation as respect to each other and a lot of microtubules, growing from them, contain the internal orchestrated system of librational water effectons. It represent a quantum system with correlated $(a \rightleftharpoons b)_{lb}^{1,2,3}$ transitions of the effectons. The resonance superradiation or absorption of a *number* of librational photons ($3q$) in the process of above transitions, is dependent on

the number of primary *lb* effectons ($q$) in the internal hollow core of a microtubule:
$$q = \frac{\pi L_{MT}^2 \times l}{V_{ef}^{lb}} \qquad 18.27$$

The value of $q$ - determines the intensity (amplitude) of coherent longitudinal librational IR photons radiation from microtubule with internal radius $L_{MT} = 7nm$ and length ($l$), for the case, when condition of standing waves (18.22) is violated.

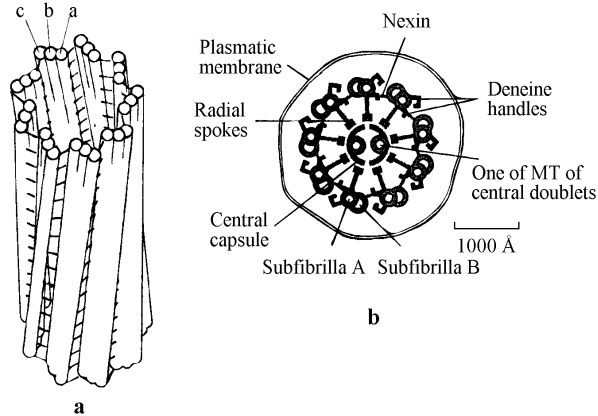

**Figure 53**. ($a$) : the scheme of centriole construction from nine triplets of microtubules. The length and diameter of cylinder are 3000 Å and 1000 Å, correspondingly. Each of triplets contain one complete microtubule and two noncomplete MT;
(b): the scheme of cross-section of cilia with number of MT doublets and MT-associated proteins (MAP): $[2 \times 9 + 2] = 20$. One of MT of periphery doublets is complete and another is noncomplete (subfibrilles A and B).

It is important that the probabilities of pair of longitudinal and two pairs of *transversal* photons, emission as a result of superradiance by primary librational effectons are equal, being the consequence of the same collective $(b \to a)_{lb}$ transition. These probabilities can be "tuned" by the electric component of electromagnetic signals, accompanied axon polarization and nerve cell excitation due to piezoelectric properties of MT.

Coherent *longitudinal* emission of IR photons from the ends of each *pair* of microtubules of two perpendicular centrioles of the *same cell* and from ends of *one* microtubule of *other cell* can form a 3D superposition of standing photons (primary deformons) as a result of 3 photons pairs superposition.

The system of such longitudinal electromagnetic deformons, as well as those formed by transversal photons, have a properties of pilotless 3D hologram. Such an electromagnetic hologram can be responsible for the following physico-chemical phenomena:

-Nonmonotonic distribution of intra-cell water viscosity and diffusion processes in cytoplasm due to corresponding nonmonotonic spatial distribution of macro-deformons;

-Regulation of spatial distribution of water activity ($a_{H_2O}$) in cytoplasm as a result of corresponding distribution of inorganic ions (especially bivalent such as $Ca^{2+}$) in the field of standing electromagnetic waves. Concentration of ions in the nodes of standing waves should be higher than that between them. Water activity ($a_{H_2O}$) varies in the opposite manner than ions concentration.

The spatial variation of ($a_{H_2O}$) means the modulation of [assembly $\Leftrightarrow$ disassembly] equilibrium of filaments of the actin and partly MTs. As a consequence, the volume and shape of cell compartments will be modulated also. The activity of numerous oligomeric allosteric enzymes can be regulated by the water activity also.

The following properties of microtubules can affect the properties of 3D standing waves, radiated by them:
a) total number of microtubules in the cell;
b) spatial distribution of microtubules in the volume of cytoplasm;
c) distribution of microtubules by their length.

The constant of $(a \Leftrightarrow b)$ equilibrium of primary librational effectons

$$(K_{a \Leftrightarrow b})_{lb} = \exp[-(E_a - E_b)/kT]_{lb} \qquad 18.28$$

and that of $(A^* \Leftrightarrow B^*)$ equilibrium of super-effectons are dependent on the structure and dynamics of $\alpha\beta$ tubulin pairs forming MT walls.

This equilibrium is interrelated, in turn, with librational photons frequency $(v_{lb})^{1,2,3}$:

$$[v_{lb} = c(\tilde{v})_{lb} = (V_b - V_a)_{lb}/h]^{1,2,3} \qquad 18.29$$

which is determined by the difference of potential and total energies between (b) and (a) states of primary effectons in the hollow core of microtubules, as far the kinetic energies of these states are equal $T_b = T_a$:

$$[V_b - V_a = E_b - E_a]_{lb}^{1,2,3} \qquad 18.30$$

$(\tilde{v})_{lb}^{1,2,3}$ is the librational band wave number.

The refraction index $(n)$ and dielectric constant of the internal water in MT depends on $[a \Leftrightarrow b]$ equilibrium of the effectons because the polarizability of water and their interaction in (a) state are higher, than that in (b) state.

The condition of mesoscopic Bose condensation (mBC) of any kind of liquid or solids is the increasing of librational or translational de Broglie wave length of molecules over the average separation between two neighboring molecules:

$$\lambda = \left[ \frac{h}{m_{H_2O} \mathbf{v}} \right] > \left[ l_m = \left( \frac{V_0}{N_0} \right)^{1/3} \right] \qquad 18.31$$

The dimensions and stability of water clusters in state of mBC in MTs are dependent on relative position of nonpolar cavities between alpha and beta tubulins, forming MTs and cavities dynamics.

The open state of nonpolar cavities provides the condition of *clusterphilic interaction* with water. This new kind of interaction was introduced by Kaivarainen (1985; 2000; 2007; http://arxiv.org/abs/physics/0102086) as the intermediate one between the hydrophilic and hydrophobic interactions. *The water clusters, in vicinity of nonpolar cavities are bigger and more stable, than the clusters of bulk water* (Fig. 55). As a consequence, the frequency of librational photons, radiated and absorbed in the process of quantum beats between the acoustic (a) and optic (b) states of such clusters is higher that of bulk water. Such phenomena makes the cytoplasmic water transparent for IR photons, radiated by the internal clusters of MTs. The tubulins, composing the walls of MTs are also transparent for these photons. So our approach do not consider MTs as a light-guids.

*This coherent photons exchange between remote water clusters, enhances in MTs (Fig.6) is the precondition of macroscopic Bose condensation and origination of quantum entanglement between big number of 'tuned' neurons.*

The transition of cavities between tubulins from the open to closed state of protein cavities is accompanied by water clusters disassembly and the destruction of non-uniform BC and macroscopic entanglement.

The quantum beats between the ground - acoustic (a) and excited - optic (b) states of primary librational effectons (mBC) of water are accompanied by super-radiation of coherent librational IR photons and their absorption (Kaivarainen, 1992). Similar idea for water in microtubules was proposed later by Jibu at al. (1994).

The number of coherent IR photons radiation ⇌ absorption is dependent on the life-time of water cluster in the open (B) state of nonpolar cleft between α and β tubulins.

*This number can be approximately evaluated.*

The life-time of the open B- state of cleft is interrelated with that of water cluster. Our computer calculations gives a value of the life-time: $(10^{-6} - 10^{-7})$ second.

The frequency of librational IR photons, equal to frequency of quantum beats between the optical and acoustic modes of these clusters is equal to product of corresponding wave number ($\tilde{\nu}_{lb} = 700\ cm^{-1}$) to the light velocity (c):

$$\nu_{lb} = \tilde{\nu}_{lb} c \simeq 700\ cm^{-1} \times 3 \cdot 10^{10}\ cm/s = 2.1 \times 10^{13} s^{-1} \qquad 18.32$$

The characteristic time of these beats, equal to period the photons is

$$\tau = 1/\nu_{lb} \simeq 0.5 \times 10^{-13} s$$

Consequently, the number of librational photons, *radiated* ⇌ *absorbed* during the life-time of water cluster is:

$$n_{ph}^{lb} = \frac{\tau_{clust}}{\tau_{ph}^{lb}} = \frac{10^{-7}}{0.5 \times 10^{-13}} = 2 \times 10^6\ photons \qquad 18.33$$

The corresponding photon exchange provides the EM interaction between water clusters in state of mesoscopic Bose condensation (mBC) in remote microtubules of remote neurons.

If the cumulative energy of this distant EM interaction between microtubules of centrioles, mediated by librational photons, exceeds *kT*, it may induce spatial reorientation - 'tuning' of pairs of centrioles in remote neuron's bodies.

## 18.9. The processes accompanied the nerve excitation

The normal nerve cell contains few dendrites, increasing the surface of cell's body. It is enable to form synaptic contacts for reception the information from thousands of other cells. Each neuron has one axon for transmitting the "resulting" signal in form of the electric impulses from the ends of axons of cells-transmitters to neuron-receptor.

The synaptic contacts, representing narrow gaps (about hundreds of angstrom wide) could be subdivided on two kinds: the *electric* and *chemical* ones. In chemical synapsis the signal from the end of axon - is transmitted by *neuromediator, i.e. acetylholine.* The neuromediator molecules are stored in *synaptic bubbles* near *presynaptic membrane.* The releasing of mediators is stimulated by ions of $Ca^{2+}$. After diffusion throw the synaptic gap mediator form a specific complexes with receptors of post synaptic membranes on the surface of neurons body or its dendrites. Often the receptors are the ionic channels like $(Na^+, K^+)$ - ATP pump. Complex - formation of different channels with mediators opens them for one kind of ions and close for the other. Two kind of mediators interacting with channels: small molecules like acetylholine., monoamines, aminoacids and big ones like set of neuropeptides are existing..

The quite different mechanism of synaptic transmission, related to stimulation of production of secondary mediator is existing also. For example, activation of adenilatcyclase by first mediator increases the concentration of intra-cell cyclic adenozin-mono-phosphate (cAMP). In turn, cAMP can activate enzymatic phosphorylation of ionic channels, changing the electric properties of cell. This secondary mediator can participate in a lot of regulative processes, including the genes expression.

In the normal state of dynamic equilibrium the ionic concentration gradient producing by ionic pumps activity is compensated by the electric tension gradient. The *electrochemical gradient* is equal to zero at this state.

The equilibrium concentration of $Na^+$ and $Cl^+$ in space out of cell is bigger than in cell, the gradient of $K^+$ concentration has an opposite sign. The external concentration of very important for regulative processes $Ca^{2+}$ (about $10^{-3}M$) is much higher than in cytosol (about $10^{-7}M$). Such a big gradient provide fast and strong increasing of $Ca^{2+}$ internal concentration after activation of corresponding channels.

At the "rest" condition of equilibrium the resulting concentration of internal anions of neurons is bigger than that of external ones, providing the difference of potentials equal to 50-100mV. As far the thickness of membrane is only about 5nm or 50Å it means that the gradient of electric tension is about:

$$100.000 \; V/sm$$

i.e. it is extremely high.

Depolarization of membrane usually is related to penetration of $Na^+$ ions into the cell. This process of depolarization could be inhibited by selected diffusion of $Cl^-$ into the cell. Such diffusion can produce even *hyperpolarization* of membrane.

*The potential of action and nerve impulse can be excited in neuron - receptor only if the effect of depolarization exceeds certain threshold.*

In accordance to our model of elementary act of consciousness (EAC) three most important consequences of neuron's body polarization can occur:

- reorganization of MTs system and change of the ionic channels activity, accompanied by short-term memorization;

-reorganization of synaptic contacts on the surface of neuron and its dendrites, leading to long-term memory;

- generation of the nerve impulse, transferring the signal to another nerve cells via axons.

The propagation of nerve signal in axons may be related to intra-cellular water activity ($a_{H_2O}$) decreasing due to polarization of membrane. As a result of feedback reaction the variation of $a_{H_2O}$ induce the [*opening/closing*] of the ionic channels, thereby stimulating signal propagation along the axons.

We put forward the hypothesis, that the periodic transition of *clusterphilic* interaction of the ordered water between inter-lipid tails in nonpolar central regions of biomembranes to hydrophobic one, following by water clusters disassembly and vice verse, could be responsible for lateral nerve signal propagation/firing via axons (Kaivarainen, 1985, 1995, 2001). *The anesthetic action can be explained by its violation of the ordered water structure in the interior of axonal membranes, thus preventing the nerve signal propagation. The excessive stabilization of the internal clusters also prevent the axonal firing.*

The change of the ionic conductivity of the axonal membranes of the axons in the process of signal propagation is a secondary effect in this explanation.

The proposed mechanism, like sound propagation, can provide distant cooperative interaction between different membrane receptors on the same cell and between remote neurons bodies without strong heat producing. The latter phenomena is in total accordance with experiments.

As far the $\alpha\beta$ pairs of tubulins have the properties of "electrets" (Debrabander, 1982), the *piezoelectric properties* of core of microtubules can be predicted (Athenstaedt, 1974; Mascarenhas, 1974).

It means that structure and dynamics of microtubules can be regulated by electric component of electromagnetic field, which accompanied the nerve excitation. In turn, dynamics of microtubules hollow core affects the properties of internal ordered water in state of mesoscopic Bose condensation (mBC).

For example, shift of the [open ⇔ closed] states equilibrium of cavity between $\alpha$ and $\beta$ tubulins to the open one in a course of excitation should lead to:

[I]. Increasing the dimensions and life-time of coherent clusters, represented by primary *lb*

effectons (mBC)

[II]. Stimulation the distant interaction between MT of different neurons as a result of increased frequency and amplitude/coherency of IR librational photons, radiated/absorbed by primary librational effectons of internal water;

[III] Turning the mesoscopic entanglement between water molecules in coherent clusters to nonuniform macroscopic entanglement.

Twisting of the centrioles of distant interacting cells and bending of MTs can occur after [gel→sol] transition. This tuning is necessary for enhancement of the number of MTs with the parallel orientation, most effective for their remote exchange interaction by means of 3D coherent IR photons and vibro-gravitational waves.

Reorganization of actin filaments and MTs system should be accompanied by corresponding changes of neuron's body and its dendrites shape and activity of certain ionic channels and synaptic contacts redistribution; This stage is responsible for long-term memory emergency.

At [sol]-state the $Ca^{2+}$ - dependent $K^+$ channels turns to the open state and internal concentration of potassium decreases. The latter oppose the depolarization and decrease the response of neuron to external stimuli. Decay of neuron's response is termed "adaptation". This *response adaptation* is accompanied by *MTs-adaptation*, i.e. their reassembly in conditions, when concentration of $Ca^{2+}$ tends to minimum. The reverse [sol→gel] transition stabilize the new equilibrium state of given group of cells.

The described hierarchic sequence of stages: from mesoscopic Bose condensation to macroscopic one, providing entanglement of big number of cells, their simultaneous synaptic reorganization and synhronization of the excitation ⇌ relaxation cycles of nerve cells, are different stages of elementary act of consciousness.

### 18.10. Possible mechanism of wave function collapsing, following from the Cycle of Mind model

A huge number of superimposed possible quantum states of any quantum system always turn to "collapsed" or "reduced" single state as a result of measurement, i.e. interaction with detector.

In accordance to "Copenhagen interpretation", the collapsing of such system to one of possible states is unpredictable and purely random. Roger Penrose supposed (1989) that this process is due to quantum gravity, because the latter influences the quantum realm acting on space-time. After certain gravity threshold the system's wave function collapsed "under its own weight".

Penrose (1989, 1994) considered the possible role of quantum superposition and wave function collapsing in synaptic plasticity. He characterized the situation of learning and memory by synaptic plasticity in which neuronal connections are rapidly formed, activated or deactivated: "Thus not just one of the possible alternative arrangements is tried out, but vast numbers, all superposed in complex linear superposition". The collapse of many cytoskeleton configuration to single one is a nonlocal process, required for consciousness.

This idea is in-line with our model of elementary acts of consciousness as a result of transitions between nonuniform macroscopic and mesoscopic Bose condensation (BC) of big number of electromagnetically tuned neurons and corresponding oscillation between their entangled and non-entangled states.

Herbert (1993) estimated the mass threshold of wave function collapse roughly as $10^6$ daltons. Penrose and Hameroff (1995) calculated this threshold as

$$\Delta M_{\text{col}} \sim 10^{19} D \qquad 18.34$$

Non-computable self-collapse of a quantum coherent wave function within the brain may fulfill the role of non-deterministic free will after Penrose and Hameroff (1995).

For the other hand, in accordance with proposed in this author model, the induced coherency between coherent water clusters (primary librational effectons - mesoscopic Bose condensate) in MTs, as a result of distant exchange of librational photons, emitted ⇌ absorbed by them, leads to formation of *macroscopic BC* in microtubules.

The increasing of the total mass of water, involved in macroscopic nonuniform BC in a big system of remote MTs and corresponding 'tuned' neuron ensembles, up to gravitational threshold may induce the wave function collapse in accordance to Penrose hypothesis.

In our approach we explain the selection of certain configurational space of huge number of 'tuned' neurons, not by structural changes of tubulins like in Hameroff-Penrose model, but by increasing of mass of water in state of macroscopic BC in brain in the process of condensation of spatially separated mesoscopic BC (coherent water clusters in MTs). The macroscopic BC is initiated by correlated shift of dynamic equilibrium ($a \rightleftharpoons b$) of nonpolar cavities, formed by pairs of tubulins, between the open ($b$) and closed ($a$) states to the open one, stabilizing water clusters. The time of development/evolution of coherence in remote neurons, accompanied by increasing of scale of macroscopic BC is much longer, than that of mesoscopic BC (about $10^{-6}$ s, equal to average period of pulsation of tubulin dimers cavity between open and close state) and can be comparable with time between axonal firing (about $1/40 = 2.5 \times 10^{-2}$ s). The time of coherence determines the period between corresponding wave function collapsing.

The corresponding structural rearrangements of tubulins and their pairs in the process of shift of open⇌ closed clefts to the right or left, do not change their mass and can not be a source of wave function collapsing "under its own weight" in contrast to increasing of mass of water in evolution of nonuniform macroscopic BC from mBC.

The dynamics of $\left[\, increasing \rightleftharpoons decreasing \,\right]$ of the entangled water mass in state of macroscopic BC is a result of correlated shift of dynamic equilibrium between primary *librational (lb) effectons* (coherent water clusters, mBC), stabilized by *open* inter-tubulins cavities and primary *translational (tr) effectons (independent water molecules)*, corresponding to closed cavities.

The correlated conversions between librational and translational effectons [$lb \rightleftharpoons tr$] of water in remote MTs, representing the association ⇌ dissociation of the entangled water clusters in state of mBC reflect, in fact, the reversible cycles of [coherence ⇌ decoherence] corresponding to cycles of mesoscopic wave function of these clusters collapsing. The relatively slow oscillations of dynamic equilibrium of [$lb \rightleftharpoons tr$] conversions are responsible for alternating contribution of macroscopic quantum entanglement in consciousness.

*Let's make some simple quantitative evaluations in proof of our interpretation of the wave function collapsing.* The mass of water in one microtubule in nerve cell body with most probable length ~ $10^5$ Å and diameter 140 Å is about

$$m_{H_2O} \sim 10^8 D$$

In accordance with our calculations for bulk water, the fraction of molecules in composition of primary *tr* effectons is about 23% and that in composition of primary *librational* effectons (mBC) is about ten times less (Figure 54) or 2.5%. In MTs due to clusterphilic interaction, stabilizing water clusters, this fraction mBC can be few times bigger.

We assume, that in MTs at least 10% of the total water mass ($10^8 D$) can be converted to primary librational effectons (coherent clusters) as a result of IR photons exchange and entanglement between mBC of the same MTs, correlated with shift of dynamic equilibrium of the clefts between tubulins to the open state. This corresponds to increasing of mass of these quasiparticles in each MT as:

$$\Delta m_{H_2O} \simeq 10^6 D \qquad 18.35$$

Such increasing of the coherent water fraction is accompanied by decreasing of water mass, involved in other types of excitations in MT.

Based on known experimental data that each nerve cell contains about 50 microtubules, we assume

that the maximum increasing of mass of primary librational effectons in one cell, using 18.35, could be:

$$\Delta M_{H_2O} \sim 50\, \Delta m_{H_2O} = 5 \times 10^7 D \qquad 18.36$$

If the true value of mass threshold, responsible for wave function collapse, $\Delta M_{col}$ is known (for example $10^{16}D$), then the number ($N_{col}$) of neurons in assemblies, required for this process is

$$N_{col} \sim (\Delta M_{col}/\Delta M_{H_2O}) = 2 \times 10^8 \qquad 18.37$$

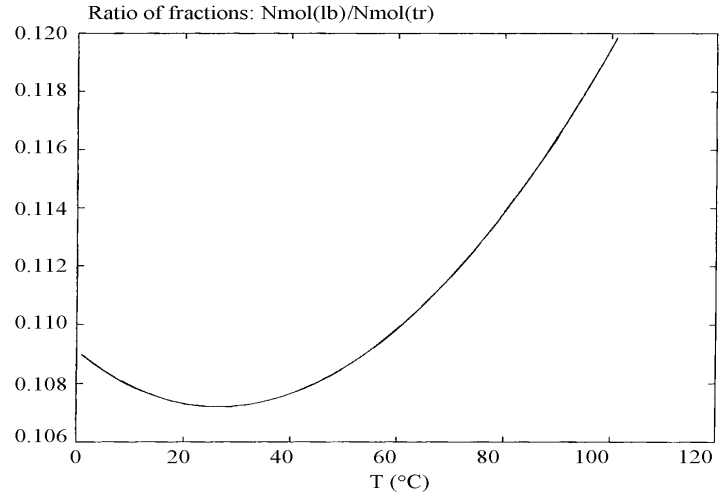

**Figure 54**. Calculated ratio of water fractions involved in primary [lb] effectons to that, involved in primary [tr] effectons for the bulk water. In microtubules this ratio can be bigger and regulated by shift of dynamic equilibrium of the inter-tubulin clefts toward the open or closed state.

The MAP– microtubules associated proteins stabilize the overall structure of MTs. They prevent the disassembly of MTs in bundles of axons and *cilia* in a course of their coherent bending. In neuron's body the concentration of MAP and their role in stabilization of MTs is much lower than in cilia.

The total number of nerve cells in human brain is about: $N_{tot} \sim 10^{11}$. The critical fraction of cells population, participating in elementary act of consciousness, following from our model, can be calculated as:

$$f_c = (N_{col}/N_{tot}) \sim 0.05 \qquad 18.38$$

This value is dependent on correct evaluation of critical mass ($\Delta M_{col}$) of collapsing.

The [gel → sol] transition, induced by coherent "collapsing" of macroscopic brain wave function and dissociation of water clusters in MTs in state BC of huge number of tuned neurons, followed by synaptic contacts reorganization, represents the elementary act of consciousness.

Our approach agree with general idea of Marshall (1989) that Bose- condensation could be responsible for "unity of conscious experience". However, our model explains how this idea can work in detail and what kind of Bose condensation is necessary for its realization.

*We can resume now, that in accordance with our Elementary Act of Consciousness or Cycle of Mind, the sequence of following interrelated stages is necessary for elementary act of perception and memory:*

1. The change of the electric component of cell's electromagnetic field as a result of neuron depolarization;

2. Shift of $A \rightleftharpoons B$ equilibrium between the closed (A) and open to water (B) states of cleft, formed by $\alpha$ and $\beta$ tubulins in microtubules (MT) to the right due to the piezoelectric effect;

3. Increasing the life-time and dimensions of coherent "flickering" water clusters, representing the 3D superposition of de Broglie standing waves of $H_2O$ molecules with properties of Bose-condensate (*effectons*) in hollow core of MT. This process is stimulated by the open nonpolar clefts between tubulin dimers in MT;

4. Spatial "tuning" of parallel MTs of distant simultaneously excited neurons due to distant electromagnetic interaction between them by means of superradiated IR photons and centrioles twisting;

5. Turning the mesoscopic BC of $H_2O$ molecules in clusters to nonuniform macroscopic BC, mediated by Virtual Replica of the clusters Multiplication in space VRM(r) and accompanied by activation of nonlocal interaction between remote clusters in big number of entangled MTs of neurons dendrites;

6. Destabilization of superimposed wave function eigenvalues of clusters (mBC) as a result of nonlinear optical effects like *bistability and self-induced transparency and superradiation;*

7. Disassembly of the actin filaments and [gel-sol] transition, decreasing strongly both - the viscosity of cytoplasm and water activity;

8. The coherent volume/shape pulsation of big group of interacting cells as a consequence *of (actin filaments+MTs) system disassembly and water activity decreasing.* The latter occur as a result of increasing of water fraction in hydration shell of actin and tubulin subunits due to increasing of their surface after disassembly. The decreasing of cytoplasmic water activity increases the passive osmoses of water from the external volume to the cell, increasing its volume.

*This stage should be accompanied by four effects:*

*(a)* Increasing the volume of the nerve cell body;

*(b)* Disrupting the (+) ends of MTs with cytoplasmic membranes, making MTs possible to bend in cell and to collective spatial tuning of huge number of MTs in the ensembles of even distant excited neurons;

(c) Origination of new MTs and microtubules associated proteins (MAP) system switch on/off the ionic channels and change the number and properties of synaptic contacts, responsible for short (MAP) and long memory;

(d) Decreasing the concentration of $Ca^{2+}$ to the limit one, when its ability to disassembly of actin filaments and MT is stopped and [gel $\rightleftharpoons$ sol] equilibrium shifts to the left again, stabilizing a new MTs and synaptic configuration.

This cyclic consequence of quantum mechanical, physico-chemical and nonlinear classical events can be considered as elementary act of memory and consciousness realization. This act can be as long as 500 ms, *i.e.* half of second, like proposed in Hamroff-Penrose model.

The elementary act of consciousness include a stage of coherent electric firing in brain (Singer, 1993) of distant neurons groups with period of about $1/40$ sec.

Probability of super-deformons and cavitational fluctuations increases after [gel→sol] transition. This process is accompanied by high-frequency (UV and visible) "biophotons" radiation due to recombination of part of water molecules, dissociated as a result of cavitational fluctuation.

The dimension of IR super-deformon edge is determined by the length of librational IR standing photon - about 10 microns. It is important that this dimension corresponds to the average microtubule length in cells confirming in such a way our idea. Another evidence in proof is that is that the resonance wave number of excitation of super-deformons, leading from our model is equal to 1200 ($1/cm$).

The experiments of Albreht-Buehner (1991) revealed that just around this frequency the response of surface extensions of 3T3 cells to weak IR irradiation is maximum. Our model predicts that IR irradiation of microtubules system *in vitro* with this frequency will dramatically increase the probability of gel → sol transition.

*Except super-radiance, two other cooperative optic effects could be involved in supercatastrophe realization: self-induced bistability and the pike regime of IR photons radiation (Bates, 1978; Andreev et al.,1988).*

The characteristic frequency of pike regime can be correlated with frequency of [gel-sol] transitions of neuronal groups in the head brain.

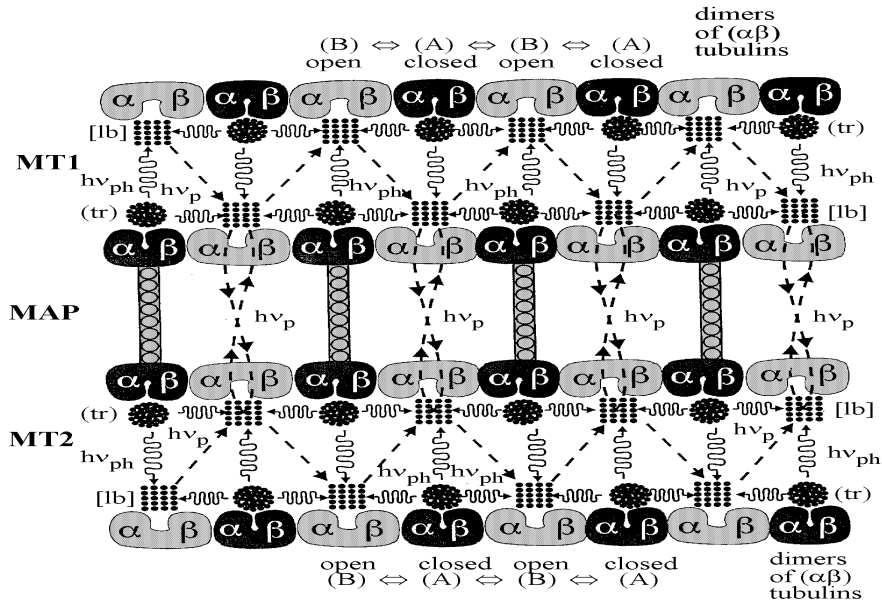

**Figure 55**. The correlation between local, conformational and distant - electromagnetic interactions between pairs of tubulins and microtubules (MT1 and MT2), connected by MAP by mean of librational IR photons exchange.

The dynamics of $\left[\,increasing \rightleftharpoons decreasing\,\right]$ of the entangled water mass in state of macroscopic BC in the process of elementary act of consciousness is a result of correlated shift of dynamic equilibrium between open and closed cavities between alpha and beta tubulins. As a result of these cavities transition from the open to closed state the primary *librational (lb) effectons* (coherent water clusters in state of mesoscopic Bose condensation - mBC) disassembly to small primary *translational (tr) effectons* (independent water molecules), induced by transition of the open states of cavities to the closed one. The nonuniform macroscopic entanglement between the remote water clusters in state of mBC is stimulated by coherent IR photons exchange and vibro-gravitational interaction between these clusters.

The MAP– microtubules associated proteins stabilize the overall structure of MTs. They prevent the disassembly of MTs in bundles of axons and *cilia* in a course of their coherent bending. In neuron's body the concentration of MAP and their role in stabilization of MTs is much lower than in cilia (Kaivarainen, 1995, 2003).

The distant electromagnetic and vibro-gravitational interactions between different MT are the consequence of IR photons and coherent vibro-gravitational waves exchange. The corresponding two types of waves are excited as a result of correlated ($a \Leftrightarrow b$) transitions of water primary librational effectons, localized in the open B- states of ($\alpha\beta$) clefts. Frequency of ($a \Leftrightarrow b$) transitions and corresponding superradiated IR photons is about $2 \times 10^{13}\ s^{-1}$. It is much higher, than frequency of transitions of clefts of $\alpha\beta$ tubulin dimers between open and closed states.

When the neighboring ($\alpha\beta$) clefts has the alternative open and closed states like on Fig 54, the general spatial structure remains straight. However, when $[A \Leftrightarrow B]$ equilibrium of all the clefts from one side of MT are shifted to the left and that from the opposite side are shifted to the right, it leads to bending of MT. Coherent bending of MTs could be responsible for [volume/shape] vibrations of the nerve cells and the cilia bending.

Max Tegmark (2000) made evaluation of decoherence time of neurons and microtubules for analyzing the correctness of Hameroff-Penrose idea of wave function collapsing as a trigger of neurons ensembles axonal firing.

The following three sources of decoherence for the ions in the act of transition of neuron between 'resting' and 'firing' are most effective:

1. Collisions with other ions
2. Collisions with water molecules
3. Coulomb interactions with more distant ions.

The coherence time of such process, calculated from this simple model appears to very short: about $10^{-20}$ s.

The electrical excitations in tubulins of microtubules, which Penrose and others have suggested may be relevant to consciousness also where analyzed. Tegmark considered a simple model of two separated but superimposed (entangled) positions of kink, travelling along the MT with speed higher than 1 m/s, as it supposed in Hameroff-Penrose (H-P) model. The life-time of such quantum state was evaluated as a result of long-range electromagnetic interaction of nearest ion with kink.

His conclusion is that the role of quantum effects and wave function collapsing in H-P model is negligible because of very short time of coherence: $10^{-13}$ s for microtubules.

Hagan, Hameroff and Tuszynski (2002) responded to this criticism, using the same formalism and kink model. Using corrected parameters they get the increasing of life-time of coherent superposition in H-P model for many orders, up to $10^{-4}$ s. This fits the model much better.

Anyway the approach, used by Tegmark for evaluation of the time of coherence/entanglement is not applicable to our model, based on oscillation between mesoscopic and macroscopic nonuniform Bose condensation. It follows from our approach that even very short life-time of oscillating semivirtual macroscopic Bose condensation can be effective for realization of entanglement.

### 18.11 Experimental data, confirming the Cycle of Mind model

There are some experimental data, which support the role of microtubules in the information processing. Good correlation was found between the learning, memory peak and intensity of tubulin biosynthesis in the baby chick brain (Mileusnic *et al.*1980). When baby rats begin their visual learning after they first open eyes, neurons in the visual cortex start to produce vast quantities of tubulin (Cronley- Dillon et. al., 1974). Sensory stimulation of goldfish leads to structural changes in cytoskeleton of their brain neurons (Moshkov *et al.*, 1992).

There is evidence for interrelation between cytoskeleton properties and nerve membrane excitability and synaptic transmission (Matsumoto and Sakai, 1979; Hirokawa, 1991). It has been shown, that microtubules can transmit electromagnetic signals between membranes (Vassilev *et al.*, 1985).

Desmond and Levy (1988) found out the learning-associated change in dendritic spine shape due to reorganization of actin and microtubules containing, cytoskeleton system. After "learning" the number of receptors increases and cytoskeleton becomes more dense.

Other data suggest that cytoskeleton regulates the genome and that signaling along microtubules occurs as cascades of phosphorylation/dephosphorylation linked to calcium ion flux (Puck, 1987; Haag et al, 1994).

The frequency of super-deformons excitations in bulk water at physiological temperature ($37^0 C$) is

around:
$$\nu_S = 3 \times 10^4 \ s^{-1}$$

The frequency of such cavitational fluctuations of water in MT, stimulating in accordance to our model cooperative disassembly of the actin and partly MT filament, accompanied by gel→ *sol* transition, could differ a bit from the above value for bulk water.

Our model predicts that if the neurons or other cells, containing MTs, will be treated by acoustic or electromagnetic field with resonance frequency of intra-MT water ($\nu_{res} \sim \nu_S^{MT} \geq 10^4 \ s^{-1}$), it can induce simultaneous disassembly of the actin filaments and destabilization of MTs system, responsible for maintaining the specific cell volume and geometry. As a result, it activates the neuron's body volume/shape pulsation.

Such external stimulation of *gel* ⇌ *sol* oscillations has two important consequences:

-*The first one* is generation of oscillating high-frequency nerve impulse, propagating via axons and exciting huge number of other nerve cells, i.e. distant nerve signal transmission in living organism;

-*The second one* is stimulation the leaning process as far long-term memory in accordance to Elementary Act of Consciousness, is related to synaptic contacts reorganization, accompanied the neuron volume/shape pulsation.

Lehardt *et al.* (1989) supposed that ultrasonic vibrations are perceptive by tiny gland in the inner ear, known as the *Saccule. It looks that Saccule may* have a dual functions of detection *gravity and auditory signals. Cohlea could be a result of Saccule evolution in mammals.*

Lenhardt and colleagues constructed the an amplitude modulated by audio-frequencies ultrasonic transmitter that operated at frequencies: (28-90) kHz. The output signal from their device was attached to the deaf people heads by means of piezoelectric ceramic vibrator. All people "heard" the modulated signal with clarity.

Our approach allows to predict the important consequence. Excitation of water super-deformons in cells, leading to gel→ *sol* transition, cell's volume/shape pulsation and generation of high-frequency nerve impulse - could be stimulated by hypersound, electromagnetic waves and coherent IR photons with frequency, corresponding to excitation energy of super-deformons.

The calculated from this assumption frequency is equal to
$$\nu_p^S = c \times \tilde{\nu}_p^S = (3 \times 10^{10}) \times 1200 = 3.6 \times 10^{13} \ c^{-1} \qquad 18.39$$

the corresponding photons wave length:
$$\lambda_p^S = c/(n_{H_2O} \times \nu_p^S) \simeq 6.3 \times 10^{-4} \ cm = 6.3\mu \qquad 18.40$$

where $\tilde{\nu}_p^S = 1200 \ cm^{-1}$ is wave number, corresponding to energy of super-deformons excitation; $n_{H_2O} \simeq 1.33$ is refraction index of water.

### *18.11.1 The additional experimental verification of the Cycle of Mind model "in vitro"*

It is possible to suggest some experimental ways of verification of Elementary Act of Consciousness using model systems. The important point of Cycle of Mind model is stabilization of highly ordered water clusters (primary librational effectons) in the hollow core of microtubules. One can predict that in this case the IR librational bands of water in the oscillatory spectra of model system, containing sufficiently high concentration of MTs, must differ from IR spectra of bulk water as follows:

- the shape of librational band of water in the former case must contain 2 components: the first one, big and broad, like in bulk water and the second one small and sharp, due to increasing coherent fraction of librational effectons. The second peak should disappeared after disassembly of MTs with specific reagents;

- the velocity of sound in the system of microtubules must be bigger, than that in disassembled

system of MTs and bulk pure water due to bigger fraction of ordered ice-like water;
 - all the above mentioned parameters must be dependent on the applied electric potential, due to piezoelectric properties of MT;
 - the irradiation of MTs system in vitro by ultrasonic or electromagnetic fields with frequency of super-deformons excitation of the internal water of MTs at physiological temperatures $(25 - 40^0 C)$ :

$$\nu_s = (2 - 4) \times 10^4 \; Hz$$

have to lead to increasing the probability of disassembly of MTs, induced by cavitational fluctuations. The corresponding effect of decreasing turbidity of MT-containing system could be registered by light scattering method.

Another consequence of super-deformons stimulation by external fields could be the increasing of intensity of radiation in visible and UV region due to emission of corresponding "biophotons" as a result of recombination reaction of water molecules:

$$HO^- + H^+ \stackrel{h\nu}{\rightleftharpoons} H_2O$$

Cavitational fluctuations of water, representing in accordance to our theory super-deformons excitations, are responsible for dissociation of water molecules, *i.e.* elevation of protons and hydroxyls concentration. These processes are directly related to sonoluminiscence phenomena.

The coherent transitions of $(\alpha\beta)$ dimers, composing MTs, between "closed" (A) and "open" (B) conformers with frequency $(\nu_{mc} \sim 10^7 \; s^{-1})$ are determined by frequency of water macroconvertons (flickering clusters) excitation, localized in cavity between $\alpha$ and $\beta$ tubulins. If the charges of (A) and (B) conformers differ from each other, then the coherent $(A \rightleftharpoons B)$ transitions generate the vibro-gravitational and electromagnetic field with the same radio-frequency. The latter component of biofield could be detected by corresponding radio waves receiver.

We can conclude that the Hierarchic Theory of condensed matter and its application to water and biosystems provide reliable models of informational exchange between different cells and correlation of their functions. The Cycle of Mind model is based on proposed quantum exchange mechanism of interactions between neurons, based on very special properties of microtubules, [gel-sol] transitions and interrelation between spatial distribution of MTs in neurons body and synaptic contacts.

The described mechanism of IR photons - mediated conversion of mesoscopic Bose condensation to macroscopic one with corresponding change of wave function spatial scale can be exploit in the construction of artificial quantum computers, using inorganic microtubules.

## RESUME

The presented in book comprehensive Hierarchic Theory of solids and liquids is verified by computer simulations on examples of ice and water. The theory unifies a substantial number of physical, optical and biological phenomena in terms of the quantum physics. Instead of big number of separate theories of these phenomena, the Hierarchic Theory suggest general quantitative approach to their description.

The discovered by computer simulations mesoscopic Bose condensation (mBC) in water and ice in form of coherent molecular clusters with dimensions of the most probable 3D de Broglie waves of molecules, is an important fundamental phenomena for understanding the quantum properties of condensed matter.

The new quantitative theories are created in the framework of the same Hierarchic concept, such as hierarchic thermodynamics, hierarchic theory of the refraction index, of Brillouin light scattering, Mössbauer effect, viscosity, self-diffusion, thermal conductivity, surface tension, vapor pressure, turbidity, superfluidity and superconductivity etc. All of these theories fit very well with the available experimental data.

*A lot of applications of Hierarchic theory to molecular biophysics, including mechanism of specific complex formation and enzyme catalytic action are presented.*

Our theory of Elementary act of consciousness (Cycle of Mind) considers the possibility of dynamic equilibrium between coherence and decoherence, mesoscopic and nonuniform macroscopic Bose condensation of water in microtubules and relatively slow oscillations of this equilibrium, providing the neurons ensembles firing and relaxation.

It follows from the mechanism proposed for heterogeneous *chemical catalysis*, including the enzymatic one, that the treatment of a reactor by an acoustic and/or electromagnetic field with frequency of quantum beats ($\omega^*$) between de Broglie waves of reacting atoms of substrate and that of active site (eq. 16.16) can accelerate the catalytic acts and make them coherent.

The treatment of medium of *crystal growth* by electromagnetic and/or acoustic field with frequency, equal to that of primary deformons or macroconvertons of crystal or with frequency, corresponding to frequency of primary electromagnetic deformons and definite polarization, should increase the *ordering of crystal structure and improve its quality*. It is a perspective way for regulation the nanotechnology processes and that of new materials technology.

# References

Aksnes, G., & Asaad, A.N. (1989). Influence of the water structure on chemical reactions in water. A study of proton-catalyzed acetal hydrolysis. *Acta Chem. Scand., 43*, 726-734.
Aksnes, G., & Libnau, O. (1991). Temperature dependence of ether hydrolysis in water. *Acta Chem. Scand.*, *45*, 463-467.
Alberts, B., Bray, D., Lewis, J., Ruff, M., Roberts, K., & Watson, J.D. (1983). *Molecular Biology of Cell. Chapter 10*. New York, London: Garland Publishing, Inc.
Albrecht-Buehler, G. (1991). Surface extensions of 3T3 cells towards distant infrared light sources. *J. Cell Biol.*, *114*, 493-502.
Albrecht-Buehler, G. (1992). Rudimentary form of cellular "vision". *Proc. Natl. Acad. Sci. USA, 89*, 8288-8292.
Aliotta, F., Fontanella, M.E., & Magazu, S. (1990). Sound propagation in thyxotropic strucures. *Phys. Chem. Liq*.
Amos, L. A., & Klug, A. (1974). Arrangements of subunits in flagellar microtubules. *J. Cell Sci.*, *14,* 523-550.
Andreev, A.V., Emeljanov, V.I., & Iljinski, Yu. A. (1988). *The cooperative phenomena in optics*, ISBN 5-02-013837-1 (in Russian). Moscow: Nauka.
Andronov, A.A., Vitt, A.A., & Haikin, S.E. (1981). *Theory of oscillations* (3rd ed.). Moscow.
Antonchenko, V.Ya. (1986). *Physics of water*. Kiev: Naukova dumka.
Ashkroft, N., & Mermin, N. (1976). *Solid state physics*. New York: Holt, Rinehart and Winston.
Atema, J. (1973). Microtubule theory of sensory transduction. *J. Theor. Biol., 38*, 181-190.
Athenstaedt, H. (1974). Pyroelectric and piezoelectric properties of vertebrates. *Ann. NY Acad. Sci.*, *238*, 68-93.
Audenaert, R., Heremans, L., Heremans, K., & Engleborghs, Y. (1989). Secondary structure analysis of tubulin and microtubules with Raman spectroscopy. *Biochim. Biophys. Acta.*, *996*, 110-115.
Babloyantz, A. (1986). *Molecules, Dynamics and Life. An introduction to self-organization of matter*. New York: John Wiley & Sons, Inc.
Bardeen, J., Cooper, L.N., & Schrieffer, J.R. (1957). *Phys. Rev.*, *108*, 1175.
Bardeen, J., & Schrieffer, J.R. (1961). Recent developments in superconductivity, in the series *Progress in Low Temperature Physics,* C. J. Gorter, editor, North Holland Publishing Co., Amsterdam, vol. 3, p. 170.
Bednorz, J.G., & Muller, K.A. (1986). *Z. Phys. B. Condensed Matter*, *64*, 189.
Bates, R. (1978). *Phys. Reports, N3,* v.35.
Begich, N. (1996). *Towards a new alchemy. The millennium science*. Anchorage, Alaska: Earth Press.
Bershadsky & Vasiliev, J.M. (1988). Cytosceleton. In P. Siekevitz (Ed.), *Cellular Organelles*. New York: Plenum Press.
Beizer, A. (1973). *Basic ideas of modern physics*. Moscow: Nauka.
Bellisent-Funel, M.C., Teixeira, J., Chen, S.H., Dorner, B., & Crespi, H.L. (1989). Low-frequency collective modes in dry and hydrated proteins. *Byphys. J.*, *56*, 713-716.
Bellfante, F.J. (1987). The Casimir effect revisited. *Am. J. Phys., 55 (2)*, 134.

Benassi, P., D'Arrigo, G., & Nardone, M. (1988). Brilouin light scattering in low temperature water-ethanol solutions. *J. Chem. Phys., 89*, 4469-4477.

Ben-Naim, A. (1980). *Hydrophobic Interaction*. New York: Plenum Press.

Benveniste, J., Aïssa, J., Jurgens, P., & Hsueh, W. (1998, April 20). Specificity of the digitized molecular signal, Presented at Experimental Biology '98 (FASEB). San Francisco.

Benveniste, J., Ducot, B., & Spira, A. (1994). "Memory of water revisited", *Nature, Letter to the Editor, 370* (6488):322.

Berezin, M.V., Lyapin, R.R., & Saletsky, A.N. (1988). Effects of weak magnetic fields on water solutions light scattering. Preprint of Physical Department of Moscow University, No.21, 4 p. (in Russian).

Bertolini, D., Cassetari, M., Grigolini, P., Salvetti, G., & Tani, A. (1989). The mesoscopic systems of water and other complex systems. *J. Mol. Liquids, 41*, 251.

Bertolini, D., Cassetari, M., Salvetti, G., Tombari, E., Veronesi, S., & Squadrito, G. (1992). *Il nuovo Cim., 14D,* 199.

Bischof, M. (1996). Vitalistic and mechanistic concepts in the history of bioelectromagnetics. In L. Beloussov, & F.A. Popp, (Eds.), *Biophotononics - non-equilibrium and coherent systems in biology, biophysics and biotechnology*. Moscow: Bioinform Services.

Blakemore, J.S. (1985). *Solid state physics*. Cambrige, N.Y. Cambridge University Press.

Bogolyubov, N.N. (1970). *Lectures on quantum statistics. Collected works. Vol.2*. Kiev.

Bohm, D. (1986, April). A new theory of the relationship of Mind and Matter. *J. Amer. Soc. for Psyhial Research, N2*, p.128.

Brändas, E.J., & Chatzdmitriou-Dreismann, C.A. (1991). *Intrn. J. Quantum Chem., 40*, 649-673.

Casimir, H.B. (1948). *Proc. K. Akad. We*t., *51*, 793.

Cantor, C.R., & Schimmel, P.R. (1980). *Biophysical Chemistry*. San Francisco: W.H. Freemen and Company.

Celegnin E., Graziano, E., & Vitello, G. (1990). Classical limit and spontaneous breakdown of symmetry as an environment effect in quantum field theory. *Physics Letters A., 145*, 1-6.

Chatzidimitriou-Dreismann, C.A., & Brändas, E.J. (1988). Coherence in disordered condensed matter. I. Dissipative structure in liquids and dynamics of far-infrared absorption. *Ber. Bundsenges. Phys. Chem., 92*, 549.

Chatzidimitriou-Dreismann, C.A., & Brändas, E.J. (1988). Coherence in disordered condensed matter. II. Size of coherent structure in condensed systems owing to luminescent transitions and detection of D- fluctuations. *Ber. Bundsenges. Phys. Chem.*, *92*, 554.

Chernikov, F.R. (1985). Lightscattering intensity oscillations in water-protein solutions. *Biofizika (USSR)*, *31*, 596.

Chernikov, F.R. (1990a). Effect of some physical factors on light scattering fluctuations in water and water biopolymer solutions. *Biofizika (USSR), 35,* 711.

Chernikov, F.R. (1990b). Superslow light scattering oscillations in liquids of different types. *Biofizika (USSR), 35,* 717.

Christiansen, P.L., Pagano, S., & Vitello, G. (1991). The lifetime of coherent excitations in Langmuir-Blodgett Scheibe aggregates. *Phys. Lett. A, 154(7,8)*, 381-384.

Clegg, J.S. (1985). On the physical properties and potential roles of intra-cellular water. *Proc. NATO Adv. Res. Work Shop*.

Clegg, J.S., & Drost-Hansen, W. (1991). On the biochemistry and cell physiology of water. In Hochachka, & Mommsen (Eds.), *Biochemistry and molecular biology of fishes* . Elsevier Science Publ. *vol.1*, pp.1-23.

Conrad, M. (1988). Proton supermobility: a mechanism for coherent dynamic computing. *J.

*Mol. Electron.*, *4*, 57-65.
Cooper, L.N. (1956). Bound Electron Pairs in a Degenerate Fermi Gas. *Phys. Rev.*, *104*, 1189.
Coffey, W., Evans, M., & Grigolini, P. (1984). *Molecular diffusion and spectra*. N.Y., Chichester, Toronto: A.Wiley Interscience Publication.
Crick, F., & Koch, C. (1990). Towards a neurobiological theory of consciousness. *Semin. Neurosci*, *2*, 263-275.
Cronley-Dillon, J., Carden, D., & Birks, C. (1974). The possible involvement of brain microtubules in memory fixation. *J. Exp. Biol.*, *61*, 443-454.
D'Aprano, A., Donato, I., & Liveri, V.T. (1990a). Molecular association of n-alcohols in nonpolar solvents. Excess volumes and viscosities of *n*- pentanol+n-octane mixtures at 0, 5, 25, 35 and $45^0$C. *J. Solut. Chem.*, *19*, 711-720.
D'Aprano, A., Donato, I., & Liveri, V.T. (1990b). Molecular interactions in 1-pentanol + 2-butanol mixtures: static dielectric constant, viscosity and refractive index investigations at 5, 25, and $45^0$C. *J. Solut. Chem.*, *18*, 785-793.
D'Aprano, A., & Donato, I. (1990c). Dielectric polarization and polarizability of 1-pentanol + n-octane mixtures from static dielectric constant and refractive index data at 0, 25 and $45^0$. *J. Solut. Chem.*, *19*, 883-892.
D'Arrigo, G., & Paparelli, A. (1988a). Sound propagation in water-ethanol mixtures at low temperatures. I. Ultrasonic velocity. *J. Chem. Phys.*, *88*, *No.1*, 405-415.
D'Arrigo G., Paparelli A. (1988b). Sound propagation in water-ethanol mixtures at low temperatures II. Dynamical properties. *J. Chem. Phys*, *88*, *No.12*, 7687-7697.
D'Arrigo, G., & Paparelli, A. (1989). Anomalous ultrasonic absorption in alkoxyethanls aqeous solutions near their critical and melting points. *J. Chem. Phys.*, *91*, *No.4*, 2587-2593.
D'Arrigo, G., & Texiera, J. (1990). Small-angle neutron scattering study of $D_2O$-alcohol solutions. *J. Chem. Faraday Trans.*, *86*, 1503-1509.
Davenas, E.F., Beauvais, J., Arnara, M., Oberbaum, B., Robinzon, A., Miadonna, A., Fortner, P., Belon, J., Sainte-Laudy, B., Poitevin & Benveniste, J. (1988). Human basophil degranulation triggered by very dilute antiserum against IgE. *Nature*, *333(6176),* 816-18.
Davydov, A.S. (1979). Solitons in molecular systems. *Phys. Scripta*, *20*, 387-394.
Davydov, A.S. (1984). Solitons in molecular systems. Kiev: Naukova dumka. (in Russian).
Del Giudice, E., Dogulia, S., Milani, M., & Vitello, G. (1985). A quantum field theoretical approach to the collective behaviour of biological systems. *Nuclear Physics*, *B251[FS13]*, 375-400.
Del Guidice, E., Doglia, S., & Milani, M. (1988). Spontaneous symmetry breaking and electromagnetic interactions in biological systems. *Physica Scripta, 38*, 505-507.
Del Guidice, E., Doglia, S., Milani, M., & Vitello, G. (1991). Dynamic Mechanism for Cytoskeleton Structures. In *Interfacial phenomena in biological systems*. New York: Marcel Deccer, Inc.
Del Guidice, E., Preparata, G., & Vitello, G. (1989). Water as a true electric dipole laser. *Phys. Rev. Lett.*, *61*, 1085-1088.
Desmond, N.L., & Levy, W.B. (1988). Anatomy of associative long-term synaptic modification. In P.W. Landfield, & S.A. Deadwyler (Eds.), *Long-Term Potentiation: From Biophysics to Behavior*.
Dicke, R.H. (1954). Coherence in spontaneous radiation processes. *Phys. Rev.,* *93*, 99-110.
Dirac, P. (1957). Principles of quantum mechanics.
Dirac, P.A.M. (1982). The Principles of Quantum Mechanics, 4th Ed. (International Series of Monographs on Physics). New York: Oxford University Press.
Dote, J.L., Kivelson, D., & Schwartz, H. (1981). *J. Phys. Chem.*, *85*, 2169.

Drost-Hansen, W. (1976). In M. Kerker (Ed.), *Colloid and Interface Science*, (p. 267). New York: Academic Press.
Drost-Hansen, W., Singleton, J., & Lin. (1992). Our aqueous heritage: evidence for vicinal water in cells. In *Fundamentals of Medical Cell Biology*, *Chemistry of the living cell* (v.3A, pp. 157-180). JAI Press, Inc.
Duhanin, V.S., & Kluchnikov, N.G. (1975). The problems of theory and practice of magnetic treatment of water. *Novocherkassk*, (pp. 70-73), (in Russian).
Dunne, B.J., & Jahn, R.G. (1992). Experiments in remote human/machine interactions. *J. Scientic Exploration*, *6*, *No 4*, 311-332.
Dzeloshinsky (Lifshiz), E.M., & Pitaevsky, L.P. (1961). *Uspekhi fizitcheskikh nauk (USSR)*, *73*, 381.
Eftink, M.R, & Hagaman, K.A. (1986). *Biophys. Chem., 25, 277.*
Einstein, A. (1965). *Collected Works*. Moscow: Nauka, (in Russian).
Eisenberg, D., & Kauzmann, W. (1969.). *The structure and properties of water*. Oxford: Oxford University Press.
Egelstaff, P.A. (1993). Static and dynamic structure of liquids and glasses. *J. Non-Crystalline solids*, *156*, 1-8.
Etzler, F.M., & Conners, J.J. (1991). Structural transitions in vicinal water: Pore size and temperature dependence of the heat capacity of water in small pores. *Langmuir*, *7*, 2293-2297.
Etzler, F.M., & White, P.J. (1987). The heat capacity of water in silica pores. *J. Colloid and Interface Sci.*, *120*, 94-99.
Farsaci, F., Fontanella, M.E., Salvato, G., Wanderlingh, F., Giordano, R., & Wanderlingh, U. (1989). Dynamical behaviour of structured macromolecular solutions. *Phys. Chem. Liq*, *20*, 205-220.
Farwell, L.A. (1996). Quantum-mechanical processes and consciousness: An empirical investigation. In S. R. Hameroff, A. Kaszniak, & A. Scott (Eds.), *Abstracts of Conference: Toward a Science of Consciousness,* (p. 162.) Tucson, USA.
Ferrario, M., Grigolini, P., Tani, A., Vallauri, R., & Zambon, B. (1985).*Adv. Chem. Phys.*, *62*, 225.
Feynman, R. (1998, January). Statistical Mechanics: A Set of Lectures (Advanced Book Classics). New York: Perseus Books Group.
Feynman, R. (1965). *The character of physical law*. London: Cox and Wyman Ltd.
Fild, R., & Burger, M. (Eds.) (1996). *Oscillations and progressive waves in chemical systems*. Moscow: Mir, (in Russian).
Fine, R.A, & Millero, F.J. (1973). Compressibility of water as a function of temperature. *J. Chem. Phys.*, *59*, 5529.
Frank, H.S., & Evans, M.W. (1945). Free volume and entropy in condensed systems III. Entropy in binary liquid mixtures. *J. Chem. Phys.*, *13*, 507.
Frank, H.S., & Wen, W. V. (1945). Ion - solvent interaction. *Disc. Faraday Soc.*, *24,* 133.
Franks, F., & Mathias, S. (Eds) (1982). *Biophysics of water*. N.Y.: John Wiley.
Franks, F. (Ed.). (1973). *Water: A comprehensieve treatise. Vols: 1-4*. New York, London: Plenum Press.
Franks, F. (1975). The hydrophobic interaction. In F. Franks (Ed.), *Water. A comprehensive treatise.* (Vol. 4, pp. 1-64). N.Y.: Plenum Press.
Frauenfelder, H. (1983). Summary and outlook. In R. Porter, M. O'Conner, & J. Wehlan, (**Eds**.), *Mobility and function in proteins and nucleic acids*. (Ciba Foundation Symposium, Vol. 93, pp. 329-339). London: Pitman.
Frauenfelder, H., & Wolynes, P.G. (1985). Rate theories and puzzles in hemoprotein

kinetics. *Science*, *229*, 337.
Frauenfelder, H., Parak, F., & Young, R.D. (1988). *Ann. Rev. Biophys. Chem.*,*17*, 451.
Frehland, E. (Ed.). (1984). *Synergetics from microscopic to macroscopic order*. Berlin: Springer.
Frontas'ev, V.P., & Schreiber, L.S. (1966). *J. Struct. Chem. (USSR), 6,* 512.
Fröhlich, H. (1950). *Phys. Rev.*, *79*, 845.
Fröhlich, H. (1968). Long-range coherence and energy storage in biological systems. *Int. J. Quantum Chem., 2*, 641-649.
Fröhlich, H. (1975). *Phys. Lett., 51,* 21.
Fröhlich, H. (1975). *Proc. Nat. Acad. Sci., USA*, *72*, 4211.
Fröhlich, H. (Ed.) (1988). *Biological coherence and response to external stimuli*. Berlin: Springer.
Fröhlich, H. (1986). In *The fluctuating enzyme,* (pp. 421-449). A Wiley-Interscience Publication.
Gavish, B., & Weber, M. (1979). Viscosity-dependent structural fluctuations in enzyme catalysis. *Biochemistry, 18*, 1269.
Gavish, B. (1986). In G. R.Welch, (Ed.), *The fluctuating enzyme,* (pp. 264-339). Wiley-Interscience Publication.
Gell-Mann, M., & Hartle, J.B. (Eds.) (1989). Quantum mechanics in the light of quantum cosmology. In Kobyashi, (Ed.), *Proc.3rd Int. Symp. Found. of Quantum Mechanics*. Tokyo: Phys. Soc. of Japan.
Genberg, L., Richard, L., McLendon, G., & Dwayne-Miller, R.V. (1991). Direct observation of global protein motion in hemoglobin and mioglobin on picosecond time scales. *Science*, *251*, 1051-1054.
Genzel, L., Kremer, F., Poglitsch, A., & Bechtold, G. (1983). Relaxation process on a picosecond time-scale in hemoglobin observed by millimeter-wave spectroscopy. *Biopolimers*, *22*, 1715-1729.
Giordano, R., Fontana, M.P., & Wanderlingh, F. (1981a). *J. Chem. Phys*., *74*, 2011.
Giordano, R., et al. (1983b). *Phys. Rev.*, *A28*, 3581.
Giordano, R., Salvato, G., Wanderlingh, F., & Wanderlingh, U. (1990). Quasielastic and inelastic neutron scattering in macromolecular solutions. *Phys. Rev. A.*, *41*, 689-696.
Glansdorf, P., & Prigogine, I. (1971). *Thermodynamic theory of structure, stability and fluctuations*. N.Y.: Wiley and Sons.
Gordeyev, G.P., & Khaidarov, T. (1983). In *Water in biological systems and their components,* (p. 3). Leningrad: Leningrad University Press, (in Russian).
Grawford, F.S. (1973) *Waves. Berkley Physics Course. Vol. 3*. N.Y.: McGraw- Hill Book Co.
Grigolini, P. (1988). *J. Chem. Phys.*, *89*, 4300.
Grundler, W., & Keilmann, F. (1983). Sharp resonance in yeast growth proved nonthermal sensitivity to microwaves. *Phys. Rev. Letts*., *51*, 1214-1216.
Guravlev, A.I., & Akopjan, V.B. (1977). *Ultrasound shining*. Moscow: Nauka.
Haag, M.M., Krystosek, A., Arenson, E., & Puck, T.T. (1994). Reverse transformation and genome exposure in the C6 glial tumor cell line. *Cancer Investigation, 2(1)*, 33-45.
Haake, F. (1991). *Quantum signatures of chaos*. Berlin: Springer.
Hagen, S., Hameroff, S. & Tuszynski, J. (2002). Quantum computation in brain microtubules: decoherence and biological feasibility. *Phys. Rev. **E65**, 061901-1-061901-11.
Haida, A., Matsuo, T., Suga, H., & Seki, S. (1972). *Proc. Japan Acad.*, *48*, 489.
Haken, H. (1983.). *Advanced synergetics*. Berlin: Springer.
Haken, H. (1988). *Information and self-organization*, Berlin: Springer.

Haken, H. (1990). *Synergetics, computers and cognition*. Berlin: Springer.
Hameroff, S. (1987.). *Ultimate computing: Biomolecular consciousness and nanotechnology*. Amsterdam: Elsevier-North Holland.
Hameroff, S. (1996, October 4-6). Cytoplasmic gel states and ordered water: Possible roles in biological quantum coherence. *Proceedings of 2nd annual advanced water science symposium*, Dallas, Texas, USA.
Hameroff, S., & Penrose, R. (1996). Orchestrated reduction of quantum coherence in brain microtubules: A model of consciousness. In S. Hameroff, A. Kaszniak, & A. Scott (Eds.), *Toward a Science of Consciusness - Tucson I,* (pp. 507-540). Cambridge, MA: MIT Press.
Hameroff, S.R., & Penrose, R. (1996). Conscious events as orchestrated spacetime selections. *Journal of Consciousness Studies, 3(1)*, 36-53.
Hameroff, S. (1998). Quantum computation in brain microtubules? The Penrose-Hameroff "Orch OR" model of consciousness. *Philos. Trans. R. Soc. London Ser., A 356*, 1869-896.
Hameroff, S., & Tuszynski, J. (2003). Search for quantum and classical modes of informational processing in microtubules: Implications of "The living state". In F. Musumeci, L. Brizhik, & Mae-Wan Ho (Eds.), *Energy and information transfer in biological systems*. World Scientific.
Huang, K. (1966). *Statistical mechanics*, (pp. 31-57). Moscow: Mir, (in Russian.)
Il'ina, S.A., Bakaushina, G.F., Gaiduk, V.I., et al. (1979). *Biofizika (USSR), 24*, 513.
Ise, N., & Okubo, T. (1980). *Accounts of Chem. Res.*, *13*, 303.
Ise, N., Matsuoka, H., Ito, K., & Yoshida, H. (1990). Inhomogenity of solute distribution in ionic systems. *Faraday Discuss. Chem. Soc., 90*, 153-162.
Ito, K., Yoshida, H., & Ise, N. (1994). Void Structure in colloid dispersion. *Science*, *263*, 66.
Jahn, R.G., & Dunne, B.J. (1987). *Margins of reality: The role of consciousness in the physical world*. San Diego, New York, London: Hacourt Brace Jovanovich.
Jahn, R.G., & Dunne, B.J. (1986). On the quantum mechanics of consciousness with application to anomalous phenomena. *Foundations of physics, 16, No. 8,* 721-772 .
Jia, Z., DeLuca, C.I., Chao, H., & Davies, P.L. (1996). Structural basis for the binding of globular antifreeze protein to ice. *Nature, Nov.21, 384:6606,* 285-288.
Jibu, M., and Yasue K. (1995). Quantum brain dynamics and consciousness - an introduction. Amsterdam: John Benjamins.
Jibu, M., Hagan, S., Hameroff, S.R., Pribram, K.H., & Yasue, K. (1994). Quantum optical coherence in cytosceletal microtubules: Implications for brain function. *Biosystems*, *32*, 195-209.
Jibu, M., & Yasue. K. (1992). The basis of quantum brain dynamics. In *Materials of the First Appalachian Conference on Behavioral Neurodynamics. September 17-20. Center for Brain Research and Informational Science*. Radford, Virginia: Radford University.
Jibu, M., & Yasue, K. (1992). Introduction to quantum brain dynamics. In E. Carvallo (Ed.), *Nature, cognition and system III*. London: Kluwer Academic.
Johri, G.K., & Roberts, J.A. (1990). Study of the dielectric response of water using a resonant microwave cavity as a probe. *J. Phys. Chem.*, *94*, 7386.
Kajava, A.V., & Lindow, S.E. (1993). A model of the three-dimensional structure of the ice nucleation proteins. *J. Mol. Biol.*, *232*, 709-717.
Kivelson, D., & Tarjus, G. (1993). Connection between integrated intensities of depolarized-light-scattering spectra and mesoscopic order in liquids. *Phys. Rev. E.*, *47(6)*, 4210-4214.
Kaivarainen, A. (1985). Solvent-dependent flexibility of proteins and principles of their function, (pp. 290). Dordrecht, Boston, Lancaster: D. Reidel Publ. Co.

Kaivarainen, A. (1989). Theory of condensed state as a hierarchical system of quasiparticles formed by phonons and three-dimensional de Broglie waves of molecules. Application of theory to thermodynamics of water and ice. *J. Mol. Liq.*, *41*, 53-60.
Kaivarainen, A. (1992). *Mesoscopic theory of matter and its interaction with light. Principles of self-organization in ice, water and biosystems*, (pp. 275). Finland: University of Turku.
Kaivarainen, A. (2000). *Hierarchic theory of condensed matter: Long relaxation, macroscopic oscillations and the effects of magnetic field*. Retreived from http://arxiv.org/abs/physics/0003107.
Kaivarainen, A. (2000). *Hierarchic theory of complex systems (biosystems, colloids): Self-organization & osmosis*. Retreived from http://arxiv.org/abs/physics/0003071.
Kaivarainen, A. (2000). *Hierarchic theory of condensed matter and its interaction with light: New theories of light refraction, Brillouin scattering and Mossbauer effect*. Retreived from http://arxiv.org/abs/physics/0003070.
Kaivarainen, A. (2001). *Hierarchic Ttheory of condensed matter: Role of water in protein dynamics, function and cancer emergence*. Retreived from http://arxiv.org/abs/physics/0003093.
Kaivarainen, A. (2001). *Mechanism of antifreeze proteins action, based on hierarchic theory of water and new "clusterphilic" interaction*. Retreived from http://arxiv.org/abs/physics/0105067.
Kaivarainen, A. (2001). *New hierarchic theory of condensed matter and its computerized application to water and ice*. Retreived from http://arxiv.org/abs/physics/0102086.
Kaivarainen A. (2001). *Hierarchic theory of matter, general for liquids and solids: ice, water and phase transitions*. American Inst. of Physics (AIP) Conference Proceedings (ed. D. Dubois), New York, vol. 573, 181-200.
Kaivarainen A. (2001). *Hierarchic model of consciousness: from molecular Bose condensation to synaptic reorganization.* CASYS-2000: International J. of Computing Anticipatory Systems, (ed. D. Dubois), Liege, v.10, 322-338.
Kaivarainen A. (2003). New Hierarchic Theory of Water and its Role in Biosystems. The Quantum Psi Problem. *Proceedings of the international conference: "Energy and Information Transfer in Biological Systems: How Physics Could Enrich Biological Understanding"*, F. Musumeci, L. S. Brizhik, M.W. Ho (Eds), World Scientific, ISBN 981-238-419-7, p. 82-147.
Kaivarainen, A. (2003). *New hierarchic theory of water and its application to analysis of water perturbations by magnetic field. Role of water in biosystems.* Retreived from http://arxiv.org/abs/physics/0207114.
Kaivarainen, A. (2004). *Hierarchic model of consciousness: From molecular Bose condensation to synaptic reorganization*. Retreived from http://arxiv.org/abs/physics/0003045.
Kaivarainen, A. (2005) Theory of bivacuum as a source of particles and particle duality as a cause of fields, in R. L. Amoroso, B. Lehnert & J-P Vigier (eds.) *Beyond The Standard Model: Searching For Unity in Physics,* pp. 199-272, Oakland: The Noetic Press.
Kaivarainen, A (2006). *Unified theory of bivacuum, particles duality, fields and time. New bivacuum mediated interaction, overunity devices, cold fusion and nucleosynthesis*. Retreived from http://arxiv.org/ftp/physics/papers/0207/0207027.pdf
Kaivarainen, A. (2007). *Hierarchic model of consciousness, including the distant and nonlocal interactions.* Toward a Science of Consciousness 2007, Budapest, Hungary, July 23-26, 2007.
Kaivarainen, A. (2007). *Origination of Elementary Particles from Bivacuum and their

*Corpuscle - Wave Pulsation.* Fifth International Conference on Dynamic Systems and Applications. May 30 - June 2, Morehouse College, Atlanta, Georgia, USA.

Kaivarainen, A. (2007). *Hierarchic models of turbulence, superfluidity and superconductivity*. In book: Superconductivity Research Horizons. Ed. Eugene H. Peterson, Nova Sci. Publ. (NY, USA). ISBN: 1-60021-510-6.

Kaivarainen, A., Fradkova, L., & Korpela, T. (1992) Separate contributions of large- and small-scale dynamics to the heat capacity of proteins. A new viscosity approach. *Acta Chem. Scand., 47*, 456-460.

Kampen, N.G., van. (1981). *Stochastic process in physics and chemistry.* Amsterdam: North - Holland.

Karachentseva, A., & Levchuk, Yu. (1989). *J. Biopolymers and Cell*, 5, 76.

Karmaker, A., & Joarder, R.N. (1993). Molecular clusters and correlations in water by x-ray and neutron diffraction. *Physical Rev. E, 47(6)*, 4215-4218.

Karplus, M., & McCammon, J.A. (1986). *Scientific American*, April, p.42.

Kauzmann, W. (1957). *Quantum Chemistry*. New York: Academic Press.

Kell, G.S. (1975). *J. Chem. and Eng. Data., 20*, 97.

Kikoin, I.K. (Ed.). (1976). *Tables of physical values*. Moscow: Atomizdat, (in Russian).

Kirschner, M., & Mitchison, T. (1986). Beyond self-assembly: From microtubules to morphogenesis. *Cell, 45*, 329-342.

Kiselevm, V.F., Saletsky, A.N., & Semikhina, L.P. (1975). *Theor. experim. khimya (USSR), 2*, 252-257.

Kittel, Ch. (1975). *Thermal physics*. N.Y: John Wiley and Sons.

Kittel, Ch. (1978). *Introduction to the solid state physics*. Moscow: Nauka, (in Russian).

Klassen, V.I. (1982). *Magnetization of the aqueous systems*. Moscow: Khimiya, (in Russian).

Kneppo, P., & Titomir, L.I. (1989). *Biomagnetic measurments*. Moscow: Energoatomizdat.

Koshland, D.E. (1962). *J. Theoret. Biol. 2*, 75.

Kozyrev, N.A. (1958). Causal or nonsymmetrical mechanics in a linear approximation. Pulkovo. Academy of Science of the USSR.

Kovacs, A.L. (1990). Hierarcical processes in biological systems. *Math. Comput. Modelling, 14*, 674-679.

Kovacs, A.L. (1991). A hierarchical model of information flow in complex systems. *Int. J. General Systems, 18*, 223-240.

Kramers, H.A. (1940). *Physika, 7*, 284.

Krinsky, V.J. (**Ed**.). (1984). *Self-organization*. Berlin: Springer.

Krtger, J. (1991). *Neuronal cooperativity*. Berlin: Springer.

Kuchino, Y., Muller, W.E.G., & Pine, P.L. (1991). Berlin, Heidelberg: Springer-Verlag.

Lagrage, P., Fontaine, A., Raoux, D., Sadoc, A., & Miglardo, P. (1980). *J. Chem. Phys., 72*, 3061

Lockwood, M. (1989). *Mind, brain and the quantum.* Basil Blackwell.

London, F. (1938). On the Bose-Einstein condensation. *Phys. Rev.*, 54, 947.

London, F. (1950). *Superfluids, (Vol. 1)*. Wiley.

Landau, L.D., & Lifshits, E.M. (1976). *Statistical physics*. Moscow: Nauka, (in Russian).

Lahoz-Beltra, R., Hameroff, S., & Dayhoff, J.E. (1993). Cytosceletal logic: A model for molecular computation via Boolean operations in microtubules and microtubule-associated proteins. *BioSystems, 29*, 1-23.

Lechleiter, J., Girard, S., Peralta, E., & Clapham, D. (1991). Spiral waves: Spiral calcium wave propagation and annihilation in Xenopus laevis oocytes. *Science, 252*, 123-126.

Lee, J.C., Field, D.J., George, H.J., & Head, J. (1986). Biochemical and chemical properties

of tubulin subspecies. *Ann. N.Y. Acad. Sci. 466*, 111-128.
Leggett, A.J. (1980). *Progr. Theor. Phys., supp. 69*, 80.
Lenhardt, M., et al., (1991). Human ultrasonic speech perception. *Science*, July 5.
Lumry, R., & Gregory, R.B. (1986). Free-energy managment in protein reactions: Concepts, complications and compensations. In *The fluctuating enzyme*. A Wiley-Interscience Publication.
Maisano, G., Majolino, D., Mallamace, F., Aliotta, F., Vasi, C., & Wanderlingh, F. (1986). Landau-Placzek ratio in normal and supercooled water. *Mol. Phys. 57*, 1083-1097.
Marshall, I.N. (1986). Consciousness and Bose-Einstein condensates. *New Ideas in Psychol. 7*, 77-83.
Magazu, S., Maisano, G., Majolino, D., Mallamace, F., Migliardo P., Aliotta, F., & Vasi, C. (1989). Relaxation process in deeply supercooled water by Mandelstam-Brilloin scattering. *J. Phys. Chem.*, *93*, 942-947.
Magazu, S., Majolino, D., Mallamace, F., Migliardo, P., Aliotta, F., Vasi, C., D'Aprano, A., & Donato, D.I. (1989). Velocity and damping of thermal phonons in isomeric alcohols. *Mol. Phys.*, 66, *No. 4*, 819-829.
Malcolm, S., Gordon, I., Ben, H., Amdur, I., & Scholander, P.F. (1962). Freezing resistance in some northern fishes. *Biol. Bull., 122*, 52-62.
Mascarenhas, S. (1974). The electret effect in bone and biopolymers and the bound water problem. *Ann. N.Y. Acad. Sci.*, *238*, 36-52..
Maslovski, V.M., & Postnikov, S.N. (1989). In *The treatment by means of the impulse magnetic field. Proceedings of the IV seminar on nontraditional technology in mechanical engineering*. Sofia-Gorky.
Miamoto, S. (1995). Changes in mobility of synaptic vesicles with assembly and disassembly of actin network. *Biochemica et Biophysica Acta*, *1244*, 85-91.
McCall, S.L., & Hahn, E.L. (1980). Self-induced transparency by pulsed coherent light. *Phys. Rev. Lett., 18*, 908-911.
Melki, R., Carlier, M.F., Pantaloni, D., & Timasheff, S.N. (1980). Cold depolimerization of microtubules to double rings: Geometric stabilization of assemblies. *Biochemistry, 28*, 9143-9152.
Mileusnic, R., Rose, S.P., & Tillson, P. (1980). Passive avoidance learning results in region specific changes in concentration of, and incorporation into, colchicine binding proteins in the chick forebrain. *Neur. Chem., 34*, 1007-1015.
Minenko, V.I. (1981). Electromagnetic treatment of water in thermoenergetics. Harkov, (in Russian).
Moshkov, D.A., Saveljeva, L.N., Yanjushina G.V., & Funticov, V.A. (1992). Structural and chemical changes in the cytosceleton of the goldfish Mauthner cells after vestibular stimulation. *Acta Histochemica Suppl-Band XLI*, 241-247.
Muallem, S., Kwiatkowska, K., Xu X, & Yin, H.L. (1995). Actin filament disassembly is a sufficient final trigger for exocytosis in nonexcitable cells. *J. Cell Biol. 128*, 589-598.
Nemethy, G., & Scheraga, H.A. (1962). *J. Chem. Phys., 36*, 3382.
Neubauer, C., Phelan, A.M., Keus, H., & Lange, D.G. (1990). Microwave irradiation of rats at 2.45 GHz activates pinocytotic uptake of tracer by capillary endothelial cells of cerebral cortex. *Bioelectromagnetics*, *11*, 261-268.
Nicolis, J.C. (1986). *Dynamics of hierarchical systems*. Berlin: Springer.
Nicolis, J.C., & Prigogine, I. (1977). *Self-organization in nonequilibrium systems. From dissipative structures to order through fluctuations*. N.Y.: Wiley and Sons.
Ohmine, I. (1992, May 31 - June 4). Water dynamics: Inherent structure, analysis for fluctuations and collective motion. In *Water - biomolecule interactions (a book of*

*abstracts)*. Palermo.
Penrose, R. (1989). *The emperor's new mind*. London: Oxford University Press.
Penrose, R. (1994). *Shadows of the mind*. London: Oxford University Press.
Peres, A. (1992, April). Classification of quantum parodoxes: Nonlocality vs. contextuality. In *Conference proceedings: "The Interpretation of Quantum Theory: Where do we stand?"*. New-York.
Peres, A. (1993). *Quantum theory: Concepts & methods*. Dordrecht: Kluwer Acad. Publ.
Peschel, G., & Adlfinger, K.H. (1970). *J. Coll. Interface Sci.*, *34* (4), 505.
Pribram, K. (1977). *Languages of the brain*, (p.123). Monterey, Calif.: Wadsworth Publishing
Pribram, K. (1991). *Brain and perception*. New Jersey: Lawrence Erlbaum.
Prigogine, I. (1980). *From being to becoming: Time and complexity in physical sciences*. San Francisco: W. H. Freeman and Company.
Prigogine, I., & Strengers, I. (1984). *Order out of chaos*. London: Hainemann.
Prokhorov, A.M. (Ed.). (1988). *Physical encyclopedia* (Vol.1-4). Moscow.
Prokhorov, A.A. (1991). Study of thermoinduced processes in liquid water within the range of associative absorption band $(\nu_2 + \nu_L)$. *J. Appl. Spectroscopy (USSR)*, *54*, 740.
Prokhorov, A.A. (1991). Sensitivity of associative absorption band $(\nu_2 + \nu_L)$ to structural changes in water and water solutions. *J. Appl. Spectroscopy (USSR). 55*, 951.
Pokorny, J., & Fiala, J. (1994). Condensed energy and interaction energy in Frolich systems. *J. Neural Network World*, *Vol. 4, N3*, 299-313.
Pokorny, J., & Fiala, J. (1994). Information content of Frolich coherent systems. *J. Neural Network World*, *Vol. 4, N3*, 315-325.
Popp, F.A., Li, K.H., et al. (1988). Physical aspects of biophotons. *Experentia*, *44*, 576-585.
Popp, F.A., Li, K.H., & Gu, Q. (1992). *Recent advances in biophoton research*. Singapore: World Scientific.
Rasmussen, S., Karampurwala, H., Vaidyanath, R., Jensen, K.S., & Hameroff, S. (1990). Computational connectionism within neurons: A model of cytosceletal automata subserving neural networks. *Physica D*, *42,* 428-449.
Rey, L. (2003). Thermoluminescence of ultra-high dilutions of lithium chloride and sodium chloride. Physica A 323 (2003) 67 – 74.
Ribary, J., Ioannides, A.A., Singh, K.D., Hasson, R., Bolton, J.P., Lado, F., Mogilner, A., & Llinas, R. (1991). Magnetic field tomography of coherent thalamcortical 40 Hz oscillations in humans. *Proc. Natl. Acad. Sci. USA*, *88*, 11037-11041.
Riccardi, L.M., & Umezawa, H. (1967). Brain and physics of many-body problems. *Kybernetik, 4*, 44-48.
Roberts, J., & Wang, F. (1993). Dielectric relaxation in water over the frequency range $13 \le f \le 18$ GHz using a resonant microwave cavity operating in the $TM_{010}$ mode. *J. Microwave Power and Electromagnetic Energy, 28*, 196-205.
Rubik, B. (1996). *Life at the edge of science,* (p. 185). The Inst. for Frontier Science.
Sadoc, A., Fountaine, A., Lagarde, D., & Raoux, D. *J. Am. Chem. Soc.*, *103*, 6287.
Samsonovich, A., Scott A., & Hameroff, S.R. (1992). Acousto-conformational transitions in cytosceletotal microtubules: Implications for neuro- like protein array devices. *Nanobiology*, 457-468.
San Biagio, P.L., & Palma, M.U. (1991). *Biophys. J.*, *60*, 508-515.
Sciortino, F., Geiger, A., & Stanley, H.E . (1990). *Phys. Rev. Lett.*, *65*, 3452-5.
Schulz, G.E., & Schirmer, R.H. (1979). *Principles of protein structure*. New York: Springer-Verlag.

Semikhina, L.P.,  Kiselev, V.F., Mogilner, A., & Llinas, R. (1988). *Izvestiya VUZov. Fizika (USSR). 5*, 13, (in Russian).

Semikhina, L.P. (1979). *Kolloidny Jurnal (USSR), 43*, 401.

Sherderake R. (1983). New Science of Life: Hypothesis of Formative Causation. Flamingo. pp.240.

Shih, Y., & Alley, C.O. (1988). *Phys. Rev. Lett. 61*, 2921.

Shnoll, S.E., Rubinshtejn, I.A., Zenchenko, K.I., et al. (2005, January 02). Experiments with rotating collimators cutting out pencil of alpha-particles at radioactive decay of Pu-239 evidence sharp anisotropy of space.  Retrieved from http://www.arXiv.org :physics/0501004.

Shnoll, S. E., Zenchenko, K. I., & Udaltsova, N.V. (2005, April 13). Cosmo-physical effects in structure of the daily and yearly periods of change in the shape of the histograms constructed by results of measurements of alpha-activity Pu-239. Retrieved from http://arXiv.org:physics/0504092.

Shoulders K.R. (1991). Method and apparatus for production and manipulation of high density charge. U.S. Patent No. 5,054,046, issued Oct. 1.

Shpinel, V.C. (1969). *Gamma-ray resonance in crystals*. Moscow: Nauka, (in Russian).

Shutilov, V.A. (1980). *Principles of ultrasound physics*. Leningrad: Leningrad University Press, (in Russian).

Singer, W. (1986). Synchronization of cortical activity and its putative role in information processing and learning. *Annu. Rev. Physiol.*, *55*, 349-374.

Singvi, K., & Sielander, A. (1962). In  Kogan Yu  (Ed.), *Mössbauer effect* (p. 193). Moscow, (in Russian).

Smith, C.W. (1994). Coherence in living biological systems. *Neural Network World, 4, N3*, 379-388.

Somogyi, B., & Damjanovich, S. (1986). In  G.R.Welch (Ed.), *The fluctuating enzyme*, (pp. 341-368). A Wiley-InterScience Publication.

Stapp, H.P. (2003). *Mind, matter and quantum mechnics*. New York: Springer - Verlag.

Strogatz, S.H., & Stewart, I. (1993). Coupled Oscillators and biological synchronization. *Scientific American*, *Dec.*, 68-75.

Tegmark, M. (2000). Importance of quantum decoherence in brain process. *Phys. Rev.* **E61**, 4194-4206.

Teixeira, J., Bellisent-Funel, M.C., Chen, S.H., & Dorner, B. (1985). Observation of new short-wavelength collective excitations in heavy water by coherent inelastic neutron scattering. *Phys. Rev. Letters*,  *Vol. 54, N.25*, 2681-2683.

Tereshkevitch, M.O., Kuprin F.V., Kuratova, T.S., & Ivishina, G.A. (1974). *J. Phys. Chem. (USSR)*, *48*, 2498.

Theiner, O., & White, K.O. (1969). *J. Opt. Soc. Amer., 59*, 181.

Timashev, S.F. ( 1985). *Doklady Academii Nauk SSSR*, *281*, 112.

Trincher, K.S. (1967). *State and role of water in biological objects*, (pp.143-149.) Moscow: Nauka.

Tsukita, S., Kobayashi, T., & Matsumoto, G. (1986). Subaxolemmal cytosceleton in squid giant axon. II. Morpological identification of microtubules and microfilament-associate domains of axolemma. *J. Cell Biol., 102*, 1710-1725.

Umezawa, H. (1993). *Advanced field fheory: Micro, macro and thermal physics*. New York: American Institute of Physics.

Umezawa, H., Matsumoto, H., & Tachiki M. (1982). *Thermo field dynamics and condensate states*. Amsterdam: North-Holland.

Vassilev, P., Kanazirska, M., & Tien, H. (1985) Intermembrane linkage mediated by tubulin. *Biochem. Biophys. Res. Comm., 126(1)*, 559-565.

Vuks, M.F. (1977). *Light scattering in gases, liquids and solutions*. Leningrad: Leningrad University Press.

Vuks, M.F. (1984). *Electrical and optical properties of molecules and condensed matter*. Leningrad: Leningrad University Press.

Wang, H., Lu, B., & Roberts, J.A. (1994). Behavior of hydrogen bonding in water over the temperature range $100 \leq T \leq 360K$. *Molecular materials*.

Watterson, J. (1987). Solvent cluster size and colligative properties. *Phys. Chem. Liq.*, *16*, 317-320.

Watterson, J. (1988). The role of water in cell architecture. *Mol. Cell. Biochem.*, *79*, 101-105.

Watterson, J. (1988). A model linking water and protein structures. *Bio Systems 22*, 51-54.

Wertheim, G.K. (1964). *Mössbauer effect*. N.Y., London: Academic Press.

Wiggins, P.M. (1972). Thermal anomalies in ion distribution in rat kidney slices and in a model system. *Clin. Exp. Pharmacol. Physiol. 2*, 171-176.

Yoshida, H., Ito, K., & Ise, N. (1991). Colloidal crystal growth. *J. Chem. Soc. Faraday Trans.*, *87(3)*, 371-378.

Yashkichev, V.I. (1980). *J. Inorganic Chem. (USSR) 25*, 327.

Zeldovitch, Ya.B., & Khlopov, M.Yu. (1988). Drama of concepts in cognition of nature. Moscow: Nauka, (in Russian).

Zhitenev, N.B. et al., (1999). Science 285, 715.

# Appendix:

## The Basic Properties of Bivacuum, Responsible for Nonlocality. Origination of Mass and Charge of Elementary Particles without Higgs Bosons

*The excerpt from paper of this author http://arxiv.org/ftp/physics/papers/0207/0207027.pdf and the next book to Nova Publishers Corp. (NY, USA)*

**A1**. **Virtual microtubules (microfilaments) formed by Bivacuum dipoles in state of quasi 1D Bose condensation**
**A2**. **Virtual multi-bilayers (membranes) formed by association of Virtual Microtubules**
**A3**. **Virtual Pressure Waves (VPW$^\pm$)**
**A4**. **Virtual Spin Waves (VirSW)**
**A5**. **Virtual Bose condensation (VirBC), as a condition of Bivacuum superfluid and nonlocal properties**
**A6**. **Three Postulates and related Conservation Rules for asymmetric Bivacuum fermions (BVF$^\updownarrow$)$_{as}$ and Bivacuum bosons (BVB$^\pm$)$_{as}$**
**A7**. **The relation between the external and internal parameters of Bivacuum fermions and their absolute external velocity**
**A8**. **The Hidden Harmony of asymmetric Bivacuum dipoles, as a background of Golden Mean**
**A9**. **The rest mass and charge origination**
**A10**. **The solution of Dirac monopole problem, following from Unified theory**
**A11**. **The elementary fermions origination from asymmetric Bivacuum dipoles: muons and tauons**
**A12**. **The energy of elementary fermions fusion from muons and tauons.**
**A13**. **New scenario of the Big Bang**

The original Bivacuum theory worked out by this author is a result of development of Dirac theory of vacuum. Both theories proceeds from assumption of equal probability of positive and negative energy of vacuum. Our theory relates the Bivacuum dipoles symmetry shift with origination of sub-elementary and elementary particles and their pulsation (beats) between Corpuscular and Wave phases. The pulsation of asymmetric Bivacuum dipoles are accompanied by energy exchange with Virtual Pressure Waves (VPW$^\pm$). The Corpuscle $\rightleftharpoons$ Wave pulsation are followed by recoil $\rightleftharpoons$ antirecoil effects, generating the electromagnetic and gravitational fields (Kaivarainen, 2005, 2006, 2007).

The concept of Bivacuum is introduced, as a dynamic matrix of the Universe, composed from non-mixing subquantum particles and antiparticles. The *subquantum particles and antiparticles* are considered, as the minimum stable vortical structures of Bivacuum with dimensions about or less than $10^{-19}m$ of opposite direction of rotation (clockwise and counter-clockwise) of zero mass and charge.

Their spontaneous collective vortical excitations produce strongly correlated pairs of donuts of positive and negative energy, mass, opposite electric and magnetic charges. These Bivacuum dipoles have shape of [torus($V^+$) + antitorus($V^-$)], separated by energetic gap.

*Three kinds of Bivacuum dipoles are named Bivacuum fermions, antifermions* (**BVF**$^\uparrow$ *and* **BVF**$^\downarrow$) *and Bivacuum bosons* (**BVB**$^\pm$).

Their tori and antitori in primordial Bivacuum have the opposite energy/mass, charge and magnetic moment, compensating each other and making Bivacuum a neutral continuum with resulting energy and charge densities equal to zero. The radius of tori and antitori, forming two generation of Bivacuum dipoles, in symmetric primordial Bivacuum are equal to each other and determined by Compton radius of muons ($\mu$) and tauons ($\tau$).

The infinitive number of Bivacuum fermions and antifermions are represented by both tori, rotating clockwise or counter-clockwise:

$$\mathbf{BVF}^\uparrow_{\mu,\tau} \equiv [V^+ \uparrow\uparrow V^-]^{\mu,\tau} \quad and \quad \mathbf{BVF}^\downarrow_{\mu,\tau} \equiv [V^+ \downarrow\downarrow V^-]^{\mu,\tau} \qquad 1$$

and Bivacuum bosons contain tori and antitori, rotating in opposite direction:

$$\mathbf{BVB}^\pm_{\mu,\tau} \equiv [V^+ \uparrow\downarrow V^-]^{\mu,\tau} \qquad 2$$

The Bivacuum bosons can be considered as intermediate state between **BVF**$^\uparrow$ and **BVF**$^\downarrow$. The continuum of Bivacuum dipoles partly is in state of virtual macroscopic Bose condensate with superfluid properties.

Symmetry shift of $\mathbf{BVF}^\uparrow_{\mu,\tau}$ and $\mathbf{BVF}^\downarrow_{\mu,\tau}$, turning them to sub-elementary fermions and sub-elementary antifermions of *muon generation*, following by their fusion of triplets originate the electrons and positrons. The similar processes with Bivacuum fermions of *tauon generation* originate the triplets of quarks: protons, neutrons and their antiparticles.

The elementary particles fusion from sub-elementary ones should be accompanied by release of huge amount of energy, determined by the mass defect.

The current quasi - symmetric Bivacuum can be considered as the Universal Reference Frame (**URF**), i.e. *Ether*, in contrast to Relative Reference Frame (**RRF**), used in special relativity (SR) theory. The elements of *Ether* - are represented by Bivacuum dipoles. It is shown in our work (Kaivarainen, 2006), that the result of Michelson - Morley experiment is a consequence of the part of Ether drug by the Earth. This special fraction of the Ether is named Virtual Replica of the Earth. Its origination and virtual hologram properties are described in our Unified Theory: http://arxiv.org/ftp/physics/papers/0207/0207027.pdf.

### A1. Virtual microtubules (microfilaments) formed by Bivacuum dipoles in state of quasi 1D Bose condensation

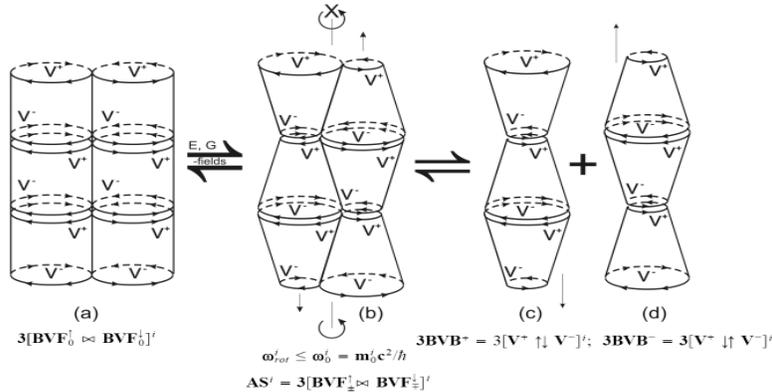

**Figure A1.** (**a**): three Cooper pairs of symmetric Bivacuum fermions in primordial

Bivacuum: $3[\mathbf{BVF}_0^\uparrow \bowtie \mathbf{BVF}_0^\downarrow]^i$ forming a double coherent Virtual microtubules ($\mathbf{VirMT}^i$). This symmetric structures do not rotate around the main common axis and their tangential velocity is zero: $\mathbf{v} = \mathbf{0}$. Only the internal rotation of torus ($\mathbf{V}^+$) and antitorus ($\mathbf{V}^-$) of Bivacuum dipoles takes a place;

(**b**): the same structure as (**a**) in strong electric and gravitational fields. The symmetric Bivacuum fermions and antifermions pairs turns to asymmetric three Cooper pairs Bivacuum fermions in secondary Bivacuum, rotating around common axis (X). These structures may be assembled to closed Virtual microtubules $\mathbf{VirMT}^i$ or nonlocal Virtual guides ($\mathbf{VirG}_{SME}^i$) of spin, momentum and energy (always open and connecting remote elementary particles of opposite spin). The $\mathbf{VirG}_{SME}^i$ are responsible for entanglement between remote 'tuned' elementary particles of similar basic frequency of $[\mathbf{C} \rightleftharpoons \mathbf{W}]$ pulsation:

$$\omega_{rot}^{\mu,\tau} \leq \omega_0^{\mu,\tau} = \mathbf{m}_0^{\mu,\tau} \mathbf{c}^2/\hbar$$

(**c**) + (**d**) are the result of dissociation of double virtual microtubules $\mathbf{VirMT}^i_{[\mathbf{BVF}_\pm^\uparrow \bowtie \mathbf{BVF}_\mp^\downarrow]}$, presented at (**b**):

$$3[\mathbf{BVF}_\pm^\uparrow \bowtie \mathbf{BVF}_\mp^\downarrow]^{\mu,\tau} \rightleftharpoons 3(\mathbf{BVB}^\pm)^{\mu,\tau} + 3(\mathbf{BVB}^\mp)^{\mu,\tau}$$

in strong electric field, to single virtual microtubules $\mathbf{VirMT}_{\mathbf{BVB}}^i$ of Bivacuum bosons of opposite polarization and resulting charge. These two single virtual microtubules propagate in opposite direction, stimulated by the resonant interaction with basic virtual pressure waves ($\mathbf{VPW}_{q=1}^\pm$) of Bivacuum. As a result of this dissociation the rotational kinetic energy of double $\mathbf{VirMT}^{\mu,\tau}_{[\mathbf{BVF}_\pm^\uparrow \bowtie \mathbf{BVF}_\mp^\downarrow]}$ turns to translational kinetic energy of single $\mathbf{VirMT}_{\mathbf{BVB}}^i$, providing the same Bivacuum dipoles symmetry shift and external velocity, corresponding to Golden Mean condition ($\mathbf{v}^2 = \boldsymbol{\phi}\mathbf{c}^2$). The latter is stimulated by resonant exchange interaction of $\mathbf{VirMT}^{\mu,\tau}_{[\mathbf{BVF}_\pm^\uparrow \bowtie \mathbf{BVF}_\mp^\downarrow]}$ and $\mathbf{VirMT}_{\mathbf{BVB}}^i$ with basic Bivacuum $(\mathbf{VPW}_{q=1}^\pm)^{\mu,\tau}$ of Compton frequency $\omega_0^{\mu,\tau} = \mathbf{m}_0^{\mu,\tau} \mathbf{c}^2/\hbar$.

**A2**. **Virtual multi - bilayers (membranes) formed by association of Virtual Microtubules**

The virtual multi-layer membranes (*VirMem*) can be formally presented as a result of infinitive number of Bivacuum fermions Cooper pairs of opposite polarization:

$$VirMem = \sum^\infty \left[ \frac{BVF^\uparrow \bowtie BVF^\downarrow}{BVF^\downarrow \bowtie BVF^\uparrow} \right]^{\mu,\tau}_{x,y} \qquad 3$$

where Bivacuum fermions and antifermions, rotating in the opposite direction are interacting with each other side-by-side. Each of such Cooper pair $BVF^\uparrow \bowtie BVF^\downarrow$ has the counterpart $BVF^\downarrow \bowtie BVF^\uparrow$. These pairs are interacting with each other by 'head-to-tail' principle, forming virtual microtubules (Fig.1a) in symmetric unperturbed by fields Bivacuum.

The oscillation of such two Cooper pairs between the symmetrically excited and the ground states are counterphase. This means that the virtual *gap oscillation* between tori and antitori of corresponding Bivacuum dipoles are also counterphase. This process may be accompanied by the exchange of pairs of Virtual Pressure Waves $\left[ \mathbf{VPW}_q^+ \bowtie \mathbf{VPW}_q^- \right]^{\mu,\tau}$ between virtual Cooper pairs. The oscillation of gap between positive and negative mass/energy, opposite charges and magnetic moments can be considered as *inter-space waves (ISW)*, in-phase with $\left[ \mathbf{VPW}_q^+ \bowtie \mathbf{VPW}_q^- \right]^{\mu,\tau}$. The new notion of positive and negative space is related to positive and negative Compton radiuses of conjugated torus and antitorus with mass and charge of muons and anti-muons, tauons and anti-tauons, forming Bivacuum dipoles:

$$\pm \mathbf{L}_{\mu,\tau} = \frac{\hbar}{\pm m_{\mu,\tau}c} \qquad 3a$$

The virtual multilayer membranes in Bivacuum can be considered also, as the assembly of huge number of virtual microtubules of Bivacuum side-by-side.

Each of the layer of virtual membranes, can pulse between the excited and ground state in counterphase with the next one, interacting with each other via dynamic exchange by pairs of virtual pressure waves $[\mathbf{VPW}^+ \bowtie \mathbf{VPW}^-]^{\mu,\tau}$. This process occur without violation of the energy conservation law because of positive and negative energy oscillation compensation. Such pulsations are accompanied by nonlocal Bivacuum gap oscillation over the space of virtual Bose condensation (BC) of Bivacuum dipoles.

The value of spatial gap between the actual and complementary torus and antitorus of Bivacuum fermions is decreasing with their excitation quantum number (**n**) increasing (Kaivarainen, 2006):

$$[\mathbf{d}_{\mathbf{v}^+ \updownarrow \mathbf{v}^-}]_n = \frac{h}{\mathbf{m}_0 \mathbf{c}(1+2\mathbf{n})} \qquad 4$$

The Bivacuum gap oscillations can be responsible for the lateral or transversal nonlocality of Bivacuum in contrast to longitudinal one, realized via virtual guides (single or double virtual microtubules): $\mathbf{VirG}_{SME}$ (Fig. 50).

The decreasing of gap between torus and antitorus of Bivacuum dipoles is in-phase with decreasing of their radius as a result of excitation and increasing of quantum number (**n**):

$$\left[ \mathbf{L}_{\mathbf{V}^\pm}^n = \frac{\pm \hbar}{\pm \mathbf{m}_0 \mathbf{c}(\frac{1}{2}+\mathbf{n})} = \frac{\mathbf{L}_0}{\frac{1}{2}+\mathbf{n}} \right] \qquad 5$$

the ratio of gap between torus and antitorus to their to radius of dipoles in primordial (symmetric) Bivacuum is a permanent value, independent on the amplitude of gap oscillation:

$$\frac{[\mathbf{d}_{\mathbf{v}^+ \updownarrow \mathbf{v}^-}]_n}{\mathbf{L}_{\mathbf{V}^\pm}^n} = \pi \qquad 5a$$

The gap and radius oscillations are accompanied by the emission and absorption of virtual clouds and virtual pressure waves excitation $[\mathbf{VPW}^+ \bowtie \mathbf{VPW}^-]$, providing exchange interaction between counterphase oscillating dipoles.

### A3. Virtual Pressure Waves ($\mathbf{VPW}^\pm$)

The emission and absorption of Virtual clouds $(\mathbf{VC}_{j,k}^+)^i$ and anti-clouds $(\mathbf{VC}_{j,k}^-)^i$ by Bivacuum dipoles of primordial Bivacuum (in the absence of matter and fields) are the result of correlated transitions between different excitation states $(j,k)$ of tori $(\mathbf{V}_{j,k}^+)^i$ and antitori $(\mathbf{V}_{j,k}^-)^i$, forming symmetric Bivacuum dipoles: $[\mathbf{BVF}^\updownarrow]^{\mu,\tau}$ and $[\mathbf{BVB}^\pm]^{\mu,\tau}$:

$$(\mathbf{VC}_q^+)^{\mu,\tau} \equiv [\mathbf{V}_j^+ - \mathbf{V}_k^+]^{\mu,\tau} \quad - \; virtual \; cloud \qquad 6$$

$$(\mathbf{VC}_q^-)^{\mu,\tau} \equiv [\mathbf{V}_j^- - \mathbf{V}_k^-]^{\mu,\tau} \quad - \; virtual \; anticloud \qquad 6a$$

where: $j > k$ are the integer quantum numbers of torus and antitorus excitation states; $q = j - k$.

The virtual clouds: $(\mathbf{VC}_q^+)^i$ and $(\mathbf{VC}_q^-)^i$ exist in form of collective excitation of *subquantum* particles and antiparticles of opposite energies, correspondingly. They can be considered as 'drops' of virtual Bose condensation of subquantum particles of positive and negative energy.

The process of [*emission* ⇌ *absorption*] of virtual clouds of positive and negative energy: $(\mathbf{VC}_q^+)^{\mu,\tau}$ and $(\mathbf{VC}_q^-)^{\mu,\tau}$ by Bivacuum fermions, antifermions and bosons is accompanied by oscillation of *virtual pressure*

$$[\mathbf{VirP}^+ \bowtie \mathbf{VirP}^-]^{\mu,\tau}$$

*and excitation of corresponding pairs of positive and negative virtual pressure waves:*

$$\left[\mathbf{VPW}_q^+ \bowtie \mathbf{VPW}_q^-\right]^{\mu,\tau}$$

In symmetric primordial Bivacuum the in-phase oscillation of energy of torus ($\mathbf{V}^+$) and antitorus ($\mathbf{V}^-$) of each Bivacuum dipole compensate each other. However, when symmetry is shifted, for example by external magnetic or electric field, corresponding polarization of Bivacuum can be accompanied by the emergence of uncompensated energy of virtual pressure waves: $\Delta\mathbf{VPW}^\pm = |\mathbf{VPW}^+ - \mathbf{VPW}^-|^{\mu,\tau}$, which can be used as a source of pure and 'free' energy.

### A4. Virtual Spin Waves (VirSW)

*The nonlocal virtual spin waves* ($\mathbf{VirSW}_{j,k}^{\pm 1/2}$), with properties of massless collective Nambu-Goldstone modes, like a real spin waves, represent the oscillation of angular momentum equilibrium of individual Bivacuum fermions or in composition of Cooper pairs with opposite spins via "flip-flop" mechanism, accompanied by origination of intermediate states - Bivacuum bosons ($\mathbf{BVB}^\pm$):

$$\mathbf{VirSW}_{j,k}^{\pm 1/2} \sim \left[\mathbf{BVF}^\uparrow(\mathbf{V}^+\!\!\upuparrows\mathbf{V}^-) \rightleftharpoons \mathbf{BVB}^\pm(\mathbf{V}^+\updownarrow\mathbf{V}^-) \rightleftharpoons \mathbf{BVF}^\downarrow(\mathbf{V}^+\!\!\downdownarrows\mathbf{V}^-)\right]^{\mu,\tau} \qquad 7$$

The $\mathbf{VirSW}_{j,k}^{+1/2}$ and $\mathbf{VirSW}_{j,k}^{-1/2}$ are excited by $(\mathbf{VC}_q^\pm)_{S=1/2}^{\circlearrowleft}$ and $(\mathbf{VC}_q^\pm)_{S=-1/2}^{\circlearrowleft}$ of opposite angular momentums, $S_{\pm 1/2} = \pm\frac{1}{2}\hbar = \pm\frac{1}{2}\mathbf{L}_0\mathbf{m}_0\mathbf{c}$ and frequency, equal to $\mathbf{VPW}_q^\pm$:

$$\mathbf{q}\omega_{\mathbf{VirSW}^{\pm 1/2}}^{\mu,\tau} = \mathbf{q}\omega_{\mathbf{VPW}^\pm}^{\mu,\tau} = \mathbf{q}\,\mathbf{m}_0^i\mathbf{c}^2/\hbar = \mathbf{q}\,\omega_0^{\mu,\tau} \qquad 8$$

minimum quantum number $\mathbf{q} = (\mathbf{j} - \mathbf{k}) = \mathbf{1}$ determines the most probable basic virtual pressure waves $\mathbf{VPW}_{q=1}^\pm$ and virtual spin waves $\mathbf{VirSW}_{q=1}^{\pm 1/2}$.

The $\mathbf{VirSW}_q^{\pm 1/2}$, like so-called torsion field, can serve as a carrier of the phase/spin (angular momentum) and information - *qubits*, but not the energy.

### A5. Virtual Bose condensation (VirBC), as a condition of Bivacuum superfluid and nonlocal properties

It follows from our model of Bivacuum, that the infinite number of Cooper pairs of Bivacuum fermions $[\mathbf{BVF}^\uparrow \bowtie \mathbf{BVF}^\downarrow]_{S=0}^i$ and their intermediate states - Bivacuum bosons $(\mathbf{BVB}^\pm)^i$, as elements of Bivacuum, have zero or very small (in presence of fields and matter) translational momentum: $\mathbf{p}_{\mathbf{BVF}^\uparrow\bowtie\mathbf{BVF}^\downarrow}^i = \mathbf{p}_{\mathbf{BVB}}^i \to 0$ and corresponding de Broglie wave length tending to infinity: $\lambda_{\mathbf{VirBC}}^i = \mathbf{h}/\mathbf{p}_{\mathbf{BVF}^\uparrow\bowtie\mathbf{BVF}^\downarrow,\mathbf{BVB}}^i \to \infty$.

This condition leads to origination of 3D system of virtual double virtual microtubules from Cooper pairs of Bivacuum fermions $[\mathbf{BVF}^\uparrow \bowtie \mathbf{BVF}^\downarrow]_{S=0}$, and single virtual microtubules, formed by Bivacuum bosons $(\mathbf{BVB}^\pm)_{S=0}$, closed or open, connecting remote coherent elementary particles.

The longitudinal momentum of Bivacuum dipoles forming virtual microfilaments and their bundles/beams can be close to zero and corresponding de Broglie wave length:

$$\lambda = \frac{h}{|\mathbf{m}_V^+ - \mathbf{m}_V^-|c} \qquad 9$$

$$\lambda \to \infty \quad \text{at} \quad \mathbf{m}_V^+ \to \mathbf{m}_V^-$$

exceeding the distance between neighboring dipoles many times.

The 3D system of these double and single microtubules (see Fig. 50) represents Bose condensate with superfluid properties.

The Bivacuum, like liquid helium, can be considered as a liquid, containing two components: the described superfluid and normal, representing fraction of Bivacuum dipoles not involved in virtual guides (VirG). The radii of VirG are determined by the Compton radii of the electrons, muons and tauons, interconnecting similar particles with opposite spins.

Their length is limited by decoherence effects, related to Bivacuum symmetry shift. In highly symmetric Bivacuum the length of **Virtual Guides** with nonlocal properties, connecting remote coherent elementary particles, may have the order of stars and galactic separation.

In some cases virtual microfilaments/microtubules (**VirMT**) may form a closed rotating rings with perimeter, determined by resulting standing de Broglie wave length of Bivacuum dipoles forming the rings. The life-time of such closed structures can be big, if they have a properties of standing and non-dissipating systems of virtual de Broglie waves of Bivacuum dipoles. Corresponding interference pattern of virtual standing waves can be a part of the Virtual Replicas (quantum hologram) of real objects. For brief description of Virtual Replica see section 18.4 of this book.

The Nonlocality can be formulated as the independence of potential energy of any elements of Bivacuum or other medium on the distance from the energy source. This formulation follows from application of the Virial theorem to systems of Cooper pairs of Bivacuum fermions $[BVF^\uparrow \bowtie BVF^\downarrow]_{S=0}$ and Bivacuum bosons ($\mathbf{BVB}^\pm$), composing virtual (and real) Bose condensate.

The described above properties of Bivacuum as the infinitive continuum and its active elements - Cooper pairs of Bivacuum fermions, is a background for development of Universe model as a quantum supercomputer.

*The idea of such supercomputer, able to calculate the most probable future and past, based on current superposition of Virtual Replicas, is based on principle of macroscopic entanglement between huge number of elementary particles (tuned de Broglie waves) of remote star systems by means of Virtual Channels, Bivacuum gap oscillation and theory of Virtual Replica of material objects (Kaivarainen, 1995; 2005-2007).*

### A6. Three postulates and related conservation rules for asymmetric Bivacuum fermions ($\mathbf{BVF}^\updownarrow)_{as}$ and Bivacuum bosons ($\mathbf{BVB}^\pm)_{as}$

*There are three basic postulates in our theory, interrelated with each other:*

**Postulate I**. The absolute values of the *internal* rotational kinetic energies of torus and antitorus are permanent equal to each other and to the half of the rest mass energy of the electrons of corresponding lepton generation, independently on the *external* rotational and translational group velocity (**v**), turning the symmetric Bivacuum fermions ($\mathbf{BVF}^\updownarrow$) to asymmetric ones:

$$[\mathbf{I}]: \quad \left( \frac{1}{2}\mathbf{m}_V^+ (\mathbf{v}_{gr}^{in})^2 = \frac{1}{2}|-\mathbf{m}_V^-|(\mathbf{v}_{ph}^{in})^2 = \frac{1}{2}\mathbf{m}_0 \mathbf{c}^2 = \mathbf{const} \right)_{in}^{,\mu,\tau} \quad 10$$

where the positive $\mathbf{m}_V^+$ and negative $-\mathbf{m}_V^- = i^2 \mathbf{m}_V^-$ are the 'actual' - inertial and 'complementary' (imaginary) - inertialess masses of torus ($\mathbf{V}^+$) and antitorus ($\mathbf{V}^-$); the $\mathbf{v}_{gr}^{in}$ and $\mathbf{v}_{ph}^{in}$ are the *internal* angular group and phase velocities of subquantum particles and antiparticles, forming torus and antitorus, correspondingly. In symmetric conditions of *primordial* Bivacuum and its virtual dipoles, when the influence of matter and fields is absent: $\mathbf{v}_{gr}^{in} = \mathbf{v}_{ph}^{in} = \mathbf{c}$.

It is proved in our theory of time (http://arxiv.org/ftp/physics/papers/0207/0207027.pdf), that the above condition means the infinitive life-time of torus and antitorus of $\mathbf{BVF}^\updownarrow$ and $\mathbf{BVB}^\pm$, independently of symmetry shift.

The *pace of time* (**dt/t**) for any closed *conservative system* is determined by the pace of its kinetic energy change $(-\mathbf{dT}/\mathbf{T}_k)_{x,y,z}$, *anisotropic* in general case (Kaivarainen, 2006, 2007):

$$\left[ \frac{\mathbf{dt}}{\mathbf{t}} = \mathbf{d}\ln\mathbf{t} = -\frac{\mathbf{dT}_k}{\mathbf{T}_k} = -\mathbf{d}\ln\mathbf{T}_k \right]_{x,y,z} \quad 11$$

since the actual kinetic energy $\mathbf{T}_k = \mathbf{m}_V^+ \mathbf{v}^2/2$, the time for conservative system has a following dependence on the resulting velocity and acceleration:

$$t = -\frac{\vec{v}}{d\vec{v}/dt}\frac{1-(v/c)^2}{2-(v/c)^2} \qquad 12$$

The conventional formula for relativistic time of system do not contain the acceleration:

$$t = \frac{t_0}{\sqrt{1-(v/c)^2}} \qquad 13$$

The *actual* (inertial) mass has the regular relativistic dependence on the external rotational and translational velocity $\mathbf{v} = \mathbf{v}^{ext}$ of Bivacuum dipoles:

$$\pm \mathbf{m}_V^+ = \frac{\mathbf{m}_0}{\pm\sqrt{1-(v/c)^2}} = \mathbf{m} \quad \text{(inertial mass)} \qquad 14$$

while the *complementary* (inertialess) mass ($\mp \mathbf{m}_V^-$) of antitorus $\mathbf{V}^-$ with sign, opposite to that of the actual one ($\pm \mathbf{m}_V^+$), has the reciprocal relativistic dependence on external velocity:

$$\mp \mathbf{m}_V^- = \mp \mathbf{m}_0 \sqrt{1-(v/c)^2} \quad \text{(inertialess mass)} \qquad 14a$$

It is important result of our approach, that in the case of nonzero external velocity of Bivacuum dipole ($\mathbf{v} > \mathbf{0}$) the difference between total energies of torus and antitorus is equal to doubled kinetic energy of dipole, anisotropic in general case:

$$\left[ \begin{array}{c} (\mathbf{m}_V^+ - \mathbf{m}_V^-)c^2 = \mathbf{m}_V^+ v^2 = \mathbf{m}_V^+ L^2 \omega^2 \\ = 2\mathbf{T}_k = \frac{\mathbf{m}_0 v^2}{\sqrt{1-(v/c)^2}} \end{array} \right]_{x,y,z}^{\mu,\tau} \qquad 15$$

The ratio of absolute values (14a) to (14) is:

$$\frac{|-\mathbf{m}_V^-|}{\mathbf{m}_V^+} = \frac{\mathbf{m}_0^2}{(\mathbf{m}_V^+)^2} = 1 - \left(\frac{v}{c}\right)^2 \qquad 16$$

**Postulate II**. The absolute *internal* magnetic moments of torus ($\mathbf{V}^+$) and antitorus ($\mathbf{V}^-$) of asymmetric Bivacuum fermions $\mathbf{BVF}_{as}^\uparrow = [\mathbf{V}^+\uparrow\uparrow \mathbf{V}^-]$ and antifermions: $\mathbf{BVF}_{as}^\downarrow = [\mathbf{V}^+\downarrow\downarrow \mathbf{V}^-]$ are permanent, equal to each other and to that of Bohr magneton ($\mu_B$):

$$[\mathbf{II}] : \left( \begin{array}{c} |\pm\mu_+| = \frac{1}{2}|e_+|\frac{|\pm\hbar|}{|\mathbf{m}_V^+|(v_{gr}^{in})_{rot}} = |\pm\mu_-| = \frac{1}{2}|-e_-|\frac{|\pm\hbar|}{|-\mathbf{m}_V^-|(v_{ph}^{in})_{rot}} = \\ = \mu_B \equiv \frac{1}{2}|e_0|\frac{\hbar}{\mathbf{m}_0 c} = \mathbf{const} \end{array} \right)^{\mu,\tau} \qquad 17$$

Consequently, the magnetic moments of torus and antitorus are independent on their internal $\left(v_{gr,ph}^{in}\right)_{rot}$ and external velocity ($\mathbf{v} > \mathbf{0}$) and mass and charge symmetry shifts.

The actual and complementary masses $\mathbf{m}_V^+$ and $|-\mathbf{m}_V^-|$, internal angular velocities ($v_{gr}^{in}$ and $v_{ph}^{in}$) and electric charges $|e_+|$ and $|e_-|$ of $\mathbf{V}^+$ and $\mathbf{V}^-$ are dependent on the external and internal velocities, however, in such a way, that their change compensate each other and

$$|\pm\mu_+| = |\pm\mu_-| = \mu_B = \mathbf{const} \qquad 17a$$

This postulate reflects the condition of the invariance of magnetic moments $|\pm\mu_\pm|$ and spin values ($\mathbf{S} = \pm\frac{1}{2}\hbar$) of torus and antitorus of Bivacuum dipoles with respect to their internal and external velocity, i.e. the absence of these parameters symmetry shifts.

One may see also that **Postulate II** means in fact that the resulting spins of Bivacuum fermion or antifermion are equal correspondingly to:

$$\mathbf{S} = \pm\frac{1}{2}\hbar \qquad 18$$

because the resulting magnetic moments of sub-elementary fermion or antifermion ($\mu_\pm$) are equal

to the Bohr magneton ($\mu_B$), like each of them:

$$|\pm\mu_+| = |\pm\mu_-| = \mu_B = \pm\frac{1}{2}\hbar\frac{e_0}{m_0 c} = S\frac{e_0}{m_0 c} \qquad 19$$

where: $e_0/m_0 c$ is gyromagnetic ratio of the electron.

*We may conclude that in fact the Postulate II reflects the permanent half-integral value of spin $\pm\frac{1}{2}\hbar$ of the fermions.*

**Postulate III**. *The equality of Coulomb attraction force between torus and antitorus of primordial Bivacuum dipoles*: $[\mathbf{V}^+ \updownarrow \mathbf{V}^-]^{\mu,\tau}$ *of $\mu$ and $\tau$ kinds (muons and tauons), providing uniform electric energy distribution in Bivacuum*:

$$[\mathbf{III}] : \mathbf{F}_0^i = \left(\frac{e_0^2}{[d_{V^+\updownarrow V^-}^2]_n}\right)^\mu = \left(\frac{e_0^2}{[d_{V^+\updownarrow V^-}^2]_n}\right)^\tau \qquad 20$$

where: $[d_{V^+\updownarrow V^-}]_n^{\mu,\tau} = \frac{\hbar}{m_0^{\mu,\tau} c(1+2n)}$ is the separation between torus and antitorus of Bivacuum dipoles at the same state of excitation $(n)$ and $\left(e_0^2 = |e_+||e_-|\right)^{\mu,\tau}$.

The important consequences of *Postulate III* are the following relations, unifying the rest mass and charges of the tori and antitori of Bivacuum dipoles - the precursors of sub-elementary fermions: muons and tauons with mass and charge of the regular electron ($m_0$ and $e_0$)$^e$:

$$(e_0 m_0)^e = (e_0 m_0)^\mu = (e_0 m_0)^\tau \qquad 21$$

$$or : (e_0 m_0)^{e,\mu,\tau} = \sqrt{|e_+ e_-||m_V^+ m_V^-|}^{e,\mu,\tau} = const \qquad 21a$$

The other forms of dependence of the charges of tori and antitori ($e_0^{\mu,\tau}$) of Bivacuum fermions on their mass ($m_0^{\mu,\tau}$) are:

$$e_0^\mu = e_0^e(m_0^e/m_0^\mu) = \frac{e_0^e}{206,7} \qquad 22$$

$$e_0^\tau = e_0^e(m_0^e/m_0^\tau) = \frac{e_0^e}{3487,28} \qquad 22a$$

where $e_0^e$ is the charge of the regular electron.

It follows from Postulate III and eqs.(22 and 22a), that the tori and antitori of symmetric $[\mathbf{V}^+ \updownarrow \mathbf{V}^-]^{\mu,\tau}$ with bigger mass: $m_0^\mu = 206,7\, m_0^e$; $m_0^\tau = 3487,28\, m_0^e$ and smaller separation (gap) have correspondingly smaller charges, providing the uniform charge density distribution in Bivacuum.

As is shown in section **A9**, just these conditions provide *the same charge symmetry shift* of Bivacuum fermions of $\mu$ and $\tau$ generations (i.e. the same charges of muon and tauon, equal to that of regular electron), notwithstanding of different mass symmetry shift between corresponding torus and antitorus (equal to the rest masses of muon and tauon), determined by Golden mean value.

*In other words, Postulate III explains why sub-elementary fermions: muons and tauons, fusing the electron/positron and proton/antiproton have same by the absolute value of electric charges, nonetheless of their big mass difference.*

### A6a. Three compensation principles, following from postulates I and II

The *Mass Compensation Principle* follows from relativistic dependencies of the actual and complementary mass of Bivacuum dipoles, reciprocal to each other (eqs 14 and 14a):

$$|\pm m_V^+||\mp m_V^-| = m_0^2 \qquad 23$$

The internal *Group and Phase Velocities Compensation Principle* for internal dynamics of torus ($v_{gr}^{in}$) and antitorus ($v_{ph}^{in}$) of Bivacuum dipoles follows from the **Postulate** (**I**) in form:

$$v_{gr}^{in}\, v_{ph}^{in} = c^2 \qquad 24$$

Similar relation is well known already for external group and phase velocities of relativistic de Broglie waves: $\mathbf{v}_{gr}^{ext} \mathbf{v}_{ph}^{ext} = \mathbf{c}^2$.

The *Charge Compensation Principle* of the same shape as previous compensation principles is a consequence of **Postulate II**:

$$|\mathbf{e}_+| \, |\mathbf{e}_-| = \mathbf{e}_0^2 \qquad 25$$

### A7. The relation between the external and internal parameters of Bivacuum fermions and their absolute external velocity

The important relativistic formula, unifying a lot of internal and external parameters of asymmetric Bivacuum fermions ($\mathbf{BVF}_{as}^{\updownarrow}$) was derived from the Postulates I and II:

$$\left(\frac{\mathbf{c}}{\mathbf{v}_{gr}^{in}}\right)^2 = \frac{1}{[1-(\mathbf{v}^2/\mathbf{c}^2)^{ext}]^{1/2}} \qquad 26$$

$$= \left(\frac{\mathbf{m}_V^+}{\mathbf{m}_V^-}\right)^{1/2} = \frac{\mathbf{v}_{ph}^{in}}{\mathbf{v}_{gr}^{in}} = \frac{\mathbf{L}_V^-}{\mathbf{L}_V^+} = \frac{\mathbf{L}_0}{(\mathbf{L}_V^+)^2} = \frac{|\mathbf{e}_+|}{|\mathbf{e}_-|} \qquad 26a$$

where the radiuses of torus ($\mathbf{L}_V^+$) and antitorus ($\mathbf{L}_V^-$), as a basis of truncated cone, as a shape of asymmetric Bivacuum fermions, have the following relativistic dependencies on their external rotational or translational group velocity ($\mathbf{v} \equiv \mathbf{v}_{gr}$):

$$\left[\mathbf{L}_V^+ = \mathbf{L}_0[1-(\mathbf{v}^2/\mathbf{c}^2)^{ext}]^{1/4}\right]^i \qquad 27$$

$$\left[\mathbf{L}_V^- = \frac{\mathbf{L}_0}{[1-(\mathbf{v}^2/\mathbf{c}^2)^{ext}]^{1/4}}\right]^i \qquad 27a$$

where: $\left[\mathbf{L}_0 = (\mathbf{L}_V^+ \mathbf{L}_V^-)^{1/2} = \hbar/\mathbf{m}_0 \mathbf{c}\right]^i$ – *Compton radius* $\qquad 27b$

*The absolute external velocity of Bivacuum dipoles squared ($v^2$) as respect to primordial Bivacuum (absolute reference frame)*, can be expressed, using 26 and 26a, as a criteria of asymmetry of these dipoles torus and antitorus, accompanied their external motion:

$$\left[\mathbf{v}^2 = \mathbf{c}^2\left(1-\frac{\mathbf{m}_V^-}{\mathbf{m}_V^+}\right) = \mathbf{c}^2\left(1-\frac{\mathbf{e}_-^2}{\mathbf{e}_+^2}\right) = \mathbf{c}^2\left(1-\frac{\mathbf{S}_+}{\mathbf{S}_-}\right)\right]_{x,y,z} \qquad 28$$

where: $\mathbf{S}_+ = \pi(\mathbf{L}_V^+)^2$ and $\mathbf{S}_- = \pi(\mathbf{L}_V^-)^2$ are the squares of cross-sections of torus and antitorus of Bivacuum dipoles as the truncated cones.

*The existence of absolute velocity in our Unified theory (anisotropic in general case) and the Universal reference frame of Primordial Bivacuum, pertinent for Ether concept, is an important difference with Special relativity theory.*

*The light velocity in Unified Theory, like sound velocity in condensed matter, is a function of Bivacuum matrix elastic properties.*

### A8. The Hidden Harmony of asymmetric Bivacuum dipoles, as a background of Golden Mean

*The formula, unifying the internal and external group and phase velocities of asymmetric Bivacuum dipoles* represents one of the multiple forms of more general expression above (26 and 26a):

$$\left(\frac{\mathbf{v}_{gr}^{in}}{\mathbf{c}}\right)^4 = 1 - \left(\frac{\mathbf{v}_{gr}^{ext}}{\mathbf{c}}\right)^2 \qquad 29$$

where: $(\mathbf{v}_{gr}^{ext}) \equiv \mathbf{v}$ is the external translational-rotational group velocity of Bivacuum dipole.

*The conditions of "Hidden Harmony"* were introduced in our approach as:
a) the equality of the internal and external group velocities and
b) the equality of the internal and external phase velocities of asymmetric Bivacuum dipoles:

$$(\mathbf{v}_{gr}^{in})_{\mathbf{V}^+}^{rot} = (\mathbf{v}_{gr}^{ext})^{tr} \equiv \mathbf{v} \qquad 30$$

$$(\mathbf{v}_{ph}^{in})_{\mathbf{V}^-}^{rot} = (\mathbf{v}_{ph}^{ext})^{tr} \qquad 30a$$

and introducing the notation:

$$\left(\frac{\mathbf{v}_{gr}^{in}}{\mathbf{c}}\right)^2 = \left(\frac{\mathbf{v}}{\mathbf{c}}\right)^2 \equiv \phi \qquad 31$$

formula (29) turns to a simple quadratic equation:

$$\phi^2 + \phi - 1 = 0, \qquad 32$$

$$\text{which has a few modes :} \quad \phi = \frac{1}{\phi} - 1 \quad or: \quad \frac{\phi}{(1-\phi)^{1/2}} = 1 \qquad 32a$$

$$or: \quad \frac{1}{(1-\phi)^{1/2}} = \frac{1}{\phi} \qquad 32b$$

The solution of (32), is equal to Golden mean: $(\mathbf{v}/\mathbf{c})^2 = \phi = 0.618$. *It is remarkable, that the Golden Mean, which plays so important role on different Hierarchic levels of matter organization: from elementary particles to galactic and even in our perception of beauty (i.e. our mentality), has so deep physical roots, corresponding to Hidden Harmony conditions* (4.12 and 4.12a).

Our theory is the first one, elucidating these roots (Kaivarainen, 1995; 2000; 2005). This important fact points, that we are on the right track searching the mechanism of mass and charge origination from Bivacuum dipoles.

## A9  The rest mass and charge origination

If we use the Golden Mean equation in shape (32b): $\frac{1}{(1-\phi)^{1/2}} = \frac{1}{\phi}$, we can see, that all the ratios in the unified formula (26 and 26a) at Golden Mean conditions: $(\mathbf{v}/\mathbf{c})^2 = \phi = 0.618$ turns to:

$$\left[\left(\frac{\mathbf{m}_V^+}{\mathbf{m}_V^-}\right)^{1/2} = \frac{\mathbf{m}_V^+}{\mathbf{m}_0} = \frac{\mathbf{v}_{ph}^{in}}{\mathbf{v}_{gr}^{in}} = \frac{\mathbf{L}^-}{\mathbf{L}^+} = \frac{|\mathbf{e}_+|}{|\mathbf{e}_-|} = \left(\frac{\mathbf{e}_+}{\mathbf{e}_0}\right)^2\right]^\phi = \frac{1}{\phi} \qquad 33$$

where the actual ($e_+$) and complementary ($e_-$) charges and corresponding mass at Golden Mean conditions are:

$$\mathbf{e}_+^\phi = \mathbf{e}_0/\phi^{1/2}; \qquad \mathbf{e}_-^\phi = \mathbf{e}_0 \phi^{1/2} \qquad 34$$

$$(\mathbf{m}_V^+)_{\mu,\tau}^\phi = \mathbf{m}_0^{\mu,\tau}/\phi; \qquad (\mathbf{m}_V^-)_{\mu,\tau}^\phi = \mathbf{m}_0^{\mu,\tau} \phi \qquad 34a$$

using (34a) it is easy to see, that the difference between the actual and complementary mass of tori and antitori of asymmetric Bivacuum fermions of two generations $[\mathbf{V}^+ \Updownarrow \mathbf{V}^-]^{\mu,\tau}$ at Golden Mean conditions is equal to the rest mass of corresponding sub-elementary fermions: muon and tauon:

$$\left[ |\Delta \mathbf{m}_V|^\phi = \mathbf{m}_V^+ - \mathbf{m}_V^- = \mathbf{m}_0(1/\phi - \phi) = \mathbf{m}_0 \right]^{\mu,\tau} \qquad 35$$

One of the form of Golden Mean equation (32a) is: $(1/\phi - \phi) = \mathbf{1}$

This is an important result, confirming our approach, that the Bivacuum fermions symmetry shift, responsible for origination of the rest mass of unstable muons and tauons before their fusion to the electrons and protons, correspondingly, is determined by the universal for all hierarchical levels of Nature - Golden mean condition, based on Hidden Harmony (eqs.30 and 30a).

*The same is true for the charge origination.* The Golden Mean symmetry shift between actual and complementary charges of Bivacuum dipoles of two generations $[\mathbf{V}^+ \Updownarrow \mathbf{V}^-]^{\mu,\tau}$ stands for elementary

charge of sub-elementary fermions or antifermions. From (34) we get:

$$[\phi^{3/2}\mathbf{e}_0 = |\Delta\mathbf{e}_\pm|^\phi = |\mathbf{e}_+ - \mathbf{e}_-|^\phi \equiv |\mathbf{e}|^\phi]^{e,\mu,\tau} \qquad 36$$

$$\text{where:} \quad [(|\mathbf{e}_+||\mathbf{e}_-|) = \mathbf{e}_0^2]^{\mu,\tau} \qquad 36a$$

where the charges of tori and antitori in symmetric Bivacuum fermions $[\mathbf{V}^+ \Updownarrow \mathbf{V}^-]^{\mu,\tau}$ are related to their mass like it follows from *Postulate III* of this theory eqs (22 and 22a).

These conditions (22 and 22a) provide the same charge symmetry shift of $[\mathbf{V}^+ \Updownarrow \mathbf{V}^-]^{\mu,\tau}$ and stand for the same uncompensated charges of muon and tauon (sub-elementary fermions), equal to that of regular electron. This is notwithstanding of different mass symmetry shift between corresponding torus and antitorus (equal to the rest masses of muon and tauon), determined by Golden mean value (35).

*This is a reason, why sub-elementary fermions: muons and tauons, creating the electrons and protons, correspondingly, have the same absolute electric charges, nonetheless of their big mass difference:* $\mathbf{m}_0^\mu = 206,7\,\mathbf{m}_0^e$ *and* $\mathbf{m}_0^\tau = 3487,28\,\mathbf{m}_0^e$.

### A10. **The solution of Dirac Monopole problem**, following from Unified theory

The Dirac theory, searching for elementary magnetic charges ($g^-$ and $g^+$), symmetric to electric ones ($e^-$ and $e^+$), named *monopoles*, leads to the following relation between the magnetic monopole and electric charge of the same signs:

$$\mathbf{ge} = \frac{n}{2}\hbar\mathbf{c} \quad \text{or:} \quad \mathbf{g} = \frac{n}{2}\frac{\hbar\mathbf{c}}{\mathbf{e}} = \frac{n}{2}\frac{\mathbf{e}}{\alpha} \qquad 37$$

$$n = 1, 2, 3 \quad \text{is the integer number}$$

where $\alpha = \mathbf{e}^2/\hbar\mathbf{c} \simeq 1/137$ is the electromagnetic fine structure constant.

It follows from this definition, that minimal magnetic charge (*at* $\mathbf{n} = 1$) is as big as $g \cong 67.7e$. The mass of monopole should be huge $\sim 10^{16}\,GeV$. All numerous and very expensive attempts to reveal the Monopoles experimentally has failed.

*From our approach it follows that the uncompensated magnetic charge (monopole) simply is absent.*

In contrast to electric and mass dipoles symmetry shifts of Bivacuum dipoles, providing electric charge and mass origination of sub-elementary fermions, the symmetry shift between the internal actual $|\mu_+|$ and complementary $|\mu_-|$ *magnetic charges* of Bivacuum dipoles is absent, as it follows from the *Postulate II* of our theory (eq.17):

$$\Delta|\mu_\pm| = (|\mu_+| - |\mu_-|) = 0 \qquad 38$$

This is a consequence of their permanent values, equal to the Bohr magneton:

$$|\mu_+| = |\mu_-| = \mu_B \equiv \frac{1}{2}|e_0|\frac{\hbar}{\mathbf{m}_0\mathbf{c}} = \mathbf{const} \qquad 39$$

In accordance to *Postulate II*, the magnetic charges of torus and antitorus are independent on the external velocity (**v**) of Bivacuum dipoles and sub-elementary particles. The equality of the actual (torus) and complementary (antitorus) magnetic moments of Bivacuum dipoles - independent on their external velocity and the absence of symmetry shift, creating uncompensated moment, explains the *absence of magnetic monopoles in Nature.*

The excitation of magnetic field in accordance to our approach (Kaivarainen, 2006; http://arxiv.org/ftp/physics/papers/0207/0207027.pdf) is a result of the equilibrium shift between Bivacuum fermions of the opposite spins: $[\mathbf{BVF}_\pm^\uparrow \rightleftharpoons \mathbf{BVF}_\mp^\downarrow]^{\mu,\tau}$ to the left or right, depending on translational and rotational dynamics of charged elementary particles in selected directions, determined by the electric current direction.

### A11. **The elementary fermions origination from asymmetric Bivacuum dipoles**: muons and tauons

The process of elementary fermions origination can be subdivided to two stages:

*The 1st stage is the creation of uncompensated mass and charge of Bivacuum dipoles, equal to mass and charge of muon and tauon.*

This process is a result of Bivacuum fermions equilibrium shift towards the torus ($\mathbf{V}^+$) or antitorus ($\mathbf{V}^-$), determined by the Golden mean condition. It involves the Cooper pairs of $[muon \bowtie antimuon]^\mu$ and $[tauon \bowtie antitauon]^\tau$ origination. This symmetry shift is a result of Bivacuum virtual Cooper pairs:

$$[\mathbf{BVF}^\uparrow_\pm \bowtie \mathbf{BVF}^\downarrow_\mp]^{\mu,\tau}_{GM} \qquad 40$$

rotation around the main common axis and the opposite relativistic dependence of torus and antitorus on the tangential velocity of rotation (eqs.14 and 14a).

The shape of each of Bivacuum fermion at Golden Mean conditions, when the rest mass and charge of muons and tauons originates, is the *truncated cone* with ratio of diameters of bases, equal to Golden mean, as it follows from eq.33:

$$D^+/D^- = \left[\frac{\mathbf{m}_0}{\mathbf{m}^+_V}\right]^{\mu,\tau} = \left(\frac{\mathbf{v}}{\mathbf{c}}\right)^2 = \phi = 0.618 \qquad 40a$$

*The 2nd stage of elementary particles formation* is the fusion of triplets: electrons, protons, neutrons and their antiparticles $< [\mathbf{F}^+_\uparrow \bowtie \mathbf{F}^-_\downarrow] + \mathbf{F}^\pm_\updownarrow >^{e,p,n}$ from Cooper pairs of muons and tauons and their antiparticles (40) and one unpaired sub-elementary fermion $\mathbf{F}^+_\updownarrow >^{e,p,n}$ or antifermion $\mathbf{F}^-_\updownarrow >^{e,p,n}$. This fusion is accompanied by strong decreasing of mass of muons and tauons (mass defect with corresponding energy release) and their conversion to sub-elementary fermions in composition of triplets.

The paired sub-elementary fermion ($\mathbf{F}^+_\uparrow$) and antifermion ($\mathbf{F}^-_\downarrow$) in $[\mathbf{F}^+_\uparrow \bowtie \mathbf{F}^-_\downarrow]^{\mu,\tau}$ have the opposite spin, charge and energy. Consequently, they compensate each other and the properties of triplets (electrons, protons and metastable in free state neutron) are determined only by unpaired sub-elementary fermion.

This stage of elementary particles origination may occur, for example, after splitting of sextet Fig.1b on two triplets, like presented at Fig.A1c,d and subsequent fusion to electron and positron or proton and antiproton. It is accompanied by huge energy release, determined by mass difference (mass defect) between muons and electrons and tauons and protons.

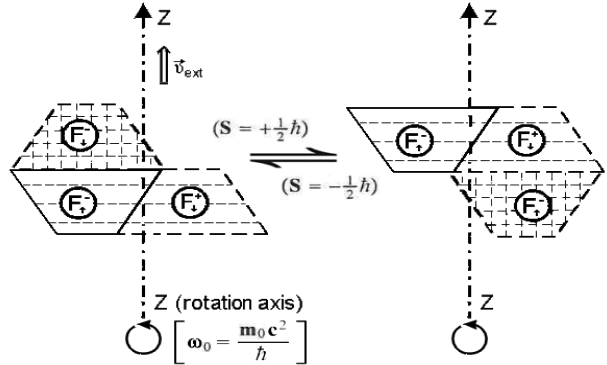

**Model of the electron, as a rotating triplet** $< [F_\uparrow^+ \bowtie F_\downarrow^-] + F_\updownarrow^- >$
**of sub-elementary fermions and antifermion, oscillating between
two spin states:** $(S = +\frac{1}{2}\hbar) \rightleftharpoons (S = -\frac{1}{2}\hbar)$.

The total resulting energy of each sub-elementary fermion in the
triplets is determined by the frequency of quantum betas between
the actual and complementary torus and antitorus:

$$E_{res} = h\nu_{C\rightleftharpoons W} = |m_V^+ - m_V^-|_{res} c^2 = m_0(\omega_0 L_0)^2 + \frac{h^2}{m_V^+ \lambda_B^2}$$

The spin state oscillation of the triplet is a result of jumping of
unpaired sub-elementary antifermion $F_\updownarrow^- >$ between two possible
positions in the triplets, corresponding to opposite phase of its
$[C \rightleftharpoons W]$ pulsation.

**Figure A2**. Model of the electron or positron, as a triplets $< [F_\uparrow^+ \bowtie F_\downarrow^-]_{x,y} + F_\updownarrow^\pm >^e$ of sub-elementary fermions. These elementary particles are the result of fusion of the unpaired Bivacuum antifermion (muon) or antifermion (antimuon) with Cooper pair $[BVF_\pm^\uparrow \bowtie BVF_\mp^\downarrow]_{GM}^{\mu,\tau}$ at Golden mean condition. Similar model is valid for proton and neutron. For explanation of neutral charge of neutron we suppose, that in the neutron the unpaired sub-elementary particle oscillate with high frequency between sub-elementary fermion and antifermion properties with opposite charges. The averaged charge is zero in this case, however spin in both isomers is $\hbar/2$.

The tangential velocity of rotating unpaired sub-elementary fermion $[F_\downarrow^-]$ in triplet around the same axis, as each of paired, is also the same $(v/c)^2 = \phi = 0.618$. This provides the origination of similar rest mass $(m_0)$ and charge $|e^\pm|$, as has each of the paired sub-elementary fermions $[F_\uparrow^+ \bowtie F_\downarrow^-]^{e,p}$ after fusion to triplet $< [F_\uparrow^+ \bowtie F_\downarrow^-]_{x,y} + F_\updownarrow^\pm >^{e,p}$. The properties of paired $[F_\uparrow^+$ and $F_\downarrow^-]^{e,p}$ totally compensate each other and the mass, charge and spin of elementary particle (triplet) is determined only by the unpaired sub-elementary fermion $F_\updownarrow^\pm >^{e,p}$. The asymmetry of each of sub-elementary fermions/antifermion in triplets and their rest mass and charge is maintained by the resonance exchange interaction with Bivacuum virtual pressure waves $(VPW_{e,p}^\pm)$ in the process of their corpuscle - wave pulsations, described in section A14.

### A12. The energy of elementary fermions fusion from muons and tauons

We suppose, that the regular electrons and positrons are the result of fusion of three *muons* and *antimuons* in relation 2:1 and 1:2, correspondingly. The protons and antiprotons are resulted from fusion of three *tauons* and *anti-tauons* in the same proportion. A single muons and tauons, as asymmetric Bivacuum fermions, are existing also, however, they have very short life-time. The experimental values of life-times of unstable muons and tauons with properties of Bivacuum fermions at Golden Mean (GM) conditions $\left(BVF_{as}^\updownarrow\right)_{GM}^{\mu,\tau}$, are very small: $2.19 \times 10^{-6} s$ and $3.4 \times 10^{-13} s$, respectively. *The stability of monomeric muons and tauons, strongly increases, as a result of their fusion to triplets of the electrons, protons and neutrons, since this process is accompanied by huge*

*energy release, determined by their mass decreasing (mass defect).*

Different superpositions of three sub-elementary fermions (former tauons) after fusion to triplets, like different combinations of three interlacing *Borromean rings* (symbol, popular in Medieval Italy) and their different dynamics, can be responsible for different properties of the protons and neutrons.

The mass of tauon and antitauon is: $\mathbf{m}_{\tau^{\pm}} = 1782(3)$ MeV. For the other hand, the mass of proton and neutron are: $\mathbf{m_p} = 938,280(3)$ MeV and $\mathbf{m_n} = 939,573(3)$ MeV, correspondingly. They are about two times less, than the mass of $\tau$-electron (tauon), equal, in accordance to our model, to mass of its unpaired sub-elementary fermion $(\mathbf{F}_\uparrow^+)^\tau$, turning to one of the quark after fusion. This mass/energy difference is close to the energy of neutral massless *gluons* (exchange bosons), stabilizing the triplets of protons and neutrons.

In the case of neutrons this difference is a bit less, providing, however, much shorter life-time of isolated neutrons (918 sec.), than that of protons ($>10^{31}$ years). We suppose that this huge difference in the life-span is determined by different dynamic structure of these two triplets, providing the positive charge of proton and neutrality of the neutron. One of possible explanation of neutrality, is fast oscillation of unpaired sub-elementary fermion between states of positive and negative charges in neutron. In contrast to conventional model in our approach the charge of quark (sub-elementary fermion in term of our approach) is supposed to be the integer number ($\mathbf{e}^{\pm}$), not the fractional one.

The additional mass defect of the paired tauons should be twice the same as one of the unpaired. The total big difference between the mass of 3 independent sub-elementary particles (tauons) and the mass of triplets (protons and neutrons), determines the sum of the gluons energy and the excessive kinetic energy - thermal energy release, as a result of these elementary particles fusion.

The mass of the regular electron is: $\mathbf{m}_{e^{\pm}} = 0,511003(1)$ MeV and the mass of $\mu$ – electron (muon) is: $\mathbf{m}_{\mu^{\pm}} = 105,6595(2)\ MeV$. The relative difference in these masses, about 200, is much higher that for protons and tauons. This provides very high stability of the electron as a triplet. This is a reason, why it is generally accepted, that the electron is a real elementary particle.

Like in the case of protons, the fusion of the electrons and positrons from muons and antimuons should be accompanied by the release of huge amount of kinetic thermal energy. Part of the fusion energy is used for *electronic – gluons* origination, similar to gluons in hadrons (protons, neutrons, etc.), responsible for strong exchange interaction, providing stability of these kind of triplets.

The electronic gluons ($e-gluons$) are responsible for the yet unknown electronic strong interaction.

*Our model of the electrons fusion from muons and antimuons in ratio 2:1 can be verified experimentally using muon + antimuon collider. The velocity of muon and antimuon in the point of colliding/scattering should correspond to Golden Mean condition:* $\mathbf{v} = \sqrt{\phi}\ \mathbf{c} = 0.786\ \mathbf{c}$. The muon collider should be designed in such a way that interception of three beams in one point will be possible.

## A13. New Scenario of the Big Bang

*This author propose a new scenario of Big Bang,* taking into account the experimental data, pointing to acceleration of the Universe expansion. It can be anticipated, that after hundred of billions years of such expansion and dying of stars, the mass and energy density of the Universe will tend to zero and the actual slightly asymmetric Bivacuum tend to primordial, symmetric one.

The process of the big collective symmetry fluctuations of vast number of Bivacuum dipoles toward the positive and negative energy simultaneously occurs in such conditions without violation of energy conservation. The spontaneous fusion of elementary particles from asymmetric Bivacuum dipoles (sub-elementary fermions and antifermions) can be considered as a kind of coherent chain - reaction, involving increasing number of asymmetric elements of Bivacuum.

*Just this energy of the avalanche reaction of elementary particles fusion from sub-elementary ones could be a source of energy of Big Bang* (Kaivarainen, 2006:

http://arxiv.org/ftp/physics/papers/0207/0207027.pdf).

The still-remaining minor asymmetry of Bivacuum before Big Fluctuation (the trace of the dyed Universe) may be responsible for small difference between probability of matter and antimatter origination, providing the relict radiation after annihilation. *So we may conclude that Big Fluctuation of Bivacuum dipoles symmetry, which may happens simultaneously in very remote region of primordial Bivacuum with nonlocal properties, is a precondition of Big Bang.*